\pdfoutput=1
\documentclass[a4paper,11pt,twoside,openright,titlepage,DIV=14,BCOR=2mm,headinclude=true,footinclude=false]{scrbook}
\KOMAoptions{headinclude,footinclude}
\usepackage{amssymb}
\usepackage[dvips]{color}
\usepackage[latin1]{inputenc}
\usepackage{graphicx}
\usepackage{gensymb}
\usepackage{amsmath}
 \usepackage{subcaption}
\usepackage{subfig}
\usepackage{relsize}
\usepackage[english]{babel}
\usepackage{color}
\usepackage{rotating}
\usepackage{xfrac}
\usepackage{mathrsfs}
\usepackage{latexsym}
\usepackage{wasysym}
\usepackage{soul}
\usepackage[toc,page]{appendix}
\usepackage{empheq}
\usepackage{tocloft}
\usepackage{bbold}
\usepackage{array}
\usepackage{amsthm}
\usepackage{mathtools}
\usepackage{enumerate}

\def\beq{\begin{equation}}
\def\eeq{\end{equation}}
\def\beqq{\begin{eqnarray}}
\def\eeqq{\end{eqnarray}}
\def\D{\partial}

\newcommand{\bdm}{\begin{displaymath}}
\newcommand{\edm}{\end{displaymath}}
\newcommand{\E}{{\rm e}}
\newcommand{\I}{{\rm i}}

\def\pmb#1{\setbox0=\hbox{$#1$}%
  \kern-.025em\copy0\kern-\wd0
  \kern.05em\copy0\kern-\wd0
  \kern-.025em\raise.0433em\box0}

\def\D{\partial}

\makeatletter
\renewcommand*{\@fnsymbol}[1]{\ensuremath{\ifcase#1\or *\or \dagger\or
    \ddagger\or 
   \mathsection\or **\or \dagger\dagger
   \or \ddagger\ddagger \else\@ctrerr\fi}}
   
   \if@twoside 
    \newlength{\textblockoffset}
\fi
\makeatother

\begin{document}
\frontmatter

   \begin{titlepage}
\vspace*{0.001cm}
\Large
\begin{center}
  \underline{Universit\`a degli Studi di Napoli ``Federico II''} \\
  \large
  \vspace{.3cm}
Dipartimento di Fisica ``Ettore Pancini''
  \vspace{8mm}

\resizebox{4cm}{!}{
  \includegraphics[scale=0.9]{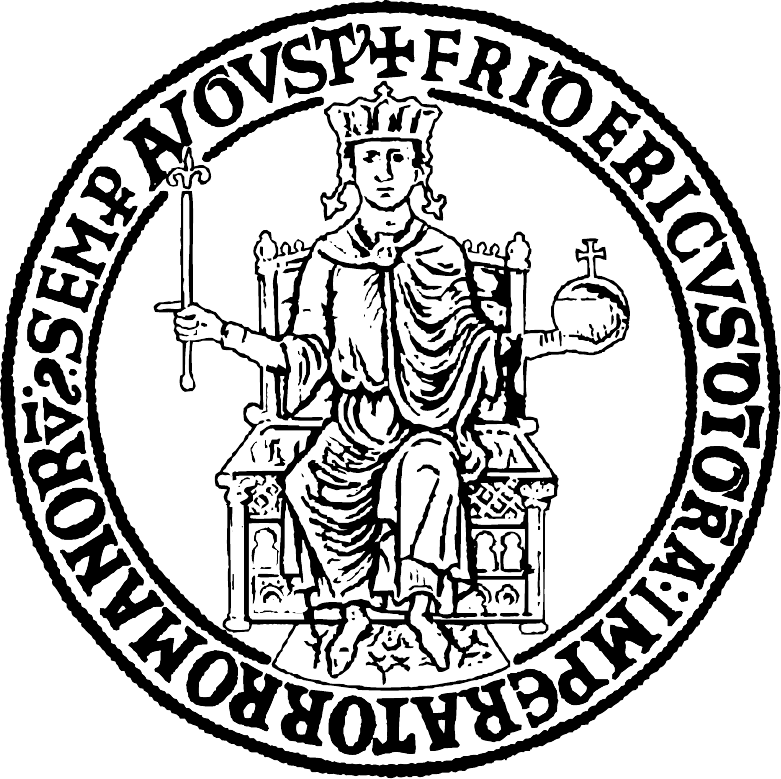}
  \vspace{3cm}
    }
\\

\vspace{0.5cm}
 {\large \bf{PhD Thesis in ``Fundamental and Applied Physics''}}
  \vspace{2cm}

\begin{center}
\bfseries\Huge{Extreme Regimes in Quantum Gravity}
\end{center}

\vspace{2cm}
\end{center}

\vspace{2cm}

\begin{minipage}[t]{7 cm}
     \Large{S}\large{upervisor:}\\
    \Large{D}\large{r.}  \Large{G}\large{iampiero} \Large{E}\large{sposito} \\
\end{minipage}
\hfill
\begin{minipage}[t]{7 cm}
\begin{flushleft}
  \Large{C}\large{andidate:}\\
   \Large{E}\large{mmanuele} \Large{B}\large{attista} \\
   \end{flushleft}
\end{minipage}

\vspace{1.7cm}

\end{titlepage}

\clearpage{\pagestyle{empty}\cleardoublepage}
\chapter*{Abstract}

The thesis is divided into two parts. In the first part the low-energy limit of quantum gravity is analysed, whereas in the second we deal with the high-energy domain. 

In the first part, by applying the effective field theory point of view to the quantization of general relativity, detectable, though tiny, quantum effects in the position of Newtonian Lagrangian points of the Earth-Moon system are found. In order to make more realistic the quantum corrected model proposed, the full three-body problem where the Earth and the Moon interact with a generic massive body and the restricted four-body problem involving the perturbative effects produced by the gravitational presence of the Sun in the Earth-Moon system are also studied. After that, a new quantum theory having general relativity as its classical counterpart is analysed. By exploiting this framework, an innovative interesting prediction involving the position of Lagrangian points within the context of general relativity is described. Furthermore, the new pattern provides quantum corrections to the relativistic coordinates of Earth-Moon libration points of the order of few millimetres.

The second part of the thesis deals with the Riemannian curvature characterizing the boosted form assumed by the Schwarzschild-de Sitter metric. The analysis of the Kretschmann invariant and the geodesic equation shows that the spacetime possesses a ``scalar curvature singularity'' within a 3-sphere and that it is possible to define what we here call ``boosted horizon'', a sort of elastic wall where all particles are surprisingly pushed away, suggesting that such 
``boosted geometries'' are ruled by a sort of ``antigravity effect''. Eventually, the equivalence with the coordinate shift method is invoked in order to demonstrate that all $\delta^2$ terms appearing in the Riemann curvature tensor give vanishing contribution in distributional sense.

\newpage   
￼\tableofcontents

\newpage

\addcontentsline{toc}{chapter}{List of Figures}
\listoffigures

\newpage

\addcontentsline{toc}{chapter}{List of Tables}
\listoftables

\chapter*{Introduction\markboth{Introduction}{}}
\addcontentsline{toc}{chapter}{Introduction}

The title of this thesis has been chosen for a specific reason. The manuscript is indeed divided into two parts. In the first part we will present the low-energy limit of the quantum theory of gravitation by employing the tools of effective field theories, whereas in the second we will outline its high-energy domain through the description of the boosting procedure involving an exact solution of Einstein field equations. 

The first reaction of Einstein to the birth of quantum mechanics was undoubtedly not positive, although his far-reaching ideas concerning the phenomena of emission and transformation of light developed in the renowned paper of 1905 ``On a heuristic point of view about the creation and conversion of light'' (ideas that, for example, were able to provide a correct interpretation of all phenomenology underlying the photoelectric effect) led to the concept of energy quanta, laying the basis of the so-called ``quantum revolution'' in physics. Celebrated are indeed the Bohr-Einstein debates of the beginning of last century and the statement Einstein wrote in 1926 in a letter to Max Born: ``I, at any rate, am convinced that He (God) does not throw dice''. This witnesses how the probabilistic interpretation underlying the new quantum framework was totally rejected by the German physicist. Nowadays, having passed about one hundred years from these renowned events, I somehow consider Einstein as prophetic. In fact, general relativity stubbornly resists any attempts to provide it with a quantum description. Moreover, it is widely accepted that all the known interactions of nature must be part of a unified theory. Electromagnetic and weak interactions have received a unified description through the Weinberg-Glashow-Salam model, whereas the inclusion of the strong interaction, as described by quantum chromodynamics, into a wider gauge theory has led to the so-called model of grand unified theories. The odd one out in this unification is gravity. In fact, the lack of a quantum pattern involving the gravitational field makes gravity stand apart from the other three forces of nature. The quantization of Einstein theory has been pursued with great vigour over the last sixty years, but a completely satisfactory quantum description of gravitational field remains elusive. Nevertheless, the major contenders for a quantum theory of gravitation are string theory and loop quantum gravity. String theory is a theoretical framework which solves in an elegant and efficient way the divergence issues arising in Einstein theory, since it demands that point-like particles are replaced by one-dimensional objects called strings. On the other hand, the picture of a granular space as formed by finite loops employed within loop quantum gravity has led to a well defined version of the Wheeler-DeWitt equation and consequently to the possibility of performing explicit computations, which however turn out to be quite distant from being testable.

In the absence of a viable theory of quantum gravity, is it possible to describe some effects involving the gravitational field at quantum level? The answer is surely affirmative. First of all, a semi-classical approximation can provide some valuable information about quantum gravity. This procedure does not represent a novel feature in theoretical physics. As an example, consider the pattern undertaken in the early days of quantum field theory involving a classical electromagnetic field interacting with quantized matter. All phenomena revealed within this approach were afterwards proved to be in accordance with the outcomes of the full theory of quantum electrodynamics. The same hybrid scheme can be adopted also in the case of Einstein theory. The regime where the gravitational field is retained as a classical background, while matter fields are quantized in the usual way, gives rise to the so-called quantum field theory in curved spaces. The most notable effect resulting from such a scheme is represented by the thermal radiation of black holes, known as Hawking radiation, which represents an example of a quantized electromagnetic field evolving in Schwarzschild background. The Hawking effect is somehow fundamental, since it has been re-derived in a number of ways, strongly reinforcing its credibility. Therefore, despite being a semi-classical result, we expect that it will represent an unavoidable feature of the unknown theory of quantum gravity. 

Another possible approach is represented by the application of the effective field theory point of view to the quantization of general relativity. Once again, this does not represent something new in the pattern of quantum field theory. In fact, the idea that the non-renormalizability, in the traditional sense, of a field theory prevents us from obtaining useful quantum predictions has been clearly demonstrated to be wrong within the context of chiral perturbation theory. Thence, the same effective field theory approach can be applied to general relativity in order to overcome its bad ultraviolet behaviour. In fact, by employing such a tool the troublesome singularities occurring in the traditional renormalization scheme of gravity can be easily absorbed into the phenomenological constants characterizing the full action of the theory. In this way, the resulting effective theory is finite and contains no singularities at any finite order of the loop expansion. Of course, treating general relativity as an effective field theory amounts to introduce a never ending set of additional higher-derivative couplings into the full Lagrangian. Then, within this framework general relativity represents only the minimal theory, whereas additional terms are related to its high-energy component. However, the low-energy domain turns out to be independent of the new couplings and hence it represents a true model-independent result of quantum gravity. 

The low-energy regime of quantum Einstein theory, analysed in the first part of this thesis, is intimately connected with its massless modes. We will see that the propagation of such massless particles in Feynman diagrams gives non-analytic contribution to the $S$ matrix. These non-analytic effects are long-ranged and, in the low-energy limit of the effective theory, dominate over the analytic contributions arising from massive degrees of freedom. Typical non-analytic terms are of type $1/\sqrt{-q^2}$ and $\log q^2$, while analytic contributions are powers series in $q$. 

By exploiting the above-mentioned set-up along with the background field quantization scheme, it is possible to derive the leading (i.e., one-loop) quantum corrections to the Newtonian potential. These result entirely from the Einstein-Hilbert sector of the full Lagrangian of gravity, which represents the lowest-order part of the theory. Depending on the definition adopted and consequently on the physical processes analysed, three different kinds of quantum corrected potential can be obtained: one-particle reducible, scattering, and bound-states potential. In the first case in fact nothing but the one-particle reducible part of the scattering amplitude is taken into account (i.e., vertex corrections and vacuum polarization diagrams are the only diagrams employed), whereas in the second the full set of Feynman diagrams constituting the scattering matrix are analysed. Finally, the third option can be derived by subtracting off the second-order Born approximation used in bound-state quantum mechanics from the definition of scattering potential. 

I have supposed that it is possible to apply the framework outlined above also to the description of the three-body problem of celestial mechanics. The second chapter of this thesis is indeed dedicated to the restricted three-body problem consisting of the Earth and the Moon as the primaries. Not only has this problem fulfilled an important role in the historical development of celestial mechanics and classical dynamics, but it has also found important applications to modern physics. For example, it has been recently discovered, by analytic and numerical methods, that there exist stable, although non-stationary, quantum states of electrons moving on circular orbits that are trapped in an effective potential well made of the Coulomb potential and the rotating electric field produced by a strong circularly polarized electromagnetic wave. 

The characterization of the restricted three-body problem follows the hybrid/semi-classical scheme described above. In fact, the gravitational interaction involving the two primaries is entirely described by employing the classical tools of Newtonian theory, whereas the motion of the planetoid, which is supposed to move in the known (background) gravitational field produced by the Earth and the Moon, is ruled by the quantum corrections resulting from the analysis of Feynman diagrams within the effective field theory approach. In the theory of gravitation, the indisputable smallness of classical and quantum corrections to the Newtonian potential had always
discouraged the investigation of their role in the restricted three-body problem. On the contrary, within this theoretical model, tiny but non-negligible  effects involving Lagrangian points are predicted. In fact, the positions of non-collinear Lagrangian points, related to a pair of fifth degree algebraic equations, are slightly modified, so that the planetoid  is no longer at equal distance from the two bodies of large mass in the  configuration of stable equilibrium, unlike in the classical Newtonian case where an equilateral triangle picture exists. Furthermore, the position of collinear Lagrangian points is described in terms of an algebraic ninth degree equation. Both for collinear and non-collinear libration points quantum corrections to the corresponding classical values turn out to be of the order of few millimetres. This represents a striking result since these predictions can be tested with the help of modern Satellite/Lunar Laser Ranging techniques. In other words, the theoretical model developed in the first part of this thesis is able to provide testable low-energy quantum gravity effects in the Earth-Moon system.  

In order to make more realistic the theoretical framework proposed in this manuscript, in the third chapter more detailed models of Newtonian theory are accounted for: the full three-body problem involving, like before, the Earth and the Moon and the restricted four-body problem, which makes it possible to consider also the perturbative effects due to the gravitational presence of the Sun in the Earth-Moon system. In the context of the full problem of three bodies, Poincar\'e theorem regarding periodic solutions is invoked to show that, even at quantum level, there may exist periodic orbits. We will see that for this purpose a fundamental role is played by the extreme smallness of Planck length. Furthermore, a scheme involving the repeated application of a $2 \times 2$ matrix of first-order linear differential operators for the resolution of the quantum corrected version of variational equations is proposed. Within the context of the restricted four-body problem, I will explain how the effective field theory pattern is able to reproduce the classical results involving the motion of a spacecraft in the vicinity of the Earth-Moon (triangular) Lagrangian points $L_4$ or $L_5$. In particular, it is shown that the gravitational presence of the Sun spoils the equilibrium condition at $L_4$ and $L_5$, in the sense that a vehicle initially placed at these libration points will not remain near them but it will escape in about two years. Thus, $L_4$ and $L_5$ can be considered as ``stable'' equilibrium points in a somewhat weak sense, i.e., only during the length of experimental observations. I will also evaluate the impulse required to cancel out the perturbing force due to the Sun in order to force the spacecraft to stay precisely at $L_4$ or $L_5$. It turns out that this value is slightly modified with respect to the corresponding Newtonian one.

In the fourth chapter I set up a scheme where the theory which is quantum corrected has as its classical counterpart the Einstein theory, instead of the Newtonian one. In other words, we will deal with a theory involving quantum corrections to Einstein gravity, rather than to Newtonian gravity. By virtue of the effective-gravity correction to the long-distance form of the potential among two point masses, all terms involving the ratio between the gravitational radius of the primary and its 
separation from the planetoid get modified. Within this framework, both relativistic and quantum corrections involving the coordinates of all Lagrangian points are once again of the order of few millimetres. Moreover, the new pattern will also allow me to show that, among all quantum coefficients occurring in the long-distance corrections to the Newtonian potential, the most suitable ones to describe the gravitational interactions involving (at least) three bodies in celestial mechanics are those connected to the bound-states potential. 

As was pointed out before, the second part of this thesis is dedicated to the high-energy limit of quantum gravity. Here the subject of gravitational waves comes into play. Similarly to what happens when one goes from Coulomb theory of electrostatics to Maxwell's electromagnetism, when we pass from Newtonian gravitation to Einstein picture the gravitational field becomes a dynamical entity: small ripples rolling across spacetime exist, i.e., gravitational waves. Gravitational waves were predicted by Einstein a century ago, in 1915, but the first indirect proof of their existence was only achieved by Hulse and Taylor nearly sixty years later, with the discovery in 1974 of the binary system ``PSR 1913+16''. After that, the scientific community has waited for further forty years until gravitational waves were finally observed by the two LIGO detectors in United States and by Virgo team, marking a discovery which represents a milestone in the history of physics and the beginning of a new era both in astrophysics and in cosmology. 

Two years before figuring out the final form of gravitational field equations, during a lecture held in Vienna in 1913 Einstein pointed out that in the linearised regime it is quite simple to prove that the action of gravitation is propagated in his theory at the speed of light, but meanwhile stuffs would have become far more complicated in the full theory, since it is governed by non-linear equations. Once again, Einstein was prophetic. By adopting a local point of view, it is indeed possible to describe small-scale ripples in the spacetime curvature propagating at the speed of light throughout the universe. Moreover, ignoring their interaction with the large-scale curvature of spacetime and their non-linear interaction with each other turns out to be conceivable within the linear domain. In this regime one can thus pretend that waves evolve in a flat Minkowski background and a wave equation catching their features arises quite simply. Globally this picture is no longer valid. In the real universe curvature is produced not only by gravitational waves, but also (and more importantly) by the material content of universe itself, such as galaxies, stars, planets, and so forth. The interaction with large-scale curvature fulfils now a significant role. As a gravitational wave propagates, its wave fronts can change shape (refraction effects), its wavelength varies (gravitational redshift), and it backscatters off the curvatures encountered during its path to some extent. 

Something similar is described in the last chapter of this thesis, where I deal with the gravitational shock-wave produced by a zero rest mass point particle moving at the speed of light. The formal method that, starting from a known exact solution of Einstein field equation, allows us to delineate such a geometry is known in the literature as boosting procedure. This process might be interpreted in an equivalent way as a pattern describing new exact solutions of general relativity equations. Gravitational shock-waves represent an example of impulsive wave, being characterized by the presence of distributional Dirac-delta-like singularities. Aichelburg and Sexl solution, for example, is an asymmetric plane-fronted wave evolving in Minkowski background, whereas the case I will handle with in this manuscript, which was first considered by Hotta and Tanaka in 1993, turns out to be a symmetric spherical wave in de Sitter background, i.e., a space having a non-vanishing cosmological constant and hence a non-zero (constant) curvature. Aichelburg and Sexl derived their framework by Lorentz-boosting to the speed of light a Schwarzschild solution, on the other hand the Hotta and Tanaka solution can be achieved by boosting through the de Sitter group transformations a Schwarzschild-de Sitter black hole.

An equivalent technique to describe ``boosting geometries'' is the so-called coordinate shift method, developed by Dray and 't Hooft. This approach is equivalent to the scissors-and-paste method introduced by Penrose in the sixties and it clearly shows how the background geometry can affect the shock-wave during its evolution. In particular, refraction effects and discontinuity phenomena occur to geodesics crossing the shock-wave.

It should now be clear that high-energy processes are intimately connected with such field configurations, since the velocities of particles involved approach the speed of light. Furthermore, these techniques have a lot of implications at quantum level: the Hotta and Tanaka solution is related to divergence issues occurring in graviton propagation in de Sitter space; the shock-wave geometry evolving in Schwarzschild background, described by Dray and 't Hooft, is connected to back-reaction and self-interaction involving Hawking particles crossing the event horizon of a black hole. For these reasons the second part of this thesis is meant to be dedicated to the high-energy domain of quantum gravity. The key point of the last chapter is represented by the Riemannian curvature of boosted Schwarzschild-de Sitter spacetime, with a particular attention to the ultrarelativistic regime. The most important features of such a ``boosting geometry'' is represented by a singularity 3-sphere where the Kretschmann invariant is not defined and by the presence of a sort of elastic wall, surrounding the above-mentioned 3-sphere, where all geodesics are pushed away, despite maintaining their completeness condition. I will call this elastic barrier ``boosted horizon''. Hence, it seems that the boosted Schwarzschild-de Sitter spacetime is ruled by a sort of antigravity effect, which seems to be in accordance with the refraction effects predicted by the coordinate shift method. Finally, the presence of the singularity 3-sphere can be ascribed to the discontinuity phenomena provided by Dray and 't Hooft picture. 

To sum up, the thesis is organized as follows: chapter 1 is dedicated to the treatment of general relativity as an effective field theory; in the following chapter I describe the restricted three-body problem in the context of effective field theories; the full three-body problem and the restricted four-body problem are analysed in chapter 3; in chapter 4 I will deal with the quantum gravitational theory having general relativity as its classical counterpart; in chapter 5 I will outline the boosting procedure and the Riemann curvature of the boosted Schwarzschild-de Sitter spacetime; finally, I will discuss conclusions and open problems.

\section*{Conventions and notations} 

In the first part of this thesis we adopt the metric signature $(+---)$, whereas in the second we employ the choice $(-+++)$. We also make use of the coordinate index notation. In those few cases in which we adopt the abstract index notation, we always stress this choice. 

The symbol $\nabla_\mu$ refers to the covariant derivative operator associated with the Levi-Civita connection, whose components in a coordinate basis are  given by the Christoffel symbols 
\begin{equation} 
\Gamma^\lambda_{ \;  \alpha \beta}= \frac{1}{2} g^{\lambda \sigma} \left(\partial_\alpha g_{\beta \sigma}+ \partial_\beta g_{\alpha \sigma} - \partial_\sigma g_{\alpha \beta } \right).  \label{Christoffel symbols}
\end{equation}
The Riemann curvature tensor is defined by the commutators
\begin{eqnarray}
& R^\mu_{\phantom{\mu} \nu \lambda \delta} \: u^\nu =\left[\nabla_\lambda,\nabla_\delta\right]u^\nu= \left(\nabla_\lambda \nabla_\delta - \nabla_\delta \nabla_\lambda \right) u^ \mu , \\
& R_{\lambda \delta \nu }^{\phantom{\lambda \delta \nu} \mu} \: v_\mu = - R^\mu_{\phantom{\mu} \nu \lambda \delta } \: v_\mu =  \left[\nabla_\lambda,\nabla_\delta\right]v_\mu=\left(\nabla_\lambda \nabla_\delta - \nabla_\delta \nabla_\lambda \right) v_\nu,
\end{eqnarray}  
for arbitrary vectors $u^\nu$ and one-forms $v_\mu$. Its coordinate expression reads as
\begin{equation}
R^\rho_{\;\; \sigma \mu \nu}=\partial_\mu \Gamma^\rho_{\; \; \nu \sigma}- \partial_\nu \Gamma^\rho_{\; \; \mu \sigma}+\Gamma^\rho_{\; \; \mu \lambda} \Gamma^\lambda_{\; \; \sigma \nu}- \Gamma^\rho_{\; \; \nu \lambda} \Gamma^\lambda_{\; \;  \mu \sigma}. 
\end{equation}
The Ricci tensor is obtained by contraction on the first and third indices, i.e.,
\begin{equation}
R_{\mu \nu}= R^\lambda_{\phantom{\lambda} \mu \lambda \nu}= \partial_\nu \Gamma^\lambda_{\; \; \mu \lambda}- \partial_\lambda \Gamma^\lambda_{\; \; \mu \nu}+\Gamma^\sigma_{\; \; \mu \lambda} \Gamma^\lambda_{\; \; \sigma \nu}- \Gamma^\sigma_{\; \; \mu \nu } \Gamma^\lambda_{\; \;  \sigma \lambda}. 
\end{equation}
The Ricci scalar (or scalar curvature) is defined as the trace of Ricci tensor, i.e., 
\begin{equation}
R= g^{\mu \nu} R_{\mu \nu}.
\end{equation}
Formulae can be changed when passing from signature $(+---)$ to $(-+++)$ by changing the sign of $g_{\mu \nu}$, $R^\rho_{\;\; \sigma \mu \nu}$, $R_{\mu \nu}$, and $T^{\mu}_{\; \nu}$, but leaving $R_{\alpha \beta \gamma \delta}$, $R^{\mu}_{\; \nu}$, $R$, and $T_{\mu \nu}$ unchanged.

Round and square brackets denote respectively symmetrisation and antisymmetrisation (including division by the number of permutations of the indices).

\mainmatter
\cftaddtitleline{toc}{chapter}{Part I: the low-energy limit}{}

\begin{titlepage}
 \vspace{1cm}
 \setlength{\parindent}{0pt}
\setlength{\parskip}{0pt}
\vspace*{\stretch{1}}
\rule{\linewidth}{1pt}
\begin{center}
\Large{\bf Part I: the low-energy limit}
\end{center}
\rule{\linewidth}{2pt}
\vspace*{\stretch{2}}

\end{titlepage}

\chapter{General relativity as an effective field theory}

\vspace{2cm}
\emph{By an application of the theory of relativity to the taste of readers, today in Germany I am called a German man of science, and in England I am represented as a Swiss Jew. If I come to be represented as a b\^{e}te noire, the descriptions will be reversed, and I shall become a Swiss Jew for the Germans and a German man of science for the English!}
\begin{flushright}
A. Einstein
\end{flushright}
                                                                 
\vspace{2cm}   

One of the most outstanding problems of modern theoretical physics is represented by the incompatibility of the two major theories of twentieth century, i.e., quantum mechanics and general relativity, which gives rise to a perturbatively non-renormalizable theory of quantum gravity. On one hand, quantum mechanics (and its offspring quantum field theory) provides an incredibly successful description of all known non-gravitational phenomena, which results in an agreement between predictions and experiment sometimes taking place at the part-per-billion level. On the other hand, general relativity is an elegant classical theory brilliantly tested within the Solar System. Despite the absence of an over-arching theoretical framework within which both successes can be accommodated, quantum predictions can be made in non-renormalizable theories by employing the techniques of effective field theory \cite{D94,D94b,D94c}. Within this scheme, calculations are organized in a systematic expansion in energy, where the high-energy effects show themselves only in the shifting of a small number of parameters which can be measured experimentally, in exactly the same way as it happens for renormalizable field theories. To any given order in the energy expansion there are only a finite number of parameters and, once predictions are expressed in terms of the measured values, it is possible to separate out the known low-energy quantum effects  from the (unknown) high-energy regime of the theory.  General relativity fits naturally into the framework of effective field theory, since gravitational interactions are proportional to energy and are easily organized into an energy expansion where the expansion scale factor is the Planck length $l_P=  \sqrt{\hbar G/c^3} \approx 1.616 \times 10^{-35} {\rm m}$. In the low-energy limit, the leading (i.e., one-loop) long-distance quantum predictions, which dominate over other quantum effects, can be isolated, since they are linked to the propagation of the massless particles of the theory and their couplings at low energy, and produce non-local/non-analytic contributions to vertex functions and propagators. These leading quantum corrections, besides being parameter free (apart from the Newton constant $G$), are entirely ruled by the Einstein-Hilbert part of the full action functional and hence represent first order modifications due to quantum mechanics. The fact that they are independent of the eventual high-energy theory of gravity makes them represent true predictions of the theory of quantum gravity. Therefore, general relativity can be considered as a well-behaved quantum field theory at ordinary energies whose predictions could be (hopefully) tested (see Sec. \ref{Laser_Ranging_Sec}).   

\section{The quantization of general relativity} \label{Sec. Quantization GR}

General relativity represents one of the most elegant and exciting theory of theoretical physics which has brought a revolutionary view point on spacetime structure and gravitation, but it suffers from a ``serious illness'': it is a classical theory, whereas it is well established that all known fundamental interactions of nature must be described by the principles of quantum theory. Thus, general relativity, despite its prominence, is not special enough to avoid this ``law''. 

\subsection{Three approaches to quantizing gravity}

There are two main reasons to develop a quantum theory of gravity \cite{Hawking79,Wald}. First of all, as we pointed out above, within the general relativity pattern the gravitational field has got  a purely classical meaning, whereas all other observed fields seem to be quantized. This crucial point is intimately connected to the nature of Einstein equations themselves, which tell us that gravity couples to $T_{\mu \nu}$, the energy-momentum tensor of matter, in a diffeomorphism-invariant way, by virtue of the tensor equations \cite{Einstein1916,Bruhat09}
\begin{equation}
R_{\mu \nu}-{1\over 2}g_{\mu \nu}R= {8 \pi G \over c^{4}}T_{\mu \nu}.
\label{Einstein_equations}
\end{equation}
When Einstein arrived at these equations, although he had already understood that the classical Maxwell theory of electromagnetic phenomena is not valid in all circumstances, the only known forms of $T_{\mu \nu}$ were classical, e.g., the energy-momentum tensor of a relativistic fluid, or even just the case of vacuum Einstein 
equations, for which $T_{\mu \nu}$ vanishes. In due course, it was realized that matter fields are quantum fields in the first place (e.g., a massive Dirac field, or spinor electrodynamics). Quantum fields are operator-valued distributions, for which a regularization and renormalization procedure is necessary and even fruitful. However, the mere replacement of $T_{\mu \nu}$ by its regularized and renormalized form $\langle T_{\mu \nu} \rangle$ on the right-hand side of Eq. (\ref{Einstein_equations}) leads to a hybrid scheme, because the classical Einstein tensor $R_{\mu \nu}-{1 \over 2}g_{\mu \nu}R$ is affected by the coupling to $\langle T_{\mu \nu} \rangle$. The question then arises whether the appropriate, full quantum theory of gravity should have field-theoretical nature or should involve, instead, other structures. This issue has led to the growth of various approaches to the quantization of the gravitational field (see below). Secondly, a number of theorems \cite{Haw-Ell} have proved that singularities appear in spacetime under very general assumptions (provided that physically realistic energy conditions hold), showing that they are true ingredient of general relativity and not a mere artifact of the high degree of symmetry of the known exacts solutions of Einstein field equations. Spacetime singularities represent a breakdown of the Einstein theory, which therefore turns out to be incomplete since it is not able to provide boundary conditions for the field equations at singular points. For this purpose, it is interesting to note that it has been demonstrated that, by relying upon different hypotheses from those adopted by Hawking and Penrose, there exists a class of global, smooth solutions to the vacuum Einstein equations looking asymptotically like the Minkowski spacetime, which in particular have the important property to be singularity free \cite{Christodoulou-Klainerman}. Anyway, driven by both the above reasons, one would like to achieve a complete framework for the theory of quantum gravity, which primarily will allow us to reach a better understanding of the early universe. There is not a well defined prescription for deriving such a theory from classical general relativity, but on general grounds we demand that the final theory be complete, consistent and agree with general relativity for macroscopic bodies and low spacetime curvatures. So far, we do not have a theory satisfying the above criteria but just an incomplete scheme whose results (e.g., Hawking radiation \cite{Hawking75}) are so compelling that we strongly believe that they will be part of the final complete picture. 

Three main approaches to quantizing gravity have been developed so far \cite{Hawking79,Wald}. The first one is the operator approach where the metric in the classical Einstein equations (\ref{Einstein_equations}) gets replaced by distribution-valued operators on some Hilbert space. This procedure has got some problems since field equations are non-polynomial and it involves products of field operators at the same spacetime point, which as we know make no sense. The other approach is represented by the canonical one \cite{DeWitt1967a}. In this case we adopt the Hamiltonian formulation of general relativity and the original framework of quantum geometrodynamics, i.e., the Arnowitt-Deser-Misner (hereafter referred to as ADM) formalism. ADM formalism enables one to re-write Einstein field equations in first-order form and with an explicit time variable dependence. For this purpose, one assumes that four-dimensional spacetime can be foliated by a family of $t =$ constant spacelike surfaces, giving rise to a $3+1$ decomposition of the original four-geometry. Then the basic ideas are to take the states of the system to be described by wave function(al)s depending on the configuration variables and to replace each momentum variable by (functional) differentiation with respect to the conjugate configuration variable. The complete quantum theory thus stems from equal-time commutation relations ruled by the Heisenberg Uncertainty Principle and from the fact that all classical constraints which are first-class are turned into operators that annihilate the wave functional, a procedure which leads to the well-known Wheeler-DeWitt equation. This approach has the advantages of being applicable to strong gravitational fields and of ensuring unitarity, but it seems to betray the whole spirit of general relativity by destroying the general covariance with a restriction of the topology of spacetime to the product of the real line with a three-dimensional manifold via the $3+1$ ADM formalism. Moreover, one would expect that quantum gravity will allow also more complicated topologies of spacetime, not only those which are products. Eventually, we should also remember that equal-time commutation relations have no precise meaning when the geometry, instead of being fixed, is quantized and obeys the Uncertainty Principle. For these reasons another approach exists which, despite presenting a lot of unsolved problems, seems to offer the best hope towards the quantization of the gravitational field, i.e., the Feynman path integral approach. In this case, instead of the state of the system or the operators, a central role is fulfilled by the probability amplitude for physical processes. In this context we define the amplitude to go from an induced three-metric $h^{\prime \prime}_{ij}$ on a spatial hypersurface $S^{\prime \prime}$ with matter fields $\phi^{\prime \prime}$ to another induced three-metric $h^\prime_{ij}$ on a spatial hypersurface $S^\prime$ with matter fields $\phi^\prime$ as the sum over all field configurations $g$ and $\phi$ which take the given values on $S^{\prime \prime}$ and $S^\prime$. More precisely, the fundamental entry of the theory is represented by the path integral 
\begin{equation}
Z = \langle h^\prime_{ij}, \phi^\prime, S^\prime \vert   h^{\prime \prime}_{ij}, \phi^{\prime \prime}, S^{\prime \prime}   \rangle = \int_{\mathcal{C}} \mathcal{D}[g,\phi] \; {\rm e}^{    ({\rm i}/ \hbar )  S[g,\phi]}, \label{Z_integral}
\end{equation}
where $\mathcal{D}[g,\phi]$ is the measure on the space of all field configurations $g$ and $\phi$ defined on the set $\mathcal{C}$ of all four-metrics $g$ and matter fields $\phi$ which coincide with $h^{\prime \prime}_{ij}$ and $\phi^{\prime \prime}$ on $S^{\prime \prime}$ and with $h^\prime_{ij}$ and $\phi^\prime$ on $S^\prime$, while $S[g,\phi]$ is the action of the fields. A purported benefit of this scheme lies in the fact that it allows the description of all those physical situations which involve a change of spatial topology by including in (\ref{Z_integral}) all spacetime metrics for which such a change occurs. This issue is inconceivable in the canonical approach since the hypothesis of global hyperbolicity (which is essential for the ADM foliation) prevents any topology change. On the other hand, the most important problem arising in this context is represented by the presence of the formal measure $\mathcal{D}[g,\phi]$, to which we are still unable to give a precise mathematical sense, except in the context of perturbation theory about a free field. Moreover, the path integral (\ref{Z_integral}), which should represent the probability amplitude at two different times $t^{\prime \prime}$ and $t^{\prime}$, turns out to be a meaningless quantity since general relativity is a parametrized theory where time is just a label treated as a dynamical variable which in lot of situations carries no physical significance.  
We should mention at this point that the three approaches discussed here represent those which are best suited for the purposes of this thesis, but they are not the only ones developed so far. In fact, in modern literature other frameworks have been proposed and we remember, among the others, the loop space representation (coming from the evolution of the canonical approach) \cite{Rovelli2004} and the string and brane theory, which is peculiar because it is not field-theoretic, spacetime points being replaced by extended structures such as strings \cite{Polchinski98,Horowitz2005}.

\subsection{Feynman rules}

In the context of Feynman path integral approach, no analytical method exists which allows us to solve the theory exactly. Therefore, perturbation theory is the method generally adopted to perform all calculations. In the following we are going to set $\hbar=c=1$, nevertheless we have to remind that, upon quantizing via the path integral, the coupling constant $\chi$ (see below) occurs only in the combination $\left( \chi^2 \hbar \right)^{-1}$ multiplying the action and hence a perturbative expansion in powers of $\hbar$ is the same as an expansion in powers of $\chi^2$ and any $L$-loop diagram always gets a factor $\hbar^{L}$.  However, we will recover the constants $\hbar$ and $c$ in the subsequent sections of this thesis in order to perform a dimensional analysis. In perturbative quantum gravity the genuine spacetime metric is separated into two parts: a fixed background  spacetime metric $\bar{g}_{\mu \nu}$ and a perturbation metric $h_{\mu \nu }$ (i.e., the one which must be quantized and not to be confused with the three-metric of the previous section) representing the fluctuation of the spacetime geometry due to quantum gravity interactions \cite{DeWitt1967b,DeWitt1967c}. One possible choice is (recall that any quantum boson field has the dimensions of a mass)
\begin{equation}
g_{\mu \nu} = \bar{g}_{\mu \nu} + \chi h_{\mu \nu }, \label{perturbed_metric_h}
\end{equation}
with $\chi =\sqrt{32 \pi G}$. The usual request $g_{\mu \lambda} \; g^{\lambda \nu}= \delta_{\mu}^{\; \nu}$ implies that
\begin{equation}
g^{\mu \nu} = \bar{g}^{\mu \nu} - \chi h^{\mu \nu } + \chi^2 h^{\mu}_{\; \lambda} h^{\lambda \nu} +{ \rm O} (\chi^3 h^3), 
\end{equation}
where it is meant that indices are always raised and lowered with the background metric. The pure gravitational field is described by the Einstein-Hilbert Lagrangian
\begin{equation}
\mathcal{L}_{grav}= \sqrt{-g} \left(\dfrac{2}{\chi^2}R \right), \label{Einstein_Lagrangian}
\end{equation}
where we have assumed for simplicity that the cosmological constant vanishes. The infinitesimal gauge transformations of the theory are represented by
\begin{equation}
\hat{x}^\mu = x^\mu + \epsilon^\mu(x),
\end{equation}
$\epsilon^\mu(x)$ being the infinitesimal generator of the transformations. Therefore, in the classical theory (i.e., before considering the quantum perturbation (\ref{perturbed_metric_h}) of $g_{\mu \nu}$) the Einstein-Hilbert Lagrangian turns out to be invariant under the infinitesimal gauge transformation of the metric tensor of the form
\begin{equation}
\hat{g}_{\mu \nu}= g_{\mu \nu}+ \epsilon^\alpha \nabla_{\alpha} g_{\mu \nu} + g_{\alpha \nu} \nabla_\mu \epsilon^\alpha +  g_{\mu \alpha } \nabla_\nu \epsilon^\alpha = g_{\mu \nu}+(\pounds_{\epsilon}g)_{\mu \nu}, \label{gauge_transf_g}
\footnote{$\nabla_\mu$ represents the covariant derivative with respect to the Levi-Civita connection of $g_{\mu \nu}$.}
\end{equation}
where $(\pounds_{\epsilon}g)_{\mu \nu}$ denotes the components of the Lie derivative along the vector field $\epsilon$ of the four-metric $g_{\mu \nu}\, {\rm d}x^\mu \otimes {\rm d}x^\nu$. In the above equation the well-known condition $\nabla_\alpha g_{\mu \nu}=0$ holds, whereas the last two terms tell us that $g_{\mu \nu}$ transforms as a tensor. Thus, bearing in mind (\ref{perturbed_metric_h}) and (\ref{gauge_transf_g}) we obtain the transformation rule for the perturbed metric $h_{\mu \nu}$
\begin{equation}
\hat{h}_{\mu \nu}= h_{\mu \nu} + 2\bar{\nabla}_{( \mu} \epsilon_{\nu)},\label{gauge_transf_h}
 \end{equation}
where $\bar{\nabla}$ denotes covariant differentiation with respect to the background metric and $\epsilon_\nu$ is such that $\epsilon_\nu {\rm d}x^\nu$ is the ghost one-form. The simplest form of matter coupled in an invariant way to gravity is a set of spinless scalar particles of mass $m$ described by the Klein-Gordon Lagrangian
\begin{equation}
\mathcal{L}_{m}=\dfrac{1}{2} \sqrt{-g} \left(g^{\mu \nu} \partial_{\mu} \phi \; \partial_{\nu} \phi -m^2 \phi^2 \right).
\label{Klein-Gordon_matter_Lagrangian}
\end{equation}
We will suppose that matter is minimally coupled to gravity. This means that the  field $\phi$ is not coupled to the scalar curvature $R$, but its only coupling to gravity is represented by the term $\sqrt{-g}$. Therefore, the gravity-scalar Lagrangian is simply given by the sum $\mathcal{L}_{grav}+\mathcal{L}_m$. The energy-momentum tensor of matter is given by
\begin{equation}
\begin{split}
T_{\mu \nu} & = -2 \dfrac{\partial \mathcal{L}_m}{\partial g^{\mu \nu}}+ g_{\mu \nu} \mathcal{L}_{m} \\
 &=  -\sqrt{-g} \partial_{\mu} \phi \; \partial_{\nu} \phi + \dfrac{1}{2} \sqrt{-g} \; g_{\mu \nu} \left( g^{\lambda \sigma} \partial_{\lambda} \phi \; \partial_{\sigma} \phi - m^2 \phi^2 \right),
\end{split} \label{energy-momentum_scalar}
\end{equation}
and represents a conserved quantity, i.e.,
\begin{equation}
\nabla^\mu T_{\mu \nu}=0.
\end{equation}
As we know, in order to have a non-singular dynamical operator on metric perturbations (the so called gauge field operator) and to avoid ``overcounting" problems in the generating functional $Z$ of the theory (and to save unitarity, too), we need to add to $\mathcal{L}_{grav}$ both a gauge and a ghost Lagrangian. Moreover, the complexity of Einstein action makes it convenient to choose a gauge leading to the simplest possible graviton propagator. We will see that the choice will fall on the de Donder gauge, the gravity analogue for the Lorenz gauge for quantum electrodynamics (QED). For this reason we first consider the gauge-fixing functional
\begin{equation}
C_{\mu}[h]= \bar{\nabla}^\nu h_{\mu \nu}-\dfrac{1}{2} \bar{\nabla}_{\mu}h, \label{gauge_fixing_Functional}
\end{equation} 
where $h\equiv  \bar{g}^{\lambda \mu}h_{\lambda \mu}= h^{\lambda}_{\; \lambda}$ (i.e., the trace of $h_{\mu \nu}$). The gauge-fixing Lagrangian will be
\begin{equation}
\begin{split}
\mathcal{L}_{gf} &=\dfrac{1}{ \alpha} \sqrt{- \bar{g}} \; C_\mu C^\mu \\
 &= \dfrac{1}{ \alpha}\sqrt{- \bar{g}} \left( \bar{\nabla}^\nu h_{\mu \nu} - \dfrac{1}{2} \bar{\nabla}_\mu h \right) \left( \bar{\nabla}_\lambda h^{\mu \lambda} - \dfrac{1}{2}\bar{\nabla}^\mu h \right), \label{gauge_fixing_lagrangian}
\end{split}
\end{equation} 
and we recover de Donder gauge with the choice $\alpha=1$. By considering the variation of the gauge-averaging functional $C_\mu[h]$ under the infinitesimal gauge transformation (\ref{gauge_transf_h}), which can be written in the form 
\begin{equation}
C_\mu [h]-C_\mu [\hat{h}]= - \left( \bar{g}^{\; \nu}_{\mu} \; \bar{g}^{\alpha \beta} \; \bar{\nabla}_{\alpha} \bar{\nabla}_{\beta}+\bar{R}^{\; \nu}_{ \mu}\right) \epsilon_\nu \equiv \mathcal{F}_{\mu}^{\; \nu} \epsilon_\nu,
\end{equation}
$ \mathcal{F}_{\mu}^{\; \nu}$ being the ghost operator acting linearly on the ghost one-form and $\bar{R}_{ \mu \nu }$ the Ricci tensor of the background geometry, we obtain the (Faddeev-Popov) ghost Lagrangian \cite{Goroff-Sagnotti}
\begin{equation}
\begin{split}
\mathcal{L}_{ghost} &= \sqrt{-\bar{g}} \; \Bigl [ -\bar{\nabla}^\nu\bar{c}^\mu \; \bar{\nabla}_\nu c_\mu - \bar{R}_{\mu \nu} \bar{c}^\mu c^\nu - \left(\bar{\nabla}^\nu\bar{c}^\mu \; \bar{\nabla}_\mu c^\rho\right) h_{\nu \rho} \\
& - \left(\bar{\nabla}^\nu\bar{c}^\mu \; \bar{\nabla}_\nu c^\rho\right) h_{\mu \rho} -\left(\bar{\nabla}^\nu\bar{c}^\mu\right) c^\rho \bar{\nabla}_\rho h_{\mu \nu} + \left(\bar{\nabla}_\mu\bar{c}^\mu \; \bar{\nabla}^\nu c^\rho\right) h_{\nu \rho} \\ 
& + \dfrac{1}{2} \left(\bar{\nabla}_\mu\bar{c}^\mu \right)  c^\rho \bar{\nabla}_\rho h \Bigr], \label{ghost Lagrangian}
\end{split}
\end{equation}
where $c^\mu$ is the spin-one anticommuting complex ghost field. \\
To further simply the calculations, we will derive Feynman rules in the case in which the background metric is represented by the flat Minkowski metric, i.e., $\bar{g}_{\mu \nu}= \eta_{\mu \nu}$. Therefore, we have
\begin{equation}
g_{\mu \nu} = \eta_{\mu \nu} + \chi h_{\mu \nu }, \label{perturbed_metric_flat}
\end{equation}
\begin{equation}
g^{\mu \nu} = \eta^{\mu \nu} - \chi h^{\mu \nu } + \chi^2 h^{\mu}_{\; \lambda} h^{\lambda \nu} +{ \rm O} (\chi^3 h^3). 
\end{equation}
In order to set up a perturbative calculation, we first need to consider the expansion of the metric determinant in powers of $h_{\mu \nu}$. By bearing in mind the property 
\begin{equation}
\det (A+B)= \det (1+A^{-1}B) \det (A),
\end{equation}
holding for generic matrices $A$ and $B$ and the Taylor expansion about zero of the logarithmic and the exponential functions up to quadratic order, we have
\begin{equation}
\begin{split}
\sqrt{-\det (g_{\mu \nu})} &=\sqrt{-g}= {\rm e}^{(1/2) \log[ -\det (\eta_{\mu \nu}+\chi h_{\mu \nu})]} =\sqrt{- \eta} \;  {\rm e}^{(1/2) \log[ \det (1+\chi\; \eta^{-1}h)]} \\
&={\rm e}^{(1/2) {\rm Tr}[ \log (1+\chi\; \eta^{-1}h)]}={\rm e}^{(1/2) {\rm Tr}[\chi \; \eta^{-1} h-(1/2) \chi^2 (\eta^{-1} h)^2+{\rm O}(\chi^3 h^3)]} \\
&= 1+ \dfrac{\chi}{2} \left[ {\rm Tr} (\eta^{-1} h)-\dfrac{\chi}{2} {\rm Tr} (\eta^{-1} h)^2 \right] \\
&+\dfrac{\chi^2}{8} \left[ {\rm Tr} (\eta^{-1} h)-\dfrac{\chi}{2} {\rm Tr} (\eta^{-1} h)^2 \right]^2 + {\rm O}(\chi^3 h^3) \\
&= 1 + \dfrac{\chi}{2} {\rm Tr} (\eta^{-1} h) -\dfrac{\chi^2}{4} {\rm Tr} (\eta^{-1} h)^2 + \dfrac{\chi^2}{8} {\rm Tr}^2 (\eta^{-1} h)+ {\rm O}(\chi^3 h^3) \\
&= 1+ \dfrac{\chi}{2} h^{\alpha}_{\; \alpha}-\dfrac{\chi^2}{4} h^{\alpha}_{\; \beta} h^{\beta}_{\; \alpha} + \dfrac{\chi^2}{8} (h^{\alpha}_{\; \alpha})^2 + {\rm O}(\chi^3 h^3).
\end{split}
\end{equation}
Next, we need the expansion of the Ricci scalar $R$. For this reason we start by considering the components of the Levi-Civita connection of the full metric $g_{\mu \nu }$ (see Eq. (\ref{Christoffel symbols})) and, after the insertion of (\ref{perturbed_metric_flat}), we end up with the relation
\begin{equation}
\begin{split}
\Gamma^{\lambda}_{\; \mu \nu} &= \dfrac{\chi}{2} \left[ \eta^{\lambda \sigma}-\chi h^{\lambda \sigma} + \chi^2 h^{\lambda}_{\; \alpha}h^{\alpha \sigma} +{ \rm O} (\chi^3 h^3) \right]  \left(\partial_\mu h_{\nu \sigma}+ \partial_\nu h_{\mu \sigma} - \partial_\sigma h_{\mu \nu } \right) \\
&= \dfrac{\chi}{2} \left(\partial_\mu h^{\lambda}_{\;\nu}+ \partial_\nu h^{\lambda}_{\;\mu} - \partial^\lambda h_{\mu \nu } \right) \\
&  - \dfrac{\chi^2}{2} h^{\lambda \sigma}\left(\partial_\mu h_{\nu \sigma}+ \partial_\nu h_{\mu \sigma} - \partial_\sigma h_{\mu \nu } \right)+ { \rm O} (\chi^3 h^3).
\end{split}
\end{equation}
The expansion for the scalar curvature $R$ then will be given by \cite{'t Hooft-Veltman74}
\begin{equation}
\begin{split}
R &= \chi \biggl(\D^\alpha \D_\alpha h- \D^\alpha \D_\beta h^{\beta}_{\; \alpha}\biggr) + \chi^2 \biggl \{ -\dfrac{1}{2} \D_{\alpha} \left(h_\mu^{\; \beta}\D^\alpha h_{\beta}^{\; \mu}\right) \\
& + \dfrac{1}{2} \D_\beta \left[ h^{\beta}_{\; \nu} \left(2 \D_\alpha h^{\nu \alpha} - \D^\nu h\right) \right] -\dfrac{1}{2} h^{\nu \alpha} \D_\nu \D_\alpha h \\
& +\dfrac{1}{4} \left(\D_{\alpha} h^{\nu}_{\; \beta}+ \D_{\beta} h^{\nu}_{\; \alpha} -\D^{\nu} h_{ \beta \alpha}\right) \left(\D^{\alpha} h^{\beta}_{\; \nu}+ \D_{\nu} h^{\beta \alpha}-\D^{\beta} h^{\alpha}_{\; \nu} \right) \\
& - \dfrac{1}{4} \left(2 \D_\alpha h^{\nu \alpha}-\D^\nu h \right)\D_{\nu} h    \\
& + \dfrac{1}{2} h^{\nu}_{\; \alpha} \D_\beta \left(\D^{\alpha} h^{\beta}_{\; \nu}+\D_\nu h^{\beta \alpha}- \D^{\beta} h^{\alpha}_{\; \nu} \right) \biggr \} + {\rm O} (\chi^3 h^3).
\end{split}
\end{equation}
At this point the above relations allow us to write down the quadratic part of the Einstein-Hilbert Lagrangian $\mathcal{L}_{grav}$
\begin{equation}
\mathcal{L}^{(2)}_{\; grav} (h^2)= \dfrac{1}{2} \left(\partial_\mu h_{\alpha \beta} \partial^\mu h^{\alpha \beta}- \dfrac{1}{2}\partial_\lambda h^{\alpha}_{\; \alpha} \partial^\lambda h^{\beta}_{\; \beta} \right) - C_\mu C^\mu.
\end{equation}
Therefore, we can now appreciate the advantages of the de Donder gauge, since we see that the addition of the gauge-fixing Lagrangian (\ref{gauge_fixing_lagrangian}) (with $\alpha=1$ and $\bar{g}_{\mu \nu}= \eta_{\mu \nu}$) to $\mathcal{L}^{(2)}_{\; grav}$ cancels out the term $-C_\mu C^\mu$, giving rise to an invertible dynamical operator on metric perturbations that turns out to be the wave operator, which in turn leads to a smooth graviton propagator that looks renormalizable (the theory however is still not renormalizable because of the derivatives that will occur in the interaction terms). Thus in the de Donder gauge and at the quadratic order in metric perturbations (and with a flat Minkowski background) we just have \cite{'t Hooft-Veltman74}
\begin{equation}
\mathcal{L}^{(2)}_{\; grav} (h^2) + \mathcal{L}_{gf} = \dfrac{1}{2}\left(\partial_{\lambda} h_{\alpha \beta} \mathcal{V}^{\alpha \beta \mu \nu} \partial^\lambda h_{\mu \nu}\right),
\end{equation} 
which, after performing partial integration and omitting total derivatives, can be written as
\begin{equation}
\mathcal{L}^{(2)}_{\; grav} (h^2) + \mathcal{L}_{gf} = - \dfrac{1}{2} h_{\alpha \beta}\left(\mathcal{V}^{\alpha \beta \mu \nu} \partial_{\lambda}  \partial^\lambda \right) h_{\mu \nu},
\end{equation} 
where the matrix
\begin{equation}
\mathcal{V}_{\alpha \beta \mu \nu}=\eta_{\alpha \mu} \eta_{\beta \nu} -\dfrac{1}{2} \eta_{\alpha \beta} \eta_{\mu \nu},
\end{equation}
is easily invertible once we symmetrize it with respect to the interchange $\alpha\leftrightarrow \beta$, $\mu \leftrightarrow \nu$ and $(\alpha \beta) \leftrightarrow (\mu \nu)$. In this way we can obtain the graviton propagator $D_{\mu \nu \rho \sigma} (k)$ quite straightforwardly by solving the tensor equation
\begin{equation}
\mathcal{V}^{\alpha \beta \mu \nu} \mathcal{P}_{\mu \nu \rho \sigma}=\dfrac{1}{2} \left( \delta^\alpha_{\; \rho} \delta^\beta_{\; \sigma}+\delta^\alpha_{\; \sigma} \delta^\beta_{\; \rho} \right), 
\end{equation}
whose solution is given by 
\begin{equation}
\mathcal{P}_{\mu \nu \rho \sigma}=\dfrac{1}{2} (\eta_{\mu \rho} \eta_{\nu \sigma} + \eta_{\mu \sigma} \eta_{\nu \rho}-\eta_{\mu \nu} \eta_{\rho \sigma}), 
\footnote{The term in brackets is reminiscent of DeWitt supermetric, which is defined as the one-parameter family of metrics on the space of rank-two tensor fields
\begin{equation}
g^{(\mu \nu)(\rho \sigma)} (\lambda)= \dfrac{1}{2}\left(g^{\mu \rho} g^{\nu \sigma}+g^{\mu \sigma} g^{\nu \rho} + \lambda \; g^{\mu \nu} g^{\rho \sigma}\right),
\end{equation} 
where the constraint $\lambda \neq -\dfrac{2}{d}$ ($d$ being the spacetime dimensions) guarantees that $g^{(\mu \nu)(\rho \sigma)}$ has an inverse.}
\end{equation}
finally obtaining
\begin{equation}
D_{\mu \nu \rho \sigma} (k)= {\rm i}\dfrac{\mathcal{P}_{\mu \nu \rho \sigma} }{k^2}. \label{graviton_propagator}
\end{equation}
It is easy to see in the non-covariant Prentki gauge (which is the counterpart of Coulomb gauge in QED and reads as $C_\mu [h]= \sum\limits_{i=1}^{3} \D^i h_{i \mu},$ with $\mu=1, \dots,4$) that there exist only two polarization states of a mass zero spin-two particle that propagate, i.e., the two helicities of the graviton \cite{'t Hooft-Veltman74}. Therefore, (\ref{graviton_propagator}) propagates a massless spin-two graviton with the speed of light (or equivalently the theory is unitary). Moreover, Eq. (\ref{graviton_propagator}) clearly shows that the graviton propagator is independent of the Newton constant $G$. We could have expected this feature from the the very beginning, since both the fact that Einstein-Hilbert Lagrangian (\ref{Einstein_Lagrangian}) depends on $G^{-1}$ and that $h_{\mu \nu}$ (as we said before) has the dimension of a mass make an $m$-point function be proportional to $G^{m/2-1}$. Higher order corrections in the perturbed metric to the pure gravity Lagrangian lead at the order $h^3$ to the three-graviton vertex, at the order $h^4$ to the four-graviton vertex, and so forth, because the terms $\sqrt{-g}$ and $g^{\mu \nu}$ in (\ref{Einstein_Lagrangian}) give rise to an infinite number of graviton vertices. Moreover, each term of this expansion always contains derivatives of second order of $h_{\mu \nu}$ because the Ricci scalar involves deriving the metric two times (this fact in turn implies that all vertices of quantum gravity are proportional to the square of the momentum unlike QED, where vertices are momentum-independent). For example, at the cubic order in metric perturbations we have \cite{Goroff-Sagnotti}
\begin{equation}
\begin{split}
\mathcal{L}^{(3)}_{\; grav} (h^3) & = -\chi\biggl\{  h^{\alpha \beta} \Bigl[  \D_\alpha h^{\gamma \delta}   \left(-\dfrac{1}{2} \D_\beta h_{\gamma \delta}+2 \D_\delta h_{\beta \gamma }\right)- \D_\alpha h \; \D_\delta h^{\delta}_{\; \beta} \\
& -\D_\delta h \; \D_\beta h^{\delta}_{\; \alpha}+\dfrac{1}{2} \D_\alpha h \; \D_\beta h +\D_\gamma h_{\alpha \beta} \left(- \D_\delta h^{\gamma \delta}+  \D^\gamma h \right) \\
&+\D_\delta h^\gamma_{\; \alpha} \left(- \D^\delta h_{\beta \gamma} +  \D_\gamma h^\delta_{\, \beta}\right) \Bigr]+\dfrac{1}{2} h \Bigl[ \D^\delta h^{\beta \gamma} \left( -\D_\gamma h_{\beta \delta} + \dfrac{1}{2} \D_\delta h_{\beta \gamma} \right) \\
& + \D_\gamma h \left( \D_\delta h^{\gamma \delta} + \dfrac{1}{2} \D^\gamma h \right) \Bigr] \biggr \},
\end{split}
\end{equation}
which gives rise to the three-graviton vertex $\tau^{\mu \nu}_{\; \; \; \alpha \beta \gamma \delta}(k,q)$ of Eq. (\ref{three-graviton_vertex}) (See Appendix \ref{Appendix_Feynman_rules} for a summary of all Feynman rules). By bearing in mind the previous equations regarding the expansion of the Einstein-Hilbert Lagrangian (\ref{Einstein_Lagrangian}) in powers of $h_{\mu \nu}$, note how a term involving $n$ graviton fields, i.e., $\left(h_{\mu \nu}\right)^n$, carries a coupling constant going as $\chi^{n-2}$.

Now we turn our attention to the vertices describing the gravity-scalar interaction. From the expansion of the Klein-Gordon Lagrangian (\ref{Klein-Gordon_matter_Lagrangian}) in terms of the perturbed metric we have at the first order
\begin{equation}
\mathcal{L}_{m}^{\; (1)}(h \phi^2)= \dfrac{\chi}{2} \left[ -h^{\mu \nu} \D_\mu \phi \D_\nu \phi  + \dfrac{1}{2} h \left( \D_\mu \phi \D^\mu \phi - m^2 \phi^2 \right) \right].
\end{equation}
If we apply the same expansion to the energy-momentum tensor of the matter (\ref{energy-momentum_scalar}) we obtain
\begin{equation}
T_{\mu \nu}= T_{\mu \nu}^{\; (0)} (\phi^2) +  T_{\mu \nu}^{\; (1)} (h \phi^2) + \dots,
\end{equation}
with
\begin{equation}
T_{\mu \nu}^{\; (0)} (\phi^2) =-\D_\mu \phi \D_\nu \phi+\dfrac{1}{2} \eta_{\mu \nu} \left(\D_\sigma \phi \D^\sigma \phi - m^2 \phi^2 \right),
\end{equation}
\begin{equation}
\begin{split}
T_{\mu \nu}^{\; (1)} (h \phi^2) & = \dfrac{\chi}{2} \biggl[- h \; \D_\mu \phi \D_\nu \phi+ \Bigl( h_{\mu \nu} \eta^{\lambda \sigma} -\eta_{\mu \nu} h^{\lambda \sigma} + \dfrac{1}{2} h \; \eta_{\mu \nu} \eta^{\lambda \sigma} \Bigr) \D_\lambda\phi \D_\sigma \phi \\
& - m^2 \phi^2 \left(h_{\mu \nu}+\dfrac{1}{2} h\; \eta_{\mu \nu} \right) \Bigr].
\end{split}
\end{equation}
Therefore, it is easy to see that the gravity-scalar interaction Lagrangian at the lowest order is given by
\begin{equation}
\mathcal{L}_{m}^{\; (1)}(h \phi^2) = \dfrac{1}{2} \chi h^{\mu \nu} T_{\mu \nu}^{\; (0)} (\phi^2),
\end{equation}
which reminds us of the QED interaction term
\begin{equation}
\mathcal{L}_{int}=-{\rm i} e J_{\mu} A^{\mu} = -{\rm i} e \bar{\Psi }\gamma_{\mu} A^{\mu} \Psi,
\end{equation}  
$e$ being the electron charge, $J_\mu$ the conserved probability four-current, $\Psi$ the Dirac spinor, $\gamma_\mu$ the Dirac matrices, $\bar{\Psi}= \Psi^\dagger \gamma^0$ the Dirac adjoint and $A^\mu$  the four-potential of the electromagnetic field generated by the electron itself. The interaction term $\mathcal{L}_{m}^{\; (1)}(h \phi^2)$ leads to the two scalar-one graviton vertex $\tau_{\mu \nu} (p,p^\prime,m)$ (\ref{2scalar-1graviton_vertex}). The expansion of $\mathcal{L}_m$ at the second order reads as 
\begin{equation}
\begin{split}
\mathcal{L}_{m}^{\; (2)}(h^2 \phi^2) &= \dfrac{\chi^2}{2} \biggl[ \left( h^{\mu}_{\;  \rho} h^{\rho \nu}-\dfrac{1}{2} h h^{\mu \nu} \right) \D_\mu \phi \D_\nu \phi \\
& + \left( \dfrac{1}{8} h^2 -\dfrac{1}{4} h^{\lambda \sigma} h_{\lambda \sigma} \right) \left( \D_\mu \phi \D^\mu \phi - m^2 \phi^2 \right) \biggr],
\end{split}
\end{equation}
which gives rise to the two scalar-two graviton vertex $\tau_{\eta \lambda\rho \sigma }(p, p^\prime, m)$ (\ref{2scalars-2gravitons vertex}). This concludes our section about the Feynman rules we will use throughout this thesis. However, we stress once again the fact that the expansions given here (and hence the Feynman rules) hold for a flat Minkowski background, whereas the most general calculation involves also curvature terms of the background geometry. Moreover, in the general case it is possible to show that the part of $\mathcal{L}_{grav}$ and $\mathcal{L}_{m}$ linear in the quantum perturbations $h_{\mu \nu}$ vanishes if the background metric obeys classical fields equations. We refer the reader to Refs. \cite{Goroff-Sagnotti,'t Hooft-Veltman74} for the general expressions.

\subsection{Ultraviolet divergences}

The birth of quantum field theory has introduced the concept of ultraviolet divergences. In hindsight, such divergences are inevitable, because they reflect the fact that in the transition from quantum mechanics to quantum field theory a change to an infinite number of degrees of freedom picture is involved and therefore we always sum over an infinite number of internal modes while performing loop integrations. Moreover, since the divergent nature of the theory probes spacetime regions at a high-energy scale (or equivalently at low distances), it witnesses our ignorance about physics at extremely high-momenta regime (so far, almost nothing is known about high-energy physics). This peculiarity has forced several generations of physicist to struggle with the topic of renormalization. A quantum field theory is said to be renormalizable if counter-terms, required to cancel divergences at each order in perturbation theory, are of the same form as those appearing in the original Lagrangian. If this is the case, the renormalization scheme leads to charge, mass and field re-definitions by means of (infinite) multiplicative factors \cite{Ryder}. The application of this procedure to the pure $SU(N)$ Yang-Mills action leads to \cite{'t Hooft-71-72}
\begin{equation}
S_{YM}=\dfrac{1}{-4 \tilde{e}^2} \int d^4x F^{\alpha}_{\; \mu \nu} F_{\alpha}^{\; \mu \nu} \rightarrow \dfrac{1}{-4\tilde{e}_R^{2}} \int d^4x \left(F^{\alpha}_{\; \mu \nu} \right)_R \left(F_{\alpha}^{\; \mu \nu}\right)_R,
\end{equation}
$\tilde{e}$ being the dimensionless coupling constant of the theory and $F^{\alpha}_{\; \mu \nu}$ the field-strength ($\alpha$ is a Lie algebra index). As we can see, the renormalized action has the same form as the original one and no new interaction terms (involving for example the gauge-covariant derivative of the field strength) need to be introduced in order to re-absorb the divergences. It is an important point to be stressed the fact that Yang-Mills theory becomes not perturbatively renormalizable when considered in curved spaces \cite{DeWitt2003}.   

By applying dimensional analysis, it is easily found that a renormalizable theory must have in $\hbar=c=1$ units a coupling constant whose mass dimension is non-negative, a condition which assures that perturbation series does not give an infinite number of different types of divergent graphs. In fact, on general grounds we know that if a quantum field theory has a coupling constant with dimension (mass)$^\delta$, then a Feynman diagram of order $N$ behaves at large momenta as $\int {\rm d}p \; p^{A-N \delta}$, where $A$ depends on the physical process considered but not on the order $N$. Therefore, interactions having $\delta <0$ are characterized by diagrams that diverge at sufficiently high order and gravity belongs to this class of theories since Newton constant $G$ has mass dimension $\delta=-2$ \cite{Weinberg79}. More precisely, since the scalar curvature $R$ contains second order derivatives of the spacetime metric, the corresponding momentum-space vertex functions (as we have shown in the previous section and in Appendix \ref{Appendix_Feynman_rules}) behave like $p^2$, and the propagator like $p^{-2}$. In $d$ dimensions each loop integral contributes $p^d$, so that with $L$ loops, $V$ vertices and $I$ internal lines, the superficial degree of divergence $\mathfrak{D}$ of a Feynman diagram is given by
\begin{equation}
\mathfrak{D}=dL+2V-2I,
\end{equation}
which, by invoking the topological relation concerning the number of independent momenta valid for any diagram
\begin{equation}
L=I-V+1,
\end{equation}
becomes
\begin{equation}
\mathfrak{D}=2+(d-2)L.
\end{equation}
In other words, $\mathfrak{D}$ increases with increasing loop order for $d=4$, so that general relativity clearly leads to a non-renormalizable theory where the cancellation of ultraviolet divergences would require the introduction of an infinite number of terms (not present in the original Lagrangian (\ref{Einstein_Lagrangian})) proportional to arbitrarily high powers of the Riemann curvature tensor and its covariant derivatives. We believe that the essence of the bad ultraviolet behaviour of Einstein theory can be enlightened by the following theorem \cite{Nieuw-Wu}:
\newtheorem{Wu}{Theorem}
\begin{Wu} \label{Wu}
The leading $L$-loop divergences of the quantum $S$ matrix for pure Einstein theory in $d$ dimensions have the form
\begin{equation}
S^{div}(L-{\rm loop},d-{\rm dim})=\dfrac{G^{L-1}}{\epsilon^L} \int {\rm d}^d x \sqrt{-g} \; \mathcal{B}(x), \label{S_div}
\end{equation}
\end{Wu}
where $\epsilon=4-d$ is the usual regularization parameter and
\vskip 0.3cm
\noindent
(i) $\mathcal{B}(x)$ is a local scalar function depending on the spacetime metric $g_{\mu \nu}(x)$ but not on the Newton constant $G$;
\vskip 0.3cm
\noindent
(ii) the fields $g_{\mu \nu}(x)$ occurring in $\mathcal{B}(x)$ are on-shell, i.e., they satisfy the vacuum Einstein equation $R_{\mu \nu}=0$;
\vskip 0.3cm
\noindent
(iii) $\mathcal{B}(x)$ is constructed from $(1/2)d+L-l-1$ Riemann tensors and $2l$ covariant derivatives.

Note that the theorem does not provide any information about non-leading divergences, but it proves the really important fact that for example no expressions of the form $R^{-1}$ or $(\nabla_\mu \nabla^\mu)^{-1}$ can appear in $S^{div}$. It is possible to sketch a proof of this theorem by means of a dimensional analysis. In fact, as explained in the previous section, a general $m$-point vertex is a function of $G^{m/2-1}$ and this explains the presence of the term $G^{L-1}$ in Eq. (\ref{S_div}). Moreover, from the fact that the $S$ matrix is dimensionless follows that $\mathcal{B}(x)$ depends on the above indicated number of Riemann tensors. The simplest application of the above theorem is obviously the case $L=1$ and $d=4$, where it forecasts a leading divergence of the form
\begin{equation}
\mathcal{B}(x)= \alpha_1 R^2 + \alpha_2 R^{\mu \nu}R_{\mu \nu} + \alpha_3 R^{\mu \nu \rho \sigma} R_{\mu \nu \rho \sigma}.
\end{equation}
In accordance with theorem \ref{Wu}, once we have imposed the on-shell condition on this expression, the first two terms vanish whereas the last one does not. This is however a dummy issue since (only) in $d=4$ dimensions there is a relation involving a topological invariant called the Euler number density which states that
\begin{equation}
\begin{split}
& R_{\alpha \beta \gamma \delta} R_{\varphi \lambda \sigma \tau} \left( \sqrt{-g}\; \epsilon^{\alpha \beta \varphi \lambda}\right) \left(\sqrt{-g}\; \epsilon^{\gamma \delta \sigma \tau} \right) \\
& = 4 \left( R_{\varphi \lambda \gamma \delta} R^{\varphi \lambda \gamma \delta} - 4 R_{\varphi \lambda} R^{\varphi \lambda} +R^2 \right)= {\rm total \; derivative},
\end{split} \label{Gauss-Bonnet}
\end{equation}
where
\begin{equation}
\epsilon_{\alpha \beta \varphi \lambda} =
\biggl \{\begin{array}{l}
 1 \; \; \; \; \; \; \; \; {\rm if} \; \; \alpha, \beta, \varphi, \lambda=1,2,3,4 \\[2mm]
 {\rm antysimmetric \; under \; exchange \; of \; any \; two \; indices.  }
 \end{array}
\end{equation}
This means that the integral defining the Euler characteristic 
\begin{equation}
\int {\rm d}^4 x \;\sqrt{-g}  \left( R_{\varphi \lambda \gamma \delta} R^{\varphi \lambda \gamma \delta} - 4 R_{\varphi \lambda} R^{\varphi \lambda} +R^2 \right) \equiv \int {\rm d}^4 x \;\sqrt{-g} \; \mathcal{G} , 
\end{equation}
vanishes, provided the right boundary conditions are engaged (Gauss-Bonnet theorem). As a consequence, it follows that
\begin{equation}
S^{div}(L=1,d=4)  =\dfrac{1}{\epsilon} \int {\rm d}^4 x \; \sqrt{-g} \left(  \alpha^{\prime}_1 R^2 + \alpha^{\prime}_2 R^{\mu \nu}R_{\mu \nu} + \alpha^{\prime}_3 \mathcal{G}\right),
\end{equation}
vanishes on-shell and hence we can conclude that the lowest order quantum corrections to the $S$ matrix of pure Einstein theory are finite. It is possible to show that the coefficients $\alpha^{\prime}_1$ and $\alpha^{\prime}_2$ are connected to the one-loop corrections to the graviton propagator (\ref{graviton_propagator}), whereas $\alpha^{\prime}_3$ comes from the analysis of the one-loop corrections to the three-graviton vertex (\ref{three-graviton_vertex}) \cite{Goroff-Sagnotti}. Unluckily, there is not another topological relation like the one of Eq. (\ref{Gauss-Bonnet}) which can shelter us from problems at two-loop level. In this case in fact theorem \ref{Wu}, jointly with the symmetries of the Riemann tensor, constrains the leading divergences of the $S$ matrix to assume the form
\begin{equation}
S^{div}(L=2,d=4)= \tilde{\alpha}\;\dfrac{G}{\epsilon^2} \int {\rm d}^4 x \; \sqrt{-g} \; R^{\alpha \beta}_{\; \; \; \gamma \delta} R^{\gamma \delta}_{\; \; \; \rho \sigma} R^{\rho \sigma}_{\; \; \; \alpha \beta}, \label{Wu_14}
\end{equation} 
where $\tilde{\alpha}$ is a constant which, at this level, we may only hope that due to miraculous cancellations will vanish in order to have a finite $S$ matrix. However, we will see below that $\tilde{\alpha}$ is a non-zero coefficient. Moreover, it is important to stress that the particular combination of three Riemann tensors occurring in Eq. (\ref{Wu_14}) is a direct consequence of a topological relation valid in the $d=6$ case, analogous to the four-dimensional Gauss-Bonnet theorem. Indeed, all topological relations derived in $2d$ dimensions become exact identities in lower dimensions, since their reduction to lower dimensions is formally equivalent to consider Riemann tensors with vanishing components outside the directions lying in the lower-dimensional space.   

The explicit form of the one-loop divergences in the theory of gravitation has been first calculated by 't Hooft and Veltman in their celebrated paper of 1974 (Ref. \cite{'t Hooft-Veltman74}). The starting point is the result of Ref. \cite{'t Hooft73}, whose generalization to our purposes allows to prove that to the Lagrangian describing the dynamics of a complex scalar field in an external gravitational field $g_{\mu \nu}$
\begin{equation}
\mathcal{L}= \sqrt{-g} \left( - g^{\mu \nu} \D_\mu \phi^* \D_\nu \phi + 2 \phi^* \mathcal{N}^\mu \D_\mu \phi + \phi^* \mathcal{M} \phi \right),\label{'tHooft_3.8}
\end{equation}
$\mathcal{N}$ and $\mathcal{M}$ being functions that  do not depend on the quantum fields $\phi$ and $\phi^*$, corresponds the counter-Lagrangian
\begin{equation}
\begin{split}
\Delta \mathcal{L} &=\dfrac{\sqrt{-g}}{8 \pi^2 \epsilon}\;  {\rm Tr} \biggl[ \dfrac{1}{12} \mathcal{Y}^{\mu \nu} \mathcal{Y}_{\mu \nu} + \dfrac{1}{2} \left( \mathcal{M}- \mathcal{N}^\mu\mathcal{N}_\mu - \nabla_\mu \mathcal{N}^\mu - \dfrac{1}{6} R \right)^2 \\
& + \dfrac{1}{60} \left( R_{\mu \nu} R^{\mu \nu} - \dfrac{1}{3} R^2 \right) \biggr],\label{'tHooft_3.35}
\end{split}
\end{equation}
which eliminates all one-loop divergencies \cite{'t Hooft-Veltman74}. Here the trace must be enforced on the ``internal'' indices labelling the scalar field and 
\begin{equation}
\mathcal{Y}_{\mu \nu} = \nabla_\mu \mathcal{N}_\nu- \nabla_\nu \mathcal{N}_\mu +\mathcal{N}_\mu  \mathcal{N}_\nu- \mathcal{N}_\nu \mathcal{N}_\mu.
\end{equation}
For a real scalar field described by the Lagrangian
\begin{equation}
\mathcal{L}= \sqrt{-g} \left( - \dfrac{1}{2} g^{\mu \nu} \D_\mu \phi \D_\nu \phi +  \phi N^\mu \D_\mu \phi +\dfrac{1}{2} \phi M \phi \right),
\end{equation}
the counter-Lagrangian assumes the form
\begin{equation}
\begin{split}
\Delta \mathcal{L} &=\dfrac{\sqrt{-g}}{8 \pi^2 \epsilon}\;  {\rm Tr} \biggl[ \dfrac{1}{24} Y^{\mu \nu} Y_{\mu \nu} + \dfrac{1}{4} \left( M- N^\mu N_\mu - \nabla_\mu N^\mu - \dfrac{1}{6} R \right)^2 \\
& + \dfrac{1}{120} \left( R_{\mu \nu} R^{\mu \nu} - \dfrac{1}{3} R^2 \right) \biggr],
\end{split}
\end{equation}
with
\begin{equation}
Y_{\mu \nu} = \nabla_\mu N_\nu- \nabla_\nu N_\mu + N_\mu  N_\nu- N_\nu N_\mu.
\end{equation}
At this point, in order to treat the gravitational field within a quantum scheme, we employ the background field method and hence we write the spacetime metric as in Eq. (\ref{perturbed_metric_h}), the scalar field as
\begin{equation}
\tilde{\phi}= \bar{\phi} + \phi,
\end{equation}
and thus the gravity-scalar Lagrangian as
\begin{equation}
\mathcal{L} = \sqrt{-g} \left( \dfrac{2}{\chi^2}R+\dfrac{1}{2} g^{\mu \nu} \D_\mu \tilde{\phi} \D^\mu \tilde{\phi} \right).\label{'tHooft_4.1}
\end{equation}
Like explained in the previous section, we have to perform an expansion of Eq (\ref{'tHooft_4.1}), with the difference that now we have to deal with the two quantum fields $h_{\mu \nu}$ and $\phi$. In this way we will obtain
\begin{equation}
\mathcal{L}= \mathcal{L}^{(0)}+ \mathcal{L}^{(1)}+ \mathcal{L}^{(2)}+ \mathcal{L}^{({\rm h.o.})}, \label{'tHooft_4.3}
\end{equation}
where $\mathcal{L}^{(0)}$ is the classical Lagrangian which has the same form as (\ref{'tHooft_4.1}) but with $\bar{g}_{\mu \nu}, \bar{R}, \bar{\phi}$ instead of $g_{\mu \nu},R,\tilde{\phi}$, $\mathcal{L}^{(1)}$ and $\mathcal{L}^{(2)}$ are linear and quadratic in the quantum fields, respectively, and $\mathcal{L}^{({\rm h.o.})}$ contains higher-order terms.  As we said before, we can dispose of $\mathcal{L}^{(1)}$ if the $c$-number quantities $\bar{g}_{\mu \nu}$ and $\bar{\phi}$ are chosen in such a way that they are on-shell, whereas we can ignore all terms contained in $\mathcal{L}^{({\rm h.o.})}$ because they give a contribution beyond the one-loop level. In other words, all one-loop divergences are determined by the quadratic term $\mathcal{L}^{(2)}$, once we have supported it with a gauge-fixing and a ghost Lagrangian. By ``extending'' the use of the de Donder gauge also to the gravity-scalar theory, the gauge-fixing functional (cf. Eq. (\ref{gauge_fixing_Functional})) reads as
\begin{equation}
C^{\alpha} [h,\phi]= \left( \D_\nu h^{\nu}_{\; \mu} - \dfrac{1}{2} \D_\mu h - \phi \D_\mu \bar{\phi} \right) t^{\mu \alpha}, \label{'tHooft_4.27}
\end{equation}
$ t^{\mu \alpha}$ being the square root of the inverse metric $\bar{g}^{\mu \nu}$, i.e.,
\begin{equation}
t^{\mu}_{\; \alpha}t^{\alpha \nu}= \bar{g}^{\mu \nu},
\end{equation}
so that we have
\begin{equation}
\begin{split}
\mathcal{L}^{(2)} + \mathcal{L}_{gf} & = \mathcal{L}^{(2)} + \sqrt{-\bar{g}} \; C^{\alpha}  C_{\; \alpha} \\
 &=\sqrt{-\bar{g}} \biggl ( \dfrac{1}{2} \partial_\mu h_{\alpha \beta} \partial^\mu h^{\alpha \beta} -\dfrac{1}{4} \D_\mu h \; \D^\mu h + \dfrac{1}{2} \; \bar{g}^{\mu \nu} \D_\mu \phi \D_\mu \phi \\
 & - h^{\alpha}_{\; \beta} X^{\beta \mu}_{\; \; \; \alpha \nu} h^{\nu}_{\; \mu} - 2 \phi Y^{\mu}_{\; \; \nu} h^{\nu}_{\; \mu} - \phi L \phi \biggr), \label{'tHooft_4.30}
\end{split}
\end{equation}
with
\begin{equation}
\begin{split}
X^{\beta \mu}_{\; \; \; \alpha \nu}  & = 2 \biggl ( -\dfrac{1}{2}\delta^\beta_{\; \nu} \bar{\nabla}^\mu \bar{\phi} \; \bar{\nabla}_\alpha \bar{\phi} + \dfrac{1}{4}  \delta^\beta_{\; \alpha} \bar{\nabla}^\mu \bar{\phi} \; \bar{\nabla}_\nu \bar{\phi} -\dfrac{1}{16} \delta^\beta_{\; \alpha} \delta^\mu_{\; \nu} \bar{\nabla}^\gamma \bar{\phi} \; \bar{\nabla}_\gamma \bar{\phi} \\
& + \dfrac{1}{8} \delta^\beta_{\; \nu} \delta^\mu_{\; \alpha} \bar{\nabla}^\gamma \bar{\phi} \; \bar{\nabla}_\gamma \bar{\phi} -\dfrac{1}{8}  \delta^\beta_{\; \alpha} \delta^\mu_{\; \nu} \bar{R} + \dfrac{1}{4}  \delta^\beta_{\; \nu} \delta^\mu_{\; \alpha} \bar{R} -\dfrac{1}{2}  \delta^\beta_{\; \nu} \bar{R}^{\mu}_{\; \alpha} \\
&+ \dfrac{1}{2}  \delta^\beta_{\; \alpha} \bar{R}^{\mu}_{\; \nu} +\dfrac{1}{2}  \bar{R}^{\beta \mu}_{\; \;  \; \alpha \nu} \biggr),
\end{split}
\end{equation}
\begin{equation}
Y^\alpha_{\; \; \beta}= \dfrac{1}{2}\delta^{\alpha}_{\; \beta} \bar{\nabla}^\gamma \bar{\nabla}_\gamma \bar{\phi} - \bar{\nabla}_\beta \bar{\nabla}^\alpha \bar{\phi},
\end{equation}
\begin{equation}
L= - \bar{\nabla}^\mu \bar{\phi} \; \bar{\nabla}_\mu \bar{\phi}.
\end{equation}
Since the second order Lagrangian (\ref{'tHooft_4.30}) has formally the same forms as the one in Eq. (\ref{'tHooft_3.8}), its counter-terms can be easily read off from (\ref{'tHooft_3.35}) \cite{'t Hooft-Veltman74}. As a final step, we need to evaluate the divergences coming from the ghost Lagrangian, which, as explained before, is obtained once we gauge-transform the gauge-breaking functional (\ref{'tHooft_4.27}). By bearing in mind that the gauge transformation law of the scalar field reads as
\begin{equation}
\phi \rightarrow \phi + \epsilon^\alpha \; \bar{\nabla}_\alpha \left( \bar{\phi} + \phi \right),
\end{equation}
whereas the one for $h_{\mu \nu}$ is given by (\ref{gauge_transf_h}), we obtain
\begin{equation}
\mathcal{L}_{ghost}=\sqrt{-\bar{g}} \;  \bar{c}^\mu \left[\D_\alpha \D^\alpha c_\mu - \bar{R}_{\alpha \mu} c^\alpha - \left( \D_\alpha \bar{\phi} \D_\mu \bar{\phi} \right)c^\alpha \right]. \label{'tHooft_5.19}
\end{equation}
The important feature according to which ghosts always appear in closed loop diagrams (i.e., they are never external) make pointless an eventual split of $c^\mu$ in a classical and a quantum part and therefore all terms containing both $h_{\mu \nu}$ and $\phi$ can be ignored in (\ref{'tHooft_5.19}) (see Eq. (\ref{ghost Lagrangian}) for a comparison). Anyway, to the ghost Lagrangian (\ref{'tHooft_5.19}) corresponds the counter-Lagrangian \cite{'t Hooft-Veltman74}
\begin{equation}
\begin{split}
\Delta \mathcal{L}_{ghost} &= -\dfrac{\sqrt{-\bar{g}}}{8 \pi^2 \epsilon} \biggl [ \dfrac{1}{6} \bar{R} \left(\bar{g}^{\mu \nu} \D_\mu \bar{\phi} \D_\nu \bar{\phi} \right) + \dfrac{17}{60} \bar{R}^2 +\dfrac{7}{30} \bar{R}_{\alpha \beta} \bar{R}^{\alpha \beta} \\
& + \bar{R}^{\alpha \beta} \left( \D_\alpha \bar{\phi} \D_\beta \bar{\phi} \right) +\dfrac{1}{2} \left(\bar{g}^{\mu \nu} \D_\mu \bar{\phi} \D_\nu \bar{\phi} \right)^2 \biggr],
\end{split}
\end{equation}
where the minus sign is due to the fermionic nature of ghost loops. At this point, by means of both quadratic and ghost contributions shown above, we are able to conclude that the one-loop counter-Lagrangian for the gravity-scalar theory reads as
\begin{equation}
\begin{split}
\Delta \mathcal{L}^{\rm (1-loop)} & = \dfrac{\sqrt{-\bar{g}}}{8 \pi^2 \epsilon} \biggl [ \dfrac{9}{720} \bar{R}^2 + \dfrac{43}{120} \bar{R}_{\alpha \beta} \bar{R}^{\alpha \beta} +\dfrac{1}{2} \left(\bar{g}^{\mu \nu} \D_\mu \bar{\phi} \D_\nu \bar{\phi} \right)^2 \\
&  - \dfrac{1}{12} \bar{R} \left(\bar{g}^{\mu \nu} \D_\mu \bar{\phi} \D_\nu \bar{\phi} \right) + 2 \left( \bar{\nabla}_{\mu} \bar{\nabla}^{\mu} \bar{\phi} \right)^2 \biggr],\label{'tHooft_5.22}
\end{split}
\end{equation}
whereas in the case of pure gravity the result is\footnote{It is possible to compute the whole set of one-loop divergencies by using the equivalent method of heat kernel expansion and zeta function regularization. A comprehensive review can be found in Ref. \cite{Hawking Comm}.}
\begin{equation}
\Delta \mathcal{L}^{\rm (1-loop)}_{grav}= \dfrac{\sqrt{-\bar{g}}}{8 \pi^2 \epsilon} \left(\dfrac{1}{120} \bar{R}^2 + \dfrac{7}{20} \bar{R}_{\alpha \beta} \bar{R}^{\alpha \beta} \right).\label{'tHooft_5.24}
\end{equation}
It is important to stress the fact that in deriving (\ref{'tHooft_5.22}) and (\ref{'tHooft_5.24}) a fundamental role is played by Gauss-Bonnet theorem. However, at this stage we can conclude that general relativity is not perturbatively renormalizable, since curvature terms different from the Ricci scalar are absent in the original Einstein-Hilbert Lagrangian.  Despite that, it is possible to transform away all one-loop divergencies by a field re-definition by replacing the Einstein-Hilbert Lagrangian $\mathcal{L}_{grav}$ by a more general one of the form
\begin{equation}
\mathcal{L}^{\prime}_{grav}=  \sqrt{-\bar{g}} \left( \dfrac{2}{\chi ^2} \bar{R} + c_1 \bar{R}^2 + c_2 \bar{R}_{\mu \nu} \bar{R}^{\mu \nu} + {\rm O} (\bar{R}^3) \right),
\end{equation}
where products of two background Riemann tensors are not considered because of the Gauss-Bonnet theorem. Therefore, from Eq. (\ref{'tHooft_5.24}) it follows immediately that the renormalized values leading to the cancellation of all one-loop divergencies of pure gravity are given by 
\begin{gather}
\begin{aligned}
c^{{\rm (ren)}}_1 = c_1 + \dfrac{1}{960 \pi^2 \epsilon},  \\
c^{{\rm (ren)}}_2 = c_2 + \dfrac{7}{160 \pi^2 \epsilon}.
\end{aligned} \label{renormalized c1-c2}
\end{gather}
We will give some information about experimental bounds on $c_1$ and $c_2$ in the next section. As we know, by adopting the classical tree-level equations of motion all divergences that are physically irrelevant will disappear, since this amounts to put all external lines of one-loop diagrams on-shell and with physical polarization states. The classical equations of motion are those that make the linear Lagrangian $\mathcal{L}^{(1)}$ in Eq. (\ref{'tHooft_4.3}) vanish and they read as
\begin{equation}
\bar{\nabla}_{\mu} \bar{\nabla}^{\mu} \bar{\phi}=0,
\end{equation}
\begin{equation}
\bar{R}_{\mu \nu} = -\dfrac{1}{2} \bar{\nabla}_{\mu} \bar{\phi} \; \bar{\nabla}_{\nu} \bar{\phi},
\end{equation}
\begin{equation}
\bar{R} = -\dfrac{1}{2} \bar{\nabla}_{\mu} \bar{\phi} \; \bar{\nabla}^{\mu} \bar{\phi}.
\end{equation}
From the above relations it follows immediately that the one-loop on-shell divergences of the gravity-scalar theory are represented by
\begin{equation}
\Delta \mathcal{L}^{\rm (1-loop)}= \dfrac{\sqrt{-\bar{g}}}{8 \pi^2 \epsilon} \dfrac{203}{80} \bar{R}^2, \label{'tHooft_6.5}
\end{equation}
which in the case of pure gravity reduce to
\begin{equation}
\Delta \mathcal{L}^{\rm (1-loop)}_{grav}= 0.  \label{'tHooft_6.6}
\end{equation}
Equations (\ref{'tHooft_6.5}) and (\ref{'tHooft_6.6}) imply two important results:
\vskip 0.3cm
\noindent
(i) the theory of gravitational field interacting with scalar particles show even at one-loop level physically meaningful divergencies that can not be re-absorbed into a field re-definition: the theory is absolutely not renormalizable;
\vskip 0.3cm
\noindent
(ii) all divergencies of pure Einstein gravity vanish on shell and thus can be disposed of by a re-definition of the background metric tensor. Therefore, we can conclude that the theory of gravitation is one-loop on-shell finite.

If topological invariants and classical field equations kindle at the one-loop level a light of hope in the path towards a renormalizable quantum theory of gravity, all these auspices are completely swept away at the following level. Two-loop divergences for the pure Einstein theory were first calculated by Goroff and Sagnotti in 1986 by using computer methods \cite{Goroff-Sagnotti} and their result was later confirmed by the author of Ref. \cite{vandeVen}\footnote{A modern application in the context of non-perturbative quantum gravity involving Weinberg asymptotic safety scenario and the results of Goroff and Sagnotti can be found in Ref. \cite{Gies}.}. In order to evaluate these divergences, the  expansion of Einstein-Hilbert Lagrangian at cubic level in quantum fields is needed. The key point of the calculation is represented by the analysis of all corrections to the three-graviton vertex (\ref{three-graviton_vertex}), which give rise to the interaction term
\begin{equation}
h^{\alpha \beta} \; \D_\alpha \D_\epsilon \D_\lambda h^{\gamma \delta} \;  \D_\beta \D_\gamma \D_\delta h^{\epsilon \lambda}. \label{Goroff_3.16}
\end{equation} 
All two-loop vertex corrections can be divided into two groups: those with a single graviton line from each vertex and those with two external lines from one vertex, with the latter which can not contribute to the structure of Eq. (\ref{Goroff_3.16}). In fact, we know that a second order derivative interaction leads to the two-graviton vertex and hence the corresponding Feynman rule contributes to graph terms with two quantum fields $h_{\mu \nu}$ leaving four, five, or six free indices coming from each vertex according to whether the external momenta are two, one, or zero, respectively. These terms would contract with another quantum field and with four, five, or six powers of its momentum (which is the only one circulating in the graph) in the pole part of the diagram and therefore a simple index-counting clearly shows that the structure of (\ref{Goroff_3.16}) can not be generated. This feature reduces the number of graphs to be calculated considerably, but it still takes a great effort to evaluate the counter-Lagrangian. By calculating all one-graviton emission vertex corrections, it is possible to prove that at the two-loop level the counter-Lagrangian for pure gravity reads as \cite{Goroff-Sagnotti}
\begin{equation}
\begin{split}
\Delta \mathcal{L} ^{\rm (2-loop)}_{grav} &= \dfrac{\chi^2}{(4 \pi)^4 \epsilon}\sqrt{-\bar{g}} \biggl[ \dfrac{209}{2880} \bar{R}^{\alpha \beta}_{\; \; \; \gamma \delta} \bar{R}^{\gamma \delta}_{\; \; \; \rho \sigma} \bar{R}^{\rho \sigma}_{\; \; \; \alpha \beta} \\ 
& - \left( \dfrac{5}{18 \epsilon} + \dfrac{5771}{4800} \right) \bar{R}^{\alpha \beta} \bar{\nabla}_{\mu} \bar{\nabla}^{\mu} \bar{R}_{\alpha \beta} \\
& + \left( \dfrac{1255}{54 \epsilon} - \dfrac{703049}{64800} \right)  \bar{R}_{\alpha \beta} \bar{R}_{\gamma \delta} \bar{R}^{\alpha \gamma \beta \delta} \\
& - \left( \dfrac{551}{27  \epsilon} - \dfrac{833}{16200} \right) \bar{R}^{\alpha}_{\; \beta} \bar{R}^{\beta}_{\; \gamma} \bar{R}^{\gamma}_{\; \alpha} \\
& + \left( \dfrac{1033}{108 \epsilon} - \dfrac{47417}{8100} \right) \bar{R}_{\alpha \beta \gamma \delta} \bar{R}^{\alpha \beta \gamma \sigma} \bar{R}^{\delta}_{\; \sigma} \\
& + {\rm terms \; involving \; the \; scalar \; curvature}\biggr],
\end{split}
\end{equation}
leading to the on-shell non-vanishing counter-term (cf. Eq. (\ref{Wu_14}))
\begin{equation}
\Delta \mathcal{L} ^{\rm (2-loop)}_{grav} = \dfrac{\chi^2}{(4 \pi)^4 \epsilon}\sqrt{-\bar{g}} \;  \dfrac{209}{2880} \bar{R}^{\alpha \beta}_{\; \; \; \gamma \delta} \bar{R}^{\gamma \delta}_{\; \; \; \rho \sigma} \bar{R}^{\rho \sigma}_{\; \; \; \alpha \beta}. \label{2loop-divergence}
\end{equation} 
The presence of this new non-removable on-shell $R^3$-type divergence arising at the two-loop level represents a clear evidence that the theory of gravitation has a bad ultraviolet behaviour, or in other words that it is not renormalizable. 

In conclusion, we have the following situation. Among the three different approaches to quantizing gravity quite intensely discussed so far, perturbation theory within Feynman path integral scheme represents the one which provides a covariant framework in which Feynman rules and radiative corrections to physical processes can be computed. Unluckily, as one would expect from simple power-counting arguments, Einstein theory turns out to be not perturbatively renormalizable, since at every order in the loop expansion divergences involve curvature invariants (and their covariant derivatives) of growing order, whose effects can not be simply absorbed into a re-definition of the original parameters of the original Lagrangian, as the explicit two-loop calculation (\ref{2loop-divergence}) shows. As a result, the theory of gravitation is only one-loop on-shell finite. Moreover, if we try to clear off all divergences through a perturbative expansion of the Einstein-Hilbert Lagrangian, then we would end up with a badly divergent mechanism, since at every loop level new invariants must be added to the Lagrangian in order to transform away the ultraviolet divergences. This attempt represents therefore a temporary dead end, but it can be nevertheless used for other purposes. Then, only at this point we can appreciate why we need to set up the formalism of effective field theory of gravity. This will represent the content of the following sections and, as we will point out, the starting point is just the perturbative expansion of pure gravity Lagrangian. 

\section{The effective field theory of gravitation}

The major difference between quantum and effective field theories is represented by the fact that in the former scheme the Lagrangian is believed to be fundamental and valid at any energy scale, whereas the latter case is based on a perturbative approach where the various terms of the action correspond to different energy scales of the theory. In this framework, every term consistent with the symmetries of the theory must be included in the action. As a result, any effective theory has by construction an infinite number of couplings  and is trivially renormalizable. Moreover, at each loop order only a finite number of terms are present in the action and hence calculations can be performed by employing standard techniques. 

Effective field theories scheme represents a widely used approach in physics and the example of chiral perturbation theory, which represents the low-energy limit of quantum chromodynamics (QCD), witnesses this trend of modern theoretical applications. It is possible to treat general relativity as an effective field theory, too. In fact, from a perturbative point of view Einstein theory has a bad behaviour since its coupling grows with energy and hence the theory is strongly coupled at high energies. Moreover, large quantum fluctuations of the spacetime metric, which may have a topology-changing nature, represent a huge problem to be dealt with in the path integral approach, an issue whose solution is yet unknown. However, low-energy fluctuations are weakly coupled and behave normally in perturbation theory. Therefore, it is natural to try to separate these low-energy quantum fluctuations from the high-energy corrections. As we said before, the tool to perform this distinction is represented by effective field theory. In this way, up to the scale of Planck energy (see below) we end up with a well-behaved quantum field theory.     

\subsection{The energy expansion of the gravitational action}

In principle, there exist two main reasons to modify the Einstein-Hilbert Lagrangian (\ref{Einstein_Lagrangian}). First of all, it is clear from previous section that a non-renormalizable theory is not predictive, since well defined predictions potentially require an infinite number of counter-terms to be added to the original Lagrangian. Therefore, due to the non-renormalizability of gravity, its validity is restricted only to the low-energy domain, i.e., to large scales, while it fails at high energy (i.e., small scales). This implies that the full unknown theory of gravity has to be invoked near or at the Planck era (i.e., the period of time going from Big-Bang to the Planck time $t_P=\sqrt{\hbar G /c^5} \approx 5.4 \times 10^{-44} {\rm s}$) and that, sufficiently far from the Planck scale ($E_P=M_P c^2 \approx 1.22 \times 10^{19} {\rm GeV}$, $M_P=\sqrt{\hbar c/G} \approx 2.18 \times 10^{-8}{\rm Kg}$ being the Planck mass) general relativity and its first loop corrections describe gravitational interactions. In this context, it makes sense to add higher order terms in the curvature invariant and non-minimal couplings between matter and gravity
to the Einstein-Hilbert Lagrangian. In fact, it is clear from Eqs. (\ref{'tHooft_6.5}), (\ref{'tHooft_6.6}), and (\ref{2loop-divergence}) how divergences introduce in Einstein theory more complicated expressions involving $R$, $R_{\mu \nu}$, and $R_{\mu \nu \rho \sigma}$. From a physical point of view, this fact implies that extra degrees of freedom, in addition to the usual spin-two graviton, need to be introduced. Besides, if the free parameters are chosen appropriately, the theory has a better ultraviolet behaviour and is asymptotically free. Secondly, it is widely known that general relativity is a gauge theory whose invariance group (i.e., the group of transformations that leaves the forms of all dynamical equations invariant) is the infinite dimensional group of general differentiable coordinate transformations known as the diffeomorphism group (which is a Lie pseudo-group). This means that the flat Minkowski spacetime of special relativity (whose invariance group is represented by the Poincar\'e group) is replaced by a curved Riemannian manifold. As a result, the action of the theory has to reflect the features of the geometry and, most important, must be invariant under the action of the diffeomorphism group. Therefore, for the aforementioned reasons the action for Einstein theory in principle can be given by
\begin{equation}
S_{grav} = \int {\rm d}^4 x \sqrt{-g} \left( -\dfrac{4}{\chi^2}\Lambda + \dfrac{2}{\chi ^2} R + c_1 R^2 + c_2 R_{\mu \nu} R^{\mu \nu} + \cdots \right), \label{full_action}
\end{equation}
where $\Lambda$ is the cosmological constant, $c_1$ and $c_2$ are the constants encountered in the last section and the ellipsis denote higher powers of $R$, $R_{\mu \nu}$, and $R_{\mu \nu \rho \sigma}$. At this point other physics principles can be invoked in order to simplify the the full action. For example, experiments tell us that the expansion of the universe is accelerating and hence it is ruled by a positive and non-zero cosmological constant whose value is given by $\Lambda \approx 10^{-52} \, {\rm m}^{-2}$ \cite{Perlmutter} (i.e., the de Sitter space, which will be discussed in chapter \ref{Boosted_chapter}), whereas experimental limits on $c_1$ and $c_2$ are very weak because curvatures are usually very small in the Solar System. In fact, it has been estimated that from the perihelion shift of Mercury it is only possible to get $c_1, c_2 \lesssim 10^{88}$ \cite{Stelle1978}. Moreover, higher powers of $R$ have essentially no experimental bounds. Then, it seems reasonable to put $\Lambda=0$ in (\ref{full_action}) but not $c_1$ nor $c_2$ if we bear in mind, for example, the quantum corrections described before. Thus, we can conclude at this point that instead of setting $c_1=c_2=0$, we can view (\ref{full_action}) as organized in energy expansion below a certain scale (i.e., the Planck scale) where reasonable values of $c_1$  and $c_2$ do not affect physics at low energies, reflecting what we actually observe in our Solar System. This amounts to treat gravity as an effective field theory. Since the loop expansion of Feynman diagrams represents a perturbative series in $\chi^2 \hbar$, the term $G \hbar \sim l_P$ is the fundamental parameter for the expansion underlying Eq. (\ref{full_action}). Now, by bearing in mind that $R$ involves second order derivatives of the metric tensor and that in the momentum-space $\I \D_\mu \sim p_\mu$, we can understand that the terms in the action with $n$ powers of the curvature are of order $p^{2n}$. Therefore, at low energies higher-order terms like $c_1 R^2$ and $c_2 R_{\mu \nu}R^{\mu \nu}$ are negligible compared to Einstein-Hilbert Lagrangian and hence we automatically recover general relativity and its well established predictions within Solar System. Thus, we can now realize why experimental bounds on $c_1$ and $c_2$ are so poor, since reasonable values of these constants give little effects at low energies. The most general gravitational action will have an infinite number of parameters such as $\chi^2$, $c_1$, $c_2$ and it would be possible to predict them once we have achieved the final theory of quantum gravity. Experiments will in principle make it possible to determine the final renormalized value of these constants, such as the ones in Eq. (\ref{renormalized c1-c2}), but at this stage our incomplete knowledge at low energy forces us to treat them as free parameters.   

Another advantage coming from viewing gravity as an effective field theory is represented by the fact that it can solve problems arising in $R^2$-theories. In fact such theories, despite being renormalizable, show negative squared mass states (i.e., the tachyons) and hence, most important, violate unitarity \cite{DeWitt1967b,Stelle1977}. The reason is that in these frameworks the propagator $D$ has the form
\begin{equation}
D(k) \propto \dfrac{1}{k^4 + A k^2}=\dfrac{1}{A} \left( \dfrac{1}{k^2}-\dfrac{1}{k^2+A} \right),
\end{equation}
and the negative sign in front of the second term spoils unitarity, whereas a negative constant $A$ makes a tachyonic state appear. However, in the regime of small curvatures, as explained before, $R^2$ can be considered as a small correction to vacuum Einstein theory leading only to a small modification of the vertices, not of the propagator. Practically speaking, the problems arise when terms like $R^2$ are comparable to $R$ or, in other words, in the case in which we treat $R^2$-theory like a fundamental theory when curvature is of order of the Planck mass squared (i.e., at high-energy scales). 

Since the key point of an effective field theory is represented by the separation between high-energy and leading (one-loop) long-range effects, it is important to mark a distinction in the contributions coming from heavy and massless particles. In fact, the Heisenberg Uncertainty Principle limits the range $\Delta r$ of virtual heavy particles according to
\begin{equation}
\Delta r \sim \dfrac{1}{m},
\end{equation}
which means that their contribution is local, as if they were described by a local Lagrangian. On the contrary, massless particles give non-local contributions because they can propagate for long distances. This difference between massive and massless modes can be caught through the analysis of their propagators. In fact a massive propagator can always be Taylor expanded about $q^2=0$, giving rise to a power series in the momentum $q$, i.e., 
\begin{equation}
\dfrac{1}{q^2-m^2}= -\dfrac{1}{m^2} - \dfrac{q^2}{m^4}-\dfrac{q^4}{m^6} + {\rm O}(q^6), \label{D94_16}
\end{equation}
whereas the same is obviously not true for the massless propagator, since it is proportional to $1/q^2$. Therefore, massive particles always give analytic contributions (near $q^2=0$) to Feynman diagrams, while massless ones generate non-analytic components.  Moreover, a direct application of Fourier analysis (see Appendix \ref{Appendix_Useful_integrals}) to Eq. (\ref{D94_16}) clearly shows that massive propagators produce only local interactions, because the term $1/m^2$ produces a delta function, while factors of $q^2$ are turned into derivatives. This means that massive particles yield a local low-energy Lagrangian when they are integrated out of a theory, producing shifts in the coefficients of the most general action (\ref{full_action}) which can be absorbed via a simple field re-definition (cf. Eq. (\ref{renormalized c1-c2})). On the other hand, typical non-analytical contributions coming from massless particles have the form 
\begin{gather}
\begin{aligned}
& \chi^2 q^2 \log \left( -q^2 \right), \\
& \chi^2 q^2 \sqrt{-\dfrac{m^2}{q^2}},
\end{aligned} 
\end{gather}
which for small enough $q^2$ dominate on the Fourier-transformed factors coming from (\ref{D94_16}). The quantum effects of massless modes are twofold: on the one side they produce, for example in the high-energy domain, local shifts in the parameters of the Lagrangian, which can not be absorbed by performing a renormalization procedure due to the non-analytic nature of their contributions, but on the other side their low-energy manifestation is not local, as we have seen before. Therefore, unlike massive particles, massless ones can not be integrated out of the theory but they must be included explicitly in quantum calculations. Since their low-energy couplings come directly from the Einstein-Hilbert Lagrangian, Eq. (\ref{Einstein_Lagrangian}) is sufficient to determine the leading low-energy quantum corrections occurring in physical phenomena. 

\subsection{The path integral}

The dynamical information about effective field theory can be obtained formally in the same way as in quantum field theory. Then, with the help of (\ref{Z_integral}) we define the generating functional
\begin{equation}
Z\left[ J, \bar{g} \right]=\int \mathcal{D}[\phi,h_{\mu \nu}] \E^{\I \tilde{S} \left(\phi,\bar{g},h,J \right)}, \label{D94_18}
\end{equation}
where 
\begin{equation}
\tilde{S}=\int {\rm d}^4 x  \; \mathcal{\tilde{L}},
\end{equation} 
is the most general covariant action which contains, as explained above, an infinite number of free parameters such as $\chi$, $c_1$, and $c_2$, taking into account the effects of the high-energy part of the true fundamental theory. The coefficients $c_1$, $c_2, \dots$ play the role of effective couplings in the action. On the other hand, as we know the low-energy degrees of freedom can not be ignored and therefore are explicitly considered in the path integral. Moreover, issues concerning the functional measure for high values of $h_{\mu \nu}$ can be completely disregarded, since we will focus only on low-energy configurations of the perturbed metric and we are free to employ any measure and regularization scheme, provided it does not violate unitarity. Another great advantage characterizing this framework is represented by the fact that (\ref{D94_18}) has a well-behaved perturbative expansion, because the coupling of the low-energy fluctuations $h_{\mu \nu}$ is weak. We also stress the fact that we have implicitly assumed that the only low-energy particles surviving the full gravity theory are the gravitons. This means that other possible massless particles, if present, must be included. 

It is possible to perform an expansion of $\mathcal{\tilde{L}}$, which in the most general case contains both a gravitational and a matter sector, in powers of the momentum transferred, in analogy with the quantum field theory case. However, at low energies only the minimally coupled Lagrangians are important. By recalling that the derivatives of the massless field essentially go as powers of momentum while those of massive field generate powers of the interacting mass, we thus have \cite{D94b}
\begin{equation}
\mathcal{\tilde{L}} =\sqrt{-\bar{g}} \left( \mathcal{\tilde{L}}_{grav} + \mathcal{\tilde{L}}_m \right),
\end{equation}
 with
 \begin{equation}
 \mathcal{\tilde{L}}_{grav}= \mathcal{\tilde{L}}_{\; grav}^{(0)}+\mathcal{\tilde{L}}_{\; grav}^{(2)}+\mathcal{\tilde{L}}_{\, grav}^{(4)}+ {\rm O}(p^6),
 \end{equation}
\begin{equation}
 \mathcal{\tilde{L}}_{m}= \mathcal{\tilde{L}}_{m}^{\; (0)}+\mathcal{\tilde{L}}_{m}^{\; (2)}+ {\rm O}(p^4),
 \end{equation}
where  the gravitational part reads as
\begin{equation}
\mathcal{\tilde{L}}_{\; grav}^{(0)}= \Lambda,
\end{equation}
\begin{equation}
\mathcal{\tilde{L}}_{\; grav}^{(2)}= \dfrac{2}{\chi^2} \bar{R},
\end{equation}
\begin{equation}
\mathcal{\tilde{L}}_{\; grav}^{(4)}= c_1 \bar{R}^2+c_2 \bar{R}_{\mu \nu}\bar{R}^{\mu \nu},
\end{equation}
whereas the matter terms are given by
\begin{equation}
 \mathcal{\tilde{L}}_{m}^{\; (0)}= \dfrac{1}{2} \left( \bar{g}^{\mu \nu} \D_\mu \phi \D_\nu \phi - m^2 \phi^2 \right),
\end{equation}
\begin{equation}
 \mathcal{\tilde{L}}_{m}^{\; (2)}= d_1\bar{R}^{\mu \nu} \D_\mu \phi \D_\nu \phi + \bar{R} \left( d_2 \D_\mu \phi \D^\mu \phi + d_3 m^2 \phi^2 \right),
\end{equation}
the parameters $c_1$, $c_2$, $d_1$, $d_2$, $d_3, \dots$ being scale-dependent coupling constants to be measured experimentally. As pointed out before, all divergences coming from the lowest-order Lagrangian are thus absorbed into the effective action, leaving us a with a finite one-loop order theory characterized by a finite number of parameters whose renormalized values must be determined experimentally. Moreover, the discussion about divergences of the previous sections has  demonstrated that loops involving low-order terms in the energy expansion always demand a renormalization procedure involving the coefficients appearing at higher order. We have already mentioned the issues concerning the experimental bounds on $\Lambda$, $c_1$, and $c_2$. About the matter Lagrangian constants $d_i$, they have the dimension of an inverse mass squared and in the presence of point particles having only gravitational interactions it has been shown that $d_i \approx {\rm O}(1/M_{P}^{2})$ \cite{D94b}.

In conclusion, we have explained so far the main properties about effective field theories of gravity because they represent the starting point of the following sections of this thesis. We conclude by observing that the great obstacle towards a full phenomenological implementation of such theories is represented by the difficulties arising in experimentally measuring the unknown coefficients of the full action (\ref{full_action}). Despite that, at this point it should be clear that the so-called leading quantum corrections are both independent of these unknown parameters and dominant at large distances over the other one-loop gravitational effects. The aforementioned leading one-loop quantum corrections will represent the heart of the first part of this thesis, dedicated to the low-energy domain of the unknown theory of quantum gravity.

\section{The leading quantum corrections to the Newtonian potential}

We have seen in the previous section how the effective field theory approach makes it possible to ignore renormalization difficulties of general relativity in the low-energy domain, because all emerging divergencies can be easily absorbed in the phenomenological constants characterizing the effective action (\ref{full_action}). Therefore, physical quantities such as the leading quantum corrections to the Newtonian potential can be calculated within this framework by exploiting, as pointed out before, the non-analytical part of the one-loop amplitude generated by the (lowest-order) Einstein-Hilbert Lagrangian. Moreover, since higher-derivative terms of (\ref{full_action}) only affect the analytical part of the one-loop amplitude, they will not contribute to the potential and hence all divergencies arising at one loop (which should be eliminated by renormalizing the parameters of these higher-derivative terms) do not constitute a problem.

\begin{figure} [htbp] 
\centering
\includegraphics[scale=0.7]{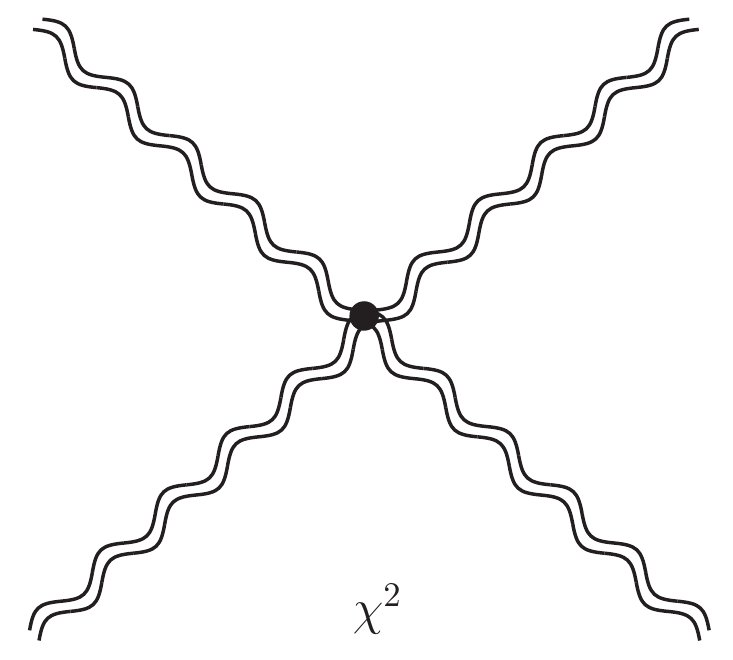}
\caption[The four-graviton vertex]{The four-graviton vertex gives a contribution of order $\chi^2 q^2$ to the scattering amplitude.} \label{4-graviton.pdf}
\end{figure}

\begin{figure} [htbp] 
\centering
\includegraphics[scale=0.7]{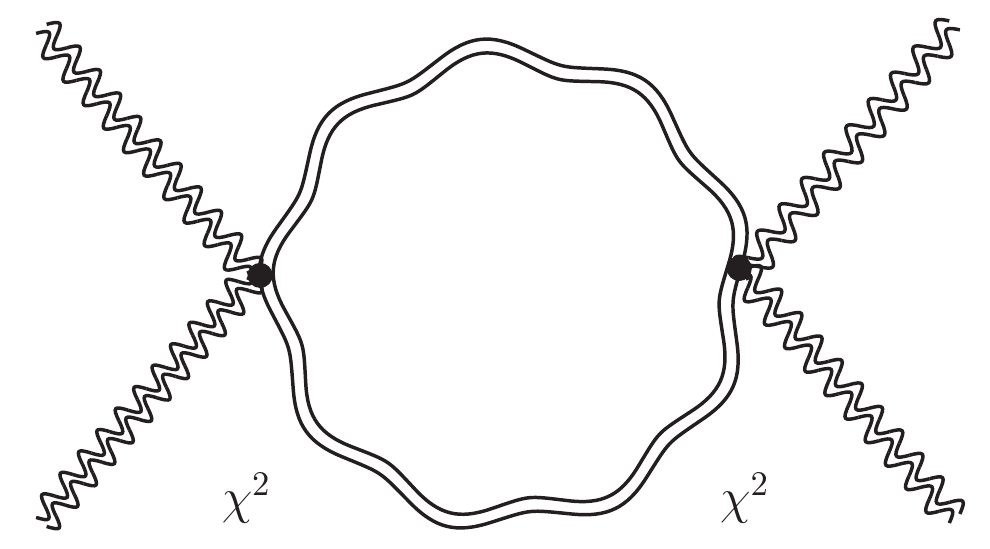}
\caption[The one-loop correction to the four-graviton vertex]{The one-loop correction to the four-graviton vertex. The contribution to the scattering amplitude is of order $\chi^4 q^4$.} \label{4-graviton-one-loop.pdf}
\end{figure}

For point-sources separated by a large distance $r$ we expect the corrections to be weak and thus they can be determined by employing perturbation theory about flat space, which we have presented in Sec. \ref{Sec. Quantization GR}. The strength of gravitational interactions at large separation is ruled by two dimensionless parameters that suggest themselves on dimensional grounds \cite{D94, D-Torma}. In effective theory the expansion parameter for quantum corrections is given by $\chi^2 q^2 \sim G/r^2$, such that at low energies/long distances higher-order loops effects are suppressed with respect to tree-diagrams and low-order loops and hence we can obtain predictions to a given order with a finite amount of calculation. For example, the three-graviton vertex (\ref{three-graviton_vertex}) goes as $\chi q^2$, while the four-graviton vertex and its one-loop correction (Figs. \ref{4-graviton.pdf} and \ref{4-graviton-one-loop.pdf}) bring a contribution of the order $\chi^2 q^2$ and $\chi^4 q^4$, respectively.
General relativity also contains the classical expansion parameter $\chi^2 m q \sim G m /r$ due to the non-linearities of the classical theory which is connected to the non-analytic terms of the form $Gq^2\sqrt{-m^2/q^2}$. However, also an expansion parameter of the form $\chi^2 m^2 \sim G m^2$ arising in the interaction with matter seems to be present. This fact can be seen directly by using Eqs. (\ref{2scalar-1graviton_vertex}) and (\ref{2scalars-2gravitons vertex}), which show that the two scalar-one graviton and the two scalar-two graviton vertices go as $\chi m^2$ and $\chi^2 m^2$, respectively. If an expansion ruled by these terms was really present, it would represent a disaster for two reasons. In fact, first of all the factor $G m^2 $ expressed as units of the Planck mass can be a very large number in all those situations where $m \gg M_p$ (as for example in the case of the Sun or the Earth); in addition if we temporarily restore $\hbar$ and $c$ this dimensionless combination goes as $G m^2/\hbar c$ and hence we would end up with the paradoxical situation where the classical limit $\hbar \rightarrow 0$ would make quantum effects dominate over classical ones. Anyway, it is possible to demonstrate that this problem is just a gauge artifact because a number of cancellations during the calculation of Feynman diagrams occurs which remove this undesirable parameter, saving in this way the utility of the energy expansion \cite{D-Torma}. Therefore, the aforementioned arguments clearly prove what we have anticipated before, i.e., the leading one-loop long-distance quantum corrections to the Newtonian potential can only be led by two dimensionless parameters having the form $G m /r c^2$ and $G \hbar/r^2 c^3=l_{P}^2/r^2$. Both of them go to zero for large distances, the first controlling the size of relativistic (post-Newtonian) corrections whereas the second the quantum ones. Thence, at one-loop level the quantum corrected Newtonian potential between two bodies of masses $m_A$ and $m_B$ will read as \cite{D94,D03,BE14a}
\begin{equation}
V_Q(r)=-{G m_{A}m_{B}\over r} \left[ 1+\left( {k_{1}\over r}+{k_{2}\over r^{2}}  \right) +{\rm O}(G^{2})  \right],
\label{1.2b}
\end{equation}
where
\begin{equation}
k_{1} \equiv \kappa_{1}{G (m_{A}+m_{B})\over c^{2}} = \kappa_1 \left(R_A+R_B \right), \label{1.3b}
\end{equation}
\begin{equation}
k_{2} \equiv \kappa_{2}{G \hbar \over c^{3}}= \kappa_{2}(l_{P})^{2}, \label{1.4b}
\end{equation}
where $R_A$ and $R_B$ are the gravitational radii of the bodies $A$ and $B$, respectively. The numbers $\kappa_1$ and $\kappa_2$ can be worked out only by a direct calculation of Feynman diagrams. We will see that they strongly depend on the different definitions we will adopt of the potential. Our first task then will be the description of such definitions.

\subsection{Three ways to define a potential} \label{three ways_sec}

\begin{figure}
\centering
\includegraphics[scale=0.7]{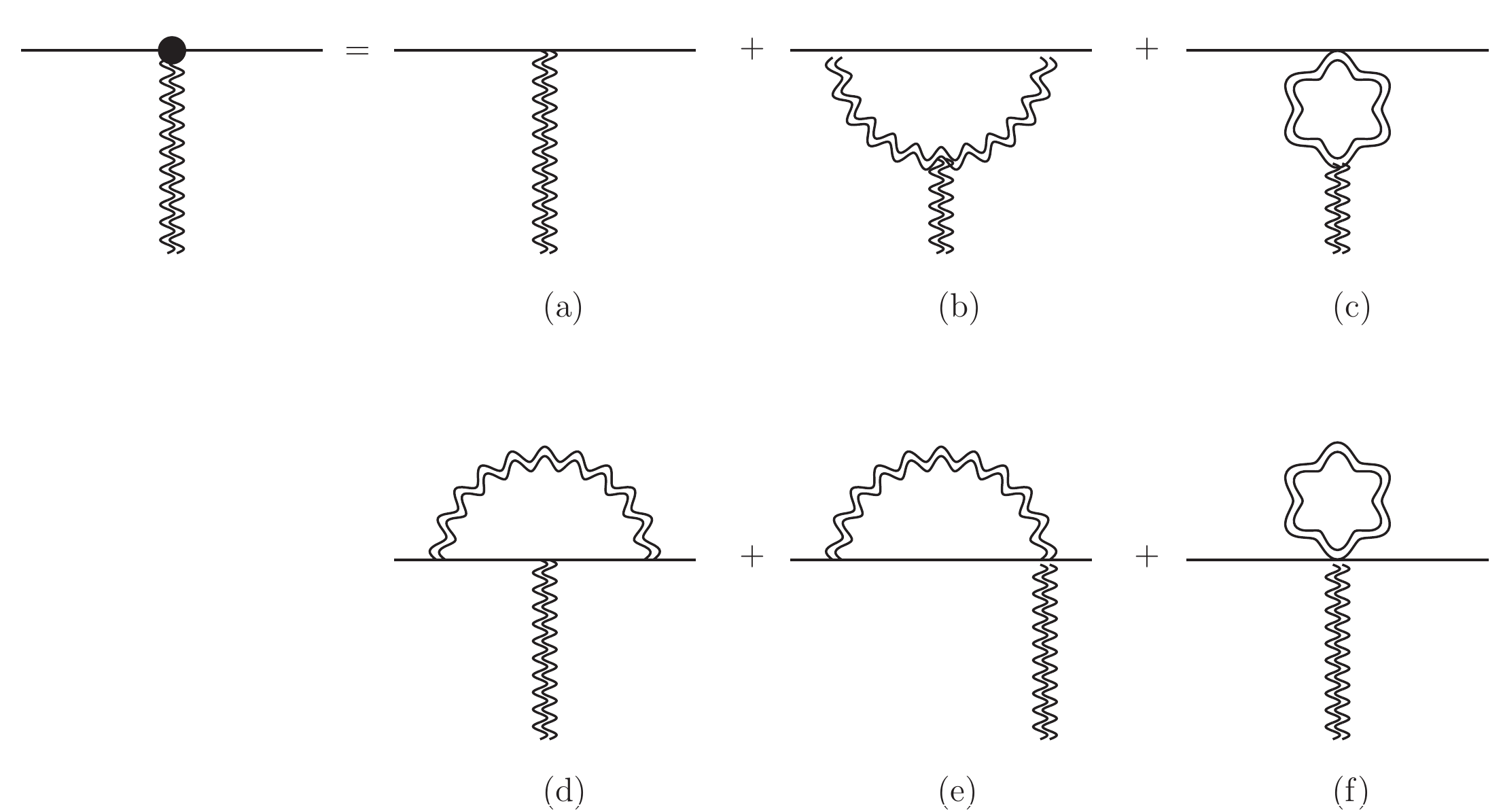}
\caption[Vertex correction diagrams]{The Feynman diagrams involved in the vertex correction. Graphs ({\rm d}), ({\rm e}), and ({\rm f}) do not have any non-analytic terms. }
\label{Fig1D94a.pdf}
\end{figure}

\begin{figure}
\centering
\includegraphics[scale=0.7]{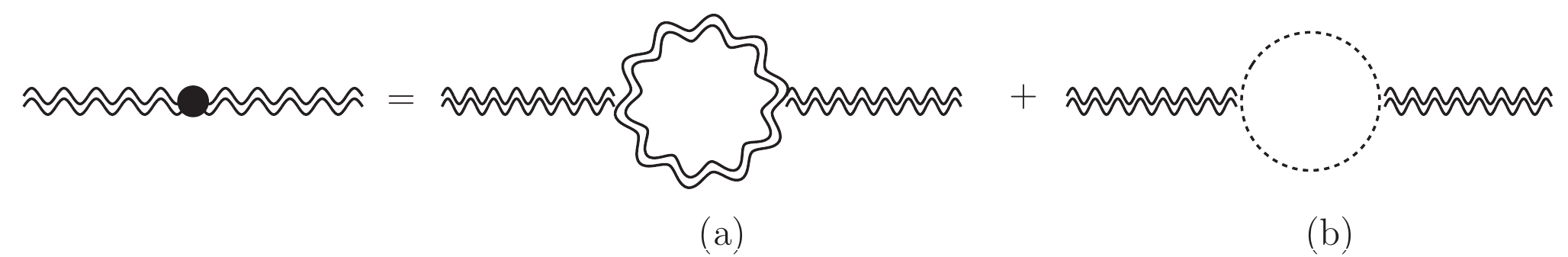}
\caption[Graviton vacuum polarization diagrams]{Graviton vacuum polarization diagrams. The dotted lines indicate the ghost fields. We do not consider any heavy particles loops because they give analytical contributions to the potential.}
\label{Fig1D94b.pdf}
\end{figure}

\begin{figure}
\centering
\includegraphics[scale=0.7]{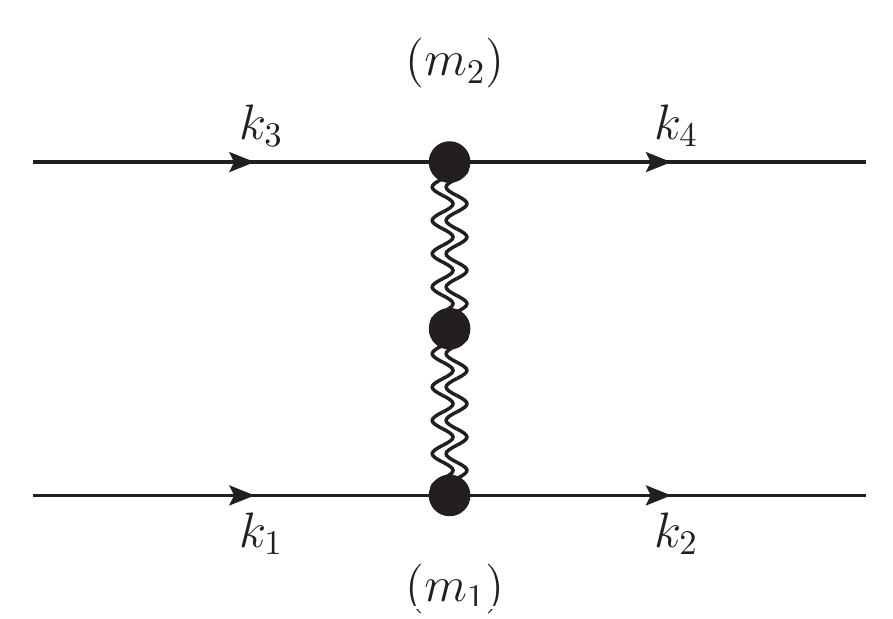} 
\caption[The set of corrections included in the one-particle reducible potential]{The set of corrections involved in the one-particle reducible potential. Dots indicate radiative corrections. The four-momentum transferred $q$ is such that $k_1-k_2=k_4-k_3=q$.}
\label{Fig2D94.pdf}
\end{figure}

The definition of a potential in a relativistic quantum field theory such as general relativity is not obvious and a lot of approaches have been discussed in the literature. Among all possible choices, we are mainly interested in three of them. For systems near the flat-space limit a natural definition of the interaction potential between slowly-moving point particles would involve their scattering amplitudes. One could for example define the potential in terms of the one-particle reducible part of the scattering amplitude, as is common done for QED and QCD. This was the choice adopted in Refs. \cite{D94,D94b}, which leads to the definition of what we will call one-particle reducible potential. The logic behind this method consists in the fact that graviton exchange dominates long-distance interactions due to the form of its propagator, which is proportional to $1/q^2$. Moreover, it has the advantage of giving rise to physically meaningful results, because the leading quantum corrections occurring in the potential can be interpreted as modifications to the Schwarzschild, Kerr-Newman \cite{BDH2003}, and Reisner-Nordstr\"om metrics \cite{DHGK2002}. Anyway, one-particle reducible graphs are not observable and so need not to form a gauge-invariant subset. As a consequence, also the potential shares the same property. The Feynman diagrams involved in the calculation are given in Figs. \ref{Fig1D94a.pdf} and \ref{Fig1D94b.pdf}, i.e., one-loop radiative corrections to the gravitational vertex and vacuum polarization graphs. The potential is eventually given by considering the Fourier transformation of the non-relativistic limit of (the non-analytical part of) the set of one-particle reducible graphs of Fig. \ref{Fig2D94.pdf}.

The great limit of the previous definition is represented by the lack of gauge invariance. Gauge theories like QCD allow a gauge invariant Wilson loop definition of the potential, but an equivalent construction in quantum gravity turns out to be quite cumbersome. Nevertheless, an approach based on the full one-loop scattering amplitude seems to represent the simplest way to define a gauge-invariant (non-relativistic) potential, as suggested in Ref. \cite{D03}. The general form of the scattering amplitude in the mixed gravity-scalar theory reads as  
\begin{equation}
\mathcal{M}(q)= \mathcal{A}(q)_{\rm analytic} + \mathcal{A}(q)_{\rm non-analytic} \equiv \mathcal{A}(q) + \mathcal{A}^\prime(q),
\end{equation}
$q$ being the transferred momentum, and 
\begin{equation}
\mathcal{A}(q) \sim A + B q^2 + {\rm O}(q^4),
\end{equation}
\begin{equation}
\mathcal{A}^\prime(q) \sim \chi^4 \dfrac{1}{q^2} + \kappa^\prime_1 \chi^4 \sqrt{-\dfrac{m^2}{q^2}}+ \kappa^\prime_2 \chi^4 \log \left( -q^2 \right) + {\rm (beyond\;one-loop\;contributions)},
\end{equation}
where $\kappa^\prime_1$ and $\kappa^\prime_2$ reduce to $\kappa_1$ and $\kappa_2$ (cf. Eqs. (\ref{1.2b})--(\ref{1.4b})), respectively, in the non-relativistic limit. $\mathcal{A}(q)$ is an analytic function of $q^2$ near $q^2=0$ giving contributions to the potential that in the coordinate-space turn out to be local, i.e., proportional to Dirac-delta function or its derivatives, and hence not dominant at low energies. All one-loop ultraviolet divergences which can be absorbed by renormalizing the couplings of the higher-derivative terms of (\ref{full_action}) are included in $\mathcal{A}(q)$, since on general grounds their contributions are polynomials in the momenta. By recalling that the $S$ matrix can be written as $S=\mathbb{1} + {\rm i} \mathcal{T}$, the full scattering amplitude $\mathcal{M}(q)$ can be related to the expectation value of the transition matrix $\mathcal{T}$ through the relation
\begin{equation}
\langle p^\prime_1,p^\prime_2,\dots \vert {\rm i}\mathcal{T} \vert p_1,p_2,\dots \rangle = \left( 2 \pi \right)^4 \delta^{(4)} (p-p^\prime) \left[ {\rm i} \mathcal{M}(q) \right], \label{scattering_amplitude}
\end{equation}
where $p$ and $p^\prime$ are the ingoing and outgoing four-momentum, respectively. The key point of this second approach consists in the fact that it is assumed that in the non-relativistic limit the matrix elements of the interaction potential within single-particle states reproduce the full field-theoretical amplitude of the scattering process according to \cite{D03}
\begin{equation}
\langle p^\prime_1,p^\prime_2 \vert {\rm i}\mathcal{T} \vert p_1,p_2 \rangle = -{\rm i }\left( 2 \pi \right)\delta (E-E^\prime) \langle p^\prime_1,p^\prime_2 \vert \tilde{V}(\mathbf{q}) \vert p_1,p_2 \rangle,
\end{equation}  
where $E-E^\prime$ is the energy difference between ingoing and outgoing states and $\tilde{V}(\mathbf{q})$ indicates the non-relativistic potential in momentum-space.
A comparison between the last two relations shows that it is possible to obtain the non-relativistic potential $V(\mathbf{r})$ in the coordinate-space by performing the Fourier transformation
\begin{equation}
V(\mathbf{r})= \int \dfrac{{\rm d}^3 q}{(2 \pi)^3}\; \E^{\I {\bf q} \cdot {\bf r}}\, \tilde{V}(\mathbf{q}) = \dfrac{1}{N} \int \dfrac{{\rm d}^3 q}{(2 \pi)^3}\; \E^{\I {\bf q} \cdot {\bf r}} \mathcal{M}(\bf{q}), \label{overallN}
\end{equation}
where by writing $\mathcal{M}(\bf{q})$ we have underlined that the non-relativistic limit of $\mathcal{M}(q)$ must be taken into account.
The overall normalization factor $N$ depends on the conventions used for the normalization of the initial and final states and it is chosen in such a way that it yields the correct Newtonian potential in the classical (i.e., tree-level) limit. It turns out that $N=1/(2 m_1 2 m_2)$ \cite{D03}. Recalling that in the low-energy domain only the non-analytical part of the scattering amplitude will give the most significant contribution, the interaction potential between two bodies of masses $m_1$ and $m_2$ can be obtained by substituting in Eq. (\ref{overallN}) $\mathcal{M}({\bf q})$ with the non-relativistic amplitude $\mathcal{A}^\prime({\bf q})$ and by putting $N=1/(2 m_1 2 m_2)$, i.e.,
\begin{equation}
V(\mathbf{r})= \dfrac{1}{2m_1}\dfrac{1}{2m_2}\int \dfrac{{\rm d}^3 q}{(2 \pi)^3}\; \E^{\I {\bf q} \cdot {\bf r}}\mathcal{A}^\prime({\bf q}). \label{D03_4}
\end{equation}
The set of Feynman diagrams contributing to $\mathcal{A}^\prime({\bf q})$ will be analysed in Sec. \ref{Sec. scattering-bound-state potential}. The potential obtained through this definition will be referred to as scattering potential.

An alternative path in the context of scattering potential can be followed. It consists in subtracting off the second-order Born approximation coming from the scattering theory of quantum mechanics according to
\begin{equation}
\begin{split}
\I \langle f \vert \mathcal{T} \vert i \rangle & = -\I (2 \pi) \delta (E-E^\prime) \biggl [ \langle f \vert \tilde{V}_{\rm bs}(\mathbf{q}) \vert i \rangle  \\
& + \sum_n \dfrac{\langle f \vert \tilde{V}_{\rm bs}(\mathbf{q})\vert n \rangle   \langle n \vert \tilde{V}_{\rm bs}(\mathbf{q}) \vert i \rangle }{E-E_n + \I \epsilon} + \dots \biggr],
\end{split}
\end{equation}
where $\tilde{V}_{\rm bs}(\mathbf{q})$ is the non-relativistic bound-state potential used in quantum mechanics evaluated in the momentum-space. Its expression in the coordinate-space is obtained by means of Eq. (\ref{D03_4}), but it turns out that it can be linked to the scattering potential by
\begin{equation}
V_{\rm bs} (r)= V(r) - \dfrac{G m_1 m_2}{r} \left[ -\dfrac{7}{2} \dfrac{G \left(m_1+m_2 \right)}{c^2 r} \right]. \label{D03_6}
\end{equation} 
The bound state potential represents the only one choice having a direct physical meaning in celestial mechanics, because in the classical limit $\hbar \rightarrow 0$ it is able to reproduce the Hamiltonian describing within the context of Einstein theory the perihelion shift of Mercury \cite{Iwasaki}. This issue represents an important point towards the construction of a consistent and correct quantum theory of gravitation, since, before predicting new quantum effects, we should first make sure that the quantum theory describes correctly classical phenomena. Therefore, the fact that the bound state potential foretells properly one of the three classical tests of general relativity surely represents an important aspect to be taken into account.

\subsection{One-particle reducible potential}

We are now ready to find out the features of the quantum corrected potential (\ref{1.2b}). We start with the one-particle reducible potential. An essential tool within this approach is represented by the concept of the form factors. By choosing the normalization convention 
\begin{equation}
\langle k^\prime \vert k \rangle = \left( 2 \pi \right)^3 2 E \, \delta^{(3)} (k-k^\prime),
\end{equation}
the general gravity-matter vertex can be written in terms of the two form factors $F_1(q^2)$ and $F_2(q^2)$ through the on-shell matrix elements of the energy-momentum tensor as
\begin{equation}
\mathcal{C}_{\mu \nu}(q)=\langle k^\prime \vert T_{\mu \nu} \vert k \rangle = F_1(q^2) \left( k_\mu k^{\prime}_{\nu}+ k^{\prime}_{\mu} k_{\nu} + q^2 \dfrac{\eta_{\mu \nu}}{2} \right) + F_2(q^2) \left(q_\mu q_\nu - \eta_{\mu \nu}q^2 \right), 
\end{equation} 
with the normalization condition $F_1(0)=1$. As an example, the two scalar-one graviton vertex (\ref{2scalar-1graviton_vertex}) reads (on-shell) as
\begin{equation}
\tau_{\mu \nu} (p,p^\prime,m) = \dfrac{-{\rm i} \chi}{2} \; \mathcal{C}_{0 \mu \nu} (p,p^\prime,m),
\end{equation}
with
\begin{equation}
\mathcal{C}_{0 \mu \nu} (p,p^\prime,m)= \left( p_\mu p^{\prime}_{\nu}+ p^{\prime}_{\mu} p_{\nu} + q^2 \dfrac{\eta_{\mu \nu}}{2} \right),
\end{equation}
the subscript $0$ indicating that no radiative corrections have been considered and where the relation $p^\prime \cdot p = m^2 - q^2/2$ coming form the momenta conservation $q+p^\prime=p$ (see Appendix \ref{Appendix_Useful_integrals}) and the on-shell condition $p^{\prime^2} = p^2 = m^2$ has been exploited. 

The energy expansion corresponds to an expansion of the form factors in powers of $q^2$, i.e., \cite{D94b}
\begin{equation}
F_1(q^2) = 1+d_1 q^2 + \chi^2 q^2 \left[ \ell_1 + \ell_2 \log \left( \dfrac{-q^2}{\mu^2} \right) + \ell_3 \sqrt{\dfrac{m^2}{-q^2}} \; \right] + \dots,
\end{equation} 
\begin{equation}
F_2(q^2)= - 4 \left(d_2 + d_3 \right) m^2 + \chi^2 m^2 \left[ \ell_4 + \ell_5 \log \left( \dfrac{-q^2}{\mu^2} \right) + \ell_6 \sqrt{\dfrac{m^2}{-q^2}} \; \right] + \dots,
\end{equation}
where ellipses denote higher powers of $q^2$ and the constant $\mu^2$ is a mass parameter. Note that no corrections of the form $\chi^2 m^2$ can be present in $F_1(q^2)$ because of the normalization condition $F_1(0)=1$. $d_i$ ($i=1,2,3$) represents the unknown effects of the true high-energy theory, whereas the coefficients $\ell_i$ ($i=1,2,\dots,6$) are related to the computation of loop diagrams. In particular, $\ell_1$ and $\ell_4$ come from the high-energy end of loop integrals and are in general divergent, while $\ell_2$, $\ell_3$, $\ell_5$, and $\ell_6$ must be finite. For $q^2 >0$ (i.e., a time-like vector) the non-analytical terms $\log \left( -q^2 \right)$ and $\sqrt{1/(-q^2)}$ pick up an imaginary part corresponding to physical (on-shell) intermediate states as described by unitarity. Since almost certainly loop integrals concerning the high-energy domain are not well represented by low-energy vertices and low-energy degrees of freedom, we should combine $\ell_1$ and $\ell_4$ with $d_i$ ($i=1,2,3$) in order to define the renormalized values
\begin{equation}
d^{(r)}_1(\mu^2) = d_1 + \chi^2 \ell_1,
\end{equation}
\begin{equation}
d^{(r)}_2(\mu^2)+ d^{(r)}_3(\mu^2) = d_2+d_3 - \chi^2 \dfrac{\ell_4}{4},
\end{equation}
which are (in principle) measurable. The dependence on $\mu^2$ indicates that the measured values depend on the choice of $\mu^2$ occurring in the logarithms, although all physical quantities are independent of such a parameter. 

\begin{figure}
\centering
\includegraphics[scale=0.7]{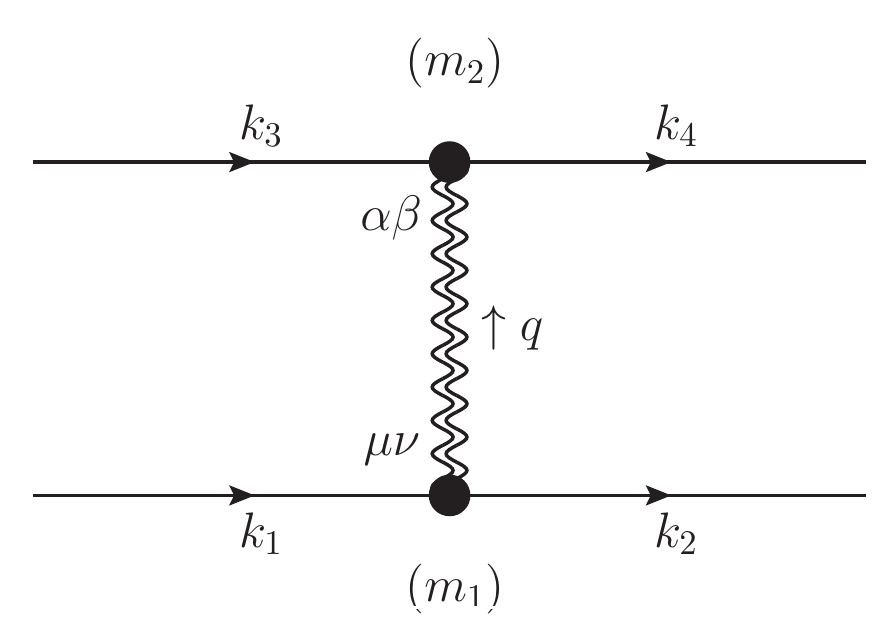} 
\caption[The gravitational interaction of two particles without vacuum polarization]{The gravitational interaction of two particles obtained from Fig. \ref{Fig2D94.pdf} without considering vacuum polarization diagrams.}\label{Fig2D94b.pdf}
\end{figure}

The gravitational interaction of two particles leading to the one-particle reducible potential is obtained, as we said before, by combining the vertices with the propagators as shown in Fig. \ref {Fig2D94.pdf}. Disregarding for a while vacuum polarization diagrams, we have (Fig. \ref{Fig2D94b.pdf})
\begin{equation}
{\rm i} \mathcal{M}(q)= \dfrac{\chi^2}{4}  \; \mathcal{C}_{\mu \nu} (-q) \; {\rm i} \dfrac{\mathcal{P}^{\mu \nu \alpha \beta}}{q^2} \; \mathcal{C}_{\alpha \beta} (q),
\end{equation}
therefore, on considering the non-relativistic limit, the scattering amplitude becomes \cite{D94b}
\begin{equation}
\begin{split}
{\rm i} \mathcal{M}({\bf q}) & \propto \chi^2 m_1 m_2 \biggl \{ \dfrac{1}{\bf{q}^2} + 2 (d_1-2 d_2 - 2d_3) \\
& + \chi^2 \Big[ (2 \ell_1-\ell_4) + (2 \ell_2 - \ell_5) \log \left( \dfrac{-{\bf q}^2}{\mu^2} \right) + (2 \ell_3 - \ell_6)\sqrt{m^2/(-{\bf q}^2)} \; \Big] \biggr \}.
\end{split} \label{D94_58}
\end{equation}
As we know, linear analytic terms in $\bf{q}^2$ lead to Dirac-delta interaction, whereas the non-analytic contributions represent the source of the power law behaviour underlying the long-distance corrections to the Newtonian potential. A similar result holds also for vacuum polarization diagrams, Fig. \ref{Fig1D94b.pdf}. By temporarily suppressing, for the sake of simplicity, Lorentz indices and constants of the order of unity, the generic form of the vacuum polarization tensor follows directly from dimensional counting \cite{D94b}
\begin{equation}
\Pi(q) \sim \chi^2 q^4 \left[ c_1 + c_2 + \ell_7 + \ell_8 \log \left( - q^2 \right) + \dots \right], \label{D94_61}
\end{equation}
such that the graviton propagator $D(q)$ can be written as
\begin{equation}
D(q) \sim\dfrac{1}{q^2}+ \dfrac{1}{q^2} \Pi(q) \dfrac{1}{q^2} + \dots = \dfrac{1}{q^2} + \chi^2 \left[ c_1 + c_2 + \ell_7 + \ell_8 \log \left( - q^2 \right) + \dots \right],
\end{equation}
$c_1$ and $c_2$ being the high-energy unknown parameters appearing in the Lagrangian (\ref{full_action}), $\ell_7$ and $\ell_8$ constants calculable from the vacuum polarization diagrams. Like before, $\ell_7$ is divergent and the combination $(c_1+c_2+\ell_7)$ forms a renormalized parameter. Again the constants in the graviton propagator lead to a $\delta^{(3)}(\bf{x})$-interaction, while the logarithm to a long-rage $1/r^3$- (quantum) effect. 

\begin{figure}
\centering
\includegraphics[scale=0.7]{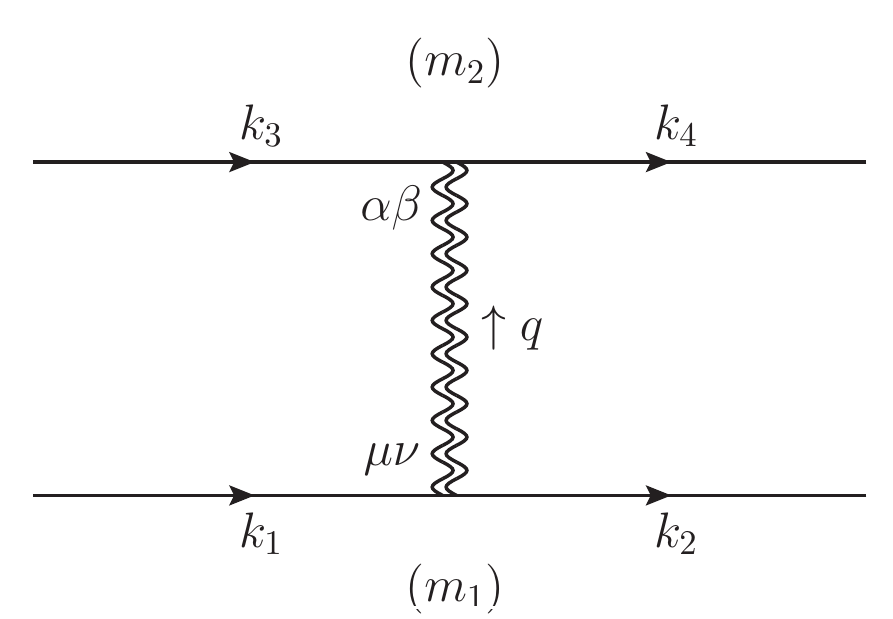} 
\caption[The Newtonian potential]{The tree-diagram giving the Newtonian potential.}\label{Newton.pdf}
\end{figure}
At the lowest order (i.e., at tree-level) the graviton exchange yields the Newtonian potential. In fact, from Fig. \ref{Newton.pdf} we have
\begin{equation}
{\rm i} \mathcal{M}(q)= -\dfrac{\chi^2}{4}  \; \mathcal{C}_{0 \mu \nu} (-q) \; {\rm i} \dfrac{\mathcal{P}^{\mu \nu \alpha \beta}}{q^2} \; \mathcal{C}_{0 \alpha \beta} (q), \label{D94_46}
\end{equation}
whose non-relativistic limit amounts to consider the relations
\begin{equation}
q^{\mu} = (0, \bf{q})
\end{equation}
\begin{equation}
k^{\mu}= (m, \bf{0}),
\end{equation}
and
\begin{equation}
\dfrac{1}{2m}\mathcal{C}_{0 \mu \nu} (q) = m \delta_{\mu 0} \delta_{\nu 0},
\end{equation}
where the factor $1/2m$ takes into account the covariant normalization. Then, it is easy to show \cite{D94b} that the Fourier transformation of the non-relativistic limit of (\ref{D94_46}) leads to the Newtonian potential
\begin{equation}
V(r)= - G \dfrac{m_1 m_2}{r},
\end{equation}
once the analytic components have been separated out. We will give the details of this calculation in the next section.

As we have already said, we are interested only in the non-analytic components of Feynman diagrams. The following example shows how this fact can simplify the calculations somewhat. Suppose we have a diagram involving the scalar particle momenta $k$ (ingoing) and $k^\prime$ (outgoing) and the graviton momenta $l$ and $q$. The conservation of momenta is such that $k - k^\prime = q$, while $l$ represents a loop momentum. All loop integrals involving any factor of $l^2$ or $(q \pm l )^2$ at the numerator gives no non-analytical contribution. In fact, consider for example the integral
\begin{equation}
\int \dfrac{{\rm d}^4 l}{(2 \pi)^4} \dfrac{l^2}{l^2 \left( q-l \right)^2 \left[ \left( k- l \right)^2 -m^2\right]},
\end{equation}
which can be written as 
\begin{equation}
\begin{split}
\int \dfrac{{\rm d}^4 l}{(2 \pi)^4} \dfrac{1}{\left( q-l \right)^2 \left[ \left( k - l \right)^2 -m^2\right]}= \int \dfrac{{\rm d}^4 l^\prime}{(2 \pi)^4} \dfrac{1}{l^{\prime^2} \left[ \left( k^\prime+ l^\prime \right)^2 -m^2\right]}= f(k^2),
\end{split}
\end{equation}
once both the shift $l^\prime=q-l$ and the momentum conservation have been considered. As we can see, we have obtained a function of $k^2$ only which does not depend on $q^2$. Therefore, this simple example allows us to conclude that any integral whose integrand would vanish if the gravitons were on-shell leads to contributions which are not non-analytical. As a result, all the components in the curly brackets of the three-graviton vertex (\ref{three-graviton_vertex}) do not contribute to our calculations and hence can be dropped (see Appendix \ref{Appendix_Useful_integrals} for further details). Another kind of simplification is given by exploiting the tensor relations
\begin{equation}
\mathcal{P}_{\alpha \beta}^{\; \; \; \; \lambda \kappa} \mathcal{P}_{\lambda \kappa \gamma \delta} = I_{\alpha \beta \gamma \delta}, \label{tensor_relation1}
\end{equation} 
\begin{equation}
I_{\alpha \beta \gamma \delta} \; t^{ \gamma \delta}= t_{\alpha \beta}, \; \; \; \;  \forall \; t_{\alpha \beta} \; {\rm symmetric}, \label{tensor_relation2}
\end{equation}
where $I_{\alpha \beta \gamma \delta} $ is defined in Eq. (\ref{I_tens}). In particular we have that
\begin{equation}
\mathcal{P}_{\alpha \beta \gamma \delta} \; t^{ \gamma \delta}= t_{\alpha \beta}-\dfrac{1}{2} \eta_{\alpha \beta} t^{\lambda}_{\; \lambda}, \; \; \; \;  \forall \; t_{\alpha \beta} \; {\rm symmetric}, 
\end{equation}
and hence for example
\begin{equation}
\mathcal{P}_{\mu \nu \alpha \beta} \; \tau^{\alpha \beta}(k,k^\prime,m)= \dfrac{-\I \chi}{2} \left(k_{\mu} k^{\prime}_{\nu} + k_{\nu} k^{\prime}_{\mu} - \eta_{\mu \nu} m^2 \right). \label{tensor_relation3}
\end{equation}
Other useful relations are (cf. Eq. (\ref{2scalars-2gravitons vertex}))
\begin{equation}
\mathcal{P}_{\alpha \beta \lambda \kappa} \mathcal{P}_{\gamma \delta \rho \sigma} \; \tau^{\lambda \kappa \rho \sigma}(k,k^\prime,m) = \tau_{\alpha \beta \gamma \delta}(k,k^\prime,m), 
\end{equation}
and 
\begin{equation}
\mathcal{P}_{\alpha \beta}^{\; \; \;  \;  \lambda \kappa} \; \tau^{\mu \nu}_{\; \; \; \lambda \kappa \gamma \delta}(k,q)  \;\mathcal{P}^{\gamma \delta}_{\; \; \; \,  \rho \sigma} = \tau^{\mu \nu}_{\; \; \; \alpha \beta \rho \sigma}(k,q), 
\end{equation}
which is valid {\it only} for terms in Eq. (\ref{three-graviton_vertex}) leading to non-analytic corrections. 

\begin{figure}
\centering
\includegraphics[scale=0.7]{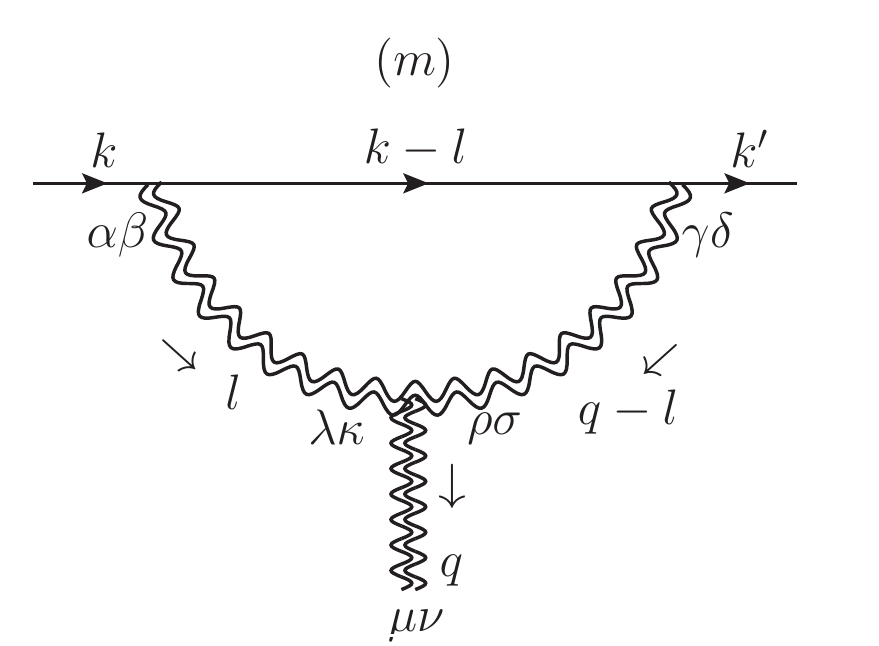} 
\caption[Vertex correction diagram 1]{The vertex correction diagram of Fig. \ref{Fig1D94a.pdf}b in detail. The momenta are such that $k-k^\prime=q$.}\label{D94b.pdf}
\end{figure}

\begin{figure}
\centering
\includegraphics[scale=0.7]{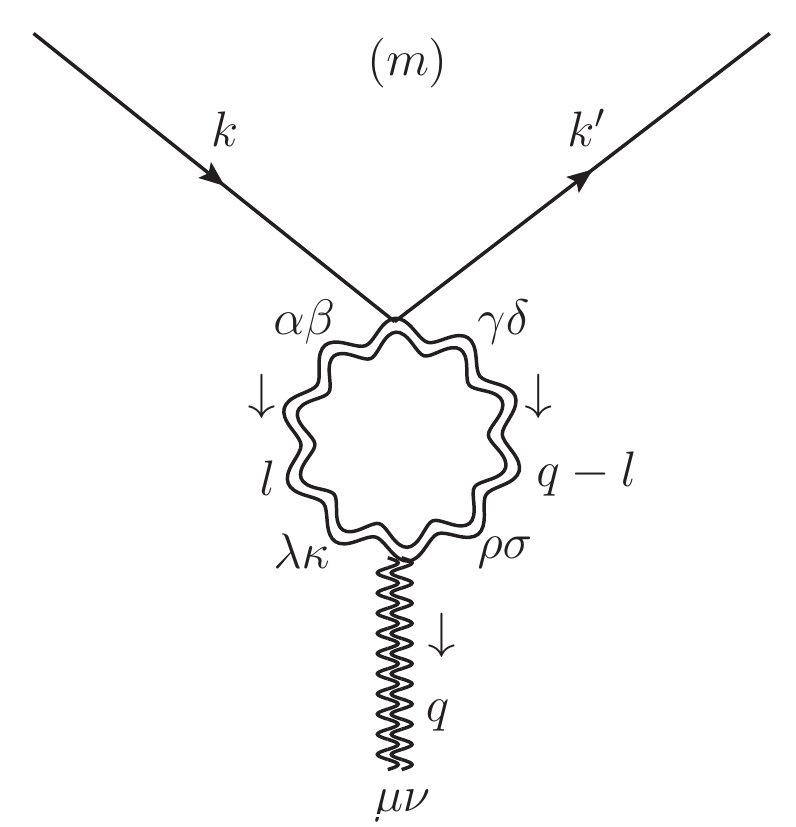} 
\caption[Vertex correction diagram 2]{The vertex correction diagram of Fig. \ref{Fig1D94a.pdf}c in detail. The momenta are such that $k-k^\prime=q$.}\label{D94c.pdf}
\end{figure}

In order to obtain the constants $\kappa_1$ and $\kappa_2$ (see Eqs. (\ref{1.2b})--(\ref{1.4b})) featuring the one-particle reducible potential, we need to evaluate Figs. \ref{Fig1D94a.pdf}b and c (as we pointed out before Figs. \ref{Fig1D94a.pdf}d, e, and f give no non-analytic contributions) and Fig. \ref{Fig1D94b.pdf}. We start with Fig. \ref{Fig1D94a.pdf}b which we have retrieved in more detail in Fig. \ref{D94b.pdf}. This diagram leads to the Green function
\begin{equation}
\begin{split}
G^{\mu \nu}(q) = \int  \dfrac{{\rm d}^4 l}{(2 \pi)^4} \biggl [ & \tau_{\alpha \beta}(k, k-l, m) \tau_{\gamma \delta}(k-l, k^\prime, m) \tau^{\mu \nu}_{\; \; \; \lambda \kappa \rho \sigma}(-l, -q)  \\
&  \dfrac{{\rm i} \mathcal{P}^{\alpha \beta \lambda \kappa}  }{l^2} \dfrac {\I \mathcal{P}^{\gamma \delta \rho \sigma}  }{\left( q-l \right)^2} \dfrac{\I}{\left(k-l \right)^2-m^2} \biggr],
\end{split}
\end{equation}
where the momentum transferred $q$ is such that $q=q(k,k^\prime)$. In terms of the form factors, the non-analytical contributions coming from Fig. \ref{D94b.pdf} are given by \cite{D94,BDH2003} \footnote{The error occurring in Ref. \cite{D94} has been corrected in Ref. \cite{BDH2003}.}
\begin{equation}
\begin{split}
F_1(q^2) & =\dfrac{\chi^2 q^2}{32 \pi^2} \left \{ \left[ \dfrac{1}{4} -2+1+0 \right] \log \left( -q^2 \right) + \left[ \dfrac{1}{16}-1+1+0 \right] \dfrac{\pi^2 m}{\sqrt{-q^2}} \right \} \\
& = \dfrac{\chi^2 q^2}{32 \pi^2} \left[ -\dfrac{3}{4} \log \left(-q^2\right) +\dfrac{1}{16}  \dfrac{\pi^2 m}{\sqrt{-q^2}} \right],
\end{split}
\end{equation}
\begin{equation}
\begin{split}
F_2(q^2) & =\dfrac{\chi^2 q^2}{32 \pi^2} \left \{ \left[ \dfrac{13}{3} -1+0-1 \right] \log \left( -q^2 \right) + \left[ \dfrac{7}{8}-1+2-1 \right] \dfrac{\pi^2 m}{\sqrt{-q^2}} \right \} \\
& = \dfrac{\chi^2 q^2}{32 \pi^2} \left[ \dfrac{7}{3} \log \left(-q^2\right) +\dfrac{7}{8}  \dfrac{\pi^2 m}{\sqrt{-q^2}} \right],
\end{split}
\end{equation}
where the sequence of numbers in the first line of each $F_i(q^2)$ refers to the four sets of terms in the square brackets of Eq. (\ref{three-graviton_vertex}), respectively. The other contribution to the potential is represented by Fig. \ref{Fig1D94a.pdf}c, whose detailed version is reported in Fig. \ref{D94c.pdf}. From this figure it follows that\footnote{For each closed graviton bubble loop, we have to include a symmetry factor equals to $\dfrac{1}{2!}$.}
\begin{equation} 
G^{\mu \nu}(q) = \dfrac{1}{2!}\int  \dfrac{{\rm d}^4 l}{(2 \pi)^4} \biggl [  \tau_{\gamma \delta \alpha \beta}(k, k^\prime)  \tau^{\mu \nu}_{\; \; \; \lambda \kappa \rho \sigma}(-l, -q)   \dfrac{{\rm i} \mathcal{P}^{\alpha \beta \lambda \kappa}  }{l^2} \dfrac {\I \mathcal{P}^{\gamma \delta \rho \sigma}  }{\left( q-l \right)^2} \biggr],
\end{equation}
and hence the non-analytical terms occurring in the form factors are \cite{D94}
\begin{equation}
F_1(q^2)  =\dfrac{\chi^2 q^2}{32 \pi^2}  \left[ 0+2+0-2 \right] \log \left( -q^2 \right)  = 0,
\end{equation}
\begin{equation}
F_2(q^2)  =\dfrac{\chi^2 q^2}{32 \pi^2}  \left[ -\dfrac{25}{3} +0+2+2 \right] \log \left( -q^2 \right)   = \dfrac{\chi^2 q^2}{32 \pi^2} \left[ -\dfrac{13}{3} \log \left(-q^2\right) \right].
\end{equation}
Therefore, the resulting non-analytical part of the form factors reads as
\begin{equation}
F_1(q^2)  =1+ \dfrac{\chi^2 q^2}{32 \pi^2}  \left[ -\dfrac{3}{4} \log \left( -q^2 \right) + \dfrac{1}{16} \dfrac{\pi^2 m}{\sqrt{-q^2}} \right] ,
\end{equation}
\begin{equation}
F_2(q^2)  =\dfrac{\chi^2 q^2}{32 \pi^2}  \left[ -2 \log \left( -q^2 \right) + \dfrac{7}{8} \dfrac{\pi^2 m}{\sqrt{-q^2}} \right].
\end{equation}
The integrals needed for the calculation of the diagrams of Figs. \ref{D94b.pdf} and \ref{D94c.pdf} are given in Appendix \ref{Appendix_Useful_integrals}.

The divergent part coming from the graviton vacuum polarization diagrams of Fig. \ref{Fig1D94b.pdf} can be directly read off from the counter-Lagrangian (\ref{'tHooft_5.24}), which in the context of effective theories can be rewritten as
\begin{equation}
\Delta \mathcal{L}^{\rm (1-loop)}_{grav}= -\dfrac{1}{16 \pi^2} \log \left( {\bf q}^2 \right) \left[\dfrac{1}{120} R^2 + \dfrac{7}{20} R_{\mu \nu} R^{\mu \nu} \right].
\end{equation}
Then, it is possible to show that the vacuum polarization tensor assumes the form (cf. Eq. (\ref{D94_61}))
\begin{equation}
\begin{split}
\Pi_{\alpha \beta \gamma \delta}(q) &= - \dfrac{\chi^2}{16 \pi^2} \log \left(-q^2 \right) \biggl[ \dfrac{21}{120} q^4 I_{\alpha \beta \gamma \delta}+\dfrac{23}{120} q^4 \eta_{\alpha \beta} \eta_{\gamma \delta} \\
& - \dfrac{23}{120} q^2 \left( \eta_{\alpha \beta}q_\gamma q_\delta + \eta_{\gamma \delta} q_\alpha q_\beta \right) \\
& -\dfrac{21}{240} q^2 \left( \eta_{\beta \gamma } q_\alpha q_\delta + \eta_{\beta \delta} q_\alpha q_\gamma   + \eta_{\alpha \delta} q_\gamma q_\beta + \eta_{\alpha \gamma } q_\delta q_\beta \right) \\
& +\dfrac{11}{30} q_\alpha q_\beta q_\gamma q_\delta \biggr] + ({\rm nonlogs}). \label{D94_71}
\end{split}
\end{equation}
Therefore, the gravitational interaction of Fig. \ref{Fig2D94.pdf} is characterized by the one-particle reducible amplitude
\begin{equation}
\I \mathcal{M}(q)= - \dfrac{\chi^2}{4} \mathcal{C}_{\mu \nu}(-q) \left[ \I D^{\mu \nu \alpha \beta}(q) + \I D^{\mu \nu \rho \sigma}(q) \I \Pi_{\rho \sigma \eta \lambda}(q) \I D^{\eta \lambda \alpha \beta}(q) \right] \mathcal{C}_{\alpha \beta} (q), \label{D94_73b}
\end{equation} 
which on taking the non-relativistic limit becomes (cf. Eq. (\ref{D94_58})) \cite{D94}
\begin{equation}
\I \mathcal{M}({\bf q})= -\I \left(4 \pi G m_1 m_2 \right) \left\{\dfrac{1}{{\bf q}^2} -\dfrac{\chi^2}{32 \pi^2} \left[-\dfrac{167 \pi}{60} \log \left( {\bf q}^2 \right)+\dfrac{\pi^2 \left(m_1+m_2 \right)}{2 \sqrt{{\bf q}^2}} \right] + {\rm const.} \right\}, \label{D94_73}
\end{equation}
thence from the non-analytical part $\mathcal{A}^\prime(\bf{q})$ of (\ref{D94_73}) evaluated in the coordinate-space (and by reinserting the constants $c$ and $\hbar$) we obtain the result \cite{D94,BDH2003}
\begin{equation}
V_Q(r)= -\dfrac{G m_1 m_2}{r} \left[1-\dfrac{G(m_1+m_2)}{r c^2} - \dfrac{167}{30 \pi} \dfrac{G \hbar}{r^2 c^3}\right].
\end{equation}
A comparison with Eqs. (\ref{1.2b})--(\ref{1.4b}) clearly shows that for the one-particle reducible potential the numbers $\kappa_1$ and $\kappa_2$ are both negative and read as
\begin{equation}
\kappa_1 = -1,
\end{equation}
\begin{equation}
\kappa_2 = -\dfrac{167}{30 \pi}.
\end{equation}

\subsection{Scattering and bound-states potential} \label{Sec. scattering-bound-state potential}

Before introducing the details of the scattering and bound-state potential, we first analyse the tree-diagram of Fig. \ref{Newton.pdf}, which in the non-relativistic domain leads to the Newtonian potential. Bearing in mind the definition (\ref{scattering_amplitude}) and the Feynman rules of Appendix \ref{Appendix_Feynman_rules}, the relation
\begin{equation}
\begin{split}
\langle & k_2,k_4 \vert {\rm i}\mathcal{T} \vert k_1,k_3 \rangle  = \\
 & \int \dfrac{{\rm d}^4 q}{(2 \pi)^4} \left[ \tau^{\mu \nu}\left(k_1,k_2,m_1\right) \I \dfrac{\mathcal{P}_{\mu \nu \alpha \beta}}{q^2} \tau^{\alpha \beta}\left(k_3,k_4,m_2\right) (2 \pi)^4 \delta^{(4)} \left(k_3+q-k_4\right)(2 \pi)^4 \delta^{(4)} \left(k_1-q-k_2\right) \right] \\
 & = (2 \pi)^4 \delta^{(4)} \left(k_1+k_3-k_2-k_4\right) \tau^{\mu \nu}\left(k_1,k_2,m_1\right) \I \dfrac{\mathcal{P}_{\mu \nu \alpha \beta}}{q^2} \tau^{\alpha \beta}\left(k_3,k_4,m_2\right) \\
 & = \left( 2 \pi \right)^4 \delta^{(4)}\left(k_1+k_3-k_2-k_4\right) {\rm i} \mathcal{M}(q),
\end{split}
\end{equation} 
implies that the full scattering amplitude associated to the process of Fig. \ref{Newton.pdf} assumes the form
\begin{equation}
\begin{split}
 {\rm i} \mathcal{M}(q) &= \tau^{\mu \nu}\left(k_1,k_2,m_1\right) \I \dfrac{\mathcal{P}_{\mu \nu \alpha \beta}}{q^2} \tau^{\alpha \beta}\left(k_3,k_4,m_2\right) \\
 & = \dfrac{\I}{q^2} \left[ \left( I_{\mu \nu \alpha \beta} -\dfrac{1}{2} \eta_{\mu \nu} \eta_{\alpha \beta} \right) \tau^{\mu \nu}\left(k_1,k_2,m_1\right)  \tau^{\alpha \beta}\left(k_3,k_4,m_2\right) \right] \\
 & = \left\{  \left[     \tau_{\mu \nu}\left(k_3,k_4,m_2\right) -\dfrac{1}{2} \eta_{\mu \nu}  \tau^{\epsilon}_{\; \epsilon}\left(k_3,k_4,m_2\right)    \right]  \tau^{\mu \nu}\left(k_1,k_2,m_1\right)   \right\} \\
 & = -\dfrac{\I \chi^2}{4 q^2}  \biggl [2 \left(k_1 \cdot k_3 \right) \left(k_2 \cdot k_4 \right) + 2 \left(k_1 \cdot k_4 \right) \left(k_2 \cdot k_3 \right) - 2 (m_2)^2 \left(k_1 \cdot k_2 \right) \\
 & + q^2 \left(k_3 \cdot k_4 \right) - 2 q^2 (m_2)^2 \biggr ] \\
 & = -\dfrac{\I \chi^2}{4 q^2}  \left[2 \left(k_1 \cdot k_3 \right)^2 + 2 \left(k_1 \cdot k_4 \right)^2 - \dfrac{q^4}{2} - 2 (m_1 m_2)^2  \right],
 \end{split}
\end{equation}
where we have exploited the tensor relations (\ref{tensor_relation2})--(\ref{tensor_relation3}) and Eqs. (\ref{momentum_scalar_product1})--(\ref{momentum_scalar_product2}). By the means of the non-relativistic expressions
\begin{equation}
\begin{split}
& q^2 \approx - {\bf q}^2, \\
& q^4 = q^2 q^2 \approx  {\bf q}^4, \\
& \left( k_1+k_3 \right)^2 \approx \left( m_1 + m_2 \right)^2, \\
&  \left( k_1-k_4 \right)^2 \approx \left( m_1 - m_2 \right)^2 + {\bf q}^2, 
\end{split}
\end{equation}
the scattering amplitude becomes
\begin{equation}
-\I \mathcal{M}({\bf q})= \I \dfrac{ \chi^2}{2 {\bf q}^2} \biggl[ (m_1 m_2)^2 - (m_1 m_2) {\bf q}^2  \biggr], 
\end{equation}
so that the non-analytical part is given by
\begin{equation}
\mathcal{A}^\prime({\bf q}) = -\dfrac{16 \pi G (m_1 m_2)^2}{{\bf q}^2},
\end{equation}
and hence by employing the definition (\ref{D03_4}) and Eq. (\ref{Fourier1}) we can easily obtain the Newtonian potential $V(r)=-Gm_1 m_2 /r$.  

\begin{figure}
\centering
\includegraphics[scale=0.7]{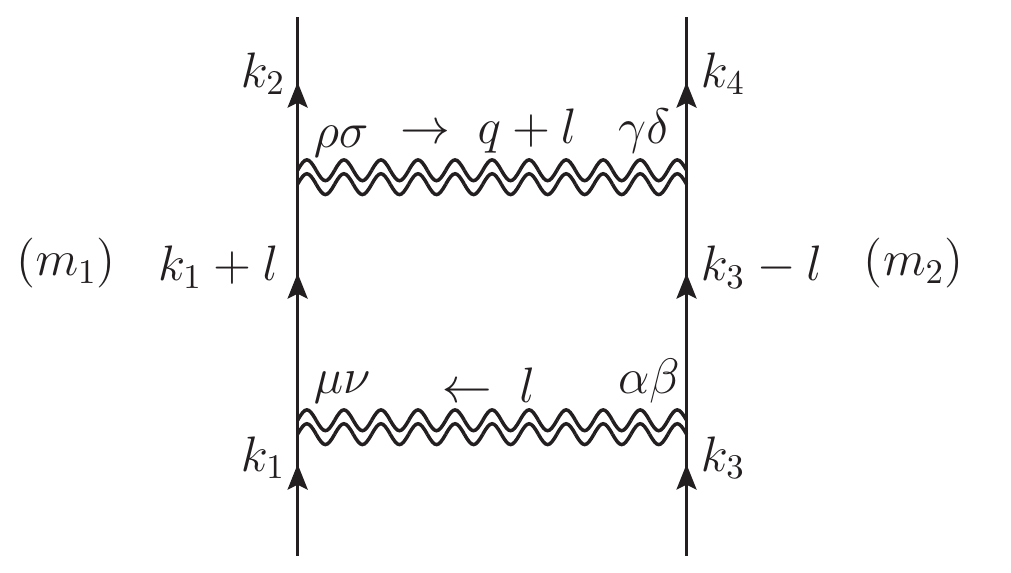} 
\caption[The box diagram]{The box diagram.}\label{box.pdf}
\end{figure}

\begin{figure}
\centering
\includegraphics[scale=0.7]{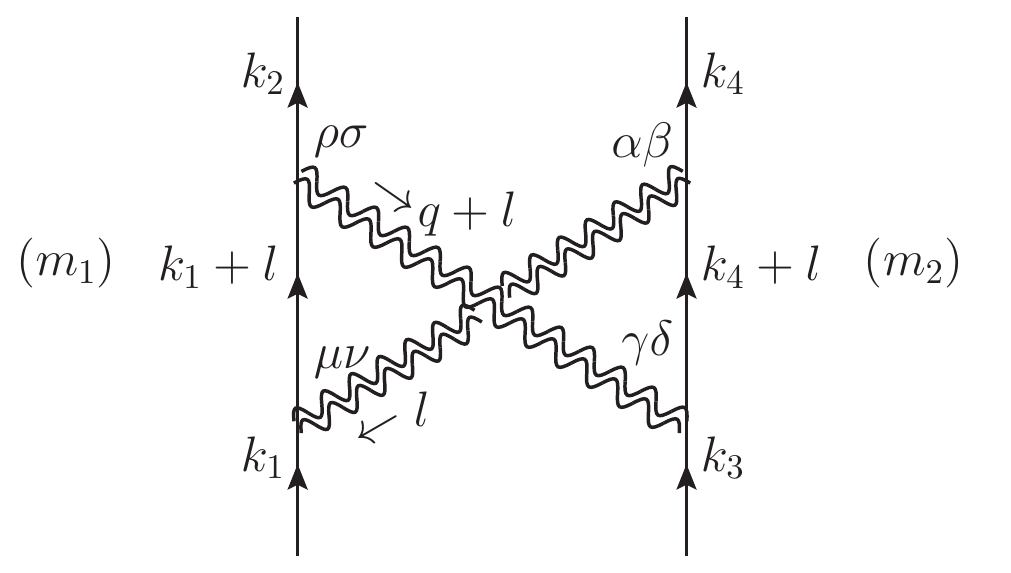} 
\caption[The crossed-box diagram]{The crossed-box diagram.}\label{crossedbox.pdf}
\end{figure}

\begin{figure}
\centering
\begin{subfigure}{.45\textwidth}
  \centering
  \includegraphics[width=1.1 \linewidth]{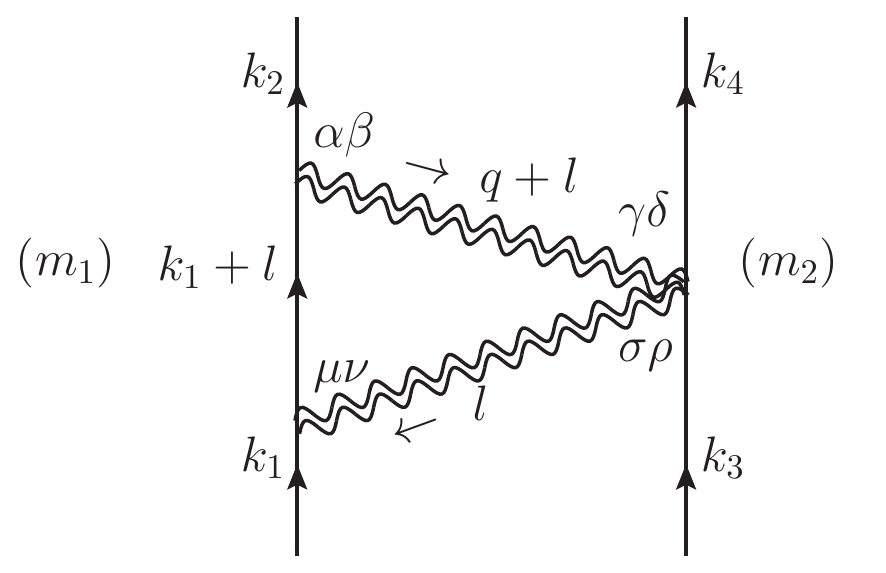}
  \caption{}
  \label{triangle1.pdf}
\end{subfigure}
\begin{subfigure}{.45\textwidth}
  \centering
  \includegraphics[width=1.1 \linewidth]{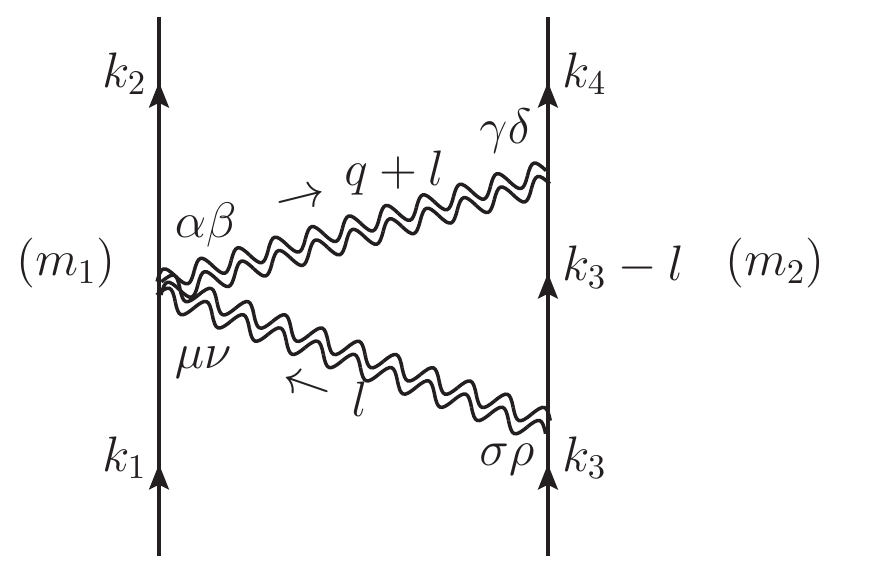}
  \caption{}
  \label{triangle2.pdf}
\end{subfigure}
\caption[The triangle diagrams]{The triangle diagrams.} \label{triangle.pdf}
\end{figure}

\begin{figure}
\centering
\includegraphics[scale=0.7]{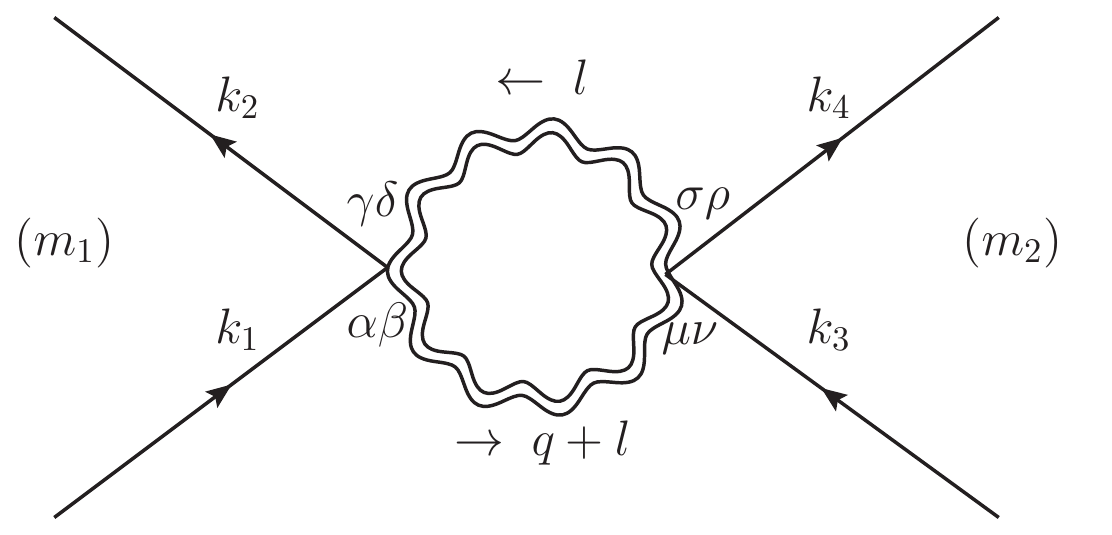} 
\caption[The double-seagull diagram]{The double-seagull diagram.}\label{seagull.pdf}
\end{figure}

\begin{figure}
\centering
\begin{subfigure}{.45\textwidth}
  \centering
  \includegraphics[width=1.05 \linewidth]{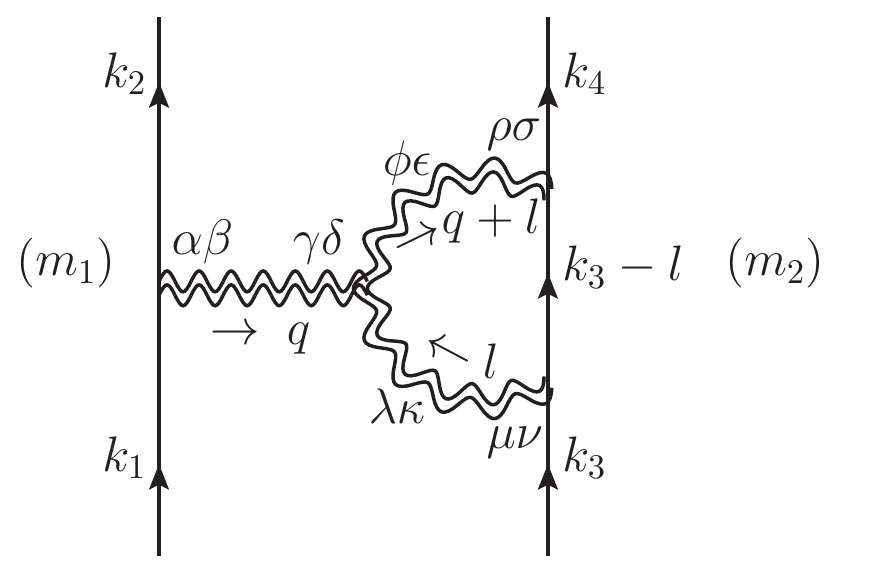}
  \caption{}
  \label{vertex1.pdf}
\end{subfigure}
\begin{subfigure}{.45\textwidth}
  \centering
  \includegraphics[width=1.05 \linewidth]{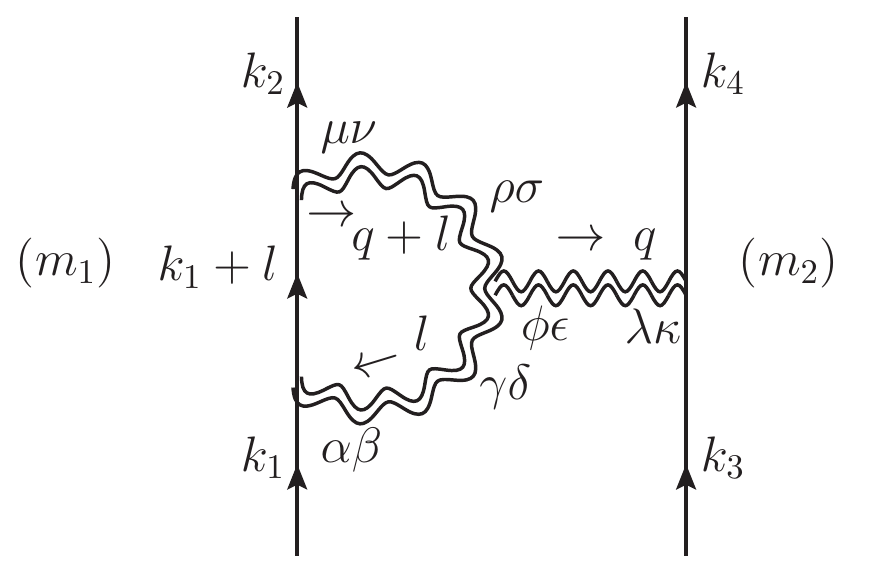}
  \caption{}
  \label{vertex2.pdf}
\end{subfigure}
\begin{subfigure}{.45\textwidth}
  \centering
  \includegraphics[width=1.05 \linewidth]{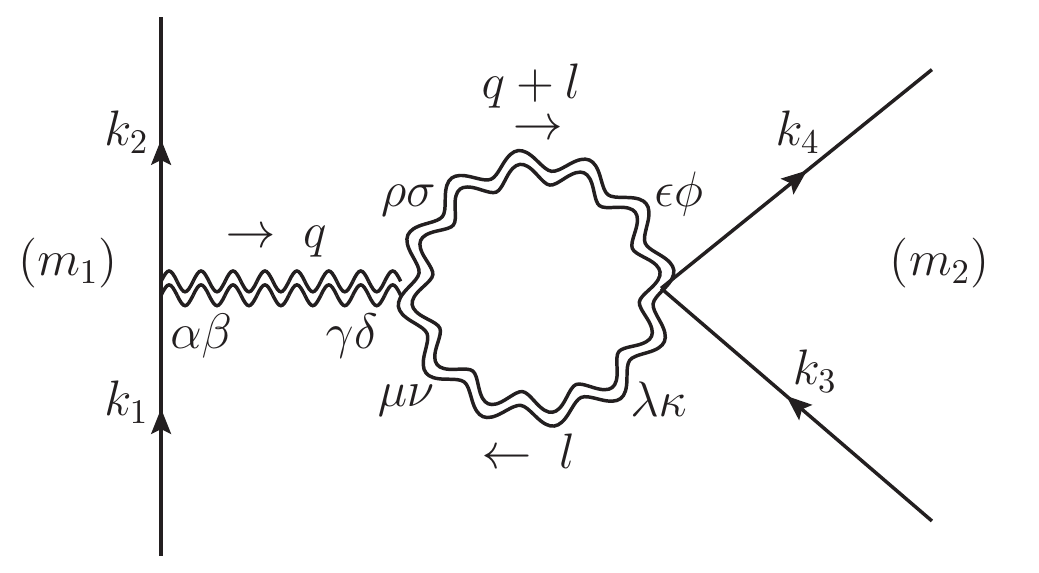}
  \caption{}
  \label{vertex3.pdf}
\end{subfigure}
\begin{subfigure}{.45\textwidth}
  \centering
  \includegraphics[width=1.05 \linewidth]{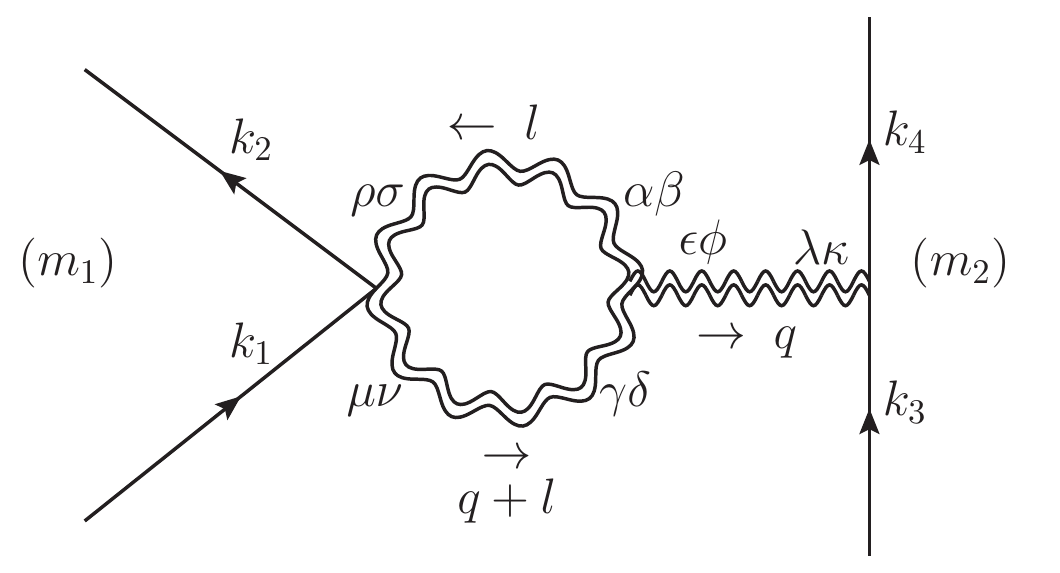}
  \caption{}
  \label{vertex4.pdf}
\end{subfigure}
\caption[The set of vertex correction diagrams contributing to the scattering potential]{The set of vertex correction diagrams contributing to the scattering potential.} \label{vertex.pdf}
\end{figure}

\begin{figure}
\centering
\begin{subfigure}{.85\textwidth}
  \centering
  \includegraphics[width=.9 \linewidth]{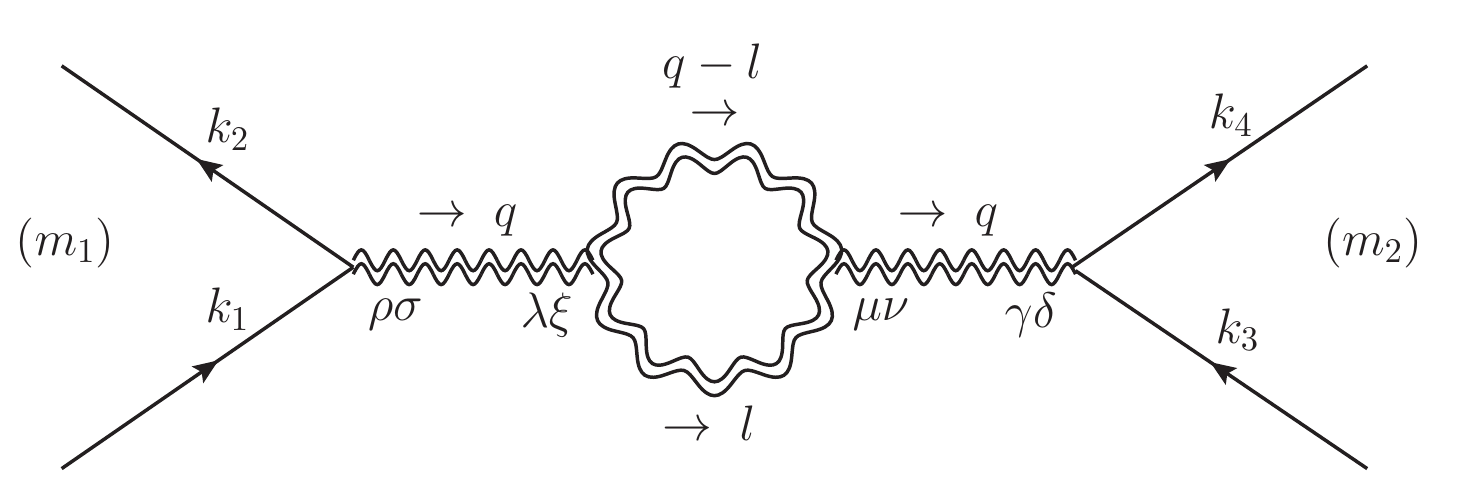}
  \caption{}
  \label{vacuum1.pdf}
\end{subfigure}
\begin{subfigure}{.85\textwidth}
  \centering
  \includegraphics[width=.9 \linewidth]{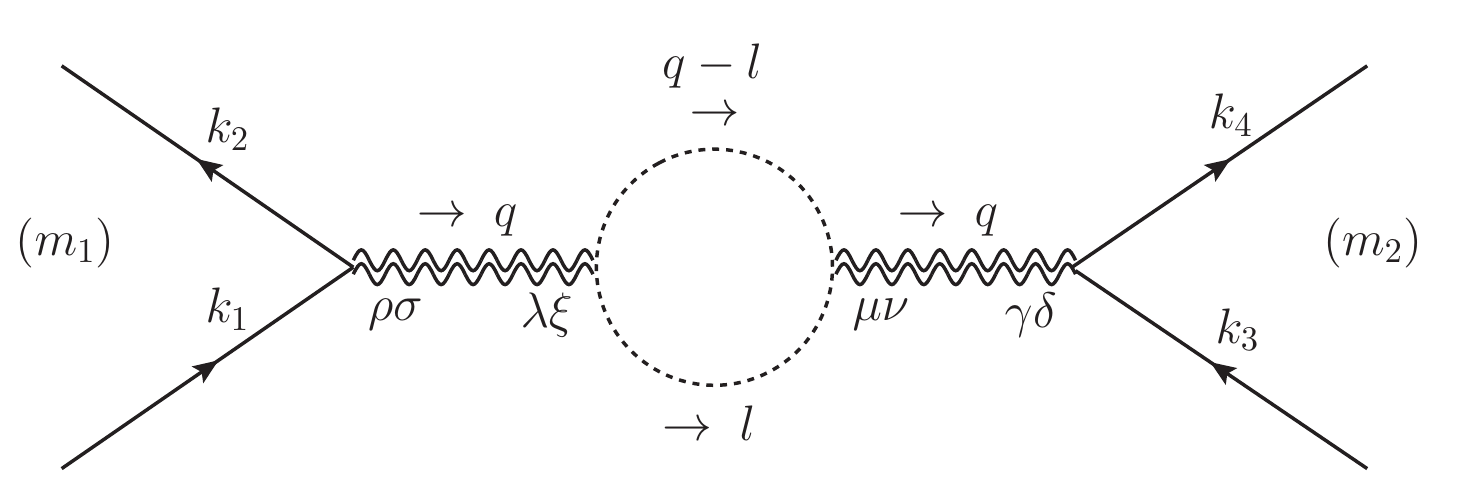}
  \caption{}
  \label{vacuum2.pdf}
\end{subfigure}
\caption[The vacuum polarization diagrams contributing to the scattering potential]{The vacuum polarization diagrams contributing to the scattering potential.} \label{vacuum.pdf}
\end{figure}

Now we turn our attention to the derivation of the scattering potential \cite{D03}. The diagrams involved are
\vskip 0.3cm
\noindent
- The box and crossed-box diagrams of Figs. \ref{box.pdf} and \ref{crossedbox.pdf}, respectively.
\vskip 0.3cm
\noindent
- The two triangle diagrams of Fig. \ref{triangle.pdf}.
\vskip 0.3cm
\noindent
- The double-seagull diagram, Fig. \ref{seagull.pdf}.
\vskip 0.3cm
\noindent
- The vertex correction (Fig. \ref{vertex.pdf}) and vacuum polarization diagrams (Fig. \ref{vacuum.pdf}). 

For all the above listed diagrams the ingoing momenta are indicated with $k_1$ and $k_3$, the outgoing ones with $k_2$ and $k_4$, while the momentum transferred $q$ is such that $k_1-k_2=k_4-k_3=q$. The integrals involved in the calculation are discussed in Appendix \ref{Appendix_Useful_integrals}. 

The contribution coming from the box diagram is given by
\begin{equation}
\begin{split}
\I \mathcal{M}(q)= \int \dfrac{{\rm d}^4 l}{(2 \pi)^4} \Biggl [ & \tau^{\mu \nu}(k_1,k_1+l,m_1) \tau^{\rho \sigma}(k_1+l,k_2,m_1) \tau^{\alpha \beta}(k_3,k_3+l,m_2) \tau^{\gamma \delta}(k_3-l,k_4,m_2) \\
& \dfrac{\I}{\left(k_1+l \right)^2-\left(m_1\right)^2}  \dfrac{\I}{\left(k_3-l \right)^2-\left(m_2\right)^2}\dfrac{\I \mathcal{P}_{\alpha \beta \mu \nu}}{l^2} \dfrac{\I \mathcal{P}_{\rho \sigma \gamma \delta}}{\left( q+l \right)^2} \Biggr ],
\end{split}
\end{equation}
whereas for the crossed-box
\begin{equation}
\begin{split}
\I \mathcal{M}(q)= \int \dfrac{{\rm d}^4 l}{(2 \pi)^4} \Biggl [ & \tau^{\mu \nu}(k_1,k_1+l,m_1) \tau^{\rho \sigma}(k_1+l,k_2,m_1)  \tau^{\gamma \delta}(k_3,k_4+l,m_2) \tau^{\alpha \beta}(k_4+l,k_4,m_2) \\
& \dfrac{\I}{\left(k_1+l \right)^2-\left(m_1\right)^2}  \dfrac{\I}{\left(k_4+l \right)^2-\left(m_2\right)^2}\dfrac{\I \mathcal{P}_{\alpha \beta \mu \nu}}{l^2} \dfrac{\I \mathcal{P}_{\rho \sigma \gamma \delta}}{\left( q+l \right)^2} \Biggr ].
\end{split}
\end{equation}
After taking the non-relativistic limit, it turns out that these two diagrams lead to \cite{D03}
\begin{equation}
\mathcal{A}^\prime({\bf q})= 4 m_1 m_2 \left( \dfrac{94}{3} G^2 m_1 m_2 \right) \log \left( {\bf q}^2 \right), 
\end{equation}
and hence from (\ref{D03_4}) the contribution to the potential is
\begin{equation}
V(r)=-\dfrac{47}{3 \pi} \dfrac{G^2 m_1 m_2}{r^3}.
\end{equation}

The two triangle diagrams of Figs. \ref{triangle1.pdf} and \ref{triangle2.pdf} give
\begin{equation}
\begin{split}
\I \mathcal{M}(q)= \int \dfrac{{\rm d}^4 l}{(2 \pi)^4} \Biggl [ & \tau^{\mu \nu}(k_1,k_1+l,m_1) \tau^{\alpha \beta}(k_1+l,k_2,m_1)  \tau^{\gamma \delta \sigma \rho}(k_3,k_4,m_2)  \\
& \dfrac{\I}{\left(k_1+l \right)^2-\left(m_1\right)^2}  \dfrac{\I \mathcal{P}_{\alpha \beta \gamma \delta}}{\left(q+l\right)^2} \dfrac{\I \mathcal{P}_{\sigma \rho \mu \nu}}{l^2} \Biggr ],
\end{split}
\end{equation}
and
\begin{equation}
\begin{split}
\I \mathcal{M}(q)= \int \dfrac{{\rm d}^4 l}{(2 \pi)^4} \Biggl [ & \tau^{\alpha \beta \mu \nu}(k_1,k_2,m_1)  \tau^{\sigma \rho}(k_3,k_3-l,m_2) \tau^{\gamma \delta}(k_3-l,k_4,m_2)   \\
& \dfrac{\I}{\left(k_3-l \right)^2-\left(m_2\right)^2}  \dfrac{\I \mathcal{P}_{\sigma \rho \mu \nu}}{l^2}  \dfrac{\I \mathcal{P}_{\alpha \beta \gamma \delta}}{\left(q+l\right)^2} \Biggr ],
\end{split}
\end{equation}
respectively. Taking the non relativistic limit we have \cite{D03}
\begin{equation}
\mathcal{A}^\prime({\bf q})=- 4 m_1 m_2  \left( 8 G^2 m_1 m_2 \right) \left[ \dfrac{7}{2} \log  \left( {\bf q}^2 \right) + \pi^2 m_1 \left( \dfrac{1}{{\bf q}} \right) \right],
\end{equation}
for Fig. \ref{triangle1.pdf} and 
\begin{equation}
\mathcal{A}^\prime({\bf q})=- 4 m_1 m_2  \left( 8 G^2 m_1 m_2 \right) \left[ \dfrac{7}{2} \log  \left( {\bf q}^2 \right) + \pi^2 m_2 \left( \dfrac{1}{{\bf q}} \right) \right],
\end{equation}
for Fig. \ref{triangle2.pdf}. This means that through a Fourier transformation we obtain for Fig. \ref{triangle.pdf} the overall result
\begin{equation}
V(r)= -4 \dfrac{G^2 m_1 m_2 \left(m_1+m_2 \right)}{r^2} + \dfrac{28}{\pi} \dfrac{G^2 m_1 m_2}{r^3}. 
\end{equation}

The scattering amplitude associated to Fig. \ref{seagull.pdf} reads as
\begin{equation}
\I \mathcal{M}(q)= \dfrac{1}{2!} \int \dfrac{{\rm d}^4 l}{(2 \pi)^4} \biggl [  \tau^{\gamma \delta \alpha \beta }(k_1,k_2,m_1)  \tau^{\sigma \rho \mu \nu}(k_3,k_4,m_2)  \dfrac{\I \mathcal{P}_{\alpha \beta \mu \nu}}{\left(q+l\right)^2} \dfrac{\I \mathcal{P}_{\sigma \rho \gamma \delta}}{l^2}  \biggr ],
\end{equation}
whose non-analytical contribution in the non-relativistic domain assumes the form \cite{D03}
\begin{equation}
\mathcal{A}^\prime({\bf q})= 4 m_1 m_2 \left( 44 G^2 m_1 m_2 \right) \log \left( {\bf q}^2 \right),
\end{equation}
giving
\begin{equation}
V(r)= -\dfrac{22}{\pi} \dfrac{G^2 m_1 m_2}{r^3}. 
\end{equation}

The vertex correction diagrams have already been analysed within the context of one-particle reducible potential. We now consider the associated scattering processes (or equivalently the associated four-point functions), Fig. \ref{vertex.pdf}. These graphs can be divided into two groups depending on whether massive loops (Figs. \ref{vertex1.pdf} and \ref{vertex2.pdf}) or pure graviton loops (Figs. \ref{vertex3.pdf} and \ref{vertex4.pdf}) are present. Massive loop diagrams give
\begin{equation}
\begin{split}
\I \mathcal{M}(q)= \int \dfrac{{\rm d}^4 l}{(2 \pi)^4} \Biggl [ & \tau^{\alpha \beta}(k_1,k_2,m_1)  \tau^{\mu \nu}(k_3,k_3-l,m_2) \tau^{\rho \sigma}(k_3-l,k_4,m_2)  \tau^{\gamma \delta}_{\; \; \; \lambda \kappa \phi \epsilon}(-l,q)  \\
&  \dfrac{\I \mathcal{P}_{\alpha \beta \gamma \delta }}{q^2}  \dfrac{\I \mathcal{P}^{\; \; \;  \lambda \kappa}_{\mu \nu}}{l^2}  \dfrac{\I \mathcal{P}^{\phi \epsilon}_{\; \; \; \rho \sigma}}{\left(q+l\right)^2} \dfrac{\I}{\left(k_3-l \right)^2-\left(m_2\right)^2} \Biggr ],
\end{split}
\end{equation}
\begin{equation}
\begin{split}
\I \mathcal{M}(q)= \int \dfrac{{\rm d}^4 l}{(2 \pi)^4} \Biggl [ & \tau^{\alpha \beta}(k_1,k_1+l,m_1)  \tau^{\mu \nu}(k_1+l,k_2,m_1) \tau^{\lambda \kappa}(k_3,k_4,m_2)  \tau^{\phi \epsilon}_{\; \; \; \gamma \delta \rho \sigma}(l,-q)  \\
&  \dfrac{\I \mathcal{P}_{\mu \nu }^{\; \; \; \, \rho \sigma}}{\left(q+l\right)^2}  \dfrac{\I \mathcal{P}^{\gamma \delta}_{\; \; \;  \alpha \beta}}{l^2}   \dfrac{\I \mathcal{P}_{\phi\epsilon \lambda \kappa  }}{q^2}   \dfrac{\I}{\left(k_1+l \right)^2-\left(m_1\right)^2} \Biggr ],
\end{split}
\end{equation}
from which it follows that \cite{D03}
\begin{equation}
\mathcal{A}^\prime({\bf q})= 4 m_1 m_2 \left( 2 G^2 m_1 m_2 \right) \left[ \dfrac{\pi^2 \left(m_1+m_2\right)}{{\bf q}} + \dfrac{5}{3} \log \left( {\bf q}^2 \right) \right],
\end{equation}
while for pure graviton ones the amplitudes
\begin{equation}
\begin{split}
\I \mathcal{M}(q)=\dfrac{1}{2!} \int \dfrac{{\rm d}^4 l}{(2 \pi)^4} \Biggl [ & \tau^{\alpha \beta}(k_1,k_2,m_1)   \tau^{\gamma \delta}_{\; \; \; \, \mu \nu \rho \sigma }(-l,q)  \tau_{\epsilon \phi \lambda \kappa}(k_3,k_4,m_2)\\
&  \dfrac{\I \mathcal{P}_{\alpha \beta \gamma \delta}}{q^2}  \dfrac{\I \mathcal{P}^{\rho \sigma \epsilon \phi }}{\left(q+ l \right)^2}   \dfrac{\I \mathcal{P}^{\lambda \kappa \mu \nu  }}{l^2}  \Biggr ],
\end{split}
\end{equation}
and 
\begin{equation}
\begin{split}
\I \mathcal{M}(q)=\dfrac{1}{2!} \int \dfrac{{\rm d}^4 l}{(2 \pi)^4} \Biggl [ & \tau^{\rho \sigma \mu \nu}(k_1,k_2,m_1)   \tau^{\epsilon \phi}_{\; \; \; \, \alpha \beta \gamma \delta }(l,-q)   \tau^{\lambda \kappa}(k_3,k_4,m_2) \\
&  \dfrac{\I \mathcal{P}^{\alpha \beta}_{ \; \; \; \, \rho \sigma }}{l^2}        \dfrac{\I \mathcal{P}_{\mu \nu}^{\; \; \, \gamma \delta}}{\left(q+ l \right)^2}   \dfrac{\I \mathcal{P}_{\epsilon \phi \lambda \kappa }}{q^2}  \  \Biggr ],
\end{split}
\end{equation}
once evaluated in the non-relativistic regime give rise to \cite{D03}
\begin{equation}
\mathcal{A}^\prime({\bf q})= -4 m_1 m_2 \left( \dfrac{52}{3} G^2 m_1 m_2 \right) \log \left( {\bf q}^2 \right).
\end{equation}
In terms of the potential we have
\begin{equation}
V(r)= \dfrac{G^2 m_1 m_2 \left( m_1 + m_2 \right)}{r^2}-\dfrac{5}{3 \pi} \dfrac{G^2 m_1 m_2 }{r^3},
\end{equation}
\begin{equation}
V(r)= \dfrac{26}{3 \pi} \dfrac{G^2 m_1 m_2 }{r^3},
\end{equation}
for massive and pure graviton loop graphs, respectively.

Lastly, we consider the vacuum polarization diagrams depicted in Fig. \ref{vacuum.pdf}. Also this set has been considered in the previous section and its resulting contribution to the scattering amplitude can be easily read from Eq. (\ref{D94_73b}), which for Figs. \ref{vacuum1.pdf} and \ref{vacuum2.pdf} reduces to
\begin{equation}
\I \mathcal{M}(q)= \tau_{\rho \sigma}(k_1,k_2,m_1) \dfrac{\I \mathcal{P}^{\rho \sigma \lambda \xi }}{q^2} \Pi_{\lambda \xi \mu \nu}(q) \dfrac{\I \mathcal{P}^{\mu \nu \gamma \delta }}{q^2} \tau_{\gamma \delta}(k_3,k_4,m_2),
\end{equation}
$\Pi_{\lambda \xi \mu \nu}(q)$ being given by (\ref{D94_71}). From this expression we have \cite{D03}
\begin{equation}
\mathcal{A}^\prime({\bf q})= 4 m_1 m_2 \left( \dfrac{43}{15} G^2 m_1 m_2 \right) \log \left( {\bf q}^2 \right),
\end{equation}
and hence
\begin{equation}
V(r)= - \dfrac{43}{30 \pi} \dfrac{G^2 m_1 m_2}{r^3}.
\end{equation}
Adding up all the corrections coming from Figs. \ref{box.pdf}--\ref{vacuum.pdf} and restoring $c$ and $\hbar$ we achieve the final expression of the scattering potential, i.e., 
\begin{equation}
V_Q(r)= -\dfrac{G m_1 m_2}{r} \left[1+3 \dfrac{G(m_1+m_2)}{r c^2} + \dfrac{41}{10 \pi} \dfrac{G \hbar}{r^2 c^3}\right],
\end{equation}
which implies that $\kappa_1$ and $\kappa_2$ are both positive and are given by (cf. (\ref{1.2b})--(\ref{1.4b}))
\begin{equation}
\kappa_1 = 3,
\end{equation}
\begin{equation}
\kappa_2 = \dfrac{41}{10 \pi}.
\end{equation}
Bearing in mind Eq. (\ref{D03_6}), the result for bound-state potential reads as
\begin{equation}
V_Q(r)= -\dfrac{G m_1 m_2}{r} \left[1-\dfrac{1}{2} \dfrac{G(m_1+m_2)}{r c^2} + \dfrac{41}{10 \pi} \dfrac{G \hbar}{r^2 c^3}\right],
\end{equation}
or in terms of $\kappa_1$ and $\kappa_2$
\begin{equation}
\kappa_1 = -\dfrac{1}{2},
\end{equation}
\begin{equation}
\kappa_2 = \dfrac{41}{10 \pi}.
\end{equation}
The values assumed by these constants for the three kinds of potential discussed so far are given in Tab. \ref{kappa_tab}. It is fair to mention that different results from the ones analysed in this thesis have been achieved in  the literature (for further details see Ref. \cite{effective_theories}). 

Now that we have obtained the ultimate structure, some remarks on the nature of the quantum corrected potential $V_Q(r)$ are essential. First of all, we note that Eq. (\ref{1.2b}) implies that, $\forall \varepsilon >0$, there exists an {\it unknown} $r_{0}$ value of $r$ such that \cite{BE14b,BEDS15}
\begin{equation}
\left | V_{Q}(r) + {G m_{A}m_{B}\over r}\left(1+{k_{1}\over r}
+{k_{2}\over r^{2}}\right)\right | < \varepsilon, \; \; \; \; \; \; \;  \forall r > r_{0},  \label{1.5b}
\end{equation}
underlying the fact that we do not have a formula for $V_Q(r)$ which is equally good at all points. Since in the course of an orbit of a celestial body around another celestial body their mutual separation may change by a non-negligible amount, we see that Eq. (\ref{1.5b}) means that $V_Q(r)$ is not apt for the characterization orbits in general but, in contrast, it is well suited for issues such as the evaluation of equilibrium points of a dynamical system or the description of displaced periodic orbits. These subjects will be investigated in the next chapter. Moreover, we also stress that the dimensionless parameter $\kappa_{1}$ depends on the dimensionless parameter $\kappa_{2}$. In other words, $k_{1}$ is a post-Newtonian term which only depends on classical physical constants, but its weight, expressed by the real number $\kappa_{1}$, is affected by the calculational procedure leading to the fully quantum term $k_{2}$, where the real number $\kappa_{2}$ weighs the Planck length squared, i.e.,
\begin{equation}
\kappa_1 = \kappa_1 (\kappa_2), \; \; \; \; \; \; \;  k_2=\kappa_2 (l_{P})^2.
\end{equation}
Thus, we are not dealing with corrections to the relativistic celestial mechanics (cf. Ref. \cite{Damour}). More precisely, the perturbative expansion (in the Poincar\'e sense, see Appendix \ref{Asymptotic exp_App}) leading to Eq. (\ref{1.2b}) involves only integer powers of Newton constant $G$: \cite{BEDS15}
\begin{equation}
V_{Q}(r) \sim -{G m_{A}m_{B}\over r} \left(1+\sum_{p=1}^{\infty}f_{p}(r)G^{p} \right)
\sim -{G m_{A}m_{B}\over r} \left(1+\sum_{n=1}^{\infty}{k_{n}\over r^{n}}\right),
\end{equation}
where the coefficients are such that $k_{n}=k_{n}(R_{A}+R_{B},(l_{P})^{2})$ (see Eqs. (\ref{1.3b}) and (\ref{1.4b})). At one loop, i.e., to linear order in $G$, where
\begin{equation}
f_{1}(r)=\kappa_{1}{(m_{A}+m_{B})\over c^{2}}{1 \over r}
+\kappa_{2}{\hbar \over c^{3}} {1 \over r^{2}},
\end{equation}
we can only have the contribution ${(R_{A}+R_{B})\over r}$ with weight equal to the real number $\kappa_{1}$, and the contribution ${(l_{P})^{2}\over r^{2}}$ with weight equal to the real number $\kappa_{2}$. Although the term ${(l_{P})^{2}\over r^{2}}$ is overwhelmed by the factor ${(R_{A}+R_{B})\over r}$, the two are inextricably intertwined because $\kappa_{1}$ is not a free real parameter but depends on $\kappa_{2}$: both $\kappa_{1}$ and $\kappa_{2}$ result from loop
diagrams, as we have shown before. Thus, the one-loop long-distance quantum correction is {\it the whole term}
\begin{equation}
f_{1}(r)G={k_{1}\over r}+{k_{2}\over r^{2}}=\kappa_{1}{(R_{A}+R_{B})\over r}
+\kappa_{2}{(l_{P})^{2}\over r^{2}},
\end{equation}
where $\kappa_{1}$ takes a certain value because there exists a non-vanishing value of $\kappa_{2}$.
\begin{table}[htbp]
\centering
\caption[The values assumed by $\kappa_1$ and $\kappa_2$ in the three different potentials]{The values assumed by $\kappa_1$ and $\kappa_2$ in the three different potentials.}
\renewcommand\arraystretch{2.2}
\begin{tabular}{|c|c|c|c|}
\hline
$\kappa_i$ & one-particle reducible & scattering & bound-states \\
\hline
$\kappa_1$ & $-1$ & $3$ & $-\dfrac{1}{2}$ \\
\hline
$\kappa_2$ & $ -\dfrac{167}{30 \pi}$ & $\dfrac{41}{10 \pi}$ &  $\dfrac{41}{10 \pi}$ \\
\hline
\end{tabular}
\label{kappa_tab}
\end{table}

We are now ready to apply the topics described so far to the context of the restricted three-body problem of celestial mechanics consisting of the Earth and the Moon as the primaries \cite{BE14a,BE14b, BEDS15,testbed}. This issue will represent the heart of the next chapter. 

\chapter{The restricted three-body problem in effective field theories of gravity} \label{Restricted_Chapter}

\vspace{2cm}
\emph{I have almost completed a treatise on analytical mechanics based uniquely on the principle of virtual velocities; but, as I do not yet know when or where I shall be able to have it printed, I am not rushing to put the finishing touches to it.}
\begin{flushright}
J. L. Lagrange
\end{flushright}

\vspace{2cm}

One of the most famous issues of classical dynamics is represented by the problem of three bodies, consisting of three particles moving in space under their mutual gravitational attraction. The problem involves finding the position of the particles at any subsequent time once their positions and velocities at the initial time $t=t_0$ are prescribed. This problem dates in substance from 1687, when Isaac Newton published his ``Principia" and has profoundly influenced classical mechanics since then. In 1887, mathematicians Heinrich Bruns and Henri Poincar\'e \cite{P1890,P1892} showed that there is no general analytical solution in terms of elementary functions for the three-body problem (this will be discussed in the next chapter, Sec. \ref{classical_integrals_Sec}), unlike the two-body one which is completely solved in the sense indicated above. In particular, it was brilliantly proved by Poincar\'e that all the series used by astronomers\footnote{We remember the achievements of the late IXX century of Delaunay, Lindstedt, Gyld\'en and Hill \cite{Marchal}.} to integrate Lagrange equations regarding the problem of three bodies were actually not convergent in a rigorous mathematical sense. In fact, for the astronomers a series was considered to be convergent if the terms they had calculated decreased rapidly, regardless of the fact they had no knowledge of the behaviour of the subsequent terms. On the other side, for the mathematicians a series is convergent only if it was rigorously proved to be so. As an example, consider the trigonometric series of the form
\begin{equation}
\sum_n A_n \sin \left(\alpha_n t \right) + \sum_n B_n \cos \left(\alpha_n t \right),
\end{equation} 
which was long considered by astronomers in the context of perturbation theory as a solution of differential equations such as
\begin{equation}
\dfrac{{\rm d}^2 x}{{\rm d}t^2}+nx^2=\Phi(x,t),
\end{equation}
where $\Phi(x,t)$ is a function expandable in powers of $x$ with coefficients which are periodic functions of $t$ and $\alpha_n$ represent some parameters which may decrease or increase indefinitely, a feature that makes the series differ from Fourier one. It was the genius of Poincar\'e that proved that such series is not absolutely convergent. In so doing, he laid the foundations for the formal definition of asymptotic series (cf. Eqs. (\ref{Acta1})--(\ref{Acta2})) which represent, as Poincar\'e himself showed \cite{P1890}, true solutions of what he called for the first time ``restricted" problem of three bodies. In his famous memoir {\it Sur le probl\`eme des trois corps et les \'equations de la dynamique} presented in the competition celebrating the 60th birthday of King Oscar II of Sweden and Norway in 1889, this problem was defined as follows:
\begin{quote}
\emph{I consider three masses: the first very large, the second small but finite, the third infinitely small; I assume that the first two describe a circle around their common center of gravity and the third moves in the plane of these circles.}
\end{quote}
Since then, the most general gravitational problem of celestial mechanics involving three masses is called full three-body problem and we will consider it in the next chapter. In this chapter indeed we are going to describe the features of the restricted three-body problem in the context of effective field theories of gravity. 

\section{Restricted three-body problem} \label{restricted_sec}

We have seen in the first chapter of this thesis that the application of the effective field theory point of view to the quantization of Einstein's general relativity can be performed by including all possible higher derivative couplings of the fields in the gravitational Lagrangian (\ref{full_action}). By doing so, any field singularities
generated by loop diagrams can be associated with some component of the action and can be absorbed through a redefinition of the coupling
constants of the theory. By treating all coupling coefficients as experimentally determined in this way, the effective field theory is finite and singularity-free at any finite order of the loop expansion, even though it remains true, as pointed out before, that Einstein's gravity is not perturbatively renormalizable and not even two-loop on-shell finite (see Eq. (\ref{2loop-divergence})). Moreover, the crucial point according to which the leading (i.e., one-loop) long-distance quantum corrections to the Newtonian potential are entirely ruled by the Einstein-Hilbert part of the full action functional (\ref{full_action}) has led to the most important outcome (for our purposes) of the effective field theory framework, i.e., the quantum corrected potential (\ref{1.2b}) (see also Eqs. (\ref{1.3b}) and (\ref{1.4b}) and Tab. \ref{kappa_tab}). We now {\it assume} that this theoretical model can be applied to long distances and macroscopic bodies occurring in celestial mechanics, starting with the restricted three-body problem. Is it possible to obtain a quantum perspective on this issue, despite the extremely small numbers involved? The question is not merely of academic interest. Indeed, on the one hand, we know already that very small quantities may produce non-trivial effects in physics. An example, among the many, is provided by the Stark effect: no matter how small is the external electric field, the Stark-effect Hamiltonian has absolutely continuous spectrum on the whole real line \cite{ReedSimon1978}, whereas the unperturbed Hamiltonian for hydrogen atom has discrete spectrum on the negative half-line. Yet another relevant example is provided by singular perturbations in quantum mechanics: if a one-dimensional harmonic oscillator is perturbed by a term proportional to negative powers of the position operator, then no matter how small is the weight coefficient one cannot recover the original Hamiltonian if the perturbation is switched off. The unperturbed Hamiltonian has in fact both even and odd eigenfunctions, whereas the singular perturbation enforces the stationary states to vanish at the origin, and the latter condition survives if the perturbation gets switched off \cite{Klauder}, so that one eventually recovers a sort of ``halved'' harmonic oscillator, with only half of the original eigenfunctions. On the other hand, by virtue of the improved technology with respect to the golden age of Poincar\'e, it becomes conceivable to send off satellites in the solar system that, within our lifetime, might become part of suitable three-body systems. This is the starting point of the modern laser ranging techniques, which will be discussed in Sec. \ref{Laser_Ranging_Sec}. Thence, the putative quantum corrected Newtonian potential (\ref{1.2b}) can be tested in circumstances which were inconceivable a century ago. 

The restricted three-body problem within the context of effective field theories represents an example of hybrid scheme in which we try to overcome our lack of knowledge about quantum gravity. This is not a novel feature in physics, since such a scheme, logically incomplete, is frequently employed in a variety of contexts as it turns out to be quite useful because the full theory is unknown or leads to equations that cannot be solved. Among the many conceivable examples of this feature, we mention the following, since they are relevant for motivating the research problem we are going to study: 
\vskip 0.3cm
\noindent
(i) The non-relativistic particle in curved spacetime \cite{DeWitt2003}, where the Schr\"odinger equation is studied, which is part of non-relativistic quantum theory, but the potential in such equation receives a contribution from spacetime curvature, which is instead defined and studied in general relativity;
\vskip 0.3cm
\noindent
(ii) Quantum field theory in curved spacetime, where the right-hand side of the Einstein equations is replaced by the expectation value of the regularized and renormalized energy-momentum tensor $\langle T_{\mu \nu} \rangle$ evaluated in a classical spacetime geometry. Only at a subsequent stage does one try to consider the back-reaction on the Einstein tensor, which, being coupled to a non-classical object like $\langle T_{\mu \nu} \rangle$, cannot remain undisturbed.

\subsection{Quantum corrected Lagrangian } \label{Quantum corrected Lagrangian_Sec}
   
The circular restricted three-body problem we are going to analyse consists of a body $A$ of mass $\alpha$ and a body $B$ of mass $\beta < \alpha$ moving under their mutual gravitational attraction and forming a two-body system in which their motion in known. $A$ and $B$ are called primaries and they can be considered as approximately spherical with a spherical symmetrical distribution of mass so that the attraction between two such bodies can be considered the same as that between two particles at their centres. We will suppose that $A$ coincides with the Earth whereas the role of $B$ is played by the Moon. The center of mass $C$ of the primaries moves uniformly in a straight line, and one can suppose it to be at rest without loss of generality. The initial conditions are such that the orbit of $B$ relative to $A$ is a circle, hence the orbit of each body relative to $C$ is a circle as well. Moreover, a third body, the planetoid $P$, moves in the plane of motion of $A$ and $B$. By hypothesis, $P$ is subjected to the quantum corrected Newtonian attraction of $A$ and $B$, but its mass $m$ is so small that it cannot affect their motion. The problem consists therefore in evaluating the motion of $P$ at any time \cite{BE14a,Pars,Szebehely67}.  
    
\begin{figure}
\centering
\includegraphics[scale=0.7]{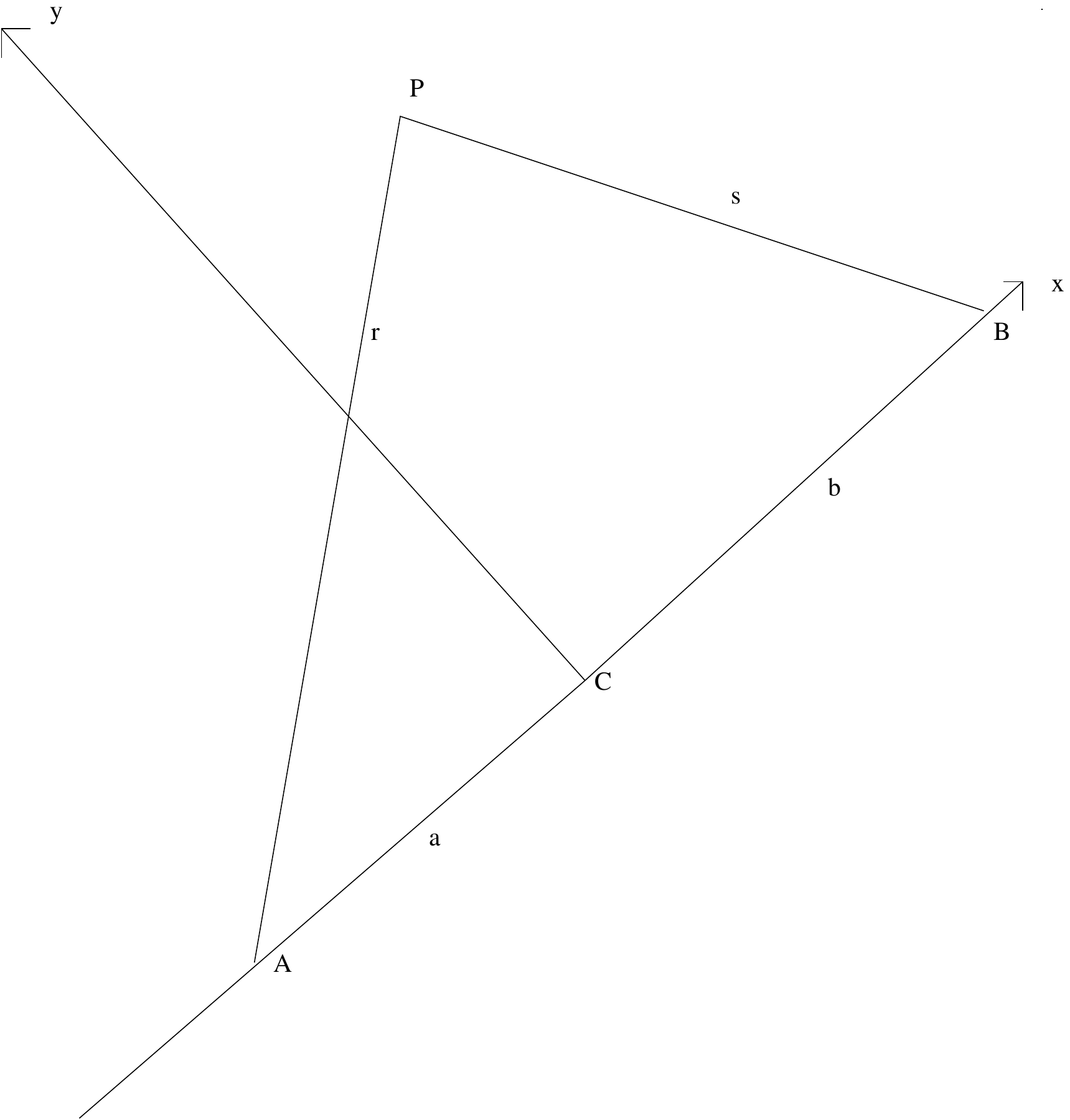}
\caption[The synodic coordinate system]{The synodic coordinate system showing the two primaries, $A$ and $B$, the center of mass $C$, and the planetoid $P$.}
\label{1a.pdf}
\end{figure}

The problem can be stated either in an inertial (fixed) coordinate system (called sidereal system) or in a rotating coordinate system (called synodic system). We will adopt the latest because it possesses the advantage that the motion of the primaries shows no explicit dependence on time. With reference to Fig. \ref{1a.pdf}, the synodic system has the center of mass $C$ as its origin, with $A$ and $B$ lying on the $x$-axis. The orbital plane normal to the total angular momentum coincides with the $x-y$ plane. Balance between the gravitational and centrifugal forces requires that
\begin{equation}
G \dfrac{\alpha \beta}{l^2} = \beta \, b \, \omega^2 = \alpha \, a \, \omega^2, \label{balance}
\end{equation}
where the length $AB$ has been denoted by $l$, $\omega$ is the common angular velocity of $A$ and $B$ and  
\begin{equation}
a={\beta \over (\alpha+\beta)}l, \; \; \; \; \; \; \; \; \; \; \; \; \; \; \;    b={\alpha \over (\alpha+\beta)}l. \label{2.2a}
\end{equation}
The quantity $\omega$ is called in celestial mechanics mean motion and it is given by
\begin{equation}
\omega=\sqrt{{G(\alpha+\beta)\over l^{3}}},
\label{2.1a}
\end{equation}
which represents the mathematical content of Kepler's third law. Note that Eq. (\ref{balance}) means that we are choosing to neglect any correction, either classical
or quantum, to the Newtonian potential between the primaries. Thus, $A$ and $B$ are permanently at rest, relative to the rotating axes, at the points of coordinates $(-a,0)$ and $(b,0)$, respectively \cite{BE14a,Pars}. The motion of the planetoid at $P(x,y)$ is the same as it would be if $A$ and $B$ were constrained to move as they do, hence the kinetic energy reads as
\begin{equation}
T={m \over 2}[({\dot x}-y \omega)^{2}+({\dot y}+x \omega)^{2}].
\end{equation}
In order to apply the effective field theories point of view to the Earth-Moon system, we need to employ the quantum corrected potential (cf. Eqs. (\ref{1.2b})--(\ref{1.4b})) in the dynamical equations describing the motion of the planetoid. Therefore, on denoting by $r$ the distance $AP$ and by $s$ the distance $BP$, i.e.,
\begin{equation}
\begin{split}
& r^{2}=(x+a)^{2}+y^{2}, \\ 
& s^{2}=(x-b)^{2}+y^{2}, 
\label{2.4a}
\end{split}
\end{equation}
the interaction potential is taken to be \cite{BE14a}
\begin{equation}
V=-{G m \alpha \over r}\left(1+{k_{1}\over r}+{k_{2}\over r^{2}}\right) -{Gm \beta \over s}\left(1+{k_{3}\over s}+{k_{4}\over s^{2}}\right),
\label{2.5a}
\end{equation}
where
\begin{equation}
k_{1}=\kappa_{1}{G(m+\alpha)\over c^{2}}= \kappa_1 \left( R_m + R_\alpha \right), \label{2.6a}
\end{equation}
\begin{equation}
k_{2}=k_{4}=\kappa_{2} (l_{P})^{2},
\end{equation}
\begin{equation}
k_{3}=\kappa_{1}{G(m+\beta)\over c^{2}}=\kappa_1 \left( R_m + R_\beta \right), \label{2.8a}
\end{equation}
and the classical Newtonian domain is recovered in the limits
\begin{equation}
k_1 \rightarrow 0, \; \; \; \; \; \; \; \; \; \; \; k_2 \rightarrow 0,\; \; \; \; \; \; \; \; \; \; \; k_3 \rightarrow 0.
\end{equation}
The quantum corrected Lagrangian underlying the dynamics of $P$ is therefore assumed to take the form \cite{BE14a} 
\begin{equation}
\begin{split}
{\mathcal{L} \over m} &= {1\over 2}({\dot x}^{2}+{\dot y}^{2})
+\omega (x{\dot y}-y{\dot x})+{1\over 2}\omega^{2}(x^{2}+y^{2})  + {G \alpha \over r}\left(1+{k_{1}\over r}+{k_{2}\over r^{2}}
\right)+{G \beta \over s}\left(1+{k_{3}\over s}
+{k_{2}\over s^{2}}\right) \\
&= T-V=T_{2}+T_{1}+T_{0}-V,
\end{split} \label{2.11a}
\end{equation}
having denoted by $T_{n}$ the part of $T$ containing $n$-th order derivatives of $x$ or $y$. In particular, $T_1$ is related to Coriolis force while $T_0$ to centrifugal force. The resulting Lagrange equations of motion read as \cite{BE14a}
\begin{equation}
{\ddot x}-2 \omega {\dot y}=G {\partial U \over \partial x}, \label{2.15a}
\end{equation}
\begin{equation}
{\ddot y}+2 \omega {\dot x}=G {\partial U \over \partial y}, \label{2.16a}
\end{equation}
having set 
\begin{equation}
U(x,y) \equiv {1\over 2}{(\alpha+\beta)\over l^{3}}(x^{2}+y^{2}) +{\alpha \over r}\left(1+{k_{1}\over r}+{k_{2}\over r^{2}} \right)+{\beta \over s}\left(1+{k_{3}\over s}
+{k_{2}\over s^{2}}\right). \label{2.14a}
\end{equation}
$U$ is usually referred to as full potential. Since the Lagrangian function (\ref{2.11a}) does not depend on time explicitly, its Jacobi integral exists. If we multiply Eq. (\ref{2.15a}) by $\dot{x}$ and (\ref{2.16a}) by $\dot{y}$ and then add the resulting equations member by member we obtain the relation
\begin{equation}
\dot{x} {\ddot x} + \dot{y} {\ddot y}= G \left( \dot{x} {\partial U \over \partial x} + \dot{y} {\partial U \over \partial y} \right) = G \dfrac{{\rm d}U}{{\rm d}t},
\end{equation}
which, once integrated with respect to $t$, gives
\begin{equation}
\dfrac{1}{2} \left({\dot x}^{2}+{\dot y}^{2}\right) = GU+\mathcal{J},
\end{equation}
$\mathcal{J}$ representing an integration constant called Jacobi integral (or Jacobi constant or integral of energy). By virtue of (\ref{2.1a}) and (\ref{2.5a}) we have
\begin{equation}
GU=T_0-V, 
\end{equation}
and hence the Jacobi integral assumes the form
\begin{equation}
\mathcal{J}=T_{2}+V-T_{0}. \label{2.12a}
\end{equation}
It is to be noted that $\mathcal{J}$ is the only conserved quantity of the circular restricted three-body problem, depending only on the initial conditions and not on time. By noting that $\mathcal{J}$ represents the energy per unit mass of P\footnote{Recall that Coriolis force is perpendicular to the trajectory.}, the total energy of the system of three bodies reads as
\begin{equation}
H_{\rm tot} =\mathcal{J} - G \dfrac{\alpha \beta}{l},
\end{equation}
which turns out to be a constant in the synodic system. By bearing in mind that from the above relations we can write $\mathcal{J}=T_{2}-GU$, one has the simple but non-trivial restriction according to which the motion of $P$ is only possible where
\begin{equation}
GU+\mathcal{J}=T_{2}>0 \Longrightarrow U > -{\mathcal{J} \over G}.
\end{equation}

Had we chosen the sidereal system $X,Y$ instead of the synodic one, Eqs. (\ref{2.15a}) and (\ref{2.16a}) would have been replaced by
\begin{equation}
{\ddot X}=G {\partial \tilde{U} \over \partial X}, 
\end{equation}
\begin{equation}
{\ddot Y}=G {\partial \tilde{U} \over \partial Y}, 
\end{equation}
where
\begin{equation}
\tilde{U}(X,Y,t) \equiv  {\alpha \over \tilde{r}}\left(1+{k_{1}\over \tilde{r}}+{k_{2}\over \tilde{r}^{2}} \right)+{\beta \over \tilde{s}}\left(1+{k_{3}\over \tilde{s}}
+{k_{2}\over \tilde{s}^{2}}\right)=-\dfrac{\tilde{V}(X,Y,t)}{G},
\end{equation}
where $\tilde{r}(X,Y,t)$ and $\tilde{s}(X,Y,t)$, being the time-dependent distances from the planetoid of $A$ and $B$, respectively, introduce the time explicitly in the equations of motion. The coordinate transformation between the sidereal and the synodic systems is given by the well-known rotation relation
\begin{equation}
\begin{split}
& X= x \cos \left( \omega t \right) - y \sin \left( \omega t \right), \\
& Y= x \sin \left( \omega t \right) + y \cos \left( \omega t \right), 
\end{split}
\end{equation}   
$\omega t$ being the angle that the $x$-axis forms with $X$-axis at time $t$, known also as the longitude of the body $A$. 

As we have just shown, in the sidereal coordinate system the potential $\tilde{U}(X,Y,t)$ contains the time variable explicitly and hence also the Lagrangian (as well as the Hamiltonian) will depend on time. As a consequence, the Jacobi integral will not exist. In fact, instead of (\ref{2.12a}) the only invariant relation of the problem is given by 
\begin{equation}
\dfrac{1}{2} \left(\dot{X}^2+ \dot{Y}^2 \right) = G \left( \tilde{U} - \int_{t_0}^{t} {\rm d}t \, \dfrac{\partial \tilde{U}}{\partial t} \right).
\end{equation}
Moreover, the energy (per unit mass) of $P$ has got the form
\begin{equation}
\mathfrak{h}=  \dfrac{1}{2} \left(\dot{X}^2+ \dot{Y}^2 \right) + \tilde{V},
\end{equation}
which is not a constant due to its time dependence, whereas, by recalling that the primaries undergo a purely classical gravitational interaction, the energy of the system $A+B$ is given by
\begin{equation}
\mathfrak{h}^\prime = \dfrac{1}{2} \omega^2 \left( \alpha \, a^2 + \beta \, b^2 \right) - G \dfrac{\alpha \beta }{l},
\end{equation} 
where, bearing in mind Eqs. (\ref{2.2a}) and (\ref{2.1a}), the kinetic term can be written as
\begin{equation}
\dfrac{1}{2} \omega^2 \left( \alpha \, a^2 + \beta \, b^2 \right) = \dfrac{1}{2}G \dfrac{\alpha \beta }{l},
\end{equation}
and hence $\mathfrak{h}^\prime$ is constant and reads as
\begin{equation}
\mathfrak{h}^\prime = -\dfrac{1}{2} G \dfrac{\alpha \beta }{l},
\end{equation} 
which means that the total energy 
\begin{equation}
\tilde{H}_{\rm tot} = m \mathfrak{h} + \mathfrak{h}^\prime,
\end{equation}
is not a constant because $\mathfrak{h}$ depends on time. The reason is due to the fact that we have chosen to neglect the effects of the planetoid on the motion of $A$ and $B$, creating a dynamical situation that exists, strictly speaking, only when $m=0$. In fact, if this condition is fulfilled we have that $H_{\rm tot}=\mathfrak{h}^\prime={\rm constant}$. We will see that for the full three-body problem the total energy of the system is conserved, since the potential energy function does not depend explicitly on time (Sec. \ref{classical_integrals_Sec}). 

\subsection{Lyapunov definition of stability} \label{Lagrangian_points_Sec} 

\begin{figure}
\centering
\includegraphics[scale=0.7]{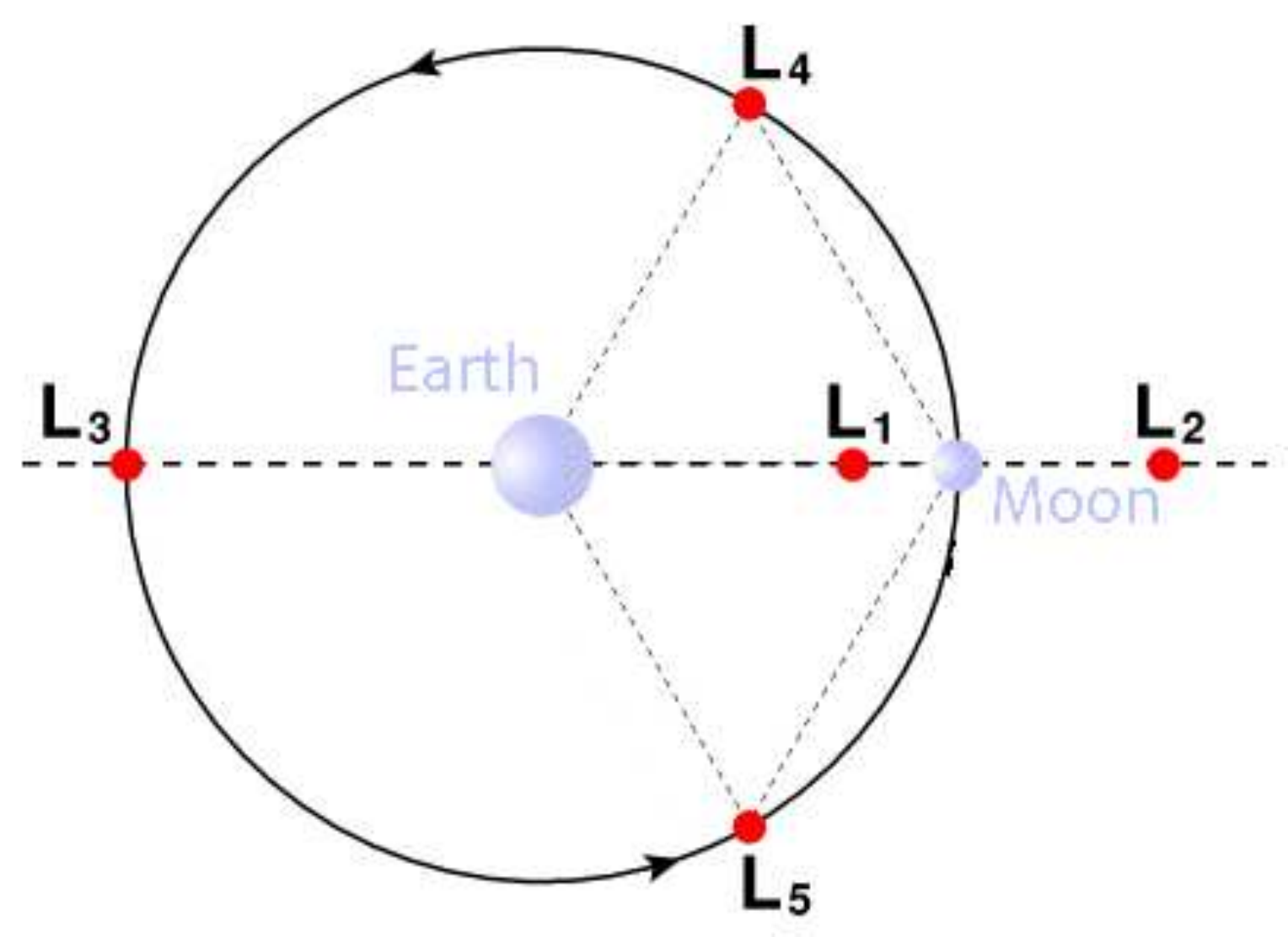}
\caption[The Lagrangian points of the Earth-Moon system in Newtonian theory]{A pictorial representation of the position of Lagrangian points in the Earth-Moon system as expected by Newtonian theory. It is possible to appreciate that $L_4$ and $L_5$ define an equilateral triangle.}
\label{Lagrangian_points.pdf}
\end{figure}

In the space surrounding two bodies that orbit about their mutual mass center there are five points where a third body will remain in equilibrium under the gravitational attraction of the other two bodies. From a physical point of view, these five equilibrium solutions represent points where the forces acting on the small planetoid in the rotating system are balanced. Since the solutions are stationary, there is no motion relative to the synodic system and hence the Coriolis force vanishes: only the gravitational and the centrifugal forces are to be considered. These equilibrium points are called Lagrangian points (or libration points) in honour of Joseph Lagrange, who discovered them in 1772 while studying the restricted problem formed by the Sun-Jupiter system. Lagrangian points can be divided into two groups: the collinear ones\footnote{Sometimes these points are referred to as Euler collinear solution.} ($L_1$, $L_2$, and $L_3$), which lie on the line joining $A$ to $B$ and turn out to be unstable and the non-collinear ones ($L_4$ and $L_5$), which instead are stable at first order (as will be proved in Sec. \ref{stability_Sec}). Lagrangian points of the Earth-Moon system within the context of Newtonian theory are depicted in Fig. \ref{Lagrangian_points.pdf}.

The discovery of the physical realization of the equilibrium points theorized by Lagrange is represented by the Trojan group of asteroids and it began only in 1906 thanks to the astronomer Max Wolf with the first-seen member of this group, called 588 Achilles, which is located near the triangular libration point of the Sun-Jupiter system. Today we know that there are $3898$ known Trojans at the triangular Lagrangian point $L_4$ and $2049$ at $L_5$ \cite{MPC}. In the sixties, simultaneously with the increased interest in space explorations, the question of existence of Lagrangian points with respect to other primaries, especially for the Earth-Moon system, arose quite naturally. In fact, if there are stable stationary solutions for various primary combinations, then from a practical point of view placing observational platforms at these points becomes feasible, especially in a really close and accessible system like the Earth-Moon system, which is also the most convenient system from an economic point of view. While the Sun-Jupiter system clearly possesses a collection of asteroids at the triangular libration points, the ability of the Earth-Moon system to collect debris or dust at the corresponding points and in what is called Kordylewski clouds is still in question (see Ref. \cite{Freitas} for further details). The major perturbing effect on the Trojans is represented by Saturn, while the stabilizing forces come from the Sun and Jupiter. The major perturbation on the Earth-Moon libration clouds is the Sun and the stabilizing effects are derived from the Earth and the Moon. This explains why the existence of accumulated material at $L_4$ or $L_5$ in the Earth-Moon system is not so obvious. Bodies at the triangular libration points of the system consisting of the Sun and another planet would face the perturbations from Jupiter; therefore, it is not surprising that the only currently known material accumulation is confined to the Sun-Jupiter system, although some asteroids were found also in the Sun-Earth system around the libration point $L_4$, as is shown by recent observations \cite{Connors}. 
As far as the collinear Lagrangian points for the Earth-Moon system are concerned, we know that $L_1$ allows comparatively easy access to Lunar and Earth orbits with minimal change in velocity and has this as an advantage to position a half-way manned space station intended to help transport cargo and personnel to the Moon and backwards, whereas $L_2$ would be a good location for a communications satellite covering the Moon's far side and would be an ideal location for a propellant depot as part of the proposed depot-based space transportation architecture \cite{Zegler}.

We now give the definition of equilibrium point of a dynamical system. Consider in the space $\mathbb{R}^n$ the system of first-order homogeneous ordinary differential equations
\begin{equation}
\dot{\bold{x}}= \bold{X}(\bold{x}), \label{RomanoA1}
\end{equation}
whose solution $\bold{x}(t)$ represents in $\mathbb{R}^n$ an integral curve of $\bold{X}(\bold{x})$ (for further details see Appendix \ref{Fundamental_App}). The system (\ref{RomanoA1}) is called autonomous due to its independence of the time variable. Recall that it is always possible to transform the set of $k$ second-order Lagrangian equations
\begin{equation}
\ddot{q}_i=Q_i\left(q_1,\dots,q_k,\dot{q}_1,\dots,\dot{q}_k \right), \; \; \; \; (i=1,\dots,k),\;(2k=n), 
\end{equation}
in the first-order system (\ref{RomanoA1}) by putting
\begin{equation}
\begin{split}
& q_i = x_i , \\
& \dot{q}_i=x_{i+k}.
\end{split}
\end{equation}
By employing the tools of mathematical analysis \cite{Whittaker-Watson}, it is possible to prove that if the vector field $\bold{X}(\bold{x}$) is continuous in the open subset $ \mathbb{U}\subseteq {\mathbb R}^n$, then, once the initial condition
\begin{equation}
\bold{x}(t_0)=\bold{x}_0, \label{RomanoA2}
\end{equation}
is assigned, at most one solution $\bold{x}(t,t_0,\bold{x}_0)$ of the Cauchy problem (\ref{RomanoA1}) and (\ref{RomanoA2}) exists at least locally, i.e., with $t \in (t_0-\delta,t_0+\delta)$, $\delta>0$. If $\bold{X}(\bold{x})$ turns out to be a local Lipschitz function in $\mathbb{U}$\footnote{$\bold{X}(\bold{x})$ represents a local Lipschitz function if it has a limited growth. Formally, the following condition must hold: $\forall \bold{x}_0 \in \mathbb{U}$ there exist an open subset $\mathbb{U}_0\subseteq \mathbb{U}$ and a constant $K>0$ such that
\begin{equation}
\vert \bold{X}(\bold{x})- \bold{X}(\bold{y}) \vert \leq K \vert \bold{x} -  \bold{y} \vert, \; \; \; \;\; \;  \forall \, \bold{x}, \bold{y} \in \mathbb{U}_0.
\end{equation}
}, then there exists locally an unique solution of the above-mentioned Cauchy problem. If the solution $\bold{x}(t,t_0,\bold{x}_0)$ is such that $t\in \mathbb{R}$, we say that we have a global solution. Moreover, since the solutions of (\ref{RomanoA1}) are independent from $t_0$, we can set $t_0=0$ and hence we can simply set hereafter $\bold{x}(t,0,\bold{x}_0) \equiv \bold{x}(t,\bold{x}_0)$. Given two solutions $\bold{x}(t,\bold{x}_{0_1})$ and $\bold{x}(t,\bold{x}_{0_2})$ of (\ref{RomanoA1}) in the interval $[0,\delta]$, it is possible to show that there exists a constant $\tilde{K}>0$ such that
\begin{equation}
\vert \bold{x}(t,\bold{x}_{0_1}) - \bold{x}(t,\bold{x}_{0_2}) \vert \leq \vert \bold{x}_{0_1}- \bold{x}_{0_2} \vert {\rm e}^{\tilde{K}t}.
\end{equation}  
In particular, the last condition implies that the solution of (\ref{RomanoA1}) is continuous with respect to the initial value. A point $\bold{x}_* \in \mathbb{U}$ is called equilibrium (or stationary or critical) point if \cite{Szebehely67}
\begin{equation}
\bold{X}(\bold{x}_*)= \bold{0},
\end{equation}
which means that $\bold{x}_*$ can be obtained from (\ref{RomanoA1}) from the condition $\dot{\bold{x}}=\bold{0}$. Therefore, according to what we have said above, $\bold{x}(t,\bold{x}_*)$ is the unique solution of the Cauchy problem (\ref{RomanoA1}) and (\ref{RomanoA2}) with initial value $\bold{x}_*$. The point $\bold{x}_*$ represents a stable equilibrium point (in the sense of Lyapunov) if $\forall \epsilon >0$, $\exists \, \delta(\epsilon)>0$ with $0 < \delta(\epsilon) < \epsilon$ such that, when the disturbances satisfy
\begin{equation}
\vert \bold{x}_0 - \bold{x}_* \vert \leq \delta(\epsilon), 
\end{equation}
then $\forall t > t_0=0$
\begin{equation}
\vert \bold{x}(t,\bold{x}_0) - \bold{x}_* \vert < \epsilon.
\end{equation}
In other words, it is always possible to define {\it a priori} a confinement of the solution. Moreover, $\bold{x}_*$ is said to be asymptotically stable if it is stable and if
\begin{equation}
\lim_{t \rightarrow + \infty} \bold{x}(t,\bold{x}_0) = \bold{x}_*.
\end{equation}
Lastly, $\bold{x}_*$ is unstable if it is not stable, i.e., $\forall \epsilon >0$, $\exists \, \delta(\epsilon)>0$ with $0 < \delta(\epsilon) < \epsilon$ and, once a neighbourhood of $\bold{x}_*$ having radius $\delta(\epsilon)$ is considered, there exist at least a point $\bold{x}_0$ and an instant $\bar{t}$ such that the corresponding solution $\bold{x}(t,\bold{x}_0)$ leaves such a neighbourhood for $t>\bar{t}$. 

It is clear from the above definitions that the analysis of the equilibrium points of (\ref{RomanoA1}) requires the knowledge of its solutions in the neighbourhood of $\bold{x}_*$. Unluckily, such a circumstance is quite unusual in physics, except for few particular cases. Thence, it becomes rather essential to establish some feasible criteria that allow the description of the stability of the system (\ref{RomanoA1}) without resorting to the explicit acquaintance of its solution. An important role for this purpose is fulfilled by the stability criterion set up by Lyapunov, which can be stated as follows \cite{Romano}:
\newtheorem*{Lyapunov1}{Lyapunov stability criterion}
\begin{Lyapunov1} 
Let $\mathfrak{V}(\bold{x}): S(\bar{\epsilon}) \rightarrow \mathbb{R} $ be a smooth function in the sphere $ S(\bar{\epsilon})$ centred in $\bold{x}_*$ and having radius equals to $\bar{\epsilon}$ such that 
\begin{equation}
\mathfrak{V}(\bold{x}_*)=0,
\end{equation}
\begin{equation}
\mathfrak{V}(\bold{x})>\mathfrak{V}(\bold{x}_*), \; \; \; \; \forall \, \bold{x} \in  S(\bar{\epsilon}).
\end{equation}
If
\begin{equation}
\dot{\mathfrak{V}}\bigl(\bold{x}(t,\bold{x}_0)\bigr) =0,
\end{equation}
for all solutions of (\ref{RomanoA1}) having initial data in $\mathbb{U}-\{\bold{x}_* \}$, then $\bold{x}_* $ represents a stable equilibrium point. Instead, if
\begin{equation}
\dot{\mathfrak{V}}\bigl(\bold{x}(t,\bold{x}_0)\bigr) <0,\; \; \; \; \forall \, \bold{x} \in  S(\bar{\epsilon}) - \{ \bold{x}_* \},
\end{equation} 
then $\bold{x}_* $ is an asymptotically stable equilibrium point. Finally, if
\begin{equation}
\dot{\mathfrak{V}}\bigl(\bold{x}(t,\bold{x}_0)\bigr) >0,\; \; \; \; \forall \, \bold{x} \in  S(\bar{\epsilon}) - \{ \bold{x}_* \},
\end{equation} 
then $\bold{x}_* $ turns out to be unstable.
\end{Lyapunov1}
The function $\mathfrak{V}$, called Lyapunov function, possesses some analogies to the potential function of classical dynamics. It is important to stress that it is possible to invoke Lyapunov stability criterion without explicit knowledge of solutions of (\ref{RomanoA1}), because from this equation it follows easily
\begin{equation}
\dot{\mathfrak{V}}\bigl(\bold{x}(t,\bold{x}_0)\bigr) = \bold{X}(\bold{x}) \cdot {\rm grad} \biggl[ \mathfrak{V}\bigl(\bold{x}(t,\bold{x}_0)\bigr)\biggr]. 
\end{equation}

Lyapunov stability criterion plays a major role in the proof of the following crucial theorem \cite{Romano}:
\newtheorem*{Dirichlet1}{Dirichlet stability criterion}
\begin{Dirichlet1}
Consider a dynamical system having a potential energy function $U$. If $\bold{x}_* $ represents a local absolute minimum for $U$, then $\bold{x}_* $ is a stable equilibrium point for the system. Instead, if $\bold{x}_* $ represents only a critical point for the function $U$, then $\bold{x}_* $ is an equilibrium point (stable or unstable or asymptotically stable).
\end{Dirichlet1}

Driven both by the Dirichlet theorem and by the fact that, as we have pointed out in Sec. \ref{Sec. scattering-bound-state potential}, the quantum corrected potential (\ref{1.2b}) is suitable for the analysis of equilibrium points but not for the description of the orbits of celestial bodies (cf. Eq. (\ref{1.5b})), the method we will adopt throughout this thesis for the analysis of the modifications occurring to the Newtonian equilibrium points of a dynamical system will consist in the evaluation of the zeroes of the gradient of its potential energy function. In the case of the quantum corrected restricted three-body problem we have just introduced, the potential  energy is given by Eq. (\ref{2.14a}), whereas in Secs. \ref{Theorethical predictions GR_Sec} and \ref{New quantum theory Sec}, where a description employing the framework of general relativity will be performed, the potential will be represented by Eqs. (\ref{4.9c}) and (implicitly) (\ref{5.11c}). The technique involving the potential energy function is the one which has been exploited in Refs. \cite{BE14a, BE14b, BEDS15, testbed}. However, other methods have been developed in the literature for the analysis of equilibrium points in the context of general relativity and we mention the one pursued by the authors of Refs. \cite{Asada09,Yamada10,Yamada,Yamada15,Asadabook}, where the analysis of Lagrangian points at the first post-Newtonian order is given. In this case the starting point is represented by Einstein-Infeld-Hoffmann equation \cite{EIH,Landau-Lif} and not by a potential energy function, as we will explain in Sec. \ref{Corrections on the position of Lagrangian points_Sec}.

\subsection{Derivatives of the full potential}

\begin{figure}
\centering
\includegraphics[scale=0.7]{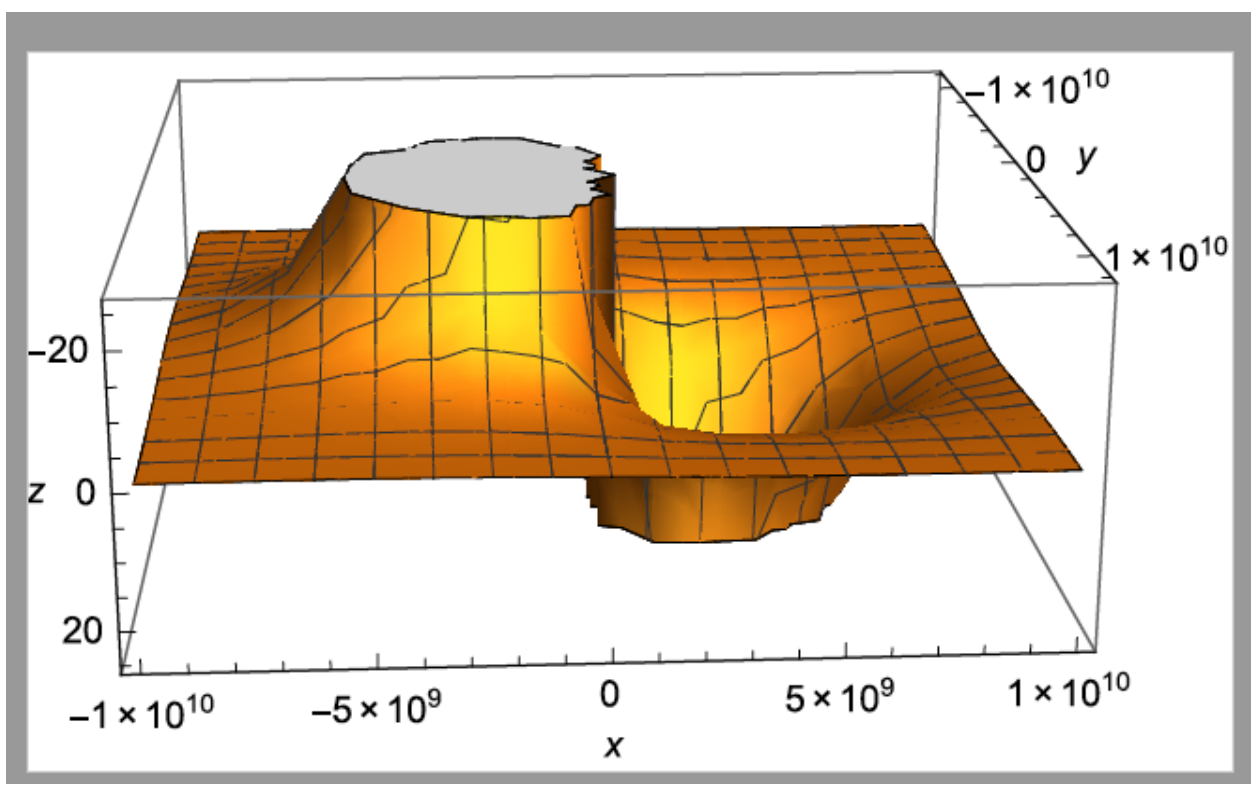}
\caption[Plot of $\left. \dfrac{\partial U(x,y)}{\partial x} \right\vert_{\lambda=0}$]{Plot of the partial derivative with respect to the
$x$-coordinate of the full potential $U(x,y)$ (see Eq. (\ref{2.14a})) obtained by setting
$\lambda=0$. The graph has been accomplished by choosing the constants $\kappa_1$ and $\kappa_2$ of the one-particle reducible potential (see Tab. \ref{kappa_tab}).} 
\label{2a.pdf}
\end{figure}

As we have seen in the last section, the equilibrium points, either stable or unstable, are points at which the full potential (\ref{2.14a}) is stationary, and hence one has
to study its first and second order partial derivatives. To begin, one finds \cite{BE14a}
\begin{equation}
{\partial U \over \partial x}=(\alpha+\beta){x \over l^{3}}
-{\alpha (x+a)\over r^{3}}\left(1+2{k_{1}\over r}
+3{k_{2}\over r^{2}}\right)
-{\beta(x-b)\over s^{3}}\left(1+2{k_{3}\over s}
+3{k_{2}\over s^{2}}\right). 
\label{3.1a}
\end{equation}
Thus, on using (\ref{2.2a}) and defining (cf. the classical formulas in Ref. \cite{Pars})
\begin{equation}
\lambda \equiv {(\alpha+\beta)\over l^{3}}
-{\alpha \over r^{3}}\left(1+2{k_{1}\over r}+3{k_{2}\over r^{2}}
\right)-{\beta \over s^{3}}\left(1+2{k_{3}\over s}
+3{k_{2}\over s^{2}}\right), \label{3.2a}
\end{equation}
one can re-express ${\partial U \over \partial x}$ in the form (see Fig. \ref{2a.pdf}) 
\begin{equation}
{\partial U \over \partial x}=\lambda x 
+{\alpha \beta l \over (\alpha + \beta)}\left[{1\over s^{3}}
\left(1+2{k_{3}\over s}+3{k_{2}\over s^{2}}\right)
-{1\over r^{3}}\left(1+2{k_{1}\over r}+3{k_{2}\over r^{2}}
\right)\right],  \label{3.3a}
\end{equation}
while, with the same notation, the other first derivative reads as
\begin{equation}
{\partial U \over \partial y}=\lambda y. \label{3.4a}
\end{equation}
For this to vanish, it is enough that either $y$ or $\lambda$ vanishes, in complete formal analogy with the classical case \cite{Pars}. When $y=0$, we find the (collinear) equilibrium points $L_1$, $L_2$, and $L_3$ lying on the line joining $A$ to $B$, while the condition $\lambda=0$ yields the libration points not lying on the line joining the two primaries, i.e., the triangular Lagrangian points $L_4$ and $L_5$. Second-order derivatives of $U$ along with their sign are important to understand
the nature of equilibrium points. For this purpose, we need the first derivatives of the function $\lambda$, which are found to be \cite{BE14a}
\begin{equation}
{\partial \lambda \over \partial x}={(x+a)\over r^{5}}\alpha
\left(3+8{k_{1}\over r}+15{k_{2}\over r^{2}}
\right)+{(x-b)\over s^{5}}\beta \left(3+8{k_{3}\over s}
+15{k_{2}\over s^{2}}\right),
\end{equation}
\begin{equation}
{\partial \lambda \over \partial y}=y \left[{\alpha \over r^{5}}
\left(3+8{k_{1}\over r}+15{k_{2}\over r^{2}}\right)
+{\beta \over s^{5}}\left(3+8{k_{3}\over s}
+15 {k_{2} \over s^{2}}\right)\right],
\end{equation}
by virtue of the identities (see Eq. (\ref{2.4a}))
\begin{equation}
\begin{split}
& {\partial r \over \partial x}={(x+a)\over r}, \\
& {\partial r \over \partial y}={y \over r}, \\
& {\partial s \over \partial x}={(x-b)\over s}, \\
& {\partial s \over \partial y}={y \over s}.
\end{split}
\end{equation}
The second order derivatives of $U$ are hence given by (see Figs. \ref{3a.pdf}, \ref{4a.pdf}, and \ref{5a.pdf}) \cite{BE14a}
\begin{equation}
{\partial^{2}U \over \partial x^{2}}=\lambda
+(x+a)^{2}{\alpha \over r^{5}}\left(3+8{k_{1}\over r}
+15 {k_{2}\over r^{2}}\right)
+(x-b)^{2}{\beta \over s^{5}}\left(3+8{k_{3}\over s}
+15 {k_{2}\over s^{2}}\right), \label{3.8a}
\end{equation}
\begin{equation}
{\partial^{2}U \over \partial x \partial y}
=y \left[{(x+a)\over r^{5}}\alpha \left(3+8{k_{1}\over r}
+15 {k_{2}\over r^{2}}\right)
+{(x-b)\over s^{5}}\beta \left(3+8{k_{3}\over s}
+15{k_{2}\over s^{2}}\right)\right], \label{3.9a}
\end{equation}
\begin{equation}
{\partial^{2}U \over \partial y^{2}}=\lambda+y^{2}\left[
{\alpha \over r^{5}}\left(3+8{k_{1}\over r}+15{k_{2}\over r^{2}}
\right)+{\beta \over s^{5}}\left(3+8{k_{3}\over s}
+15{k_{2}\over s^{2}}\right)\right]. \label{3.10a}
\end{equation}
\begin{figure}
\centering
\includegraphics[scale=0.7]{3a.pdf}
\caption[Plot of $\left. \dfrac{\partial^2 U(x,y)}{\partial x^2} \right\vert_{\lambda=0}$]{Plot of the partial derivative $U_{,xx}(x,y)$ obtained by setting
$\lambda=0$. The graph has been obtained by choosing the constants $\kappa_1$ and $\kappa_2$ of the scattering potential (see Tab. \ref{kappa_tab}).} 
\label{3a.pdf}
\end{figure}
\begin{figure}
\centering
\includegraphics[scale=0.7]{4a.pdf}
\caption[Plot of $\left. \dfrac{\partial^2 U(x,y)}{\partial x \partial y} \right\vert_{\lambda=0}$]{Plot of the partial derivative $U_{,xy}(x,y)$ obtained by setting
$\lambda=0$. The graph has been obtained by choosing the constants $\kappa_1$ and $\kappa_2$ of the bound-states potential (see Tab. \ref{kappa_tab}).} 
\label{4a.pdf}
\end{figure}
\begin{figure}
\centering
\includegraphics[scale=0.7]{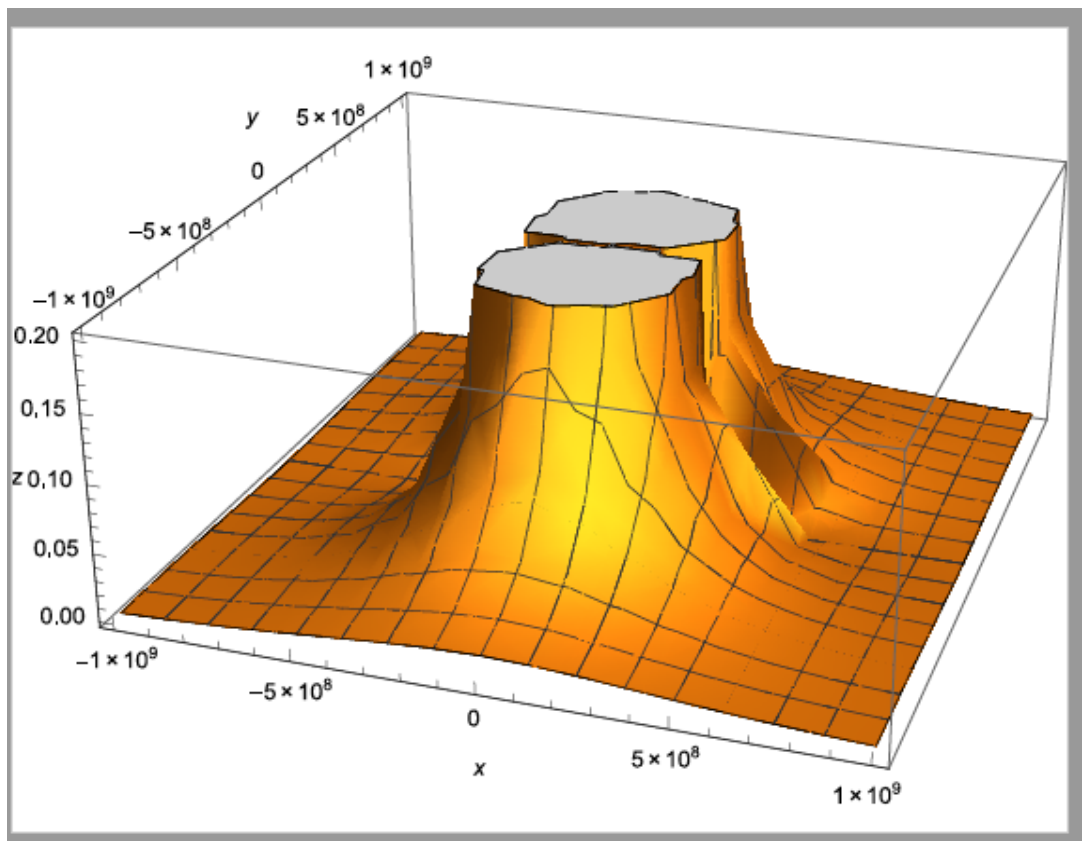}
\caption[Plot of $\left. \dfrac{\partial^2 U(x,y)}{\partial y^2} \right\vert_{\lambda=0}$]{Plot of the partial derivative $U_{,yy}(x,y)$ obtained by setting
$\lambda=0$. The graph has been obtained by choosing the constants $\kappa_1$ and $\kappa_2$ of the one-particle reducible potential (see Tab. \ref{kappa_tab}).} 
\label{5a.pdf}
\end{figure}

The analysis of the behaviour of the potential and its derivatives concerning the collinear Lagrangian points is intriguing and, as we will shortly see, sheds some light on the role that effective field theories could fulfil in celestial mechanics. To begin, we know that the line joining $A$ to $B$ is an axis having equation $y=0$, and it can be divided into 3 regions:
\begin{equation}
\begin{split}
&{\mathbb R}_{1}: \; x \in (-\infty,-a),  \\
&{\mathbb R}_{2}: \; x \in (-a,b),  \\
& {\mathbb R}_{3}: \; x \in (b,+\infty). 
\end{split}
\end{equation}
From Eq. (\ref{2.4a}) and the condition $y=0$ one has $r=|x+a|$, $s=|x-b|$, and hence Eqs. (\ref{3.2a}) and (\ref{3.8a}) yield \cite{BE14a}
\begin{equation}
\left . {\partial^{2}U \over \partial x^{2}} \right |_{y=0}
=\left[{(\alpha+\beta)\over l^{3}}+2{\alpha \over r^{3}}
+2{\beta \over s^{3}}\right]
+2{\alpha \over r^{4}}\left(3 k_{1}+6{k_{2}\over r}\right)
+2{\beta \over s^{4}}\left(3k_{3}+6{k_{2}\over s}\right). \label{4.1a}
\end{equation}
In Newtonian theory, since all terms in square brackets in (\ref{4.1a}) are positive, one concludes that $U_{,xx}$ is always positive on $y=0$. However, by virtue of (\ref{2.5a})--(\ref{2.8a}) and Tab. \ref{kappa_tab}, this may no longer be true in our case, if one adopts the one-particle reducible or the bound-states potential. In fact, the sufficient condition for preservation of the sign in Newtonian theory reads as \cite{BE14a}
\begin{equation}
\left(3k_{1}+6{k_{2}\over r}\right)+{\beta \over \alpha}
\left({r \over s}\right)^{4}\left(3k_{3}+6{k_{2}\over s}\right)>0,
\label{4.2a}
\end{equation}
which is however violated with the choice of the $\kappa_1$ and $\kappa_2$ coming from the one-particle reducible potential. Since the contribution of $k_2$ is overwhelmed by that of $k_1$ and $k_3$, it remains true that (\ref{4.2a}) is violated also by adopting the bound-states potential. 
\begin{figure}
\centering
\includegraphics[scale=0.7]{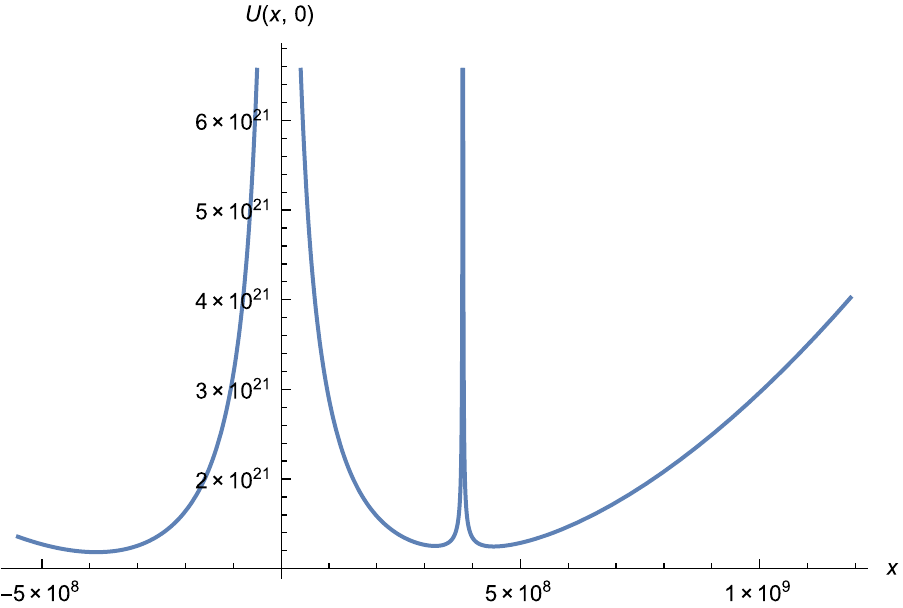}
\caption[Plot of the potential $U(x,0)$]{Plot of the potential $U(x,0)$. The three minima in $\mathbb{R_1}$, $\mathbb{R_2}$, and $\mathbb{R_3}$ correspond to $L_3$, $L_1$, and $L_2$, respectively. The vertical asymptote having negative $x$-coordinate represents the position of the body $A$, whereas the one with $x>0$ represents $B$. The three kinds of potential of Tab. \ref{kappa_tab} give no appreciable differences.} 
\label{6a.pdf}
\end{figure}

Note that the function $U(x,0)$ has, from (\ref{2.14a}), the limiting behavior (see Fig. \ref{6a.pdf})
\begin{equation}
\lim_{x \to -a}U(x,0)=\lim_{x \to b}U(x,0)=+\infty,
\end{equation}
\begin{equation}
\lim_{x \to -\infty}U(x,0)=\lim_{x \to +\infty}U(x,0)=+\infty.
\end{equation}
Moreover, $U_{,x}$ passes just once through the $x$-axis in each of the three regions ${\mathbb R}_{1}$, ${\mathbb R}_{2}$, and ${\mathbb R}_{3}$, which implies that $U$ has three minima and hence there exist three equilibrium points on $AB$, i.e., the Lagrangian points $L_{1}(x=l_{1})$, $L_{2}(x=l_{2})$, and $L_{3}(x=l_{3})$. To study the location of the equilibrium points, we note that
\begin{equation}
{r\over (x+a)}=(-1,1,1), \; \; \; \; \; \;\; \; \;   {s \over (x-b)}=(-1,-1,1),
\end{equation}
the three values on the right-hand side referring to ${\mathbb R}_{1}$, ${\mathbb R}_{2}$, and ${\mathbb R}_{3}$, respectively, so that in ${\mathbb R}_{1}$ for example (see (\ref{3.3a}))
\begin{equation}
{\partial U \over \partial x}=(\alpha+\beta){x \over l^{3}}
+{\alpha \over r^{2}}\left(1+2{k_{1}\over r}+3{k_{2}\over r^{2}}
\right)+{\beta \over s^{2}}\left(1+2{k_{3}\over s}
+3{k_{2}\over s^{2}}\right). \label{4.6a}
\end{equation}
At the point $x=-a-l$ one has $r=l,s=2l$, and from (\ref{2.2a}) and (\ref{4.6a}) one finds \cite{BE14a}
\begin{equation}
\left . {\partial U \over \partial x}\right |_{x=-a-l}
=-{7\over 4}{\beta \over l^{2}}
+{1\over l^{3}}\left[\alpha \left(2k_{1}+3{k_{2}\over l}\right)
+{\beta \over 4}\left(k_{3}+{3\over 4}{k_{2}\over l}\right)
\right]. \label{4.7a}
\end{equation}
In Newtonian theory, the sum in square brackets in (\ref{4.7a}) is absent and one can say that $U_{,x}$ is negative and hence $L_{3}$ lies between $x=-a-l$ and $x=-a$. In our model, for this to remain true, one should impose the sufficient condition \cite{BE14a}
\begin{equation}
2k_{1}+3{k_{2}\over l}+{\beta \over 4 \alpha}\left(k_{3}
+{3\over 4}{k_{2}\over l}\right) <0,
\label{4.8a}
\end{equation}
which is violated in the context of effective field theories if we adopt the scattering potential. In fact, the inequality (\ref{4.8a}) clearly matches with the one-particle reducible potential and, since $k_1$ and $k_3$ weigh more than $k_2$, it is in accordance also with the bound-states potential, despite in this latest case $\kappa_2 >0$. Similarly, to understand whether the equilibrium point $L_{1}$ lies between $C$ and $B$, one has to evaluate $U_{,x}$ at $C$,
where $r=a$, $s=b$, $x=y=0$, which yields, from (\ref{3.3a}), \cite{BE14a}
\begin{equation}
\left . {\partial U \over \partial x} \right |_{C}
=-(\alpha^{3}-\beta^{3}){(\alpha+\beta)^{2}\over 
\alpha^{2}\beta^{2}l^{2}}
-\left[{\alpha \over a^{3}}\left(2k_{1}+3{k_{2}\over a}
\right)+{\beta \over b^{3}}\left(2k_{3}+3{k_{2}\over b}\right)
\right].
\label{4.9a}
\end{equation}
In Newtonian theory, the sum in square brackets in (\ref{4.9a}) does not occur, and hence $\left . {\partial U \over \partial x} \right |_{C}$ is always negative. For this to hold true in our model, one has to impose the sufficient condition \cite{BE14a}
\begin{equation}
k_{1}+{3\over 2}{k_{2}\over a}+{\beta \over \alpha}
\left({a\over b}\right)^{3}\left(k_{3}+{3\over 2}
{k_{2}\over b}\right) >0,
\label{4.10a}
\end{equation}
which instead is not in accordance neither with the one-particle reducible potential nor with the bound-states one.

At this stage, despite the incompleteness of our analysis, we have already proved a simple but non-trivial result: not only can our model be used to discriminate among competing theories of effective gravity, but there exists no choice of signs for the coefficients $\kappa_1$ and $\kappa_2$ of the quantum corrected potential (\ref{1.2b}) for which all qualitative features of the restricted three-body problem in Newtonian theory remain unaffected. As far as we can see, this means that either we reject effective theories of gravity or we should expect them to be able to lead to testable effects. We will explain in the course of this thesis why we have good confidence that our model could predict measurable outcomes in the Earth-Moon system.  

Furthermore, from (\ref{3.10a}) we find \cite{BE14a}
\begin{equation}
 \left . {\partial^{2}U \over \partial y^{2}}
\right |_{L_{3}}=\lambda={\alpha \beta l \over (\alpha+\beta)}
{1\over x}\left({1\over r^{3}}-{1\over s^{3}}\right) +
 {1\over x}\left[2 \left({k_{1}\over r^{4}}-{k_{3}\over s^{4}}
\right)+3k_{2}\left({1\over r^{5}}-{1\over s^{5}}\right)\right].
\label{4.11a}
\end{equation}
The classical theory does not involve the sum of terms in square brackets in (\ref{4.11a}), and hence one points out that, since at $L_{3}$ $x$ is negative and $r<s$, the second derivative of $U$ at $L_{3}$ is negative \cite{Pars}. With similar arguments, we can arrive at the same conclusion also if we evaluate this derivative at $L_2$. In the quantum regime, however, the sufficient condition for this to be still valid, i.e., \cite{BE14a}
\begin{equation}
\left({k_{1}\over r^{4}}-{k_{3}\over s^{4}}\right)
+{3\over 2}k_{2}\left({1\over r^{5}}-{1\over s^{5}} \right)>0,
\label{4.12a}
\end{equation}
can be violated with the parameters characterizing the one-particle and the bound-states potentials. We note also that at $L_{1}$, where $r=x+a$ and $s=-(x-b)$, one has from (\ref{3.10a}) \cite{BE14a}
\begin{equation}
\left . {\partial^{2}U \over \partial y^{2}} \right |_{L_{1}}
={(\alpha +\beta)\over l^{3}}-{\alpha \over r^{3}}
-{\beta \over s^{3}}
-\left[2 \left(\alpha {k_{1}\over r^{4}}
+\beta {k_{3}\over s^{4}}\right)
+3k_{2}\left({\alpha \over r^{5}}+{\beta \over s^{5}}
\right)\right].
\label{4.13a}
\end{equation}
Since the Newtonian theory consists only of the first three terms in (\ref{4.13a}), one finds that $U_{,yy}$ is negative at $L_{1}$, because in ${\mathbb R}_{2}$ both $r$ and $s$ are less than $l$. In order to keep this feature also in the context of effective field theories, the following sufficient condition should hold \cite{BE14a}: 
\begin{equation}
\alpha{k_{1}\over r^{4}}+\beta{k_{3}\over s^{4}}
+{3\over 2}k_{2}\left({\alpha \over r^{5}}
+{\beta \over s^{5}}\right)>0,
\label{4.14a}
\end{equation}
which is respected only in the case of the scattering potential.

On reverting now to the graph of $U(x,0)$, as we have already pointed out, there are minima at $L_{1}$, $L_{2}$, and $L_{3}$, and we would like to determine at which of these three points $U(x,0)$ has the greatest value, and at which it has instead the least value. In Newtonian theory, one finds that $U(l_{1})>U(l_{2})>U(l_{3})$ \cite{Pars}. To establish the counterpart in our model, let $Q_{3}(x=q_{3})$ be the point of ${\mathbb R}_{3}$ whose distance from $B$ is equal to the distance of $L_{1}$ from $B$, i.e., $L_{1}B=BQ_{3}=j$. Thus, following patiently a number of cancellations, we find \cite{BE14a}
\begin{equation}
\begin{split}
 U(l_{1})-U(q_{3})  & =U(x=b-j,y=0,r=l-j,s=j) - U(x=b+j,y=0,r=l+j,s=j) \\
&= 2\alpha j \left[{1\over (l-j)^{2}}-{1\over l^{2}}\right]
+ {2 \alpha j \over (l-j)^{2}(l+j)^{2}}
\left[2k_{1}l +{k_{2}(j^{2}+3l^{2})\over (l^{2}-j^{2})}
\right].
\label{4.15a}
\end{split}
\end{equation}
In the classical theory, we have that $U(l_{1})-U(q_{3})>0$ because only the first term of (\ref{4.15a}) is present. In the quantum domain, for this to remain true, one should impose the following sufficient condition \cite{BE14a}
\begin{equation}
k_{1}+{1\over 2}k_{2}{(j^{2}+3l^{2})\over l(l^{2}-j^{2})}>0,
\label{4.16a}
\end{equation}
which fails to comply with the features of both the one-particle reducible and the bound-states potential. Lastly, let $Q_{1}(x=q_{1})$ be the point of ${\mathbb R}_{1}$ whose distance from $C$ is equal to the distance of $L_{2}$ from $C$, i.e., $Q_{1}C=CL_{2}=f$. Then we find \cite{BE14a}
\begin{equation}
\begin{split}
 & U(l_{2})-U(q_{1})  =U(x=f,y=0,r=x+a,s=x-b) - U(x=-f,y=0,r=x-a,s=-x+b)   \\
 &  = {2\alpha \beta l (b^{2}-a^{2})\over (\alpha+\beta)  (f^{2}-a^{2})(f^{2}-b^{2})} + {2\alpha \beta l \over (\alpha+\beta)(f^{2}-a^{2})^{2}
(f^{2}-b^{2})^{2}} \\
&\times \left\{ 2f \Bigr[k_{3}(f^{2}-a^{2})^{2}  
. -k_{1}(f^{2}-b^{2})^{2}\Bigr] 
 + {k_{2}\Bigr[(b^{2}+3f^{2})(f^{2}-a^{2})^{3}
-(a^{2}+3f^{2})(f^{2}-b^{2})^{3}\Bigr]
\over (f^{2}-a^{2})(f^{2}-b^{2})}  \right\}.
\label{4.17a}
\end{split}
\end{equation}
In Newtonian theory, the sum of terms in curly brackets in (\ref{4.17a}) does not occur, and one finds $U(l_{2})>U(q_{1})$. For this inequality to be saved in the quantum corrected case, one should impose the sufficient condition \cite{BE14a}
\begin{equation}
2f \Bigr[k_{3}(f^{2}-a^{2})^{2}-k_{1}(f^{2}-b^{2})^{2}\Bigr]
+{k_{2}\Bigr[(b^{2}+3f^{2})(f^{2}-a^{2})^{3}
-(a^{2}+3f^{2})(f^{2}-b^{2})^{3}\Bigr] \over
(f^{2}-a^{2})(f^{2}-b^{2})}>0.
\label{4.18a}
\end{equation}
This is more involved than (\ref{4.16a}), and it is not {\it a priori} so obvious whether a choice of signs of $\kappa_1$ and $\kappa_2$ leads always to its fulfilment.

To sum up, the seven {\it sufficient} conditions (\ref{4.2a}), (\ref{4.8a}), (\ref{4.10a}), (\ref{4.12a}), (\ref{4.14a}), (\ref{4.16a}), and (\ref{4.18a}) coming from the analysis of collinear libration points imply that some changes of qualitative features are unavoidable with respect to Newtonian theory, regardless of the choice of parameters made in (\ref{1.2b}) (or equivalently in Eq. (\ref{2.5a})), although five out of seven inequalities are fulfilled with the choice of the scattering potential.

\subsection{Non-collinear Lagrangian points} \label{noncoll_sec}

When the equilibrium points do not lie on the line joining $A$ to $B$, the coordinate $y$ is different from zero and hence the first derivative (\ref{3.4a}) vanishes because $\lambda=0$. On the other hand, the first derivative (\ref{3.3a}) should vanish as well, which then implies, by virtue of $\lambda=0$,
\begin{equation}
{1\over r^{3}}\left(1+2{k_{1}\over r}+3{k_{2}\over r^{2}}\right)
={1\over s^{3}}\left(1+2{k_{3}\over s}+3{k_{2}\over s^{2}}
\right).
\label{5.1a}
\end{equation}
Unlike Newtonian theory where $k_1=k_2=k_3=0$ \cite{Pars}, this equation is no longer solved by $r=s$. The definition (\ref{3.2a}), jointly with (\ref{5.1a}), makes it now possible to express the condition $\lambda=0$ in the form \cite{BE14a,BE14b,BEDS15}
\begin{equation}
{1\over l^{3}}={1\over r^{3}}+2{k_{1}\over r^{4}}
+3{k_{2}\over r^{5}}.
\end{equation}
This is an algebraic equation of fifth degree in the variable
\begin{equation}
w \equiv {1\over r},
\end{equation}
and we divide both sides by $3k_{2}$ and exploit the definitions (\ref{2.6a})-(\ref{2.8a}) to write it in the form 
\begin{equation}
\sum_{k=0}^{5}\zeta_{k}w^{k}=0, 
\label{5.4a}
\end{equation}
where \cite{BE14a,BE14b,BEDS15}
\begin{equation}
\zeta_{5} \equiv 1,
\label{5.5a}
\end{equation}
\begin{equation}
\zeta_{4} \equiv {2\over 3}{\kappa_{1}\over \kappa_{2}}
\dfrac{R_m + R_\alpha}{(l_{P})^2},
\label{5.6a}
\end{equation}
\begin{equation}
\zeta_{3} \equiv {1\over 3 \kappa_{2}}{1\over (l_{P})^{2}},
\label{5.7a}
\end{equation}
\begin{equation}
\zeta_{2}=\zeta_{1} \equiv 0,
\end{equation}
\begin{equation}
\zeta_{0} \equiv -{1\over 3 \kappa_{2}}{1\over (l_{P})^{2}l^{3}}.
\label{5.9a}
\end{equation}
Moreover, by defining
\begin{equation}
r(l) \equiv \dfrac{1}{w_{+}(l)}, \label{r(l)}
\end{equation}
$w_{+}(l)$ being the positive roots of (\ref{5.4a}), one can  evaluate $s(l)=s(r(l))$ from Eq. (\ref{5.1a}), which can be viewed as an algebraic equation of fifth degree in the variable 
\begin{equation}
u \equiv {1\over s},
\end{equation}
i.e., (cf. Eq. (\ref{5.4a}))
\begin{equation}
\sum_{k=0}^{5}{\widetilde \zeta}_{k}u^{k}=0,
\label{5.11a}
\end{equation}
where \cite{BE14a,BE14b,BEDS15}
\begin{equation}
{\widetilde \zeta}_{k}=\zeta_{k}, \; \; \; \; \;\; \forall k=0,1,2,3,5,
\label{5.12a}
\end{equation}
\begin{equation}
{\widetilde \zeta}_{4} \equiv {2\over 3}
{\kappa_{1}\over \kappa_{2}}
{R_m+R_\beta\over (l_{P})^{2}},
\label{5.13a}
\end{equation}
and similarly to Eq. (\ref{r(l)}), we define
\begin{equation}
s(l) \equiv \dfrac{1}{u_{+}(l)}. \label{s(l)}
\end{equation}
Since Eqs. (\ref{5.4a}) and (\ref{5.11a}) are of odd degree with real coefficients, the fundamental theorem of algebra guarantees the existence of at least one real solution, despite the lack of a general algebraic solution algorithm for all polynomial equations of degree greater than four (Abel-Ruffini theorem, see next section). Moreover, by virtue of the small term ${G\over c^{2}}$ appearing in the definition of $R_m$, $R_\alpha$, and $R_\beta$, the coefficients $\zeta_{4}$ and $\tilde{\zeta}_{4}$ play a negligible role both in the Earth-Moon system and in many other conceivable toy models of the restricted three-body problem, as is confirmed by detailed numerical checks. However, since the left-hand side of Eqs. (\ref{5.4a}) and (\ref{5.11a}) is a fairly simple polynomial function, the basic rules for studying functions of a real variable provide already a valuable information. For example, from (\ref{5.4a}) one has \cite{BE14b}
\begin{equation}
f'(w)=w^{2}(3 \zeta_{3}+4 \zeta_{4}w+5w^{2}),
\end{equation}
which therefore vanishes either at $w=0$ or at \cite{BE14b}
\begin{equation}
w_{1}=-{2 \over 5}\zeta_{4}+{1 \over 5}\sqrt{4(\zeta_{4})^{2}
-15 \zeta_{3}},
\end{equation}
\begin{equation}
w_{2}=-{2 \over 5}\zeta_{4}-{1 \over 5}\sqrt{4(\zeta_{4})^{2}
-15 \zeta_{3}}.
\end{equation}
By virtue of (\ref{5.6a}) and (\ref{5.7a}), such roots are real provided that \cite{BE14b}
\begin{equation}
(\kappa_{1})^{2} \geq   {45 \over 16} \left[ {c^{2} l_{P}\over G (m+\alpha)} \right]^2 \kappa_{2} , \label{A11f}
\end{equation}
which is satisfied in the Earth-Moon system by virtue of the small value of the Planck length, regardless of the choice of the potential. In particular, the roots $w_{1}$ and $w_{2}$ are both negative in the case of the scattering potential and both positive for the bound-states one (see Tab. \ref{kappa_tab}). By adopting the one-particle reducible potential we have
\begin{equation}
\begin{split}
& w_1= 84.6 \; {\rm m}^{-1},\\
& w_2 = -5.11 \times 10^{66} \; {\rm m}^{-1},
\end{split}
\end{equation}
while in the case of scattering potential we have found that
\begin{equation}
\begin{split}
& w_1=-28.2 \; {\rm m}^{-1},\\
& w_2 =-2.08 \times 10^{67} \; {\rm m}^{-1},
\end{split}
\end{equation}
finally the bound-states potential gives
\begin{equation}
\begin{split}
& w_1= 3.47 \times 10^{66} \; {\rm m}^{-1},\\
& w_2 = 169.1 \; {\rm m}^{-1}.
\end{split}
\end{equation}
Moreover, the second derivative of $f(w)$ reads as
\begin{equation}
f''(w)=2w(3 \zeta_{3}+6 \zeta_{4}w +10 w^{2}) \equiv 2w g(w).
\end{equation}
The point $w=0$ is therefore a flex point, while the sign of $f''(w)$ at $w_{1}$ and $w_{2}$, and hence maxima or minima of $f(w)$, is governed by the sign of the second degree polynomial $g(w) \equiv 3 \zeta_{3}+6 \zeta_{4} w +10 w^{2}$. Furthermore, Descartes's rule of signs, which states that the number of positive roots of an algebraic equation either equals to that of sign changes in its coefficients or is less than it by a multiple of two, can be applied to Eqs. (\ref{5.4a}) and (\ref{5.11a}). We have two, one or three sign changes in these equations according to whether the one-particle reducible, the scattering, or the bound-states potential is adopted, respectively. In fact, we have numerically checked that Eqs. (\ref{5.4a}) and (\ref{5.11a}) give two, one, and three positive roots in these three different cases. 

The Cartesian coordinates $(x,y)$ of the equilibrium points not lying along $AB$ can be found from the general formulas (\ref{2.4a}), i.e., 
\begin{equation}
r^{2}(l)=x^{2}+y^{2}+2ax+a^{2},
\label{5.15a}
\end{equation}
\begin{equation}
s^{2}(l)=x^{2}+y^{2}-2bx+b^{2}.
\label{5.16a}
\end{equation}
Subtraction of Eq. (\ref{5.16a}) from Eq. (\ref{5.15a}) yields \cite{BE14a,BE14b,BEDS15}
\begin{equation}
x(l) \equiv {(r^{2}(l)-s^{2}(l)+b^{2}-a^{2})\over 2(a+b)},
\label{5.17a}
\end{equation}
while $y(l)$ can be obtained from (\ref{5.15a}) in the form
\begin{equation}
y_{\pm}(l) \equiv \pm \sqrt{r^{2}(l)-x^{2}(l)-2ax(l)-a^{2}}.
\label{5.18a}
\end{equation}
Thus, the two non-collinear libration points of the Earth-Moon system assume coordinates
\begin{equation}
\begin{split}
& L_{4}(x(l),y_{+}(l)), \\ 
& L_{5}(x(l),y_{-}(l)).
\label{5.19a}
\end{split}
\end{equation}
In Newtonian theory, where $r=s$, the formula (\ref{5.19a}) reduces to the familiar \cite{Pars}
\begin{equation}
L_{4}\left({(\alpha-\beta)\over (\alpha+\beta)}{l\over 2},
{\sqrt{3}\over 2}l \right), \;
L_{5}\left({(\alpha-\beta)\over (\alpha+\beta)}{l\over 2},
-{\sqrt{3}\over 2}l \right), 
\label{5.20a}
\end{equation}
by virtue of (\ref{2.2a}), giving numerically \cite{BEDS15}
\begin{equation}
\begin{split}
& x_{cl}= 1.875281488022488 \times 10^{8} \; {\rm m}, \\
& y_{cl}=  \pm 3.329001652147382  \times 10^{8} \; {\rm m},
\label{5.20a_ter}
\end{split}
\end{equation}
and
\begin{equation}
r_{cl}=s_{cl}=3.844000000000000 \times 10^{8} \; {\rm m}.
\label{5.20a_bis}
\end{equation}
The geometric interpretation of the above formulas is simple but it has a non-trivial consequence: since at quantum level we have $r(l) \neq s(l)$, at the points $L_{4}$ and $L_{5}$ the planetoid is not at the same distance from the primaries, unlike Newtonian theory. Therefore, we can assert that to the equilateral libration points of Newtonian celestial mechanics there correspond points no longer exactly at the vertices of an equilateral triangle. In fact our quantum corrected model predicts a very tiny displacement from the case $r=s$, which can be expressed by the differences \cite{BE14a}
\begin{equation}
\begin{split}
& \bar{\delta}_{1}(l) \equiv x(l)-{(\alpha-\beta)\over (\alpha+\beta)}
{l\over 2}, \\
& \bar{\delta}_{2}(l) \equiv y_{+}(l)-{\sqrt{3}\over 2}l, \\
& \bar{\delta}_{3}(l) \equiv y_{-}(l)+{\sqrt{3}\over 2}l.
\label{5.21a}
\end{split}
\end{equation}
We have solved numerically Eqs. (\ref{5.4a}) and (\ref{5.11a}) by adopting the three different kinds of potential \cite{BE14a,BE14b,BEDS15}. The details are given in Tab. \ref{noncoll_details_tab}, whereas the quantum corrections on classical values are summarized in Tab. \ref{noncoll_corrections_tab}. In particular, the quantum differences between the distances of the planetoid from the primaries read as
\begin{equation}
\begin{split}
& r_Q-s_Q= -2.92 \; {\rm mm}, \\
& r_Q-s_Q=  8.76 \; {\rm mm}, \\
& r_Q-s_Q=  -1.46 \; {\rm mm},
\end{split}
\end{equation}
for the one-particle reducible, scattering, and bound-states potential, respectively. 
\begin{table}
\centering
\caption[Quantum details concerning non-collinear Lagrangian points]{The quantum values of the distances from the Earth and of the coordinates of the non-collinear Lagrangian points obtained by solving numerically Eqs. (\ref{5.4a}) and (\ref{5.11a}) for the three different types of potential.}
{\relsize{-2.49}
\renewcommand\arraystretch{2.0}
\begin{tabular}{|c|c|c|c|}
\hline
\multicolumn{4}{|c|}{Quantum details of non-collinear Lagrangian points}\\
\hline
\; \;  $L_i$ \; \;  &  One-particle reducible &  Scattering & Bound-states \\
\cline{1-4}
& $r_4= 3.843999999970434 \times 10^8 \; {\rm m}$& $r_4= 3.844000000088697 \times 10^8 \;  
{\rm m}$ & $r_4= 3.843999999985217    \times 10^8 \; {\rm m} $ \\ 
\cline{2-4}
$L_4$ & $x_4 = 1.875281487993286 \times 10^8 \; {\rm m} $  & $x_4 = 1.875281488110093  \times 10^8 \;  
{\rm m}$   & $x_4 = 1.875281488007887  \times 10^8 \;  {\rm m}$  \\
& $y_4= 3.329001652130102  \times 10^8 \; {\rm m} $ &   $y_4= 3.329001652199221 \times 10^8 \; {\rm m} $ &  $y_4= 3.329001652138742 \times 10^8 \; {\rm m} $ \\
\hline
& $r_5= 3.843999999970434 \times 10^8 \; {\rm m}$& $r_5= 3.844000000088697 \times 10^8 \;  
{\rm m}$ & $r_5= 3.843999999985217    \times 10^8 \; {\rm m} $ \\ 
\cline{2-4}
$L_5$ & $x_5 = 1.875281487993286 \times 10^8 \; {\rm m} $  & $x_5 = 1.875281488110093  \times 10^8 \;  
{\rm m}$   & $x_5 = 1.875281488007887  \times 10^8 \;  {\rm m}$  \\
& $y_5=- 3.329001652130102  \times 10^8 \; {\rm m} $ &   $y_5=- 3.329001652199221 \times 10^8 \; {\rm m} $ &  $y_5=- 3.329001652138742 \times 10^8 \; {\rm m} $ \\
\hline
\end{tabular} 
\label{noncoll_details_tab}
}
\end{table} 
\begin{table}
\centering
\caption[Quantum corrections on the position of Newtonian non-collinear Lagrangian points]{Quantum corrections on the position of Newtonian non-collinear Lagrangian points obtained by solving numerically Eqs. (\ref{5.4a}) and (\ref{5.11a}) for the three different types of potential.}
{
\renewcommand\arraystretch{2.0}
\begin{tabular}{|c|c|c|c|}
\hline
\multicolumn{4}{|c|}{Quantum corrections on Newtonian non-collinear Lagrangian points}\\
\hline
\; \;  $L_i$ \; \;  &  One-particle reducible &  Scattering & Bound-states \\
\cline{1-4}
& $r_Q-r_{cl}=-2.96  \; {\rm mm}$& $ r_Q-r_{cl}= 8.87 \; {\rm mm} $ & $r_Q-r_{cl}= -1.48 \; {\rm mm} $ \\ 
\cline{2-4}
$L_4$ & $x_Q-x_{cl}=  -2.92 \; {\rm mm}$  & $x_Q-x_{cl}=8.76  \; {\rm mm}$   & $x_Q-x_{cl}= -1.46  \; {\rm mm} $  \\
  & $y_Q-y_{cl}= -1.73 \; {\rm mm} $ &  $y_Q-y_{cl}= 5.18 \; {\rm mm} $ & $y_Q-y_{cl}=  -0.864 \; {\rm mm}$ \\
\hline
& $r_Q-r_{cl}= -2.96 \; {\rm mm}$& $ r_Q-r_{cl}=8.87  \; {\rm mm} $ & $r_Q-r_{cl}= -1.48  \; {\rm mm} $ \\ 
\cline{2-4}
$L_5$ & $x_Q-x_{cl}=-2.92  \; {\rm mm}$  & $x_Q-x_{cl}=  8.76 \; {\rm mm}$   & $x_Q-x_{cl}= -1.46  \; {\rm mm} $  \\
  & $y_Q-y_{cl}= 1.73 \; {\rm mm} $ &  $y_Q-y_{cl}= -5.18 \; {\rm mm} $ & $y_Q-y_{cl}= 0.864  \; {\rm mm}$ \\
\hline
\end{tabular} 
\label{noncoll_corrections_tab}
}
\end{table} 

In conclusion, our quantum model provides corrections on the position of non-collinear Lagrangian points which are of the order of few millimetres. Since sub-centimetre effects are accessible to modern laser ranging techniques (discussed in Sec. \ref{Laser_Ranging_Sec}), such modifications could be testable, but, for the sake of honesty, we should stress that a lot of difficulties due to several perturbations, of gravitational and non-gravitational nature, and to engineering issues regarding the positioning of satellites make this experiment hard, but not impossible, to accomplish. 

\subsection{Alternative route to quintic equations}\label{Alternative_route_5_Sec}

The results spelled out in the previous section have proved that the position of non-collinear libration points is ruled by a pair of fifth degree equations which have been solved by means of numerical tools. In order to double check our theoretical predictions, we now adopt another route to solve Eqs. (\ref{5.4a}) and (\ref{5.11a}) by employing the rich mathematical theory of quintic equations and their roots \cite{Hermite,Birkeland1924}.  In fact, since throughout this thesis we are going to need the roots of our algebraic equations up to the fifteenth or sixteenth decimal digit, the method exposed in this section will turn out to be very useful, because it leads to exact formulas for the roots of the quintic which are {\it then} evaluated numerically, which is possibly better than solving numerically the quintic from the beginning.  

Algebraic equations up to the fourth order can be solved with a finite number of radicals. This is no longer possible if the equation is of higher degree than four. In fact thanks to the joint work of Ruffini (in 1799) and Abel (in 1824), we know from the so-called Abel-Ruffini theorem (also known as Abel's impossibility theorem) that there is no general algebraic solution, i.e., solution in terms of radicals, to polynomial equations of degree five or higher with arbitrary coefficients\footnote{This result can be stated in an equivalent and elegant way by using Galois theory \cite{Artin}.}.  Therefore, the solution of such equations should be given in terms of more involved functions.
The first step in this direction was taken by Hermite in 1858 \cite{Hermite}, who, on exploiting the results achieved by Bring and Jerrard \cite{Bring,Jerrard}, proved that  
the quintic equation of the form
\begin{equation}
X^5-X-c=0,
\end{equation}
can be solved in terms of elliptic functions. However, in this section we will not employ the method outlined by Hermite, but we will exploit Birkeland theorem \cite{Birkeland1920}, according to which the roots of any quintic can be re-expressed through generalized hypergeometric functions. This wonderful theorem can be stated, in its general form, as follows:
\newtheorem*{Birkeland1}{Birkeland theorem}
\begin{Birkeland1} 
Consider the algebraic equation of the form
\begin{equation}
X^n=\mathfrak{g}\,X^{n^\prime} + \mathfrak{b}, \label{Birkeland1}
\end{equation}
where $n$ and $n^\prime$ are positive integer numbers, prime to each other, such that $n> n^\prime$ and $\mathfrak{g}$ and $\mathfrak{b}$ are constants. If the roots of (\ref{Birkeland1}) are considered as functions of the variable 
\begin{equation}
\sigma = (-1)^{n-n^\prime} \dfrac{n^n}{(n^\prime)^{n^\prime}(n-n^\prime)^{n-n^\prime}} \dfrac{\mathfrak{b}^{n-n^\prime}}{\mathfrak{g}^n}, \label{Birkeland3}
\end{equation}
they turn out to be integrals of the higher hypergeometric differential equation of order $n-1$
\begin{equation}
\left[\sigma^{n-2}(\sigma-1){{\rm d}^{n-1}\over {\rm d} \sigma^{n-1}} +\sigma^{n-3}(A_{1}\sigma-B_{1}){{\rm d}^{n-2}\over {\rm d}\sigma^{n-2}}
+ \dots +(A_{n-2}\sigma-B_{n-2}){{\rm d}\over {\rm d}\sigma} +\tilde{C} \right]\Lambda=0,\label{Birkeland2}
\end{equation}
where the quantities $A_i$, $B_i$ and $\tilde{C}$ are constants.
\end{Birkeland1}
The functions solving (\ref{Birkeland2}) are the above-mentioned generalized hypergeometric functions (or higher hypergeometric functions), which are defined as
\begin{equation}
F: \sigma \rightarrow F(\sigma) \equiv F \left(\begin{matrix}
a_{1}, & a_{2}, & ..., & a_{n-2}, & a_{n-1} \cr
b_{1}, & b_{2}, & ..., & b_{n-2}, & \sigma
\end{matrix}\right)
=\sum_{j=0}^{\infty}C_{j} \sigma^{j},
\label{2.19b}
\end{equation}
with the coefficients evaluated according to the rules
\begin{equation}
C_{0} \equiv 1, \; C_{j} \equiv 
{(a_{1},j)(a_{2},j)...(a_{n-2},j)(a_{n-1},j) \over (1,j)(b_{1},j)...(b_{n-3},j)(b_{n-2},j)},
\end{equation}
and where the symbol $(a_i,j)$ means
\begin{equation}
(a_i,j) \equiv a_i (a_i+1)(a_i+2)...(a_i+j-1).
\end{equation}
The constants $A_i$, $B_i$ and $\tilde{C}$ appearing in (\ref{Birkeland2}) are determined by the quantities $a_i$ and $b_i$. The higher hypergeometric functions are characterized by the property that the ratio of any two coefficients $C_{j+1}$ and $C_{j}$ is a rational function of $j$ with numerator and denominator having fixed degree independent of $j$ (and in fact they have degree equals to $n-1$). 

Let $i_0$ and $i_1$ be the integer numbers ($i_0 \leq n$, $i_1<n^\prime$) given by
\begin{equation}
(i_0-1)n^\prime +1=i_1 n,
\end{equation}
and let
\begin{equation}
a_i= \dfrac{i-1}{n}-\dfrac{1}{n(n-n^\prime)}, \; \; \; \; \; \; \; \; \; \; \; \; \; \; \; (i=1,2,\dots,i_0-1),
\end{equation}
\begin{equation}
a_i= \dfrac{i}{n}-\dfrac{1}{n(n-n^\prime)}, \; \; \; \; \; \; \; \; \; \; \; \; \; \; \; (i=i_0,i_0+1,\dots,n-1),
\end{equation}
\begin{equation}
b_i= \dfrac{i}{n^\prime}-\dfrac{1}{n^\prime(n-n^\prime)}, \; \; \; \; \; \; \; \; \; \; \; \; \; \; \; (i=1,2,\dots,i_1-1),
\end{equation}
\begin{equation}
b_i= \dfrac{i+1}{n^\prime}-\dfrac{1}{n^\prime(n-n^\prime)}, \; \; \; \; \; \; \; \; \; \; \; \; \; \; \; (i=i_1,i_1+1,\dots,n^\prime-1),
\end{equation}
\begin{equation}
b_i = \dfrac{i-n^\prime+1}{n-n^\prime}, \; \; \; \; \; \; \; \; \; \; \; \; \; \; \; (i=n^\prime,n^\prime+1,\dots,n-2),
\end{equation}
and also
\begin{equation}
F_0(\sigma )= F \left(\begin{matrix}
a_{1}, & a_{2}, & ..., & a_{n-2}, & a_{n-1} \cr
b_{1}, & b_{2}, & ..., & b_{n-2}, & \sigma
\end{matrix}\right),
\end{equation}
\begin{equation}
F_{n-i-1}(\sigma )= F \left(\begin{matrix}
a_{1}+1-b_i, & a_{2}+1-b_i, & ..., & a_{n-2}+1-b_i, & a_{n-1}+1-b_i \cr
2-b_{i}, &b_1+1- b_{i}, & ..., & b_{n-2}+1-b_i, & \sigma
\end{matrix}\right),
\end{equation}
\begin{equation}
\begin{split}
& \hat{\nu} = {\rm e}^{2 \pi \I/(n-n^\prime)}, \\
& \hat{\delta}= {\rm e}^{2 \pi \I/n^\prime}. 
\end{split}
\end{equation}
Let us suppose for simplicity $\mathfrak{g}=1$ in Eq. (\ref{Birkeland1}). Then, depending on the value assumed by the variable $\sigma$ defined in Eq. (\ref{Birkeland3}), from Birkeland theorem it follows that the roots of (\ref{Birkeland1}) are found through the following procedure \cite{Birkeland1924}: 
\begin{description}
\item[ $\vert \sigma \vert < 1 \; \; {\rm or} \; \; \sigma=1$ ] In this case the $n-n^\prime$ roots of (\ref{Birkeland1}) are given by
\begin{equation}
X_i = F_0 (\sigma) + \sum_{k=1}^{n-n^\prime - 1} \left[ \theta_k \, \hat{\nu}^{\I (1-k n^\prime)} \sigma^{n^\prime-b_{k} } F_{k} (\sigma)\right], \;\;\;\;\;(i=1,2,\dots,n-n^\prime), 
\label{Birkeland_root1}
\end{equation}
where the coefficients $\theta_k$ are numerical quantities. The remaining $n^\prime$ roots $X_{n-n^\prime+1}, \dots, X_n,$ (with $n^\prime <n-1$) are instead
\begin{equation}
\begin{split}
X_{n-n^\prime+i} = - \dfrac{n-n^\prime}{n^\prime} \theta_{k_0} \sigma^{n^\prime-b_{k_0}}F_{k_0}(\sigma) + \sum_{h=1}^{n^\prime-1} \left[ \hat{\Delta}_{n^\prime-h} \hat{\delta}^{\I (1-h n )} \sigma^{1-b_h} F_{n-h-1} (\sigma)\right], \\
  (i=1,2,\dots,n^\prime),
\end{split}
\end{equation}
where $\hat{\Delta}_{n^\prime-h}$ are constants and $k_0$ is defined by the congruence relation\footnote{In modular arithmetic we define a congruence relation in the following way \cite{Apostol}:\\
for a positive integer $n$, two integers $i$ and $j$ are said to be congruent modulo $n$, written
\begin{equation}
i \equiv j \; \; ({\rm mod}\; n), \nonumber
\end{equation}
if their difference $i-j$ is an integer multiple of $n$ (or equivalently $n$ divides $i-j$). The number $n$ is called the modulus of the congruence. For example,
\begin{equation}
44 \equiv 20 \; \; ({\rm mod}\; 12), \nonumber
\end{equation}
because $44-20=24$, which is a multiple of $12$.
}
\begin{equation}
n^\prime k_0 - 1 \equiv 0 \; \; \; \left( {\rm mod} \left(n-n^\prime \right) \right).
\label{Birkeland_root2}
\end{equation} 
\item[$\vert \sigma \vert >1$] In this second possibility we have the $n$ roots
\begin{equation}
X_i= \mathfrak{b}^{1/n} \sum_{s=1}^{n-n^\prime} \left[d_s \, \hat{\epsilon}^{\I \left[1+\left(s-1\right)n^\prime\right]} \, \sigma^{-a_s} \, \Psi_{s-1} \left( \dfrac{1}{\sigma} \right) \right], \; \; \; \; \; \; \; \; \; \; \; \;  (i=1,2,\dots,n),
\end{equation}
 where $\hat{\epsilon}={\rm e}^{2 \pi \I /n}$, $d_s$ are numerical constants and
\begin{equation}
\Psi_{s-1} \left( \dfrac{1}{\sigma} \right) = F \left(\begin{matrix}
a_{s}, & a_{s}+1-b_1, & ..., & a_{s}+1-b_{n-3}, & a_{s}+1-b_{n-2} \cr
a_{s}+1-a_1, & a_{s}+1-a_2, & ..., & a_{s}+1-a_{n-1}, & \dfrac{1}{\sigma}
\end{matrix}\right).
\end{equation}
\end{description}
When the $n$ roots of (\ref{Birkeland1}) are considered as functions of $\mathfrak{b}$, we can define the so-called critical points. They are given by those values of $\mathfrak{b}$ which make $\sigma=1$. Thus, we have the $n-n^\prime+1$ critical points
\begin{equation}
\mathfrak{b}_0=0, \; \; \mathfrak{b}_1=-\mathfrak{C} \, \hat{\nu}^{n^\prime}, \; \; \mathfrak{b}_2=-\mathfrak{C} \, \hat{\nu}^{2n^\prime}, \dots, \; \; \mathfrak{b}_{n-n^\prime} = -\mathfrak{C}, 
\label{Birkelandcritical1}
\end{equation}
where
\begin{equation}
\mathfrak{C}= \dfrac{n-n^\prime}{n^\prime} \left(\dfrac{n^\prime}{n} \right)^{\dfrac{n}{n-n^\prime}}>0.
\label{Birkelandcritical2}
\end{equation}
Critical points play an important role within this scheme. In fact, by defining how the roots of (\ref{Birkeland1}) vary and interchange around them, we automatically determine the symmetry (Lie) group of the differential equation (\ref{Birkeland2}), i.e., the continuous transformation group that takes each solution curve of (\ref{Birkeland2}) into another. To fix ideas, let us consider the case $\vert \sigma \vert <1$. Consider for each critical point a small cut which does not intersect the other points and which is not crossed when $\mathfrak{b}$ varies continuously. Then, it is possible to prove the following law for the permutation of roots \cite{Birkeland1924,Birkeland1920}:
\begin{equation}
\begin{split}
& X_1  \xrightarrow{\makebox[0.5cm]{$\mathfrak{b}_1$}}  X_n, \; \; X_2  \xrightarrow{\makebox[0.5cm]{$\mathfrak{b}_2$}} X_n, \dots , X_{n-n^\prime}  \xrightarrow{\makebox[0.5cm]{$\mathfrak{b}_{n-n^\prime}$}}  X_n, \\
& X_{n-n^{\prime}+i}  \xrightarrow[\text{$+$}]{\makebox[0.5cm]{$\mathfrak{b}_0$}} X_{n-n^{\prime}+i+1}, \; \; \; \;\; \; \; \;\; \; \; \;\; \; \; \; \; \; \; \; \; \; \; \;\; \; \; \;\; \; \; \;(i=1,2,\dots,n^\prime), 
\end{split}
\label{Birkelandpermutation1}
\end{equation} 
whereas 
\begin{equation}
\begin{split}
& X_i  \xrightarrow{\makebox[0.5cm]{$\mathfrak{b}_h$}} X_i, \; \; \; \;\; \; (i\lessgtr h, \; i \; {\rm and}\; h \leq n-n^\prime), \\
& X_i  \xrightarrow{\makebox[0.5cm]{$\mathfrak{b}_0$}} X_i, \; \; \; \;\; \; \, (i=1,2,\dots,n-n^\prime),
\end{split}
\label{Birkelandpermutation2}
\end{equation}
where the symbol $X_1  \xrightarrow{\makebox[0.5cm]{$\mathfrak{b}_1$}}  X_n$ denotes that $X_1$ changes into $X_n$ when $\mathfrak{b}$ describes a small closed contour about $\mathfrak{b}_1$ and $\xrightarrow[\text{$+$}]{\makebox[0.5cm]{$\mathfrak{b}_0$}} $ means that the (closed) contour is defined in a positive direction about $\mathfrak{b}_0=0$. Thence, through the permutation scheme defined above we can say that have found the symmetry group for the higher hypergeometric differential equation (\ref{Birkeland2}).

At this stage, we note that when an algebraic equation is characterized by the presence of three (or more) non-vanishing coefficients as in Eqs. (\ref{5.4a}) and (\ref{5.11a}), Birkeland theorem leads to too many (for numerical purposes) hypergeometric functions in the general expansion of roots. Furthermore, the set of linear partial differential equations obeyed by the roots when viewed as functions of all coefficients does not lead easily to their explicit form \cite{Sturmfels}. Nevertheless, it is possible to achieve a more feasible form of Birkeland's results by transforming our quintic equations (\ref{5.4a}) and (\ref{5.11a}) in their Bring-Jerrard form (i.e., the form with $n=5$ and $n^\prime=1$). This can be done by means of the so-called Tschirnhaus transformations, which we are going to define.

In 1683 Ehrenfried Walther von Tschirnhaus developed a method for solving an algebraic equation of degree $n$ by exploiting a polynomial transformation (nowadays called Tschirnhaus transformation) that, upon removing the intermediate terms, transforms the starting equation into another having a simpler form \cite{T}. Tschirnhaus demonstrated the utility of this method by applying it to the resolution of the cubic equation. This pattern was later developed further by Bring \cite{Bring} and Jerrard \cite{Jerrard}.  
Before describing the details of this procedure, we introduce a nomenclature which is quite common in the context of quintic theory. The different forms of a fifth degree equation are defined as follows:
\begin{equation}
\begin{split}
& X^{5}+a_{4}X^{4}+a_{3}X^{3}+a_{2}X^{2}+a_{1}X+a_{0}=0, \; \; \; \; \, \,{\rm General \; quintic },\\
& X^{5}+b_{3}X^{3}+b_{2}X^{2}+b_{1}X+b_{0}=0, \; \; \; \; \; \; \; \; \; \; \; \; \; \; \; \; \; \; \; \, {\rm Reduced \; quintic },\\
& X^{5}+c_{2}X^{2}+c_{1}X+c_{0}=0, \; \; \; \; \; \; \; \; \; \; \; \; \; \; \; \; \; \; \; \;  \; \; \; \; \;  \; \; \; \; \; \; \;  {\rm Principal \; quintic },\\
& X^{5}+d_{1}X+d_{0}=0,  \; \; \; \; \; \; \; \; \; \; \; \; \; \; \; \; \; \; \; \;  \; \; \; \; \;  \; \; \; \; \; \; \; \; \; \; \; \; \; \; \; \; \; \; \;  {\rm Bring-Jerrard \; quintic }. \label{quintic}
\end{split}
\end{equation} 
To fix ideas, let us apply the above definitions to our case. Consider for example Eq. (\ref{5.4a}), which can be written in dimensionless units by defining
\begin{equation}
w={1 \over r} \equiv {\gamma \over l_{P}}, \label{2.5b}
\end{equation}
where $\gamma$ is a real number to be determined. The quintic equation obeyed by $\gamma$ is therefore
\begin{equation}
\gamma^{5}+\bar{\rho}_{4}\gamma^{4}+\bar{\rho}_{3}\gamma^{3}+\bar{\rho}_{0}=0,
\label{2.6b}
\end{equation}
where $\bar{\rho}_{4}$, $\bar{\rho}_{3}$, $\bar{\rho}_{0}$ are all dimensionless and read as (see Eqs. (\ref{5.5a})--(\ref{5.9a}))
\begin{equation}
\bar{\rho}_{4} \equiv \zeta_{4}l_{P}={2 \over 3}{\kappa_{1}\over \kappa_{2}}
{G(m+\alpha)\over c^{2}l_{P}},
\end{equation}
\begin{equation}
\bar{\rho}_{3} \equiv \zeta_{3}l_{P}^{2}={1 \over 3 \kappa_{2}},
\end{equation}
\begin{equation}
\bar{\rho}_{0} \equiv \zeta_{0}l_{P}^{5}=-{1 \over 3 \kappa_{2}}
\left({l_{P}\over l}\right)^{3}
=-\bar{\rho}_{3} \left({l_{P}\over l}\right)^{3}.
\end{equation}
Therefore, the Bring-Jerrard form of (\ref{2.6b}) is
\begin{equation}
\gamma^{5}+d_{1}\gamma+d_{0}=0.
\label{2.10b}
\end{equation}
Similarly, for (\ref{5.11a}) we can write
\begin{equation}
\Gamma^{5}+d_{1}\Gamma+d_{0}=0, 
\label{2.10bb}
\end{equation}
where we have defined
\begin{equation}
u= \dfrac{1}{s}= \dfrac{\Gamma}{l_P}. \label{2.32b}
\end{equation}
At this stage, we are ready to see how Tschirnhaus procedure works and how it will allow us to transform (\ref{2.6b}) into (\ref{2.10b}) (the pattern that starting from (\ref{5.11a}) leads to (\ref{2.10bb}) is exactly the same). First of all, denote the roots of Eq. (\ref{quintic}a) by $X_{i}$ ($i=1,...,5$) and let
\begin{equation}
S_{n}=S_{n}(X_{k}) \equiv \sum_{k=1}^{5}(X_{k})^{n}, \label{A2b}
\end{equation}
be the sum of the $n$-th powers of such roots. By virtue of the Newton power-sum formula, a general representation of $S_{n}$ is
\begin{equation}
S_{n}=-n a_{5-n}-\sum_{j=1}^{n-1}S_{n-j}a_{5-j}, \label{A3b}
\end{equation}
with the understanding that $a_{j}=0$ for $j<0$. For the lowest values of $n$, Eq. (\ref{A3b}) yields
\begin{equation}
\begin{split}
& S_{1}(X_{k})=-a_{4}, \\
& S_{2}(X_{k})=(a_{4})^{2}-2a_{3}, \\
& S_{3}(X_{k})=-(a_{4})^{3}+3a_{3}a_{4}-3a_{2}, \\
& S_{4}(X_{k})=(a_{4})^{4}-4a_{3}(a_{4})^{2}+4 a_{2}a_{4}+2 (a_{3})^{2}-4 a_{1}, \\
& S_{5}(X_{k})=-(a_{4})^{5}+5\Bigr[a_{3}(a_{4})^{3}-a_{2}(a_{4})^{2}-(a_{3})^{2}a_{4}+a_{1}a_{4}-a_{0}
+a_{2}a_{3}\Bigr]. \label{A5b}
\end{split}
\end{equation}
A systematic way to proceed involves two steps, i.e., first a quadratic Tschirnhaus transformation \cite{T}
\begin{equation}
Y_{k}=(X_{k})^{2}+\mu X_{k}+\nu,
\label{A6b}
\end{equation}
between the roots $X_{k}$ of Eq. (\ref{quintic}a) and the roots $Y_{k}$ of the principal quintic (\ref{quintic}c), supplemented \cite{Adamchik} by the evaluation of $S_{1}(Y_{k}),\dots,S_{5}(Y_{k})$ to obtain through radicals $\mu$, $\nu$, $c_{0}$, $c_{1}$, $c_{2}$, and eventually a quartic Tschirnhaus transformation \cite{T}
\begin{equation}
Z_{k}=(Y_{k})^{4}+u_{1}(Y_{k})^{3}+u_{2}(Y_{k})^{2}+u_{3}Y_{k}+u_{4},
\label{A8b}
\end{equation}
between the roots $Y_{k}$ of Eq. (\ref{quintic}c) and the roots $Z_{k}$ of the Bring-Jerrard form (\ref{quintic}d). The power sums for the principal quintic form are indeed 
\begin{equation}
\begin{split}
& S_{1}(Y_{k})=S_{2}(Y_{k})=0, \\
& S_{3}(Y_{k})=-3c_{2}, \\
& S_{4}(Y_{k})=-4 c_{1}, \\
& S_{5}(Y_{k})=-5 c_{0}. \label{B1b}
\end{split}
\end{equation}
On the other hand, we can evaluate $S_{1}(Y_{k})$ and $S_{2}(Y_{k})$ by using the quadratic transformation (\ref{A6b}) and exploiting the identities
\begin{equation}
S_{1}(Y_{k})=S_{2}(X_{k})+\mu S_{1}(X_{k})+5 \nu,
\end{equation}
\begin{equation}
S_{2}(Y_{k})=S_{4}(X_{k})+2 \mu S_{3}(X_{k})+(\mu^{2}+2 \nu)S_{2}(X_{k})
+2 \mu \nu S_{1}(X_{k})+5 \nu^{2},
\end{equation} 
obtaining therefore the following equations for $\mu$ and $\nu$:
\begin{equation}
\mu a_{4}-5 \nu +2 a_{3}-(a_{4})^{2}=0, \label{B4b}
\end{equation}
\begin{equation}
\mu^{2}a_{3}-10 \nu^{2}+\mu (3 a_{2}-a_{3}a_{4})+2a_{1}-2a_{2}a_{4}+(a_{3})^{2}=0, \label{B5b}
\end{equation}
where, in the course of arriving at Eq. (\ref{B5b}), we have re-expressed repeatedly $(a_{4})^{2}$ from Eq. (\ref{B4b}). This system is quadratic with respect to $\mu$ and $\nu$, and hence leads to two sets of coefficients. For the case studied in Eq. (\ref{2.6b}), they reduce to (here $a_{3}=\bar{\rho}_{3}$, $a_{4}=\bar{\rho}_{4}$) \cite{BEDS15}
\begin{equation}
\mu_{\pm}={a_{4}[13a_{3}-4 (a_{4})^{2}] \pm \sqrt{60 (a_{3})^{3}-15 (a_{3}a_{4})^{2}} \over
2 [5a_{3}-2(a_{4})^{2}]},
\end{equation}
\begin{equation}
\nu_{\pm}={\mu_{\pm}\over 5}a_{4}+{2 \over 5}a_{3}-{1 \over 5}(a_{4})^{2}.
\end{equation}
There is complete freedom to choose either of these. After finding $\mu$ and $\nu$ in such a way, one can use Eq. (\ref{B1b}) to obtain $c_{0}$, $c_{1}$, $c_{2}$. One finds explicitly, in general, \cite{BEDS15}
\begin{equation}
\begin{split}
c_{0}&= -\nu^{5}-\mu \nu^{4}S_{1}(X_{k})-(2 \mu^{2}\nu^{3}+\nu^{4})S_{2}(X_{k})
- \left(2\mu^{3}\nu^{2}+4\mu \nu^{3}\right)S_{3}(X_{k}) \\
& - \left(\mu^{4}\nu +6\mu^{2}\nu^{2}+2\nu^{3}\right)S_{4}(X_{k})
-\left({\mu^{5}\over 5}+4\mu^{3}\nu+6 \mu \nu^{2}\right)S_{5}(X_{k}) \\
&- (\mu^{4}+6 \mu^{2}\nu+2 \nu^{2})S_{6}(X_{k})
-(2\mu^{3}+4 \mu \nu) S_{7}(X_{k})-(2 \mu^{2}+\nu)S_{8}(X_{k})  \\
&- \mu S_{9}(X_{k})-{1 \over 5}S_{10}(X_{k}),
\end{split}
\end{equation}
\begin{equation}
\begin{split}
c_{1}&= -{5 \over 4}\nu^{4}-\mu \nu^{3}S_{1}(X_{k})
-\left({3 \over 2}\mu^{2}\nu^{2}+\nu^{3}\right)S_{2}(X_{k}) - (\mu^{3}\nu+3 \mu \nu^{2})S_{3}(X_{k}) \\
& -\left({\mu^{2}\over 4}+3 \mu^{2}\nu+{3 \over 2}\nu^{2}\right)S_{4}(X_{k}) -(\mu^{3}+3 \mu \nu)S_{5}(X_{k}) - \left({3 \over 2}\mu^{2}+\nu \right)S_{6}(X_{k}) \\
& -\mu S_{7}(X_{k})-{1 \over 4}S_{8}(X_{k}),
\end{split}
\end{equation}
\begin{equation}
\begin{split}
c_{2} &= -{5 \over 3}\nu^{3}-\mu \nu^{2}S_{1}(X_{k})-(\mu^{2}\nu+\nu^{2})S_{2}(X_{k})
-{\mu \over 3}(\mu^{2}+6 \nu)S_{3}(X_{k})  \\
&- (\mu^{2}+\nu)S_{4}(X_{k})-\mu S_{5}(X_{k})-{1 \over 3}S_{6}(X_{k}).
\end{split}
\end{equation}
By virtue of the Newton formulas (\ref{A3b}), the power sums for (\ref{quintic}d) are
\begin{equation}
\begin{split}
& S_{1}(Z_{k})=S_{2}(Z_{k})=S_{3}(Z_{k})=0, \\
& S_{4}(Z_{k})=-4d_{1}, \\
& S_{5}(Z_{k})=-5 d_{0}.
\label{B12b}
\end{split}
\end{equation}
Assuming now, following Bring \cite{Bring}, that the roots $Z_{k}$ of (\ref{quintic}d) are related by the quartic transformation (\ref{A8b}) to the roots $Y_{k}$ of the principal quintic (\ref{quintic}c), we can substitute Eq. (\ref{A8b}) into Eq. (\ref{B12b}). This leads to a system of five equations with six unknown variables. More precisely, from the equation
\begin{equation}
S_{1}(Z_{k})=5 u_{4}-4c_{1}-3 u_{1}c_{2}=0,
\end{equation}
one finds 
\begin{equation}
u_{4}={4\over 5}c_{1}+{3 \over 5}c_{2}u_{1}.
\end{equation} 
The second equation \cite{Adamchik}
\begin{equation}
\begin{split}
S_{2}(Z_{k})&= -10 u_{1}u_{2}c_{0}-4 (u_{2})^{2}c_{1}
+{4 \over 5}(c_{1})^{2}+8 c_{0}c_{2}+{46 \over 5}u_{1}c_{1}c_{2}  \\
&+ \left[{6 \over 5}(u_{1})^{2}+6 u_{2}\right](c_{2})^{2}
-2 u_{3}(5c_{0}+4 u_{1}c_{1}+3 u_{2}c_{2})=0,
\end{split}
\label{B15b}
\end{equation}
obtained from the identities
\begin{equation}
\begin{split}
S_{2}(Z_{k})&= S_{8}(Y_{k})+2 u_{1}S_{7}(Y_{k})+[(u_{1})^{2}+2 u_{2}]S_{6}(Y_{k})
+2(u_{1}u_{2}+u_{3})S_{5}(Y_{k})  \\
&+ [(u_{2})^{2}+2 u_{4}+2 u_{1}u_{3}]S_{4}(Y_{k})
+2(u_{2}u_{3}+u_{1}u_{4})S_{3}(Y_{k})+5(u_{4})^{2},
\end{split}
\end{equation}
\begin{equation}
\begin{split}
& S_{6}(Y_{k})=3(c_{2})^{2}, \\
& S_{7}(Y_{k})=7 c_{1}c_{2}, \\
& S_{8}(Y_{k})=8 c_{0}c_{2}+4 (c_{1})^{2},
\end{split}
\end{equation}
relates $u_{2}$ and $u_{3}$. The clever idea of the Bring-Jerrard method lies in choosing $u_{2}$ in such a way that the coefficient of $u_{3}$ in Eq. (\ref{B15b}) vanishes. By inspection one finds immediately
\begin{equation}
u_{2}=-{5 \over 3}{c_{0}\over c_{2}} -{4 \over 3}{c_{1}\over c_{2}}u_{1}.
\end{equation}
Thus, Eq. (\ref{B15b}) now depends only on $u_{1}$ and is a quadratic, i.e., \cite{Adamchik}
\begin{equation}
\begin{split}
S_{2}(Z_{k}) &= \Bigr[27 (c_{2})^{4}-160 (c_{1})^{3}+300 c_{0}c_{1}c_{2} \Bigr](u_{1})^{2}
+\Bigr[27 c_{1}(c_{2})^{3}-400 c_{0}(c_{1})^{2}
+375 (c_{0})^{2}c_{2} \Bigr]u_{1}  \\
&+ 18 (c_{1}c_{2})^{2}-45c_{0}(c_{2})^{3}-250 (c_{0})^{2}c_{1}=0.
\end{split}
\end{equation}
Lastly, by setting the sum of the cubes of (\ref{A8b}) to zero by virtue of (\ref{B12b}), a cubic equation for $u_{3}$ is obtained, by virtue of the identity
\begin{equation}
S_{3}(Z_{k})=5(u_{4})^{3}+\sum_{l=2}^{12}b_{l}S_{l}(Y_{k}),
\end{equation}
where (recall that we already know $S_{1}(Y_{k}),\dots,S_{8}(Y_{k})$) \cite{BEDS15}
\begin{equation}
\begin{split}
& b_{2}=3 u_{2}(u_{4})^{2}, \\
& b_{3}=(u_{3})^{3}+3 u_{1}(u_{4})^{2}+6 u_{2}u_{3}u_{4}, \\
& b_{4}=3(u_{2})^{2}u_{4}+3(u_{4})^{2}+3 u_{2}(u_{3})^{2} +6 u_{1}u_{3}u_{4}, \\
& b_{5}=3(u_{2})^{2}u_{3}+3 u_{1}(u_{3})^{2} +6u_{4}(u_{3}+u_{1}u_{2}), \\
& b_{6}=(u_{2})^{3}+3(u_{1})^{2}u_{4}+3(u_{3})^{2} +6 u_{2}(u_{4}+u_{1}u_{3}), \\
& b_{7}=3(u_{1})^{2} u_{3}+3(u_{2})^{2}u_{1} +6(u_{1}u_{4}+u_{2}u_{3}), \\
& b_{8}=3 u_{4}+3(u_{1})^{2}u_{2}+3(u_{2})^{2} +6 u_{1}u_{3}, \\
& b_{9}=(u_{1})^{3}+3 u_{3}+6 u_{1}u_{2}, \\
& b_{10}=3 u_{2}+3(u_{1})^{2}, \\
& b_{11}=3 u_{1}, \\
& b_{12}=1, \\
& S_{9}(Y_{k})=9c_{0}c_{1}-3(c_{2})^{3}, \\
& S_{10}(Y_{k})=5(c_{0})^{2}-10c_{1}(c_{2})^{2}, \\
& S_{11}(Y_{k})=-11c_{0}(c_{2})^{2}-11(c_{1})^{2}c_{2}, \\
& S_{12}(Y_{k})=-24c_{0}c_{1}c_{2}-4(c_{1})^{3}+3(c_{2})^{4}. 
\end{split}
\end{equation}
All intermediate quantities for reduction to the Bring-Jerrard form can be therefore found in terms of radicals. As we can see, this procedure is conceptually clear, although rather lengthy, and the joint effect of inverting (\ref{A8b}) and then (\ref{A6b}) to find $X_{k}=X_{k}(Y_{j}(Z_{l}))$ leads to twenty candidate roots, which is not very helpful if one is interested in the numerical values of such roots, as indeed we are. Rather than feeling in despair, at this stage we point out that, since in our original quintic (\ref{2.6b}) two coefficients vanish, i.e., with the notations of (\ref{quintic}a) we have $a_{2}=a_{1}=0$, it is more convenient to use what is normally ruled out in the generic case \cite{Adamchik}, i.e., a cubic Tschirnhaus transformation between the roots $X_{k}$ of Eq. (\ref{quintic}a) and the roots $Y_{k}$ of Eq. (\ref{2.10b}) \cite{BEDS15}:
\begin{equation}
Y_{k}=(X_{k})^{3}+\lambda_{1}(X_{k})^{2}+\lambda_{2}(X_{k})+\lambda_{3}.
\label{A9b}
\end{equation}
By virtue of Eqs. (\ref{2.10b}) and (\ref{A2b})--(\ref{A5b}), we find 
\begin{equation}
S_{1}(Y_{k})=S_{2}(Y_{k})=S_{3}(Y_{k})=0, 
\label{A10b}
\end{equation}
and
\begin{equation}
\begin{split}
& S_{4}(Y_{k})=-4 d_{1}, \\
& S_{5}(Y_{k})=-5 d_{0}.
\label{A11b}
\end{split}
\end{equation}
Upon assuming the cubic relation (\ref{A9b}), Eq. (\ref{A10b}) become a non-linear algebraic system leading to the numerical evaluation of $\lambda_{1}$, $\lambda_{2}$, $\lambda_{3}$. More precisely, from $S_{1}(Y_{k})=0$ we find \cite{BEDS15}
\begin{equation}
5\lambda_{3}+\lambda_{2}S_{1}(X_{k})+\lambda_{1}S_{2}(X_{k})+S_{3}(X_{k})=0,
\label{A12b}
\end{equation}
while from $S_{2}(Y_{k})=0$ we obtain \cite{BEDS15}
\begin{equation}
\begin{split}
& 5(\lambda_{3})^{2}+2 \lambda_{2}\lambda_{3}S_{1}(X_{k})
+[(\lambda_{2})^{2}+2 \lambda_{1}\lambda_{3}]S_{2}(X_{k})
+2(\lambda_{1}\lambda_{2}+\lambda_{3})S_{3}(X_{k}) \\
&+ [(\lambda_{1})^{2}+2 \lambda_{2}]S_{4}(X_{k})
+2 \lambda_{1}S_{5}(X_{k})+S_{6}(X_{k})=0.
\label{A13b}
\end{split}
\end{equation}
Lastly, from the vanishing of $S_{3}(Y_{k})$ we get \cite{BEDS15}
\begin{equation}
\begin{split}
& 5 (\lambda_{3})^{3}+3 \lambda_{2}(\lambda_{3})^{2}S_{1}(X_{k})
+3 (\lambda_{2})^{2}\lambda_{3}S_{2}(X_{k})+(\lambda_{2})^{3}S_{3}(X_{k}) + 3 (\lambda_{1})^{2}\lambda_{3}S_{4}(X_{k}) \\
& +[3(\lambda_{1})^{2}\lambda_{2}+6 \lambda_{1}\lambda_{3}]S_{5}(X_{k})
+[(\lambda_{1})^{3}+3 \lambda_{3}+6 \lambda_{1}\lambda_{2}]S_{6}(X_{k}) + 3[(\lambda_{1})^{2}+\lambda_{2}]S_{7}(X_{k}) \\
& +3 \lambda_{1}S_{8}(X_{k})+S_{9}(X_{k})=0.
\label{A14b}
\end{split}
\end{equation}
The system (\ref{A12b})--(\ref{A14b}) cannot be solved by radicals because, if one expresses for example $\lambda_{1}$ as a linear function of $\lambda_{2}$ and $\lambda_{3}$ from Eq. (\ref{A12b}), and then solves the resulting quadratic equation for $\lambda_{2}=\lambda_{2}(\lambda_{3})$ or $\lambda_{3}=\lambda_{3}(\lambda_{2})$ from Eq. (\ref{A13b}), one discovers that Eq. (\ref{A14b}) is not a polynomial in $\lambda_{3}$ (respectively, $\lambda_{2}$). Nevertheless, {\it for numerical purposes}, the system (\ref{A12b})--(\ref{A14b}) can be solved, as was indeed first done in Ref. \cite{BEDS15}. Lastly, from Eq. (\ref{A11b}) we find the coefficients $d_{1}$ and $d_{0}$ in the Bring-Jerrard form of the quintic, according to the formulas \cite{BEDS15}
\begin{equation}
d_{1}=-{1 \over 4}S_{4}(Y_{k})=\sum_{i=0}^{12}b_{1i}S_{i}(X_{k}),
\end{equation}
\begin{equation}
d_{0}=-{1 \over 5}S_{5}(Y_{k})=\sum_{i=0}^{15}b_{0i}S_{i}(X_{k}),
\end{equation}
We have evaluated all $b_{1i}$ and $b_{0i}$ coefficients by applying patiently the Tschirnhaus transformation (\ref{A9b}) and the definition (\ref{A2b}). We find therefore six triplets of possible values for $\lambda_{1}$, $\lambda_{2}$, $\lambda_{3}$ (see Tables \ref{lambda1_tab}--\ref{lambda2_tab}), which lead always to the same values of $d_{1}$ and $d_{0}$ (this is a crucial consistency check), i.e.,
\begin{table}
\centering
\caption[Values of $\lambda_3$, $\lambda_2$, and $\lambda_1$ for Eq. (\ref{5.4a}) in the case of one-particle reducible potential]{The six triplets of values of $\lambda_3$, $\lambda_2$, and $\lambda_1$ for the $1/r$-equation (\ref{5.4a}) in the case of one-particle reducible potential.}
{\relsize{-1.5}
\renewcommand\arraystretch{2.0}
\begin{tabular}{|c|c|c|c|}
\hline
\multicolumn{4}{|c|}{Equation (\ref{5.4a}), one-particle reducible potential}\\
\hline
$n$th triplet & $\lambda_3$  & $\lambda_2$  &   $\lambda_1$   \\
\hline
$n$=1 &  $3.67 \times 10^{-45}- \I \,  7.61 \times 10^{-56}  $ &  $ -0.19 + \I \,  3.90 \times10^{-12}  $ &  $1.03 \times 10^{32} +\I \, 3.78   \times 10^{-44} $ \\
\hline
$n$=2 &   $3.67 \times 10^{-45}+ \I \,  7.61 \times 10^{-56}  $ &  $ -0.19 - \I \,  3.90 \times10^{-12}  $ &  $1.03 \times 10^{32} -\I \, 3.78   \times 10^{-44} $ \\
\hline
$n$=3 &  $-1.84 \times 10^{-45}- \I \,  3.18 \times 10^{-45}  $ &  $ -0.19 + \I \,  6.31 \times10^{-12}  $ &  $1.03 \times 10^{32} +\I \, 6.11   \times 10^{-44} $ \\
\hline
$n$=4 &  $-1.84 \times 10^{-45}+ \I \,  3.18 \times 10^{-45}  $ &  $ -0.19 - \I \,  6.31 \times10^{-12}  $ &  $1.03 \times 10^{32} -\I \, 6.11   \times 10^{-44} $ \\
\hline
$n$=5 & $-1.84 \times 10^{-45}- \I \,  3.18 \times 10^{-45}  $ &  $ -0.19 + \I \,  2.41 \times10^{-12}  $ &  $1.03 \times 10^{32} +\I \, 2.34   \times 10^{-44} $ \\
\hline
$n$=6 &  $-1.84 \times 10^{-45}+ \I \,  3.18 \times 10^{-45}  $ &  $ -0.19 - \I \,  2.41 \times10^{-12}  $ &  $1.03 \times 10^{32} -\I \, 2.34   \times 10^{-44} $ \\
\hline
\end{tabular}
\label{lambda1_tab}
}
\end{table} 
\begin{table}
\centering
\caption[Values of $\lambda_3$, $\lambda_2$, and $\lambda_1$ for Eq. (\ref{5.4a}) in the case of scattering potential]{The six triplets of values of $\lambda_3$, $\lambda_2$, and $\lambda_1$ for the $1/r$-equation (\ref{5.4a}) in the case of scattering potential.}
{\relsize{-1.5}
\renewcommand\arraystretch{2.0}
\begin{tabular}{|c|c|c|c|}
\hline
\multicolumn{4}{|c|}{Equation (\ref{5.4a}), scattering potential}\\
\hline
$n$th triplet & $\lambda_3$  & $\lambda_2$  &   $\lambda_1$   \\
\hline
$n$=1 &  $-4.98 \times 10^{-45}- \I \, 3.10 \times 10^{-55}  $ &  $ 0.26 + \I \,   1.59\times10^{-11}  $ &  $4.21 \times 10^{32} +\I \,  3.78 \times 10^{-44} $ \\
\hline
$n$=2 &  $ -4.98 \times 10^{-45} + \I \, 3.10 \times 10^{-55} $ &  $ 0.26 -\I \,   1.59 \times 10^{-11} $ & $4.21\times10^{32} - \I \,  3.78 \times 10^{-44} $ \\
\hline
$n$=3 & $2.49 \times 10^{-45} + \I \,  4.32\times 10^{-45} $ &  $ 0.26 + \I \,  2.57 \times 10^{-11} $&   $ 4.21\times 10^{32} + \I \,  6.11 \times 10^{-44}  $\\
\hline
$n$=4 &  $ 2.49\times 10^{-45} -\I \,  4.32\times 10^{-45} $ & $ 0.26 - \I \, 2.57 \times 10^{-11} $ &  $ 4.21\times 10^{32} - \I \,  6.11 \times 10^{-44} $ \\
\hline
$n$=5 & $2.49 \times 10^{-45} + \I \, 4.32 \times 10^{-45} $ &  $0.26 + \I \,   9.82 \times 10^{-12} $ & $ 4.21 \times 10^{32} + \I \,   2.34 \times 10^{-44} $ \\
\hline
$n$=6 &  $ 2.49 \times 10^{-45} - \I \,  4.32 \times 10^{-45} $&  $0.26 - \I \,  9.82 \times 10^{-12} $ &  $4.21\times 10^{32} - \I \,  2.34\times 10^{-44} $\\
\hline
\end{tabular}
}
\end{table} 
\begin{table}
\centering
\caption[Values of $\lambda_3$, $\lambda_2$, and $\lambda_1$ for Eq. (\ref{5.4a}) in the case of bound-states potential]{The six triplets of values of $\lambda_3$, $\lambda_2$, and $\lambda_1$ for the $1/r$-equation (\ref{5.4a}) in the case of bound-states potential.}
{\relsize{-1.5}
\renewcommand\arraystretch{2.0}
\begin{tabular}{|c|c|c|c|}
\hline
\multicolumn{4}{|c|}{Equation (\ref{5.4a}), bound-states potential}\\
\hline
$n$th triplet & $\lambda_3$  & $\lambda_2$  &   $\lambda_1$   \\
\hline
$n$=1 &  $-4.98 \times 10^{-45}- \I \, 5.17   \times 10^{-56}  $ &  $ 0.26+ \I \, 2.65  \times10^{-12}  $ &  $ -7.01 \times 10^{31} - \I \, 3.78 \times 10^{-44} $ \\
\hline
$n$=2 &   $-4.98 \times 10^{-45}+ \I \, 5.17   \times 10^{-56}  $ &  $ 0.26- \I \, 2.65  \times10^{-12}  $ &  $ -7.01 \times 10^{31} + \I \, 3.78 \times 10^{-44} $ \\
\hline
$n$=3 &  $2.49 \times 10^{-45}- \I \, 4.32   \times 10^{-45}  $ &  $ 0.26+ \I \, 4.29  \times10^{-12}  $ &  $ -7.01 \times 10^{31} - \I \, 6.11 \times 10^{-44} $ \\
\hline
$n$=4 &  $2.49 \times 10^{-45}+ \I \, 4.32   \times 10^{-45}  $ &  $ 0.26- \I \, 4.29  \times10^{-12}  $ &  $ -7.01 \times 10^{31} + \I \, 6.11 \times 10^{-44} $ \\
\hline
$n$=5 & $2.49 \times 10^{-45}- \I \, 4.32   \times 10^{-45}  $ &  $ 0.26 + \I \, 1.64  \times10^{-12}  $ &  $ -7.01 \times 10^{31} - \I \, 2.34 \times 10^{-44} $ \\
\hline
$n$=6 &  $2.49 \times 10^{-45} + \I \, 4.32   \times 10^{-45}  $ &  $ 0.26 - \I \, 1.64  \times10^{-12}  $ &  $ -7.01 \times 10^{31} + \I \, 2.34 \times 10^{-44} $ \\
\hline
\end{tabular}
}
\end{table} 
\begin{table}
\centering
\caption[Values of $\lambda_3$, $\lambda_2$, and $\lambda_1$ for Eq. (\ref{5.11a}) in the case of one-particle reducible potential]{The six triplets of values of $\lambda_3$, $\lambda_2$, and $\lambda_1$ for the $1/s$-equation (\ref{5.11a}) in the case of one-particle reducible potential.}
{\relsize{-1.5}
\renewcommand\arraystretch{2.0}
\begin{tabular}{|c|c|c|c|}
\hline
\multicolumn{4}{|c|}{Equation (\ref{5.11a}), one-particle reducible potential}\\
\hline
$n$th triplet & $\lambda_3$  & $\lambda_2$  &   $\lambda_1$   \\
\hline
$n$=1 &  $3.67 \times 10^{-45}- \I \, 9.37 \times 10^{-58}  $ &  $ -0.19 + \I \, 4.80  \times10^{-14}  $ &  $1.27 \times 10^{30} +\I \, 3.78   \times 10^{-44} $ \\
\hline
$n$=2 &   $3.67 \times 10^{-45}+ \I \, 9.37 \times 10^{-58}  $ &  $ -0.19 - \I \, 4.80  \times10^{-14}  $ &  $1.27 \times 10^{30} -\I \, 3.78   \times 10^{-44} $ \\
\hline
$n$=3 &  $-1.84 \times 10^{-45}- \I \, 3.18 \times 10^{-45}  $ &  $ -0.19 + \I \, 7.77  \times10^{-14}  $ &  $1.27 \times 10^{30} +\I \, 6.11   \times 10^{-44} $ \\
\hline
$n$=4 &  $-1.84 \times 10^{-45}+ \I \, 3.18 \times 10^{-45}  $ &  $ -0.19 - \I \, 7.77  \times10^{-14}  $ &  $1.27 \times 10^{30} -\I \, 6.11   \times 10^{-44} $ \\
\hline
$n$=5 & $-1.84 \times 10^{-45} - \I \, 3.18 \times 10^{-45}  $ &  $ -0.19 + \I \, 2.97  \times10^{-14}  $ &  $1.27 \times 10^{30} + \I \, 2.34   \times 10^{-44} $ \\
\hline
$n$=6 &  $-1.84 \times 10^{-45} + \I \, 3.18 \times 10^{-45}  $ &  $ -0.19 - \I \, 2.97  \times10^{-14}  $ &  $1.27 \times 10^{30} - \I \, 2.34   \times 10^{-44} $ \\
\hline
\end{tabular}
}
\end{table} 
\begin{table}
\centering
\caption[Values of $\lambda_3$, $\lambda_2$, and $\lambda_1$ for Eq. (\ref{5.11a}) in the case of scattering potential]{The six triplets of values of $\lambda_3$, $\lambda_2$, and $\lambda_1$ for the $1/s$-equation (\ref{5.11a}) in the case of scattering potential.}
{\relsize{-1.5}
\renewcommand\arraystretch{2.0}
\begin{tabular}{|c|c|c|c|}
\hline
\multicolumn{4}{|c|}{Equation (\ref{5.11a}), scattering potential}\\
\hline
$n$th triplet & $\lambda_3$  & $\lambda_2$  &   $\lambda_1$   \\
\hline
$n$=1 &   $-4.98\times 10^{-45} -  \I \, 3.82\times 10^{-57} $& $ 0.26 +  \I \,  1.96\times 10^{-13} $&  $ 5.17\times 10^{30} +  \I \,  3.78\times 10^{-44} $ \\
\hline
$n$=2 & $-4.98\times 10^{-45} +  \I \,  3.82\times 10^{-57}  $ &  $ 0.26 - \I \,  1.96\times 10{-13}  $ & $ 5.17\times 10^{30} - \I \,  3.78\times 10^{-44}  
$ \\
\hline
$n$=3 & $2.49\times 10^{-45} + \I \,  4.32\times 10^{-45}  $ & $0.26 + \I \,  3.16\times 10^{-13}  $ & $5.17\times 10^{30} +  \I \,  6.11\times10^{-44} $\\
\hline
$n$=4 & $2.49\times 10^{-45} -\I \,   4.32\times 10^{-45} $ & $ 0.26 - \I \,  3.16\times 10^{-13} $ & $ 5.17 \times 10^{30} -  \I \,  6.11\times 10^{-44} $ \\
\hline
$n$=5 & $ 2.49\times 10^{-45} +  \I \,  4.32\times 10^{-45}  $ &  $ 0.26 + \I \,  1.21\times 10^{-13} $ & $5.17\times 10^{30} + \I \,   2.34\times10^{-44} $\\
\hline
$n$=6 & $ 2.49\times 10^{-45} -  \I \,  4.32\times 10^{-45} $ & $0.26 -  \I \,   1.21\times 10^{-13} $ & $5.17\times 10^{30} - \I \,   2.34\times 10^{-44}   $ \\
\hline
\end{tabular}
}
\end{table} 
\begin{table}
\centering
\caption[Values of $\lambda_3$, $\lambda_2$, and $\lambda_1$ for Eq. (\ref{5.11a}) in the case of bound-states potential]{The six triplets of values of $\lambda_3$, $\lambda_2$, and $\lambda_1$ for the $1/s$-equation (\ref{5.11a}) in the case of bound-states potential.}
{\relsize{-1.5}
\renewcommand\arraystretch{2.0}
\begin{tabular}{|c|c|c|c|}
\hline
\multicolumn{4}{|c|}{Equation (\ref{5.11a}), bound-states potential}\\
\hline
$n$th triplet & $\lambda_3$  & $\lambda_2$  &   $\lambda_1$   \\
\hline
$n$=1 &  $ -4.98 \times 10^{-45}- \I \, 6.36  \times 10^{-58}  $ &  $0.26 + \I \, 3.26  \times10^{-14}  $ &  $ -8.62 \times 10^{29} -\I \, 3.78 \times 10^{-44} $ \\
\hline
$n$=2 &   $ -4.98 \times 10^{-45}+ \I \, 6.36  \times 10^{-58}  $ &  $0.26 - \I \, 3.26  \times10^{-14}  $ &  $ -8.62 \times 10^{29} +\I \, 3.78 \times 10^{-44} $ \\
\hline
$n$=3 &  $ 2.49 \times 10^{-45} - \I \, 4.32  \times 10^{-45}  $ &  $0.26 + \I \, 5.27  \times10^{-14}  $ &  $ -8.62 \times 10^{29} -\I \, 6.11 \times 10^{-44} $ \\
\hline
$n$=4 &  $ 2.49 \times 10^{-45} + \I \, 4.32  \times 10^{-45}  $ &  $0.26 - \I \, 5.27  \times10^{-14}  $ &  $ -8.62 \times 10^{29} +\I \, 6.11 \times 10^{-44} $ \\
\hline
$n$=5 & $ 2.49 \times 10^{-45} - \I \, 4.32  \times 10^{-45}  $ &  $0.26 + \I \, 2.01  \times10^{-14}  $ &  $ -8.62 \times 10^{29} -\I \, 2.34 \times 10^{-44} $ \\
\hline
$n$=6 &  $ 2.49 \times 10^{-45} + \I \, 4.32  \times 10^{-45}  $ &  $0.26 - \I \, 2.01  \times10^{-14}  $ &  $ -8.62 \times 10^{29} +\I \, 2.34 \times 10^{-44} $ \\
\hline
\end{tabular}
\label{lambda2_tab}
}
\end{table} 
\begin{equation}
\begin{split}
& d_{1}(w)= 8.17 \times 10^{-177} - \I \, 6.78 \times 10^{-187},  \\
& d_{1}(u)= 8.17 \times 10^{-177}+ \I \, 8.34 \times 10^{-189},
\label{A17b}
\end{split}
\; \; \; \; \; \; \; \; {\rm (one-particle \; reducible \; potential) }
\end{equation}
\begin{equation}
\begin{split}
& d_{1}(w)=2.78 \times 10^{-176}+{\rm i} \, 6.91 \times 10^{-186}, \\
& d_{1}(u)=2.78 \times 10^{-176}-{\rm i} \, 8.51 \times 10^{-188},
\end{split}
\; \; \; \; \; \; \; \; {\rm (scattering \; potential) }
\end{equation}
\begin{equation}
\begin{split}
& d_{1}(w)= 2.78 \times 10^{-176}+ {\rm i} \ 1.15 \times 10^{-186},  \\
& d_{1}(u)= 2.78 \times 10^{-176}- {\rm i} \ 1.42 \times 10^{-188},
\end{split}
\; \; \; \; \; \; \; \; {\rm (bound-states \; potential) }
\end{equation}
and 
\begin{equation}
\begin{split}
& d_{0}(w)= 3.60 \times 10^{-221} - \I \, 3.73 \times 10^{-231},  \\
& d_{0}(u)=  3.60 \times 10^{-221} + \I \, 4.59 \times 10^{-233}, 
\end{split}
\; \; \; \; \; \; \; \; {\rm (one-particle \; reducible \; potential) }
\end{equation}
\begin{equation}
\begin{split}
& d_{0}(w)=1.66 \times 10^{-220}-{\rm i} \, 5.17  \times 10^{-230}, \\
& d_{0}(u)=-1.66 \times 10^{-220}+{\rm i}  \, 6.36 \times 10^{-232},
\end{split}
\; \; \; \; \; \; \; \; {\rm (scattering \; potential) }
\end{equation}
\begin{equation}
\begin{split}
& d_{0}(w)=  -1.66 \times 10^{-220} - \I \, 8.62 \times 10^{-231}, \\
& d_{0}(u)= -1.66 \times 10^{-220} + \I \, 1.06 \times 10^{-232}, 
\label{A18b}
\end{split}
\; \; \; \; \; \; \; \; {\rm (bound-states \; potential) }
\end{equation}
where $w$ and $u$ are the variables defined in (\ref{2.5b}) and (\ref{2.32b}), respectively. 

At this point, all the coefficients appearing in (\ref{2.10b}) and (\ref{2.10bb}) are known and we are ready to invoke Birkeland theorem to solve them. This means that for our purposes we need to consider $n=5$ and $n^\prime=1$ (cf. Eq. (\ref{Birkeland1})) and hence Eq. (\ref{Birkeland3}) becomes
\begin{equation}
\sigma \equiv {3125 \over 256}{(-d_{0})^{4}\over (-d_{1})^{5}}.
\label{2.11b}
\end{equation}
By focusing our attention on (\ref{2.10b}) (the procedure involving Eq. (\ref{2.10bb}) is exactly the same), we can further simplify it by rescaling $\gamma$ according to\footnote{The variable $\chi$ has not to be confused with the constant appearing in (\ref{perturbed_metric_h}).}
\begin{equation}
\gamma=\chi {\tilde \gamma}.
\end{equation}
The quintic for ${\tilde \gamma}$ is then
\begin{equation}
{\tilde \gamma}^{5}+{d_{1}\over \chi^{4}}{\tilde \gamma}+{d_{0}\over \chi^{5}}=0. 
\label{2.13b}
\end{equation}
One can choose $\chi$ in such a way that \cite{BEDS15}
\begin{equation}
-{d_{1}\over \chi^{4}}=1 \Longrightarrow \chi=\chi(d_{1})=(-d_{1})^{1 \over 4},
\end{equation}
so that the Bring-Jerrard quintic (\ref{2.13b}) becomes
\begin{equation}
{\tilde \gamma}^{5}-{\tilde \gamma}-{\tilde \beta}=0, \label{2.13bb}
\end{equation}
where
\begin{equation}
{\tilde \beta} \equiv -{d_{0}\over \left[\chi(d_{1})\right]^{5}}
\end{equation}
while the corresponding $\sigma$ of (\ref{2.11b}) reads as
\begin{equation}
{\tilde \sigma}={3125 \over 256}\left(-{d_{0}\over (\chi(d_{1}))^{5}}\right)^{4}=\sigma.
\end{equation}
By virtue of the numerical results (\ref{A17b})--(\ref{A18b}), $\vert \tilde \sigma \vert <1$ for the three types of potential and hence, according to Eqs. (\ref{Birkeland_root1})--(\ref{Birkeland_root2}), the five roots of (\ref{2.13bb}) are given in terms of hypergeometric functions of order four, i.e., \cite{BEDS15,Birkeland1924}
\begin{equation}
\left(\begin{matrix}
{\tilde \gamma}_{1} \cr {\tilde \gamma}_{2} \cr {\tilde \gamma}_{3} \cr 
{\tilde \gamma}_{4} 
\end{matrix}\right)
= \left(\begin{matrix}
{\rm i} & {{\tilde \beta}\over 4} & {5 \over 32}{\rm i}{\tilde \beta}^{2} 
& -{5 \over 32}{\tilde \beta}^{3} \cr
-1 & {{\tilde \beta}\over 4} & {5 \over 32}{\tilde \beta}^{2} 
& {5 \over 32}{\tilde \beta}^{3} \cr
-{\rm i} & {{\tilde \beta}\over 4} & -{5 \over 32}{\rm i}{\tilde \beta}^{2} 
& -{5 \over 32}{\tilde \beta}^{3} \cr
1 & {{\tilde \beta}\over 4} & -{5 \over 32}{\tilde \beta}^{2} 
& {5 \over 32}{\tilde \beta}^{3} 
\end{matrix}\right)
\left(\begin{matrix}
F_{0}({\tilde \sigma}) \cr F_{1}({\tilde \sigma}) \cr
F_{2}({\tilde \sigma}) \cr F_{3}({\tilde \sigma})
\end{matrix}\right),
\end{equation}
\begin{equation}
{\tilde \gamma}_{5}=-{\tilde \beta}F_{1}({\tilde \sigma}),
\end{equation}
where
\begin{equation}
F_{0}({\tilde \sigma}) \equiv F \left(\begin{matrix}
-{1 \over 20}, & {3 \over 20}, & {7 \over 20}, & {11 \over 20} \cr
{1 \over 4}, & {1 \over 2}, & {3 \over 4}, & {\tilde \sigma}
\end{matrix}\right),
\end{equation}
\begin{equation}
F_{1}({\tilde \sigma}) \equiv F \left(\begin{matrix}
{1 \over 5}, & {2 \over 5}, & {3 \over 5}, & {4 \over 5} \cr
{1 \over 2}, & {3 \over 4}, & {5 \over 4}, & {\tilde \sigma}
\end{matrix}\right),
\end{equation}
\begin{equation}
F_{2}({\tilde \sigma}) \equiv F \left(\begin{matrix}
{9 \over 20}, & {13 \over 20}, & {17 \over 20}, & {21 \over 20} \cr
{3 \over 4}, & {5 \over 4}, & {3 \over 2}, & {\tilde \sigma}
\end{matrix}\right),
\end{equation}
\begin{equation}
F_{3}({\tilde \sigma}) \equiv F \left(\begin{matrix}
{7 \over 10}, & {9 \over 10}, & {11 \over 10}, & {13 \over 10} \cr
{5 \over 4}, & {3 \over 2}, & {7 \over 4}, & {\tilde \sigma}
\end{matrix}\right).
\end{equation}
Moreover, the differential equation (\ref{Birkeland2}) obeyed by the roots assumes now the form \cite{BEDS15,Birkeland1924}
\begin{equation}
\left[\sigma^{3}(\sigma-1){{\rm d}^{4}\over {\rm d} \sigma^{4}}
+\sigma^{2}(A_{1}\sigma-B_{1}){{\rm d}^{3}\over {\rm d}\sigma^{3}}
+\sigma(A_{2}\sigma-B_{2}){{\rm d}^{2}\over {\rm d}\sigma^{2}}
+(A_{3}\sigma-B_{3}){{\rm d}\over {\rm d}\sigma}+\tilde{C} \right]\Lambda=0,
\label{2.26b}
\end{equation}
and the critical points turn out to be (see Eqs. (\ref{Birkelandcritical1}) and (\ref{Birkelandcritical2}))
\begin{equation}
\begin{split}
& {\tilde \beta}_{1}=-{\rm i} \mathfrak{C}, \\
& {\tilde \beta}_{2}=\mathfrak{C}, \\
&{\tilde \beta}_{3}={\rm i} \mathfrak{C}, \\ 
&{\tilde \beta}_{4}=-\mathfrak{C}, 
\end{split}
\label{2.27b}
\end{equation}
where $\mathfrak{C} \equiv {1024 \over 3125}$. Thence, bearing in mind these relations jointly with (\ref{Birkelandpermutation1}) and (\ref{Birkelandpermutation2}), we have the permutation scheme
\begin{equation}
\tilde{\gamma}_1  \xrightarrow{\makebox[0.5cm]{$\tilde{\beta}_1$}}  \tilde{\gamma}_5, \; \; \tilde{\gamma}_2  \xrightarrow{\makebox[0.5cm]{$\tilde{\beta}_2$}} \tilde{\gamma}_5, \; \;  \tilde{\gamma}_3  \xrightarrow{\makebox[0.5cm]{$\tilde{\beta}_3$}}  \tilde{\gamma}_5, \; \; \tilde{\gamma}_4 \xrightarrow{\makebox[0.5cm]{$\tilde{\beta}_4$}} \tilde{\gamma}_5. 
\end{equation} 
Thus, the symmetry group of the forth-order linear differential equation (\ref{2.26b}) has the property that the root ${\tilde \gamma}_{k}$ is changed into ${\tilde \gamma}_{5}$, for all $k=1,2,3,4$, when ${\tilde \beta}$ describes a small closed contour about the critical point ${\tilde \beta}_{k}$ defined by Eq. (\ref{2.27b}). Eventually, the roots $\gamma_{i}$ of Eq. (\ref{2.10b}) are given by
\begin{equation}
\gamma_{i}=\chi(d_{1}){\tilde \gamma}_{i}, \; \; \; \; \forall i=1,2,3,4,5.
\end{equation}

The final step towards the solution of our original quintic (\ref{2.6b}) consists in solving Eq. (\ref{A9b}) for $X_{k}=X_{k}(Y_{k})$, with the help of the solution algorithm for the cubic equation. This means that we first re-express (\ref{A9b}) in the form
\begin{equation}
h(X_{k}) \equiv (X_{k})^{3}+\vartheta_{2}(X_{k})^{2}+\vartheta_{1}X_{k}+\vartheta_{0}=0,
\label{A19b}
\end{equation}
where $\vartheta_{2} \equiv \lambda_{1}, \vartheta_{1} \equiv \lambda_{2}, \vartheta_{0} \equiv \lambda_{3}-Y_{k}$. We then define the new variable
\begin{equation}
B_{k} \equiv X_{k}+{\vartheta_{2} \over 3}=X_{k}+{\lambda_{1} \over 3},
\label{A20b}
\end{equation}
in terms of which Eq. (\ref{A19b}) is mapped into its canonical form
\begin{equation}
(B_{k})^{3}+pB_{k}+q=0, \; \; \; \; \; \; \; \;\; \; \; \;\; \; \; \;\; \; \; \;  p \equiv h' \left(-{\vartheta_{2}\over 3}\right), \;\; \; \; 
q \equiv h \left(-{\vartheta_{2}\over 3}\right).
\label{A21b}
\end{equation}
It is possible to solve (\ref{A21b}) once again by means of Birkeland theorem. By bearing in mind Eq. (\ref{Birkeland1}), in the case $n=3$ and $n^\prime=1$ it follows from (\ref{Birkeland3}) that
\begin{equation}
\hat{\sigma} \equiv -{27 \over 4}{q^{2}\over p^{3}}.
\label{A22b}
\end{equation}
Thus, the theorem states that if (\ref{A22b}) is such that $|\hat{\sigma}|<1$ or $\hat{\sigma}=1$, the three roots of (\ref{A21b}) can be expressed through the Gaussian or ordinary hypergeometric functions in the form \cite{Birkeland1924}
\begin{equation}
(B_{k})_{j}=\sqrt{-p}\left[(-1)^{3j}F \left(-{1 \over 6},{1 \over 6}, {1 \over 2}; \hat{\sigma} \right)
+{1 \over 3}\sqrt{\hat{\sigma} \over 3}F \left({1 \over 3},{2 \over 3},{3 \over 2}; \hat{\sigma} \right)\right],
\; \; \; \; \;\; \; \; \; \; (j=1,2), 
\label{A23b}
\end{equation}
\begin{equation}
(B_{k})_{3}=-{2 \over 3}\sqrt{-{p \hat{\sigma} \over 3}} \; 
F \left({1 \over 3},{2 \over 3},{3 \over 2}; \hat{\sigma} \right),
\label{A24b}
\end{equation}
whereas if $|\hat{\sigma}|> 1$ we have \cite{Birkeland1924}
\begin{equation}
\begin{split}
(B_{k})_{j} & =\sqrt{-\dfrac{p}{3}} \left[ \hat{\phi}^{\,j}\, 2^{1/3} \,\hat{\sigma}^{1/6} \, F \left(-\dfrac{1}{6},\dfrac{1}{3},\dfrac{2}{3};\dfrac{1}{\hat{\sigma}} \right) + \hat{\phi}^{\,2j}\, 2^{-1/3} \,\hat{\sigma}^{-1/6}   \,  F \left(-\dfrac{1}{6},\dfrac{2}{3},\dfrac{4}{3};\dfrac{1}{\hat{\sigma}} \right) \right] , \\
& \; \; \; \; \;\; \; \; \; \;\; \; \; \; \;\; \; \; \; \;\; \; \; \; \;\; \; \; \; \;\; \; \; \; \;\; \; \; \; \;\; \; \; \; \;\; \; \; \; \;\; \; \; \; \;\; \; \; \; \;\; \; \; \; \;\; \; \; \; \;\; \; \; \; \;\; \; \; \; \;\; \; \; \; \;\; \; \; \; \;\; \; \; \; \;\; \; \; \; \;\; \;  (j=1,2,3), 
 \end{split}
\end{equation}
where $\hat{\phi}$ represents any root of the algebraic equation $\hat{\phi}^{\,3}=1$. For our purposes and for the three different potentials, we always have $|\hat{\sigma}| <1$. As is clear from Eqs. (\ref{A20b}), (\ref{A23b}), and (\ref{A24b}), our method yields eventually fifteen candidate roots, and by insertion into the original quintic (\ref{2.6b}) we have found the five effective solutions. Obviously, the same procedure has been applied also to (\ref{2.10bb}). The results we have found following this scheme are in perfect agreement with those of Tabs. \ref{noncoll_details_tab} and \ref{noncoll_corrections_tab}\footnote{A little mistake occurred in Ref. \cite{BEDS15} which has led to wrong corrections only on the $y$-coordinates of $L_4$ and $L_5$. On the other side, the values involving the $x$-coordinate written in Eq. (2.38) of Ref. \cite{BEDS15} are correct.}.

To sum up, the pattern described in this section has allowed us to transform, through the cubic Tschirnhaus transformation (\ref{A9b}), the original quintic (\ref{2.6b}) into its Bring-Jerrard counterpart (\ref{2.10b}), which, once rescaled to the form (\ref{2.13bb}), has been solved by employing Birkeland theorem. Therefore, the five roots of the starting algebraic equation (\ref{2.6b}) are found by inverting (\ref{A9b}). In this way we have confirmed the outcomes of Sec. \ref{noncoll_sec}, which were obtained by adopting numerical tools from the very beginning.

Yet another valuable solution algorithm is available, i.e., the method in Ref. \cite{King}, which expresses the roots of the quintic (\ref{quintic}a) through two infinite series, i.e., the Jacobi nome and the theta series, for which fast convergence is obtained, but the need to evaluate the roots with a large number of decimal digits makes it problematic, as far as we can see, to deal with such series. 

\subsection{Collinear Lagrangian points} \label{coll_Sec}

From the theoretical point of view, it is equally important to work out how collinear Lagrangian points get affected by the one-loop long-distance quantum corrections to the Newtonian potential \cite{BEDS15}. On the side of the applications, their importance is further strengthened, since satellites (e.g., the Wilkinson Microwave Anisotropy Probe) have been sent so far to the points $L_{1}$, $L_{2}$, and $L_{3}$ of some approximate three-body configurations in the solar system. 

As we have pointed out before, when the libration points are collinear, the coordinate $y$ vanishes, which ensures the vanishing of (\ref{3.4a}) as well. On the other hand, bearing in mind Eq. (\ref{2.4a}a), the vanishing of $y$ implies that $x$ obeys the algebraic equation
\begin{equation}
x^{2}+2ax+a^{2}-r^{2}=0,
\label{3.4b}
\end{equation}
which is solved by the two roots
\begin{equation}
x=\varepsilon r -a=\varepsilon r -{\beta l \over (\alpha+\beta)}, \;\; \; \; \; \; \;\;\; \; \; \; \; \; ( \varepsilon= \pm 1).
\label{3.5b}
\end{equation}
In particular, when $\varepsilon=1$ we obtain the solution describing the coordinates of $L_1$ and $L_2$, while the choice $\varepsilon=-1$ is connected to the position of $L_3$. Furthermore, the geometry of the problem yields also
\begin{equation}
x={(r^{2}-s^{2})\over 2l}+{1 \over 2}{(\alpha-\beta)\over (\alpha+\beta)}l,
\end{equation}
which implies, by comparison with Eq. (\ref{3.5b}),
\begin{equation}
s^{2}=(r -\varepsilon l)^{2} \Longrightarrow s= \pm (r -\varepsilon l),
\label{3.7b}
\end{equation}
where both signs should be considered, since $(r-\varepsilon l)$ may be negative. Note now that the insertion of (\ref{3.5b}) into Eq. (\ref{3.1a}) yields
\begin{equation}
{\partial U \over \partial x}={\beta \over s^{3}}\left(1+2{k_{3}\over s}
+3{k_{2}\over s^{2}}\right)(l- \varepsilon r)
-{\alpha \varepsilon \over r^{2}}\left(1+2{k_{1}\over r}+3{k_{2}\over r^{2}}
\right)+{(\alpha+\beta)\over l^{3}}\left(\varepsilon r -{\beta l \over (\alpha+\beta)}\right)=0.
\label{3.8b}
\end{equation}
Moreover, we consider first the solution $s=r - \varepsilon l$ in Eq. (\ref{3.7b}). This turns Eq. (\ref{3.8b}) into the form
\begin{equation}
{\beta \over (r-\varepsilon l)^{2}}+{2k_{3}\beta \over (r-\varepsilon l)^{3}}
+{3 k_{2}\beta \over (r-\varepsilon l)^{4}}
+{\alpha \over r^{2}}+{2 k_{1}\alpha \over r^{3}}+{3 k_{2} \alpha \over r^{4}}
-{(\alpha+\beta)r \over l^{3}}+{\beta \varepsilon \over l^{2}}=0.
\label{3.9b}
\end{equation}
This form of the equation to be solved for $r={\overline {AP}}$ suggests multiplying  both sides by $(r-\varepsilon l)^{4}r^{4}$, which makes it clear that we end up by studying a nonic algebraic equation \cite{BEDS15}. Moreover, it is now convenient to adopt dimensionless units. For this purpose, since the length parameters $k_1$ and $k_3$ appearing in the quantum corrected potential (\ref{2.5a}) are a linear combination of the gravitational radii $R_{\alpha}$, $R_{\beta}$ of primaries and $R_{m}$ of the planetoid, which in turn represent a fraction of the distance $l$ between $A$ and $B$ (cf. Eqs. (\ref{2.6a}) and (\ref{2.8a})), we can set
\begin{equation}
\begin{split}
& \psi \equiv {r \over l}, \\
& \rho \equiv {\beta \over \alpha}, \\ 
& \rho_{\alpha} \equiv {\bar{R}_{\alpha}\over l}, \\ 
& \rho_{\beta} \equiv {\bar{R}_{\beta}\over l}, \\
& \rho_{P} \equiv {l_{P}\over l},
\label{3.13b}
\end{split}
\end{equation}
where
\begin{equation}
\begin{split}
& \bar{R}_\alpha= R_\alpha + R_m, \\
& \bar{R}_\beta= R_\beta + R_m.
\end{split}
\end{equation}
In light of (\ref{3.13b}), we find the following dimensionless form of the nonic resulting from Eq. (\ref{3.9b}) \cite{BEDS15}:
\begin{equation}
\sum_{n=0}^{9}A_{n}\psi^{n}=0,
\label{3.14b}
\end{equation}
where 
\begin{equation}
\begin{split}
A_{0} \equiv & -3 (1+\rho)^{-1}\kappa_{2}(\rho_{P})^{2}, \\
 A_{1} \equiv & -2(1+\rho)^{-1}\Bigr[\kappa_{1}\rho_{\alpha}-6 \varepsilon \kappa_{2}(\rho_{P})^{2}\Bigr], \\
 A_{2} \equiv & -(1+\rho)^{-1}\Bigr[1-8 \varepsilon \kappa_{1}\rho_{\alpha} +18 \kappa_{2} (\rho_{P})^{2}\Bigr], \\
 A_{3} \equiv & 4(1+\rho)^{-1}\Bigr[\varepsilon -3 \kappa_{1}\rho_{\alpha} +3 \varepsilon \kappa_{2}(\rho_{P})^{2}\Bigr], \\
 A_{4} \equiv & -(1+\rho)^{-1}\Bigl\{ \left[6+(1+\varepsilon)\rho\right] -2 \kappa_{1}(4 \rho_{\alpha}+\rho_{\beta}\rho)\varepsilon +3(1+\rho)\kappa_{2}(\rho_{P})^{2}\Bigr\}, \\
 A_{5} \equiv & (1+\rho)^{-1}\Bigr[(1+4 \varepsilon) +(5+ 2 \varepsilon)\rho -2 \kappa_{1}(\rho_{\alpha}+\rho_{\beta}\rho)\Bigr], \\
 A_{6} \equiv & -(1+\rho)^{-1}\Bigr[(1+4 \varepsilon)+(10 \varepsilon+1)\rho\Bigr], \\
 A_{7} \equiv & 2(1+\rho)^{-1}(3+5 \rho), \\
 A_{8} \equiv & -(1+\rho)^{-1}(4+5 \rho)\varepsilon, \\ 
 A_{9} \equiv & 1.
\end{split}
\end{equation}
If we take instead the root $s=-(r-\varepsilon l)$ in Eq. (\ref{3.7b}) and insert it into Eq. (\ref{3.8b}), we find, with analogous procedure, the nonic equation \cite{BEDS15}
\begin{equation}
\sum_{n=0}^{9}B_{n}\psi^{n}=0,
\label{3.25b}
\end{equation}
where
\begin{gather}
\begin{aligned}
B_{k}= & A_{k}, \; \; \; \; \; \; {\rm if} \; \;  k=0,1,2,3,7,8,9, \\
 B_{4} \equiv & (1+\rho)^{-1}\Bigr\{[-6+(1-\varepsilon)\rho]
+2 \kappa_{1}\varepsilon (4 \rho_{\alpha}+\rho_{\beta}\rho)
+3(\rho-1)\kappa_{2}(\rho_{P})^{2}\Bigr\}, \\
 B_{5} \equiv & (1+\rho)^{-1}\Bigr[(1+4 \varepsilon)+(5 -2 \varepsilon)\rho 
-2 \kappa_{1}(\rho_{\alpha}+\rho_{\beta}\rho)\Bigr], \\
 B_{6} \equiv & -(1+\rho)^{-1}\Bigr[(1+4 \varepsilon)+(10 \varepsilon -1)\rho].
\end{aligned}
\end{gather}

In Newtonian theory, the collinear libration points are ruled instead by a quintic equation, as is clear by setting $k_{1}=k_{2}=k_{3}=0$ in Eq. (\ref{3.8b}) and multiplying the resulting equation by $(r-\varepsilon l)^{2}r^{2}$. By virtue of the two choices of sign in Eq. (\ref{3.7b}) one gets, if $s=r -\varepsilon l$, the quintic \cite{BEDS15}
\begin{equation}
\sum_{n=0}^{5}C_{n}\psi^{n} = \psi^{5}-{(2+ 3 \rho)\over (1+\rho)}\varepsilon \psi^{4}
+{(1+3 \rho)\over (1+\rho)}\psi^{3}
-{[1+(1+ \varepsilon)\rho]\over (1+\rho)}\psi^{2}  
+ {2 \varepsilon \over (1+\rho)}\psi -{1 \over (1+\rho)}=0,
\label{3.30b}
\end{equation}
while $s=-(r- \varepsilon l)$ leads to the quintic \cite{BEDS15}
\begin{equation}
\sum_{n=0}^{5}D_{n}\psi^{n} = \psi^{5}-{(2+ 3 \rho)\over (1+\rho)}\varepsilon \psi^{4}
+{(1+3 \rho)\over (1+\rho)}\psi^{3}
-{[1-(1- \varepsilon)\rho]\over (1+\rho)}\psi^{2} + {2 \varepsilon \over (1+\rho)}\psi -{1 \over (1+\rho)}=0.
\label{3.31b}
\end{equation}
In this case, the coefficients are related by
\begin{equation}
C_{k} =  D_{k},    \; \; \; \; \; \; {\rm if} \; \; k=0,1,3,4,5, 
\end{equation}
and
\begin{equation}
\begin{split}
C_{2,-} = & -(1+\rho)^{-1}=D_{2,+}, \\
C_{2,+}=& -(1+\rho)^{-1}(1+2 \rho) \not = D_{2,-}=-(1+\rho)^{-1}(1-2 \rho).
\end{split}
\end{equation}
\begin{table}
\centering
\caption[Quantum details concerning collinear Lagrangian points]{The quantum values of the distances from the Earth and of the coordinates of the collinear Lagrangian points obtained by solving numerically Eqs. (\ref{3.14b}) and (\ref{3.25b}) for the three different types of potential.}
{\relsize{-2.49}
\renewcommand\arraystretch{2.0}
\begin{tabular}{|c|c|c|c|}
\hline
\multicolumn{4}{|c|}{Quantum details of collinear Lagrangian points}\\
\hline
\; \;  $L_i$ \; \;  &  One-particle reducible &  Scattering & Bound-states \\
\cline{1-4}
& $r_1= 3.263762881726546  \times 10^8 \; {\rm m}$& $r_1= 3.263762881775874 \times 10^8 \;  
{\rm m}$ & $r_1= 3.263762881732712    \times 10^8 \; {\rm m} $ \\ 
\cline{2-4}
$L_1$ & $x_1 = 3.217044369749034   \times 10^8 \; {\rm m} $  & $x_1 =  3.217044369798362   \times 10^8 \;  
{\rm m}$   & $x_1 =   3.217044369755200  \times 10^8 \;  {\rm m}$  \\
& $y_1= 0 \; {\rm m} $ &   $y_1= 0 \; {\rm m} $ &  $y_1= 0 \; {\rm m} $ \\
\hline
& $r_2= 4.489205600336842 \times 10^8 \; {\rm m}$& $r_2= 4.489205600368175 \times 10^8 \;  
{\rm m}$ & $r_2=  4.489205600340759    \times 10^8 \; {\rm m} $ \\ 
\cline{2-4}
$L_2$ & $x_2 = 4.442487088359330  \times 10^8 \; {\rm m} $  & $x_2 = 4.442487088390662  \times 10^8 \;  
{\rm m}$   & $x_2 =  4.442487088363246   \times 10^8 \;  {\rm m}$  \\
& $y_2= 0 \; {\rm m} $ &   $y_2= 0  \; {\rm m} $ &  $y_2= 0  \; {\rm m} $ \\
\hline
& $r_3= 3.816747156909998  \times 10^8 \; {\rm m}$& $r_3= 3.816747157028501  \times 10^8 \;  
{\rm m}$ & $r_3=  3.816747156924810    \times 10^8 \; {\rm m} $ \\ 
\cline{2-4}
$L_3$ & $x_3 = -3.863465668887510  \times 10^8 \; {\rm m} $  & $x_3 = -3.863465669006013    \times 10^8 \;  
{\rm m}$   & $x_3 =  -3.863465668902323  \times 10^8 \;  {\rm m}$  \\
& $y_3= 0 \; {\rm m} $ &   $y_3= 0  \; {\rm m} $ &  $y_3= 0  \; {\rm m} $ \\
\hline
\end{tabular} 
\label{coll_details_tab}
}
\end{table} 
\begin{table}
\centering
\caption[Quantum corrections on the position of Newtonian collinear Lagrangian points]{Quantum corrections on the position of Newtonian collinear Lagrangian points obtained by solving numerically Eqs. (\ref{3.14b}) and (\ref{3.25b}) for the three different types of potential.}
{
\renewcommand{\arraystretch}{2.0}
\begin{tabular}{|c|c|c|c|}
\hline
\multicolumn{4}{|c|}{Quantum corrections on Newtonian collinear Lagrangian points}\\
\hline
\; \;  $L_i$ \; \;  &  One-particle reducible &  Scattering & Bound-states \\
\cline{1-4}
$L_1$ & $r_Q-r_{cl}=-1.23   \; {\rm mm}$& $ r_Q-r_{cl}= 3.70  \; {\rm mm} $ & $r_Q-r_{cl}= -0.617 \; {\rm mm} $ \\ 
 & $x_Q-x_{cl}=  -1.23 \; {\rm mm}$  & $x_Q-x_{cl}= 3.70 \; {\rm mm}$   & $x_Q-x_{cl}= -0.617  \; {\rm mm} $  \\
\hline
$L_2$ & $r_Q-r_{cl}= -0.783  \; {\rm mm}$& $ r_Q-r_{cl}= 2.35  \; {\rm mm} $ & $r_Q-r_{cl}= -0.392  \; {\rm mm} $ \\ 
 & $x_Q-x_{cl}=  -0.783 \; {\rm mm}$  & $x_Q-x_{cl}=  2.35 \; {\rm mm}$   & $x_Q-x_{cl}= -0.392  \; {\rm mm} $  \\
 \hline
$L_3$ & $r_Q-r_{cl}= -2.96 \; {\rm mm}$& $ r_Q-r_{cl}= 8.89 \; {\rm mm} $ & $r_Q-r_{cl}=  -1.48 \; {\rm mm} $ \\ 
& $x_Q-x_{cl}= 2.96 \; {\rm mm}$  & $x_Q-x_{cl}=   -8.89 \; {\rm mm}$   & $x_Q-x_{cl}= 1.48  \; {\rm mm} $  \\
\hline
\end{tabular} 
\label{coll_corrections_tab}
}
\end{table} 
In light of Eqs. (\ref{3.30b}) and (\ref{3.31b}), the distances of the planetoid from the Earth in Newtonian theory are given by \cite{BEDS15}
\begin{equation}
\begin{split}
& r_{1,cl}= 3.263762881738878 \times 10^8 \; {\rm m}, \\
& r_{2,cl}= 4.489205600344675  \times 10^8 \; {\rm m}, \\
& r_{3,cl}= 3.816747156939623  \times 10^8 \; {\rm m}, 
\end{split}
\label{dist_coll_Newton}
\end{equation}
at $L_1$, $L_2$, and $L_3$, respectively. The coordinates of such points within the classical theory read as
\begin{equation}
\begin{split}
&  L_1  \left( 3.217044369761366 \times 10^8,  0 \right) \, {\rm m}, \\
& L_2  \left( 4.442487088367163 \times 10^8,  0 \right) \, {\rm m}, \\
& L_3  \left(-3.863465668917136  \times 10^8,  0 \right) \, {\rm m}.
\end{split}
\label{coord_coll_Newton}
\end{equation}
The contribution coming from the numerical resolution of the nonic equations (\ref{3.14b}) and (\ref{3.25b}) is written in Tabs. \ref{coll_details_tab} and \ref{coll_corrections_tab}. Interestingly, the order of magnitude of quantum corrections to the location of collinear libration points coincides with that involving $L_{4}$, $L_{5}$, as a comparison between Tabs. \ref{noncoll_corrections_tab} and \ref{coll_corrections_tab} shows. In particular, the scattering potential gives modifications to classical values which in modulus result to be always huger than those obtained through the other two potentials. This may not have any practical consequence, since $L_{1}$, $L_{2}$, $L_{3}$ are points of unstable equilibrium (as we will see in Sec \ref{stability_Sec}), but the detailed analysis performed in this section and in Sec. \ref{noncoll_sec} adds evidence in favour of our evaluation of quantum corrections to {\it all} Lagrangian points in the Earth-Moon system being able to predict effects of order of millimetre. However, the main perturbations of such a scheme may result from the Sun. If one then considers a restricted four-body problem where the Earth and the Moon move in circular orbits around their center of mass, which in turn moves in a circular orbit about the Sun\footnote{The Sun's effect on the planetoid is much larger than the Sun's effect on the Moon.}, one finds that $L_{4}$ and $L_{5}$ are no longer points of stable equilibrium. This issue will be described in details in Sec. \ref{Restricted4b_Sec}.

\section{Motion in the neighbourhood of a given motion} \label{Neighbourhood_Sec}

The fact that many dynamical problems are intractable, in the sense that we are unable to display an analytical solution, represents a quite common issue in physics. Nevertheless, there are some classes of problems for which the solution turns out to be easier due to the presence of some simplifying factors. An example is provided by the description of a motion which is near a known solution of the original problem, such as the motion near an equilibrium point. The tool adopted in classical mechanics for such a situation is represented by the so-called variational equations, which describe the time evolution of the deviation vector between two solutions, i.e., the undisturbed (known) solution and the disturbed (unknown) one.  

\subsection{Variational equations}

We have already pointed out in Sec. \ref{Lagrangian_points_Sec} that the autonomous system of differential equations (\ref{RomanoA1}) governing the dynamics of a point particle in the $n$-fold space $(x_1,x_2,\dots,x_n)$ is such that its Cauchy problem (see Eq. (\ref{RomanoA2})) is in general not feasible. Despite that, in many astronomical applications we can find approximations to the solution with a high degree of accuracy.

Consider the situation in which we know the solution of the Cauchy problem (\ref{RomanoA1}) and (\ref{RomanoA2}) with a particular initial value $\bold{x}_0=\boldsymbol{\varrho}$, but not its behaviour in some ranges of values of $\boldsymbol{\varrho}$. Then, we try to determine, or to determine approximately, the solution (or characteristic) originating in a point $\boldsymbol{\varrho} + \boldsymbol{\eta}$ in the neighbourhood of $\boldsymbol{\varrho}$. In other words, we aim to describe a characteristic in the vicinity of a known characteristic \cite{BE14b,BEDS15,P1890,P1892,Pars,Szebehely67}. Let the known solution (or undisturbed characteristic) be $\bold{x}$. Thus, we have $x_i(t=0)=\varrho_i$ and $\dot{x}_i=X_i(x_1,x_2,\dots,x_n),$ $(i=1,2,\dots,n)$. Moreover, suppose that the $n$ functions $X_i$ are of class $C^1$ in a certain domain of $\bold{x}$. Let the neighbouring solution (or disturbed characteristic) be $\bold{x}+\bold{y}$, $\bold{y}$ representing the displacement from the undisturbed solution. Thence,
\begin{equation}
\dot{x}_i+ \dot{y}_i=X_i(x_1+y_1,x_2+y_2,\dots,x_n+y_n), \; \; \; \; \; \; \;\; \; \; \; \; \; \; (i=1,2,\dots,n),
\end{equation}
with $y_i(t=0)=\eta_i$, $(i=1,2,\dots,n)$. Therefore, we can say that the $n$ displacement functions $y_i$ satisfy the $n$ differential equations
\begin{equation}
\dot{y}_i=Y_i(y_1,y_2,\dots,y_n;t), \; \; \; \; \; \; \;\; \; \; \; \; \; \; (i=1,2,\dots,n), \label{Pars2314}
\end{equation}
where
\begin{equation}
Y_i(y_1,y_2,\dots,y_n;t)= X_i(x_1+y_1,x_2+y_2,\dots,x_n+y_n) - X_i(x_1,x_2,\dots,x_n), \; \; \; \; \; \; \;\; (i=1,2,\dots,n). \label{Pars2315}
\end{equation}
Note how the system (\ref{Pars2314}), unlike (\ref{RomanoA1}), is obviously not autonomous and that the symbols $x_1,x_2,\dots,x_n$ appearing on the right-hand side of (\ref{Pars2315}) denote known functions of $t$, i.e., the values at time $t$ of the undisturbed characteristic. If we suppose that $\vert \boldsymbol{\eta} \vert$ small, also $\vert \bold{y} \vert$ will be small (at least for a sufficiently small range of values of $t$), since the solution of (\ref{RomanoA1}) varies continuously with its initial value. We refer to all those particular cases in which $\vert \bold{y} \vert$ remains small for all values of $t$ as stable displaced orbits. The characterization of regions of stability and instability of displaced orbits is a fascinating theoretical problem (well described for example by Levi-Civita in Ref. \cite{LeviCivita56}) whose solution might have far-reaching consequences for example for the design of space missions. We will not treat such a problem, but instead of studying the exact equations (\ref{Pars2314}) we will consider their linear approximation. By expanding the right-hand side of Eq. (\ref{Pars2315}) up to terms of first order in the $y_i$'s, we obtain the above-mentioned linear approximation, i.e.,
\begin{equation}
\dot{\xi}_i = \sum_{k=1}^{n} a_{ik} \xi_k, \; \; \; \; \; \; \;\; (i=1,2,\dots,n),\label{Pars2316}
\end{equation}
where we have written $\xi_i$ in place of $y_i$ to make a clear distinction between $\bold{y}$, which satisfies the exact equations (\ref{Pars2314}) and $\boldsymbol{\xi}$, which instead satisfies the linear approximation (\ref{Pars2316}). Moreover, we have
\begin{equation}
a_{ik}=a_{ik}(t) \equiv\left.  \dfrac{\partial X_i}{\partial x_k} \right\vert_{\bold{x}}, \label{aik_coeff}
\end{equation}
meaning that the coefficients $a_{ik}$ are known functions of $t$ whose value is represented by the derivatives $\partial X_i/\partial x_k$ evaluated at the point $\bold{x}=(x_1,x_2,\dots,x_n)$ occupied by the particle at time $t$ on the undisturbed solution. Equations (\ref{Pars2316}) are called (linear) variational equations (sometimes called also variational equations of Jacobi or of Poicar\'e) and their matrix form reads as
\begin{equation}
\dot{\boldsymbol{\xi}}= \bold{A} \boldsymbol{\xi}, 
\label{Pars2317}
\end{equation}
where $\boldsymbol{\xi}$ denotes the column matrix with entries the functions $(\xi_1,\xi_2,\dots,\xi_n)$ and $\bold{A}$ the $n \times n$ matrix whose elements are defined in Eq. (\ref{aik_coeff}). We can think of solutions to linear variational equations as representing the tangent vectors along the undisturbed trajectory which linearly approximate its difference with nearby trajectories. We will see that if the solution $\boldsymbol{\xi}$ of (\ref{Pars2317}) is such that $\vert \boldsymbol{\xi} \vert$ remains small for all the time when $\vert \boldsymbol{\eta} \vert$ is small (with $\boldsymbol{\xi}(t=0)=\boldsymbol{\eta}$), we say that the undisturbed characteristic has first-order stability (or that it is infinitesimally stable). 

Some special circumstances must be mentioned at this stage. First of all, there are two special cases in which the elements (\ref{aik_coeff}) occurring in the definition of the matrix $\bold{A}$ are constants. The first is represented by the motion in the neighbourhood of a singular point, which includes the special case of small oscillations near a stable equilibrium point. The second involves the cases of steady motion. The description of a steady motion involves {\it in primis} the definition of a gyroscopic system, i.e., a system characterized by the presence of $k$ Lagrangian coordinates $q_1, q_2, \dots, q_k$ which are ignorable while the remaining $n-k$ ($n$ being the number of degrees of freedom) $q_{k+1}, q_{k+2}, \dots, q_n$ are non-ignorable (or palpable) \cite{Romano,Pars}. Therefore, we have $k+1$ first integrals of the system, i.e., the $k$ momenta corresponding to the $k$ ignorable coordinates and the integral of energy. We define steady motion to be the one in which both the velocities $\dot{q}_1, \dot{q}_2, \dots, \dot{q}_k$ and the palpable coordinates have constant values. However, for such a motion the non-constant ignorable coordinates do not occur in the definition (\ref{aik_coeff}) and hence $\bold{A}$ ends up to be constant. Obviously, when the matrix $\bold{A}$ turns out to be constant the system (\ref{Pars2317}) is autonomous. Eventually, another situation of interest is represented by the one in which the functions $a_{ik}(t)$ in Eq. (\ref{aik_coeff}) are periodic functions of $t$ so that we need to study a periodic matrix $\bold{A}(t)$ \cite{BE14b,P1890,P1892}. We will describe these issues in the next sections.  

\subsection{Solution of variational equations}

Let 
\begin{equation}
\boldsymbol{\xi}(t=0)=\boldsymbol{\eta}.
\label{xi(0)}
\end{equation}
The solution of the Cauchy problem (\ref{Pars2317}) and (\ref{xi(0)}) is suggested by the method of successive approximation (see also Appendix \ref{Fundamental_App}) \cite{Pars}. Consider the (principal fundamental) matrix $\bold{R}(t)$
\begin{equation}
\bold{R}(t)= \bold{D}_0 + \bold{D}_1(t)+\bold{D}_2(t)+ \dots, \label{Pars2321}
\end{equation}
where
\begin{equation}
\bold{D}_0= \mathbb{1},
\end{equation}
while for the successive members
\begin{equation}
\dot{\bold{D}}_{j+1}= \bold{A} \bold{D}_{j},
\end{equation}
with 
\begin{equation}
\bold{D}_{j+1}(t=0) \equiv \bold{D}_{j+1}(0)= \bold{0}.
\end{equation}
Then, since
\begin{equation}
\bold{R}(0)= \bold{D}_{0}= \mathbb{1},
\end{equation}
and 
\begin{equation}
\dot{\bold{R}}(t)= \bold{A} \bold{R}(t),
\end{equation}
the matrix $\bold{R}(t) \boldsymbol{\eta}$ assumes the value $\boldsymbol{\eta}$ at $t=0$ and satisfies (\ref{Pars2317}), the required solution being linear in $\boldsymbol{\eta}$ (as should be expected), i.e., 
\begin{equation}
\boldsymbol{\xi}(t)=\bold{R}(t) \boldsymbol{\eta}.
\label{Pars2327}
\end{equation} 
Since the elements $d_{ij}^{\,(h)}$ of $\bold{D}_{h}(t)$ are such that, for some interval $0 \leq t \leq t_1$, \cite{Pars}
\begin{equation}
n \vert d_{ij}^{\,(h)} \vert < \dfrac{\left(nKt\right)^h}{h!},
\end{equation}
$K$ being a positive number such that, for all the $n^2$ elements $a_{ij}(t)$ of $\bold{A}$,
\begin{equation}
\vert a_{ij} (t) \vert < K,
\end{equation}
then the infinite series defining the elements of $\bold{R}$ is majorized by an exponential series having constant terms and hence (\ref{Pars2321}) is uniformly convergent for $0 \leq t \leq t_1$. 
\begin{description}
\item[{\it Example: Newtonian orbit.}] Consider a point particle of unit mass subjected to the Newtonian attraction of a second body. Let the origin of the coordinate frame be the center of the gravitational force and choose as Lagrangian coordinates the polar coordinates $r$, $\theta$ of the first body. The Lagrangian of the system will be
\begin{equation}
\mathcal{L}= \dfrac{1}{2} \left(\dot{r}^2 + r^2 \dot{\theta}^2 \right) + \dfrac{g}{r},
\end{equation}
$g$ being the product of the Newton constant $G$ and the mass of the second body. Thus, the Hamiltonian will be given by
\begin{equation}
H =  \dfrac{1}{2} \left( p_{r}^{2}+ \dfrac{1}{r^2} p_{\theta}^{2} \right) - \dfrac{g}{r},
\end{equation} 
and the Hamiltonian equation of motion reads as
\begin{equation}
\renewcommand{\arraystretch}{2.0}
\begin{dcases}
&\dot{r}=p_r,\\
& \dot{\theta}= \dfrac{1}{r^2} p_{\theta}, \\
& \dot{p}_r = - \dfrac{g}{r^2}+ \dfrac{1}{r^3} p_{\theta}^{2},\\
& \dot{p}_{\theta}=0, 
\end{dcases} \\ [2em]
\label{Pars23226}
\end{equation}
where in particular $\dfrac{1}{r^3} p_{\theta}^{2}$ represents the centrifugal force, $p_{\theta}$ being a constant which we indicate with $\tilde{p}$. By a comparison with Eq. (\ref{RomanoA1}), it follows easily that $(r, \theta, p_r, p_\theta)=(x_1,x_2,x_3,x_4)=\bold{x}$ and
\begin{equation}
\bold{X}(\bold{x})= \left( 
\renewcommand{\arraystretch}{2.0}
\begin{matrix}
p_r \cr
 \dfrac{1}{r^2} p_{\theta} \cr
 - \dfrac{g}{r^2}+ \dfrac{1}{r^3} p_{\theta}^{2} \cr
 0
\end{matrix} \right),
\end{equation}
Thus, bearing in mind Eq. (\ref{aik_coeff}), the matrix $\bold{A}$ is given by 
\begin{equation}
\bold{A}(t)= \left( 
\renewcommand{\arraystretch}{2.0}
\begin{matrix}
0 & 0 & 1 & 0 \cr
-\dfrac{2 \tilde{p}}{r^3} & 0 & 0 & \dfrac{1}{r^2} \cr
\dfrac{2 g}{r^3}-\dfrac{3 \tilde{p}^2}{r^4} & 0 & 0 & \dfrac{2 \tilde{p}}{r^3} \cr
0 &  0 & 0 & 0 \cr
\end{matrix} \right),
\end{equation}
where $r$ indicates the value of the radial distance at time $t$ on the undisturbed characteristic. If the undisturbed orbit is an ellipse characterized by a certain period, the elements $a_{ik}$ of $\bold{A}(t)$ are known periodic function of $t$ with the same period. On the other hand, in the special case when the undisturbed solution is represented by an uniform circular motion, $\bold{A}$ is a constant matrix. In this latter circumstance, from the request that that centrifugal force balances in Eq. (\ref{Pars23226}c) the Newtonian one, we obtain the value of the angular velocity $\omega$ needed for a circular orbit at any altitude, i.e., 
\begin{equation}
\omega= \sqrt{\dfrac{g}{a^3}},
\end{equation}
\end{description}
$a$ being the radius of the circular trajectory. Thus, we can write $\tilde{p}=a^2 \omega$ and $\bold{A}$ becomes the constant matrix
\begin{equation}
\bold{A}= \left( 
\renewcommand{\arraystretch}{2.0}
\begin{matrix}
0 & 0 & 1 & 0 \cr
-\dfrac{2 \omega}{a} & 0 & 0 & \dfrac{1}{a^2} \cr
-\omega^2  & 0 & 0 & \dfrac{2 \omega}{a} \cr
0 &  0 & 0 & 0 \cr
\end{matrix} \right).
\end{equation}
This is clearly an example of steady motion where $\theta$ represents an ignorable coordinate and both the velocity $\dot{\theta}=\omega$ and the radial coordinate $r=a$ assume constant values. The variational equations will be (see eq. (\ref{Pars2317}))
\begin{equation}
\begin{dcases}
& \dot{\xi}_1=\xi_3,\\ 
 & \dot{\xi}_2=-\dfrac{2 \omega}{a} \xi_1 + \dfrac{1}{a^2}\xi_4, \\ 
 & \dot{\xi}_3=-\omega^2 \xi_1 + \dfrac{2 \omega}{a} \xi_4, \\ 
 & \dot{\xi}_4=0,
\end{dcases} \\ [2em]
\end{equation}
with solution given by
\begin{equation}
\begin{dcases}
& \xi_1=\eta_1 \cos \left(\omega t\right)+\dfrac{\eta_3}{\omega} \sin \left(\omega t\right) + \dfrac{2}{a \omega} \eta_4 \left[1- \cos \left(\omega t\right) \right], \\
& \xi_2 = - \dfrac{2}{a} \eta_1 \sin \left(\omega t\right)+ \eta_2-\dfrac{2 \eta_3}{a \omega} \left[1- \cos \left(\omega t\right) \right]- \dfrac{\eta_4}{a^2 \omega} \left[3 \omega t - 4 \sin \left(\omega t\right) \right], \\
& \xi_3= - \omega \eta_1 \sin \left(\omega t\right) + \eta_3 \cos \left(\omega t\right) + \dfrac{2}{a} \eta_4 \sin \left(\omega t\right), \\
& \xi_4= \eta_4.
\end{dcases} \\ [2em]
\end{equation}
Note that $\vert \xi_1 \vert$, $\vert \xi_3 \vert$, and $\vert \xi_4 \vert$ remain small for all time if $\vert \boldsymbol{\eta} \vert$ is small, whereas 
\begin{equation}
\lim_{t \rightarrow + \infty} \vert \xi_2 \vert = - \infty,
\end{equation}
unless $\eta_4=0$. A special case is that in which $\boldsymbol{\eta}=(0,0,\eta,0)$. The system is then stable at first order, the solution of variational equations being
\begin{equation}
\begin{dcases}
& \xi_1=\dfrac{\eta}{\omega} \sin \left(\omega t\right), \\
& \xi_2 = -\dfrac{2 \eta}{a \omega} \left[1- \cos \left(\omega t\right) \right], \\
& \xi_3=  \eta \cos \left(\omega t\right), \\
& \xi_4= 0.
\end{dcases} \\ [2em]
\end{equation}

\subsection{The case of constant coefficients and the concept of first-order stability}

In the special case in which the elements $a_{ik}$ of Eq. (\ref{aik_coeff}) are constants, also the matrix $\bold{A}$ is constant and 
\begin{equation}
\bold{D}_{h}=\left(\dfrac{t^h}{h!} \right)\bold{A}^h,
\end{equation}
so that Eq. (\ref{Pars2321}) becomes
\begin{equation}
\bold{R}= \mathbb{1} + t \bold{A} + \left(\dfrac{t^2}{2!}\right) \bold{A}^2 +  \left(\dfrac{t^3}{3!}\right) \bold{A}^3 + \dots = {\rm e}^{t \bold{A}}.
\end{equation}
Therefore, the solution (\ref{Pars2327}) assumes the simple form
\begin{equation}
\boldsymbol{\xi}=  {\rm e}^{t \bold{A}} \, \boldsymbol{\eta } . \label{Pars2332}
\end{equation}
If $\bold{A}$ is a diagonal matrix, it becomes more evident that the variational equations (\ref{Pars2316}) get completely separated into the $n$ independent equations
\begin{equation}
\dot{\xi}_i = \lambda_i \xi_i, \; \; \; \; \; \; \;\; \; \; \; \; \; \;\; \; \; \; \; \; \; (i=1,2,\dots,n), \label{Pars2333}
\end{equation}
where $\lambda_i=a_{ii}$ represent the eigenvalues of $\bold{A}$. Then, the solution will be
\begin{equation}
\xi_i= \eta_i {\rm e}^{t \lambda_i}, \; \; \; \; \; \; \;\; \; \; \; \; \; \;\; \; \; \; \; \; \; (i=1,2,\dots,n), \label{Pars2334}
\end{equation}
because the matrix ${\rm e}^{t \bold{A}}$ appearing in Eq. (\ref{Pars2332}) is the diagonal matrix
\begin{equation}
\left( 
\renewcommand{\arraystretch}{1.5}
\begin{matrix}
{\rm e}^{t \lambda_1} & 0 & \dots & 0 \cr
0 & {\rm e}^{t \lambda_2} & \dots & 0 \cr
. & . & \dots & . \cr
0 & 0 & \dots &  {\rm e}^{t \lambda_n} \cr
\end{matrix}
\right).
\end{equation}

When $\bold{A}$ can not be reduced to diagonal form, the solution of (\ref{Pars2333}) contains terms of the form $\xi \sim t^{\mu} {\rm e}^{\lambda t}$, where we suppose that $\mu$ is a number such that $\mu>0$ (i.e., $t^{\mu}$ is a secular term).  

At this stage, we can describe the stability proprieties of the linearised system (\ref{Pars2333}) . In fact, we can distinguish three different cases: 
\begin{enumerate}
   \item The characteristic equation of $\bold{A}$ has complex roots. We have the following properties:
   \begin{itemize}
     \item If all the $\lambda$'s have negative real parts, we have {\it first-order asymptotic stability}. In fact, in such a case $\underset{t \to +\infty}{\lim} \vert \boldsymbol{\xi} \vert = 0$. This is true also when multiple roots are present.
     \item If some or all of the characteristic roots of $\bold{A}$ have positive real parts, the linearised system is characterized by {\it first-order instability}, since the displacement vector $\boldsymbol{\xi}$ occurring in Eq. (\ref{Pars2333}) does not remain small. This circumstance remains true also when some of the roots are multiple.
   \end{itemize}
   \item The characteristic equation of $\bold{A}$ has pure imaginary roots. We have two cases:
   \begin{itemize}
   \item If all the roots turn out to be simple, then the solution contains only sines and cosines of multiples of $t$ and hence is oscillatory. Thence, $\vert \boldsymbol{\xi} \vert $ remains small if it is initially small. We have {\it first-order stability}, but not asymptotic stability.  
   \item If multiple roots are present, both periodic and secular terms of the form $t^{\mu} \cos \left( k t \right)$, $t^{\mu} \sin \left( k t \right)$ ($k$ being a real number) are present in the solution. The linearised system has {\it first-order instability}.
   \end{itemize}
    \item The characteristic equation of $\bold{A}$ has real roots. We have two different situations:
    \begin{itemize}
    \item The roots are all negative. The solution has {\it first-order stability}. This stays true also when multiple roots are present.
    \item Some of the roots are positive. Then, even when multiple roots are present, we have {\it first-order instability}. 
    \end{itemize}
\end{enumerate}
The above considerations can be straightforwardly applied to the analysis of the stability (at first order) of the equilibrium points of a dynamical system, since, as we pointed out before, in such a case the matrix $\bold{A}$ turns out to be a constant matrix. 

An important comment on these results is called for at this point. In fact, we stress that first-order stability does not imply stability in general. In other words, if the linearised system (\ref{Pars2333}) turns out to be stable at first order, this does not necessary mean that the exact system (\ref{Pars2314}) is stable.

\subsection{The case of periodic coefficients} \label{periodic_coeff_sec}

Consider the case in which the undisturbed characteristic is represented by a periodic orbit with period $T$ \cite{BE14b,BEDS15,P1890,P1892,Pars,Eastham}. In other words, we can say that Eq. (\ref{RomanoA1}) admits the periodic solution 
\begin{equation}
x_i=\varphi_i(t), \; \; \; \; \; \; (i=1,2,\dots,n), \label{undisturbed1}
\end{equation}
which is such that
\begin{equation}
\varphi_i(t+T)=\varphi_i(t), \; \; \; \; \; \; \forall \, t. 
\end{equation}
This means that the elements $a_{ik}$ of the linear variational equations (\ref{Pars2316}) are periodic functions of $t$ sharing the same period $T$ of the undisturbed solution (\ref{undisturbed1}), so that
\begin{equation}
\bold{A}(t)= \bold{A}(t+T), \; \; \; \; \; \; \forall \, t,
\end{equation}
and hence if
\begin{equation}
\bold{F}(t) = \left[
\begin{matrix}
\boldsymbol{\xi}_1(t) & \boldsymbol{\xi}_2(t) & \dots & \boldsymbol{\xi}_n(t) \cr
\end{matrix}
\right]=
\left[
\begin{matrix}
\xi_{11}(t) & \dots & \xi_{1n}(t) \cr
\vdots &\ddots & \vdots \cr
\xi_{n1}(t) & \dots & \xi_{nn}(t) \cr
\end{matrix}
\right],
\label{Fundamental_periodic}
\end{equation}
is a fundamental matrix (where $\xi_{ij}$ indicates the $i$-th component of the $j$-th linearly independent vector, so that, for example, $\xi_{12}$ represents the first component of $\boldsymbol{\xi}_2$) satisfying the condition (\ref{Fund25}), which can now be written as
\begin{equation}
\dot{\bold{F}}(t)=\bold{A}(t) \bold{F}(t),
\end{equation}
then also $\bold{F}(t+T)$ represents a fundamental matrix. Therefore, there exists a non-singular constant matrix $\bold{M}$ such that \cite{Eastham}
\begin{equation}
\bold{F}(t+T)= \bold{F}(t)\bold{M}. \label{Pars2344}
\end{equation}
The matrix $\bold{M}$ is called monodromy matrix of the fundamental matrix $\bold{F}(t)$. It expresses the important fact that the solution of variational equation is in general not periodic, unlike the undisturbed characteristic. Since $\bold{M}$ is time-independent, it can be computed by setting $t = 0$ in Eq.(\ref{Pars2344}), yielding
\begin{equation}
\bold{M}=\bold{F}(0)^{-1}\bold{F}(T),
\end{equation}
so that in the case in which $\bold{F}(t)$ turns out to be a principal fundamental matrix, from the last condition we simply have
\begin{equation}
\bold{M}=\bold{F}(T).
\end{equation}
Thence, the monodromy matrix of the principal fundamental matrix (\ref{Pars2321}) is $\bold{R}(T)$. Consider the fundamental matrix $\bold{G}(t)=\bold{F}(t)\bold{C}$ ($\bold{C}$ being a constant matrix). The monodromy matrix $\bold{N}$ of $\bold{G}(t)$ will be given by
\begin{equation}
\bold{N}= \bold{C}^{-1} \bold{M}\bold{C}, \label{Pars2346}
\end{equation}
$\bold{M}$ being the monodromy matrix of $\bold{F}(t)$. This follows at once from the condition
\begin{equation}
\bold{G}(t+T)=\bold{F}(t+T)\bold{C}=\bold{F}(t)\bold{M}\bold{C}=\bold{G}(t)\bold{C}^{-1}\bold{M}\bold{C}. 
\end{equation}
Therefore, if two fundamental matrices are related by the condition $\bold{G}(t)=\bold{F}(t)\bold{C}$, the corresponding monodromy matrices $\bold{N}$ and $\bold{M}$ are similar, as witnessed by Eq. (\ref{Pars2346}). Thus, all monodromy matrices have the same eigenvalues and hence can be reduced to the same Jordan normal form (see Appendix \ref{Fundamental_App}). The eigenvalues $\mu_1,\mu_2,\dots,\mu_n$ of $\bold{M}$ are called characteristic multipliers. None of them vanishes, since
\begin{equation}
\mu_1\mu_2\dots\mu_n = \det \left( \bold{M} \right) \neq 0. 
\end{equation}
By virtue of the similarity condition (\ref{Pars2346}), the characteristic multipliers are an intrinsic property of variational equations and do not depend on the choice of the fundamental matrix. 

If $\bold{M}$ is the monodromy matrix of the fundamental matrix $\bold{F}(t)$, we can find a matrix $\bold{K}$ (not always a real matrix) such that \cite{Eastham}
\begin{equation}
\bold{M}= {\rm e}^{T \bold{K}}.\label{Pars2348}
\end{equation}
If the eigenvalues of $\bold{K}$ are $\alpha_1, \alpha_2, \dots, \alpha_n$, those of $\bold{M}$ are ${\rm e}^{T \alpha_1},{\rm e}^{T \alpha_2}, \dots,{\rm e}^{T \alpha_n}$, i.e.,
\begin{equation}
\mu_{k}={\rm e}^{T \alpha_k}, \; \; \; \; \; \; \;\; \; \; \; \; \; \; (k=1,2,\dots,n). 
\end{equation}
The numbers $\alpha_1, \alpha_2, \dots, \alpha_n$ are called characteristic exponents or Floquet exponents \cite{P1890,Eastham} of the given periodic orbit. Note that the characteristic exponents are not unique, since if $\mu_{j}={\rm e}^{T \alpha_j}$, then the same $\mu_j$ can also be written as $\mu_{j}={\rm e}^{\left(\alpha_j+2\pi \I/T \right)T}$.

The most important property of variational equations (\ref{Pars2317}) in the case of periodic coefficients is enlightened by the following theorem \cite{Eastham}: 
\newtheorem*{Floquet1}{Floquet Theorem}
\begin{Floquet1}
Let 
\begin{equation}
\dot{\boldsymbol{\xi}}(t)=\bold{A}(t) \boldsymbol{\xi}(t), \label{variational_periodic}
\end{equation}
represent the linear variational equations (\ref{Pars2317}) in the case in which $\bold{A}(t)$ is an $n \times n$ periodic matrix with period $T$. Then any fundamental matrix solution $\bold{F}(t)$ can be expressed in the form
\begin{equation}
\bold{F}(t)=\bold{S}(t) {\rm e}^{t \bold{K}}, \label{Pars23413}
\end{equation}
where $\bold{S}(t)$ is a non-singular periodic matrix whose elements are continuous periodic functions of $t$ with period $T$ and $\bold{K}$ is the constant matrix appearing in Eq. (\ref{Pars2348}). The representation (\ref{Pars23413}) is called Floquet normal form of the fundamental matrix $\bold{F}(t)$.
\end{Floquet1}
\begin{proof}
First of all, note that
\begin{equation}
\bold{F}(t+T)=\bold{F}(t)\bold{M}=\bold{F}(t) {\rm e}^{T \bold{K}}.
\nonumber
\end{equation}
Then, by writing
\begin{equation}
\bold{S}(t)= \bold{F}(t) {\rm e}^{-t \bold{K}},
\nonumber
\end{equation}
it follows at once that $\bold{S}(t)$ is periodic, since we have
\begin{equation}
\bold{S}(t+T)=\bold{F}(t+T) {\rm e}^{-(t+T) \bold{K}}= \bold{F}(t) {\rm e}^{T \bold{K}} {\rm e}^{-T \bold{K}} {\rm e}^{-t \bold{K}}= \bold{F}(t) {\rm e}^{-t \bold{K}}=\bold{S}(t).
\nonumber
\end{equation}
Finally, since $\bold{F}(t)$ and ${\rm e}^{t \bold{K}}$ are non-singular, also $\bold{S}(t)$ is non-singular and hence we can write
\begin{equation}
\bold{F}(t)=\bold{S}(t) {\rm e}^{t \bold{K}}.
\nonumber
\end{equation}
\end{proof}
The Floquet normal form (\ref{Pars23413}) has the great advantage of giving rise to a time-dependent change of coordinates 
\begin{equation}
\tilde{\boldsymbol{\xi}} = \bold{S}^{-1}(t) \boldsymbol{\xi},
\end{equation}
under which our original system (\ref{variational_periodic}) becomes a linear system with real constant coefficients having form
\begin{equation}
\dfrac{{\rm d}}{{\rm d}t}\,\tilde{\boldsymbol{\xi}}= \bold{K} \tilde{\boldsymbol{\xi}}.
\end{equation}
Note also that if we write
\begin{equation}
\bold{F}(t)=\left({\rm e}^{2 \pi \I t/T} \right)\bold{S}(t) {\rm e}^{t \bold{K}},
\end{equation}
the function $\left({\rm e}^{2 \pi \I t/T} \right)\bold{S}(t)$ is still periodic and hence the fact that the characteristic exponents are not unique does not alter our results.

Bearing in mind (\ref{Fundamental_periodic}), from Eq. (\ref{Pars23413}) it easily follows that, in all those cases in which the monodromy matrix is diagonalizable and admits $n$ distinct eigenvalues, the components of the $n$ linearly independent solutions $(\boldsymbol{\xi}_1, \boldsymbol{\xi}_2,\dots,\boldsymbol{\xi}_n)$ of the variational equations can be written in the form
\begin{equation}
\begin{split}
 \boldsymbol{\xi}_1 &= (S_{11},S_{21},\dots,S_{n1}){\rm e}^{t \alpha_1}, \\
 \boldsymbol{\xi}_2 & = (S_{12},S_{22},\dots,S_{n2}){\rm e}^{t \alpha_2}, \\
& \vdots \\
\boldsymbol{\xi}_n & = (S_{1n},S_{2n},\dots,S_{nn}){\rm e}^{t \alpha_n},
\end{split}
\end{equation}
or equivalently
\begin{equation}
\xi_{ij}(t)=S_{ij}(t){\rm e}^{t \alpha_j}, \; \; \; \; \; \; \; \; \; ({i,j=1,2,\dots,n}),
\end{equation}
(no summation over repeated indices). Therefore, bearing in mind the above relations jointly with the results of Appendix \ref{Fundamental_App} (see in particular Eq. (\ref{Fund18})), the solution
\begin{equation}
\boldsymbol{\xi}(t)= 
\left[
\begin{matrix}
\xi_1(t) \cr
\xi_2(t) \cr
\vdots \cr
\xi_n(t) \cr
\end{matrix}
\right], \label{variational_periodic_sol}
\end{equation}
of Eq. (\ref{variational_periodic}) can be written as ($C_j$ being constants)
\begin{equation}
\xi_i(t)=\sum_{j=1}^{n}C_j S_{ij}(t) {\rm e}^{t \alpha_j } \equiv \sum_{j=1}^{n}S_{ij}(t) {\rm e}^{t \alpha_j } , \; \; \; \; \; \; \; \; \; ({i=1,2,\dots,n}),
\label{variational_periodic_sol_2}
\end{equation}
which simply expresses the fact that the $i$th component of the solution (\ref{variational_periodic_sol}) of (\ref{variational_periodic}) can be expressed as the sum of the $i$th components of the $n$ linearly independent vectors appearing in the fundamental matrix (\ref{Fundamental_periodic}). In particular, (\ref{variational_periodic_sol_2}) does not need to be periodic, as we pointed out before. Note also that in the case in which the undisturbed characteristic is represented by a stationary solution (e.g., an equilibrium point), the functions $S_{ij}(t)$ become constant and we recover (\ref{Pars2334}). Then, variational equations are always linear differential equations whose coefficients can be constant or can be periodic functions of time, depending on whether the undisturbed solution is stationary or not.  

Thence, in this special case (i.e., $\bold{M}$ is diagonalizable and possesses $n$ distinct eigenvalues) we can distinguish the following circumstances:
\begin{itemize}
\item[-] If all the characteristic exponents have negative real parts, we have {\it first-order asymptotic stability}.
\item[-] If all the characteristic exponents have positive real parts, we have {\it first-order instability}.
\item[-] If all the characteristic exponents are real and negative, we have {\it first-order stability}. If any of the roots are positive we have {\it first-order instability}.
\item[-] If all the characteristic exponents are pure imaginary numbers, we have {\it first-order stability}.
\end{itemize}
On the contrary, if the monodromy matrix $\bold{M}$ is diagonalizable but it does not provide us with $n$ different characteristic exponents (i.e., the characteristic equation of $\bold{M}$ has multiple roots), all but one the above-mentioned cases remain unaffected: we always have {\it first-order instability} when all the characteristic exponents are pure imaginary. In fact, when the characteristic multipliers are not distinct, the solution of variational equations can no longer be placed in form (\ref{variational_periodic_sol_2}), but it can be written as
\begin{equation}
\xi_i \sim t^{\mu} S_{ij} {\rm e}^{t \alpha_j},
\label{variational_general_case}
\end{equation}
$t^{\mu}$ ($\mu>0$) being a secular term. As an example, one can easily show that when two characteristic exponents are equal, then the solution will contain secular terms linear in $t$, while if three characteristic exponents turn out to be the same, the secular terms will be quadratic in $t$. Thus, if the characteristic exponents are complex but all of them have negative real parts, we have {\it first-order asymptotic stability}, since in this case Eq. (\ref{variational_general_case}) reduces to the form
\begin{equation}
\xi_i \sim t^{\mu}  {\rm e}^{-k\,t }, \;\;\;\;\;(k>0),
\end{equation}
so that $\lim\limits_{t \to +\infty} \xi_i =0$. On the other side, it is easy to realize that when some exponent has positive real part we have {\it first-order instability}. The case with all characteristic exponents which are real is trivial. Finally, when the characteristic exponents are pure imaginary numbers but some of them are equal, we have, as anticipated before, {\it first-order instability}, since in this case Eq. (\ref{variational_general_case}) will contain both trigonometric and secular terms. This situation is reminiscent of that of constant coefficients.

The results of Appendix \ref{Fundamental_App} have showed that in all those circumstances in which a matrix is {\it not} diagonalizable, we can achieve its best (and somehow unique) ``closest-to-diagonal'' form by employing its Jordan normal form.  Therefore, we can find the explicit solution of (\ref{variational_periodic}) in the general case by introducing a transformation which reduces $\bold{K}$ to its Jordan normal form. Let $\bold{L}$ be a non-singular matrix such that
\begin{equation}
\bold{L}^{-1} \bold{K} \bold{L}=\bold{J}.
\label{Pars23414}
\end{equation} 
We write the Jordan normal form $\bold{J}$ as
\begin{equation}
\bold{J}= 
\left(
 \begin{matrix}
\bold{J}_0 & 0 & \dots & 0 &  0 \cr
0 & \bold{J}_1 & \dots & 0 & 0 \cr
0 & 0 & \bold{J}_2 & \dots & 0 \cr
\vdots & \vdots & \vdots & \ddots & \vdots \cr
0 & 0 & \dots & 0 & \bold{J}_k 
\end{matrix}
\right),
\end{equation}
$\bold{J}_0$ being a $q\times q$ diagonal matrix with entries $\alpha_1,\alpha_2,\dots,\alpha_q$ not necessarily different, $\bold{J}_1,\bold{J}_2,\dots,\bold{J}_k$ the Jordan blocks. Bearing in mind that $\bold{K}$ represents an $n \times n$ matrix, we have
\begin{equation}
n = q + \sum_{i=1}^{k} p_i,
\end{equation}
$p_i$ being the size of the $i$th block. From Eqs. (\ref{Pars23413}) and (\ref{Pars23414}) it follows at once that
\begin{equation}
\bold{F}(t)=\bold{S}(t) {\rm e}^{t\bold{L}\bold{J}\bold{L}^{-1}}=\bold{S}(t)\bold{L}{\rm e}^{t \bold{J}}\bold{L}^{-1},
\end{equation}
so that the fundamental matrix $\bold{F}(t)\bold{L}$ has the form
\begin{equation}
\bold{S}(t)\bold{L}{\rm e}^{t \bold{J}} \equiv \bold{Q}(t) {\rm e}^{t \bold{J}}, \label{Pars23416}
\end{equation}
$\bold{Q}(t)$ being, obviously, periodic. In order to evaluate the term $ {\rm e}^{t \bold{J}}$, note that the matrix $ {\rm e}^{t \bold{J_0}}$ is just the diagonal matrix whose entries are ${\rm e}^{t \alpha_1},{\rm e}^{t \alpha_2},\dots,{\rm e}^{t \alpha_q}$. As far as the Jordan blocks $\bold{J}_i$ ($i=1,2,\dots,k$) are concerned, by noticing that
\begin{equation}
\bold{J}_i= \alpha_{q+i}\, \mathbb{1}_{p_i}+\bold{B}_i, \; \; \; \;\; \; \; \;\; \; \; \; (i=1,2,\dots,k),
\end{equation}
$\bold{B}_i$ being the matrix whose non-vanishing elements are placed on the super-diagonal and are equal to one (cf. Eq. (\ref{Jordan19})), it is possible to understand that we need to evaluate the term ${\rm e}^{t \bold{B}_i }$ if we want to discover the form of ${\rm e}^{t \bold{J} }$. It is easy to show that \cite{Pars}
\begin{equation}
{\rm e}^{t \bold{B}_i }= 
\left[
\renewcommand{\arraystretch}{2.2}
\begin{matrix}
1 & t & \dfrac{t^2}{2!} & \dots & \dfrac{t^{p_i-2}}{(p_i-2)!}& \dfrac{t^{p_i-1}}{(p_i-1)!} \cr
0 & 1 & t & \dots &  \dfrac{t^{p_i-3}}{(p_i-3)!}  & \dfrac{t^{p_i-2}}{(p_i-2)!} \cr
. & . & . & \dots & . & . \cr
0 & 0 & 0 & \dots & 1 & t \cr
0 & 0 & 0 & \dots & 0 & 1 \cr
\end{matrix}
\right].
\label{Pars23317}
\end{equation}
Now
\begin{equation}
{\rm e}^{t \bold{J}_i }= {\rm e}^{t \alpha_{q+i}} {\rm e}^{t \bold{B}_i }, \; \; \; \;\; \; \; \;\; \; \; \; (i=1,2,\dots,k).
\end{equation}
and hence all the components of the exponential map
\begin{equation}
{\rm e}^{t\bold{J}}= 
\left(
 \begin{matrix}
{\rm e}^{t\bold{J}_0} & 0 & \dots & 0 &  0 \cr
0 & {\rm e}^{t\bold{J}_1} & \dots & 0 & 0 \cr
0 & 0 & {\rm e}^{t\bold{J}_2} & \dots & 0 \cr
\vdots & \vdots & \vdots & \ddots & \vdots \cr
0 & 0 & \dots & 0 & {\rm e}^{t\bold{J}_k }
\end{matrix}
\right),
\label{Pars23313}
\end{equation}
occurring in (\ref{Pars23416}) have been evaluated. 

Therefore, in the most general situation where the monodromy matrix is {\it not} diagonalizable, although the solution of (\ref{variational_periodic}) is more involved than before, we can achieve the same conclusions as the case in which $\bold{M}$ is diagonalizable but with some multiple eigenvalues and hence, in particular, even if all the characteristic exponents are purely imaginary, we can never have first-order stability, since the exponential map (\ref{Pars23313}) contains secular terms, as can be easily seen from Eq. (\ref{Pars23317}). In fact, in this case the solution of (\ref{variational_periodic}) contains terms of the form $f(t) t^k \cos\left(u t\right)$, $f(t) t^k \sin\left(u t\right)$ (where $f(t)$ is a periodic function with period $T$ and $k,u \in \mathbb{R}$). 

We conclude this section with an important remark. In any problem of variation from a periodic orbit (but not in the problem of variation from an equilibrium point) one of the characteristic exponents is always vanishing (or equivalently, one characteristic multiplier is equal to one). In fact, the undisturbed periodic motion satisfies the autonomous system (\ref{RomanoA1}), which written in components becomes
\begin{equation}
\dot{x}_i = X_i (x_1,x_2,\dots,x_n), \;\; \; \; \; \; \;\; \; \; \; \; \;\; \; \; \; \; (i=1,2,\dots,n).
\end{equation}
Thus,
\begin{equation}
\ddot{x}_i = \sum_{j=1}^{n} \dfrac{\partial X_i}{\partial x_j} \dot{x}_j=\sum_{j=1}^{n} a_{ij}(t) \dot{x}_j,  \; \;\; \; \; \; \; \;\; \; \; \; \; (i=1,2,\dots,n),
\end{equation}
and the variational equations are satisfied by
\begin{equation}
\xi_i = \dot{x}_i,  \;\; \; \; \; \; \;\; \; \; \; \; \;\; \; \; \; \; (i=1,2,\dots,n),
\end{equation}
or in other words the variational equations have a purely periodic solution with period $T$. This can only happen if one of the characteristic exponents vanishes. Note how in this analysis the fact that the original system (\ref{RomanoA1}) is autonomous represents an absolutely necessary requisite. Moreover, when a dynamical system possesses an integral of motion another characteristic exponent is zero (unless all partial derivatives of this integral vanish identically for all points of the periodic solution) \cite{P1892}. Therefore, when in general the autonomous system (\ref{RomanoA1}) has $k$ integrals of motion, then $k+1$ characteristic exponents will be zero. On the other hand, when the original system (\ref{RomanoA1}) depends explicitly on time (i.e., it is not autonomous) and has $j$ first integrals, then $j$ characteristic exponents will be zero.  

It should now be clear that the approach described so far is essentially a linear analysis. The underlying idea is very clear. Since it is very improbable that, in any application, the initial conditions of (\ref{RomanoA1}) are exactly those which generate the periodic solution (\ref{undisturbed1}), it will result more likely that the actual motion differs from the periodic (undisturbed) one very little. In other words, we are interested in a solution differing very little from (\ref{undisturbed1}) and having form
\begin{equation}
\tilde{x}_i=\varphi_i(t) + \xi_i, \; \; \; \; \; \; (i=1,2,\dots,n), \label{disturbed1}
\end{equation}
where $\xi_i$'s satisfy the linear variational equations (\ref{variational_periodic}). Thus, we must interpret the coordinates (\ref{disturbed1}) as those of a body on his actual motion and the coordinates (\ref{undisturbed1}) as those that the same body would have in the periodic (hypothetical) motion. Moreover, we have so far supposed that the $\xi_i$'s are such that we can neglect, in first approximation, their squares, because the difference between the coordinates (\ref{disturbed1}) and (\ref{undisturbed1}) remain, within this framework, always very small.

\subsection{The equation defining the characteristic exponents}

We have seen in the previous section that when the undisturbed characteristic is represented by the periodic function (\ref{undisturbed1}), the solution of (\ref{variational_periodic}) is represented by (\ref{variational_periodic_sol_2}), provided that the monodromy matrix is diagonalizable and admits $n$ different characteristic multipliers. 

Consider the initial condition for the variation \cite{P1892}
\begin{equation}
\xi_i(0)= \gamma_i,  \;\; \; \; \; \; \;\; \; \; \; \; \;\; \; \; \; \; (i=1,2,\dots,n), \label{Sze_39}
\end{equation}
and consequently (cf. Eq. (\ref{disturbed1}))
\begin{equation}
\tilde{x}_i(0)= \varphi_i(0)+ \gamma_i, \;\; \; \; \; \; \;\; \; \; \; \; \;\;  \; (i=1,2,\dots,n).
\end{equation}
A period later the variations assume the value
\begin{equation}
\xi_i(T)=  \gamma_i + \psi_i \neq \xi_i(0), \;\; \; \; \; \; \;\; \; \; \; \; \;\;  \; (i=1,2,\dots,n), \label{Sze_40}
\end{equation}
since, as we said before, the solution of variational equations is not periodic in general. By bearing in mind the obvious fact that if $\gamma_i=0$ then $\psi_i=0$ (from (\ref{Sze_39}) it is clear that $\gamma_i$'s determine the solution (\ref{variational_periodic_sol_2}) and hence a given set of $\gamma_i$'s is connected to the set of $\psi_i$'s), the expansion of $\psi_i=\psi_i(\gamma_j)$ around $\gamma_j=0$ gives
\begin{equation}
\begin{split}
\psi_i &=\sum_{j=1}^{n}  \left. \dfrac{\partial \psi_i}{\partial \gamma_j} \right \vert_{\gamma_j=0} \gamma_j+ \sum_{j,k=1}^{n} \left. \dfrac{\partial^2 \psi_i}{\partial \gamma_j \partial \gamma_k} \right \vert_{\gamma_j=0} \dfrac{\gamma_j \gamma_k}{2!} + \dots  \\
& \equiv \sum_{j=1}^{n}  \dfrac{\partial \psi_i}{\partial \gamma_j} \gamma_j+ \sum_{j,k=1}^{n} \dfrac{\partial^2 \psi_i}{\partial \gamma_j \partial \gamma_k} \dfrac{\gamma_j \gamma_k}{2!} + \dots, \; \; \; \; \; \;\;  \; (i=1,2,\dots,n),
\end{split} 
\end{equation}
which, in the case in which the disturbed characteristic differs very little from the undisturbed periodic solution so that besides the squares of $\xi_i$'s we can neglect also those of $\gamma_i$'s, becomes simply
\begin{equation}
\psi_i=\sum_{j=1}^{n}  \dfrac{\partial \psi_i}{\partial \gamma_j} \gamma_j,  \;\; \; \; \; \; \;\; \; \; \; \; \;\;  \; (i=1,2,\dots,n). \label{Poincare_lemma}
\end{equation}
The above expansion is referred to as Poincar\'e lemma \cite{P1892}.

At this stage, we can describe a powerful method of determining the $j$th characteristic exponent occurring in (\ref{variational_periodic_sol_2}). The solution corresponding to this exponent is
\begin{equation}
\xi_i (t)= S_{ij}(t) {\rm e}^{t \alpha_j}, \;\; \; \; \; \; \;\; \; \; \; \; \;\;  \; (i=1,2,\dots,n), \label{Sze_42}
\end{equation}
($j$ fixed). We can establish a relation between $\gamma_i$ and $\psi_i$, since, by bearing in mind that (see Eq. (\ref{Sze_39}))
\begin{equation}
S_{ij}(T)=S_{ij}(0)=\xi_i(0)=\gamma_i,
\end{equation}
we have (cf. Eqs. (\ref{Sze_40}) and (\ref{Sze_42}))
\begin{equation}
\gamma_i + \psi_i = \xi_i(T) = S_{ij}(T) {\rm e}^{T \alpha_j} = \gamma_i {\rm e}^{T \alpha_j}, 
\end{equation}
(no summation over $j$) and hence
\begin{equation}
\gamma_i \left(1- {\rm e}^{T \alpha_j}  \right) + \psi_i=0, \; \; \; \; \; \;\; \; \; \; \; \;\;  \; (i=1,2,\dots,n). \label{Sze_43}
\end{equation}
By substituting (\ref{Poincare_lemma}) in (\ref{Sze_43}) we obtain for a {\it fixed} value of $j$
\begin{equation}
\gamma_i \left(1- {\rm e}^{T \alpha_j}  \right) + \sum_{k=1}^{n}  \dfrac{\partial \psi_i}{\partial \gamma_k} \gamma_k =0,  \;\; \; \; \; \; \;\; \; \; \; \; \;\;  \; (i=1,2,\dots,n),
\end{equation}
which represents a homogeneous system where $\gamma_i $'s can be regarded as the unknowns. Such a system will possess a solution different from the trivial one provided that its associated determinant vanishes, i.e.,
\begin{equation}
\det \left(
\renewcommand{\arraystretch}{2.2}
\begin{matrix}
\dfrac{\partial \psi_1}{\partial \gamma_1} + 1 - {\rm e}^{T \alpha_j} & \dfrac{\partial \psi_1}{\partial \gamma_2} & \dfrac{\partial \psi_1}{\partial \gamma_3} & \dots & \dfrac{\partial \psi_1}{\partial \gamma_n} \cr 
\dfrac{\partial \psi_2}{\partial \gamma_1} & \dfrac{\partial \psi_2}{\partial \gamma_2} + 1 - {\rm e}^{T \alpha_j} & \dfrac{\partial \psi_2}{\partial \gamma_3} & \dots & \dfrac{\partial \psi_2}{\partial \gamma_n} \cr 
 \vdots & \vdots &\vdots &  & \vdots \cr
 \dfrac{\partial \psi_n}{\partial \gamma_1} &  \dfrac{\partial \psi_n}{\partial \gamma_2} &  \dfrac{\partial \psi_n}{\partial \gamma_3} & \dots & \dfrac{\partial \psi_n}{\partial \gamma_n} + 1 - {\rm e}^{T \alpha_j} \cr
\end{matrix}
\right)=0. \label{Poincare_charactertic_determ}
\end{equation}
As you can see, the terms on the diagonal of the determinant are
\begin{equation}
\dfrac{\partial \psi_i}{\partial \gamma_i} + 1 - {\rm e}^{T \alpha_j},
\end{equation}
(no summution over $i$), while those on the $i$th row and $k$th column ($i \neq k$) are
\begin{equation}
\dfrac{\partial \psi_i}{\partial \gamma_k}.
\end{equation}
The determinant, when expanded, leads to a $n$th-order algebraic equation for ${\rm e}^{T \alpha_j}$, which gives the value of $\alpha_j$ (since $T$ is known) if the derivatives $\partial \psi_i/\partial \gamma_k$ are known. We can then appreciate how crucial the role played by the functional or Jacobian determinant
\begin{equation}
\underline{\Delta}= \dfrac{\partial \psi_i}{\partial \gamma_k}, \label{Poincare_functional_det}
\end{equation}
is.

\subsection{An important theorem by Poincar\'e}\label{Important_poincare_theorem_sec}

Suppose that Eq. (\ref{RomanoA1}) is such that the vector field occurring on the right-hand side depends not only on the coordinate variables but also explicitly on time (or, in other words, the system is not autonomous) and on an arbitrary parameter $\rho_{\ell}$, i.e.,
\begin{equation}
\dot{\tilde{x}}_i = X_i(\tilde{x}_1,\tilde{x}_2,\dots,\tilde{x}_n;t;\rho_{\ell}),  \; \;\; \; \; \; \; \;\;  \; (i=1,2,\dots,n). \label{Poincare_33}
\end{equation}  
Let $X_i(\tilde{x}_1,\tilde{x}_2,\dots,\tilde{x}_n;t;\rho_{\ell})$ ($i=1,2,\dots,n$) be periodic functions of time with period $T$. When $\rho_{\ell}=0$, (\ref{Poincare_33}) becomes
\begin{equation}
\dot{x}_i = X_i(x_1,x_2,\dots,x_n;t),  \; \;\; \; \; \; \; \;\;  \; (i=1,2,\dots,n), \label{Poincare_35}
\end{equation} 
and we will assume that they admit one and only one periodic solution of period $T$
\begin{equation}
x_i= \varphi_i(t), \; \;\; \; \; \; \; \;\;  \; (i=1,2,\dots,n), \label{Poincare_34}
\end{equation} 
in such a way that
\begin{equation}
\varphi_i(0)=\varphi_i(T), \; \;\; \; \; \; \; \;\;  \;(i=1,2,\dots,n).
\end{equation}
We can generalize the concepts developed in last sections and define the characteristic exponents also when we deal with systems such as the one of Eq. (\ref{Poincare_35}). In fact, Eq. (\ref{Poincare_35}) represents linear first-order differential equations whose coefficients are periodic functions of $t$ and hence it is possible to express (\ref{Poincare_34}) by means of characteristic exponents. 

At this stage, we would like to find a family of periodic solutions parametrized by $\rho_{\ell}$ that agrees with (\ref{Poincare_34}) when $\rho_{\ell}=0$. In other words, we want to investigate under which circumstances (\ref{Poincare_33}) will have a periodic solution of period $T$ when $\rho_{\ell}$ is no longer zero, but very small \cite{P1892}. When $\rho_{\ell} \neq 0$, the solution of (\ref{Poincare_33}) will be slightly modified with respect to the undisturbed solution (\ref{Poincare_34}) and hence it can be written as in (\ref{disturbed1}), i.e., 
\begin{equation}
\tilde{x}_i=\varphi_i(t) + \xi_i,  \; \; \; \; (i=1,2,\dots,n). \label{Poincare_36}
\end{equation}
Thus, it should be now clear that the regime $\rho_{\ell} \neq 0$ can be described by employing the tool of variational equations, the functions $\xi_i$ occurring in (\ref{Poincare_36}) being solutions of the linear variational equations of (\ref{Poincare_35}). Therefore, in complete analogy with the previous sections we set (see Eq. (\ref{Sze_39}))
\begin{equation}
\begin{split}
& \tilde{x}_i(0)=\varphi_i(0)+ \xi_i(0), \\
& \tilde{x}_i(T)=\varphi_i(T)+ \gamma_i + \psi_i, 
\end{split}
\; \; \; \; \; \; \; \; (i=1,2,\dots,n).
\label{Poincare34c}
\end{equation}
Poincar\'e has demonstrated \cite{P1892} that the $\psi_i$'s are analytical functions of $\rho_{\ell}$ and $\gamma_i$'s and vanish if 
\begin{equation}
\rho_{\ell}= \gamma_1=\gamma_2=\dots=\gamma_n=0.
\end{equation}
It is then obvious that the solution (\ref{Poincare_36}) will be periodic if 
\begin{equation}
\psi_i(\rho_{\ell},\gamma_1,\gamma_2,\dots,\gamma_n)=0, \; \; \; \; \; \; \; (i=1,2,\dots,n). \label{Poincare34b}
\end{equation}
If the functional determinant (\ref{Poincare_functional_det}) is not zero for $\rho_{\ell}= \gamma_i=0$ ($i=1,2,\dots,n$), then we can solve the $n$ implicit equations (\ref{Poincare34b}) with respect to the $\gamma_i$'s and find
\begin{equation}
\gamma_i = \tau_i (\rho_{\ell}), \; \; \; \; \; \; \; (i=1,2,\dots,n), \label{Poincare34d}
\end{equation}
$\tau_i (\rho_{\ell})$ being developable in powers of $\rho_{\ell}$ and vanishing with it \cite{P1892}. Therefore, if the Jacobian (\ref{Poincare_functional_det}) is such that
\begin{equation}
\underline{\Delta} \neq 0,
\end{equation}
than the system (\ref{Poincare_33}) admits a periodic solution for small but non-vanishing values of $\rho_{\ell}$. The periodic solution in fact will be obtained by simply substituting (\ref{Poincare34d}) (which assures that (\ref{Poincare34b}) holds) in (\ref{Poincare34c}). In other words, we have a periodic solution if, for $\rho_{\ell}=0$, the system (\ref{Poincare34b}) admits
\begin{equation}
\gamma_1=\gamma_2=\dots=\gamma_n=0,
\end{equation}
as a (simple) solution.

Nevertheless, by bearing in mind Eq. (\ref{Poincare_charactertic_determ}), it follows that if $\underline{\Delta} =0$, then one of the characteristic exponents is zero. On the contrary, the condition $\underline{\Delta} \neq 0$ implies that all the characteristic exponents differ from zero. Recall also that we have to calculate the determinant $\underline{\Delta}$ for $\rho_{\ell}=0$. We can therefore state the following theorem \cite{P1892}:
\newtheorem*{Poincare1}{Poincar\'e Theorem (non-autonomous systems)}
\begin{Poincare1} 
If Eqs. (\ref{Poincare_33}), which depend on a parameter $\rho_{\ell}$, admit for $\rho_{\ell}=0$ the periodic solution (\ref{Poincare_34}) for which all the characteristic exponents are different from zero (i.e., $\underline{\Delta} \neq 0$), they will admit in addition a periodic solution for small but non-vanishing values of $\rho_{\ell}$. 
\end{Poincare1}
Note that the hypothesis for which the system (\ref{Poincare_35}) is such that all the characteristic exponents are non-vanishing is consistent with the fact that it is not autonomous and has no integral of motion.

We have seen in the previous sections that when we deal with autonomous systems, then one characteristic exponent vanishes. If there is only one vanishing characteristic exponent when $\rho_{\ell}=0$, we still have a periodic solution for small but non-vanishing values of $\rho_{\ell}$. Thus, the following theorem follows \cite{P1892}:
\newtheorem*{Poincare2}{Poincar\'e Theorem (autonomous systems)}
\begin{Poincare2} 
If Eqs. (\ref{Poincare_33}), which depend on a parameter $\rho_{\ell}$, are such that time does not appear explicitly and, besides, if they admit for $\rho_{\ell}=0$ a periodic solution, then one characteristic exponent will vanish. If no other of these exponents is equal to zero, then there will still exist a periodic solution for small but non-vanishing values of $\rho_{\ell}$. 
\end{Poincare2}

The above theorems will turn out to be very useful when we will describe the quantum corrected full three-body problem (Chapter \ref{more detailed newtonian models_Chapter}, Sec. \ref{Full-3B_Sec}). 

\subsection{Variation from Hamiltonian equations}\label{Variation_Hamiltonian_Sec}

If the original equations of motion are of Hamiltonian form and there is a periodic solution, two of the characteristic exponents vanish. In fact, the first one is zero because we suppose that the system is autonomous, a condition which means that the Hamiltonian function itself is a constant of motion and this represents the reason why the second characteristic exponent equals zero. Moreover, if $\mu_k$ is an eigenvalue of the monodromy matrix, then also $1/\mu_k$ and the complex conjugate $\mu_{k}^{*}$ are eigenvalues. Thus, if $\alpha_{k}$ is a complex characteristic exponent, other characteristic exponents are $-\alpha_{k}$, $\alpha_{k}^{*}$, and $-\alpha_{k}^{*}$. On the other hand, if $\alpha_{k}$ is real or pure imaginary, then also $-\alpha_{k}$ is a characteristic exponent. In other words, the characteristic exponents are equal in pairs and of opposite sign \cite{P1892}. For example, in the case of the restricted three-body problem, the degrees of freedom are two and the characteristic exponents can be either $(0,0,\alpha,-\alpha)$ or $(0,0,\I \alpha,-\I \alpha)$. 

We proceed to establish these important results \cite{Pars}. Consider the system of Hamiltonian equations for $n$ degrees of freedom
\begin{equation}
\begin{dcases}
& \dot{q}_i = \dfrac{\partial H}{\partial p_i},\\
& \dot{p}_i =- \dfrac{\partial H}{\partial q_i}, \\
\end{dcases} \\ [2.5 em]
\; \;\; \; \; \; \; \;\;  \;\; \; \; \; \; \;(i=1,2,\dots,n),
\label{Pars2361}
\end{equation}
where $H=H(q_1,q_2\dots,q_n;p_1,p_2\dots,p_n)$. Suppose (\ref{Pars2361}) admit a periodic solution of period $T$
\begin{equation}
q_i = \varphi_i(t),  \;\;  \;\; \; \;\;\;  \;\; \; \;\;\;  \;\; \; \; p_i=\phi_i (t), \; \;\; \; \; \; \; \;\;  \;\; \; \; \; \; \;(i=1,2,\dots,n). 
\end{equation}
By writing the Lagrangian coordinates and the momenta in the varied orbit as
\begin{equation}
\tilde{q}_i= \varphi_i(t)+ \xi_i,  \;\;  \;\; \; \;\;\;  \;\; \; \;\;\;  \;\; \; \; \tilde{p}_i= \phi_i (t)+ \eta_i, \; \;\; \; \; \; \; \;\;  \;\; \; \; \; \; \;(i=1,2,\dots,n),
\end{equation}
the variational equations assume the form
\begin{equation}
\begin{dcases}
& \dot{\xi}_i = \sum_{j=1}^{n}\left[ H_{,p_{i}q_{j}} \, \xi_j +  H_{,p_{i}p_{j}} \, \eta_j \right],  \\
& \dot{\eta}_i =-\sum_{j=1}^{n}\left[ H_{,q_{i}q_{j}} \, \xi_j +  H_{,q_{i}p_{j}} \, \eta_j \right],\\ 
\end{dcases} \\ [2.5 em]
\; \;\; \; \; \; \; \;\;  \;\; \; \; \; \; \;(i=1,2,\dots,n),
\label{Pars2362}
\end{equation}
where a subscript consisting of a comma followed by a variable denotes partial derivatives with respect to that variable. Note also that the values of the $q$'s and $p$'s in the original motion have been substituted after differentiation. Equations (\ref{Pars2362}) are of Hamiltonian form. In fact the $\xi$'s can be considered as the Lagrangian coordinates and the $\eta$'s as the canonical momenta, while the Hamiltonian function is
\begin{equation}
\mathcal{H}=  \sum_{j,k=1}^{n} \left[ \dfrac{1}{2} H_{,q_{j}q_{k}} \xi_j \xi_k + H_{,q_{j}p_{k}} \xi_j \eta_k +\dfrac{1}{2} H_{,p_{j}p_{k}} \eta_j \eta_k \right]. 
\label{Pars2364}
\end{equation}
Thus, variational equations (\ref{Pars2362}) can be written as
\begin{equation}
\dot{\boldsymbol{\varsigma}}= \bold{Z}  \mathcal{H}_{\boldsymbol{\varsigma}}=\bold{Z} \bold{P}(t) \boldsymbol{\varsigma}, \label{Pars2365}
\end{equation}
where
\begin{equation}
\boldsymbol{\varsigma} = 
\left(
\begin{matrix}
\xi_1 \cr
\xi_2 \cr
\vdots \cr
\xi_n \cr
\eta_1 \cr
\eta_2 \cr
\vdots \cr
\eta_n \cr
\end{matrix}
\right),
\end{equation}
\begin{equation}
\bold{Z}=
\left(
\begin{matrix}
\bold{0} & \mathbb{1}_n \cr
- \mathbb{1}_n & \bold{0} \cr
\end{matrix}
\right),
\end{equation}
\begin{equation}
\mathcal{H}_{\boldsymbol{\varsigma}} = 
\left(
\renewcommand{\arraystretch}{2.2}
\begin{matrix}
\dfrac{\partial \mathcal{H}}{\partial \varsigma_1} \cr
\dfrac{\partial \mathcal{H}}{\partial \varsigma_2} \cr
\vdots \cr
\dfrac{\partial \mathcal{H}}{\partial \varsigma_{2n}} \cr
\end{matrix}
\right),
\end{equation}
and $\bold{P}(t)$ is a symmetric periodic $2 n \times 2 n$ matrix. It is possible to prove that if $\boldsymbol{\varsigma}^\prime$ represents another solution of (\ref{Pars2365}) (independent from $\boldsymbol{\varsigma}$), then \cite{P1890}
\begin{equation}
\dfrac{{\rm d}}{{\rm d}t}\left[\boldsymbol{\varsigma}^{\rm T} \, \bold{Z}\,\boldsymbol{\varsigma}^{\prime} \right] = \sum_{j=1}^{n} \dfrac{{\rm d}}{{\rm d}t} \left[ \xi_j \eta^{\prime}_{j}-\eta_{j} \xi^{\prime}_{j} \right]=0.
\end{equation}
From this conditions it follows that the monodromy matrix $\bold{R}(T)$ of the principal fundamental matrix (\ref{Pars2321}) now satisfies the condition
\begin{equation}
\bold{R}(T)^{\rm T} \, \bold{Z} \,\bold{R}(T) = \bold{Z}.
\label{Pars23613}
\end{equation}
Any matrix having the property exhibited in Eq. (\ref{Pars23613}) is called symplectic matrix. 

Recall that all monodromy matrices have the same eigenvalues. Let $\mu_k$ be a characteristic multiplier, then it satisfies the characteristic equation
\begin{equation}
\det \left[ \bold{R}(T)^{\rm T} - \mu_k \mathbb{1}_{2n} \right]=0,
\end{equation}
or
\begin{equation}
\det \left[ \bold{R}(T)^{\rm T} \, \bold{Z} \,\bold{R}(T)  - \mu_k \bold{Z} \,\bold{R}(T)  \right]=0,
\end{equation}
and hence, by virtue of (\ref{Pars23613}),
\begin{equation}
\det \left[  \bold{Z}   - \mu_k \bold{Z} \,\bold{R}(T)  \right]=0.
\end{equation}
It follows that
\begin{equation}
\det \left[ \bold{R}(T)   - \dfrac{1}{\mu_k}  \mathbb{1}_{2n} \right]=0,
\end{equation}
i.e., if $\mu_k$ is an eigenvalue of the monodromy matrix $\bold{R}(T)$, the same holds also for $1/\mu_k$. Since $\mu_k={\rm e}^{T \alpha_k}$, the properties stated at the beginning of this section follow easily. Moreover, one characteristic exponent vanishes because we are still dealing with autonomous systems, but the non-vanishing $\alpha$'s are paired (i.e., $\alpha_k$ and $-\alpha_k$), therefore also another characteristic exponent must be zero. Thence, the number of vanishing $\alpha$'s is equal to two, since the total number of characteristic exponents is even. Furthermore, since the monodromy matrix is real, if $\mu_k$ is an eigenvalue, so is its complex conjugate $\mu_{k}^{*}$. Then, if $\alpha_k$ is a characteristic exponent that is neither real nor pure imaginary, so is $\alpha_k^{*}$.

As we know, the sign of the real part of characteristic exponents determines the stability (in the linear sense) of the periodic solution. Since, as we have just demonstrated, the $\alpha$'s occur in pairs for Hamiltonian systems, the existence of a characteristic exponent with non-zero real part immediately indicates instability. Thus, the necessary condition for stability of a periodic orbit is that all the characteristic exponents must be pure imaginary, as we have already explained. 

As we said before, since the Hamiltonian equations (\ref{Pars2361}) do not depend on time explicitly and always admit the {\it vis viva} integral
\begin{equation}
H= {\rm const},
\end{equation}
two of the characteristic exponents vanish. If in addition they admit $k$ independent integrals of motion which are in involution (i.e., the Poisson bracket of any pair of constants of motion vanishes), then $2k+2$ characteristic exponents vanish, unless all the functional determinants of the $k+1$ integrals of motion with respect to arbitrary $k+1$ of the variables $q_i$ and $p_i$ are zero at the same time at all points of the periodic solution \cite{P1892}. 

\subsection{Stability analysis of Lagrangian points} \label{stability_Sec}

In the previous section we have seen that the Earth-Moon system is characterized by the presence of five Lagrangian points, which represent the equilibrium positions for the planetoid. Not only are these libration points solutions of the equation of motion, but near these points other families of solutions do exist \cite{Szebehely67}.

A rather important question is whether the positions of equilibrium are stable. In the affirmative case, the planetoid would therefore remain permanently near the point of stable equilibrium. It should be clear from the previous section that in order to study this issue, we need to employ the tool of variational equations. Thus, on denoting by $(x_{0},y_{0})$ one of the points $L_{1}$, $L_{2}$, $L_{3}$, $L_{4}$, $L_{5}$, one writes in the equations of motion (\ref{2.15a}) and (\ref{2.16a})
\begin{equation}
x=x_{0}+\xi, \; \; \; \; \; \;  y=y_{0}+\eta.
\end{equation} 
As usual, by expanding the right-hand sides in powers of $\xi$ and $\eta$, and retaining only terms of first order, one obtains the linear approximation \cite{BE14a}
\begin{equation}
\begin{split}
& {\ddot \xi} -2 \omega {\dot \eta}=G(\mathcal{A}\xi + \mathcal{B} \eta),\\
& {\ddot \eta} +2 \omega {\dot \xi}=G(\mathcal{B}\xi + \mathcal{C} \eta),
\label{6.2a}
\end{split}
\end{equation}
having defined
\begin{equation}
\begin{split}
& \mathcal{A} \equiv \left . {\partial^{2}U \over \partial x^{2}} 
\right |_{x_{0},y_{0}} , \\
& \mathcal{B} \equiv \left . {\partial^{2}U \over \partial x \partial y}
\right |_{x_{0},y_{0}} , \\
& \mathcal{C} \equiv \left . {\partial^{2}U \over \partial y^{2}} 
\right |_{x_{0},y_{0}}.
\label{6.4a}
\end{split}
\end{equation}
It is clear that the coupled set of ordinary differential equations (\ref{6.2a}) represents an example of linear variational equations. By virtue of (\ref{6.4a}), the coefficients of such equations are constant and hence the solution can be written in the form (cf. (\ref{Pars2334}))
\begin{equation}
\begin{split}
& \xi=\xi_0 {\rm e}^{\sigma t}, \\
& \eta=\eta_0 {\rm e}^{\sigma t}.
\end{split}
\end{equation}
This leads to the linear homogeneous system of algebraic equations \cite{BE14a}
\begin{equation}
\left(\sigma^{2}-G\mathcal{A} \right)\xi
- \left(2 \omega  \sigma +G \mathcal{B} \right)\eta=0,
\end{equation}
\begin{equation}
\left(2 \omega   \sigma -G \mathcal{B} \right)\xi
+\left(\sigma^{2}-G \mathcal{C} \right)\eta=0.
\end{equation}
Non-trivial solutions exist if and only if the determinant of the matrix of coefficients vanishes. Such a condition is expressed by the algebraic equation of fourth degree
\begin{equation}
 \sigma^{4}-[G(\mathcal{A}+\mathcal{C})-4 \omega^{2}]\sigma^{2}
+G^{2}(\mathcal{A}\mathcal{C}-\mathcal{B}^{2})=0.
\label{6.8a}
\end{equation}
This is a quadratic equation for $\sigma^2$ and, in order to accomplish first-order stability criterion, its roots must be real and negative. From the standard theory of algebraic equations of second degree, one finds that
\begin{equation}
 \sigma^{2}={1\over 2}[G(\mathcal{A}+\mathcal{C})-4 \omega^{2}] 
\pm {1\over 2}\sqrt{[G(\mathcal{A}+\mathcal{C})-4 \omega^{2}]^{2}
-4G^{2}(\mathcal{A}\mathcal{C}-\mathcal{B}^{2})}.
\label{6.9a}
\end{equation}

In Newtonian theory, $(\mathcal{A}\mathcal{C}-\mathcal{B}^{2})$ is negative at $L_{1}$, $L_{2}$, and $L_{3}$. This is clear at once by considering, at each collinear libration points, Eqs. (\ref{3.8a})--(\ref{3.10a}) in the limit $k_1 \rightarrow 0$, $k_2 \rightarrow 0$, and $k_3 \rightarrow 0$. Then, only half of the ${\sigma^{2}}$ values are negative, which implies that the criterion for first-order stability is not satisfied \cite{BE14a,Pars}. In other words, collinear Lagrangian points are unstable within classical theory. In the quantum regime, it remains true, from (\ref{3.9a}), that our $\mathcal{B}$ vanishes at $L_{1}$, $L_{2}$, and $L_{3}$, and we express our $\mathcal{A}$  at $L_{1}$, $L_{2}$, and $L_{3}$ from (\ref{4.1a}), our $\mathcal{C}$ at $L_{2}$ and $L_{3}$ from (\ref{4.11a}), and our $\mathcal{C}$ at $L_{1}$ from (\ref{4.13a}). Thus, provided that the sufficient conditions (\ref{4.2a}), (\ref{4.12a}), and (\ref{4.14a}) hold, which are in turn guaranteed, as we know, from the choice of scattering potential, it is always true that $(\mathcal{A}\mathcal{C}-\mathcal{B}^{2})<0$, and the points $L_{1}$, $L_{2}$, and $L_{3}$ remain points of unstable equilibrium even in the presence of quantum corrections obtained from an effective-gravity picture \cite{BE14a}.

As far as non-collinear Lagrangian points are concerned, the vanishing of $\lambda$ (see Eq. (\ref{3.2a})) simplifies the evaluation of $\mathcal{A}$ and $\mathcal{C}$ from (\ref{3.8a}) and (\ref{3.10a}), and we find (with the understanding that $r=r(l)$,  
$s=s(l)$ and $y=y(l)$ as in Sec. \ref{restricted_sec}) \cite{BE14a}
\begin{equation}
\mathcal{A}={\alpha(r^{2}-y^{2})\over r^{5}}
\left(3+8{k_{1}\over r}+15{k_{2}\over r^{2}}\right)
+{\beta(s^{2}-y^{2})\over s^{5}}
\left(3+8{k_{3}\over s}+15{k_{2}\over s^{2}}\right),
\end{equation}
\begin{equation}
\mathcal{C}={\alpha y^{2}\over r^{5}}
\left(3+8{k_{1}\over r}+15{k_{2}\over r^{2}}\right)
+{\beta y^{2}\over s^{5}}
\left(3+8{k_{3}\over s}+15{k_{2}\over s^{2}}\right),
\end{equation}
\begin{equation}
\begin{split}
\mathcal{B}^{2}&= {\alpha^{2}y^{2}(r^{2}-y^{2})\over r^{10}}
\left(3+8{k_{1}\over r}+15{k_{2}\over r^{2}}\right)^{2}
+{\beta^{2}y^{2}(s^{2}-y^{2})\over s^{10}}
\left(3+8{k_{3}\over s}+15{k_{2}\over s^{2}}\right)^{2}
 \\
&+ {2\alpha \beta y^{2}\over r^{5}s^{5}}
\left(3+8{k_{1}\over r}+15{k_{2}\over r^{2}}\right)
\left(3+8{k_{3}\over s}+15{k_{2}\over s^{2}}\right)
\left[x^{2}+(a-b)x-ab\right].
\end{split}
\end{equation}
In the evaluation of $(\mathcal{A}\mathcal{C}-\mathcal{B}^{2})$ we find therefore exact cancellation of the two pairs of terms involving $\alpha^{2}$ and $\beta^{2}$. Moreover, on exploiting from (\ref{2.4a}) the identity
\begin{equation}
r^{2}+s^{2}=2(x^{2}+y^{2})+2(a-b)x+a^{2}+b^{2},
\end{equation}
we obtain, bearing in mind that $(a+b)=l$,
\begin{equation}
(\mathcal{A}\mathcal{C}-\mathcal{B}^{2})={\alpha \beta y^{2}l^{2}\over r^{5}s^{5}}
\left(3+8{k_{1}\over r}+15{k_{2}\over r^{2}}\right)
\left(3+8{k_{3}\over s}+15{k_{2}\over s^{2}}\right).
\label{6.14a}
\end{equation}
This is all we need, because (\ref{6.14a}) is clearly positive if the scattering potential in employed. This fact ensures that all values of $\sigma^{2}$ 
from the solution formula (\ref{6.9a}) are negative (a result further confirmed by numerical analysis), and hence, in full agreement with the criterion for 
first-order stability of equilibrium points, in the quantum regime $L_4$ and $L_5$ are stable at first order \cite{BE14a}. The same condition holds also within the classical theory, for which, at equilateral libration points, Eq. (\ref{6.8a}) becomes \cite{Pars}
\begin{equation}
\sigma^4 + \left[\left(\alpha + \beta \right)\dfrac{G}{l^3} \right] \sigma^2+ \dfrac{27}{4}\left(\dfrac{G}{l^3}\right)^2 \alpha \beta=0,
\label{Pars2875}
\end{equation}
since the classical limit of (\ref{3.8a})--(\ref{3.10a}) gives at the classical coordinates (\ref{5.20a}) of $L_4$ and $L_5$
\begin{equation}
\begin{split}
& \mathcal{A}_{cl}= \dfrac{3}{4}\dfrac{(\alpha+\beta)}{l^3}, \\
& \mathcal{B}_{cl}= \pm \dfrac{3\sqrt{3}}{4}\dfrac{(\alpha-\beta)}{l^3}, \\
& \mathcal{C}_{cl}= \dfrac{9}{4}\dfrac{(\alpha+\beta)}{l^3},
\end{split}
\end{equation}
the upper sign referring to $L_4$ and the lower to $L_5$. From the general solution algorithm for (\ref{Pars2875})
\begin{equation}
\sigma^2 = - \dfrac{1}{2} \left(\alpha + \beta \right)\dfrac{G}{l^3} \pm \dfrac{1}{2} \sqrt{\left[\left(\alpha + \beta \right)\dfrac{G}{l^3} \right]^2-27\left(\dfrac{G}{l^3}\right)^2 \alpha \beta},
\end{equation}
it follows at once that the roots for $\sigma^2$ are real and negative if the quantities appearing under the radical sign are positive, i.e.,
\begin{equation}
\alpha^2  -25 \alpha \beta + \beta^2>0,
\end{equation}
which is satisfied if the ratio $\alpha/\beta=\rho^{-1}$ is greater than the larger root of
\begin{equation}
x^2 - 25 x + 1 =0,
\end{equation}
which is about $24.96$. Thus, in Newtonian theory, there is first-order stability at the non-collinear libration points if $\alpha$ is greater than about $25$ times $\beta$, a condition amply fulfilled in the Earth-Moon system, as well as, for the system consisting of the Sun and one of the planets of the Solar System. 

\subsection{Displaced periodic orbits for a solar sail} 

Over several years, by exploiting the tools provided by Newtonian theory, much progress has been made on modern models of planetoids and their periodic orbits at linear order (or, by using a specific engineering term, displaced periodic orbits) around the Earth-Moon Lagrangian points. In particular, a modern version of planetoid is a solar sail, which is propelled by reflecting solar photons and therefore can transform the momentum of photons into a propulsive force \cite{Simo1}. Although solar sailing has been considered as a practical means of spacecraft propulsion only relatively recently, the fundamental ideas had been already developed towards the end of the previous century and we refer the reader to Ref. \cite{McInnes1999} for further details.

Solar sailing technology appears as a promising form of advanced spacecraft propulsion, which can enable exciting new space-science mission concepts such as Solar System exploration and deep space observation. Furthermore, they can also be used for highly non-Keplerian orbits, such as closed orbits displaced high above
the ecliptic plane \cite{Waters07}, since they can apply a propulsive force continuously. This makes it possible to consider some exciting and unique trajectories. In such trajectories, a sail can be used as a communication satellite for high latitudes. For
example, the orbital plane of the sail can be displaced above that of the Earth, so that the sail can stay fixed above our planet at some distance, provided that the orbital periods are equal. Orbits around the collinear points of the Earth-Moon system are also of great interest because their unique positions are advantageous
for several important applications in space mission design \cite{Szebehely67,Roy}. 

Over the last few decades, several authors have tried to determine more accurate approximations of such equilibrium points. Such (quasi)halo orbits were first studied in Ref. \cite{Farquhar73}. Halo orbits near the collinear libration
points in the Earth-Moon system represent a prominent issue, in particular around the $L_{1}$ and $L_{2}$ points, because of their unique positions. However, as was pointed out before, a linear analysis shows that the collinear libration points $L_{1}$, $L_{2}$, and $L_{3}$ are of the type saddle$\times$center$\times$center, leading to an instability in their vicinity, whereas the equilateral equilibrium points $L_{4}$ and $L_{5}$ are stable, in that they are of the type center$\times$center$\times$center. Although the libration points $L_{4}$ and $L_{5}$ are naturally stable and require a small acceleration, the disadvantage is the longer communication path length from the lunar pole to the sail. On the other hand, if the orbit maintains visibility from Earth, a spacecraft near $L_{2}$ can be used to provide communications between the equatorial regions of the Earth and the polar regions of the Moon. The establishment of a bridge for radio communications is crucial for forthcoming space missions, which plan to use the lunar poles. Displaced non-Keplerian orbits near the Earth-Moon libration points have been investigated in Refs. \cite{Simo1,McInnes1993}. 

All the above-mentioned points make it clear that the analysis of orbits around libration points does not belong just to the history of celestial mechanics, but plays a crucial role in modern investigations of space mission design. For this reason, we believe that the quantum corrected description of displaced orbits for the Earth-Moon system involving a solar sail could represent an important task for future developments in space technology.
\begin{figure}
\centering
\includegraphics[scale=0.7]{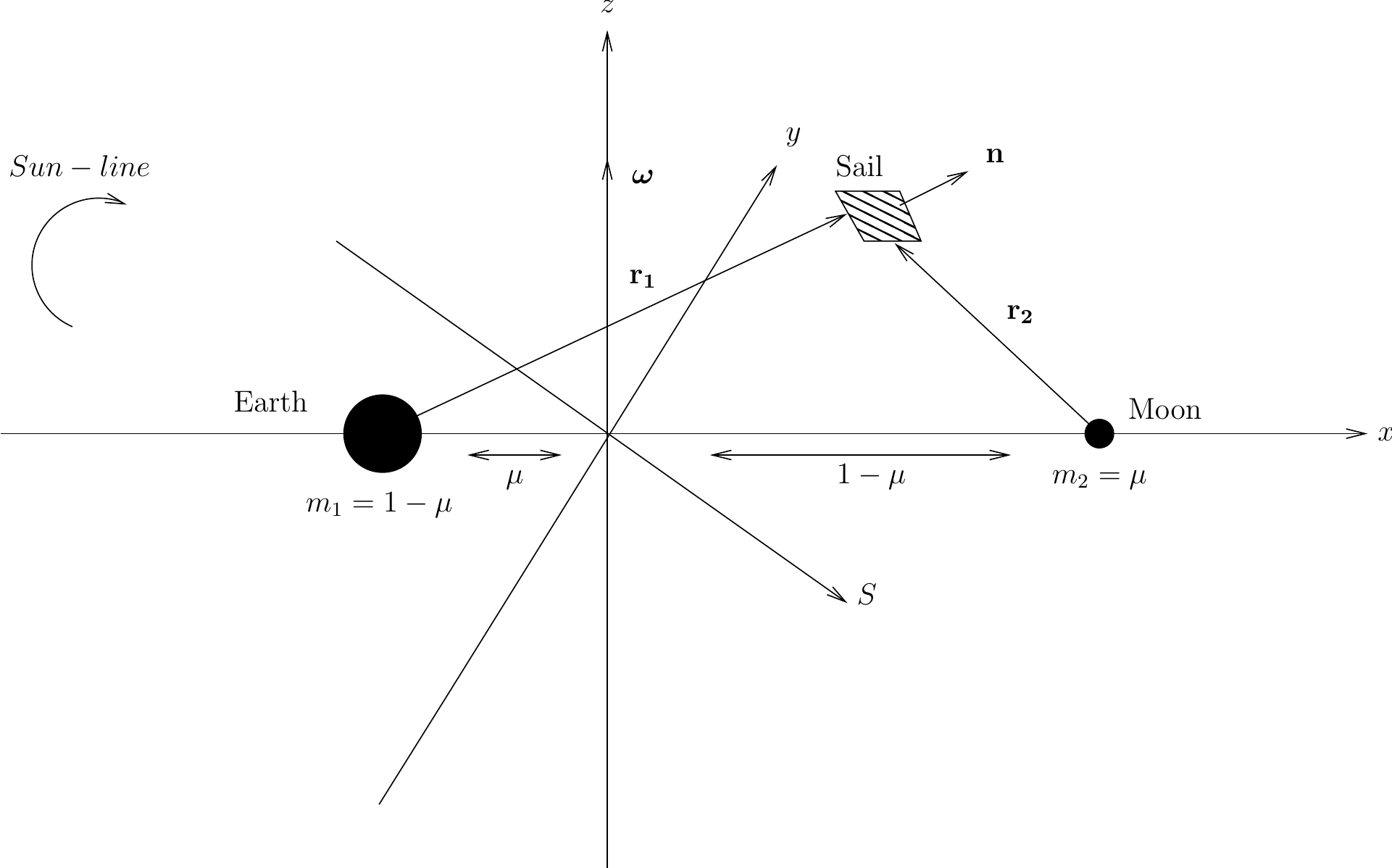}
\caption[Schematic geometry of the restricted three-body problem involving a solar sail]{Schematic geometry of the Earth-Moon restricted three-body problem when the planetoid is a solar sail. Despite the notations of this figure, the distances of the planetoid from the Earth and the Moon will be always indicated, like in the previous sections, with $r$ and $s$, respectively.}
\label{2b.pdf}
\end{figure}

The vector dynamical equation for the sail in the rotating frame of reference (Fig. \ref{2b.pdf}) is given by
\begin{equation}
\dfrac{{\rm d}^2 \bold{r}}{{\rm d}t^2}+ 2 \boldsymbol{\omega} \times \dfrac{{\rm d} \bold{r}}{{\rm d}t} - G \nabla U (\bold{r})=\boldsymbol{a},
\label{Simo_1}
\end{equation}
where $\bold{r}=(x,y,z)$ is the position vector of the sail relative to the mass center of the primaries, $\boldsymbol{\omega}$ its angular velocity and $ U (\bold{r})$ is the quantum corrected potential (see Eq. (\ref{2.14a})). For the sake of simplicity, let us consider, {\it only} for this section, units where the sum of the masses of the primaries is set to one, as well as their distance and the Newton constant. Then, the solar radiation pressure acceleration of the sail is defined by
\begin{equation}
\boldsymbol{a}= a_0 \left( \boldsymbol{S} \cdot \boldsymbol{n} \right)^2 \boldsymbol{n},
\end{equation} 
where $a_0$ is the magnitude of the solar radiation pressure force exerted on the sail, while
\begin{equation}
\begin{split}
& \boldsymbol{n}= \left[ \cos\left( \varphi\right) \cos\left( \omega_{\star}t \right), -\cos\left( \varphi\right) \sin\left( \omega_{\star}t \right), \sin\left( \varphi\right)  \right],\\
& \boldsymbol{S}= \left[ \cos\left( \omega_{\star}t \right), -\sin\left( \omega_{\star}t \right), 0\right],
\end{split}
\end{equation}
represent the unit normal to the sail and the Sun line direction vector, respectively. Furthermore, $\omega_{\star}=0.923$ is the angular rate of the Sun line in the co-rotating frame expressed in dimensionless units and $\varphi$ indicates the pitch angle relative to the Sun line, which describes the sail altitude. In order to analyse the dynamics of the solar sail in the neighbourhood of libration points, we adopt the formalism of variational equations. 

Let us first consider non-collinear Lagrangian points. Thus, by letting the components $x,y,z$ of the position vector of the sail at each libration point change by the infinitesimal amount $\xi,\eta,\zeta$, respectively, and, by retaining only first-order terms in $\xi,\eta,\zeta$ in the equation of motion (\ref{Simo_1}), one finds the following linear variational equations of motion for the libration points $L_{4}$ and $L_{5}$ describing stable equilibrium \cite{BEDS15}: 
\begin{equation}
\ddot \xi -2  \dot \eta=U_{xx}^{0}\xi + U_{xy}^{0}\eta+a_{\xi},
\label{5.1b}
\end{equation}
\begin{equation}
\ddot \eta +2  \dot \xi=U_{xy}^{0}\xi + U_{yy}^{0}\eta+a_{\eta},
\label{5.2b}
\end{equation}
\begin{equation}
\ddot \zeta =U_{zz}^{0}\zeta +a_{\zeta},
\label{5.3b}
\end{equation}
where the auxiliary variables $a_{\xi}$, $a_{\eta}$, $a_{\zeta}$ characterizing the solar sail acceleration are given by
\begin{equation}
\begin{split}
a_{\xi} & = a_0  \cos\left( \omega_{\star}t \right) \cos^3\left( \varphi\right), \\
a_{\eta} & = -a_0 \sin \left( \omega_{\star}t \right) \cos^3\left( \varphi\right), \\
a_{\zeta} & =  a_0 \sin \left( \varphi\right) \cos^2 \left( \varphi\right),
\end{split}
\end{equation}
whereas $U_{xx}^{0}$, $U_{yy}^{0}$, $U_{zz}^{0}$, $U_{xy}^{0}$ are the partial derivatives of the gravitational potential (\ref{2.14a}) evaluated at $L_{4}$ or $L_{5}$. Note how the in-plane motion, enlightened by Eqs. (\ref{5.1b}) and (\ref{5.2b}), is decoupled by the out-of-plane motion (\ref{5.3b}). Let us assume now that a solution of the linearised equations of motion (\ref{5.1b})--(\ref{5.3b}) is periodic of the form
\begin{equation}
\xi(t)=A_{\xi}\cos(\omega_{\star}t)+B_{\xi}\sin(\omega_{\star}t),
\label{5.4b}
\end{equation}
\begin{equation}
\eta(t)=A_{\eta}\cos(\omega_{\star}t)+B_{\eta}\sin(\omega_{\star}t),
\label{5.5b}
\end{equation}
where $A_{\xi}$, $A_{\eta}$, $B_{\xi}$, and $B_{\eta}$ are parameters to be determined. They just represent the amplitude of the displaced periodic orbit. By substituting Eqs. (\ref{5.4b}) and (\ref{5.5b}) in the differential equations (\ref{5.1b})-(\ref{5.3b}), we obtain the following linear non-homogeneous system in $A_{\xi}$, $A_{\eta}$, $B_{\xi}$, and $B_{\eta}$ \cite{BEDS15}:
\begin{equation}
\begin{dcases}
& -(\omega_{\star}^{2}+U_{xx}^{0})B_{\xi}+2 \omega_{\star}A_{\eta}-U_{xy}^{0}B_{\eta}=0, \\
& -U_{xy}^{0}A_{\xi}+2 \omega_{\star}B_{\xi}-(\omega_{\star}^{2}+U_{yy}^{0})A_{\eta}=0, \\
& -(\omega_{\star}^{2}+U_{xx}^{0})A_{\xi}-U_{xy}^{0}A_{\eta}-2 \omega_{\star}B_{\eta}=a_{0}\cos^{3}(\varphi), \\
& -2 \omega_{\star}A_{\xi}-U_{xy}^{0}B_{\xi}-(\omega_{\star}^{2}+U_{yy}^{0})B_{\eta} =-a_{0}\cos^{3}(\varphi).
\end{dcases} \\ [2em]
\label{5.9b}
\end{equation}
The system (\ref{5.9b}) can be solved to find the coefficients $A_{\xi}$, $B_{\xi}$, $A_{\eta}$, $B_{\eta}$, here arranged in the four rows of a column vector $\bold{P}$, while $\bold{b}$ is the column vector whose four rows are the right-hand sides of (\ref{5.9b}). Let $\bold{T}$ be the 
$4 \times 4$ matrix
\begin{equation}
\bold{T}= \left(\begin{matrix} \bold{A}_{1} & \bold{B}_{1} \cr \bold{C}_{1} & \bold{D}_{1}
\end{matrix}\right),
\end{equation}
whose $2 \times 2$ sub-matrices are represented by \cite{BEDS15}
\begin{equation}
\bold{A}_{1}=\left(\begin{matrix} 
0 & -\omega_{\star}^{2}-U_{xx}^{0} \cr
-U_{xy}^{0} & 2 \omega_{\star} 
\end{matrix}\right),
\end{equation}
\begin{equation}
\bold{B}_{1}=\left(\begin{matrix}
2 \omega_{\star} & -U_{xy}^{0} \cr
-\omega_{\star}^{2}-U_{yy}^{0} & 0
\end{matrix}\right),
\end{equation}
\begin{equation}
\bold{C}_{1}=\left(\begin{matrix}
-\omega_{\star}^{2}-U_{xx}^{0} & 0 \cr
-2 \omega_{\star} & -U_{xy}^{0} 
\end{matrix}\right),
\end{equation}
\begin{equation}
\bold{D}_{1}=\left(\begin{matrix}
-U_{xy}^{0} & -2 \omega_{\star} \cr
0 & -\omega_{\star}^{2}-U_{yy}^{0}
\end{matrix}\right).
\end{equation}
With this matrix notation, the solution of our linear system (\ref{5.9b}) reads as \cite{BEDS15}
\begin{equation}
(\bold{P})^{i}=\sum_{j=1}^{4} \left[ (\bold{T}^{-1})_{\; j}^{i} \; (\bold{b})^{j} \right], \;\;\;\;\;\;\;\;\; \forall i=1,2,3,4.
\end{equation}

Since the out-of-plane motion is independent of the in-plane one, the solution of (\ref{5.3b}) with initial values $\zeta(t=0)=\zeta_0$ and $\dot{\zeta}_0=0$ can be easily obtained and it is given by \cite{BEDS15}
\begin{equation}
\zeta(t)= \theta(t) a_{0}(\cos^{2}\varphi) (\sin \varphi)|U_{zz}^{0}|^{-1} + \cos(\omega_{\zeta}t)\Bigr[\zeta_0-a_{0}(\cos^{2}\varphi) (\sin \varphi)
|U_{zz}^{0}|^{-1}\Bigr],
\end{equation}
where $\theta(t)$ is the step function 
\begin{equation}
\theta(t)=
\begin{dcases}
& 1 \; \; \; \; \; {\rm if} \; t>0, \\
& 0 \; \; \; \; \;  {\rm if} \; t<0,
\end{dcases} \\ [2em]
\end{equation} 
and the dimensionless frequency $\omega_{\zeta}$ is defined as
\begin{equation}
\omega_{\zeta}= \vert U_{zz}^{0} \vert^{1/2}.
\end{equation}
Thus, the required sail acceleration for a fixed distance can be given by \cite{BEDS15}
\begin{equation}
a_{0}={\zeta_0|U_{zz}^{0}| \over (\cos^{2}\varphi) (\sin \varphi)}.
\end{equation}
Furthermore, the out-of-plane distance can be maximized by an optimal choice of the sail pitch angle determined by
\begin{equation}
\left. \dfrac{{\rm d}}{{\rm d}\varphi} \cos^2(\varphi) \sin(\varphi)\right\vert_{\varphi=\varphi_{\star}}=0,
\end{equation}
which gives
\begin{equation}
\varphi_{\star} = 35.264\degree.
\end{equation}
\begin{figure}
\centering
\includegraphics[scale=0.7]{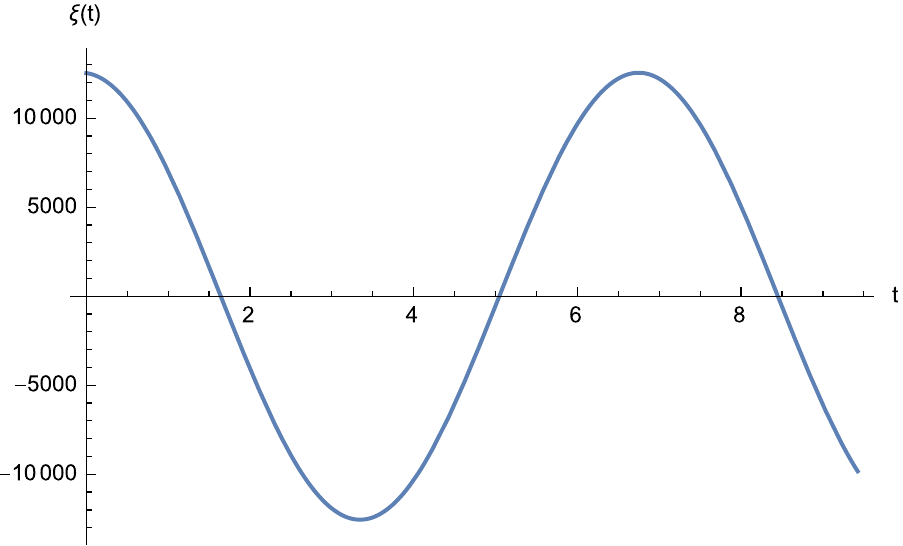}
\caption[Time evolution of the function $\xi(t)$ defined in Eq. (\ref{5.4b}) for $L_{4}$ in the Newtonian case]{Time evolution of the function $\xi(t)$ defined in Eq. (\ref{5.4b}) for $L_{4}$ in the Newtonian case.}
\label{3b.pdf}
\end{figure}
\begin{figure}
\centering
\includegraphics[scale=0.7]{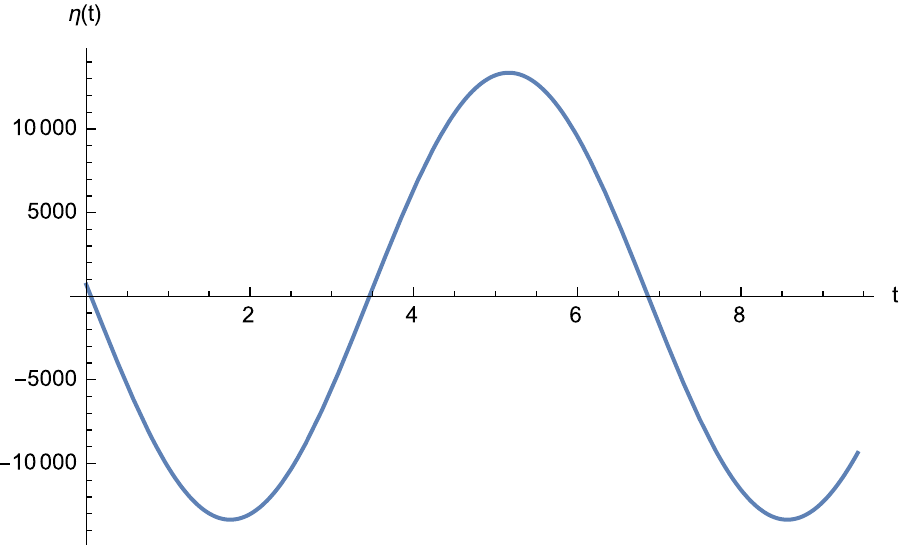}
\caption[Time evolution of the function $\eta(t)$ defined in Eq. (\ref{5.5b}) for $L_{4}$ in the Newtonian case]{Time evolution of the function $\eta(t)$ defined in Eq. (\ref{5.5b}) for $L_{4}$ in the Newtonian case.}
\label{4b.pdf}
\end{figure}
\begin{figure}
\centering
\includegraphics[scale=0.7]{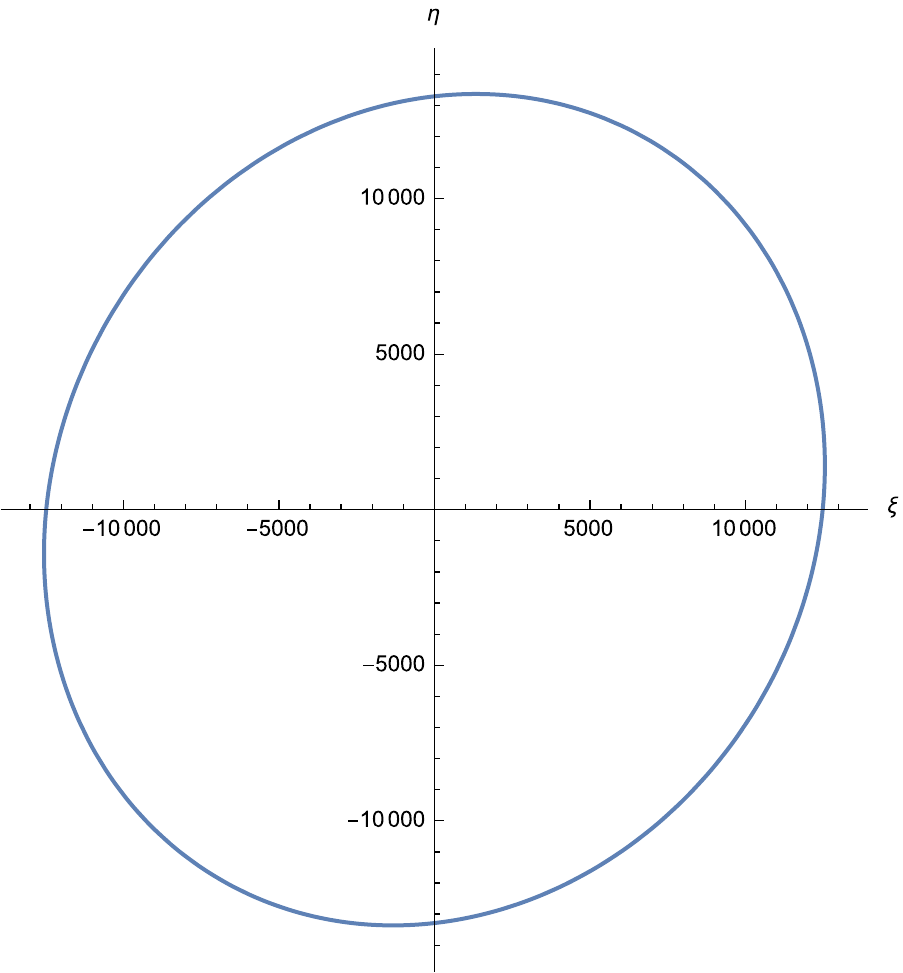}
\caption[Periodic orbits at linear order around the Lagrangian point $L_{4}$ in Newtonian theory]{Periodic orbits at linear order around the Lagrangian point $L_{4}$ in Newtonian theory.}
\label{5b.pdf}
\end{figure}
\begin{figure}
\centering
\includegraphics[scale=0.7]{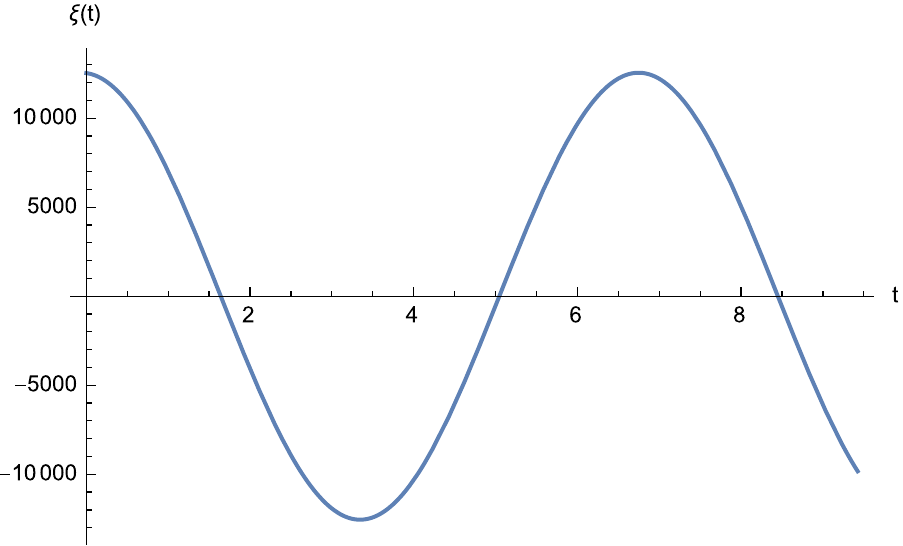}
\caption[Time evolution of the function $\xi(t)$ defined in Eq. (\ref{5.4b}) for $L_{4}$ in the quantum corrected model]{Time evolution of the function $\xi(t)$ defined in Eq. (\ref{5.4b}) for $L_{4}$ in the quantum corrected model.}
\label{6b.pdf}
\end{figure}
\begin{figure}
\centering
\includegraphics[scale=0.7]{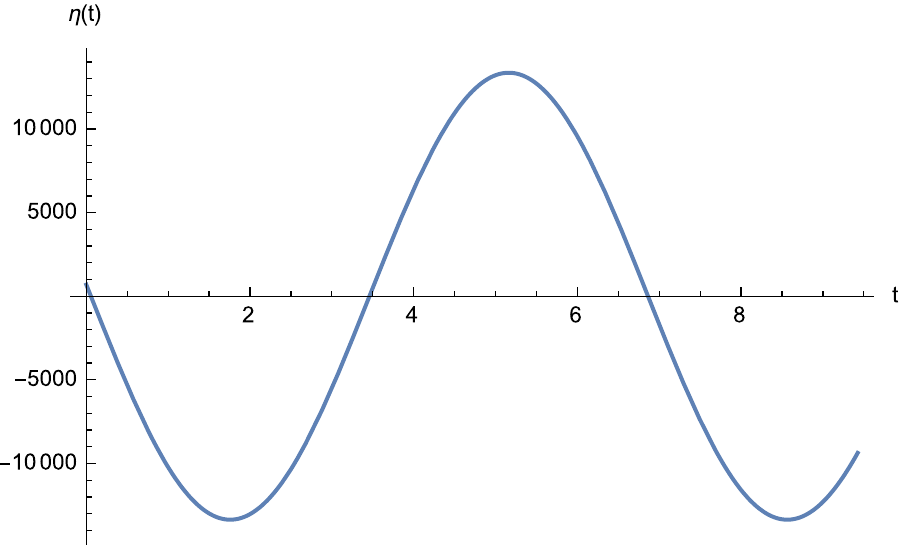}
\caption[Time evolution of the function $\eta(t)$ defined in Eq. (\ref{5.5b}) for $L_{4}$ in the quantum corrected model]{Time evolution of the function $\eta(t)$ defined in Eq. (\ref{5.5b}) for $L_{4}$ in the quantum corrected model.}
\label{7b.pdf}
\end{figure}
\begin{figure}
\centering
\includegraphics[scale=0.7]{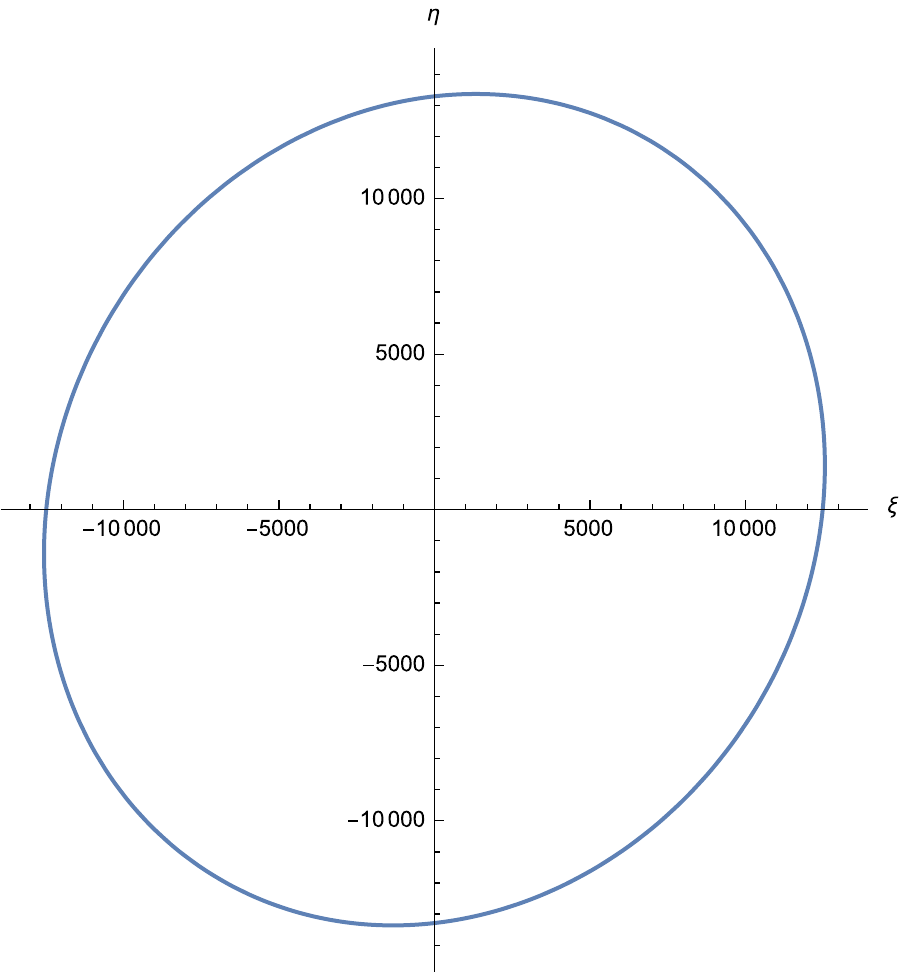}
\caption[Periodic orbits at linear order around the Lagrangian point $L_{4}$ in the quantum corrected model]{Periodic orbits at linear order around the Lagrangian point $L_{4}$ in the quantum corrected model. The periodic orbit is elliptic as in the Newtonian case displayed in Fig. \ref{5b.pdf}.}
\label{8b.pdf}
\end{figure}
The findings for displaced periodic orbits both in Newtonian theory and in the quantum corrected regime are well summarized in Figs. \ref{3b.pdf}--\ref{8b.pdf} (obtained by restoring the usual units for the distances, the masses and the Newtonian gravitational constant), which show clearly that our calculation is of interest because it proves that even at quantum level there exist periodic solutions in the neighbourhood of stable equilibrium points \cite{BEDS15}. In particular, the trajectory displayed in Fig. \ref{8b.pdf} is an ellipse centred on $L_4$, in analogy to what happens in the classical case (Fig. \ref{5b.pdf}). Furthermore, we have found that the period of such orbits is about $28$ days, i.e., the synodic lunar month. Note also that in deriving the above-mentioned figures, we have set the angle $\varphi=\varphi_{\star}$ and an initial out-of-plane distance $\zeta_0=100$ km. Moreover, the starting value of $\zeta$ has been increased gradually to reach $2500$ km.

\begin{figure}
\centering
\includegraphics[scale=0.7]{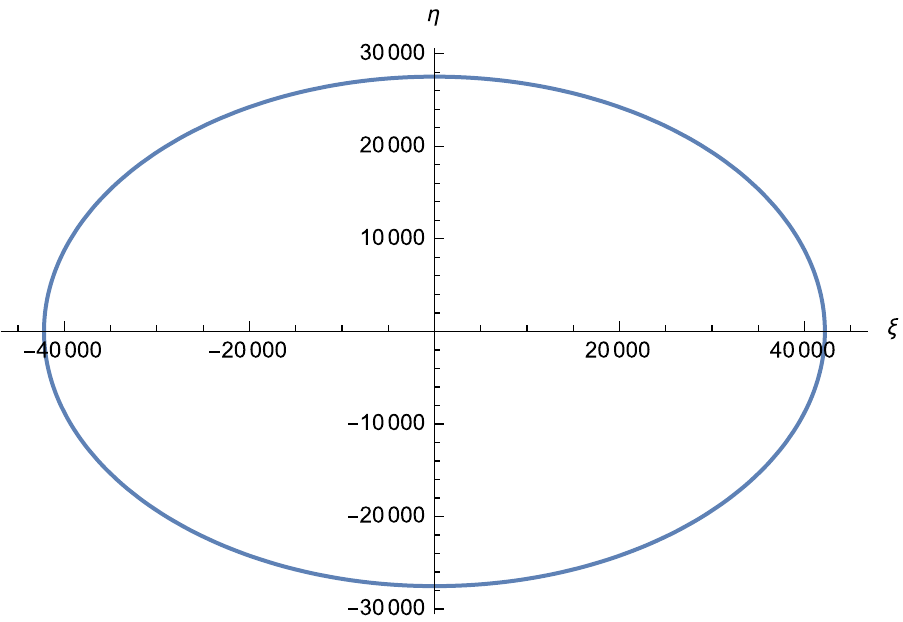}
\caption[Periodic orbits at linear order around the Lagrangian point $L_{2}$ in Newtonian theory]{Periodic orbits at linear order around the Lagrangian point $L_{2}$ in Newtonian theory.}
\label{9b.pdf}
\end{figure}

\begin{figure}
\centering
\includegraphics[scale=0.7]{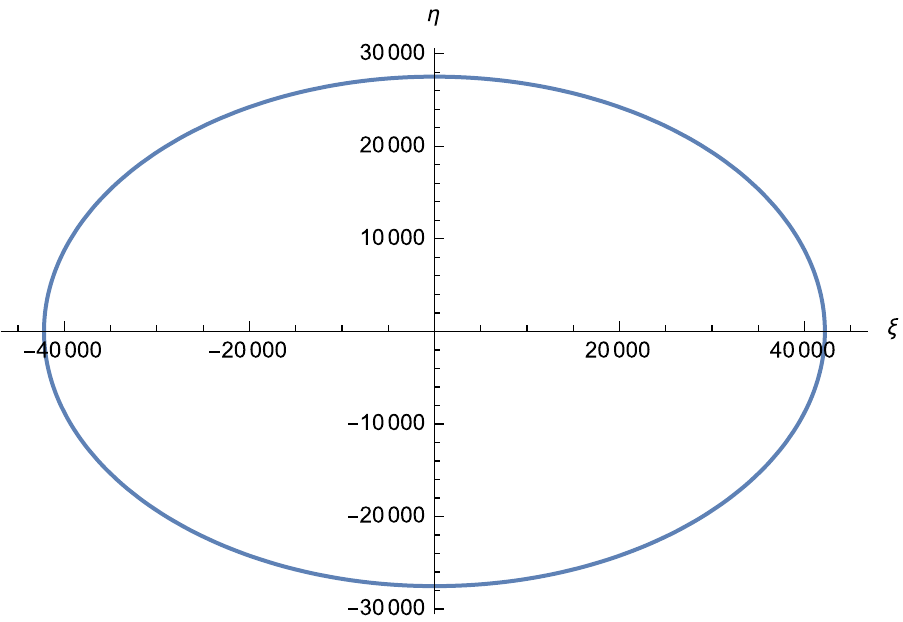}
\caption[Periodic orbits at linear order around the Lagrangian point $L_{2}$ in the quantum corrected model]{Periodic orbits at linear order around the Lagrangian point $L_{2}$ in the quantum corrected model. The periodic orbit is elliptic as in the Newtonian case displayed in Fig. \ref{9b.pdf}.}
\label{10b.pdf}
\end{figure}

Our quantum corrected model predicts the presence of displaced periodic orbits also around collinear Lagrangian points. In this case, the linear variational equations are given by
\begin{equation}
\ddot \xi -2  \dot \eta=U_{xx}^{0}\xi +a_{\xi},
\label{Simo13}
\end{equation}
\begin{equation}
\ddot \eta +2  \dot \xi= U_{yy}^{0}\eta+a_{\eta},
\end{equation}
\begin{equation}
\ddot \zeta =U_{zz}^{0}\zeta +a_{\zeta},
\label{Simo15}
\end{equation}
whose solution assumes the form
\begin{equation}
\xi(t)=\xi_0 \cos \left( \omega_{\star} t \right),
\label{Simo22}
\end{equation}
\begin{equation}
\eta(t)=\eta_0 \sin \left( \omega_{\star} t \right).
\label{Simo23}
\end{equation}
Like before, we insert Eqs. (\ref{Simo22}) and (\ref{Simo23}) into Eqs. (\ref{Simo13})--(\ref{Simo15}) and, on solving the resulting linear non-homogeneous system, we find that the amplitudes $\xi_0$ and $\eta_0$ are given by
\begin{equation}
\xi_0=a_0 \dfrac{\left(U_{yy}^{0} - \omega_{\star}^{2}-2\omega_{\star} \right) \cos^3 (\varphi)}{\left(U_{xx}^{0}-\omega_{\star}^{2}\right)\left(U_{yy}^{0}-\omega_{\star}^{2}\right)-4\omega_{\star}^{2}},
\end{equation}
\begin{equation}
\eta_0=-a_0 \dfrac{\left(U_{xx}^{0} - \omega_{\star}^{2}-2\omega_{\star} \right) \cos^3 (\varphi)}{\left(U_{xx}^{0}-\omega_{\star}^{2}\right)\left(U_{yy}^{0}-\omega_{\star}^{2}\right)-4\omega_{\star}^{2}}.
\label{Simo26}
\end{equation}
The trajectories around $L_2$ obtained through Eqs. (\ref{Simo13})--(\ref{Simo26}) are given in Figs. \ref{9b.pdf} and \ref{10b.pdf}.

We conclude this section by stressing that the solar sail model is an interesting possibility considered over the last few decades, but is not necessarily better than alternative models of planetoid. For example, the large structure and optical nature of solar sails can create a considerable challenge. If the structure and mass distribution of the sail is complicated, one has to resort to suitable approximations.


\chapter{Quantum description of more detailed Newtonian models} \label{more detailed newtonian models_Chapter}

\vspace{2cm}
\emph{It is the simple hypotheses of which one must be most wary, because these are the ones that have the most chances of passing unnoticed.}
\begin{flushright}
H. Poincar\'e
\end{flushright}

\vspace{2cm}

The restricted three-body problem represents a simplified version of the most general three-body problem. The aim of this chapter consists in adding all features that would contribute to make our quantum corrected model as close as possible to reality, in order to encourage the launch of future space missions that could verify it. For this reason, we will describe, once again within the context of effective field theories of gravity, the full three-body problem involving, like before, the Earth and the Moon, and the restricted four-body problem consisting of the Sun, the Earth, and the Moon as the primaries.

\section{Full three-body problem in effective field theories of gravity} \label{Full-3B_Sec}

As was stressed by Poincar\'e in his landmark work on the (restricted) three-body problem \cite{P1890}, the main aim of celestial mechanics is not the one of evaluating the astronomical ephemeris\footnote{In astronomy and celestial navigation, an ephemeris gives the positions of naturally occurring astronomical objects, as well as artificial satellites in the sky at a given time. Historically, positions were given as printed tables of values, given at regular intervals of date and time.}, but rather to ascertain whether Newtonian theory remains the most appropriate tool for investigating celestial gravity \cite{P1892}, at least (we would say) within the Solar System. With hindsight, this statement is not completely superseded by current developments in gravitational theories, provided in its formulation one replaces Newtonian theory by Einstein's general relativity, which has been challenged over the years by several competing theories (e.g., Brans-Dicke, $f(R)$, and so forth), to be tested both in the Solar System and on extra-galactic scales. Thus, in this context it makes sense to go one step further by assessing the full three-body problem of celestial mechanics by employing the hybrid scheme described in the previous chapters, where the Newtonian potential receives classical and quantum corrections from the calculational recipes of effective field theories of gravity \cite{BE14b}. 

\subsection{The classical integrals}\label{classical_integrals_Sec}

In the settings described before, three bodies $A_{1}$, $A_{2}$, $A_{3}$ having masses $m_{1}$, $m_{2}$, $m_{3}$, respectively, move in space under the action of their mutual gravitational attraction. The coordinates and velocities of the three bodies at $t=0$ are prescribed, and the full three-body problem consists in determining their position at any subsequent time. The differences with the restricted problem are easily recognized. In fact, in the latter the masses of only two particles are arbitrary, because the mass of the planetoid must be smaller than the other two. Moreover, the general problem allows any sets of initial conditions, while the restricted one demands circular orbits for the primaries.

Following Refs. \cite{BE14b,Pars}, we take fixed rectangular axes and denote the coordinates of $A_{r}$ at time $t$ by $x_{r},y_{r},z_{r}$. The
coordinates of the center of mass $D$ of the three bodies are instead represented by block capital letters $X,Y,Z$, so that, on denoting by $M \equiv m_{1}+m_{2}+m_{3}$ the mass of the whole system, one can write
\begin{equation}
\begin{split}
& MX=\sum_{r=1}^{3}m_{r}x_{r}, \\
& MY=\sum_{r=1}^{3}m_{r}y_{r}, \\
& MZ=\sum_{r=1}^{3}m_{r}z_{r}.
\end{split}
\end{equation}
We also set
\begin{equation}
\begin{split}
& x_r = X + \alpha_r, \\
& y_r = Y + \beta_r, \\
& z_r = Z + \gamma_r,
\end{split}
\end{equation}
so that $\alpha_r$, $\beta_r$, and $\gamma_r$ are the coordinates of $A_r$ relative to the system having $D$ as its origin and axes with the same directions as those of the fixed frame.  

The kinetic energy function $T$ of the system can be expressed by means of the relation \cite{BE14b,Pars}
\begin{equation}
\begin{split}
T &= \dfrac{1}{2} \sum_{r=1}^{3} m_r \left(\dot{x}^{2}_{r}+\dot{y}^{2}_{r}+\dot{z}^{2}_{r} \right) \\
& =\dfrac{1}{2} M ({\dot X}^{2}+{\dot Y}^{2}+{\dot Z}^{2})
+{1\over 2}\sum_{r < s}{m_{r}m_{s}\over M}v_{rs}^{2},
\end{split}
\end{equation}
where $v_{rs}$ is the speed of $A_{s}$ relative to $A_{r}$, i.e.,
\begin{equation}
v_{rs}^{2}=({\dot x}_{s}-{\dot x}_{r})^{2}
+({\dot y}_{s}-{\dot y}_{r})^{2}
+({\dot z}_{s}-{\dot z}_{r})^{2}=({\dot \alpha}_{s}-{\dot \alpha}_{r})^{2}
+({\dot \beta}_{s}-{\dot \beta}_{r})^{2}
+({\dot \gamma}_{s}-{\dot \gamma}_{r})^{2}.
\end{equation}

The potential energy function of this system is $-U$, and in the classical regime we have
\begin{equation}
U=G \left({m_{2}m_{3}\over r_{1}}+{m_{3}m_{1}\over r_{2}}
+{m_{1}m_{2}\over r_{3}}\right),
\label{Pars2919}
\end{equation}
where $r_1$ is the distance between $A_2$ and $A_3$, and so forth. 

Since the system is a holonomic dynamical system with nine degrees of freedom, we need nine Lagrangian equations, or eighteen Hamiltonian equations of motion. In the former case we have \cite{Pars}
\begin{equation}
\begin{split}
& m_r {\ddot x}_r={\partial U \over \partial x_r}, \\
& m_r {\ddot y}_r={\partial U \over \partial y_r}, \\
& m_r {\ddot z}_r={\partial U \over \partial z_r},
\end{split}
\; \; \; \; \; \; \; \; \; \; \; \;\; \; \; \;  (r=1,2,3),
\end{equation}
whereas for the latter 
\begin{equation}
\begin{dcases}
m_r {\dot x}_r = \xi_r, \\ 
m_r {\dot y}_r = \eta_r, \\ 
m_r {\dot z}_r = \zeta_r, \\   
\dot{\xi}_r = \dfrac{\partial U}{\partial x_r}, \\
\dot{\eta}_r = \dfrac{\partial U}{\partial y_r}, \\
\dot{\zeta}_r = \dfrac{\partial U}{\partial z_r}, \\
\end{dcases} \\ [2em]
\; \; \; \; \; \; \; \; \; \; \; \;\; \; \; \;  (r=1,2,3),
\label{Pars29112}
\end{equation}
where we have indicated, for the time being, the components of the canonical momenta with $\xi_r$, $\eta_r$, $\zeta_r$. The Lagrangian and Hamiltonian functions are given by
\begin{equation}
\mathcal{L}=T+U,
\end{equation} 
\begin{equation}
\mathcal{H}= \dfrac{1}{2}\sum_{r=1}^{3} \dfrac{\left(\xi_{r}^{2}+ \eta_{r}^{2}+\zeta_{r}^{2}\right)}{m_r}-U.
\end{equation}
Note how the fact that we are dealing with a eighteenth-order system explains the complex nature of the problem, especially if we make a comparison with the fourth-order system describing the restricted case\footnote{We can reduce the system made up of Eqs. (\ref{2.15a}) and (\ref{2.16a}) to a third-order system by employing the Jacobi integral (\ref{2.12a}). Moreover, a further reduction to a second-order system can be achieved by elimination of the time variable \cite{Szebehely67}.} (cf. Eqs. (\ref{2.15a}) and (\ref{2.16a})). 

The general problem of three bodies admits ten independent {\it algebraic} integrals of motion, called classical integrals. First of all, since no external forces act on the system, the center of mass $D$ moves on a straight line with constant velocity, giving six constants of motion (three regarding the components of the position of $D$ and three the components of its velocity, or equivalently the three components of the total momentum of the system). Other three integrals of motion are represented by the components of the angular momentum about the origin, and the last one is the integral of energy $T-U$ (the potential (\ref{Pars2919}) does not depend explicitly on time, as was anticipated at the end of Sec. \ref{Quantum corrected Lagrangian_Sec}). Among the classical integrals, only six of them, besides being independent, are also in involution: the three components of the total momentum, the square of the angular momentum, its third component\footnote{Recall that for the three components of the angular momentum $\vec{\ell}$ we have
\begin{equation}
\{\ell_i, \ell_j \} = \varepsilon_{ijk} \ell_k,
\end{equation}
\begin{equation}
\{\vert \vec{\ell}\vert^2, \ell_i \}=0, \; \; \; \; \forall \, i \in \{1,2,3\}, 
\end{equation}
$\{,\}$ being the Poisson bracket and $\varepsilon_{ijk}$ the total antisymmetric Levi-Civita symbol.} and the total energy. The existence of these six integrals of motion allows the reduction of the original eighteenth-order system (\ref{Pars29112}) to a sixth-order one, corresponding to a dynamical system with only three ``true'' degrees of freedom \cite{Pars,Szebehely67}.

As was pointed out before, the classical integrals are algebraic functions of the canonical coordinates. There are no further algebraic integrals independent of those already found. This remarkable result was proved by Bruns in 1887 and the general form of his theorem reads as follows \cite{Szebehely67}:
\newtheorem*{Bruns}{Bruns Theorem}
\begin{Bruns}
In the problem of $N$ bodies the only integrals of motion involving the coordinates and the momenta algebraically, and which do not involve time explicitly, are composed of the integrals of the center of mass of the system, the total angular momentum and the energy. 
\end{Bruns}  
In other words, any algebraic integrals of the problem of three bodies is merely a combination of the ten classical integrals. 

Bruns theorem was first generalized by Painlev\'e, who demonstrated that even in the case in which we consider integrals which are algebraic functions of the canonical momenta and analytic functions of the coordinates, the result does not change: the only independent constants of motion are once again given by the ten classical  integrals \cite{Szebehely67}. After that, the most general result was achieved by Poincar\'e \cite{P1892}, who proved that, besides energy, angular momentum and linear momentum, there are no other {\it analytic} functions on phase space which Poisson commute with the Hamiltonian. In other words, any constant of motion is necessarily a function of the classical integrals and hence the three-body problem does not give rise to a completely integrable system. Recall that a completely integrable Hamiltonian system having $n$ degrees of freedom is said to be completely integrable (in the Liouville sense) if it admits $n$ independent Poisson commuting integrals of motion. In fact, for completely integrable system the constants of motion determine a regular foliation of the phase space (called Lagrangian foliation), which is thus decomposed as a collection of lesser dimensional sub-manifolds so that the dynamics is automatically reduced to one with less degrees of freedom \cite{Romano}.  

To be more precise, the proof of the above-mentioned theorem given by Poinacar\'e only holds in the parameter region where one of the masses of the bodies dominates the other two. It is still possible that for very special masses the system is integrable, as we will see in Sec. \ref{Periodic_solutions_Sec}. Before Poincar\'e, mathematicians and, in particular, astronomers spent much energy in the search for sequences of changes of variables which made the system ``more and more integrable''. Poincar\'e realized that the series defining their transformations were divergent (hence his interest in divergent series, as outlined at the beginning of Chapter \ref{Restricted_Chapter}). This divergence problem is connected to the ``small denominators problem'' and getting around it by considering a number of theoretical conditions on frequencies appears as the heart of the Kolmogorov-Arnol'd-Moser theorem \cite{Romano}.

Finally, we conclude this section by mentioning two important solutions of the full three-body problem found by Lagrange in which the lengths $r_1$, $r_2$, and $r_3$ remain constant throughout the motion. They are the collinear solution, where the particles always line up, and the triangle one, where the bodies lay at the vertices of an equilateral triangle of invariable size \cite{Pars}. Moreover, by studying solutions where the particles describe orbits which are invariant only in shape but not in size, Lagrange discovered a family of solutions where the particles move forming an equilateral triangle and describing a conic with the center of mass (which is fixed in space) representing one focus. The conics all have the same eccentricity and, in the particular case in which they are ellipses, the motion turns out to be periodic \cite{Pars}.  

\subsection{Reduced form of the equations of motion}

In the previous section we have seen that in Newtonian theory the center of mass $D$ of the system moves uniformly on a straight line. This means that we could also suppose, as a special case, that it is at rest, but we will not follow this alternative. Furthermore, we also consider the most general problem in which the particles move in the space and are not constrained in a plane. 

\begin{figure} [htbp] 
\centering
\includegraphics[scale=0.7]{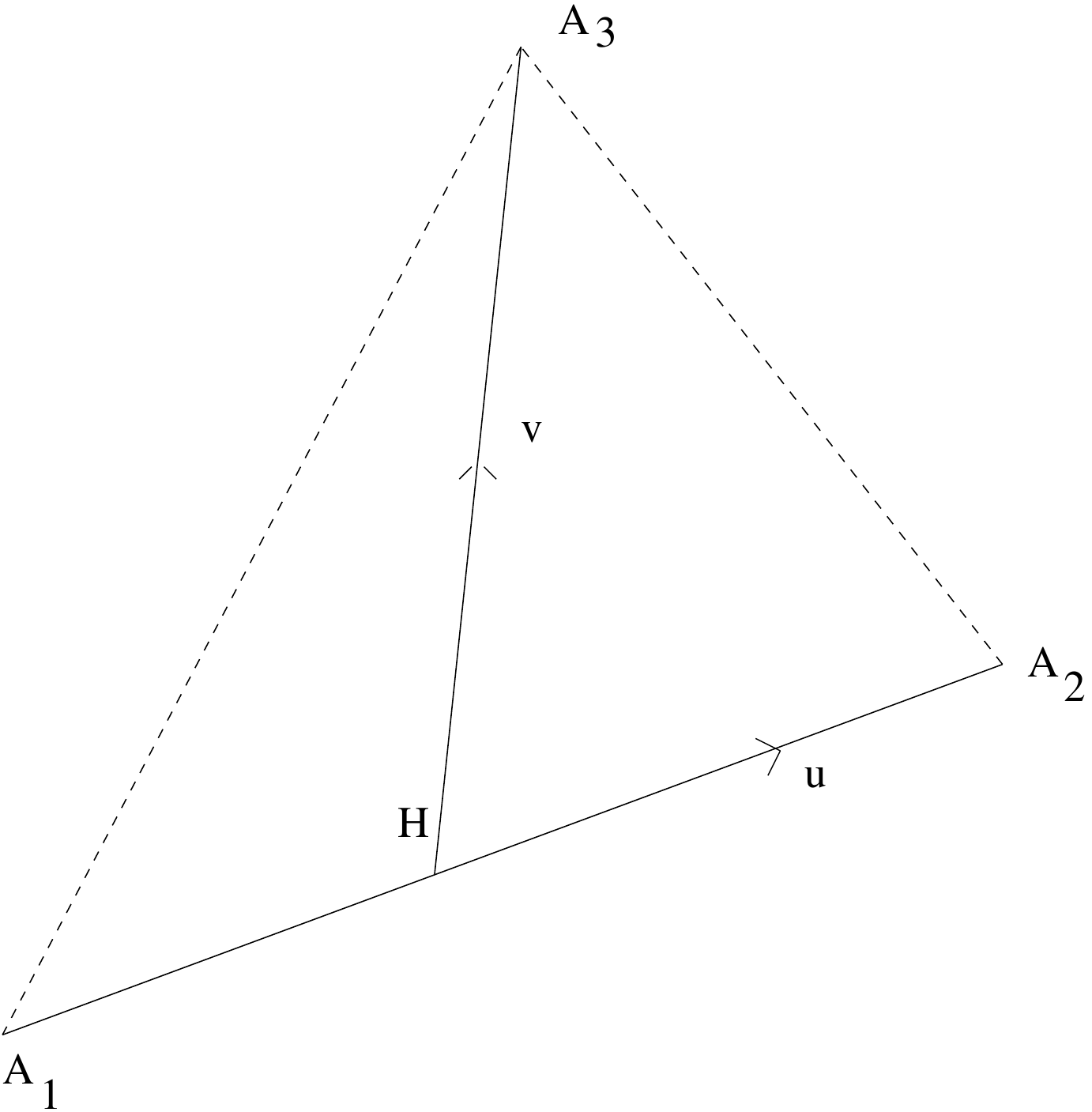}
\caption[Schematic set-up of the full three-body problem]{Schematic set-up of the full three-body problem: the three  bodies $A_{1}$, $A_{2}$, and $A_{3}$, the center of mass $H$ of $A_{1}$ and $A_{2}$, the vector $\vec u$ joining $A_{1}$ to $A_{2}$, and the vector $\vec v$ joining $H$ to $A_{3}$.} 
\label{1f.pdf}
\end{figure}

Bearing in mind the above premisses, let the vector ${\overrightarrow {A_{1}A_{2}}}$ be ${\vec u}$, and let the vector ${\overrightarrow {HA_{3}}}$ ($H$ being the center of mass of $A_{1}$ and $A_{2}$) be ${\vec v}$ (Fig. \ref{1f.pdf}). Thus, by defining the parameters
\begin{equation}
\alpha_{1} \equiv {m_{1}\over (m_{1}+m_{2})}, \;\; \; \;\;\; \; \;   \alpha_{2} \equiv 1-\alpha_{1},
\end{equation}
the vector ${\overrightarrow {A_{2}A_{3}}}$ is $(-\alpha_{1}{\vec u}+{\vec v})$,
while the vector ${\overrightarrow {A_{1}A_{3}}}$ is $(\alpha_{2}{\vec u}+{\vec v})$. Hereafter, we denote by $(x,y,z)$ the components of ${\vec u}$, and by $(\xi,\eta,\zeta)$ the components of ${\vec v}$. 

On defining the ``reduced masses''
\begin{equation}
\begin{split}
& m \equiv {m_{1}m_{2}\over (m_{1}+m_{2})}, \\
& \mu \equiv {(m_{1}+m_{2})m_{3}\over (m_{1}+m_{2}+m_{3})},
\end{split}
\end{equation}
the $x$-terms in $T$ arising from the motion relative to $D$ give \cite{BE14b,Pars}
\begin{equation}
{1\over 2M}\Bigr[m_{2}m_{3}(-\alpha_{1}{\dot x}
+{\dot \xi})^{2}+m_{3}m_{1}(\alpha_{2}{\dot x}+{\dot \xi})^{2}
+m_{1}m_{2}{\dot x}^{2}\Bigr] 
={m\over 2}{\dot x}^{2}+{\mu \over 2}{\dot \xi}^{2}.
\end{equation}
One has now to add the corresponding formulas for $y$ and $z$, which yields the neat result
\begin{equation}
T={M \over 2}\Bigr({\dot X}^{2}+{\dot Y}^{2}+{\dot Z}^{2}\Bigr)
+{m\over 2}({\dot x}^{2}+{\dot y}^{2}+{\dot z}^{2})
+{\mu \over 2}({\dot \xi}^{2}+{\dot \eta}^{2}+{\dot \zeta}^{2}).
\label{2.5f}
\end{equation}

As was pointed out before, in Newtonian theory one proceeds by assuming at this stage a potential of the form \cite{Pars}
\begin{equation}
U=G \left({m_{2}m_{3}\over r_{1}}+{m_{3}m_{1}\over r_{2}}
+{m_{1}m_{2}\over r_{3}}\right),
\label{2.6f}
\end{equation}
the distances being defined in this case as
\begin{equation}
(r_{1})^{2} \equiv (-\alpha_{1}{\vec u}+{\vec v}) \cdot
(-\alpha_{1}{\vec u}+{\vec v})
=(-\alpha_{1}x+\xi)^{2}+(-\alpha_{1}y+\eta)^{2}
+(-\alpha_{1}z+\zeta)^{2},
\label{2.7f}
\end{equation}
\begin{equation}
(r_{2})^{2} \equiv (\alpha_{2}{\vec u}+{\vec v}) \cdot
(\alpha_{2}{\vec u}+{\vec v})
=(\alpha_{2}x+\xi)^{2}+(\alpha_{2}y+\eta)^{2}
+(\alpha_{2}z+\zeta)^{2},
\end{equation}
\begin{equation}
(r_{3})^{2} \equiv {\vec u} \cdot {\vec u}
=x^{2}+y^{2}+z^{2}.
\label{2.9f}
\end{equation}
In the quantum regime, although we keep using the classical concepts of kinetic energy and center of mass, we depart from classical Newtonian theory by assuming that $U$ can be a more general function of $r_{1}$, $r_{2}$, $r_{3}$, i.e.,
\begin{equation}
U=U(r_{1},r_{2},r_{3})=\sum_{k=1}^{3} U_{k}(r_{k}).
\label{2.10f}
\end{equation}
We will first derive the equations of motion resulting from the general choice (\ref{2.10f}), and we will eventually look for explicit solutions with a choice of $U$ inspired by the issues described in the previous chapters (see Sec. \ref{Periodic_solutions_Sec}, Eq. (\ref{3.1f})).

By virtue of (\ref{2.5f}) and (\ref{2.10f}), the Lagrangian equations of motion read as
\begin{equation}
\begin{split}
& M {\ddot X}={\partial U \over \partial X}, \\
& m {\ddot x}={\partial U \over \partial x}, \\
& \mu {\ddot \xi}={\partial U \over \partial \xi},
\label{2.11f}
\end{split}
\end{equation}
supplemented by the corresponding second-order equations for $(Y,y,\eta)$ and $(Z,z,\zeta)$. Since, from (\ref{2.10f}), $U$ is independent of $X,Y,Z$, one has
\begin{equation}
{\ddot X}={\ddot Y}={\ddot Z}=0,
\end{equation}
which means that the center of mass $D$ moves uniformly in a straight line, like in the Newtonian case. We may even assume that $D$ remains at rest without losing generality, and the remaining equations for $m \ddot x$ and $\mu \ddot \xi$ in (\ref{2.11f}) can be obtained by setting
\begin{equation}
U_{,r_{j}} \equiv {\partial U \over \partial r_{j}}, \; \; \;\; \; \; \forall j=1,2,3,
\label{2.13f}
\end{equation}
and writing patiently the partial derivatives
\begin{equation}
{\partial U \over \partial x}=U_{,r_{1}}
{\partial r_{1}\over \partial x}
+U_{,r_{2}}{\partial r_{2}\over \partial x} 
+U_{,r_{3}}{\partial r_{3}\over \partial x},
\end{equation}
\begin{equation}
{\partial U \over \partial \xi}=U_{,r_{1}}
{\partial r_{1}\over \partial \xi}
+U_{,r_{2}}{\partial r_{2}\over \partial \xi}.
\end{equation}
In light of (\ref{2.7f})--(\ref{2.9f}), one arrives therefore
at the formulas
\begin{equation}
{\partial U \over \partial x}=-Ax+B \xi ,
\end{equation}
\begin{equation}
{\partial U \over \partial \xi}=Bx-C \xi,
\end{equation}
where we have defined \cite{BE14b}
\begin{equation}
A \equiv -{\alpha_{1}^{2}\over r_{1}}U_{,r_{1}}
-{\alpha_{2}^{2}\over r_{2}}U_{,r_{2}}
-{1\over r_{3}}U_{,r_{3}},
\label{2.18f}
\end{equation}
\begin{equation}
B \equiv {\alpha_{2}\over r_{2}}U_{,r_{2}}
-{\alpha_{1}\over r_{1}}U_{,r_{1}},
\end{equation}
\begin{equation}
C \equiv -{1\over r_{1}}U_{,r_{1}}-{1\over r_{2}}U_{,r_{2}}.
\label{2.20f}
\end{equation}
After writing the corresponding equations for $(y,\eta)$ and $(z,\zeta)$ one obtains eventually, bearing in mind that 
${\vec u}$ has components $(x,y,z)$, while ${\vec v}$ has components $(\xi,\eta,\zeta)$, the equations of motion in matrix form \cite{BE14b}:
\begin{equation}
\left(\begin{matrix}
m {{\rm d}^{2}\over {\rm d}t^{2}}+A & -B \cr
-B & \mu {{\rm d}^{2}\over {\rm d}t^{2}}+C
\end{matrix}
\right)
\left(\begin{matrix}
{\vec u} \cr {\vec v} 
\end{matrix}\right)=0.
\label{2.21f}
\end{equation}
Such a scheme tells us that the full three-body problem is equivalent to a system of two particles, i.e., a particle of mass $m$ at $(x,y,z)$ and a particle of mass $\mu$ at $(\xi,\eta,\zeta)$, in perfect analogy with the classical case \cite{BE14b,Pars}. The system (\ref{2.21f}) represents the so-called reduced form of the equations of motion for the system of three particles. 

The integrals of angular momentum are found to take the form \cite{BE14b}
\begin{equation}
M(Y {\dot Z}-Z{\dot Y})+m(y{\dot z}-z{\dot y})
+\mu (\eta {\dot \zeta}-\zeta {\dot \eta})=a,
\label{2.22f}
\end{equation} 
\begin{equation}
M(Z {\dot X}-X{\dot Z})+m(z{\dot x}-x{\dot z})
+\mu (\zeta {\dot \xi}-\xi {\dot \zeta})=b,
\label{2.23f}
\end{equation} 
\begin{equation}
M(X {\dot Y}-Y{\dot X})+m(x{\dot y}-y{\dot x})
+\mu (\xi {\dot \eta}-\eta {\dot \xi})=c.
\label{2.24f}
\end{equation} 
Since, as outlined before, the center of mass moves uniformly on a straight line, the terms $M(Y{\dot Z}-Z {\dot Y})$ and
$[m(y{\dot z}-z{\dot y})+\mu (\eta {\dot \zeta}-\zeta{\dot \eta})]$ in (\ref{2.22f}) are separately constant, and similarly in Eqs. (\ref{2.23f}) and (\ref{2.24f}).
Indeed, one finds from Eq. (\ref{2.11f})
\begin{equation}
{{\rm d}\over {\rm d}t}[m(y{\dot z}-z{\dot y})
+\mu(\eta {\dot \zeta}-\zeta{\dot \eta})]
=\left(y{\partial \over \partial z}-z {\partial \over \partial y}
+\eta {\partial \over \partial \zeta}
-\zeta {\partial \over \partial \eta}\right)U,
\label{(2.25)}
\end{equation}
which vanishes, because $U$ depends on $r_{1},r_{2},r_{3}$ separately, according to Eq. (\ref{2.10f}), and the following identity holds \cite{BE14b}:
\begin{equation}
\left(y{\partial \over \partial z}
-z{\partial \over \partial y}
+\eta {\partial \over \partial \zeta}
-\zeta {\partial \over \partial \eta}\right)r_{k}=0, \;\; \; \; \; 
\forall \, k=1,2,3.
\end{equation}
The forces are not in the line joining the particles, but their moment about the origin is
\begin{equation}
{\vec u} \times (-A {\vec u}+B {\vec v})
+{\vec v} \times (B {\vec u}-C {\vec v}),
\end{equation}
which vanishes by virtue of the skew-symmetry of the vector product. Hence the angular momentum about the origin remains constant as in Newtonian theory \cite{BE14b}.

\subsection{Periodic solutions} \label{Periodic_solutions_Sec}

After having written the equations of motion in a rather general form, we cannot attempt any integration without an explicit form of the potential function. For this purpose, we now investigate the implications of assuming that the classical potential (\ref{2.6f}) can be replaced by a quantum corrected potential according to the recipes considered in the previous chapters. This means that the general formula (\ref{2.10f}) can take the form \cite{BE14b}
\begin{equation}
\begin{split}
U(r_{1},r_{2},r_{3})&=
{G m_{2}m_{3}\over r_{1}}
\left(1+\kappa_{23} {G \over c^{2}}
{(m_{2}+m_{3})\over r_{1}}
+\kappa {l_{P}^{2} \over (r_{1})^{2}}\right)\\
&+{G m_{1}m_{3}\over r_{2}}
\left(1+\kappa_{13}  {G \over c^{2}}
{(m_{1}+m_{3})\over r_{2}}
+\kappa {l_{P}^{2} \over (r_{2})^{2}}\right) \\
&+{G m_{1}m_{2}\over r_{3}}
\left(1+\kappa_{12}  {G \over c^{2}}
{(m_{1}+m_{2})\over r_{3}}
+\kappa {l_{P}^{2} \over (r_{3})^{2}}\right),
\end{split}
\label{3.1f}
\end{equation}
where $U(r_{1},r_{2},r_{3})$ has been defined following the structure of Eqs. (\ref{1.2b})--(\ref{1.4b}), whereas the dimensionless parameters $\kappa$, $\kappa_{12}$, $\kappa_{23}$, and $\kappa_{13}$ can be easily read off with the help of Tab. \ref{kappa_tab}. Furthermore, it should be clear that $\kappa_{12}$, $\kappa_{23}$, and $\kappa_{13}$ depend on $\kappa$, since they are part of a calculational recipe that yields, at the same time, a post-Newtonian term and a fully quantum term (see the discussion at the end of Sec. \ref{Sec. scattering-bound-state potential}).

The first derivatives of such a potential, to be used in the definitions (\ref{2.18f})--(\ref{2.20f}) of the functions $A,B,C$ read therefore as
\begin{equation}
U_{,r_{1}}=-{Gm_{2}m_{3}\over (r_{1})^{2}}
\left(1+2 \kappa_{23}  {G \over c^{2}}
{(m_{2}+m_{3})\over r_{1}}
+3 \kappa {l_{P}^{2}\over (r_{1})^{2}}\right),
\label{3.2f}
\end{equation}
\begin{equation}
U_{,r_{2}}=-{Gm_{1}m_{3}\over (r_{2})^{2}}
\left(1+2 \kappa_{13}  {G \over c^{2}}
{(m_{1}+m_{3})\over r_{2}}
+3 \kappa {l_{P}^{2}\over (r_{2})^{2}}\right),
\end{equation}
\begin{equation}
U_{,r_{3}}=-{Gm_{1}m_{2}\over (r_{3})^{2}}
\left(1+2 \kappa_{12}  {G \over c^{2}}
{(m_{1}+m_{2})\over r_{3}}
+3 \kappa {l_{P}^{2}\over (r_{3})^{2}}\right).
\label{3.4f}
\end{equation}
The equations of motion (\ref{2.21f}) are Lagrangian second-order equations of motion. They can be re-expressed as a coupled set of twelve first-order Hamiltonian equations as follows:
\begin{equation}
\begin{dcases}
{{\rm d} \over {\rm d}t}x=p_{x}, \\
{{\rm d} \over {\rm d}t}y=p_{y},\\
{{\rm d} \over {\rm d}t}z=p_{z},\\
{{\rm d} \over {\rm d}t}\xi=p_{\xi},\\
{{\rm d} \over {\rm d}t}\eta=p_{\eta}, \\
{{\rm d} \over {\rm d}t}\zeta=p_{\zeta},\\
{{\rm d}\over {\rm d}t}p_{x}=-{1\over m}(Ax-B \xi),\\
{{\rm d}\over {\rm d}t}p_{y}=-{1\over m}(Ay-B \eta),\\
{{\rm d}\over {\rm d}t}p_{z}=-{1\over m}(Az-B \zeta),\\
{{\rm d}\over {\rm d}t}p_{\xi}=-{1\over \mu}(C \xi-Bx),\\
{{\rm d}\over {\rm d}t}p_{\eta}=-{1\over \mu}(C \eta-By),\\
{{\rm d}\over {\rm d}t}p_{\zeta}=-{1\over \mu}(C \zeta-Bz).\\
\end{dcases} \\ [2em]
\label{4.1f}
\end{equation}
We need therefore twelve initial conditions to integrate these equations of motion. Hereafter it is convenient to introduce the $6$-tuple of position variables
\begin{equation}
x_{i} \equiv (x,y,z,\xi,\eta,\zeta) 
\equiv (x_{1},\dots,x_{6}),
\label{4.13f}
\end{equation}
and the $6$-tuple of momentum variables
\begin{equation}
y_{i} \equiv (p_{x},p_{y},p_{z},p_{\xi},p_{\eta},p_{\zeta})
\equiv (p_{1},\dots,p_{6}).
\label{4.14f}
\end{equation}
Equation (\ref{4.1f}) can be therefore further re-expressed in the canonical form through the {\it autonomous} system \cite{BE14b,P1890,P1892}
\begin{equation}
\begin{dcases}
{{\rm d}\over {\rm d}t}x_{i}={\partial F \over \partial y_{i}},\\
 {{\rm d}\over {\rm d}t}y_{i}=-{\partial F \over \partial x_{i}},
\end{dcases} \\ [2em]
\label{4.15f}
\end{equation}
where the Hamiltonian function $F$ is given by
\begin{equation}
F(x_{1},\dots,x_{6},y_{1},\dots,y_{6})=\sum_{i=1}^{6}
{y_{i}^{2}\over 2}+f(x_{1},\dots,x_{6}),
\label{4.16f}
\end{equation}
and $f$ solves the second half of the Hamiltonian equations (\ref{4.1f}), i.e., \cite{BE14b}
\begin{equation}
\begin{split}
& {\partial f \over \partial x}={1\over m}(Ax-B\xi),\\
& {\partial f \over \partial y}={1\over m}(Ay-B\eta),\\
& {\partial f \over \partial z}={1\over m}(Az-B\zeta),\\
& {\partial f \over \partial \xi}={1\over \mu}(C \xi-Bx),\\
& {\partial f \over \partial \eta}={1\over \mu}(C \eta-By),\\
& {\partial f \over \partial \zeta}={1\over \mu}(C \zeta-Bz),\\
\end{split}
\label{4.17f}
\end{equation}
the functions $A(x_{1},\dots,x_{6})$, $B(x_{1},\dots,x_{6})$, $C(x_{1},\dots,x_{6})$ being defined by (\ref{2.18f})--(\ref{2.20f}), supplemented by (\ref{2.7f})--(\ref{2.9f}) and (\ref{3.2f})--(\ref{3.4f}).

At this stage, we can exploit the fundamental Poincar\'e theorem on autonomous systems enunciated at the end of Sec. \ref{Important_poincare_theorem_sec}. According to this theorem and bearing in mind the results of Sec. \ref{Variation_Hamiltonian_Sec} regarding the number of vanishing characteristic exponents for a Hamiltonian system, we can say that if Eq. (\ref{4.15f}), which depends on a parameter $\rho$, possesses for $\rho=0$ a periodic solution for which only {\it two} characteristic exponents vanish, we have again a periodic solution for small, but non-vanishing, values of $\rho$ \cite{BE14b}. In our case, the small parameter $\rho$ is the Planck length $l_{P}$, and when $\rho=0$ we revert to the three-body problem in post-Newtonian mechanics, for which, in the circular restricted case, one knows from recent work \cite{HuangWu} that orbits may be unstable, or bounded chaotic, or bounded regular. Moreover, in Ref. \cite{Imai} has been proved that the Newtonian full $N$-body problem admits a special class of  solutions, called choreographic solutions, at the first post-Newtonian order. In celestial mechanics a solution is called choreographic if every massive particles move periodically in a single closed orbit. Bearing in mind that in general relativity the periastron shift prohibits a binary system from orbiting in a single closed curve, the authors of Ref. \cite{Imai} have computed relativistic corrections to initial conditions so that an orbit for a three-body system can be choreographic and define an eight-shaped curve. This means that the stunning solution of Newtonian mechanics first discovered by Moore in 1993 \cite{Moore1993} and re-discovered with its existence proof by Chenciner and Montgomery in 2000 \cite{Chenciner00}, survives also in the context of general relativity at the first post-Newtonian order\footnote{This result holds also at the second post-Newtonian order, as shown in Ref. \cite{L-Nakano}.}. This particular solution of the classical three-body problem consists in the fact that three bodies of equal mass move periodically on the plane along the same curve. The periodic orbit has zero angular momentum, and the three bodies chase each other around a fixed eight-shaped curve. Such an orbit visits in turn every Euler configuration in which one of the bodies sits at the midpoint of the segment defined by the other two.

Therefore, by virtue of the above-mentioned Poincar\'e theorem on periodic solutions and of the extreme smallness of the Planck length, we have found a simple but non trivial result consisting in the fact that  also our quantum corrected potential (\ref{3.1f}) may lead to periodic solutions \cite{BE14b}. This is a novel perspective on a smooth matching between classical and quantum-corrected three-body problems.

\subsection{General solution of the quantum corrected variational equations}

Bearing in mind Sec. \ref{Neighbourhood_Sec}, let us now revert to Eq. (\ref{4.15f}), and assume that a periodic solution
\begin{equation}
x_{i}=\varphi_{i}(t), \;\; \; \; \; \; \;  y_{i}=\phi_{i}(t),
\label{5.1f}
\end{equation}
has been found. We now investigate an algorithm for the evaluation of characteristic exponents \cite{BE14b}. For this purpose, we consider small disturbances of such periodic solutions, written as
\begin{equation}
{\tilde x}_{i}=\varphi_{i}(t)+\xi_{i}, \;\; \; \; \; \; \; \; {\tilde y}_{i}=\phi_{i}(t)+\eta_{i},
\end{equation}
and we form the variational equations resulting from the linearised approximation (cf. Eq. (\ref{Pars2362})), i.e.,
\begin{equation}
{{\rm d}\over {\rm d}t}\xi_{i}
=\sum_{k=1}^{6}\Bigr[F_{,y_{i}x_{k}}\xi_{k}
+F_{,y_{i}y_{k}}\eta_{k}\Bigr],
\label{5.3f}
\end{equation}
\begin{equation}
{{\rm d}\eta_{i}\over {\rm d}t}
=-\sum_{k=1}^{6}\Bigr[F_{,x_{i}x_{k}}\xi_{k}
+F_{,x_{i}y_{k}}\eta_{k}\Bigr].
\label{5.4f}
\end{equation}
We now try to integrate Eqs. (\ref{5.3f}) and (\ref{5.4f}) by setting \cite{BE14b}
\begin{equation}
\begin{split}
& \xi_{i}={\rm e}^{\alpha t}S_{i}, \\
& \eta_{i}={\rm e}^{\alpha t}T_{i},
\end{split}
\label{5.5f}
\end{equation}
$S_{i}$ and $T_{i}$ being unknown periodic functions of $t$ with the same period of the unperturbed solution (\ref{5.1f}) and $\alpha$ the characteristic exponent. Next, we assume that the Hamiltonian function $F$ admits the Poincar\'e asymptotic expansion (see Appendix \ref{Asymptotic exp_App})
\begin{equation}
F \sim F_{0}+\rho F_{1}+\rho^{2}F_{2}+{\rm O}(\rho^{3}),
\label{5.9f}
\end{equation}
and we also suppose that $F_0$ depends only on the coordinates $x_i$, while it is independent of the momenta $y_i$, which therefore are ignorable coordinates. In other words, for $\rho=0$ the system under investigation is completely integrable, since it admits six independent and Poisson commuting integrals of motion. As explained in Sec. \ref{Variation_Hamiltonian_Sec}, this means that for $\rho=0$ all the characteristic exponents are zero. Poincar\'e has demonstrated that, in such a situation, for small, but non-vanishing values of $\rho$, it is possible to expand $\alpha$, $S_{i}$, and $T_{i}$ in powers of $\sqrt{\rho}$, i.e., \cite{P1892}\footnote{This framework is complementary to the one mentioned in Sec. \ref{Periodic_solutions_Sec}, where we exploited the Poincar\'e theorem on the persistence
of periodic solutions at small $\rho$. That theorem does not imply an expansion for the Hamiltonian function like the one in Eq. (\ref{5.9f}), and, in addiction, it assumes that for $\rho=0$ there are only two vanishing characteristic exponents.}
\begin{equation}
\alpha \sim \sum_{j=1}^{N}\alpha_{j}\rho^{{j \over 2}},
\label{5.6f}
\end{equation}
\begin{equation}
S_{i} \sim \sum_{l=0}^{N}S_{i}^{l}\rho^{{l \over 2}},
\end{equation}
\begin{equation}
T_{i} \sim \sum_{l=0}^{N}T_{i}^{l}\rho^{{l \over 2}}.
\label{5.8f}
\end{equation}
The ``extended version'' of Cauchy theorem provided by Poincar\'e in Refs. \cite{P1892,Pseries} represents the starting point for the demonstration that the asymptotic series (\ref{5.6f})--(\ref{5.8f}) exist. We briefly describe this result. Let
\begin{equation}
\begin{split}
& \dfrac{{\rm d}}{{\rm d}t}x = h(x,y,t,\nu),\\
& \dfrac{{\rm d}}{{\rm d}t}y = g(x,y,t,\nu), 
\end{split}
\label{Pdiff}
\end{equation}
be two differential equations where the functions $h$ and $g$ are expandible is terms of the unknown functions $x$ and $y$, the variable $t$, and an arbitrary parameter $\nu$.  Unlike Cauchy, who limited his attention to series developable with respect to the independent variable $t$ only, Poincar\'e considered expansions also in terms of $\nu$ and the initial data $x_0$, $y_0$. In this way, he showed that (\ref{Pdiff}) is satisfied by some series
\begin{equation}
\begin{split}
& x=f_{1}(t,x_0,y_0,\nu),\\
& y=f_{2}(t,x_0,y_0,\nu),
\end{split}
\label{Pdiff2}
\end{equation}
which can be developed in terms of increasing powers of $t$, $x_0$, $y_0$, $\nu$ and reduce, respectively, to $x_0$ and $y_0$ for $t=0$. Furthermore, it is proved that (\ref{Pdiff2}) converges for any value of the variable $t$, provided that $\vert \nu \vert$ is sufficiently small \cite{P1892,Pseries}. 

At this stage, after having employed the expansion (\ref{5.9f}), we insert formulas (\ref{5.5f}) and (\ref{5.6f})--(\ref{5.8f}) into the linear variational equations (\ref{5.3f}) and (\ref{5.4f}), yielding \cite{BE14b}
\begin{equation}
{{\rm d}\over {\rm d}t}\xi_{i} \sim {\rm e}^{\alpha t}
\left[{{\rm d}S_{i}^{0}\over {\rm d}t}
+\left(\alpha_{1}S_{i}^{0}+{{\rm d}S_{i}^{1}\over {\rm d}t}
\right)\sqrt{\rho}
+\left(\alpha_{1}S_{i}^{1}+\alpha_{2}S_{i}^{0}
+{{\rm d}S_{i}^{2}\over {\rm d}t}\right)\rho
+{\rm O}(\rho^{{3\over 2}})\right],
\label{5.10f}
\end{equation}
\begin{equation}
{{\rm d}\over {\rm d}t}\eta_{i} \sim {\rm e}^{\alpha t}
\left[{{\rm d}T_{i}^{0}\over {\rm d}t}
+\left(\alpha_{1}T_{i}^{0}+{{\rm d}T_{i}^{1}\over {\rm d}t}
\right)\sqrt{\rho}
+\left(\alpha_{1}T_{i}^{1}+\alpha_{2}T_{i}^{0}
+{{\rm d}T_{i}^{2}\over {\rm d}t}\right)\rho
+{\rm O}(\rho^{{3\over 2}})\right].
\label{5.11f}
\end{equation}
so that comparison of coefficients of equal powers of $\rho$ gives for all $i=1,\dots,6$, and up to first order in $\rho$, the equations \cite{BE14b}
\begin{equation}
{{\rm d}S_{i}^{0}\over {\rm d}t}=\sum_{k=1}^{6}
\Bigr({F_{0}}_{,y_{i}x_{k}}S_{k}^{0}
+{F_{0}}_{,y_{i}y_{k}}T_{k}^{0}\Bigr),
\label{5.12f}
\end{equation}
\begin{equation}
\alpha_{1}S_{i}^{0}+{{\rm d}S_{i}^{1}\over {\rm d}t}
=\sum_{k=1}^{6}\Bigr({F_{0}}_{,y_{i}x_{k}}S_{k}^{1}
+{F_{0}}_{,y_{i}y_{k}}T_{k}^{1}\Bigr),
\label{5.13f}
\end{equation}
\begin{equation}
\alpha_{1}S_{i}^{1}+\alpha_{2}S_{i}^{0}
+{{\rm d}S_{i}^{2}\over {\rm d}t}
=\sum_{k=1}^{6}\Bigr({F_{0}}_{,y_{i}x_{k}}S_{k}^{2}
+{F_{1}}_{,y_{i}x_{k}}S_{k}^{0}
+{F_{0}}_{,y_{i}y_{k}}T_{k}^{2}
+{F_{1}}_{,y_{i}y_{k}}T_{k}^{0}\Bigr),
\end{equation}
\begin{equation}
{{\rm d}T_{i}^{0}\over {\rm d}t}=-\sum_{k=1}^{6}
\Bigr({F_{0}}_{,x_{i}x_{k}}S_{k}^{0}
+{F_{0}}_{,x_{i}y_{k}}T_{k}^{0}\Bigr),
\label{5.15f}
\end{equation}
\begin{equation}
\alpha_{1}T_{i}^{0}+{{\rm d}T_{i}^{1}\over {\rm d}t}
=-\sum_{k=1}^{6}\Bigr({F_{0}}_{,x_{i}x_{k}}S_{k}^{1}
+{F_{0}}_{,x_{i}y_{k}}T_{k}^{1}\Bigr),
\label{5.16f}
\end{equation}
\begin{equation}
\alpha_{1}T_{i}^{1}+\alpha_{2}T_{i}^{0}
+{{\rm d}T_{i}^{2}\over {\rm d}t}
=-\sum_{k=1}^{6}\Bigr({F_{0}}_{,x_{i}x_{k}}S_{k}^{2}
+{F_{1}}_{,x_{i}x_{k}}S_{k}^{0}
+{F_{0}}_{,x_{i}y_{k}}T_{k}^{2}
+{F_{1}}_{,x_{i}y_{k}}T_{k}^{0}\Bigr).
\label{5.17f}
\end{equation}
To begin, one should solve Eqs. (\ref{5.12f}) and (\ref{5.15f}) for $S_{i}^{0}$ and $T_{i}^{0}$, and insert them into (\ref{5.13f}) and (\ref{5.16f}) to find $S_{i}^{1}$ and $T_{i}^{1}$, and iterate the procedure to find $S_{i}^{2}$, $T_{i}^{2},\dots$, as well as $\alpha_{1},\alpha_{2},\dots$. 

In our model, the potential function (\ref{3.1f}) contains only a part of zero-th order in $\rho \equiv l_{P}$ and a part of second order in $\rho$, and the same holds for the Hamiltonian function $F$ defined in Eq. (\ref{4.16f}). This means that in our framework the terms $F_1$ and ${\rm O}(\rho^3)$ appearing in Eq. (\ref{5.9f}) vanish identically and on defining
\begin{equation}
\gamma_{1}(r_{1}) \equiv -{G m_{2}m_{3}\over (r_{1})^{2}}
\left(1+2 \kappa_{23}{G \over c^{2}}{(m_{2}+m_{3})\over r_{1}}\right),
\label{6.1f}
\end{equation}
\begin{equation}
\gamma_{2}(r_{2}) \equiv -{G m_{1}m_{3}\over (r_{2})^{2}}
\left(1+2 \kappa_{13}{G \over c^{2}}{(m_{1}+m_{3})\over r_{2}}\right),
\end{equation}
\begin{equation}
\gamma_{3}(r_{3}) \equiv -{G m_{1}m_{2}\over (r_{3})^{2}}
\left(1+2 \kappa_{12}{G \over c^{2}}{(m_{1}+m_{2})\over r_{3}}\right),
\label{6.3f}
\end{equation}
we find that $A$, $B$, and $C$ in (\ref{2.18f})--(\ref{2.20f}) take the form \cite{BE14b}
\begin{equation}
A=A_{0}+\rho^{2}A_{2}, \;\; \; \;  B=B_{0}+\rho^{2}B_{2}, \;\; \; \; C=C_{0}+\rho^{2}C_{2},
\end{equation}
where \cite{BE14b}
\begin{equation}
A_{0}=-(\alpha_{1})^{2}{\gamma_{1}(r_{1})\over r_{1}}
-(\alpha_{2})^{2}{\gamma_{2}(r_{2})\over r_{2}}-{\gamma_{3}(r_{3})\over r_{3}},
\label{6.5f}
\end{equation}
\begin{equation}
A_{2}=3G \kappa \left[(\alpha_{1})^{2}{m_{2}m_{3}\over (r_{1})^{5}}
+(\alpha_{2})^{2}{m_{1}m_{3}\over (r_{2})^{5}}
+{m_{1}m_{2}\over (r_{3})^{5}}\right],
\end{equation}
\begin{equation}
B_{0}=\alpha_{2}{\gamma_{2}(r_{2})\over r_{2}}
-\alpha_{1}{\gamma_{1}(r_{1})\over r_{1}},
\label{6.7f}
\end{equation}
\begin{equation}
B_{2}=3G \kappa \left[\alpha_{1}{m_{2}m_{3}\over (r_{1})^{5}}
-\alpha_{2}{m_{1}m_{3}\over (r_{2})^{5}}\right],
\end{equation}
\begin{equation}
C_{0}=-{\gamma_{1}(r_{1})\over r_{1}}
-{\gamma_{2}(r_{2})\over r_{2}},
\label{6.9f}
\end{equation}
\begin{equation}
C_{2}=3G \kappa \left[{m_{1}m_{3}\over (r_{2})^{5}}
+{m_{2}m_{3}\over (r_{1})^{5}}\right].
\end{equation}
At this stage, the coupled system (\ref{4.17f}) can be re-expressed in the form \cite{BE14b}
\begin{equation}
\begin{split}
& {\partial f \over \partial x_{1}}={1\over m}(A_{0}x_{1}-B_{0}x_{4})
+{1 \over m}(A_{2}x_{1}-B_{2}x_{4})\rho^{2},\\
& {\partial f \over \partial x_{2}}={1\over m}(A_{0}x_{2}-B_{0}x_{5})
+{1 \over m}(A_{2}x_{2}-B_{2}x_{5})\rho^{2},\\
& {\partial f \over \partial x_{3}}={1\over m}(A_{0}x_{3}-B_{0}x_{6})
+{1 \over m}(A_{2}x_{3}-B_{2}x_{6})\rho^{2},\\
& {\partial f \over \partial x_{4}}={1\over \mu}(C_{0}x_{4}-B_{0}x_{1})
+{1 \over \mu}(C_{2}x_{4}-B_{2}x_{1})\rho^{2},\\
& {\partial f \over \partial x_{5}}={1\over \mu}(C_{0}x_{5}-B_{0}x_{2})
+{1 \over \mu}(C_{2}x_{5}-B_{2}x_{2})\rho^{2},\\
& {\partial f \over \partial x_{6}}={1\over \mu}(C_{0}x_{6}-B_{0}x_{3})
+{1 \over \mu}(C_{2}x_{6}-B_{2}x_{3})\rho^{2},\\
\end{split}
\label{6.11f}
\end{equation}
where the left-hand sides can be further re-expressed upon writing
\begin{equation}
f(x_{1},\dots,x_{6})=f_{0}(x_{1},\dots,x_{6})+f_{2}(x_{1},\dots,x_{6})\rho^{2}.
\label{6.17f}
\end{equation}
Furthermore, from (\ref{4.16f}) and (\ref{6.17f}) we obtain straightforwardly
\begin{equation}
\begin{split}
& F_{0,x_{i}y_{k}}=f_{0,x_{i}y_{k}}=0, \\
& F_{0,y_{i}x_{k}}=(y_{i})_{,x_{k}}=0, \\
& F_{0,y_{i}y_{k}}=\delta_{ik}, \\
& F_{0,x_{i}x_{k}}=f_{0,x_{i}x_{k}}.
\end{split}
\label{6.18f}
\end{equation}
On the other hand, from Eqs. (\ref{6.11f}) and (\ref{6.17f}), we find immediately the $6 \times 6$ matrix of partial
derivatives 
\begin{equation}
M_{ik}^{0} \equiv f_{0,x_{i}x_{k}},
\label{6.19f}
\end{equation}
whose entries are given by \cite{BE14b}
\begin{equation}
M_{11}^{0}={1 \over m}(x_{1}A_{0,1}+A_{0}-x_{4}B_{0,1}),
\end{equation}
\begin{equation}
M_{12}^{0}={1 \over m}(x_{1}A_{0,2}-x_{4}B_{0,2}),
\end{equation}
\begin{equation}
M_{13}^{0}={1 \over m}(x_{1}A_{0,3}-x_{4}B_{0,3}),
\end{equation}
\begin{equation}
M_{14}^{0}={1 \over m}(x_{1}A_{0,4}-x_{4}B_{0,4}-B_{0}),
\end{equation}
\begin{equation}
M_{15}^{0}={1 \over m}(x_{1}A_{0,5}-x_{4}B_{0,5}),
\end{equation}
\begin{equation}
M_{16}^{0}={1 \over m}(x_{1}A_{0,6}-x_{4}B_{0,6}),
\end{equation}
\begin{equation}
M_{21}^{0}={1 \over m}(x_{2}A_{0,1}-x_{5}B_{0,1}),
\end{equation}
\begin{equation}
M_{22}^{0}={1 \over m}(x_{2}A_{0,2}+A_{0}-x_{5}B_{0,2}),
\end{equation}
\begin{equation}
M_{23}^{0}={1 \over m}(x_{2}A_{0,3}-x_{5}B_{0,3}),
\end{equation}
\begin{equation}
M_{24}^{0}={1 \over m}(x_{2}A_{0,4}-x_{5}B_{0,4}),
\end{equation}
\begin{equation}
M_{25}^{0}={1 \over m}(x_{2}A_{0,5}-x_{5}B_{0,5}-B_{0}),
\end{equation}
\begin{equation}
M_{26}^{0}={1 \over m}(x_{2}A_{0,6}-x_{5}B_{0,6}),
\end{equation}
\begin{equation}
M_{31}^{0}={1 \over m}(x_{3}A_{0,1}-x_{6}B_{0,1}),
\end{equation}
\begin{equation}
M_{32}^{0}={1 \over m}(x_{3}A_{0,2}-x_{6}B_{0,2}),
\end{equation}
\begin{equation}
M_{33}^{0}={1 \over m}(x_{3}A_{0,3}+A_{0}-x_{6}B_{0,3}),
\end{equation}
\begin{equation}
M_{34}^{0}={1 \over m}(x_{3}A_{0,4}-x_{6}B_{0,4}),
\end{equation}
\begin{equation}
M_{35}^{0}={1 \over m}(x_{3}A_{0,5}-x_{6}B_{0,5}),
\end{equation}
\begin{equation}
M_{36}^{0}={1 \over m}(x_{3}A_{0,6}-x_{6}B_{0,6}-B_{0}),
\end{equation}
\begin{equation}
M_{41}^{0}={1 \over \mu}(x_{4}C_{0,1}-x_{1}B_{0,1}-B_{0}),
\end{equation}
\begin{equation}
M_{42}^{0}={1 \over \mu}(x_{4}C_{0,2}-x_{1}B_{0,2}),
\end{equation}
\begin{equation}
M_{43}^{0}={1 \over \mu}(x_{4}C_{0,3}-x_{1}B_{0,3}),
\end{equation}
\begin{equation}
M_{44}^{0}={1 \over \mu}(x_{4}C_{0,4}+C_{0}-x_{1}B_{0,4}),
\end{equation}
\begin{equation}
M_{45}^{0}={1 \over \mu}(x_{4}C_{0,5}-x_{1}B_{0,5}),
\end{equation}
\begin{equation}
M_{46}^{0}={1 \over \mu}(x_{4}C_{0,6}-x_{1}B_{0,6}),
\end{equation}
\begin{equation}
M_{51}^{0}={1 \over \mu}(x_{5}C_{0,1}-x_{2}B_{0,1}),
\end{equation}
\begin{equation}
M_{52}^{0}={1 \over \mu}(x_{5}C_{0,2}-x_{2}B_{0,2}-B_{0}),
\end{equation}
\begin{equation}
M_{53}^{0}={1 \over \mu}(x_{5}C_{0,3}-x_{2}B_{0,3}),
\end{equation}
\begin{equation}
M_{54}^{0}={1 \over \mu}(x_{5}C_{0,4}-x_{2}B_{0,4}),
\end{equation}
\begin{equation}
M_{55}^{0}={1 \over \mu}(x_{5}C_{0,5}+C_{0}-x_{2}B_{0,5}),
\end{equation}
\begin{equation}
M_{56}^{0}={1 \over \mu}(x_{5}C_{0,6}-x_{2}B_{0,6}),
\end{equation}
\begin{equation}
M_{61}^{0}={1 \over \mu}(x_{6}C_{0,1}-x_{3}B_{0,1}),
\end{equation}
\begin{equation}
M_{62}^{0}={1 \over \mu}(x_{6}C_{0,2}-x_{3}B_{0,2}),
\end{equation}
\begin{equation}
M_{63}^{0}={1 \over \mu}(x_{6}C_{0,3}-x_{3}B_{0,3}-B_{0}),
\end{equation}
\begin{equation}
M_{64}^{0}={1 \over \mu}(x_{6}C_{0,4}-x_{3}B_{0,4}),
\end{equation}
\begin{equation}
M_{65}^{0}={1 \over \mu}(x_{6}C_{0,5}-x_{3}B_{0,5}),
\end{equation}
\begin{equation}
M_{66}^{0}={1 \over \mu}(x_{6}C_{0,6}+C_{0}-x_{3}B_{0,6}),
\end{equation}
where, in order to ease the notation, we have adopted a convention for which a subscript like ${}_{,k}$ denotes partial derivative with respect to $x_{k}$, for all $k=1,\dots,6$. Now, a patient application of (\ref{4.16f}), (\ref{6.17f}), and (\ref{6.18f}) to the Eqs. (\ref{5.12f})--(\ref{5.17f}) reveals that, for all $i=1,\dots,6$ (exploiting the vanishing of $F_{1}$ in our model) \cite{BE14b}
\begin{equation}
\sum_{k=1}^{6}
\left(\begin{matrix}
\delta_{ik} {{\rm d}\over {\rm d}t} & -\delta_{ik} \cr
M_{ik}^{0} & \delta_{ik} {{\rm d}\over {\rm d}t}
\end{matrix}\right)
\left(\begin{matrix}
S_{k}^{0} \cr
T_{k}^{0}
\end{matrix}\right)=0,
\label{6.20f}
\end{equation}
while, for higher-order terms, we find the inhomogeneous equations
\begin{equation}
\sum_{k=1}^{6}
\left(\begin{matrix}
\delta_{ik}{{\rm d}\over {\rm d}t} & -\delta_{ik} \cr
M_{ik}^{0} & \delta_{ik} {{\rm d}\over {\rm d}t}
\end{matrix}\right)
\left(\begin{matrix}
S_{k}^{n} \cr
T_{k}^{n}
\end{matrix}\right)
=-\sum_{l=0}^{n-1}\alpha_{n-l}
\left(\begin{matrix}
S_{i}^{l} \cr
T_{i}^{l}
\end{matrix}\right).
\label{6.21f}
\end{equation}
The above relations represent a recursive algorithm for the solution of the quantum corrected variational equations (\ref{5.3f}) and (\ref{5.4f}) involving a repeated application of a $2 \times 2$ matrix of linear first-order differential operators \cite{BE14b}. For example, for the equations where $\alpha_{1}$ and $\alpha_{2}$ occur we find
\begin{equation}
\sum_{k=1}^{6}
\left(\begin{matrix}
\delta_{ik}{{\rm d}\over {\rm d}t} & -\delta_{ik} \cr
M_{ik}^{0} & \delta_{ik} {{\rm d}\over {\rm d}t}
\end{matrix}\right)
\left(\begin{matrix}
S_{k}^{1} \cr
T_{k}^{1}
\end{matrix}\right)
=-\alpha_{1}
\left(\begin{matrix}
S_{i}^{0} \cr
T_{i}^{0}
\end{matrix}\right),
\end{equation}
\begin{equation}
\sum_{k=1}^{6}
\left(\begin{matrix}
\delta_{ik}{{\rm d}\over {\rm d}t} & -\delta_{ik} \cr
M_{ik}^{0} & \delta_{ik} {{\rm d}\over {\rm d}t}
\end{matrix}\right)
\left(\begin{matrix}
S_{k}^{2} \cr
T_{k}^{2}
\end{matrix}\right)
=-\alpha_{2}
\left(\begin{matrix}
S_{i}^{0} \cr
T_{i}^{0}
\end{matrix}\right)
-\alpha_{1}
\left(\begin{matrix}
S_{i}^{1} \cr
T_{i}^{1}
\end{matrix}\right).
\end{equation}

Of course, it is at least equally important to study the case when the characteristic exponents do not vanish at $\rho=0$ \cite{P1890,P1892}. In such a case, we assume that the asymptotic expansion (\ref{5.6f}) can be generalized by adding the term $\alpha_{0}$, i.e., \cite{BE14b}
\begin{equation}
\alpha \sim \sum_{l=0}^{N} \alpha_{l}\rho^{l \over 2}.
\end{equation}
Thus, the scheme described before leads eventually to equations that generalize (\ref{6.20f}) and (\ref{6.21f}) upon adding $\alpha_{0}$ to the linear differential operator ${{\rm d}\over {\rm d}t}$, i.e., \cite{BE14b}
\begin{equation}
\sum_{k=1}^{6}
\left(\begin{matrix}
\delta_{ik} \left({{\rm d}\over {\rm d}t}+\alpha_{0}\right) & -\delta_{ik} \cr
M_{ik}^{0} & \delta_{ik} \left({{\rm d}\over {\rm d}t}+\alpha_{0}\right)
\end{matrix}\right)
\left(\begin{matrix}
S_{k}^{0} \cr
T_{k}^{0}
\end{matrix}\right)=0,
\label{6.25f}
\end{equation}
\begin{equation}
\sum_{k=1}^{6}
\left(\begin{matrix}
\delta_{ik} \left({{\rm d}\over {\rm d}t}+\alpha_{0}\right) & -\delta_{ik} \cr
M_{ik}^{0} & \delta_{ik} \left({{\rm d}\over {\rm d}t}+\alpha_{0}\right)
\end{matrix}\right)
\left(\begin{matrix}
S_{k}^{n} \cr
T_{k}^{n}
\end{matrix}\right)
=-\sum_{l=0}^{n-1}\alpha_{n-l}
\left(\begin{matrix}
S_{i}^{l} \cr
T_{i}^{l}
\end{matrix}\right).
\label{6.26f}
\end{equation}

Finally, we see that our computational recipes are of little help unless we say what sort of periodic solutions we have in mind. As it should be clear from Sec. \ref{Neighbourhood_Sec}, the periodic solutions alluded to in Eq. (\ref{5.1f}) are solutions of Eqs. (\ref{4.15f}) when $\rho=0$. Thus, with the notation in Eqs. (\ref{6.1f})--(\ref{6.3f}), (\ref{6.5f}), (\ref{6.7f}), and (\ref{6.9f}), the matrix (\ref{6.19f}) should be therefore evaluated along solutions of the coupled equations
\begin{equation}
{{\rm d}x_{i}\over {\rm d}t}=y_{i} \; \; \; \; \;\forall i=1,\dots,6 ,
\label{6.27f}
\end{equation}
\begin{equation}
{{\rm d}y_{i}\over {\rm d}t}=-{1 \over m}(A_{0}x_{i}-B_{0}x_{i+3}) \; \; \;\; \; \forall i=1,2,3 ,
\end{equation}
\begin{equation}
{{\rm d}y_{i}\over {\rm d}t}=-{1 \over \mu}(C_{0}x_{i}-B_{0}x_{i-3}) \; \; \; \; \;\forall i=4,5,6 .
\label{6.29f}
\end{equation}
The desired periodic solutions, whose existence is a special rather than generic property (see the remarks at the end of Sec. \ref{periodic_coeff_sec}), can be written in the form \cite{BE14b}
\begin{equation}
x_{i}=\sum_{l=0}^{\infty}D_{il} \sin(\omega_{il}t+\varphi_{il}),
\end{equation}
\begin{equation}
y_{i}=\sum_{l=0}^{\infty}E_{il} \sin(\omega_{il}t+\gamma_{il}).
\label{6.31f}
\end{equation}
When we insert such Fourier expansions into the system (\ref{6.27f})--(\ref{6.29f}), we have to bear in mind that $A_{0}$, $B_{0}$, and $C_{0}$ in (\ref{6.5f}), (\ref{6.7f}), (\ref{6.9f}) depend on $x_{1},\dots,x_{6}$ because Eqs. (\ref{2.7f})--(\ref{2.9f}) can be re-expressed in the form
\begin{equation}
\begin{split}
& (r_{1})^{2}=\sum_{k=1}^{3} (\alpha_{1}x_{k}-x_{k+3})^{2}, \\
& (r_{2})^{2}=\sum_{k=1}^{3} (\alpha_{2}x_{k}+x_{k+3})^{2}, \\
& (r_{3})^{2}=\sum_{k=1}^{3}(x_{k})^{2}.
\end{split}
\end{equation}

\subsection{A scheme for the resolution of variational equations}

In the previous section we have arrived at a broad framework for the resolution of the quantum corrected variational equations (\ref{5.3f}) and (\ref{5.4f}) that presents formidable technical difficulties, which have prevented us from showing a solution of such equations. For this purpose, one should solve completely the following
problems \cite{BE14b}:
\vskip 0.3cm
\noindent
(i) First, how to find periodic solutions of the Hamiltonian equations (\ref{4.15f}) when $\rho=0$. From Eqs. (\ref{6.27f})-(\ref{6.31f}), this means having to solve the infinite system of equations
\begin{equation}
\sum_{l=0}^{\infty}D_{il}\omega_{il}\cos(\omega_{il}t+\varphi_{il})
=\sum_{l=0}^{\infty}E_{il}\sin(\omega_{il}t+\gamma_{il}), \; \; \;\; \;  \forall i=1,\dots,6,
\label{7.1f}
\end{equation}
\begin{equation}
\begin{split}
\sum_{l=0}^{\infty}E_{il}\omega_{il}\cos(\omega_{il}t+\gamma_{il}) & = -{A_{0}\over m}\sum_{l=0}^{\infty}D_{il}\sin(\omega_{il}t+ \gamma_{il})  \\
&+  {B_{0}\over m}\sum_{l=0}^{\infty}D_{i+3,l}\sin(\omega_{i+3,l}t+ \gamma_{i+3,l}), \; \; \;\; \;  \forall i=1,2,3,
\end{split}
\end{equation}
\begin{equation}
\begin{split}
\sum_{l=0}^{\infty}E_{il}\omega_{il}\cos(\omega_{il}t+\gamma_{il}) &= -{C_{0}\over \mu}\sum_{l=0}^{\infty}D_{il}\sin(\omega_{il}t+ \gamma_{il}) \\
&+  {B_{0}\over \mu}\sum_{l=0}^{\infty}D_{i-3,l}\sin(\omega_{i-3,l}t+ \gamma_{i-3,l}), \; \; \;\; \; \forall i=4,5,6.
\end{split}
\label{7.3f}
\end{equation}
\vskip 0.3cm
\noindent
(ii) Second, how to solve variational equations through Eqs. (\ref{6.20f}) and (\ref{6.21f}), or (\ref{6.25f}) and (\ref{6.26f}), when the matrix $M_{ik}^{0}$ is evaluated along a solution of Eqs. (\ref{7.1f})--(\ref{7.3f}). In Refs. \cite{P1890,P1892}, Poincar\'e obtained an algebraic equation of third degree for the square of $\alpha_{1}$, which was the hardest part of the calculation, but we do not see an analogous equation for the square of $\alpha_{1}$ in our quantum corrected model.
\vskip 0.3cm
\noindent
(iii) Third, what is the counterpart, if any, of the variety of periodic and asymptotic solutions found by Poincar\'e \cite{P1890,P1892}, i.e., more precisely:
\begin{itemize}
\item[(a)] Periodic solutions of the Hamiltonian equations (\ref{4.15f}) with non-vanishing values of $\rho$, e.g.,
\begin{equation}
x_{l}(t)=\phi_{l}^{0}(t)+(\rho-\rho_{0})^{1 \over 2}\phi_{l}^{(1)}(t)
+(\rho-\rho_{0})\phi_{l}^{(2)}(t)
+(\rho-\rho_{0})^{3 \over 2}\phi_{l}^{(3)}(t)+\dots,
\end{equation}
where $\phi_{l}^{0}(t)$ has period $T$, while $\phi_{l}^{(1)}(t),\phi_{l}^{(2)}(t),\phi_{l}^{(3)}(t)$
have period equal to an integer multiple of $T$.
\item[(b)] Asymptotic solutions of Eqs. (\ref{4.15f}) of the first kind, for which
\begin{equation}
x_{i}(t)=\varphi_{i}(t)+A{\rm e}^{-\alpha t}\theta_{i}^{(1)}(t)
+A^{2}{\rm e}^{-2 \alpha t}\theta_{i}^{(2)}(t)
+A^{3}{\rm e}^{-3 \alpha t}\theta_{i}^{(3)}(t)+\dots,
\label{7.5f}
\end{equation}
where $\varphi_{i}(t)$ is an unstable periodic solution, $A$ is an arbitrary integration constant,
$\alpha$ is a positive characteristic exponent, $\theta_{i}^{(1)}(t),\theta_{i}^{(2)}(t), \dots$ have
period $T$. At sufficiently large positive values of $t$ such series are convergent. As
$t \rightarrow \infty$, such solutions approach asymptotically the unstable periodic solution
$\varphi_{i}(t)$.
\item[(c)] Asymptotic solutions of Eqs. (\ref{4.15f}) of the second kind, for which
\begin{equation}
x_{i}(t)=\varphi_{i}(t)+B{\rm e}^{\alpha t}\omega_{i}^{(1)}(t)
+B^{2}{\rm e}^{2 \alpha t}\omega_{i}^{(2)}(t)
+B^{3}{\rm e}^{3 \alpha t}\omega_{i}^{(3)}(t)+\dots,
\label{7.6f}
\end{equation}
where $B$ is a new integration constant,
$\alpha$ is again the positive characteristic exponent, and the functions $\omega$ are of
the same functional form as the functions $\theta$ occurring in (\ref{7.5f}).  
At sufficiently large negative values of $t$ such series are convergent. As
$t \rightarrow - \infty$, such solutions approach asymptotically the unstable periodic solution
$\varphi_{i}(t)$.
\item[(d)] Doubly asymptotic (or homoclinic) solutions which are represented by (\ref{7.6f}) if $t<0$ and $\vert t \vert$ is very large,
and by (\ref{7.5f}) if $t>0$ and $\vert t \vert$ is very large. The corresponding (chaotic) orbit, which initially differs
slightly from the unstable periodic solution, departs gradually from it at first, and after having
departed significantly from it ends up by approaching asymptotically the unstable periodic solution.
At finite values of $t$, there exist intervals of this time variable where neither (\ref{7.5f}) nor (\ref{7.6f})
converges in Newtonian physics \cite{P1890,P1892}.
\end{itemize}

At this point, it should be clear why the resolution of the quantum variational equations (\ref{5.3f}) and (\ref{5.4f}) represents a really demanding task.

\section{Restricted four-body problem}\label{Restricted4b_Sec}

The last step towards a more realistic model concerning the quantum description of the Earth-Moon system is represented by the characterization of the restricted four-body problem, which involves also the perturbations due to the gravitational presence of the Sun. In other words, we have to face up a system consisting of the Sun, the Earth, and the Moon as the three primaries and a a spacecraft, a solar sail, or a particular satellite aimed at experimental measurements and called laser-ranged test mass (see Sec. \ref{Laser_Ranging_Sec}) as the planetoid. 

We have seen that in the restricted three-body problem the motion of the two primaries is exactly described by the equations of motion governing the two-body problem, because it is assumed that the planetoid has an infinitesimal mass and hence can not affect the motion of the other two bodies. Therefore, we may generalize this problem first by solving the dynamical equations describing the motion of the three primaries and then by finding the motion of the planetoid in the presumably known gravitational field produced by the them. Since Poincar\'e has demonstrated that no closed-form solution is known for the full three-body problem, this generalization to the case of four masses is rather difficult. A practicable possibility consists in assuming the motions of the three primaries and, without attempting to establish the exact solution of the equations governing these motions, accept an approximate solution. Such an approximation may be, for instance, that the Earth and the Moon move in elliptic orbits around their mass center and that the mass center of the Earth-Moon system, in turn, moves in elliptic orbit around the Sun. The plane of the orbit of the mass center of the Earth-Moon system, which is called the plane of ecliptic, is inclined relative to the plane containing the orbits of the Earth and the Moon. A simpler approximation would consist in neglecting the eccentricity of all orbits, i.e., assuming that the Earth, the Moon and their mass center have circular orbits. 

The first who dealt with the restricted problem of four bodies (within Newtonian theory) by employing the circular orbits hypothesis were the authors of Ref. \cite{Tapley}. In fact, although it is widely accepted that, with the introduction of the Sun, the points $L_4$ and $L_5$ of the Earth-Moon system cease to be 
equilibrium points, in Ref. \cite{Tapley} it is showed that stable motion may be possible in a region around these non-collinear libration points, provided that we change the meaning of the word ``stable''. In this context in fact the term ``stable'' indicates that the planetoid will remain within a certain region only for the period of time during which the motion is studied. We will see that this feature holds also in the context of effective field theories of gravity \cite{testbed}. 

\subsection{Equations of motion}

\begin{figure} [htbp] 
\centering
\includegraphics[scale=0.7]{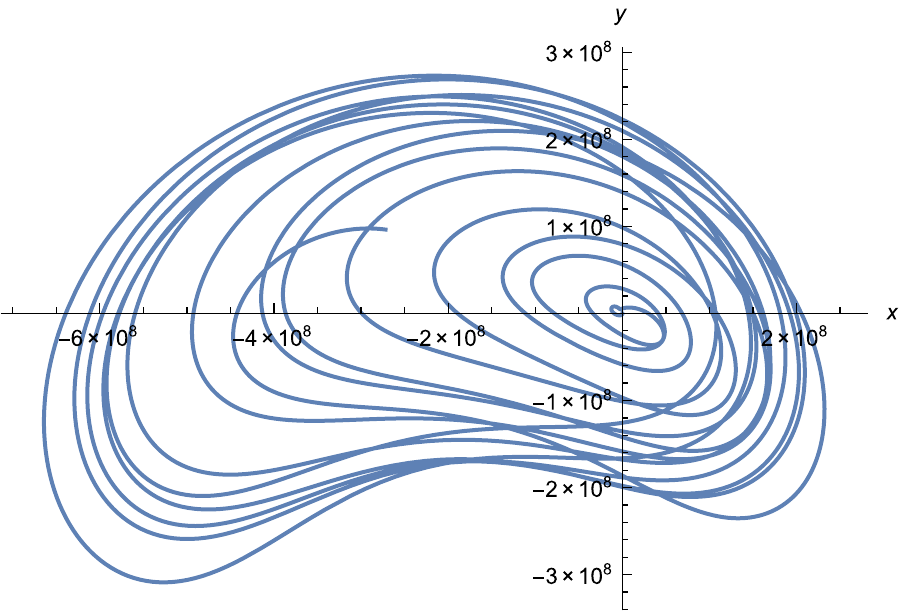}
\caption[Parametric plot of the spacecraft motion about $L_4$ in the classical restricted four-body problem]{Parametric plot of the spacecraft motion about $L_4$ resulting from zero initial displacement and velocity in the 
classical case. The quantities appearing on the axes are measured in meters and the time interval considered is about $4 \times 10^7 \; {\rm s}$.}
\label{1c.pdf}
\end{figure}

\begin{figure} [htbp] 
\centering
\includegraphics[scale=0.7]{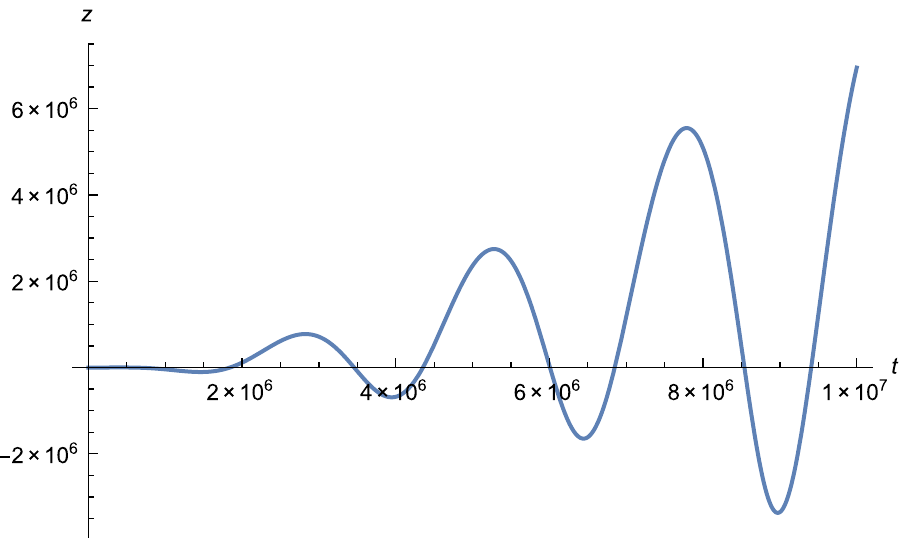}
\caption[Parametric plot of the spacecraft motion about $L_4$ in the $z$-direction in the classical restricted four-body problem]{Plot of the spacecraft motion about $L_4$ in the $z$-direction resulting from zero initial displacement and 
velocity in the classical case. The quantities on the axes are measured in meters and in seconds.}
\label{2c.pdf}
\end{figure}

We start by introducing the classical dynamical equations governing the motion of the planetoid in the gravitational field of the Earth, the Moon, and the Sun \cite{testbed,Szebehely67,Tapley}. As pointed out before, we suppose that the Earth and the Moon move in circular orbits around their mass center and  the mass center, in turn, moves in circular orbit about the Sun. The Earth-Moon orbit plane is inclined at an angle $i=5^{\degree} 9^{\prime}$ to the plane of the ecliptic. We introduce the rotating coordinate system $\xi,\eta,\zeta $ with the Earth-Moon mass center as its origin and characterized by the fact that the $\xi$ axis lies along 
the Earth-Moon line, the $\eta$-axis lies in the Earth-Moon orbit plane and the $\zeta$-axis points in the direction of the angular velocity vector of the Earth-Moon configuration. The $\xi, \eta$-axes rotate about the $\zeta$-axis with the angular velocity $\omega$ of the Earth-Moon line. If the vector $\vec{\mathcal{R}}=(\xi,\eta,\zeta)$ indicates in this coordinate system the position of a spacecraft of infinitesimal mass, the vector dynamical equation describing its motion is \cite{testbed}
\begin{equation}
\ddot{\vec{\mathcal{R}}}   + \vec{\omega} \times (2\; \dot{\vec{\mathcal{R}}} + \dot{\vec{\omega}} 
\times \vec{\mathcal{R}}) = -\vec{\nabla}_{\mathcal{R}}V+\vec{\nabla}_{\mathcal{R}}U+ \vec{S}, 
\label{vecc}
\end{equation}
where
\begin{equation}
\begin{split}
&V \equiv \dfrac{Gm_1}{\rho_1}+\dfrac{Gm_2}{\rho_2}, \\
&U \equiv Gm_3 \left[\dfrac{1}{\rho_3}-\dfrac{\vec{\mathcal{R}} \cdot 
\vec{\mathcal{R}_3}}{(\mathcal{R}_{3})^{3}} \right],
\end{split}
\end{equation}
with $G$ being as usual the universal gravitation constant; $m_1$, $m_2$, and $m_3$ the mass of the Earth, the Moon, and the Sun, respectively; $\rho_1$, $\rho_2$, and $\rho_3$ the distances from the planetoid of the Earth, the Moon, and the Sun, respectively; $\mathcal{R}_3$ the distance of the Sun from the Earth-Moon mass center; lastly, $\vec{S}$ describes the solar radiation pressure. Written in components, Eq. (\ref{vecc}) becomes
\begin{equation}
\ddot{\xi}-2\omega \dot{\eta}-\omega^2 \xi = -\dfrac{\partial V}{\partial \xi}
+\dfrac{\partial U}{\partial \xi}+S_{\xi}, 
\label{1ac}
\end{equation}
\begin{equation}
\ddot{\eta}+2 \omega \dot{\xi}-\omega^2 \eta =  -\dfrac{\partial V}{\partial \eta}
+\dfrac{\partial U}{\partial \eta}+S_{\eta},  
\end{equation}
\begin{equation}
\ddot{\zeta} =  -\dfrac{\partial V}{\partial \zeta}+\dfrac{\partial U}{\partial \zeta}+S_{\zeta}. 
\label{1cc}
\end{equation}
We can write Eqs. (\ref{1ac})--(\ref{1cc}) in what we denote by $x,y,z$ system, which is the rotating non-inertial coordinate frame of reference centred at one of the two non-collinear Lagrangian points, e.g., $L_4$. If we use the transformations
\begin{equation}
\begin{split}
& \xi =x+\xi_p,  \\
& \eta =y+\eta_p,    \\
& \zeta = z, 
\end{split}
\label{2.7c}
\end{equation}
where $\xi_p$ and $\eta_p$ are the constant coordinates of the libration point $L_4$ in the $\xi,\eta,\zeta$ system, then Eqs. (\ref{1ac})--(\ref{1cc}) become \cite{testbed}
\begin{equation}
\ddot{x} =2 \omega \dot{y} + (x+\xi_p)\omega^2-(x_3+\xi_p)(\Omega_{\omega})^2 +
S_x + \sum_{i=1}^3 \dfrac{Gm_i}{\rho_{i}^{3}}(x_i-x), 
\label{4ac}
\end{equation}
\begin{equation}
\ddot{y}=-2 \omega \dot{x} + (y+\eta_p)\omega^2-(y_3+\eta_p)(\Omega_{\omega})^2 +
S_y + \sum_{i=1}^3 \dfrac{Gm_i}{\rho_{i}^{3}}(y_i-y), 
\end{equation}
\begin{equation}
\ddot{z}=-z_3(\Omega_{\omega})^2 + S_z + \sum_{i=1}^3 \dfrac{Gm_i}{\rho_{i}^{3}}(z_i-z), 
\label{4cc}
\end{equation}
where $\Omega_{\omega}$ is the angular velocity of the Earth-Moon mass center around the Sun, and the relation 
$Gm_3/(\mathcal{R}_3)^3=(\Omega_{\omega})^2$ has been exploited. Moreover, the distances $\rho_i$ are given by
\begin{equation}
(\rho_{i})^2=(x_i-x)^2+(y_i-y)^2+(z_i-z)^2, \; \;  \; \; \; \; \; \; \; (i=1,2,3),
\end{equation}
where the coordinates $(x_1,y_1)$ and $(x_2,y_2)$ of the Earth and the Moon respectively are deduced from 
(\ref{2.7c}) once the coordinates $(\xi_p,\eta_p)$ of $L_4$ are known (remember we have $z_1=z_2=0$), whereas the 
coordinates of the Sun are given by the relations \cite{testbed,Tapley}
\begin{equation}
\begin{split}
& x_3 = \mathcal{R}_3 \left( \cos \psi \cos \theta + \cos i \sin \psi \sin \theta \right) -\xi_p, \\
& y_3 = -\mathcal{R}_3 \left( \cos \psi \sin \theta - \cos i \sin \psi \cos \theta \right) -\eta_p, \\
& z_3 = \mathcal{R}_3 \sin \psi \sin i, 
\end{split}
\end{equation}
where $\psi$ is the angular position of the Sun with respect to the vernal equinox and measured in the plane of the ecliptic, and $\theta$ describes the position of the Earth-Moon line with respect to the vernal equinox measured in the Earth-Moon orbit plane. The relations defining these angles are
\begin{equation}
\begin{split}
& \psi = \Omega_{\omega} t + \psi_0,  \\
& \theta = \Omega_{\omega} t +\theta_0,
\end{split}
\end{equation}
where $\psi_0$ and $\theta_0$ are the initial values of $\psi$ and $\theta$, respectively. For our computation we have used the following 
numerical values \cite{testbed}: 
\begin{equation}
\begin{split}
& \Omega_{\omega}=1.99082 \times 10^{-7} \; {\rm rad/s}, \\
& \omega=2.665075637 \times 10^{-6} \; {\rm rad/s}, \\
& \psi_0 = \theta_0 = 0.
\end{split}
\end{equation}
In particular, the last condition implies that the initial position of the Sun will be on the extended Earth-Moon line, with the Moon in between Earth and Sun. Furthermore, the classical values of $\xi_p$ and $\eta_p$ are given by Eq. (\ref{5.20a_ter}). If we set $\vec{S}= \vec{0}$ from the very beginning in Eq. (\ref{vecc}), we obtain that the spacecraft proceeds on a trajectory around $L_4$ for at least $700$ days before the solar influence causes it to move through wide departure from the Lagrangian point, as is shown in Figs. \ref{1c.pdf} and  \ref{2c.pdf}, obtained after having integrated Eqs. (\ref{4ac})--(\ref{4cc}). 
As can be noticed from Fig. \ref{1c.pdf}, the irregular initial motion damps out and there is an approximate one-month 
periodicity associated with the motion. Moreover, Fig. \ref{2c.pdf} shows that the amplitude of the motion increases 
with time and that the period of motion is about $27.6$ days, a value really near to the $29.53$ days of the synodical month. Furthermore, from the analysis of the plots it does not appear that, after $700$ days, a limiting value for the envelope is approached. All these results indicate that the spacecraft will ultimately escape from the equilibrium point $L_4$ (or equivalently $L_5$) or, in other words, the perturbing presence of the Sun makes the points $L_4$ and $L_5$ cease to be equilibrium points, but they are ``stable'' in the sense indicated before \cite{testbed,Tapley}.

All these considerations are valid within the classical scheme, whereas in the quantum corrected regime we have learned that the Newtonian potential is corrected by a Poincar\'e asymptotic expansion involving integer powers of $G$ only, so that Eq. (\ref{vecc}) can be replaced by the vector dynamical equation \cite{testbed}
\begin{equation}
\ddot{\vec{\mathcal{R}}}   + \vec{\omega} \times (2\; \dot{\vec{\mathcal{R}}} + \dot{\vec{\omega}} 
\times \vec{\mathcal{R}}) = -\vec{\nabla}_{\mathcal{R}}V_{q}+\vec{\nabla}_{\mathcal{R}}U_{q}+ \vec{S}, 
\label{vecq}
\end{equation}
with \cite{testbed}
\begin{equation}
V_{q}=\dfrac{Gm_1}{\rho_1} \left[ 1+\dfrac{k_1}{\rho_1}+\dfrac{k_2}{(\rho_1)^2} \right] 
+\dfrac{Gm_2}{\rho_2}\left[ 1+\dfrac{k^{\prime}_{1}}{\rho_2}+\dfrac{k_2}{(\rho_2)^2} \right], 
\end{equation}
\begin{equation}
U_{q}=\dfrac{Gm_3}{\rho_3} \left[ 1+\dfrac{k^{\prime \prime}_{1}}{\rho_3}+\dfrac{k_2}{(\rho_3)^2} \right]
-G m_3 \dfrac{\vec{\mathcal{R}} \cdot \vec{\mathcal{R}_3}}{(\mathcal{R}_{3})^{3}} 
\left[1+\dfrac{2 k^{\prime \prime}_{1}}{\mathcal{R}_{3}} +\dfrac{3 k_2}{(\mathcal{R}_{3})^2}\right],
\end{equation}
and (cf. Tab. \ref{kappa_tab})
\begin{equation}
\begin{split}
& k_1=\kappa_1 \dfrac{ G m_1}{c^2}, \\
& k^{\prime}_{1}=\kappa_1 \dfrac{ G m_2}{c^2}, \\
& k^{\prime \prime}_{1}=\kappa_1\dfrac{ G m_3}{c^2}, \\
& k_2=\kappa_2(l_{P})^{2}.
\end{split}
\end{equation}

\begin{figure}  
\centering
\includegraphics[scale=0.7]{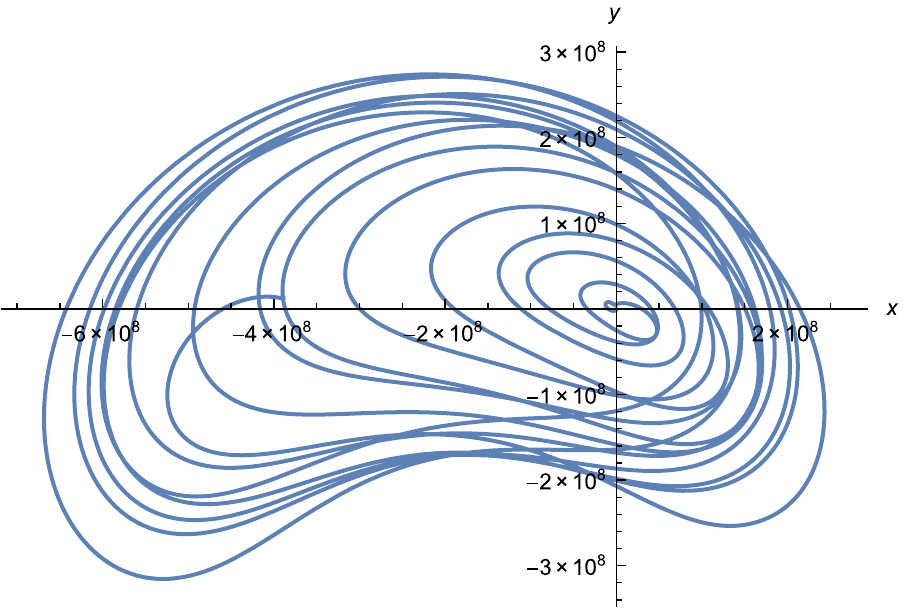}
\caption[Parametric plot of the spacecraft motion about $L_4$ in the quantum corrected restricted four-body problem]{Parametric plot of the spacecraft motion about $L_4$ resulting from zero initial displacement and velocity in the quantum case. The quantities appearing on the axes are measured in meters and the time interval considered is about $4 \times 10^7 \;  {\rm s}$.}
\label{3c.pdf}
\end{figure}

\begin{figure}  
\centering
\includegraphics[scale=0.7]{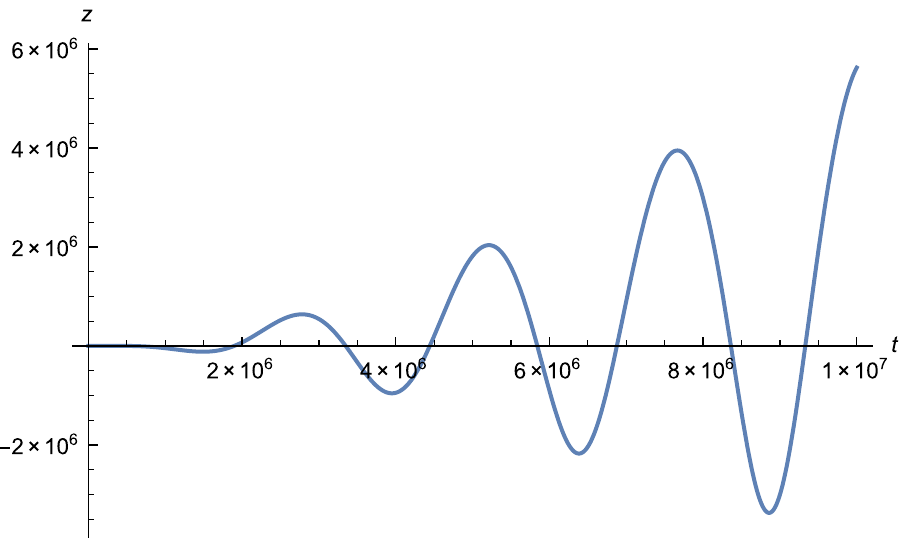}
\caption[Parametric plot of the spacecraft motion about $L_4$ in the $z$-direction in the quantum corrected restricted four-body problem]{Plot of the spacecraft motion about $L_4$ in the $z$-direction resulting from zero initial displacement and velocity in the quantum case. The quantities on the axes are measured in meters and in seconds.}
\label{4c.pdf}
\end{figure}

In the $x,y,z$ system, instead of Eqs. (\ref{4ac})--(\ref{4cc}), 
Eq. (\ref{vecq}), written in components, gives rise to the system \cite{testbed}
\begin{equation}
\begin{split}
\ddot{x} &= \; 2 \omega \dot{y} + (x+\xi_p)\omega^2-(x_3+\xi_p)(\Omega_{\omega})^2 
\left[1+\dfrac{2 k^{\prime \prime}_{1}}
{\mathcal{R}_{3}} +\dfrac{3 k_2}{(\mathcal{R}_{3})^2}\right]  +\dfrac{G m_1 (x_1-x)}{(\rho_1)^3} 
\left[ 1+\dfrac{2 k_1}{\rho_1}+\dfrac{3 k_2}{(\rho_1)^2} \right]  \\
& +  \dfrac{G m_2 (x_2-x)}{(\rho_2)^3} \left[ 1+\dfrac{2 k^{\prime}_{1}}{\rho_2}+\dfrac{3 k_2}{(\rho_2)^2} \right]  
+ \dfrac{G m_3 (x_3-x)}{(\rho_3)^3} \left[1+\dfrac{2 k^{\prime \prime}_{1}}{\rho_{3}} 
+\dfrac{3 k_2}{(\rho_{3})^2}\right]+ S_x, 
\end{split}
\label{4aq}
\end{equation}
\begin{equation}
\begin{split}
\ddot{y} &= -2 \omega \dot{x} + (y+\eta_p)\omega^2-(y_3+\eta_p)(\Omega_{\omega})^2 
\left[1+\dfrac{2 k^{\prime \prime}_{1}}{\mathcal{R}_{3}} +\dfrac{3 k_2}{(\mathcal{R}_{3})^2}\right]  
+\dfrac{G m_1 (y_1-y)}{(\rho_1)^3} \left[ 1+\dfrac{2 k_1}{\rho_1}+\dfrac{3 k_2}{(\rho_1)^2} \right]  \\
&+ \dfrac{G m_2 (y_2-y)}{(\rho_2)^3} \left[ 1+\dfrac{2 k^{\prime}_{1}}{\rho_2}+\dfrac{3 k_2}{(\rho_2)^2} \right]  
+ \dfrac{G m_3 (y_3-y)}{(\rho_3)^3} \left[1+\dfrac{2 k^{\prime \prime}_{1}}{\rho_{3}} 
+\dfrac{3 k_2}{(\rho_{3})^2}\right]+ S_y , 
\end{split}
\end{equation}
\begin{equation}
\begin{split}
\ddot{z} &= -z_3 (\Omega_{\omega})^2 \left[1+\dfrac{2 k^{\prime \prime}_{1}}{\mathcal{R}_{3}} 
+\dfrac{3 k_2}{(\mathcal{R}_{3})^2}\right]  -\dfrac{G m_1 z}{(\rho_1)^3} \left[ 1+\dfrac{2 k_1}{\rho_1}
+\dfrac{3 k_2}{(\rho_1)^2} \right]  - \dfrac{G m_2 z}{(\rho_2)^3} \left[ 1+\dfrac{2 k^{\prime}_{1}}{\rho_2}
+\dfrac{3 k_2}{(\rho_2)^2} \right]  \\
&+ \dfrac{G m_3 (z_3-z)}{(\rho_3)^3} \left[1+\dfrac{2 k^{\prime \prime}_{1}}{\rho_{3}} 
+\dfrac{3 k_2}{(\rho_{3})^2}\right]+ S_z, 
\label{4cq} 
\end{split}
\end{equation}
where we have used the fact that $z_1=z_2=0$. Setting $\vec{S}=\vec{0}$, we have integrated Eqs. (\ref{4aq})--(\ref{4cq}) 
and we have discovered that the situation is almost the same as in the classical case (see Figs. \ref{3c.pdf} and \ref{4c.pdf}), 
i.e., the planetoid is destined to run away from the triangular libration points in about $700$ days. This means that, 
also within the quantum corrected scheme, the gravitational effect of the Sun spoils the equilibrium condition at $L_4$ and $L_5$.  

\subsection{The solar radiation pressure and the linear stability at $L_4$}

\begin{figure}  
\centering
\includegraphics[scale=0.7]{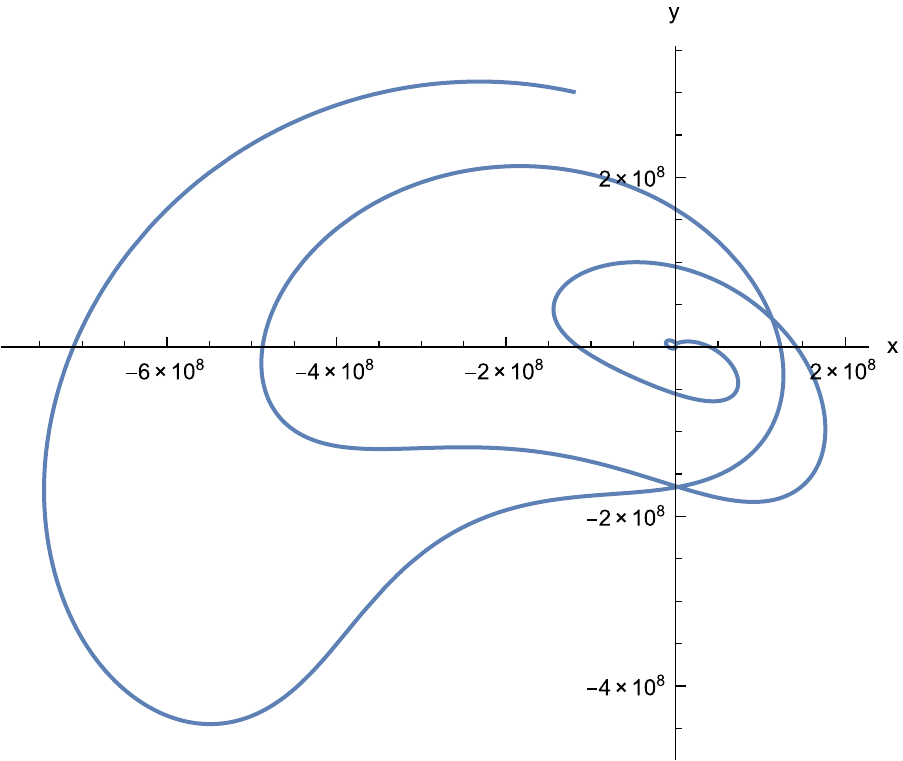}
\caption[Parametric plot in the classical restricted four-body problem of the spacecraft motion about $L_4$ in the presence of the solar radiation 
pressure]{Parametric plot in the classical regime of the spacecraft motion about $L_4$ in the presence of the solar radiation 
pressure and considering $A/m=0,159\; {\rm m^2/Kg}$. The initial displacement and velocity are zero. The quantities appearing 
on the axes are measured in meters and the time interval considered is about $1 \times 10^7 \; {\rm s}$.}
\label{5c.pdf}
\end{figure}

At this stage, we assume the presence of the radiation pressure both in the classical equations (\ref{4ac})--(\ref{4cc}) and in the quantum ones (\ref{4aq})--(\ref{4cq}). The solar radiation pressure vector is given by
\begin{equation}
\vec{S}= -K \dfrac{A}{m (\rho_3)^3} \vec{\rho}_3, \label{solar_radiation}
\end{equation}  
where $A$ is the cross-sectional area normal to $\vec{\rho}_3$, $m$ is the planetoid mass, and $K$ is a constant. 
Inspired by Refs. \cite{testbed,Tapley}, we use the value $K=2,048936 \times 10^{17}\; {\rm N}$. We have integrated the classical equations (\ref{4ac})--(\ref{4cc}) and we have found that the presence of the solar radiation pressure causes the vehicle to move further away from $L_4$ in a given time, as one can see from Fig. \ref{5c.pdf}. In particular, the larger the ratio $A/m$ is, the larger the envelope of the motion turns out to be \cite{testbed}. 

\begin{figure}  
\centering
\includegraphics[scale=0.7]{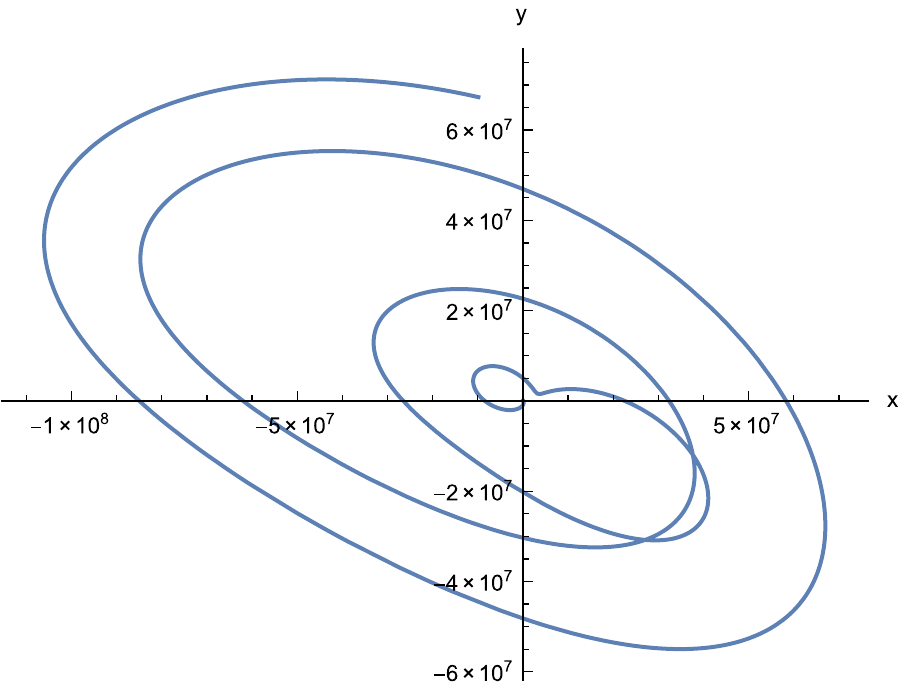}
\caption[Parametric plot in the quantum corrected restricted four-body problem of the spacecraft motion about $L_4$ in the presence of the solar radiation 
pressure]{Parametric plot in the quantum regime of the spacecraft motion about $L_4$ in the presence of the solar radiation 
pressure and considering $A/m=0,159\; {\rm m^2/Kg}$. The initial displacement and velocity are zero. The quantities appearing 
on the axes are measured in meters and the time interval considered is about $1 \times 10^7 \; {\rm s}$.}
\label{6c.pdf}
\end{figure}

Interestingly, in the quantum case ruled by effective gravity the situation is a little bit different. Unlike the classical regime, the presence of the solar radiation pressure in Eqs. (\ref{4aq})--(\ref{4cq}) does not show itself through the fact that the spacecraft goes away from the triangular libration points more rapidly, but it results in a less chaotic and irregular motion about $L_4$, which ultimately make the planetoid escape from $L_4$, like in the classical case. These effects are clearly visible from Fig. \ref{6c.pdf}.\footnote{The different scale adopted in Fig. \ref{6c.pdf} with respect to the one of Fig. \ref{5c.pdf} allows us to better appreciate its features.}

\begin{figure}
\begin{subfigure}{.5\textwidth}
\centering
  \includegraphics[width=.9\linewidth]{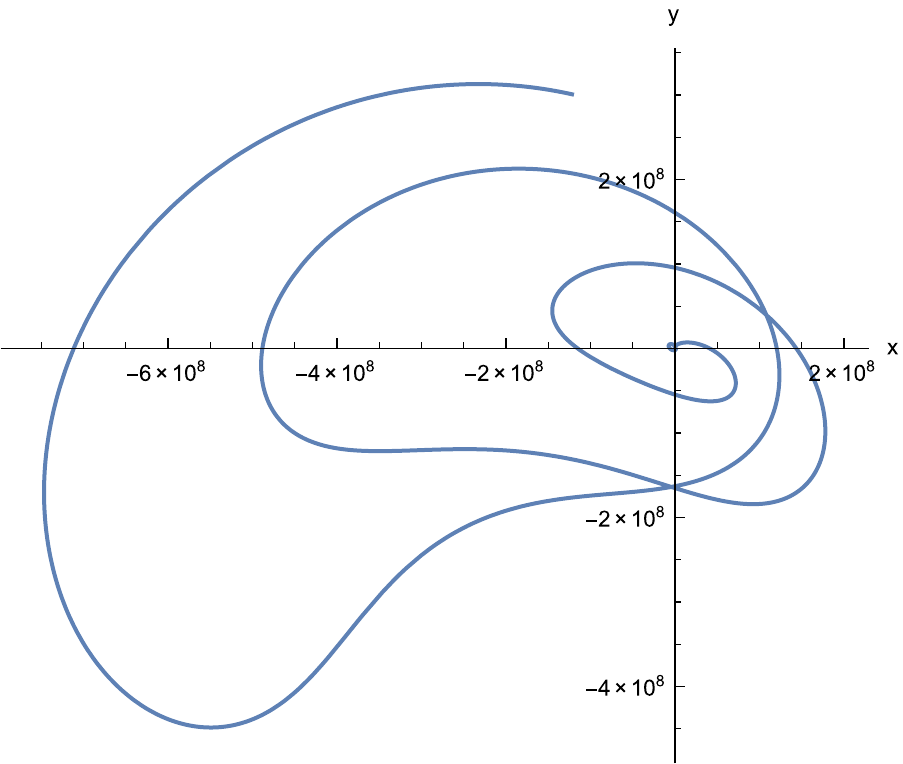}
  \caption{Spacecraft motion about $L_4$ with an initial velocity of $3\; {\rm m/s}$ at $60^{\degree}$.}
\end{subfigure}
\begin{subfigure}{.55\textwidth}
\centering
    \includegraphics[width=.9\linewidth]{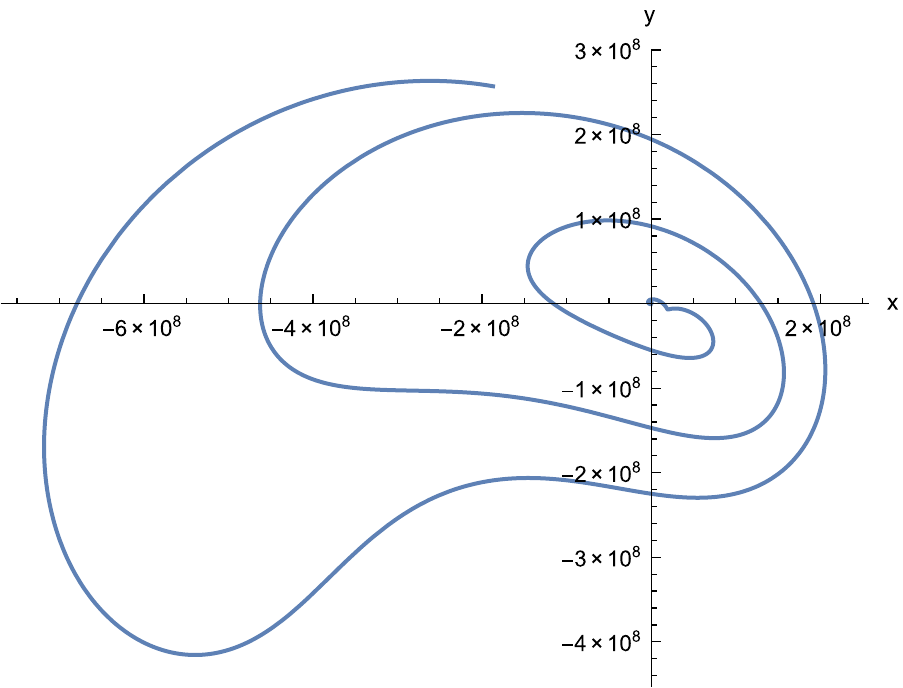}
  \caption{Spacecraft motion about $L_4$ with an initial velocity of $3\; {\rm m/s}$ at $150^{\degree}$.}
 \end{subfigure}
\begin{subfigure}{.5\textwidth}
\centering
  \includegraphics[width=.9\linewidth]{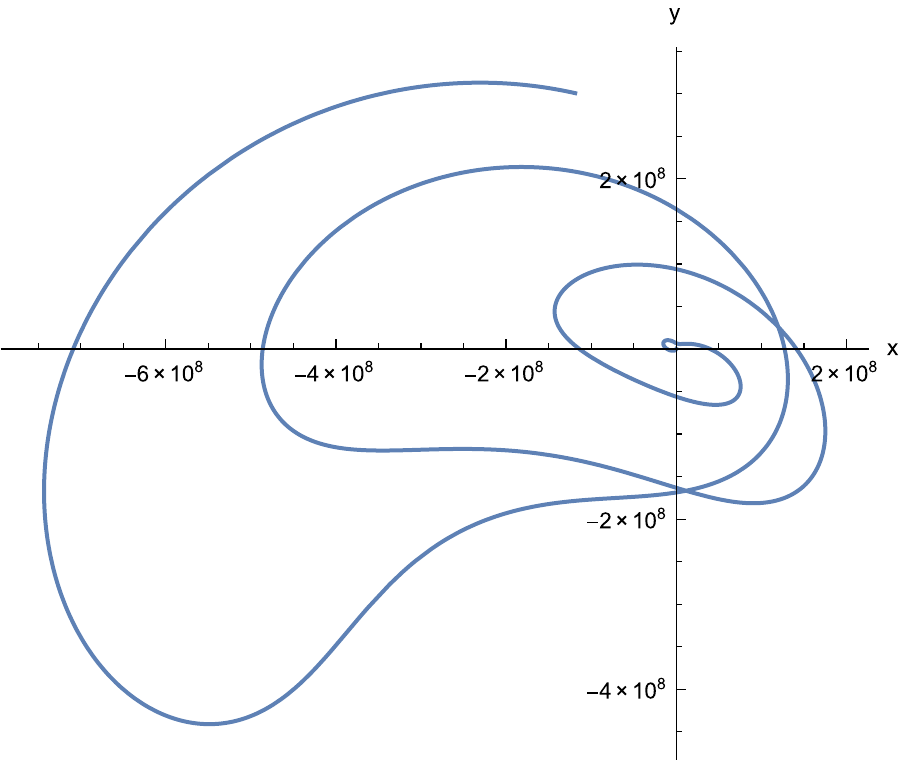}
  \caption{Spacecraft motion about $L_4$ with an initial velocity of $3\; {\rm m/s}$ at $240^{\degree}$.}
  \label{7cc.pdf}
\end{subfigure}
\begin{subfigure}{.5\textwidth}
\centering
  \includegraphics[width=.9\linewidth]{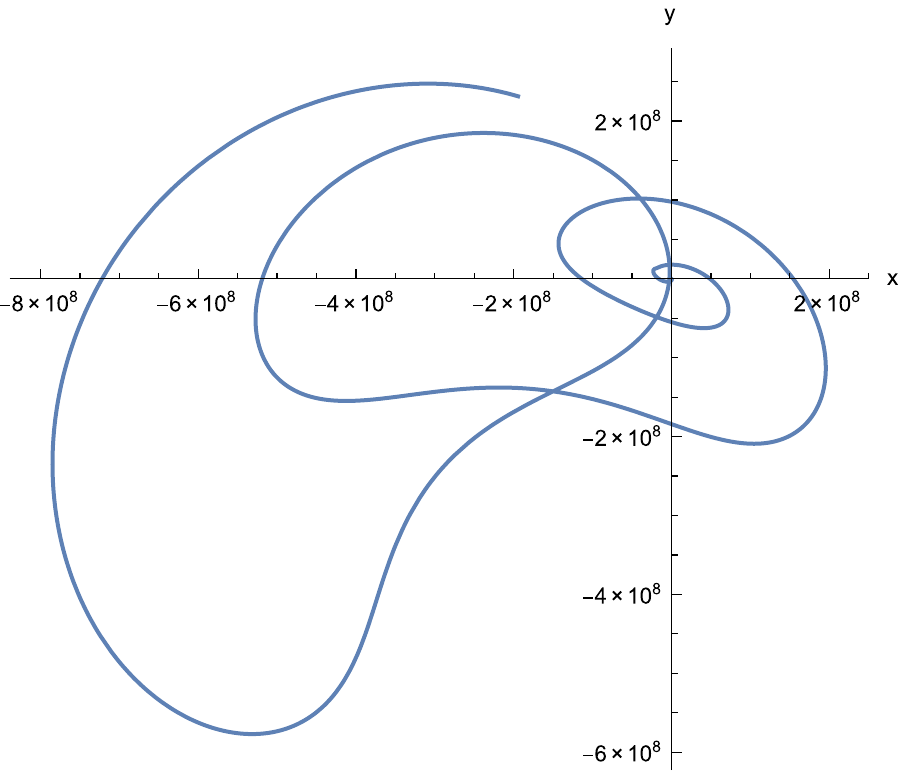}
  \caption{Spacecraft motion about $L_4$ with an initial velocity of $3\; {\rm m/s}$ at $330^{\degree}$.}
\end{subfigure}
\caption[Plots of the spacecraft motion about $L_4$ in the classical restricted four-body problem and with different initial velocities]{Plots of the spacecraft motion about $L_4$ in the classical regime and with initial velocities: (a) directed away from the Earth-Moon mass center, (b) normal to the Earth-Moon mass center-$L_4$ line and in the direction of rotation of the coordinate system, (c) directed towards the Earth-Moon mass center, (d) normal to the Earth-Moon mass center-$L_4$ line and in the opposite direction of rotation of the coordinate system.}
\label{7c.pdf}
\end{figure}

It is also possible to find the best set of initial conditions which leads to the smallest envelope of the motion of the planetoid. We have studied several sets of initial conditions both in the classical case and in the quantum one. In the classical regime, we completely agree with the results of Ref. \cite{Tapley}. We have found, in fact, that the amplitude of the spacecraft's motion depends strongly on the position of the Sun (i.e., on the values assumed by $\theta_0$ and $\psi_0$) and 
on its initial position and velocity. For example, Fig. \ref{7c.pdf} shows the motion resulting from an initial zero displacement and different initial velocity (with, like before, $\theta_0=\psi_0=0$), and a time interval of about $1 \times 10^7 \; {\rm s}$.
We have numerically checked that the envelope of the motion in Fig. \ref{7cc.pdf} is smaller at any time than the envelope of the motion shown in Fig. \ref{5c.pdf}, although at first sight it is difficult to realize this point.  

\begin{figure}
\begin{subfigure}{.5\textwidth}
  \centering
  \includegraphics[width=.9\linewidth]{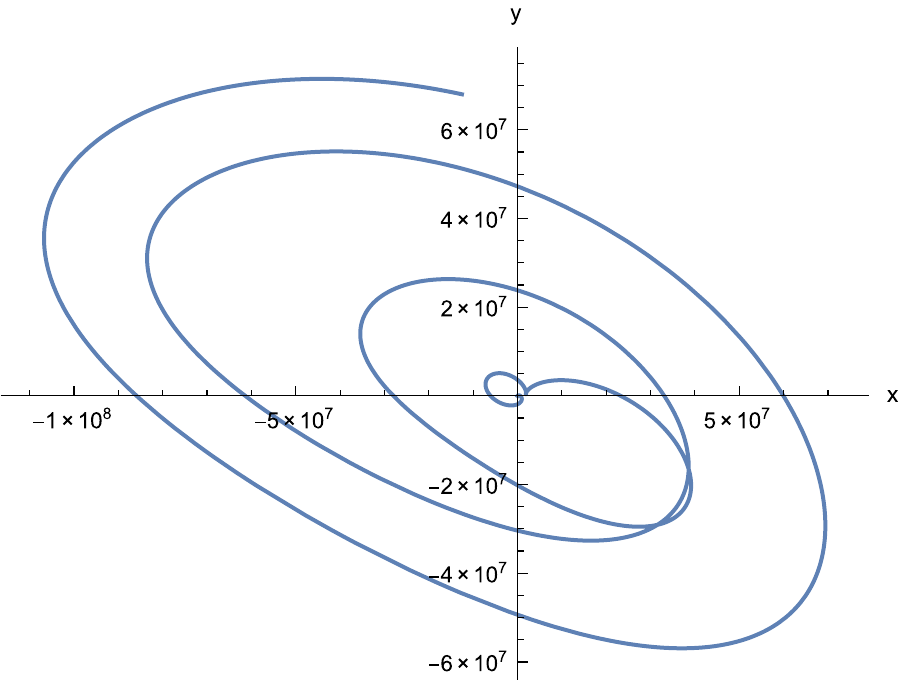}
  \caption{Spacecraft motion about $L_4$ with an initial velocity of $3\; {\rm m/s}$ at $60^{\degree}$.}
\end{subfigure}
\begin{subfigure}{.48\textwidth}
  \centering
  \includegraphics[width=.9\linewidth]{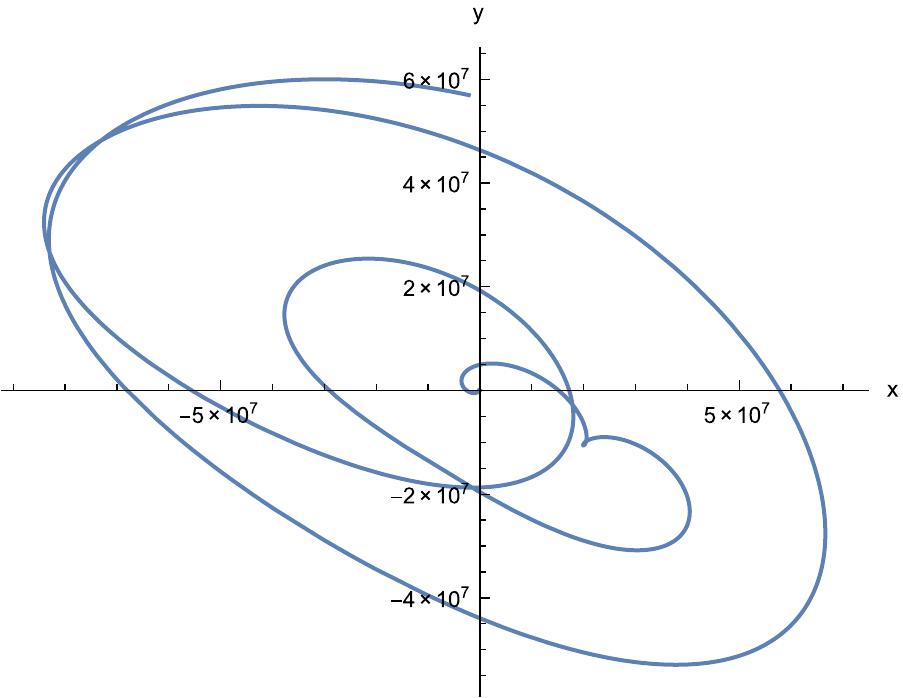}
  \caption{Spacecraft motion about $L_4$ with an initial velocity of $3\; {\rm m/s}$ at $150^{\degree}$.}
  \label{8cb.pdf}
\end{subfigure}
\begin{subfigure}{.5\textwidth}
  \centering
  \includegraphics[width=.9\linewidth]{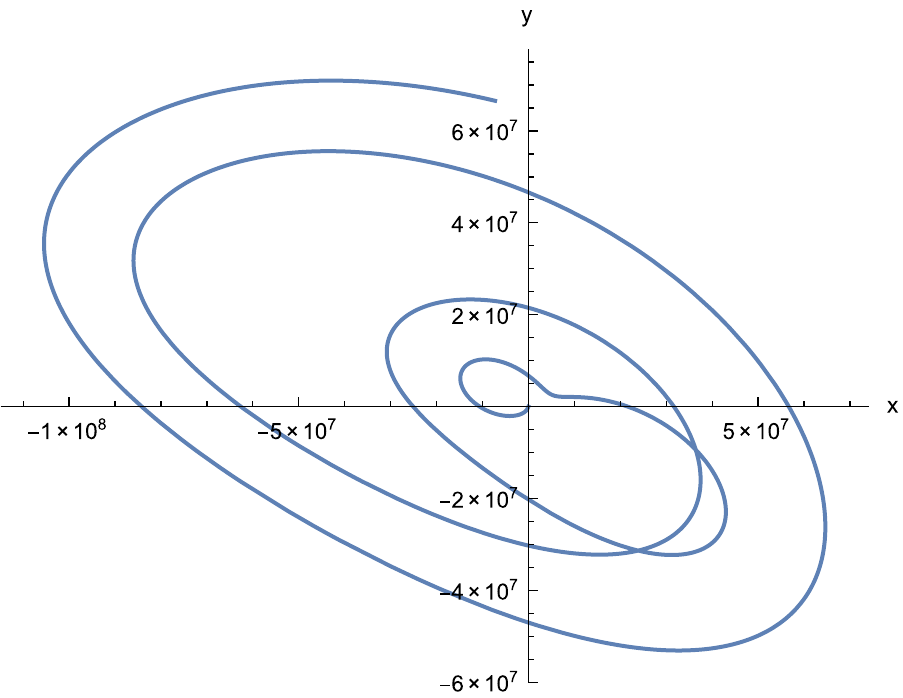}
  \caption{Spacecraft motion about $L_4$ with an initial velocity of $3\; {\rm m/s}$ at $240^{\degree}$.}
\end{subfigure}
\begin{subfigure}{.52\textwidth}
  \centering
  \includegraphics[width=.9\linewidth]{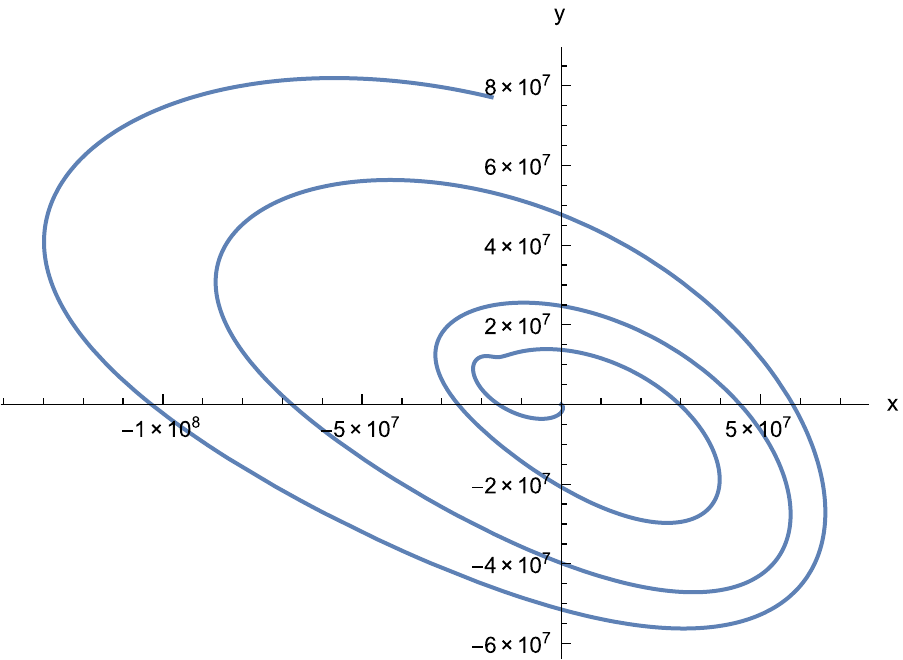}
  \caption{Spacecraft motion about $L_4$ with an initial velocity of $3\; {\rm m/s}$ at $330^{\degree}$.}
\end{subfigure}
\caption[Plots of the spacecraft motion about $L_4$ in the quantum corrected restricted four-body problem and with different initial velocities]{Plots of the spacecraft motion about $L_4$ in the quantum regime and with initial velocities: (a) directed away from the Earth-Moon mass center, (b) normal to the Earth-Moon mass center-$L_4$ line and in the direction of rotation of the coordinate system, (c) directed towards the Earth-Moon mass center, (d) normal to the Earth-Moon mass center-$L_4$ line and in the opposite direction of rotation of the coordinate system.}
\label{8c.pdf}
\end{figure}

The situation is fairly the same in the quantum regime (Fig. \ref{8c.pdf}), where we have discovered that one set of 
initial conditions (having $\theta_0=\psi_0=0$) exists, which results in a smaller envelope of the 
spacecraft motion at any given time, as one can see from Fig. \ref{8cb.pdf} \cite{testbed}. This fact can be understood with a comparison between Figs. \ref{6c.pdf} 
and \ref{8cb.pdf}. The interesting difference with respect to the classical case consists in the fact that 
the reduction of the envelope of the planetoid motion produced by a non-zero initial velocity becomes more evident 
in the quantum regime. Moreover, this reduction effect is achieved in the two regimes through different initial conditions: in the classical framework the initial velocity is directed towards the Earth-Moon mass center, while in the quantum one it results to be normal to the Earth-Moon mass center-$L_4$ line. By inspection of Figs. \ref{7c.pdf} and \ref{8c.pdf} we discover a strong dependence on the initial conditions
of the planetoid trajectories both in the classical and quantum regime. This suggests that, from an experimental
point of view, it might be useful to drop off two or more satellites close to the Lagrangian points $L_{4}$
and $L_{5}$ with slightly different initial conditions for position and velocity. Measurements of the satellite
differential positions, together with the measurement of the single orbits, could make it possible to discriminate
between classical and quantum regime, without depending on the absolute knowledge of Lagrangian points' location \cite{testbed}.  

\begin{figure}
\begin{subfigure}{.5\textwidth}
  \centering
  \includegraphics[width=.9\linewidth]{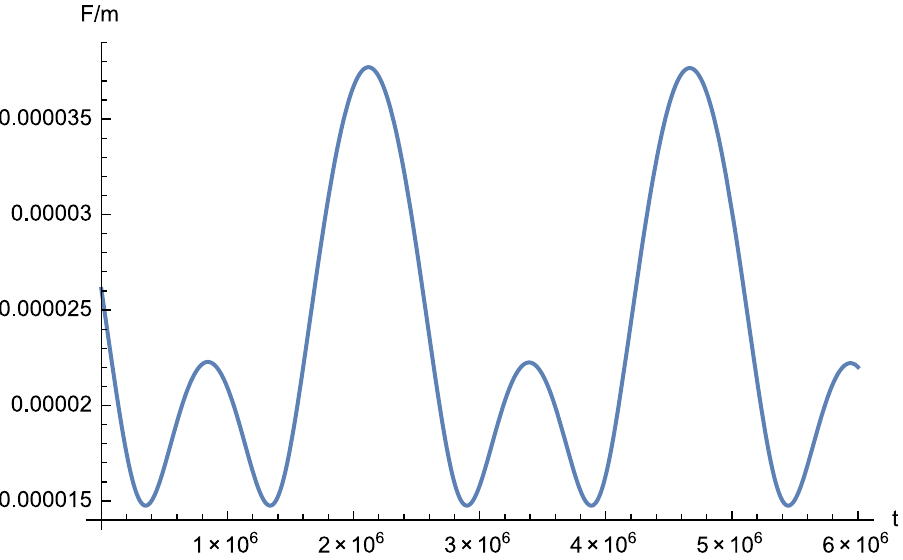}
  \caption{Force per unit mass required to induce stability at $L_4$ in the classical regime.}
  \label{impulsec}
\end{subfigure}
\begin{subfigure}{.5\textwidth}
  \centering
  \includegraphics[width=.9\linewidth]{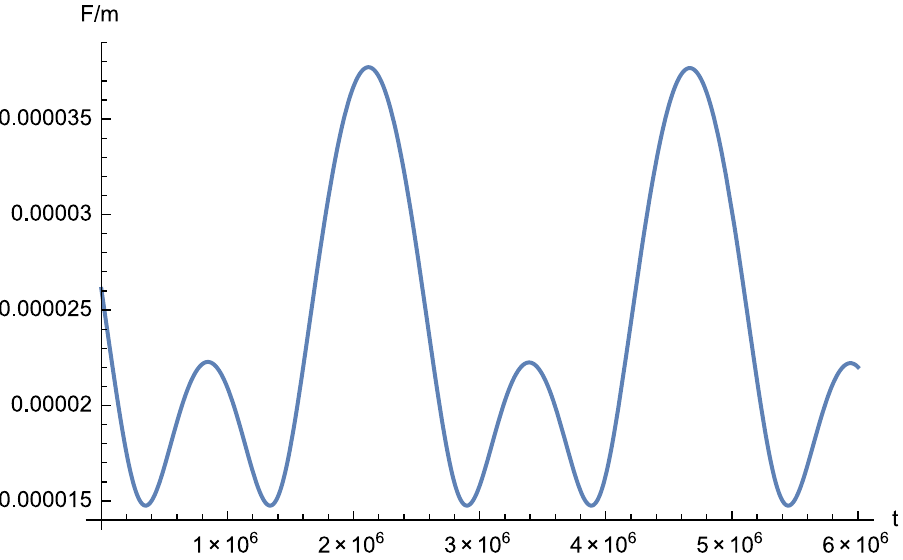}
    \caption{Force per unit mass required to induce stability at $L_4$ in the quantum regime.}
  \label{impulseq}
\end{subfigure}
\caption[Plot of the force required to induce stability at $L_4$ both in the classical and in the quantum corrected restricted four-body problem]{Plot of the force per unit mass as a function of time required to induce stability at the Lagrangian point $L_4$.}
\label{impulse}
\end{figure}

If we want to force the particle to stay precisely at $L_4$, we have to set aside the perturbing force due 
to the Sun by the application of a continuous force (see Fig. \ref{impulse}). Therefore, we have to study the following stability equation (in the $\xi,\eta,\zeta$ system) \cite{testbed}: 
\begin{equation}
-\vec{\nabla}_{\mathcal{R}}V+\vec{\nabla}_{\mathcal{R}}U+ \vec{S}+\dfrac{\vec{F}}{m}=\vec{0}, \label{vecstabilityc}
\end{equation}
which becomes in the quantum case 
\begin{equation}
-\vec{\nabla}_{\mathcal{R}}V_q+\vec{\nabla}_{\mathcal{R}}U_q+ \vec{S}+\dfrac{\vec{F_q}}{m}=\vec{0}, \label{vecstabilityq}
\end{equation}
where $m$ is the mass of the planetoid and $\vec{F}$ 
(respectively, $\vec{F}_q$) represents the force to be 
applied to the spacecraft in order to make it stay precisely at $L_4$ in the 
classical (respectively, quantum) regime. 
If we consider Eqs. (\ref{vecstabilityc}) and (\ref{vecstabilityq}) in the $x,y,z$ coordinate system, we can exploit 
the simplification resulting from the fact that the planetoid must be at the position $x=y=z=0$; hence Eq. 
(\ref{vecstabilityc}), written in components, becomes \cite{testbed}
\begin{equation}
\xi_p \omega^2-(x_3+\xi_p)(\Omega_{\omega})^2 + S_x 
+ \sum_{i=1}^3 \dfrac{Gm_i}{\rho_{i}^{3}} x_i+ \dfrac{F_x}{m}=0, \label{stabilityac}
\end{equation}
\begin{equation}
\eta_p \omega^2-(y_3+\eta_p)(\Omega_{\omega})^2 + S_y 
+ \sum_{i=1}^3 \dfrac{Gm_i}{\rho_{i}^{3}} y_i + \dfrac{F_y}{m}=0, 
\end{equation}
\begin{equation}
-z_{3}(\Omega_{\omega})^2 + S_z + \dfrac{Gm_3}{\rho_{i}^{3}}z_3 + \dfrac{F_z}{m}=0, \label{stabilitycc}
\end{equation}
whereas from Eq. (\ref{vecstabilityq}) we obtain \cite{testbed}
\begin{equation}
\begin{split}
  \xi_p & \omega^2-(x_3+\xi_p)(\Omega_{\omega})^2 
\left[1+\dfrac{2 k^{\prime \prime}_{1}}{\mathcal{R}_{3}} 
+\dfrac{3 k_2}{(\mathcal{R}_{3})^2}\right]  +\dfrac{G m_1 x_1}{(\rho_1)^3} 
\left[ 1+\dfrac{2 k_1}{\rho_1}+\dfrac{3 k_2}{(\rho_1)^2} \right]  \\
&+ \dfrac{G m_2 x_2}{(\rho_2)^3} \left[ 1+\dfrac{2 k^{\prime}_{1}}{\rho_2}+\dfrac{3 k_2}{(\rho_2)^2} \right]  
+ \dfrac{G m_3 x_3}{(\rho_3)^3} \left[1+\dfrac{2 k^{\prime \prime}_{1}}{\rho_{3}} +\dfrac{3 k_2}{(\rho_{3})^2}\right]
+ S_x +  \dfrac{F_{q_x}}{m}=0, 
\end{split}
\label{stabilityaq} 
\end{equation}
\begin{equation}
\begin{split}
\eta_p & \omega^2-(y_3+\eta_p)(\Omega_{\omega})^2 \left[1+\dfrac{2 k^{\prime \prime}_{1}}{\mathcal{R}_{3}} 
+\dfrac{3 k_2}{(\mathcal{R}_{3})^2}\right]  +\dfrac{G m_{1} y_{1}}{(\rho_1)^3} 
\left[ 1+\dfrac{2 k_1}{\rho_1}+\dfrac{3 k_2}{(\rho_1)^2} \right] \\
&+ \dfrac{G m_2 y_2}{(\rho_2)^3} \left[ 1+\dfrac{2 k^{\prime}_{1}}{\rho_2}+\dfrac{3 k_2}{(\rho_2)^2} \right]  
+ \dfrac{G m_3 y_3}{(\rho_3)^3} \left[1+\dfrac{2 k^{\prime \prime}_{1}}{\rho_{3}} 
+\dfrac{3 k_2}{(\rho_{3})^2}\right]+ S_y+\dfrac{F_{q_y}}{m}=0, 
\end{split}
\end{equation}
\begin{equation}
-z_{3} (\Omega_{\omega})^2 \left[1+\dfrac{2 k^{\prime \prime}_{1}}{\mathcal{R}_{3}} 
+\dfrac{3 k_2}{(\mathcal{R}_{3})^2}\right]  
+  \dfrac{G m_3 z_3}{(\rho_3)^3} \left[1+\dfrac{2 k^{\prime \prime}_{1}}{\rho_{3}} 
+\dfrac{3 k_2}{(\rho_{3})^2}\right]+ S_z + \dfrac{F_{q_z}}{m}=0. 
\label{stabilitycq}
\end{equation}
Equations (\ref{stabilityac})--(\ref{stabilitycc}) and (\ref{stabilityaq})--(\ref{stabilitycq}) 
make it possible for us to evaluate 
both the classical and the quantum force needed for stability and therefore the impulse per 
unit mass which the planetoid must be subjected to in order to stay in equilibrium exactly at $L_4$. 
Bearing in mind that  the impulse is defined as the integral of a force over the time interval for which it acts, 
and on considering a time interval of one year, we have found in the Newtonian regime \cite{testbed}
\begin{equation}
I_{cl} /m = 747,608255 \; {\rm N \;s /Kg},
\label{3.10c}
\end{equation}
whereas in the quantum context \cite{testbed}
\begin{equation}
\begin{split}
I_q /m = 747,608236 \; {\rm N \;s /Kg},  \\
I_q /m = 747,608315 \; {\rm N \; s /Kg},\\
I_q /m = 747,608245 \; {\rm N \; s /Kg},
\end{split}
\label{3.10cbis}
\end{equation}
for the one-particle reducible, the scattering, and the bound-states potential, respectively. We also note that this calculation suggests a gedanken experiment in which
two satellites are sent to $L_{4}$ and $L_{5}$, respectively.
If the first satellite receives the impulse $I_{cl}$ while the second receives the impulse $I_{q}$,
one might try to check, by direct comparison, which value is better suited for stabilizing the
Lagrangian point, gaining support for classical or, instead, quantum theory.
However, this configuration is merely ideal because, in light of the very small relative difference
of the impulse in the two cases, it looks practically impossible to keep all the experimental
conditions (satellite mass, actuator and readout calibration, initial conditions, solar
radiation pressure, and so forth) identical within the required accuracy (less than 0.1 parts
per million) \cite{testbed}. 

Finally, we stress that the results achieved in the last two sections hold for all the three different choices of the potential, Figs. \ref{1c.pdf}--\ref{8c.pdf} giving imperceptible differences in the three cases likewise. 

\chapter{Towards a new quantum theory}

\vspace{2cm}
\emph{It seems clear that the present quantum mechanics is not in its final form. Some further changes will be needed, just about as drastic as the changes made in passing from Bohr's orbit theory to quantum mechanics. Some day a new quantum mechanics, a relativistic one, will be discovered, in which we will not have these infinities occurring at all. It might very well be that the new quantum mechanics will have determinism in the way that Einstein wanted.}
\begin{flushright}
P. A. M. Dirac
\end{flushright}

\vspace{2cm}

In the previous chapters we have considered a framework where the theory which is quantum corrected has as its classical counterpart the Newtonian model. In fact, the quantum corrected potential (\ref{1.2b}) is characterized by the one-loop corrections to the Newtonian one. On the other hand, one has to consider that general relativity is currently the most successful gravitational theory describing the nature of space and time, and well confirmed by observations. In fact, it has been brightly confirmed by all the so-called ``classical'' tests, i.e., the perihelion shift of Mercury, the deflection of light, and the Shapiro time delay, and it has also gone through the systematic test offered by the binary pulsar system ``PSR 1913 + 16'', since its orbit decay is perfectly in accordance with the expected theoretical decay due to the emission of gravitational waves, as predicted by general relativity. Moreover, an astonishing result has been recently achieved by the scientific community: the first direct observation of gravitational waves \cite{LIGO}, as we will see in Sec. \ref{The boosting procedure_Sec}. Furthermore, Lagrangian points have recently attracted renewed interests for relativistic astrophysics \cite{Asada09,Yamada10,Yamada,Yamada15,Asadabook}, where the position and the stability of Lagrangian points is described within the post-Newtonian regime. For all these reasons, we believe that our model is incomplete without a comparison with the Einstein theory. Then, by taking 
seriously into account the important role fulfilled by general relativity within this context, we now describe a {\it new} quantum corrected regime where the underlying classical theory is represented by Einstein theory, rather than the Newtonian model \cite{testbed}. Eventually, this scheme represents a further test of both general relativity and effective field theories of gravity in the Earth-Moon system which has never been studied before. 

\section{Theoretical predictions of general relativity} \label{Theorethical predictions GR_Sec}

In order to establish the most accurate classical counterpart of the putative quantum framework that we are going to set up, we first need to describe the restricted three-body problem involving the Earth and the Moon within the context of general relativity.

\subsection{Post-Newtonian approximation} \label{Post-newtonian_approx_Sec}

Post-Newtonian approximation arises from a linearization of the Einstein field equation (\ref{Einstein_equations}) under the assumptions of weak fields (i.e., the spacetime metric is nearly flat) and low velocities as compared to the speed of light $c$. This approximation yields, in a natural coordinate system, equations of motion which in form resemble the corresponding Newtonian equations modified by corrections terms of order $1/c^2$. This is an excellent approximation except for phenomena involving gravitational collapse and black holes and phenomena dealing with the large scale structure of Universe \cite{Wald,Haw-Ell,MTW}.

Within the post-Newtonian pattern, the $n$-body Lagrangian describing the gravitational interaction among massive particles has been known since the beginning of last century. The first who dealt with such a topic was Levi-Civita \cite{LeviCivita1941}. In fact, by appealing to the geodesic principle for the motion of each celestial body and to the post-Newtonian approximation, the famous Italian mathematician derived the most general $n$-body Lagrangian of celestial mechanics, which, unlike the Einstein-Infeld-Hoffmann Lagrangian (see Eq. (\ref{EIH_Lagr})), takes into account also the dimensions of the bodies. In an inertial coordinate system $x_0 \equiv ct,x_1,x_2,x_3$, this (dimensionless) Lagrangian function reads as \cite{LeviCivita1941}
\begin{equation}
\mathcal{L}_{{\rm LC}}= \mathcal{N} + \mathcal{D},
\end{equation} 
where $\mathcal{N}$ is the Lagrangian of a material element in Newtonian mechanics
\begin{equation}
\mathcal{N}= \sum_{i=1}^{n}\dfrac{1}{2c^2}v_{i}^2+ \mathfrak{V}_{N},
\end{equation}
where
\begin{equation}
\mathfrak{V}_{N} \equiv {G \over c^{2}}\int_{S}{\mu \over r}{\rm d}S = {G\over c^{2}} \sum_{h=1}^{n}\int_{h}{\mu \over r}{\rm d}C_{h},
\end{equation}
$S$ being the region occupied by all bodies $C_{h}$ ($h=1,\dots,n$) and $\mu$ the function representing the local density, whereas $\mathcal{D}$ is the Einstein modification
\begin{equation}
\begin{split}
\mathcal{D} &= \dfrac{1}{2} \mathcal{N}^2 -\left( \mathfrak{V}_{N} \right)^2 -{G\over c^{2}} \int_{S}{\mu \mathfrak{V}_{N} \over r}{\rm d}S +{3 \over 2}{G\over c^{4}}
\sum_{i=1}^{n} \int_{S}{\mu v_{i}^2 \over r}   {\rm d}S +{1 \over 2}{G\over c^{2}} {\partial^{2}\over \partial (x^{0})^{2}}\int_{S}\mu r {\rm d}S \\
& + \mathfrak{V}_{N} \sum_{i=1}^{n} \dfrac{v_{i}^2}{c^2} - 4 {G\over c^{4}}\sum_{i=1}^{n} \left(  v_{i}\int_{S} {\mu v_{i}\over r}{\rm d}S \right) + {\rm O}(c^{-5}).
\end{split}
\end{equation}
The great contribution given by Levi-Civita consisted in proving that on passing from Newtonian to relativistic celestial mechanics, a sort of cancellation principle for which, in first approximation, the physical dimensions of bodies can be ignored, still holds, provided that the following assumptions on celestial bodies are made \cite{LeviCivita1941}:
\vskip 0.3cm
\noindent
{\bf A1} The center of gravity of each body is {\it substantial}, i.e., it always adheres to the same material element. Furthermore, the center of gravity is always a center of gravitation. The latter condition means the center of gravity of each body coincides with its mass center. 
\vskip 0.3cm
\noindent
{\bf A2} The body performs a quasi-translational motion. Indeed, in a translational motion, all points of the body have, at any instant $t$, the same vector speed, e.g., the speed ${\vec v}_{g}$ of the center of gravity. We can still regard as a translation every motion for which, defining
\begin{equation}
\left | \bigtriangleup {\vec v}(t) \right | \equiv
\left | {\vec v}_{P_{i}}(t)-{\vec v}_{P_{j}}(t) \right | \; 
\forall P_{i},P_{j} \in C,
\end{equation}
one has always
\begin{equation}
{{\left | \bigtriangleup {\vec v} \right |}\over 
{\left | {\vec v}_{g} \right |}} <<1.
\end{equation}
This is precisely what happens for planetary motions. Their deformations are initially negligible and they behave, as a consequence, as essentially rigid bodies. Their motion is actually a composition of translation and rotation. However, for every point of the body, the speed resulting from rotation attains only a few percent of the common speed of translation. For example, in the case of the Earth, one has
\begin{equation}
{{\left | \bigtriangleup {\vec v} \right |}\over 
{\left | {\vec v}_{g} \right |}} \approx 3 \cdot 10^{-2}.
\end{equation}
\vskip 0.3cm
\noindent
{\bf A3} The maximal dimension of all bodies is negligible if compared to the Euclidean distance between them.

The above assumptions are well satisfied by all the known celestial bodies of Solar System.

If we relax the hypothesis regarding the dimension of the bodies by supposing from the very beginning that they are represented by point-like masses, we obtain the Einstein-Infeld-Hoffmann Lagrangian. This Lagrangian describes in fact the dynamics of a system of $n$ particles subjected to their mutual gravitational attractions including general relativity effects within the first post-Newtonian approximation. Note that a system of gravitating bodies can be correctly described by a Lagrangian up to terms of order $1/c^4$ (in the absence of an electromagnetic field, for which a Lagrangian exists in general only up to terms of second order). In fact, a system of interacting bodies loses its energy in the form of radiation of gravitational waves, but this effect appears only at the order $1/c^5$ and hence up to second-order terms in the post-Newtonian approximation the energy of the system is constant. Einstein-Infeld-Hoffmann Lagrangian is given by (in the de Donder Gauge) \cite{Landau-Lif}
\begin{equation}
\begin{split}
\mathcal{L}_{\rm EIH} &= \dfrac{1}{2} \sum_{i=1}^{n} m_{i}v_{i}^{2} + \sum_{i=1}^{n} \sum_{j \neq i} \dfrac{G m_i m_j}{2 r_{ij}}+\dfrac{3}{2c^2} \sum_{i=1}^{n} \sum_{j \neq i} \dfrac{G m_i m_j}{r_{ij}} v_{i}^{2} + \dfrac{1}{8c^2} \sum_{i=1}^{n} m_{i}v_{i}^{4} \\
& - \dfrac{1}{4c^2} \sum_{i=1}^{n} \sum_{j \neq i} \dfrac{G m_i m_j}{r_{ij}}  \left[7 \left(\bold{v}_{i} \cdot \bold{v}_{j} \right) +\left( \bold{v}_{i} \cdot \bold{n}_{ij} \right)\left( \bold{v}_{j} \cdot \bold{n}_{ij} \right) \right] \\
&  - \dfrac{1}{2c^2}\sum_{i=1}^{n} \sum_{j \neq i} \sum_{k \neq i} G^2\dfrac{m_i m_j m_k}{r_{ij} r_{ik}} + {\rm O}(c^{-4}), 
\end{split}
\label{EIH_Lagr}
\end{equation}
whereas the Euler-Lagrange equations resulting from (\ref{EIH_Lagr}) are known as Einstein-Infeld-Hoffmann equations and read as \cite{EIH}
\begin{equation}
\begin{split}
\dfrac{{\rm d} \bold{v}_k}{{\rm d}t} & = \sum_{i \neq k} \bold{r}_{ik} \dfrac{G m_i}{r_{ik}^{3}}+\dfrac{1}{c^2} \sum_{i \neq k} \bold{r}_{ik} \dfrac{G m_i}{r_{ik}^{3}} \Biggl [ -4 \sum_{j \neq k} \dfrac{G m_k}{r_{jk}} - \sum_{h \neq i} \dfrac{G m_h}{r_{ih}} \left( 1- \dfrac{\bold{r}_{ik} \cdot \bold{r}_{ih}}{2 r_{ih}^{2}} \right) \\
&+v_{k}^2+2v_{i}^{2}   - 4 \bold{v}_i \cdot \bold{v}_k  - \dfrac{3}{2} \left( \bold{v}_{i} \cdot \bold{n}_{ik} \right)^2 \Biggr] 
- \dfrac{1}{c^2}\sum_{i \neq k} \left( \bold{v}_i - \bold{v}_k \right) \dfrac{G m_i \bold{n}_{ik} \cdot \left( 3 \bold{v}_i - 4\bold{v}_k \right)}{r_{ik}^{2}} \\
& + \dfrac{7}{2c^2} \sum_{i \neq k} \sum_{h \neq i} \bold{r}_{ih} \dfrac{Gm_i m_h }{r_{ik}r_{ih}^{3}} + {\rm O}(c^{-4}), \; \; \; \; \; \; \;\; \; \; \; \; \; \;\; \; \; \; \;  (k=1,2,\dots,n),
\end{split}
\label{EIH_equation}
\end{equation}
$\bold{v}_k$ being the vector velocity measured in an inertial frame of the $k$-th body having mass $m_k$, $\bold{r}_{ij}=\bold{r}_i-\bold{r}_j$ the distance among such masses and $\bold{n}_{ij}= \bold{r}_{ij}/r_{ij}$. Note how in the limit $c \rightarrow + \infty$ we recover Newtonian dynamics. For a two-body system we have
\begin{equation}
\begin{split}
\mathcal{L}_{{\rm EIH},2} & = \dfrac{1}{2} \sum_{i=1}^{2} m_{i}v_{i}^{2} +\dfrac{G m_1 m_2}{r}+\dfrac{3}{2c^2} \dfrac{G m_1 m_2}{r} \left(v_{1}^{2}+v_{2}^{2}\right)+\dfrac{1}{8c^2} \sum_{i=1}^{2} m_{i}v_{i}^{4} \\
& -\dfrac{1}{2c^2} \dfrac{G m_1 m_2}{r} \left[7 \left(\bold{v}_{1} \cdot \bold{v}_{2} \right) +\left( \bold{v}_{1} \cdot \bold{n} \right)\left( \bold{v}_{2} \cdot \bold{n} \right) \right]-\dfrac{1}{2c^2} \dfrac{G^2 m_1 m_2 \left(m_1 + m_2 \right)}{r^2},
\end{split}
\label{EIH_Lagr_2}
\end{equation}
$r$ being the distance between $m_1$ and $m_2$. By employing (\ref{EIH_Lagr_2}) it is possible to calculate, for example, the secular shift of the perihelion of the orbit of two gravitating bodies of comparable mass \cite{Landau-Lif,Robertson1938}
\begin{equation}
\Delta \varphi = \dfrac{6 \pi G (m_1+m_2)}{c^2 a (1-e^2)},
\end{equation}
where $a$ and $e$ represent the semi-major axis and the eccentricity of the elliptic orbit, respectively. In particular, in the case of a particle in a centrally symmetric gravitational field generated by the mass $M$ we obtain the well-known formula
\begin{equation}
\Delta \varphi = \dfrac{6 \pi G M}{c^2 a (1-e^2)},
\end{equation}
which in the case of the Sun-Mercury system gives the famous $43.0''$ per century. Finally, for a three-body interaction (\ref{EIH_Lagr}) gives
\begin{equation}
\begin{split}
\mathcal{L}_{{\rm EIH},3} & = \dfrac{1}{2} \sum_{i=1}^{3} m_{i}v_{i}^{2} +\dfrac{G m_1 m_2}{r_{12}}+\dfrac{G m_1 m_3}{r_{13}}+\dfrac{G m_2 m_3}{r_{23}} \\
& +\dfrac{3}{2c^2} \left[ \dfrac{G m_1 m_2}{r_{12}} \left(v_{1}^{2}+v_{2}^{2} \right)+\dfrac{G m_1 m_3}{r_{13}}\left(v_{1}^{2}+v_{3}^{2} \right)+\dfrac{G m_2 m_3}{r_{23}} \left(v_{2}^{2}+v_{3}^{2} \right) \right]+\dfrac{1}{8c^2} \sum_{i=1}^{3} m_{i}v_{i}^{4} \\
& -\dfrac{1}{2c^2} \biggl\{ \dfrac{G m_1 m_2}{r_{12}} \left[7 \left(\bold{v}_{1} \cdot \bold{v}_{2} \right) +\left( \bold{v}_{1} \cdot \bold{n}_{12} \right)\left( \bold{v}_{2} \cdot \bold{n}_{12} \right) \right]+\dfrac{G m_1 m_3}{r_{13}} [7 \left(\bold{v}_{1} \cdot \bold{v}_{3} \right) \\
& +\left( \bold{v}_{1} \cdot \bold{n}_{13} \right)\left( \bold{v}_{3} \cdot \bold{n}_{13} \right) ] +\dfrac{G m_2 m_3}{r_{23}} \left[7 \left(\bold{v}_{2} \cdot \bold{v}_{3} \right) +\left( \bold{v}_{2} \cdot \bold{n}_{23} \right)\left( \bold{v}_{3} \cdot \bold{n}_{23} \right) \right] \biggr\} \\
& -\dfrac{1}{2c^2} \biggl[ \dfrac{G^2 m_1 m_2 \left(m_1 + m_2 \right)}{r_{12}^{2}}  + \dfrac{G^2 m_1 m_3 \left(m_1 + m_3 \right)}{r_{13}^{2}} + \dfrac{G^2 m_2 m_3 \left(m_2 + m_3 \right)}{r_{23}^{2}} \\
& +2G^2 m_1 m_2 m_3 \left(\dfrac{1}{r_{12}r_{13}} + \dfrac{1}{r_{12}r_{23}} +\dfrac{1}{r_{13}r_{23}} \right) \biggr].
\end{split}
\label{EIH_Lagr_3}
\end{equation}
Note how in the last line of Eq. (\ref{EIH_Lagr_3}) coupling terms proportional to a product of three masses have appeared.

\subsection{Corrections on the position of Lagrangian points} \label{Corrections on the position of Lagrangian points_Sec}

The analysis of the previous chapters relies on the simple but non-trivial {\it assumption} that, since effective gravity modifies the long-distance Newtonian potential among bodies of masses $m_{A}$ and $m_{B}$ according to (\ref{1.2b}) for all values of $r$ greater than a suitably large $r_{0}$ (see Eq. (\ref{1.5b})), the resulting modification of Newtonian dynamics can be obtained by considering a classical Lagrangian where the Newtonian potential is replaced by $V_Q(r)$, while all other terms, i.e., kinetic energy, centrifugal and Coriolis forces, remain unaffected (cf. Sec. \ref{Quantum corrected Lagrangian_Sec} or chapter \ref{more detailed newtonian models_Chapter}). Although it would be inappropriate to use the quantum effective action to study the low-energy effects resulting from the expansion (\ref{1.2b}), the above assumption is a short-cut to describe
a theory lying in between classical gravity and full quantum gravity. For this reason, it becomes important to study the predictions of classical gravity when general relativity is instead assumed.

As was pointed out in Sec. \ref{Lagrangian_points_Sec}, our analysis is performed by exploiting Dirichlet stability criterion, which in turn allows us to find the position of equilibrium points by evaluating the zeros of the gradient of the potential energy function (see below, Eq. (\ref{4.9c})) governing the dynamics of the planetoid in the Earth-Moon gravitational field \cite{BE14a,BE14b,BEDS15,testbed}, but other  methods have been developed in the literature, like the one exploited in Refs. \cite{Asada09,Yamada10,Yamada,Yamada15,Asadabook}. In this latter case, the position of Lagrangian points is obtained by considering an ansatz for the solution of Eq. (\ref{EIH_equation}) in  the case of $n=3$ bodies. In fact, in Ref. \cite{Yamada10} it is shown that, after some manipulations, Eq. (\ref{EIH_equation}) leads to post-Newtonian corrections regarding the position of collinear Lagrangian points which turn out to be governed by the seventh degree algebraic equation \cite{Yamada10}
\begin{equation}
\sum_{k=0}^{7}\mathfrak{a}_k \, \mathfrak{g}^k =0, \label{Yamada1}
\end{equation} 
where the unknown $\mathfrak{g}$ is such that
\begin{equation}
\mathfrak{g} \equiv \dfrac{r_{23}}{r_{12}} \Rightarrow r_{13}= (1+\mathfrak{g}) r_{12},
\label{Yamada2}
\end{equation}
and the coefficients $\mathfrak{a}_k$ are functions of the ratios between the masses of the bodies (see Ref. \cite{Yamada10} for their detailed form). In the case of the restricted three-body problem, Eq. (\ref{Yamada1}) becomes more feasible if we modify (\ref{Yamada2}) through the ansatz
\begin{equation}
\mathfrak{g} = \mathfrak{g}_{\rm Newton} (1+ \varepsilon), 
\label{Yamada3}
\end{equation}
$\varepsilon$ being an infinitesimal quantity. On the other hand, the triangular libration points are evaluated only in the restricted case and their corrections with respect to the Newtonian values $r_{ij}=l$ are given by
\begin{equation}
r_{ij}= l (1+ \varepsilon_{ij}),
\end{equation}
where $\varepsilon_{ij}$ are again functions of the ratios between the masses of the bodies and subjected to the constraint
\begin{equation}
\varepsilon_{12}+\varepsilon_{23}+\varepsilon_{31}=0.
\end{equation}
The explicit form of $\varepsilon_{ij}$ is given in Ref. \cite{Yamada}. 

Let us apply our analysis first to non-collinear Lagrangian points. When our attention is confined to the restricted planar three-body problem in the post-Newtonian limit, from the general Lagrangian (\ref{EIH_Lagr}) we obtain\footnote{The Lagrangian function (\ref{5.6c}) can be written in the equivalent form
\begin{equation}
\mathcal{L}_{GR}=\dfrac{1}{2} c^2 \left(\dfrac{{\rm d}s}{{\rm d}x^{0} }\right)^2= \dfrac{1}{2} \left(\dfrac{{\rm d}s}{{\rm d}t} \right)^2,
\end{equation}
where ${\rm d}s^2$ represents the square of the length of an arc of curve in the spacetime. The corresponding geodesics are the lines which make the variation of $\mathcal{L}_{GR}$ vanish. In other words, the Lagrange equations associated to (\ref{5.6c}) lead to the well-known geodesic equation
\begin{equation}
\ddot{x}^{\lambda} + \Gamma^{\lambda}_{\; \mu  \nu} \dot{x}^{\mu} \dot{x}^{\nu},
\end{equation}
$\Gamma^{\lambda}_{\; \mu  \nu}$  being the Christoffel symbols of the metric. Nevertheless, alternative forms of the Lagrangian are found in the literature \cite{Landau-Lif,LeviCivita1941}, such as
\begin{equation}
\mathcal{L}^{\prime}= \sqrt{\left(\dfrac{{\rm d}s}{{\rm d}x^{0} }\right)^2}= \dfrac{\sqrt{2}}{c} \sqrt{\mathcal{L}_{GR}}.
\end{equation}
These two forms of the Lagrangian function (corresponding to two different forms of the principle of least action) are conditionally equivalent \cite{Hadamard1923}. In fact, by adopting $\mathcal{L}^{\prime}$, the time variable $t$ is considered as an arbitrary parameter, whereas the motion defined by (\ref{5.6c}) is not arbitrary in time, but it admits as an integral of motion the total energy (i.e., the Hamiltonian function). If we take into account this condition, the dynamics described by $\mathcal{L}_{GR}$ and $\mathcal{L}^{\prime}$ is in general equivalent.} \cite{Wanex}
\begin{equation}
\mathcal{L}_{GR}={1 \over 2} \sum_{\mu,\nu=0}^{3}
g_{\mu \nu}{{\rm d}x^{\mu}\over {\rm d}t}{{\rm d}x^{\nu}\over {\rm d}t},
\label{5.6c}
\end{equation}
the metric tensor components in a co-rotating frame $x^{0}=ct$, $x^{1}=\xi$, $x^{2}=\eta$, $x^{3}=\zeta$ being given by \cite{testbed,Brumberg1972,Krefetz1967}
\begin{equation}
\begin{split}
g_{00} &=1-2{R_{\alpha}\over r}-2{R_{\beta}\over s}-{\Omega^{2}\over c^{2}}(\xi^{2}+\eta^{2})
+2\left[\left({R_{\alpha}\over r}\right)^{2}+\left({R_{\beta}\over s}\right)^{2}\right] \\
&- 2 {(R_{\alpha}+R_{\beta})\over l^{3}}
\left({R_{\alpha}\over r}+{R_{\beta}\over s}\right)(\xi^{2}+\eta^{2})
+4{R_{\alpha}\over r}{R_{\beta}\over s}  \\
&+ {(2-\rho)\over (1+\rho)}{R_{\alpha}\over r}{R_{\beta}\over l}
+{(2\rho-1)\over (1+\rho)}{R_{\beta}\over s}{R_{\alpha}\over l}
-7{\xi \over l^{2}}\left({R_{\alpha}\over r}R_{\beta}-{R_{\beta}\over s}R_{\alpha}\right) \\
&+ (1+\rho)^{-1}{\eta^{2}\over l}\left[\rho \left({R_{\alpha}\over r}\right)^{3}
{R_{\beta}\over (R_{\alpha})^{2}}
+\left({R_{\beta}\over s}\right)^{3}{R_{\alpha}\over (R_{\beta})^{2}}\right],
\end{split}
\label{5.1c}
\end{equation}
\begin{equation}
2c g_{01}=\left(1+2{R_{\alpha}\over r}+2{R_{\beta}\over s}\right)2 \Omega \eta,
\end{equation}
\begin{equation}
2c g_{02}=-\left(1+2{R_{\alpha}\over r}+2{R_{\beta}\over s}\right)2 \Omega \xi
-8{\Omega^{2}l \over (1+\rho)}\left(\rho {R_{\alpha}\over r}-{R_{\beta}\over s}\right),
\end{equation}
\begin{equation}
g_{03}=0,
\end{equation}
\begin{equation}
g_{ij}=-\left(1+2{R_{\alpha}\over r}+2{R_{\beta}\over s}\right)\delta_{ij}, \;\; \; \; \; \; \; \; \;  i,j=1,2,3
\label{5.5c}
\end{equation}
where, like in chapter \ref{Restricted_Chapter}, we are considering primaries of masses $\alpha$ (i.e., the Earth) and $\beta$ (i.e., the Moon) separated by a distance $l$, with gravitational radii $R_{\alpha} \equiv {G \alpha \over c^{2}}$ and $R_{\beta} \equiv {G \beta \over c^{2}}$, mass ratio $\rho \equiv {\beta \over \alpha} <1$ and distances from the planetoid represented by $r$ and $s$, respectively. Furthermore, note how in Eqs. (\ref{5.1c})--(\ref{5.5c}) the classical angular frequency (or pulsation) $\omega \equiv \sqrt{G(\alpha+\beta)\over l^{3}}$ gets replaced by \cite{testbed,Bhat}
\begin{equation}
\Omega \equiv \omega \left[1-{3 \over 2}{(R_{\alpha}+R_{\beta})\over l}
\left(1-{1 \over 3}{\rho \over (1+\rho)^{2}}\right)\right].
\label{4.4c}
\end{equation}
The Lagrangian equations obtained from (\ref{5.6c}) describing in the synodic frame $\xi,\eta$\footnote{This coordinate system is analogous to that of Fig. \ref{1a.pdf}.} the motion of the planetoid in the gravitational field generated by the Earth and the Moon assume the form \cite{testbed,Bhat}
\begin{equation}
\ddot \xi -2 \Omega {\dot \eta}={\partial W \over \partial \xi}
-{{\rm d}\over {\rm d}t}\left( {\partial W \over \partial {\dot \xi}}\right),
\label{4.5c}
\end{equation}
\begin{equation}
\ddot \eta +2 \Omega {\dot \xi}={\partial W \over \partial \eta}
-{{\rm d}\over {\rm d}t}\left( {\partial W \over \partial {\dot \eta}}\right),
\label{4.6c}
\end{equation}
where the effective potential $W$ reads as \cite{testbed}
\begin{equation}
\begin{split}
W(\xi,\eta) &= {\Omega^{2}\over 2}(\xi^{2}+\eta^{2})+c^{2}\left[{R_{\alpha}\over r}
+{R_{\beta}\over s}-{1 \over 2}\left({(R_{\alpha})^{2}\over r^{2}}
+{(R_{\beta})^{2}\over s^{2}}\right)\right]  \\
&+ {1 \over 8c^{2}}f^{2}(\xi,\eta,{\dot \xi},{\dot \eta})
+{3 \over 2}\left({R_{\alpha}\over r}+{R_{\beta}\over s}\right)
f(\xi,\eta,{\dot \xi},{\dot \eta}) \\
&+ {R_{\beta}\over (1+\rho)}\Omega l \left(4 {\dot \eta}
+{7 \over 2}\Omega \xi \right)\left({1 \over r}-{1 \over s}\right)\\
&+ {R_{\beta}\over (1+\rho)}\Omega^{2}l^{2} \left[-{\eta^{2}\over 2(1+\rho)}
\left({\rho \over r^{3}}+{1 \over s^{3}}\right)-{l \over rs}
+{(\rho-2)\over 2(1+\rho)}{1 \over r}
+{(1-2 \rho)\over 2(1+\rho)}{1 \over s}\right],
\end{split}
\label{4.9c}
\end{equation}
where 
\begin{equation}
f(\xi,\eta,{\dot \xi},{\dot \eta}) \equiv {\dot \xi}^{2}+{\dot \eta}^{2}
+2 \Omega (\xi {\dot \eta}-\eta {\dot \xi})
+\Omega^{2}(\xi^{2}+\eta^{2}).
\end{equation}
At all equilibrium points, the first and second time derivatives of coordinates $(\xi,\eta)$ should vanish, which implies, as we said before, that it is enough to evaluate the zeros of the gradient of $W(\xi,\eta)$, because \cite{Krefetz1967}
\begin{equation}
{{\rm d}\over {\rm d}t}\left({\partial W \over \partial {\dot \xi}}\right)
={{\rm d}\over {\rm d}t}\left({\partial W \over \partial {\dot \eta}}\right)=0, \;\; \; \; \; \;  
{\rm if} \; \; {\dot \xi}={\dot \eta}={\ddot \xi}={\ddot \eta}=0.
\end{equation}
By virtue of 
\begin{equation}
\begin{split}
& r^{2}=\left(\xi + {\rho l \over (1+\rho)}\right)^{2}+\eta^{2},\\
& s^{2}=\left(\xi - {l \over (1+\rho)}\right)^{2}+\eta^{2},
\end{split}
\label{4.7c}
\end{equation}
we find \cite{testbed} 
\begin{equation}
{\partial \over \partial \xi}(r^{-p})=-pr^{-p-2}\left(\xi+{\rho l \over (1+\rho)}\right),
\end{equation}
\begin{equation}
{\partial \over \partial \xi}(s^{-p})=-ps^{-p-2}\left(\xi-{l \over (1+\rho)}\right),
\end{equation}
\begin{equation}
{\partial \over \partial \eta}(r^{-p})=-pr^{-p-2},
\end{equation}
\begin{equation}
{\partial \over \partial \eta}(s^{-p})=-ps^{-p-2}.
\end{equation}
By computing the above formulas with $p=1,2,3$, the two components of ${\rm grad}\,W$ can be expressed in the form \cite{testbed}
\begin{equation}
{\partial W \over \partial \xi}=W_{1}(\xi,\eta,r)+W_{2}(\xi,\eta,s)
+{R_{\beta}l^{3}\over (1+\rho)}{\Omega^{2}\over rs}
\left[\xi \left({1 \over r^{2}}+{1 \over s^{2}}\right)
+{l \over (1+\rho)}\left({\rho \over r^{2}}-{1 \over s^{2}}\right)\right],
\label{4.12c}
\end{equation}
\begin{equation}
{\partial W \over \partial \eta}=\eta \left[W_{3}(\xi,\eta,r)+W_{4}(\xi,\eta,s)
+{R_{\beta}l^{3}\over (1+\rho)}{\Omega^{2}\over rs}
\left({1 \over r^{2}}+{1 \over s^{2}}\right)\right],
\label{4.13c}
\end{equation}
where the functions $W_{1},\dots,W_{4}$ are defined by \cite{testbed}
\begin{equation}
\begin{split}
W_{1}(\xi,\eta,r)& \equiv  \xi \Omega^{2}+{\Omega^{4}\xi (\xi^2+ \eta^2)\over 2c^{2}}
+3 \Omega^{2}\xi {R_{\alpha}\over r}
+{7 \over 2}{R_{\beta}l \Omega^{2}\over (1+\rho)}{1 \over r} \\
&+ \left(\xi +{\rho l \over (1+\rho)}\right){1 \over r^{3}}
\biggr \{ c^{2}R_{\alpha} \left({R_{\alpha}\over r}-1 \right)+\Omega^{2}
\biggr[-{3 \over 2}R_{\alpha}(\xi^{2}+\eta^{2}) \\
&- {7 \over 2}{R_{\beta}l \xi \over (1+\rho)}
+{3 \over 2}{\rho \over (1+\rho)^{2}}{R_{\beta}l^{2}\eta^{2}\over r^{2}}
+{(2-\rho)\over 2(1+\rho)^{2}}R_{\beta}l^{2}\biggr ] \biggr \} ,
\end{split}
\end{equation}
\begin{equation}
\begin{split}
W_{2}(\xi,\eta,s)& \equiv  3 \Omega^{2}\xi {R_{\beta}\over s}
-{7 \over 2} {R_{\beta}l \Omega^{2}\over (1+\rho)}{1 \over s} \\
&+ \left(\xi -{l \over (1+\rho)}\right){1 \over s^{3}}
\biggr \{ c^{2}R_{\beta} \left({R_{\beta}\over s}-1 \right)+\Omega^{2}
\biggr[-{3 \over 2}R_{\beta}(\xi^{2}+\eta^{2}) \\
&+ {7 \over 2}{R_{\beta}l \xi \over (1+\rho)}
+{3 \over 2}{1 \over (1+\rho)^{2}}{R_{\beta}l^{2}\eta^{2}\over s^{2}}
+{(2\rho -1)\over 2(1+\rho)^{2}}R_{\beta}l^{2}\biggr ] \biggr \} ,
\end{split}
\end{equation}
\begin{equation}
\begin{split}
W_{3}(\xi,\eta,r)& \equiv  \Omega^{2}+{\Omega^{4}\over 2c^{2}}(\xi^{2}+\eta^{2})
+{c^{2}R_{\alpha}\over r^{3}}\left({R_{\alpha}\over r}-1 \right)
+3 \Omega^{2}{R_{\alpha}\over r} \\
&- {3 \over 2}\Omega^{2}(\xi^{2}+\eta^{2}){R_{\alpha}\over r^{3}}
-{7 \over 2}{R_{\beta}l \xi \Omega^{2} \over (1+\rho)}{1 \over r^{3}} \\
&+ {R_{\beta}l^{2}\Omega^{2}\over 2(1+\rho)^{2}}{\rho \over r^{3}}
\left(3{\eta^{2}\over r^{2}}-2 \right)
+{R_{\beta}l^{2}\Omega^{2}\over 2(1+\rho)^{2}}
{(2-\rho)\over r^{3}},
\end{split}
\end{equation}
\begin{equation}
\begin{split}
W_{4}(\xi,\eta,s) & \equiv  {c^{2}R_{\beta}\over s^{3}}\left({R_{\beta}\over s}-1 \right)
+3 \Omega^{2}{R_{\beta}\over s}
- {3 \over 2}\Omega^{2}(\xi^{2}+\eta^{2}){R_{\beta}\over s^{3}}
+{7 \over 2}{R_{\beta}l \xi \Omega^{2} \over (1+\rho)}{1 \over s^{3}} \\
&+ {R_{\beta}l^{2}\Omega^{2}\over 2(1+\rho)^{2}}{1 \over s^{3}}
\left(3{\eta^{2}\over s^{2}}-2 \right)
+{R_{\beta}l^{2}\Omega^{2}\over 2(1+\rho)^{2}}
{(2 \rho-1)\over s^{3}}.
\end{split}
\label{A8c}
\end{equation}
Thus, it is clear from Eqs. (\ref{4.12c})--(\ref{A8c}) that, unlike the case of Chapter \ref{Restricted_Chapter}, when the gradient of $W$ is set to zero with $\eta \not=0$, one does not get two different algebraic equations for $r$ and $s$ (cf. Sec. \ref{noncoll_sec} and in particular Eqs. (\ref{5.1a})--(\ref{s(l)})). Since we are interested in numerical solutions of such an enlarged algebraic system with (at least) ten decimal digits, we set $r \equiv \gamma l$, $s=\Gamma l$, and we study the coupled algebraic equations for the real numbers $\gamma$ and $\Gamma$ obtained from
\begin{equation}
\gamma^{5}\Gamma^{5}{\partial W \over \partial \xi}=0,
\label{4.14c}
\end{equation}
\begin{equation}
\gamma^{5}\Gamma^{5}{1 \over \eta}{\partial W \over \partial \eta}=0,
\label{4.15c}
\end{equation}
where the fifth powers of $\gamma$ and $\Gamma$ are suggested by the occurrence of terms proportional to $\gamma^{-5}$ and $\Gamma^{-5}$ in the derivatives ${\partial W \over \partial \xi}$ and ${\partial W \over \partial \eta}$. We can write Eqs. (\ref{4.14c}) and (\ref{4.15c}) in a more concise way, i.e., \cite{testbed}
\begin{equation}
\gamma^{5}\Gamma^{5}{\partial W \over \partial \xi} =\sum_{n=0}^{5}A_n(\Gamma^{j})\gamma^n=0, 
\; \; \; \; \; \;\; \; \; \, j \in \{0,1,2,3,4,5 \}, 
\label{4.16c}
\end{equation}
\begin{equation}
\gamma^{5}\Gamma^{5}{1 \over \eta}{\partial W \over \partial \eta} = \sum_{n=0}^{5}B_n(\Gamma^{j})\gamma^n=0, 
\; \; \;\; \; \; j \in \{0,1,2,3,4,5 \}, 
\label{4.17c}
\end{equation}
where the coefficients $A_n(\Gamma^j)$ are given by \cite{testbed}
\begin{equation}
\begin{split}
A_5(\Gamma^j) & \equiv  \Gamma^5 \left[ 1+\dfrac{\Omega^2}{2c^2} \left(\eta^2+ \xi^2 \right) \right] \xi \Omega^2 
+ \Gamma^4 \left[ \dfrac{3 \xi}{l}-\dfrac{7}{2 \left(1+\rho \right)}\right]R_{\beta} \Omega^2 \\
& +  \Gamma^2 \left\{ \frac{1}{2} \Omega^2 \left[ \dfrac{7 l \xi}{\left(1+ \rho \right)} 
+ \dfrac{l^2 \left(2 \rho-1 \right)}{\left(1+ \rho \right)^2} -3 \left(\eta^2 + \xi^2 \right) \right] 
- c^2  \right\} \left(\xi - \dfrac{l}{(1+ \rho)} \right) \dfrac{R_{\beta}}{l^3} \\
& + \left[  \Gamma   \left(\dfrac{c R_{\beta} }{l^2} \right)^2  
+ \dfrac{3}{2} \dfrac{R_{\beta} \eta^2}{\left(1+ \rho \right)^2}\dfrac{\Omega^2}{l^3} \right]  
\left( \xi - \dfrac{l}{(1+ \rho)} \right),
\end{split}
\end{equation}
\begin{equation}
A_4(\Gamma^j) \equiv \Gamma^5 \left( \dfrac{3 \xi R_{\alpha} \Omega^2}{l} \right) 
+ \Gamma^2 \left[  \xi \left( 1 + \rho \right) - l  \right] \dfrac{\Omega^2 R_{\beta}}{l \left( 1+ \rho \right)^2},
\end{equation}
\begin{equation}
A_3(\Gamma^j)=0,
\end{equation}
\begin{equation}
\begin{split}
A_2(\Gamma^j) & \equiv  - \Gamma^5 \biggl\{ 2c^2 R_{\alpha} \left( 1+ \rho \right)^2 
+ \Omega^2 \bigl[ 7 l R_{\beta} \xi \left(1 + \rho \right) + 3 R_{\alpha} \left( \eta^2 
+ \xi^2 \right) \left(1+ \rho \right)^2 \\
& + l^2 R_{\beta} \left(\rho-2 \right) \bigr] \biggr\}  \dfrac{\left[ \xi + \rho \left( l+ \xi \right) \right] }  {2 l^3 \left(1+ \rho \right)^3} 
+ \Gamma^4 \left[ 2 l^2 R_{\beta} (1+ \rho) \right]   \dfrac{\Omega^2 \left[ \xi 
+ \rho \left( l+ \xi \right) \right] }  {2 l^3 \left(1+ \rho \right)^3},
\end{split}
\end{equation}
\begin{equation}
A_1(\Gamma^j) \equiv \Gamma^5 \left( \xi + l \dfrac{\rho}{(1+\rho)} \right) \left(\dfrac{c R_{\alpha}}{l^2} \right)^2,
\end{equation}
\begin{equation}
A_0(\Gamma^j) \equiv \Gamma^5 \left[ \xi + \left( l + \xi \right) \rho \right] 
\dfrac{3 R_{\alpha}\rho \eta^2 \Omega^2 }{2 l^3 \left(1+\rho \right)^3},
\end{equation}
whereas the coefficients $B_n(\Gamma^j)$ are defined by \cite{testbed}
\begin{table}
\centering
\caption[General relativity corrections on the position of Newtonian non-collinear Lagrangian points]{General relativity corrections on the position of Newtonian non-collinear Lagrangian points obtained by solving numerically Eqs. (\ref{4.16c}) and (\ref{4.17c}).}
{
\renewcommand\arraystretch{2.0}
\begin{tabular}{|c|c|}
\hline
\multicolumn{2}{|c|}{General relativity corrections on Newtonian non-collinear Lagrangian points}\\
\hline
\; \;  $L_i$ \; \;  &  Corrections\\
\cline{1-2} 
& $r_{GR}-r_{cl}=-0.0139 \; {\rm mm}$ \\
\cline{2-2}
 $L_4$ & $\xi_{GR}-\xi_{cl}=2.74\; {\rm mm}$ \\
& $\eta_{GR}-\eta_{cl}=-1.60\; {\rm mm}$ \\
\hline
& $r_{GR}-r_{cl}=-0.0139 \; {\rm mm}$ \\
\cline{2-2}
 $L_5$ & $\xi_{GR}-\xi_{cl}=2.74\; {\rm mm}$ \\
& $\eta_{GR}-\eta_{cl}=1.60\; {\rm mm}$ \\
\hline
\end{tabular}
\label{noncoll_GRcorrections_tab}
}
\end{table}
\begin{equation}
\begin{split}
B_5(\Gamma^j) & \equiv  \Gamma^5 \left[ 2 c^2 + \left( \eta^2 + \xi^2 \right) \Omega^2 \right] 
\dfrac{\Omega^2}{2 c^2} + \Gamma^4 \left( \dfrac{3 R_{\beta}\Omega^2}{l} \right)  +  \Gamma^2 \biggl\{ - \left( 1 + \rho \right)^2 [ 2 c^2 \\
&  + 3 \Omega^2 \left( \eta^2  + \xi^2 \right) ] + \Omega^2 \left[ 7 l \xi \left( 1 + \rho \right) 
+l^2 \left( 2 \rho - 3 \right) \right] \biggr\}  \dfrac{R_{\beta}}{2l^3 \left(1+\rho \right)^2} \\
& + \Gamma \left(\dfrac{c R_{\beta}}{l^2} \right)^2
+\dfrac{3 R_{\beta}\eta^2 \Omega^2}{2 l^3 \left(1+ \rho \right)^2},
\end{split}
\end{equation}
\begin{equation}
B_4(\Gamma^j) \equiv \Gamma^5 \left( \dfrac{3 R_{\alpha} \Omega^2}{l} \right) 
+ \Gamma^2 \left[ \dfrac{R_{\beta} \Omega^2}{l (1+ \rho)} \right],
\end{equation}
\begin{equation}
B_3(\Gamma^j)=0,
\end{equation}
\begin{equation}
\begin{split}
B_2(\Gamma^j)& \equiv  -\Gamma^5 \left\{\dfrac{c^2 R_{\alpha}}{l^3}+ \Omega^2 \left[ \dfrac{7 R_{\beta}\xi}{2l^2(1+\rho)} 
+ \dfrac{3 R_{\alpha}\left( \eta^2 + \xi^2 \right)}{2 l^3}- \dfrac{R_{\beta}}{l (1+\rho)^2} 
+  \dfrac{3 R_{\beta} \rho}{2 l (1+ \rho)^2} \right] \right\}  \\
& +  \Gamma^4 \left[ \dfrac{\Omega^2 R_{\beta}}{l (1+\rho)} \right],
\end{split}
\end{equation}
\begin{equation}
B_1(\Gamma^j) \equiv \Gamma^5 \left( \dfrac{c R_{\alpha}}{l^2} \right)^2,
\end{equation}
\begin{equation}
B_0(\Gamma^j) \equiv \Gamma^5 \left[ \dfrac{3}{2} \dfrac{R_{\beta} \rho \eta^2 \Omega^2}{l^3 (1+\rho^2)} \right].
\end{equation}
The planetoid coordinates are eventually expressed, from the definition (\ref{4.7c}), in the form
\begin{equation}
\begin{split}
& \xi={l \over 2}\left[(\gamma^{2}-\Gamma^{2})+{(1-\rho)\over (1+\rho)}\right], \\
& \eta=\pm l \sqrt{\gamma^{2}-{1 \over 4}(\gamma^{2}-\Gamma^{2}+1)^{2}}.
\end{split}
\end{equation}
By numerical analysis of Eqs. (\ref{4.16c}) and (\ref{4.17c}) we have found that, in the Earth-Moon system, the only solution where both $\gamma$ and $\Gamma$
are different from zero is given by
\begin{equation}
\begin{split}
& \gamma=0.99999999999996386756, \\
& \Gamma=0.99999999999284192083.
\end{split}
\end{equation}
These values lead to a tiny departure from the equilateral triangle picture of Newtonian theory \cite{testbed}. This effect was first predicted in Ref. \cite{Krefetz1967} and it has been showed that it reflects the expected emission of gravitational radiation in Ref. \cite{Brumberg}. Thus, the resulting values of the distance from the Earth and the Moon of the planetoid turn out to be
\begin{equation}
\begin{split}
& r_{GR}=\gamma l=3.8439999999998611069 \times 10^{8} {\rm m}, \\
& s_{GR}=\Gamma l=3.8439999999724843437 \times 10^{8} {\rm m},
\end{split}
\end{equation}
and
\begin{equation}
 r_{GR} - s_{GR}= 2.74 \; {\rm mm}.
\end{equation}
The corrections with respect to the corresponding Newtonian values (see Eqs. (\ref{5.20a})--(\ref{5.20a_bis})) are written in Tab. \ref{noncoll_GRcorrections_tab}. At this stage, we can compare corrections of Tab. \ref{noncoll_GRcorrections_tab} with those obtained through the method adopted by the authors of 
Ref. \cite{Yamada} and that we have outlined at the beginning of this section. We have found that the correction on the $\xi$-coordinate has got the same sign and the 
same magnitude as the one obtained with the pattern followed in Ref. \cite{Yamada}, while the correction on the $\eta$-coordinate has got only the same sign, because the magnitude is three times bigger. Anyway, it is interesting to note the fact that two different approaches give exactly the same correction of the $\xi$-coordinate.

Now we turn our attention to the collinear Lagrangian points. The position $L_1$, $L_2$, and $L_3$ is described by the system of equations
\begin{equation}
\begin{dcases}
& \dfrac{\partial W}{\partial \xi}=0, \\
& \eta=0.
\end{dcases}
\end{equation}
Bearing in mind the outcomes of Sec. \ref{coll_Sec} (see Eqs. (\ref{3.4b})--(\ref{3.7b})), 
we know that the vanishing of the $\eta$-coordinate implies that
\begin{equation}
\xi = \epsilon r - l \dfrac{\rho}{(1+\rho)}, \; \; \; \; (\epsilon= \pm 1), \label{4.36c}
\end{equation}
which in turn leads to the condition 
\begin{equation}
s= \pm (r- \epsilon l). \label{4.37c}
\end{equation}
\begin{table}
\centering
\caption[General relativity details of Lagrangian points]{The values of the distances from the Earth and of the coordinates of the Lagrangian points obtained within the context of general relativity.}
{
\renewcommand\arraystretch{2.0}
\begin{tabular}{|c|c|}
\hline
\multicolumn{2}{|c|}{General relativity details of Lagrangian points}\\
\hline
\; \;  $L_i$ \; \;  &  Details   \\
\cline{1-2}
&  $r_1=3.263762881740760 \times 10^8 \; {\rm m}$ \\ 
\cline{2-2}
$L_1$ & $\xi_1 =   3.217044369763247  \times 10^8 \; {\rm m} $    \\
& $\eta_1= 0 \; {\rm m} $  \\
\cline{1-2}
&  $r_2= 4.489205600341480\times 10^8 \; {\rm m}$ \\ 
\cline{2-2}
$L_2$ & $\xi_2 =  4.442487088363968  \times 10^8 \; {\rm m} $    \\
& $\eta_2= 0 \; {\rm m} $  \\
\cline{1-2}
&  $r_3= 3.816747156939217 \times 10^8 \; {\rm m}$ \\ 
\cline{2-2}
$L_3$ & $\xi_3 = -3.863465668916729   \times 10^8 \; {\rm m} $    \\
& $\eta_3= 0 \; {\rm m} $  \\
\cline{1-2}
&  $r_4=  3.843999999999861 \times 10^8 \; {\rm m}$ \\ 
\cline{2-2}
$L_4$ & $\xi_4 = 1.875281488049864      \times 10^8 \; {\rm m} $    \\
& $\eta_4= 3.329001652131416    \times 10^8 \; {\rm m} $  \\
\cline{1-2}
&  $r_5=  3.843999999999861 \times 10^8 \; {\rm m}$ \\ 
\cline{2-2}
$L_5$ & $\xi_5 = 1.875281488049864      \times 10^8 \; {\rm m} $    \\
& $\eta_5=- 3.329001652131416    \times 10^8 \; {\rm m} $  \\
\hline
\end{tabular} 
}
\label{GRdetails_tab}
\end{table} 
\begin{table}
\centering
\caption[General relativity corrections on the position of Newtonian collinear Lagrangian points]{General relativity corrections on the position of Newtonian collinear Lagrangian points obtained by solving numerically Eqs. (\ref{4.38c}) and (\ref{4.50c}).}
{
\renewcommand\arraystretch{2.0}
\begin{tabular}{|c|c|}
\hline
\multicolumn{2}{|c|}{General relativity corrections on Newtonian collinear Lagrangian points}\\
\hline
\; \;  $L_i$ \; \;  &  Corrections\\
\cline{1-2} 
$L_1$ & $r_{GR}-r_{cl}=0.188 \; {\rm mm}$ \\
& $\xi_{GR}-\xi_{cl}=0.188 \; {\rm mm}$ \\
\hline
$L_2$ & $r_{GR}-r_{cl}=-0.320 \; {\rm mm}$ \\
& $\xi_{GR}-\xi_{cl}=-0.320 \; {\rm mm}$ \\
\hline
$L_3$ & $r_{GR}-r_{cl}=-0.0406 \; {\rm mm}$ \\
& $\xi_{GR}-\xi_{cl}= 0.0406 \; {\rm mm}$ \\
\hline
\end{tabular}
\label{coll_GRcorrections_tab}
}
\end{table}
If we substitute relations (\ref{4.36c}) and (\ref{4.37c}) into Eq. (\ref{4.12c}) and initially adopt the choice $s=(r- \epsilon l)$, we obtain an algebraic tenth degree equation where the only unknown is the distance $r$ of the planetoid from the Earth. By setting, as before, $r=\gamma l$, this equation can be written down as \cite{testbed}
\begin{equation}
\sum_{n=0}^{10} C_n \gamma^n = 0, \label{4.38c}
\end{equation} 
where \cite{testbed}
\begin{equation}
C_{10} \equiv 1,
\end{equation}
\begin{equation}
C_{9} \equiv - \dfrac{(7 \rho +4)}{\epsilon (1+ \rho)},
\end{equation}
\begin{equation}
C_{8} \equiv \dfrac{2 c^2}{\Omega^2 l^2} + \dfrac{3 (7 \rho^2 + 8 \rho +2)}{(1+\rho)^2},
\end{equation}
\begin{equation}
\begin{split}
C_{7}  & \equiv  - \dfrac{1}{\epsilon (1+ \rho)}  \bigg \{ \dfrac{c^2}{l^3 \Omega^2} \left[ 2l (5 \rho +4)
-3 \epsilon (1+ \rho) (R_{\alpha}+R_{\beta}) \right] \\
&  +\dfrac{1}{(1+ \rho)^2} \left[ \rho^2 (13 \rho + 12)+ 2 (1 + \rho)^2 (11 \rho +2) \right] \bigg  \},
\end{split}
\end{equation}
\begin{equation}
\begin{split}
C_{6} & \equiv  \dfrac{c^2}{\epsilon \Omega^2 l^3} \left \{ 12 (l \epsilon -R_{\alpha} - R_{\beta})
+ \rho \left[ 20 l \epsilon - 12  (R_{\alpha}+R_{\beta}) \right] \right \}   \\
&+ \dfrac{1}{(1+ \rho)^3} \left[ 4 \rho^3 + (1 + \rho) (31 \rho^2 + 14 \rho + 1) \right],
\end{split}
\end{equation}
\begin{equation}
\begin{split}
C_{5} & \equiv  -2\dfrac{c^4  (R_{\alpha}+R_{\beta}) }{l^5 \Omega^4} + \dfrac{c^2}{l^3 \Omega^2 (1+ \rho)^2} 
\bigg [ - 4 l \epsilon (1 + \rho) (5 \rho +2) + 3 R_{\alpha} (5 \rho^2 +12 \rho +6)  \\
& +  R_{\beta} (18 \rho^2 + 44 \rho + 23) \bigg ] -\dfrac{3 \epsilon \rho}{(1+ \rho)^3} (7 \rho^2 + 6 \rho +1), 
\end{split}
\end{equation}
\begin{equation}
\begin{split}
C_{4} & \equiv  \dfrac{2 c^2}{\epsilon l^3 \Omega^2} \bigg \{ [ 2l (R_{\beta}-2R_{\alpha}) 
- \epsilon ((R_{\alpha})^{2}+(R_{\beta})^{2}) ] 
\left( \dfrac{-c^2}{l^3 \Omega^2} \right) + \dfrac{R_{\beta}}{(1+\rho)^2} 
[ 5 \rho^2 + 2 \rho + 5 \\
&  -2 \epsilon (1 + \rho ) ]  +  \dfrac{1}{(1+\rho)^2} [ l \epsilon (1+\rho)(1+5 \rho) - 6 R_{\alpha} (1+2 \rho) ] \bigg \} 
+ \dfrac{\rho^2}{(1+\rho)^3} (7 \rho + 3),
\end{split}
\end{equation}
\begin{equation}
\begin{split}
C_{3} & \equiv  -\dfrac{2 c^4}{l^6 (1+\rho)^3 \Omega^4 \epsilon} \bigl[ R_{\beta}(R_{\beta}+l \epsilon) 
+ R_{\alpha}(4 R_{\alpha}+ 6 l \epsilon) \bigr] -\dfrac{c^2}{l^3 (1+ \rho)^2 \Omega^2 \epsilon} \bigl\{ 2 l \rho (1+ \rho) \\ 
&+ 3 R_{\alpha} \epsilon (5 \rho^2 -2 \rho -1) + R_{\beta} [10(1+ \rho) -  \epsilon (3 \rho^2 
+ 44 \rho + 18)] \bigr\} -\dfrac{\rho^3}{(1+\rho)^3 \epsilon},
\end{split}
\end{equation}
\begin{equation}
C_{2} \equiv \dfrac{c^2}{l^3 \Omega^2} \left \{ \dfrac{4 c^2 R_{\alpha}}{l^3 \Omega^2} (3 R_{\alpha} 
+ 2 l \epsilon) + \dfrac{1}{(1+\rho)^2} \{ 12 \epsilon R_{\alpha}\rho^2 
-8 R_{\beta} [ \epsilon (1+ 3 \rho) -(1+\rho)] \} \right \}, 
\end{equation}
\begin{equation}
C_{1} \equiv -\dfrac{c^2 \epsilon}{l^3\Omega^2 (1+\rho)^2} \left \{ [2 c^2 R_{\alpha} (4 R_{\alpha}+l \epsilon)(1+\rho)^2] 
\dfrac{\epsilon}{l^3 \Omega^2}+3 \rho^2 R_{\alpha} + 2 R_{\beta} [\epsilon (1+\rho)-(1+3\rho)] \right \},
\end{equation}
\begin{equation}
C_{0} \equiv 2\left(\dfrac{ c^2 R_{\alpha}}{l^3 \Omega^2}\right)^2,
\end{equation}
whereas in the other case, i.e., $s=-(r-\epsilon l)$, we end up with the algebraic equation \cite{testbed}
\begin{equation}
\sum_{n=0}^{10} D_n \gamma^n = 0, \label{4.50c}
\end{equation} 
with \cite{testbed}
\begin{equation}
D_k=C_k \; \; \;\; \; \;   {\rm if} \; \;  k=10,9,8,0,
\end{equation}
\begin{equation}
D_7 \equiv C_7 - 6 \dfrac{c^2 R_{\beta}}{l^3 \Omega^2},
\end{equation}
\begin{equation}
D_6 \equiv C_6 + 24 \dfrac{c^2 R_{\beta}}{l^3 \Omega^2 \epsilon},
\end{equation}
\begin{equation}
D_5 \equiv C_5 - \dfrac{2 c^2 R_{\beta}}{l^5 \Omega^4 (1+\rho)^2}
[l^2 \Omega^2 (18 \rho^2+38 \rho+21)-2c^2 (1+\rho)^2],
\end{equation}
\begin{equation}
D_4 \equiv C_4 - \dfrac{2 c^2 R_{\beta}}{l^5 \Omega^4 (1+\rho)^2 \epsilon} \{ 4 c^2 (1+\rho)^2 
+ 2 \Omega^2 l^2 [2 \epsilon (1+\rho)-(6 \rho^2+14 \rho+9)] \},
\end{equation}
\begin{equation}
D_3 \equiv C_3 + \dfrac{2 c^2 R_{\beta}}{l^3 \Omega^2} \left [\dfrac{2 c^2 }{l^2 \Omega^2}+\dfrac{10 }
{(1+\rho) \epsilon}-\dfrac{1}{ (1+\rho)^2} (3 \rho^2 + 8 \rho +6) \right], 
\end{equation}
\begin{equation}
D_2 \equiv C_2 -\dfrac{16 c^2 R_{\beta}}{l^3 \Omega^2 (1+\rho)},
\end{equation}
\begin{equation}
D_1 \equiv C_1+\dfrac{4 c^2 R_{\beta}\epsilon}{l^3 \Omega^2 (1+ \rho)}.
\end{equation}
The details about all Lagrangian points are summarized in Tab. \ref{GRdetails_tab}. In particular, the values of the distance of the planetoid from the Earth at the libration points $L_1$, $L_2$, and $L_3$, obtained through the solution of Eqs. (\ref{4.38c}) and (\ref{4.50c}), are given by
\begin{equation}
\begin{split}
& r_{1,GR}= 3.2637628817407598555 \times 10^8 \;  {\rm m},\\
& r_{2,GR}=  4.4892056003414800050  \times 10^8 \;  {\rm m},\\
& r_{3,GR}= 3.8167471569392170594 \times 10^8 \;  {\rm m},
\end{split}
\label{4.59c}
\end{equation}
respectively. The corrections with respect to the corresponding classical values (\ref{dist_coll_Newton}) and (\ref{coord_coll_Newton}) are written in Tab. \ref{coll_GRcorrections_tab}.

Interestingly, the correction on the position of the Lagrangian point $L_1$ is exactly the same as the one calculated  with the method of Ref. \cite{Yamada10} described at the beginning of this section\footnote{As shown in Ref. \cite{Yamada10}, the general
relativity corrections to $L_{1},L_{2}$ may be of order $30$ meters in the Sun-Jupiter system (we will show the corrections resulting from our model in the concluding remarks of this thesis). However, compared to the Earth-Moon system, a mission to test this effect at Jupiter would be exceedingly more expensive
and complex to realize and could not even benefit from the use of accurate, direct laser ranging (Sec. \ref{Laser_Ranging_Sec}) from Earth due
to the large distance. The effect of the extremely harsh Jupiter radiation environment on the test spacecraft
(planetoid) should also be considered to evaluate its impact on the integrity of the spacecraft and, therefore,
the duration of the positioning measurements.}. We believe that, according to the definitions involving the ratio 
of the distances of the planetoid from the primaries given in Ref. \cite{Yamada10} (see Eqs. (\ref{Yamada1})--(\ref{Yamada3})), the equations resulting from the 
application of the method developed by the authors of Ref. \cite{Yamada10} are well suited to describe only the position of $L_1$, and the agreement with the corrections presented here is a clue supporting our opinion. 

\section{The new quantum theory} \label{New quantum theory Sec}

The analysis of the previous section prepares the ground for a more appropriate definition and evaluation of quantum corrections of Lagrangian points, when the underlying classical theory of gravity is Einstein's general relativity \cite{testbed}. 

\subsection{Quantum effects on Lagrangian points}

It should be clear from the analysis of previous sections that the Lagrangian function underlying the classical limit of the quantum theory we are going to set up is represented by Eq. (\ref{5.6c}). Then, this Lagrangian function represents the staring point of our new quantum framework.

Consider the quantum corrected potential (\ref{1.2b}). By dividing it by the product of the mass of one of the two bodies and the square of the speed of light, we obtain straightforwardly (cf. Eqs. (\ref{1.3b}) and (\ref{1.4b}))
\begin{equation}
 {V_{Q}(r) \over c^{2}m_{B}} =
-{R_{A}\over r}\left[ 1+ \left(\kappa_{1}{(R_{A}+R_{B})\over r}
+\kappa_{2}{(l_{P})^{2}\over r^{2}} \right) +{\rm O}(G^{2})\right],
\label{4.1c}
\end{equation}
the dimensionless ratio $R_{A}/r$ representing obviously a classical term, i.e., the Newtonian potential. We now bear in mind that, in light of (\ref{4.1c}), the dimensionless ratio
\begin{equation}
U_{\alpha}(r) \equiv {R_{\alpha}\over r}=U_{\alpha},
\label{5.7c}
\end{equation}
where, as we know, $R_{\alpha} \equiv {G \alpha \over c^{2}}$ is the gravitational radius of the of the Earth, gets replaced by (or mapped into)
\begin{equation}
\begin{split}
{V}_{\alpha}(r) & = 
\left[1+\kappa_{2}{(l_{P})^{2}\over r^{2}}\right]U_{\alpha}(r)
+\kappa_{1}\left(1+{R_{m}\over R_{\alpha}}\right)(U_{\alpha}(r))^{2}
+{\rm O}(G^{3}) \\
& \sim  \left[1+\kappa_{2}{(l_{P})^{2}\over r^{2}}\right]U_{\alpha}(r)
+\kappa_{1}(U_{\alpha}(r))^{2},
\end{split}
\label{5.8c}
\end{equation}
because the gravitational radius $R_{m}$ of the planetoid or laser ranging test mass (see Sec. \ref{Laser_Ranging_Sec}) is indeed much smaller than $R_{\alpha}$. The same holds for the dimensionless ratio
\begin{equation}
U_{\beta}(s) \equiv {R_{\beta}\over s}=U_{\beta},
\end{equation}
and its effective-gravity counterpart
\begin{equation}
\begin{split}
{V}_{\beta}(s) & = \left[1+\kappa_{2}{(l_{P})^{2}\over s^{2}}\right]U_{\beta}(s)
+\kappa_{1}\left(1+{R_{m}\over R_{\beta}}\right)
(U_{\beta}(s))^{2}+{\rm O}(G^{3}) \\
& \sim  \left[1+\kappa_{2}{(l_{P})^{2}\over s^{2}}\right]U_{\beta}(s)
+\kappa_{1}(U_{\beta}(s))^{2},
\end{split}
\label{5.10c}
\end{equation}
where $R_{\beta} \equiv {G \beta \over c^{2}}$. By inserting the effective-gravity map defined by Eqs. (\ref{5.7c})--(\ref{5.10c}) into the Lagrangian (\ref{5.6c}), we are led to consider the effective-gravity Lagrangian \cite{testbed}
\begin{equation}
\begin{split}
\mathcal{L}_{V}&= {c^{2}\over 2} \biggr \{ 1-2({V}_{\alpha}+{V}_{\beta})
-{\Omega^{2}\over c^{2}}(\xi^{2}+\eta^{2})
+2 \left[({V}_{\alpha})^{2}+({V}_{\beta})^{2}\right]- 2{(R_{\alpha}+R_{\beta})\over l^{3}}(\xi^{2}+\eta^{2})
({V}_{\alpha}+{V}_{\beta})\\
& +4{V}_{\alpha}{V}_{\beta} + {(2-\rho)\over (1+\rho)}{R_{\beta}\over l}{V}_{\alpha}
+{(2 \rho-1)\over (1+\rho)}{R_{\alpha}\over l}{V}_{\beta}
-7{\xi \over l^{2}}(R_{\beta}{V}_{\alpha}
-R_{\alpha}{V}_{\beta}) \\
&+ (1+\rho)^{-1}{\eta^{2}\over l} \left[\rho {R_{\beta}\over (R_{\alpha})^{2}}
({V}_{\alpha})^{3}
+{R_{\alpha}\over (R_{\beta})^{2}}({V}_{\beta})^{3}\right] \biggr \} \\
&- {1 \over 2}\Bigr({\dot \xi}^{2}+{\dot \eta}^{2}+{\dot \zeta}^{2}\Bigr)
\Bigr[1+2({V}_{\alpha}+{V}_{\beta})\Bigr]
+\Omega \eta {\dot \xi}\Bigr[1+2({V}_{\alpha}+{V}_{\beta})\Bigr]  \\
&- \Omega \xi {\dot \eta} \Bigr[1+2({V}_{\alpha}+{V}_{\beta})\Bigr]
-4{\Omega^{2}l \over (1+\rho)}{\dot \eta}(\rho {V}_{\alpha}-{V}_{\beta}),
\end{split}
\label{5.11c}
\end{equation}
and the only non-trivial Euler-Lagrange equations for the planar restricted three-body problem are
\begin{equation}
\begin{split}
& {{\rm d}\over {\rm d}t}\left({\partial \mathcal{L}_{V}\over \partial {\dot \xi}}\right)
-{\partial \mathcal{L}_{V}\over \partial \xi}=0, \\
& {{\rm d}\over {\rm d}t}\left({\partial \mathcal{L}_{V}\over \partial {\dot \eta}}\right)
-{\partial \mathcal{L}_{V}\over \partial \eta}=0.
\end{split}
\label{5.12c}
\end{equation}
\begin{table}
\centering
\caption[Distances from the Earth and planar coordinates of the planetoid at all Lagrangian points in the new quantum regime]{Distances $r_i$ from the Earth and planar coordinates $(\xi_i,\eta_i)$ of the planetoid at all Lagrangian points $L_{i}$ in the new quantum regime obtained through the Lagrangian $\mathcal{L}_V$ (\ref{5.11c}) for the three different potentials.}
{\relsize{-2.49}
\renewcommand\arraystretch{2.0}
\begin{tabular}{|c|c|c|c|}
\hline
\multicolumn{4}{|c|}{Quantum details of Lagrangian points}\\
\hline
\; \;  $L_i$ \; \;  &  One-particle reducible &  Scattering & Bound-states \\
\cline{1-4}
& $r_1= 3.263762881728428  \times 10^8 \; {\rm m}$& $r_1= 3.263762881777756 \times 10^8 \;  
{\rm m}$ & $r_1= 3.263762881734594 \times 10^8 \; {\rm m} $ \\ 
\cline{2-4}
$L_1$ & $\xi_1 = 3.217044369750916  \times 10^8 \; {\rm m} $  & $\xi_1 = 3.217044369800243  \times 10^8 \;  
{\rm m}$   & $\xi_1 = 3.217044369757081 \times 10^8 \;  {\rm m}$  \\
& $\eta_1= 0\; {\rm m} $ &   $\eta_1=0\; {\rm m} $ &  $\eta_1=0\; {\rm m} $ \\
\hline
& $r_2= 4.489205600333647   \times 10^8 \; {\rm m} $ & $r_2= 4.489205600364979 \times 10^8 \;  
{\rm m}$ & $r_2= 4.489205600337563 \times 10^8 \; {\rm m}$ \\ 
\cline{2-4}
$L_2$ & $\xi_2 = 4.442487088356134  \times 10^8 \; {\rm m}$  & $\xi_2 = 4.442487088387467 \times 10^8 \;  
{\rm m} $  & $\xi_2 = 4.442487088360051  \times 10^8 \;  {\rm m}$  \\
& $\eta_2= 0\; {\rm m} $ &   $\eta_2= 0\; {\rm m} $&  $\eta_2= 0\; {\rm m} $ \\
\hline
& $r_3= 3.816747156909591 \times 10^8 \; {\rm m}$ & $r_3= 3.816747157028094 \times 10^8 \;  
{\rm m}$ & $r_3= 3.816747156924404 \times 10^8 \; {\rm m}$ \\ 
\cline{2-4}
$L_3$ & $\xi_3 =-3.863465668887104 \times 10^8 \; {\rm m}$  & $\xi_3 =-3.863465669005607 \times 10^8 \; 
{\rm m} $  & $\xi_3 = -3.863465668901917 \times 10^8 \; {\rm m}$  \\
& $\eta_3= 0\; {\rm m} $ &   $\eta_3= 0\; {\rm m} $ &  $\eta_3= 0\; {\rm m} $ \\
\hline
& $r_4= 3.843999999970295  \times 10^{8} {\rm m}$ & $r_4= 3.84400000008856  \times 10^{8} 
{\rm m}$ & $r_4 = 3.843999999985078  \times 10^{8} {\rm m}$ \\ 
\cline{2-4}
$L_4$ & $\xi_4 = 1.875281488020662  \times 10^8 \;  {\rm m} $  & $\xi_4 =1.875281488137470 \times 10^8 \;  {\rm m}$  & $\xi_4 = 1.875281488035263 \times 10^8 \;  {\rm m} $  \\
& $\eta_4=  3.329001652114136 \times 10^8 \;  {\rm m} $  &   $\eta_4=3.329001652183255 \times 10^8 \;  
{\rm m} $ &  $\eta_4= 3.329001652122776 \times 10^8 \;  {\rm m} $ \\
\hline
& $r_5= 3.843999999970295 \times 10^{8} {\rm m}$ & $r_5= 3.84400000008856 \times 10^{8} 
{\rm m}$ & $r_5= 3.843999999985078 \times 10^{8} {\rm m}$ \\ 
\cline{2-4}
$L_5$ & $\xi_5 = 1.875281488020662  \times 10^8 \;  {\rm m} $  & $\xi_5 = 1.875281488137470 \times 10^8 \;  
{\rm m}$  & $\xi_5 =  1.875281488035263 \times 10^8 \;  {\rm m}$  \\
& $\eta_5= -3.329001652114136 \times 10^8 \;  {\rm m} $ &   $\eta_5=-3.329001652183255 \times 10^8 \;  
{\rm m}$ &  $\eta_5= - 3.329001652122776 \times 10^8 \;  {\rm m} $ \\
  \hline 
\end{tabular} 
\label{details_new_quantum_tab}
}
\end{table} 

An important issue must be stressed at this point. In fact, in the previous chapters we have inserted the effective-gravity map
\begin{equation}
(U_{\alpha},U_{\beta}) \rightarrow ({V}_{\alpha},{V}_{\beta}),
\label{5.12bisc}
\end{equation}
in the Lagrangian of Newtonian gravity for the restricted planar three-body problem, whereas we are here inserting the same map in the Lagrangian of general relativity for the restricted three-body problem, i.e., Eq. (\ref{5.6c}). The metric tensor with components (\ref{5.1c})--(\ref{5.5c}) describes, within the framework of general relativity, a tiny departure from the Newtonian treatment of the restricted planar three-body problem. At that stage, one can recognize that many
Newtonian-potential terms, written as $R_{\alpha}/r$ and $R_{\beta}/s$, occur therein; for each of them, we apply the map (\ref{5.12bisc}) to find what we call a quantum-corrected Lagrangian. 
\begin{table}
\centering
\caption[Quantum corrections on the relativistic position of Lagrangian points]{Quantum corrections on the relativistic position of Lagrangian points for three different types of potential.}
{
\renewcommand\arraystretch{2.0}
\begin{tabular}{|c|c|c|c|}
\hline
\multicolumn{4}{|c|}{Quantum corrections on Lagrangian points}\\
\hline
\; \;  $L_i$ \; \;  &  One-particle reducible &  Scattering & Bound-states \\
\cline{1-4} 
& $r_{Q}-r_{GR}=-1.23 \; {\rm mm}$ & $r_{Q}-r_{GR}=3.70 \; {\rm mm}$ & $r_{Q}-r_{GR}= -0.617\; {\rm mm}$  \\
\cline{2-4}
 $L_1$ & $\xi_{Q}-\xi_{GR}=-1.23 \; {\rm mm}$ & $\xi_{Q}-\xi_{GR}=3.70\; {\rm mm}$ & $\xi_{Q}-\xi_{GR}=-0.617 \; {\rm mm}$ \\
\cline{1-4} 
& $r_{Q}-r_{GR}= -0.783\; {\rm mm}$ & $r_{Q}-r_{GR}=2.35 \; {\rm mm}$ & $r_{Q}-r_{GR}=-0.392 \; {\rm mm}$  \\
\cline{2-4}
 $L_2$ & $\xi_{Q}-\xi_{GR}=-0.783 \; {\rm mm}$ & $\xi_{Q}-\xi_{GR}=2.35\; {\rm mm}$ & $\xi_{Q}-\xi_{GR}=-0.392\; {\rm mm}$ \\
\cline{1-4} 
& $r_{Q}-r_{GR}=-2.96 \; {\rm mm}$ & $r_{Q}-r_{GR}= 8.89 \; {\rm mm}$ & $r_{Q}-r_{GR}=-1.48 \; {\rm mm}$  \\
\cline{2-4}
 $L_3$ & $\xi_{Q}-\xi_{GR}=2.96\; {\rm mm}$ & $\xi_{Q}-\xi_{GR}=-8.89\; {\rm mm}$ & $\xi_{Q}-\xi_{GR}=1.48\; {\rm mm}$ \\
\cline{1-4} 
& $r_{Q}-r_{GR}=-2.96 \; {\rm mm}$ & $r_{Q}-r_{GR}= 8.87 \; {\rm mm}$ & $r_{Q}-r_{GR}=-1.48 \; {\rm mm}$  \\
\cline{2-4}
 $L_4$ & $\xi_{Q}-\xi_{GR}=-2.92\; {\rm mm}$ & $\xi_{Q}-\xi_{GR}=8.76 \; {\rm mm}$ & $\xi_{Q}-\xi_{GR}=-1.46\; {\rm mm}$ \\
& $\eta_{Q}-\eta_{GR}=-1.73\; {\rm mm}$ & $\eta_{Q}-\eta_{GR}=5.18\; {\rm mm}$ & $\eta_{Q}-\eta_{GR}=-0.864\; {\rm mm}$ \\
\cline{1-4} 
& $r_{Q}-r_{GR}= -2.96\; {\rm mm}$ & $r_{Q}-r_{GR}= 8.87 \; {\rm mm}$ & $r_{Q}-r_{GR}= -1.48 \; {\rm mm}$  \\
\cline{2-4}
 $L_5$ & $\xi_{Q}-\xi_{GR}=-2.92 \; {\rm mm}$ & $\xi_{Q}-\xi_{GR}=8.76 \; {\rm mm}$ & $\xi_{Q}-\xi_{GR}=-1.46\; {\rm mm}$ \\
& $\eta_{Q}-\eta_{GR}=1.73 \; {\rm mm}$ & $\eta_{Q}-\eta_{GR}=-5.18 \; {\rm mm}$ & $\eta_{Q}-\eta_{GR}=0.864\; {\rm mm}$ \\
\hline
\end{tabular}
\label{new_quantum_corrections_tab}
}
\end{table}

Note however that in Ref. \cite{BDH2003}, where the authors derive quantum corrections to some known exact solutions in general relativity, they find that
these metrics differ from the classical metrics only for an additional term proportional to $(l_{P})^{2}$. Within such a framework, the running of $G$ at large $r$ has a universal character independent of masses, and there is no room left for $\kappa_{1}$ in the quantum-corrected Lagrangian.
The two schemes are conceptually different: quantum corrections to known {\it exact} solutions of
general relativity do not necessarily have the same nature as quantum corrections of metrics which represent 
solutions of the {\it linearized}  Einstein equations and which are used in turn to derive equations of motion 
of interacting bodies. The insertion of the map (\ref{5.12bisc}) in the general relativity Lagrangian (\ref{5.6c}) leads to
other terms quadratic in $U_{\alpha}$ and $U_{\beta}$, which are of the same order of those 
already present, and hence the resulting Euler-Lagrange equations (\ref{5.12c}) will lead to predictions
affected by $\kappa_{1}$. 

Now we set to zero all time derivatives of $\xi$ and $\eta$ in Eqs. (\ref{5.12c}), we define the real numbers $\gamma$ and $\Gamma$ as in Sec. \ref{Corrections on the position of Lagrangian points_Sec} and solve numerically the resulting algebraic system for such numbers. The values obtained through this method are written explicitly in Tabs. \ref{details_new_quantum_tab} and \ref{new_quantum_corrections_tab}. 

An important issue concerning both non-collinear and collinear Lagrangian points, consists in the fact that  we have checked numerically that the corrections of Tab. \ref{new_quantum_corrections_tab} do not change if we set $\kappa_2=0$ in the Euler-Lagrange equations (\ref{5.12c}), because $\kappa_{2}$ weighs the dimensionless ratios ${(l_{P})^{2}\over r^{2}}$ and ${(l_{P})^{2}\over s^{2}}$, which are extremely small at large values of $r$ and $s$.

Finally, we stress that, within this new scheme, quantum corrections on {\it Newtonian} quantities can be easily obtained through the algebraic sum of quantum corrections to general relativity (Tab. \ref{new_quantum_corrections_tab}) and general relativity corrections to Newtonian theory (Tabs. \ref{noncoll_GRcorrections_tab} and \ref{coll_GRcorrections_tab}) \cite{testbed}. 

\subsection{A possible choice of the quantum potential}

So far we have evaluated quantum corrections to Lagrangian points by employing the three different sets of quantum coefficients $\kappa_1$ and $\kappa_2$ occurring in the long-distance corrections to the Newtonian potential (\ref{1.2b}) (cf. Tab. \ref{kappa_tab}). Is it possible to recognize, within our new quantum framework, the most suitable choice of the quantum corrected potential to describe gravitational interactions involving (at least) three bodies in celestial mechanics? In this section we describe our proposal about this fundamental issue \cite{testbed}.

Equations (\ref{5.8c}) and (\ref{5.10c}) can be rewritten in the form
\begin{equation}
{V}_{\alpha} \sim U_{\alpha}+ \kappa_1 (U_{\alpha})^{2}
+{\rm O}(G^{2}),
\label{5.15c}
\end{equation}
\begin{equation}
{V}_{\beta} \sim U_{\beta}+\kappa_1 (U_{\beta})^{2}
+{\rm O}(G^{2}),
\label{5.16c}
\end{equation}
since
\begin{equation}
\begin{split}
& \kappa_{2}{(l_{P})^{2}\over r^{2}}U_{\alpha}={\rm O}(G^{2}),\\
& \kappa_{2}{(l_{P})^{2}\over s^{2}}U_{\beta}={\rm O}(G^{2}).
\end{split}
\end{equation}
If we insert the map (\ref{5.15c}) and (\ref{5.16c}) in the Lagrangian of {\it Newtonian} gravity for the restricted planar three-body problem (which can be obtained, as we pointed out before, from Eq. (\ref{2.11a}) in the limit $k_1 \rightarrow 0$, $k_2 \rightarrow 0$, $k_3 \rightarrow 0$), we find, with the notation adopted in this chapter, the effective potential \cite{testbed}
\begin{equation}
W^{\prime}={\omega^{2}\over 2}(\xi^{2}+\eta^{2})+c^{2}\Bigr[(U_{\alpha}+U_{\beta})
+\kappa_{1}((U_{\alpha})^{2}+(U_{\beta})^{2})\Bigr]+{\rm O}(G^{2}),
\label{5.13c}
\end{equation}
whereas we have seen that general relativity pattern yields the effective potential (\ref{4.9c}), expressible in the form
\begin{equation}
\begin{split}
W&={\Omega^{2}\over 2}(\xi^{2}+\eta^{2})+c^{2}\Bigr[(U_{\alpha}+U_{\beta})
-{1\over 2}((U_{\alpha})^{2}+(U_{\beta})^{2})\Bigr]+{\rm O}(G^{2})  \\
& \sim  {\omega^{2}\over 2}(\xi^{2}+\eta^{2})+c^{2}\Bigr[(U_{\alpha}+U_{\beta})
-{1\over 2}((U_{\alpha})^{2}+(U_{\beta})^{2})\Bigr]+{\rm O}(G^{2}),
\end{split}
\label{5.14c}
\end{equation}
because $\omega^{2}={c^{2}\over l^{3}}(R_{\alpha}+R_{\beta})={\rm O}(G)$ and, by virtue of (\ref{4.4c}), $\Omega^{2} \sim \omega^{2}+{\rm O}(G^{2})$\footnote{Now it is also possible to understand why our predictions are strongly affected by the value of $\kappa_1$ and why the map (\ref{5.12bisc}) leads to coordinates of Lagrangian points pretty close to those reported in Tabs. \ref{noncoll_details_tab} and \ref{coll_details_tab}.}. Thus, from a comparison between Eqs. (\ref{5.13c}) and (\ref{5.14c}), we can easily realize that $W^{\prime}$ and $W$ are equal, up to second order terms, if and only if $\kappa_1=-1/2$, a condition which matches exactly with the parameters characterizing the bound-states potential. Thence, the effective gravity map (\ref{5.12bisc}) is such that, once it is inserted into the Newtonian Lagrangian for the restricted planar three-body problem, it reproduces the relativistic potential $W$ up to second order terms in $G$, but this happens if and only if we consider long-distance quantum corrections resulting from the bound-states potential. Recall that considering, in a perturbative expansion, physical quantities up to the linear order in the Newtonian gravitational constant $G$ amounts to consider the tree-level of our quantum framework (cf. Eq. (\ref{1.2b})), therefore we can state, in an equivalent manner, that the bound-states potential makes the classical limit of our new quantum theory coincide with (\ref{5.13c}) at tree-level. As far as we can see, these arguments, along with the considerations regarding the perihelion shift of Mercury exposed at the end of Sec. \ref{three ways_sec}, add evidences in favour of considering the bound-states potential as the most apt choice in the context of quantum corrected relativistic mechanics \cite{testbed}.

\section{Laser ranging techniques} \label{Laser_Ranging_Sec}

\begin{figure} [htbp] 
\centering
\includegraphics[scale=0.7]{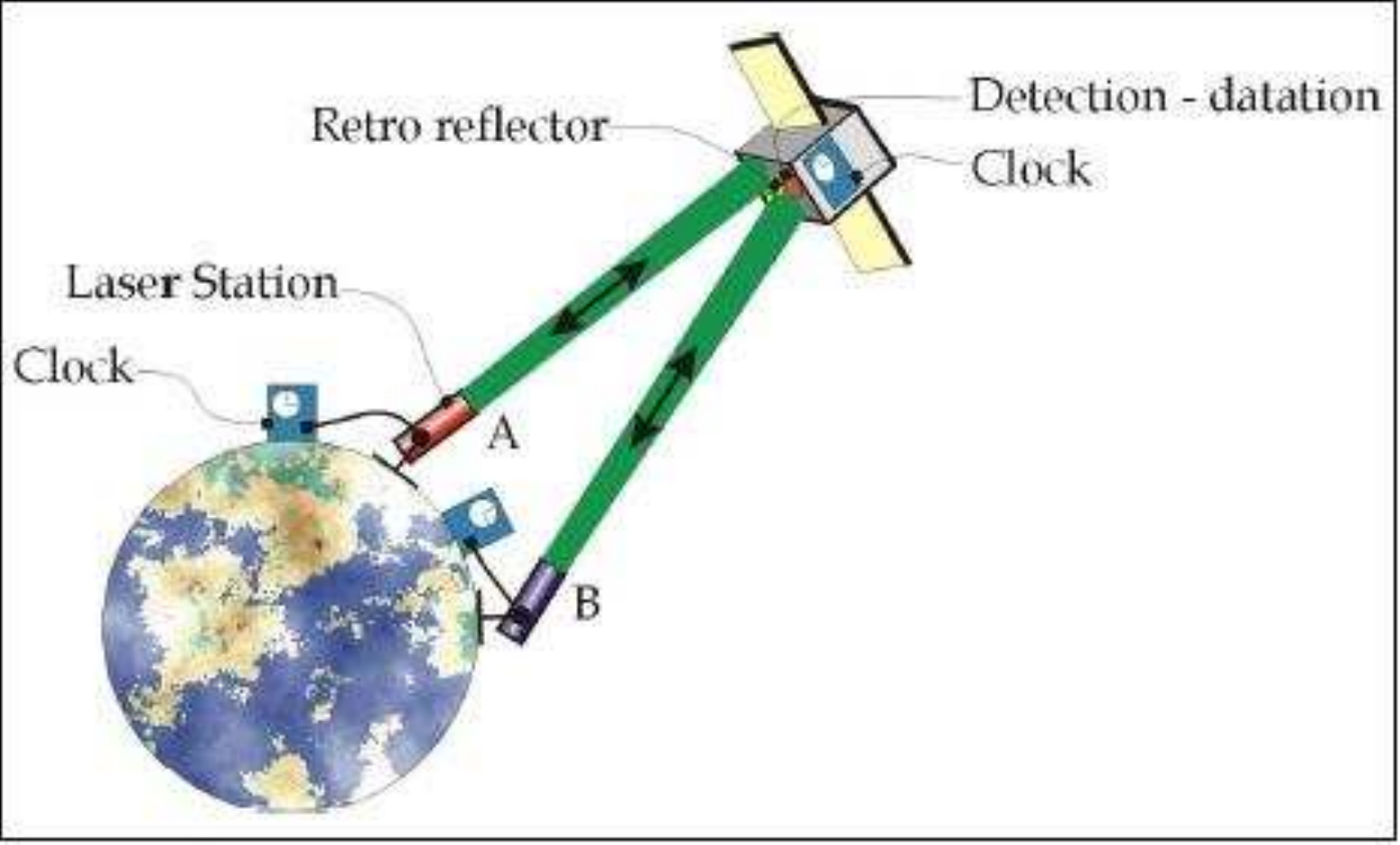}
\caption[Schematic set-up of Satellite/Lunar Laser Ranging technique]{Schematic set-up of Satellite/Lunar Laser Ranging technique.} 
\label{SLR.pdf}
\end{figure}

At this stage, it remains to be seen whether the present techniques in space sciences make it possible to realize a satellite that approaches the Earth-Moon Lagrangian points so that our tiny corrections start making themselves manifest. Remarkably, the values predicted in this thesis are very accessible in light of the advances of the modern technology. In fact, they can be studied with the technique of Satellite/Lunar Laser Ranging and a laser-ranged test mass equipped with cube corner retro-reflectors, to be designed {\it ad hoc} for this purpose. This advanced technique is conceptually very simple (Fig. \ref{SLR.pdf}). First of all, it involves firing a very short pulse of light towards satellite's cube corner retro-reflectors. Then, since the value of the speed of light is known, by measuring the two-way time of flight, it is possible to calculate the distance between the laser station and the satellite with sub-centimetre accuracy \cite{BEDS15,testbed}. 
This kind of assessment is performed by the International Laser Ranging Service, which recently celebrated the 50-th anniversary of the first successful measurement, which occurred at the Goddard Geophysical and Astronomical Observatory on October 31, 1964. 

Detecting the tiny departures from classical gravity described so far is a challenging task, which requires precise positioning in space at Lagrangian points in absolute terms, i.e., with respect to an appropriately chosen coordinate reference system. One potential choice is the International Terrestrial Reference System, which is established with several geodesy techniques, including Satellite/Lunar Laser Ranging. The latter provides almost uniquely the metrological definition of the Earth's center of mass (geo-center) and the origin of the above-mentioned reference system, as  well as, together with very long baseline interferometry, the absolute scale of length in space in Earth's orbit. Given the similarity of performing accurate positioning metrology with Lunar Laser Ranging, another option for the coordinate frame is the Solar System barycentre. In fact, the distance, for example, of $L_{4}$ and $L_{5}$ from the ground laser stations of the International Laser Ranging Service is very close to their distance to the laser retro-reflector arrays deployed by the Apollo and Lunokhod missions, which, over the last 45 years, were used in some of the best precision tests of general relativity (see Refs. \cite{Martini,Daa,Marcha,Marchb}). The Solar System barycentre is particularly apt for the purpose, since it is used for general relativity tests 
carried out with Lunar Laser Ranging data analysis by means of the orbit software package PLANETARY EPHEMERIS PROGRAM (PEP) since 1980s and up the present \cite{Daa,Martini}. PEP has been developed by the Harvard-Smithsonian Center for Astrophysics. A review of Lunar Laser Ranging data taking and analysis can be found in Ref. \cite{Martini}.

A laser ranging test mass can be designed with a dedicated effort, by exploiting the experience of Lunar Laser Ranging data taking and analysis described above, and especially by taking advantage of existing capabilities for a detailed pre-launch characterization of any kind of laser retro-reflector arrays and/or test
mass for Solar System exploration \cite{BEDS15,Dab,Dac,Currie,Dad}. Some of the key performance indicators that must be taken into account to design an appropriate laser ranging test mass for the signature of new physics described in this thesis are as follows.
\vskip 0.3cm
\noindent
(i) Adequate laser return signal (lidar optical cross section) from Lagrangian 
points.
\vskip 0.3cm
\noindent
(ii) Acceptable rejection of the unavoidable non-gravitational perturbations at all Lagrangian points, which {\it any} chosen test mass and/or 
test spacecraft will experience, whose complexity scales with the complexity of the structure of the test mass and/or test spacecraft itself.
\vskip 0.3cm
\noindent
(iii) Optimization/minimization of the value of the surface-to-mass ratio. This is a critical key performance indicator, since all non-gravitational perturbations related to the Sun radiation pressure and thermal effect are proportional
to this factor (see for example Ref. \cite{Vok}). Compared to other test spacecrafts and/or test masses, a laser ranging test mass has the advantage of the simplicity of 
geometrical shape (for example, spherical) and mechanical structure. To date, Apollo/Lunokhod are demonstrating a lifetime of at least 45 years.
\vskip 0.3cm
\noindent
(iv) Time-durability of the test mass to prolonged measurements. This key performance indicator favours laser ranging test mass over other types of any active test masses and/or spacecrafts, since the former is passive and maintenance free.

The above key performance indicators can be characterized at the dedicated laboratory described by Refs. \cite{Dab,Dac} (see also Ref. \cite{INFN}). From the experimental point of view of laser ranging investigations, arguments reported in this section can be applied to $L_{3}$, $L_{4}$, and $L_{5}$ . They do not apply to $L_{2}$, since such a position is not visible from International Laser Ranging Service stations. The distance of 
$L_{1}$ from Earth is shorter than for $L_{3}$, $L_{4}$, and $L_{5}$, which would make the laser return signal
from a laser ranging test mass in $L_{1}$ higher than from $L_{3}$, $L_{4}$, and $L_{5}$ (by a purely geometric factor equal
to the fourth power of the ratio of the distances of $L_{3}$ and $L_{1}$ from any given International Laser Ranging Service station; see, for example, Ref. \cite{Dab}). Given the relative proximity of $L_{1}$ to the Moon, gravitational effects on a laser ranging test mass in $L_{1}$ related to the non-point-like structure of the Moon (felt in $L_{1}$) should be evaluated to determine their influence, if any, on the conclusions of the previous sections. This influence is expected to be negligible for a laser ranging test mass in $L_{3}$, $L_{4}$, and $L_{5}$, since they are much more distant from the Moon than $L_{1}$.  

Finally, a last remark must be mentioned. We have seen throughout this thesis that both relativistic and quantum corrections on Lagrangian points are of the order of few millimetres. This magnitude is comparable with the instrumental accuracy of point-to-point laser time-of-flight measurements
in space of Satellite/Lunar Laser Ranging techniques. The full positioning error budget of the orbits of satellites equipped with retro-reflectors depends also on other sources of uncertainty (related to the specific orbit, satellite and
retro-reflector arrays), in addition to the pure point-to-point laser time-of-flight instrumental accuracy (related to the network of laser ranging ground stations of the International Laser Ranging Service). The full positioning error budget can be larger than millimetres. However, we believe that the most astonishing result of the first part of this manuscript consists in having found quantum gravity corrections, occurring in a familiar and close system like the Earth-Moon one, which have the opportunity to be tested. This is a novel feature in the quantum theory of the gravitational field, because all other models are so far unable to produce testable effects \cite{Rovelli2004,Polchinski98,Horowitz2005,DeWitt1967c}.

\cftaddtitleline{toc}{chapter}{Part II: the high-energy limit}{}

\begin{titlepage}
 \vspace{1cm}
 \setlength{\parindent}{0pt}
\setlength{\parskip}{0pt}
\vspace*{\stretch{1}}
\rule{\linewidth}{1pt}
\begin{center}
\Large{\bf Part II: the high-energy limit}
\end{center}
\rule{\linewidth}{2pt}
\vspace*{\stretch{2}}

\end{titlepage}

\chapter{Boosted spacetimes} \label{Boosted_chapter}

\vspace{2cm}
\emph{Arc, amplitude, and curvature sustain a similar relation to each other as time, motion, and velocity, or as volume, mass, and density.}
\begin{flushright}
C. F. Gauss
\end{flushright}
                                                                 
\vspace{2cm}  

The ultrarelativistic boosting procedure had been applied in theoretical physics in order to map known exact solutions of Einstein field equations into a class of spacetimes characterized by the presence of gravitational shock-waves. We can thus interpret this {\it modus operandi} has a formal method that allows us to describe such geometries. Although this pattern has a completely classical nature, it has many implications at quantum level. For instance, the first non-trivial gravitational effects to be seen in particle-particle interaction at extreme energies may be due to such fields. Thus, our understanding of quantum gravity can surely be helped by considering these field configurations.  

The principal purpose of this chapter consists in evaluating the features of the Riemann curvature associated to the boosted Schwarzschild-de Sitter metric \cite{BEST}. For the sake of simplicity, we will adopt units $G=c=1$. The metric signature is $(-+++)$.

\section{The boosting procedure} \label{The boosting procedure_Sec}

\begin{figure}
\centering
\includegraphics[scale=0.7]{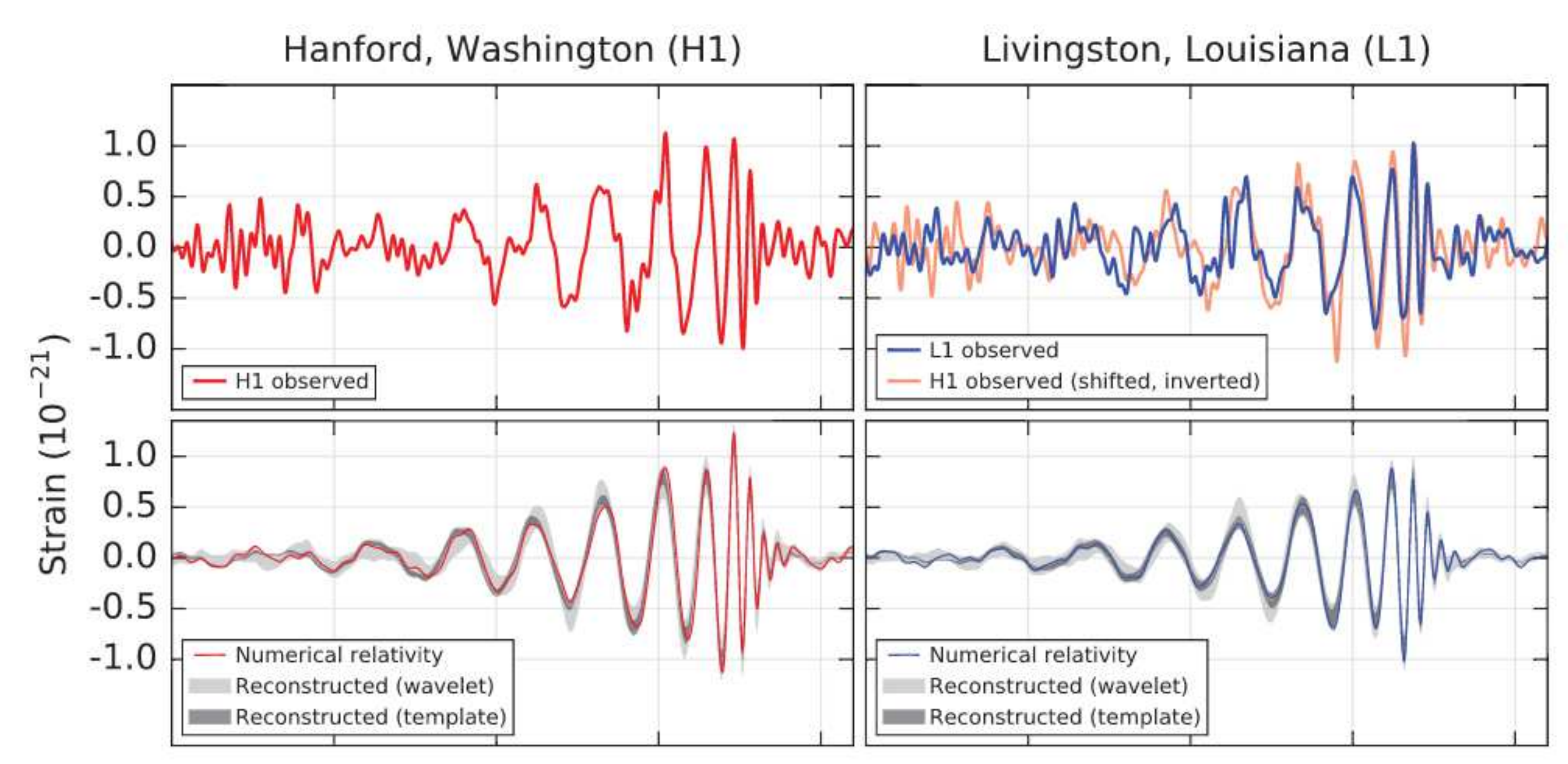}
\caption[LIGO detection of the gravitational wave event ``GW150914"]{LIGO detection of the gravitational wave event ``GW150914" at the Livingston (left) and Hanford (right) detectors, compared to the theoretical predicted values. This is the first direct detection of gravitational waves and the first observation of a binary black hole merger.}
\label{LIGO.pdf}
\end{figure}

The subject of gravitational fields generated by sources which move at the speed of light has been extensively studied in the literature because of its close connection to the topic of gravitational waves.  Predicted by Albert Einstein in 1916, it took the scientific community one hundred years to achieve the first {\it direct} experimental observation of gravitational waves (an {\it indirect} proof of the existence of gravitational waves is represented by the effects observed in 1974 by Hulse and Taylor in the binary pulsar system ``PSR 1913 + 16''). On February 11, 2016, in fact, the Laser Interferometer Gravitational-Wave Observatory (LIGO) and Virgo Collaboration teams announced the first observation of gravitational waves resulting from the merging of two black holes (with masses of about 29 and 36 times the mass of the Sun, as evaluated in the source frame) occurring about 1.3 billion light years away \cite{LIGO}. The event, called ``GW150914", has radiated in gravitational waves an energy of about $5 \times 10^{47}\, {\rm J}$ and it has been detected by the LIGO Hanford (Washington) and Livingston (Louisiana) observatories on September 14, 2015 at 09:50:45 UTC, as shown in Fig. \ref{LIGO.pdf}. A new era both for classical and quantum cosmology has just begun. 

\subsection{Aichelburg and Sexl method}

The first who dealt with the subject of gravitational waves generated by sources moving at the speed of light was Tolman in 1934 \cite{Tolman1934}, who studied the 
gravitational field of light beams and pulses in the linearised Einstein theory. But it was only in 1971 that Aichelburg and Sexl \cite{AS1971} developed a method to describe the gravitational field associated to a zero rest mass point particle moving at the speed of light in Minkowski background (i.e., the gravitational 
field from a single photon). In fact, in Ref. \cite{AS1971} the authors first derive this field by solving the linearised Einstein field equations for a particle with rest mass $m$ moving uniformly with velocity $v$. Then, they take the limit $v \to 1$ while the mass of the particle tends to zero in such a way that its energy remains finite. After that, the full non-linear Einstein theory is employed. In particular, starting from the Schwarzschild metric (the exact metric describing a particle at rest), which written in isotropic coordinates $(t,x,y,z)$ reads as 
\begin{equation}
{\rm d}s^2 = -\dfrac{(1-A)^2}{(1+A)^2}{\rm d}t^2 +(1+A)^4 ({\rm d}x^2+{\rm d}y^2+{\rm d}z^2) \label{isotropic_Schwarzchild},
\end{equation}
with $A=m/2r$ and $r^2 = x^2 + y^2 + z^2$, the Lorentz transformation in the $x$-direction 
\begin{equation}
\bar{t}  =  (1 - v^2)^{-1/2} (t+vx), \label{ASLorentztransformation1}
\end{equation}
\begin{equation}
\bar{x}  =  (1 - v^2)^{-1/2} (x+vt), 
\end{equation}
\begin{equation}
\bar{y}  =  y,
\end{equation}
\begin{equation}
\bar{z}  =  z \label{ASLorentztransformation2},
\end{equation}
is applied to (\ref{isotropic_Schwarzchild}) in order to obtain the gravitational field as seen by an observer moving uniformly with velocity $v$ relative to the mass. Once the limits $v \rightarrow 1$ and $m \rightarrow 0$ are considered, Aichelburg and Sexl obtained the remarkable result that both the linearised solution and the 
exact solution agree completely. 

The method first developed by Aichelburg and Sexl is called in the literature ``the boost of a metric''. With this procedure it is possible to 
show that the gravitational field of a null source moving in Minkowski space is non-vanishing only on a plane containing the particle itself and orthogonal to the direction of 
motion, i.e., (asymmetric) plane-fronted gravitational shock-waves, representing a special case of impulsive waves. The Riemann curvature tensor is zero everywhere except on this plane, where it assumes a 
distributional nature. The intriguing fact is that the boosted metric in the ultrarelativistic regime ($v \rightarrow 1$) has a new type of 
singularity, i.e., a distributional (Dirac-delta-like) singularity. The boosted ultrarelativistic metric obtained in Ref. \cite{AS1971} reads indeed as
\begin{equation}
{\rm d}s^2 =- {\rm d}\bar{t \;}^2 +{\rm d}\bar{x}^2  + {\rm d}\bar{y}^2  + {\rm d}\bar{z}^2 
+ 4 p \left \{ \left(\lvert \bar{t}-\bar{x} \rvert\right)^{-1} -2 \delta \left(\bar{t \;}^2 - \bar{x}^2\right) \log\sqrt{\bar{y} +\bar{z}} \;  \right\} 
({\rm d}\bar{t} - {\rm d}\bar{x})^2, \label{ASmetric}
\end{equation}
with 
\begin{equation}
p \equiv \dfrac{m}{\sqrt{1-v^2}} >0
\label{AS2.5}
\end{equation}
The role played by the parameter (\ref{AS2.5}) is fundamental. In fact, since the energy of the particle diverges as $v \rightarrow 1$ because of its finite rest mass $m$, in order to get around this issue $p$ is kept constant when the limit $v \rightarrow 1$ is evaluated. In other words, the total energy $p$ of the particle is kept constant while its rest mass goes to zero, as anticipated before.   

From Eq. (\ref{ASmetric}) it is possible to realize that the gravitational field travels with the particle, being zero everywhere except at the hypersurface $\bar{t}=\bar{x}$. In other words, the boosting method is such that as $v \to 1$ the gravitational field turns out to be compressed in the direction of motion of the particle and dilated in the orthogonal direction, sharing therefore the same characteristics as the electromagnetic field. Moreover, as pointed out before, the Riemann tensor of (\ref{ASmetric}) is zero everywhere except on the hypersurface $\bar{t}=\bar{x}$ and has non-vanishing components given by \cite{AS1971}
\begin{equation}
R_{0202}=4p \; \delta (\bar{t}-\bar{x}) \left[\dfrac{\bar{y}^2-\bar{z}^2}{(\bar{y}^2+\bar{z}^2)^2} + \pi \delta(\bar{y})\delta(\bar{z})\right], \label{ASRiemann1}
\end{equation}
\begin{equation}
R_{0303}=4p \; \delta (\bar{t}-\bar{x}) \left[\dfrac{\bar{y}^2-\bar{z}^2}{(\bar{y}^2+\bar{z}^2)^2} - \pi \delta(\bar{y})\delta(\bar{z})\right],
\end{equation}
\begin{equation}
R_{0203}=-4p \; \delta (\bar{t}-\bar{x})\dfrac{2\bar{y}\bar{z}}{(\bar{y}^2+\bar{z}^2)^2},\label{ASRiemann2}
\end{equation}
with the other components related to the ones given above by symmetry. An important remark should be made at this point. In fact, unlike what authors claim in the literature regarding this topic, the Riemann tensor is perfectly defined since it contains the tensor product of Dirac's $\delta$ distributions (and not their multiplications) \cite{BEST}. The only elements which are ``poorly defined'' in (\ref{ASRiemann1})--(\ref{ASRiemann2}) are represented by the functions
\begin{equation}
\dfrac{\bar{y}^2-\bar{z}^2}{(\bar{y}^2+\bar{z}^2)^2},
\end{equation}
and
\begin{equation}
\dfrac{2\bar{y}\bar{z}}{(\bar{y}^2+\bar{z}^2)^2},
\end{equation}
which are not locally integrable on the $(y,z)$-plane and, therefore, do not define, a priori, any distribution. Of course, their ``regularization'' (in the way of Gel'fand, see for example Ref. \cite{Gelfand}) is straightforward: the integration is understood in such a way that we first integrate over the set $y^2 + z^2 > \epsilon$ and then pass to the limit $\epsilon \rightarrow 0$ \cite{BEST}.

In order to give a precise meaning to expressions (\ref{ASRiemann1})--(\ref{ASRiemann2}), instead of boosting from the very beginning the Schwarzschild metric, the authors of Ref. \cite{AS1971} have applied the Lorentz transformations (\ref{ASLorentztransformation1})--(\ref{ASLorentztransformation2}) directly to the components of the Riemann tensor of (\ref{isotropic_Schwarzchild}) and then they have investigated the regime $v \rightarrow 1$. In this way, with the help of tetrad formalism, Aichelburg and Sexl have obtained relations which are valid only for those spacetime points where $\bar{y}^2+\bar{z}^2 \neq 0$. In particular, they obtain again the relations (\ref{ASRiemann1})--(\ref{ASRiemann2}), but without the terms $\delta(\bar{y})\delta(\bar{z})$ which vanish because of the condition $\bar{y}^2+\bar{z}^2 \neq 0$. This fact shows that on the hypersurface $\bar{x}=\bar{t}$ the Riemann tensor has a $\delta$-like singularity and is exactly of Petrov type $N$ (i.e., all four principal null directions of the Weyl spinor, describing the Weyl conformal curvature, coincide)\footnote{In Ref. \cite{AS1971} it is brilliantly explained how the original Petrov type $D$ field (i.e., two double principal null directions exist) is transformed in pure radiation.}. 

As noted at the beginning of this chapter, the boosting procedure is intimately connected with quantum theory. In fact, Aichelburg and Sexl results can provide important hints about gravitational interactions among particles at high energies. At extreme energies indeed interactions due to shock-waves dominate over all other field theoretical interactions. If we generalize the flat result of Aichelburg and Sexl by considering, for example, a massless particle moving in Schwarzschild background (or, to be more precise, along its event horizon), the resulting (spherical) shock-wave geometry can give information about the back-reaction or self-interaction to which the black hole is subjected to when matter enters or leaves the black hole itself (Hawking radiation) \cite{Dray}. Moreover, when two massless particles, along with their shock-waves, collide, the result of such an impact will be represented by curved shock-waves, a phenomenon that can be considered as a limiting case of the general problem of black hole encounters \cite{DEath}, an issue now in the limelight thanks to the recent first direct observation of gravitational waves. 

Eventually, it should be noted that, years after the work by Aichelburg and Sexl, more general impulsive waves were obtained by boosting other
black hole spacetimes with rotation, charge, and a cosmological constant \cite{FP1990,HT1993,ES2007}. However, our main contributions are devoted to the boosted Schwarzschild-de Sitter metric \cite{BEST}, which we will introduce in the next section.  
 
\subsection{Boosted Schwarzschild-de Sitter solution}

\begin{figure}
\centering
\includegraphics[scale=0.40]{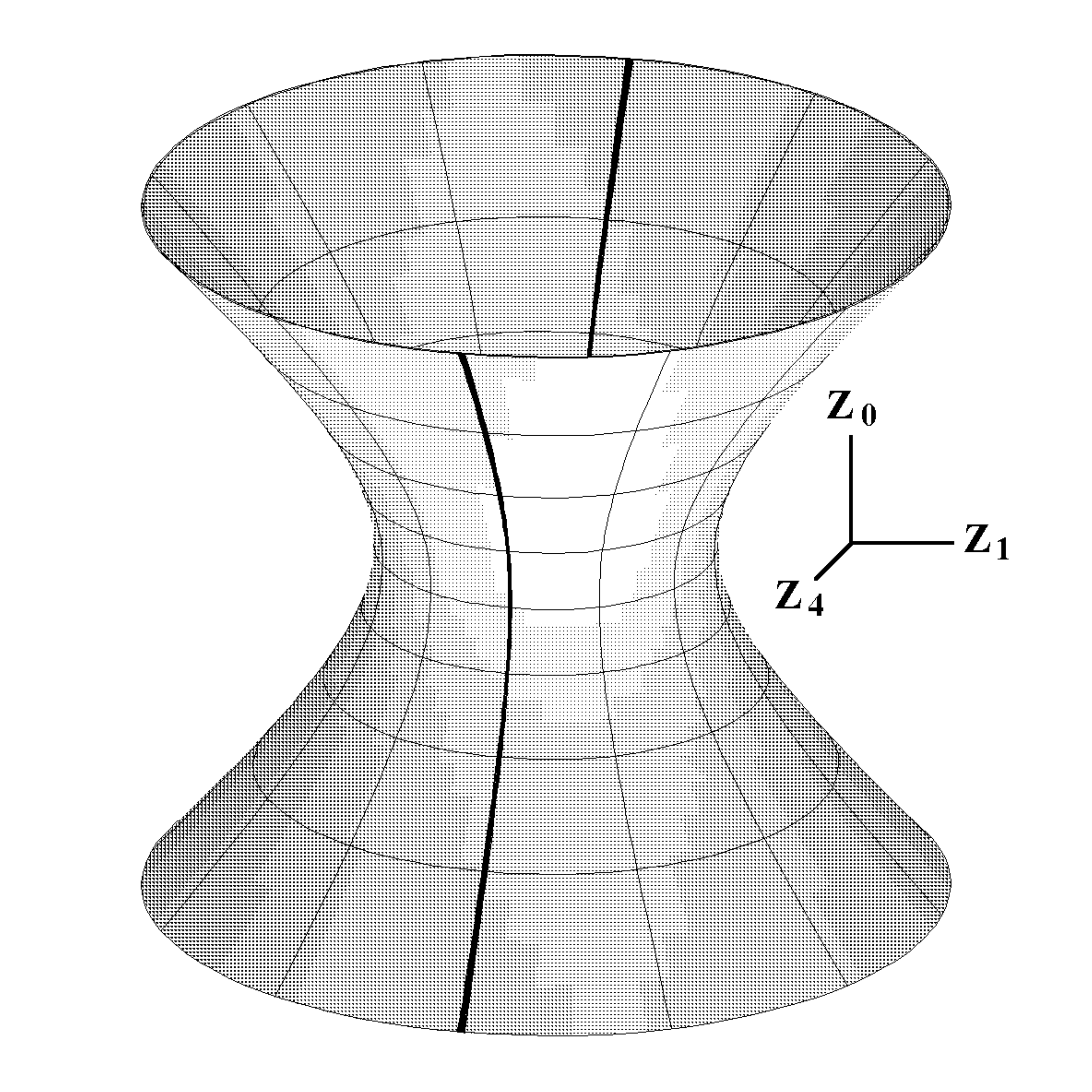}
\caption[Hyperboloid illustrating de Sitter space]{de Sitter space represented as a hyperboloid embedded in a five-dimensional flat space (two dimensions have been suppressed). The two world lines visible from the figure correspond to Eq. (\ref{worldline1}).}
\label{deSitter.pdf}
\end{figure}

The procedure involving the boost of the Schwarzschild-de Sitter solution by the means of the de Sitter group transformations\footnote{Recall that the background geometry provides us with a natural notion of boost as being its associated isometries.} was first employed by Hotta and Tanaka in 1993 \cite{HT1993}. Motivated by the analysis of quantum effects involving gravitons in de Sitter spacetime, the solution found by Hotta and Tanaka represents an example of spherical shock-wave geometry in de Sitter background (i.e., a background with a non-vanishing cosmological constant $\Lambda$), which generalizes the results of Aichelburg and Sexl, since it reduces to the latter when $\Lambda$ vanishes. 

de Sitter spacetime in four dimensions can be expressed as a four-dimensional hyperboloid 
of radius $a$ (with $\Lambda=3/a^2$) embedded in five-dimensional Minkowski spacetime having metric
\begin{equation}
{\rm d}s^{2}_{{\rm M}}= - {\rm d}Z^{2}_{0}+ {\rm d}Z_{1}^{2}+{\rm d}Z_{2}^{2}+{\rm d}Z_{3}^{2}+{\rm d}Z_{4}^{2},
\end{equation}
with coordinates satisfying the hyperboloid constraint 
\begin{equation}
a^2 = -(Z_{0})^2+ (Z_{1})^{2}+(Z_{2})^{2}+(Z_{3})^{2}+(Z_{4})^{2},
\label{Z hyperboloid constraint}
\end{equation}
as shown in Fig. \ref{deSitter.pdf}.

This represents a special case of a more general result, according to which every analytical four-dimensional spacetime can be considered, at least locally, as a surface embedded in a flat space having no more than ten dimensions. The proof of this theorem was first given by Levi-Civita \cite{LeviCivita1925}\footnote{The theorem was also demonstrated by Janet and Cartan \cite{J-C}.}, who limited his attention only to Riemannian manifolds and then was generalized by Friedman to pseudo-Riemannian ones \cite{Fried}.

de Sitter space represents the unique maximally symmetric, vacuum solution of Einstein's field equations with a positive cosmological constant. In fact, it has the same degrees of freedom of a four-dimensional Minkowski space, being characterized by the presence of ten Killing vectors. From the form of Eq. (\ref{Z hyperboloid constraint}) indeed it is easy to realize that the isometry group of de Sitter space is the ten-dimensional group $O(1,4)$ of homogeneous ``Lorentz transformations'' in the five-dimensional embedding space, called de Sitter group. Moreover, all the recent data from cosmological observations clearly indicate that in order to explain the properties of the presently observed Universe (the value of the Hubble constant, the anisotropies of the cosmic microwave background, and so forth) in the framework of inflationary cosmology, a non-vanishing repulsive cosmological constant has to be invoked \cite{Krauss}. For all these reasons, de Sitter space represents one of the most studied spacetimes in the literature.

By exploiting the relations between the $Z_i$ ($i=0,1,2,3,4$) coordinates and the spherical static coordinates $(t,r,\theta,\phi)$ \cite{BEST}
\begin{equation}
Z_0 \equiv  \sqrt{a^2 - r^2} \sinh(t/a), \label{Z0}
\end{equation}
\begin{equation}
Z_1 \equiv  r \cos \theta, 
\end{equation}
\begin{equation}
Z_2 \equiv  r \sin \theta \cos \phi , 
\end{equation}
\begin{equation}
Z_3 \equiv  r \sin \theta \sin \phi,  
\end{equation}
\begin{equation}
Z_4 \equiv  \pm  \sqrt{a^2 - r^2} \cosh(t/a), \label{Z3}
\end{equation}
de Sitter metric can be written as
\begin{equation}
{\rm d}s^{2}=-\left(1-{r^{2}\over a^{2}}\right){\rm d}t^{2}
+{{\rm d}r^{2}\over \left(1-{r^{2}\over a^{2}}\right)}
+r^{2}({\rm d}\theta^{2}+\sin^{2}\theta \;  {\rm d}\phi^{2}).  \label{dS metric}
\end{equation}

In Ref. \cite{HT1993} the Schwarzschild-de Sitter line element was interpreted as a first-order perturbation of de Sitter, i.e.,
\begin{equation}
{\rm d}s^{2} \approx -\left(1-{2m \over r}-{r^{2}\over a^{2}}\right){\rm d}t^{2} + \left( 1- \dfrac{r^2}{a^2} \right)^{-1} \left[1+ \left( 1- \dfrac{r^2}{a^2} \right)^{-1} \,\dfrac{2m}{r} \right] +r^{2}({\rm d}\theta^{2}+\sin^{2}\theta \;  {\rm d}\phi^{2}).
\label{S-dS approx}
\end{equation}
On the contrary, we have decided  to perform an exact analysis and hence, following Refs. \cite{BEST,ES2007}, we start with the standard form of the metric for a Schwarzschild-de Sitter spacetime
\begin{equation}
{\rm d}s^{2}=-\left(1-{2m \over r}-{r^{2}\over a^{2}}\right){\rm d}t^{2}
+{{\rm d}r^{2}\over \left(1-{2m \over r}-{r^{2}\over a^{2}}\right)}
+r^{2}({\rm d}\theta^{2}+\sin^{2}\theta \;  {\rm d}\phi^{2}), \label{S-dS metric}
\end{equation}
motivated by the fact that the work of Ref. \cite{ES2007} has demonstrated that both (\ref{S-dS approx}) and (\ref{S-dS metric}) lead to the same results. However, we stress that by adopting the exact approach our point of view has not changed: the background geometry is still represented by de Sitter space. Furthermore, note that from Eqs. (\ref{Z0})--(\ref{Z3}) it follows that the source of (\ref{S-dS metric}), which is located at $r=0$, corresponds to two world lines
\begin{equation}
(Z_0,Z_1,Z_2,Z_3,Z_4)=(a \sinh (t/a),0,0,0,\pm a \cosh(t/a)), \label{worldline1}
\end{equation}
propagating along the hyperboloid of Fig. \ref{deSitter.pdf}. In other words, Eq. (\ref{worldline1}) describes the source of (\ref{S-dS metric}) in the limit $m\to 0$.

At this stage, we are ready to show how the ultrarelativistic boosted form of the Schwarzschild-de Sitter metric can be obtained. First of all, we need to express (\ref{S-dS metric}) through the $Z_i$ coordinates. Then, bearing in mind that
\begin{equation}
r^2 = (Z_1)^2+ (Z_2)^2+ (Z_3)^2,
\label{r^2_def}
\end{equation}
and on defining 
\begin{equation}
f^2 \equiv  a^2 - r^2 = (Z_{4})^2 - (Z_{0})^{2}, 
\end{equation}
\begin{equation}
F_m \equiv   1 - \dfrac{2 a^2 m}{f^2 r}-\dfrac{a^2/r^2}{\left( 1- \dfrac{2 a^2 m}{f^2 r}\right)}, 
\end{equation}
\begin{equation}
Q \equiv   1+ \dfrac{2 (Z_{0})^{2}}{f^2},
\end{equation}
we can express the Schwarzschild-de Sitter metric (\ref{S-dS metric}) in the form \cite{BEST,ES2007}
\begin{equation}
{\rm d}s^2 = h_{00}{\rm d}Z_{0}^{2}+ h_{44}{\rm d}Z_{4}^{2}+2h_{04}{\rm d}Z_{0}{\rm d}Z_{4}+ {\rm d}Z_{1}^{2}
+{\rm d}Z_{2}^{2}+{\rm d}Z_{3}^{2},
\label{Z S-dS metric}
\end{equation} 
where
\begin{equation}
h_{00} \equiv -\dfrac{1}{2} \left(Q-1 \right)F_m - \left(1-\dfrac{2 a^2 m}{f^2 r} \right) - \dfrac{(Z_{0})^{2}}{r^2}, \\
\end{equation}
\begin{equation}
h_{44} \equiv  -\dfrac{1}{2} \left(Q+1 \right)F_m + \left(1-\dfrac{2 a^2 m}{f^2 r} \right) - \dfrac{(Z_{4})^{2}}{r^2}, \\
\end{equation}
\begin{equation}
h_{04} \equiv  \dfrac{Z_0 Z_4}{f^2} F_m+ \dfrac{Z_0 Z_4}{r^2}.  
\end{equation}
It is possible to interpret Eq. (\ref{Z S-dS metric}) as the the geometry produced by the two point sources having (\ref{worldline1}) as their world lines. 

At this point, we introduce a boost in the $Z_1$-direction\footnote{A boost can be performed in any direction orthogonal to $Z_0$, since a boost in the $Z_0$-direction represents simply a time shift.} (which represents, as we said before, an element of the de Sitter group $O(1,4)$) by defining a new set of coordinates independent of $v$, i.e., the $Y_i$ coordinates, such that (hereafter $\gamma \equiv  1/ \sqrt{1-v^2} \; $)
\begin{equation}
Z_0 = \gamma \left( Y_0 + v Y_1 \right),  \label{boost 1}
\end{equation}
\begin{equation}
Z_1 = \gamma \left( v Y_0 + Y_1 \right), 
\end{equation}
\begin{equation}
Z_2 = Y_2, \; \;  \; Z_3=Y_3 , \; \;  \; Z_4=Y_4. \label{boost 2}
\end{equation}
We also set 
\begin{equation}
m \equiv p \sqrt{1-v^2},
\label{HT14}
\end{equation}
$p$ being the same parameter as the energy of the black hole solution in Minkowski background (see Eq. (\ref{AS2.5})). Thus, starting from (\ref{Z S-dS metric}) jointly with (\ref{boost 1})--(\ref{boost 2}) we eventually obtain the boosted Schwarzschild-de Sitter metric \cite{BEST,ES2007}
\begin{equation}
\begin{split}
{\rm d}s^2 =& \gamma^2 \left(h_{00}+v^2\right){\rm d}Y_{0}^{2}+\gamma^2 \left(1+ v^2 h_{00}\right)
{\rm d}Y_{1}^{2}+{\rm d}Y_{2}^{2}+{\rm d}Y_{3}^{2}+h_{44}{\rm d}Y_{4}^{2}   \\
& + 2v\gamma^2 \left(1+h_{00}\right){\rm d}Y_0 {\rm d}Y_1 + 2 \gamma h_{04} {\rm d}Y_0 {\rm d}Y_4 
+ 2 v \gamma h_{04} {\rm d}Y_1 {\rm d}Y_4, \label{boosted metric}
\end{split}
\end{equation}
whose singular ultrarelativistic limit ($v \to 1$ and $p$ fixed) is expressed by \cite{ES2007}
\begin{equation}
\begin{split}
{\rm d}s^{2}=&-{\rm d}Y_{0}^{2}+{\rm d}Y_{1}^{2}+{\rm d}Y_{2}^{2}+{\rm d}Y_{3}^{2}+{\rm d}Y_{4}^{2}  \\
&+ 4 p \left[-2+{Y_{4}\over a}
\log \left({{a+Y_{4}}\over {a-Y_{4}}}\right)\right]
\delta(Y_{0}+Y_{1})({\rm d}Y_{0}+{\rm d}Y_{1})^{2}. \label{ultrarelativistic boosted metric}
\end{split}
\end{equation}
Thence, we can interpret (\ref{boosted metric}) as the low-velocity limit of (\ref{ultrarelativistic boosted metric}). Moreover, from Eq. (\ref{ultrarelativistic boosted metric}) it easily follows that the our ``boosted geometry'' differs from de Sitter spacetime only by the inclusion of an impulsive wave. In fact, the first line of Eq. (\ref{ultrarelativistic boosted metric}) describes de Sitter space viewed as a four-dimensional hyperboloid of radius $a$ having equation
\begin{equation}
(Y_{0})^{2}=-a^{2}+(Y_{1})^{2}+(Y_{2})^{2}+(Y_{3})^{2}+(Y_{4})^{2},  \label{hyperboloid constrain}
\end{equation}
embedded into flat five-dimensional space, while the second line describes a spherical shock-wave singularity located on the null
hypersurface having equations
\begin{equation}
Y_{0}+Y_{1}=0, \label{null hypersurface 1}
\end{equation}
\begin{equation}
(Y_{2})^{2}+(Y_{3})^{2}+(Y_{4})^{2}-a^{2}=0, \label{null hypersurface 2}
\end{equation}
Eq. (\ref{null hypersurface 2}) being obtained by the joint effect of the hyperboloid
constraint (\ref{hyperboloid constrain}) and the Dirac-delta condition (\ref{null hypersurface 1}) \cite{BEST}. Equivalently, it may be noted that the impulsive wave is represented by the evolving 2-sphere (\ref{null hypersurface 2}) in the five-dimensional Minkowski space at any time $Z_0$. In addiction, the two null point sources of the shock-wave are located at the points \begin{equation}
Y_2=Y_3=0,
\label{Podolsy2.14a}
\end{equation} 
\begin{equation}
Y_4= \pm a,
\label{Podolsy2.14b}
\end{equation}
of this sphere \cite{HT1993}. They result from the boost of the singularities described by the world lines (\ref{worldline1}) that were originally located at $r=0$. Thus, we can physically interpret the procedure outlined above as the boost of the source of (\ref{S-dS metric}), located at the singular point $r=0$, in the limit in which $v \to 1$ and $m \to 0$ in such a way that the energy $p$, defined in Eq. (\ref{HT14}), remains constant. It may seem surprising that the Schwarzschild-de Sitter metric (\ref{S-dS metric}) does not have a single source both in the low-velocity and in the ultrarelativistic limit. However, it must be noted that the static coordinates $(t,r,\theta,\phi)$ do not span the complete spacetime whose analytic extension contains both the black hole and white hole parts \cite{Wald,MTW}. 

\subsection{four-dimensional form of the boosted metric}

The spacetime metric (\ref{boosted metric}) is apparently expressed by a $5 \times 5$ matrix while the original metric 
(\ref{S-dS metric}) is expressed through $4$ local coordinates $(t,r,\theta,\phi)$. Hence also the metric  (\ref{boosted metric}) 
should be eventually expressed through $4$ coordinates only, if
one wants to arrive at a formula for the curvature, since our reference spacetime
remains four-dimensional. To restore the usual four-dimensional form of the metric, we have to exploit the constraint 
(\ref{Z hyperboloid constraint}) expressed in terms of $Y_i$ coordinates, i.e., Eq. (\ref{hyperboloid constrain}). By virtue of 
this condition we can write \cite{BEST}
\begin{equation}
Y_{0}=\sqrt{-a^{2}+(Y_{1})^{2}+(Y_{2})^{2}+(Y_{3})^{2}+(Y_{4})^{2}} \equiv \sqrt{\sigma(Y_{\mu })}, \label{Y0}
\end{equation}
\begin{equation}
dY_{0}={\sum_{\mu=1}^{4}Y_{\mu}dY_{\mu }\over \sqrt{\sigma(Y_{\mu })}}, \label{dYo}
\end{equation}
and eventually, using (\ref{Y0}) and (\ref{dYo}), we obtain the manifestly four-dimensional form of the boosted metric 
(\ref{boosted metric}), which can be expressed by the relations \cite{BEST}
\begin{equation}
g_{11} =  \dfrac{\gamma^2 \left(h_{00}+v^2\right)}{\sigma} (Y_{1})^{2}+\gamma^2 \left(1+v^2h_{00}\right)+\dfrac{2v\gamma^2  
\left(1+h_{00}\right)}{\sqrt{\sigma}} Y_1 \label{g11} , 
\end{equation}
\begin{equation}
g_{22} = \dfrac{\gamma^2 \left(h_{00}+v^2\right)}{\sigma} (Y_{2})^{2} +1, 
\end{equation}
\begin{equation}
g_{33} = \dfrac{\gamma^2 \left(h_{00}+v^2\right)}{\sigma} (Y_{3})^{2} +1, 
\end{equation}
\begin{equation}
g_{44} = \dfrac{\gamma^2 \left(h_{00}+v^2\right)}{\sigma} (Y_{4})^{2} +h_{44}+ \dfrac{2 \gamma h_{04}}{\sqrt{\sigma}} Y_4 , 
\end{equation}
\begin{equation}
g_{12} = \dfrac{\gamma^2 \left(h_{00}+v^2\right)}{\sigma} Y_{1}Y_{2} + \dfrac{v \gamma^2 \left(1+h_{00} \right)}{\sqrt{\sigma}}Y_2, 
\end{equation}
\begin{equation}
g_{13} = \dfrac{\gamma^2 \left(h_{00}+v^2\right)}{\sigma} Y_{1}Y_{3} + \dfrac{v \gamma^2 \left(1+h_{00} \right)}{\sqrt{\sigma}}Y_3, 
\end{equation}
\begin{equation}
g_{14} = \dfrac{\gamma^2 \left(h_{00}+v^2\right)}{\sigma} Y_{1}Y_{4} + \dfrac{v \gamma^2 \left(1+h_{00} \right)}{\sqrt{\sigma}}Y_4 
+ \dfrac{\gamma h_{04}}{\sqrt{\sigma}}+v \gamma h_{04}, 
\end{equation}
\begin{equation}
g_{23} = \dfrac{\gamma^2 \left(h_{00}+v^2\right)}{\sigma} Y_{2}Y_{3} , 
\end{equation}
\begin{equation}
g_{24} = \dfrac{\gamma^2 \left(h_{00}+v^2\right)}{\sigma} Y_{2}Y_{4} + \dfrac{\gamma h_{04}}{\sqrt{\sigma}} Y_2 , 
\end{equation}
\begin{equation}
g_{34} = \dfrac{\gamma^2 \left(h_{00}+v^2\right)}{\sigma} Y_{3}Y_{4} + \dfrac{\gamma h_{04}}{\sqrt{\sigma}} Y_3. \label{g34}
\end{equation}

\subsection{Coordinate transformations} \label{coordinate-transformation}

For future purposes, it is crucial to derive the transformations relating the spherical coordinates $(t,r,\theta,\phi)$ and the boost coordinates $(Y_1,Y_2,Y_3,Y_4)$ characterizing the metric tensor components (\ref{g11})--(\ref{g34}). 

We start by inverting (\ref{boost 1})--(\ref{boost 2}), yielding easily
\begin{equation}
Y_0 = \gamma \left( Z_0 - v Z_1 \right),  \label{boost_Y0}
\end{equation}
 \begin{equation}
Y_1 = \gamma \left( Z_1-v Z_0  \right), 
\end{equation}
\begin{equation}
Y_2 = Z_2, \; \;  \; Y_3=Z_3 , \; \;  \; Y_4=Z_4. \label{boost_Y4}
\end{equation}
By using (\ref{Z0})--(\ref{Z3}), jointly with (\ref{boost_Y0})--(\ref{boost_Y4}), we obtain that \cite{BEST}
\begin{equation}
Y_0=\gamma \left( \sqrt{a^2-r^2} \sinh (t/a)-vr\cos \theta \right), 
\end{equation}
and
\begin{equation}
Y_1=\gamma \left( r\cos \theta-v\sqrt{a^2-r^2} \sinh (t/a) \right), \label{syst1}
\end{equation}
\begin{equation}
Y_2=r \sin \theta \cos \phi,\label{syst2}
\end{equation}
\begin{equation}
Y_3=r \sin \theta \sin \phi, \label{syst3}
\end{equation}
\begin{equation}
Y_4=\sqrt{a^2-r^2} \cosh(t/a).\label{syst4}
\end{equation}
Thus, bearing in mind that Eq. (\ref{hyperboloid constrain}) allows us to get rid of the $Y_0$ coordinate, if we want to obtain $(t,r,\theta,\phi)$ coordinates occurring in Schwarzschild-de Sitter metric (\ref{S-dS metric}) as functions of $(Y_1,Y_2,Y_3,Y_4)$, then we have to invert relations (\ref{syst1})--(\ref{syst4}). First of all, by exploiting Eqs. (\ref{boost 1})--(\ref{boost 2}), the condition (\ref{r^2_def}) becomes
\begin{equation}
r^2=\gamma^2 (v \sqrt{\sigma}+Y_1)^2+(Y_2)^2+(Y_3)^2, \label{r^2_1}
\end{equation}
whereas on using (\ref{syst2}) and (\ref{syst3}) we obtain
\begin{equation}
r^2= \dfrac{(Y_2)^2+(Y_3)^2}{\sin^2 \theta},\label{r^2_2}
\end{equation}
so that a comparison between (\ref{r^2_1}) and (\ref{r^2_2}) yields \cite{BEST}
\begin{equation}
\sin^2 \theta = \dfrac{(Y_2)^2+(Y_3)^2}{\gamma^2 (v \sqrt{\sigma}+Y_1)^2+(Y_2)^2+(Y_3)^2},
\end{equation}
whose solutions are given by \cite{BEST}
\begin{equation}
\theta = \mp \arcsin \left(   \sqrt{\dfrac{(Y_2)^2+(Y_3)^2}{\gamma^2 (v \sqrt{\sigma}+Y_1)^2+(Y_2)^2+(Y_3)^2}} \right) + 2\pi n, \;\; \; (n\; {\rm integer}), \label{theta-Y_1}
\end{equation}
 \begin{equation}
\theta = \pi \mp \arcsin \left(   \sqrt{\dfrac{(Y_2)^2+(Y_3)^2}{\gamma^2 (v \sqrt{\sigma}+Y_1)^2+(Y_2)^2+(Y_3)^2}} \right) + 2\pi n, \;\; \; (n\; {\rm integer}). \label{theta-Y_2}
\end{equation}
Therefore, at this stage from (\ref{syst4}) we straightforwardly obtain the relations for $t$, i.e., \cite{BEST}
\begin{equation}
t= a \left[ \mp {\rm arccosh}\left(\dfrac{Y_4}{\sqrt{a^2-r^2}} \right) +2 \pi \I n   \right], \; \;\; (n\; {\rm integer}), \label{t-Y}
\end{equation}
and eventually from (\ref{syst3}) we get \cite{BEST}
\begin{equation}
\phi = \arcsin \left( \dfrac{Y_3}{r \sin \theta} \right) + 2 \pi n, \; \;\; (n\; {\rm integer}), \label{phi-Y_1}
\end{equation}
\begin{equation}
\phi = \pi-\arcsin \left( \dfrac{Y_3}{r \sin \theta} \right) + 2 \pi n, \; \;\; (n\; {\rm integer}). \label{phi-Y_2}
\end{equation}
Thus, Eqs. (\ref{r^2_1}), (\ref{theta-Y_1})--(\ref{phi-Y_2}), represent the relations we were looking for, because they link $(t,r,\theta,\phi)$ to $(Y_1,Y_2,Y_3,Y_4)$ coordinates.

\section{Riemann curvature of the boosted Schwarzschild-de Sitter spacetime} \label{Riemann curvature of the boosted Schwarzschild-de Sitter spacetime_Sec}

The great revolution introduced by Einstein's theory \cite{Einstein1916} consists in viewing the gravitational field as the 
curvature of spacetime. Such a curvature is directly coupled to the energy and momentum of whatever matter and radiation are present, as 
specified by the Einstein field equations (\ref{Einstein_equations}), whose content states that ``the matter and the energy say to the spacetime how to curve, and the 
curvature of spacetime says to the matter how to move'' \cite{MTW}. Thus, one of the most important objects of the theory of the gravitational field is the Riemann tensor, since it represents an intrinsic object that catches in an elegant and covariant way the features of spacetime curvature by formally measuring the extent to which the metric tensor is not locally isometric to that of flat Minkowski space. Therefore, it would be of great physical importance to evaluate the effects of shock-wave geometries (i.e., the ``boosted geometries'') on curvature.

Since ``gravitation is a manifestation of spacetime curvature, and curvature shows up in the deviation of one geodesic from a nearby 
geodesic'' \cite{MTW}, the concept of spacetime curvature is directly related to the geodesic completeness of spacetime, as we say that 
a spacetime manifold is geodesically complete if any geodesic can be extended to arbitrary values of the affine parameter (see Sec. \ref{b-completeness}). Thus, knowledge of the Riemann curvature tensor is an essential step towards the description of topological features of spacetime and this motivates the effort 
we made in calculating the Riemann tensor for the boosted Schwarzschild-de Sitter metric \cite{BEST}. Nevertheless, we stress that the usual general relativity definitions regarding curvature (see Eqs. (\ref{definition Riemann})--(\ref{geodesic deviation})) are given in terms of objects that, unlike the ones 
we will handle, have no distributional singularities (cf. (\ref{ultrarelativistic boosted metric})). Thus, we are interested 
in a sort of generalization of the usual concept of Riemann tensor, which enlarges the notion of curvature, i.e., what we call the ``boosted 
Riemann tensor'', with a particular interest in the ultrarelativistic regime, where distributional singularities show up.

\subsection{The Riemann curvature tensor} \label{Appendix Riemann}

We start by recalling some basic properties of pseudo-Riemannian geometry. 

The Riemann tensor can be defined in various alternative (and equivalent) ways \cite{Wald,Nakahara}. First, given the covariant derivative operator $\nabla$ associated with the Levi-Civita connection, the Riemann curvature tensor can defined as the map 
\begin{equation}
R : \mathcal{X} (M)\otimes\mathcal{X} (M)\otimes\mathcal{X} (M) \rightarrow \mathcal{X} (M),
\end{equation}
$\mathcal{X} (M)$ being the set of all vector fields defined on the manifold $M$, such that
\begin{equation}
R(X,Y,Z) \equiv \nabla_{X} \nabla_{Y} Z -\nabla_{Y} \nabla_{X} Z- \nabla_{[X,Y]}Z, \label{definition Riemann}
\end{equation}
where $[X,Y]$ denotes the Lie bracket of the vector fields $X$ and $Y$. In the case in which $[X,Y]=0$, the previous formula reduces to
\begin{equation}
R(X,Y,Z) \equiv \nabla_{X} \nabla_{Y} Z -\nabla_{Y} \nabla_{X} Z.
\end{equation}
Therefore, we can obtain the well-known result according to which the Riemann tensor measures the failure of successive operations of differentiation to commute when applied 
to a dual vector field $\omega \in \chi^{*}(M)$ (a condition which can be interpreted as the integrability obstruction for the existence of an isometry with Minkowski space), i.e.,
\begin{equation}
\nabla_a \nabla_b \; \omega_c - \nabla_b \nabla_a \; \omega_c =- R^{d}_{\;\, cab} \;  \omega_{d},
\end{equation}
where we have employed the abstract index notation. Moreover, the failure of a vector to return to its original value when parallel transported around a small closed loop is directly connected to the Riemann tensor, which is in this way related to the path dependence of parallel transport underlying the pseudo-Riemannian geometry. We can easily construct a small closed loop at $p\in M$ by choosing a two-dimensional surface $\mathcal{S}$ through $p$ and the coordinates $t$ and $s$ on it. Next, we construct the loop by moving of a quantity $\Delta t$ along the curve $s=0$, followed by moving $\Delta s$ along the curve 
$t= \Delta t$ and then reverting by $\Delta t$ and $\Delta s$. If we consider the vector $v^a$ at $p$ and parallel transport it around 
the closed loop we have just constructed, the change $\delta v^a$ to second order in the displacements $\Delta t$ and $\Delta s$ that we register 
when we move back to the starting point involves once again the Riemann tensor, because we have \cite{Wald}
\begin{equation}
\delta v^a = \Delta t \Delta s \; v^d \; T^c \; S^b \; R^{a}_{\;\, dcb},
\end{equation}
where $T^c$ and $S^b$ indicate the tangent to the curves of constant $s$ and $t$, respectively. Finally, the Riemann tensor appears also in 
the geodesic deviation equation, i.e., the equation measuring the tendency of geodesics to accelerate toward or away from each other. 
If $\gamma_s (t)$ denotes a smooth one-parameter family of geodesics such that for each $s \in \mathbb{R}$ the curve $\gamma_s$ is a geodesic parametrized by the affine parameter $t$, the geodesic deviation equation reads as \cite{Wald,MTW}
\begin{equation}
a^c \equiv T^a \nabla_a (T^b \nabla_b X^c) = R^{c}_{ \; \,def} T^d T^e X^f, \label{geodesic deviation}
\end{equation}
where $a^c$ represents the relative acceleration of an infinitesimally nearby geodesic in the family,  $X^a = \partial x^a (s,t) /\partial s $ 
is the deviation vector ($x^a(s,t)$ being the coordinates of one of the geodesics belonging to the family $\gamma_s (t)$) and $ T^a 
=  \partial x^a (s,t) /\partial t$ represents the tangent vector to the geodesic. Therefore, Eq. (\ref{geodesic deviation}) 
states that, if the curvature does not vanish, some initially parallel geodesics will fail to remain parallel: in the presence of a 
gravitational field the fifth postulate of Euclidean geometry is no longer valid.

The components of the Levi-Civita connection in a non-coordinate basis $\{\bold{e}_a\}$ are  given by the Riemann-Christoffel symbols \cite{MTW}
\begin{equation}
\Gamma_{abc}= \dfrac{1}{2} \left( g_{ab,c}+g_{ac,b}-g_{bc,a}+c_{abc}+c_{acb}-c_{bca} \right), \label{Gamma}
\end{equation}
where the ``commutation coefficients''  $c_{abc}$ are defined by
\begin{equation}
[\bold{e}_{b},\bold{e}_{c}] \equiv c_{bc}^{\; \; \; a} \; \bold{e}_{a}.  \label{c_abc}
\end{equation}
Then, the components of the Riemann tensor read as
\begin{equation}
R^{a}_{\; \; bcd} = \Gamma^{a}_{\; \;  bd,c} - \Gamma^{a}_{\; \;  bc,d} + \Gamma^{e}_{\; \;  bd} \Gamma^{a}_{\; \;  ec}
-\Gamma^{e}_{\; \;  bc} \Gamma^{a}_{\; \;  ed}- \Gamma^{a}_{\; \;  be}c_{cd}^{\; \; \; e}. \label{Riemann tensor} 
\end{equation}

Therefore, at this point we note that since we have obtained the formulas (\ref{g11})--(\ref{g34}) expressing the manifestly four-dimensional form of (\ref{boosted metric}), we can evaluate the Riemann-Christoffel symbols and consequently the Riemann curvature tensor of the boosted Schwarzschild-de Sitter metric by using the relations of classical general relativity outlined above. However, we can somewhat simplify Eqs. (\ref{Gamma}) and (\ref{Riemann tensor}) in the case in which $ \left \{ 
\dfrac{\partial}{\partial Y_{\mu}} \right \}$ ($\mu$ being a coordinate index such that $\mu=1,2,3,4$) is a coordinate basis. As we know, the static spherical basis $(t,r,\theta,\phi)$ is indeed a coordinate basis. Bearing in mind definitions (\ref{Z0})--(\ref{Z3}), the Jacobian of the transformation between the spherical coordinates and the   $ \left \{ \dfrac{\partial}{\partial Z_{\mu}} \right \}$ 
is expressed by \cite{BEST}
\begin{equation}
J_{\mu} ^{\; \; \lambda} =
\renewcommand{\arraystretch}{2.0}
\begin{pmatrix}
0 & \cos \theta & -r \sin \theta & 0 \\
0 & \sin \theta \cos \phi &  r \cos \theta \cos \phi &  -r \sin \theta \sin \phi \\
0  & \sin \theta \sin \phi  & r \cos \theta \sin \phi & r \sin \theta \cos \phi  \\
\dfrac{\sqrt{a^2-r^2}}{a} \sinh (t/a) &  \dfrac{-r}{\sqrt{a^2-r^2} }\cosh (t/a) & 0 & 0 
\end{pmatrix} , 
\label{Jacobian}
\end{equation}
while the inverse Jacobian reads as
\begin{equation}
 (J^{-1})_{\lambda}^{\; \; \mu} = 
 \renewcommand{\arraystretch}{2.0}
  \begin{pmatrix}
\dfrac{a \; r \cos \theta \coth (t/a)}{(a^2 - r^2)} &   \dfrac{a \; r \cos \phi  \sin \theta \coth (t/a)}{(a^2 - r^2)} 
&  \dfrac{a \; r \sin \theta \sin \phi \coth (t/a)}{(a^2 - r^2)}  & \dfrac{a \left(\sinh (t/a)\right)^{-1}}{\sqrt{a^2-r^2}} \\
\cos \theta & \cos \phi \sin \theta & \sin \theta \sin \phi & 0 \\
-\sin \theta /r & \cos \theta \cos \phi /r &   \cos \theta \sin \phi /r &  0\\
0  & - \dfrac{\sin \phi}{r \sin \theta} & \dfrac{\cos \phi}{r \sin \theta}&  0
\end{pmatrix} . 
\label{inverse jacobian}
\end{equation}

By virtue of (\ref{Jacobian}) and (\ref{inverse jacobian}), if we adopt the concise notation $x_{\lambda} \equiv 
(t,r,\theta,\phi)$ we can write 
\begin{equation}
\dfrac{\partial }{\partial Z_{\mu}}= (J^{-1})_{\lambda}^{\; \; \mu}  \dfrac{\partial }{\partial x_{\lambda}},
\end{equation}
and, by exploiting the fact that $ \left \{ \dfrac{\partial}{\partial x_{\lambda}} \right \}$ is a coordinate basis, after a 
lengthy calculation we arrive at the conclusion that also the basis $ \left \{ \dfrac{\partial}{\partial Z_{\mu}} \right \}$ is a 
coordinate basis, or in other words we have that 
\begin{equation}
\left[ \dfrac{\partial}{\partial Z_{\mu}} , \dfrac{\partial}{\partial Z_{\lambda}} \right]=0 . 
\end{equation}
The relations (\ref{boost 1})--(\ref{boost 2}) for the boost show that the transformations between $Z_{\mu}$ and $ Y_{\mu}$ are linear, 
therefore we can easily conclude that 
\begin{equation}
\left[ \dfrac{\partial}{\partial Y_{\mu}} , \dfrac{\partial}{\partial Y_{\lambda}} \right]=0 ,
\end{equation}
hence the basis $\left \{ \dfrac{\partial}{\partial Y_{\mu}} \right \}$ is a coordinate basis as well \cite{BEST}. This means that we can evaluate the Riemann-Christoffel symbols and the Riemann curvature tensor for the boosted spacetime metric 
(\ref{g11})--(\ref{g34}) by setting $c_{abc}=0$ in Eqs. (\ref{Gamma}) and (\ref{Riemann tensor}). Nevertheless, these relations are 
still too complicated to be computed analytically, and therefore a numerical calculation has been necessary. Formulas 
(\ref{g11})--(\ref{g34}) show indeed that we are dealing with a spacetime metric represented by a $4 \times 4$ matrix whose elements 
are given by some complicated non-vanishing functions of the $Y_{\mu}$ coordinates. That is why we first tried to compute the Riemann 
curvature tensor analytically in terms of tetrads (see Appendix \ref{Tetrad}) before realizing that even this solution was far too 
complicated. Thus, the only way we had to compute the Riemann-Christoffel symbols and the Riemann tensor was represented by numerical 
calculations. In this way we can evaluate the behavior of spacetime curvature also in the ultrarelativistic regime, which is the one 
we are mainly interested in, by letting the velocity $v$, defined by the boost relations (\ref{boost 1})--(\ref{boost 2}), approach gradually 
the speed of light \cite{BEST}. 

In the following sections we discuss the results of our computation mainly by studying curvature invariants and the behavior of geodesics in our 
reference spacetime, since we believe that these features represent the best tools to describe physically the concept of spacetime curvature. However, before going on, a little digression on the topic of singularities in general relativity turns out to be essential.

\subsection{Spacetime singularities}\label{b-completeness}

Intuitively, a spacetime singularity is a ``place'' where the curvature ``blows up'' \cite{Wald} or, by analogy with electrodynamics, 
a point where the metric tensor is either not defined or not suitably differentiable \cite{Haw-Ell}. Regrettably, both these statements 
are not rigorous definitions that can characterize the concept of spacetime singularity. First of all, since in general relativity we do not know the 
manifold and the metric structure in advance (they are solutions of Einstein field equations), we are not able to give a physical sense to the 
notion of an event until we solve Einstein equations, and hence the idea of a singularity as a ``place'' has not a satisfactory meaning. 
Moreover, also the notion of curvature becoming larger and larger as a general criterion for singularities has pathological problems. In fact, 
the bad behavior of components or derivatives of the Riemann tensor could be ascribed to the coordinate or tetrad basis employed. To avoid this 
problem, one might examine scalar curvature invariants constructed from the Riemann tensor or its covariant derivatives, which in some cases 
can completely characterize the spacetime (see Ref. \cite{CHP} for further details). However, even if the value of some scalar invariants 
is unbounded, curvature might blow up only ``as one goes to infinity'', a case that we would interpret as a singularity-free spacetime 
\cite{Wald}. Furthermore, spacetimes may be singular without any bad behavior of the curvature tensor (the so-called ``conical singularities'' 
\cite{Wald}). Lastly, the bad behaviour of the metric tensor at some spacetime points cannot be a way to define singularities, as one could 
always cut out such points and hence the remaining manifold, representing the whole spacetime, would turn out to be non-singular. 

A more satisfactory idea to define singularities consists in using the notion of incompleteness of timelike geodesics, i.e., geodesics which are 
inextensible in at least one direction and hence characterized only by a finite range of the affine parameter. This has the immediate physical interpretation 
that there exist freely moving observers or particles whose histories did not exist after (or before) a finite interval of proper time. 
Although the physical meaning of affine parameter on null geodesics is different from the case of timelike ones, we could also regard 
null geodesic incompleteness as a good criterion to define spacetime singularities.  Thus, timelike and null geodesic completeness are minimum 
conditions for spacetime to be considered singularity-free \cite{Haw-Ell}. However, since there are examples of geodesically complete spacetimes 
which contain an inextensible timelike curve of bounded acceleration and finite length \cite{Geroch}, we should generalize the concept of 
affine parameter to all $C^1$ curves, no matter whether they are geodesics or not. This fact is linked to the concept of bundle completeness (b-completeness), which we shortly describe following Refs. \cite{Haw-Ell,Sch}.

The b-boundary construction is a device to attach to any spacetime a set of boundary points. Such a boundary point can be considered as 
an equivalence class of inextensible curves in a spacetime, whose affine length is finite \cite{Haw-Ell,Sch}.

Let $\lambda(t)$ be a $C^1$ curve through a point $p$ of a manifold $M$ and let $\{ E_{\mu} \}$ (like before $\mu=1,2,3,4$) be a basis 
for the tangent vector space at $p$ to the manifold $M$, $T_p M$. We can propagate $\{ E_{\mu} \}$ along $\lambda(t)$ to obtain a basis 
for $T_{\lambda (t)}M,  \; \forall t$. Then any $V = \left( \partial / \partial t \right)_{\lambda(t)} \in T_{\lambda (t)}M$ can be 
expressed as $V = V^\mu (t) E_\mu$ and we can define a generalized affine parameter $u$ on the curve $\lambda(t)$ by \cite{Haw-Ell}
\begin{equation}
u = \int_{p}  \left( \sum_{\mu} V_{\mu} V^{\mu} \right)^{1/2} {\rm d}t.
\end{equation}
Let $\{ E_{\mu^\prime} \}$ be another basis of $T_p M$. Then there exists some non-singular matrix $A^{\mu}_{\; \; \nu}$ such that
\begin{equation}
E_{\nu} = \sum_{\mu^\prime} A^{\mu^{\prime}}_{\; \; \nu} E_{\mu^\prime}.
\end{equation}
As $\{ E_{\mu^\prime} \}$ and $\{ E_{\mu} \}$ are parallely transported along $\lambda(t)$, this relation is valid with constant 
$A^{\mu}_{\; \; \nu}$ and hence we have
\begin{equation}
V^{\mu^\prime} (t)= \sum_{\nu} A^{\mu^\prime}_{\; \; \nu} V^{\nu}(t).
\end{equation}
Since $A^{\mu}_{\; \; \nu}$ is non-singular, there exists some constant $C>0$ such that \cite{Haw-Ell}
\begin{equation}
C \sum_{\mu} V_\mu V^\mu \leq \sum_{\mu^\prime} V_{\mu^\prime} V^{\mu^\prime} \leq C^{-1} \sum_{\mu} V_\mu V^\mu. 
\end{equation}
Thus, the length of a curve $\lambda$ is finite in the parameter $u$ if and only if it is finite in the parameter $u^\prime$. 
If $\lambda$ is a geodesic then $u$ becomes its affine parameter, but the definition given above is still valid since it has been formulated 
in terms of a general parameter $u$ defined on any $C^1$ curve. Therefore, we say that a spacetime $(M,g)$ is b-complete if there exists 
an endpoint for every $C^1$ curve of finite length as measured by a generalized affine parameter. We have that b-completeness implies 
g-completeness (short for geodesic completeness), but the converse is not true. Therefore, we can define a spacetime to be singularity-free 
if it is b-complete. Thence, we recover the fundamental property outlined before according to which g-completeness represents the minimum condition for a spacetime to be considered singularity-free.

Therefore, we can classify a singularity represented by the presence of at least one incomplete geodesic according to whether \cite{Wald}
\begin{enumerate}
\item a curvature invariant blows up along a geodesic (``scalar curvature singularity''),
\item a component of the Riemann tensor or its covariant derivatives in a parallelly propagated tetrad blows up along a geodesic 
(``parallely propagated curvature singularity''),
\item no such invariant or component blows up (``non-curvature singularity'').
\end{enumerate}

\subsection{The Kretschmann invariant}

The review of the previous section clearly shows the important role fulfilled by scalar curvature invariants in the analysis of spacetime singularities. Being coordinate 
independent, they can provide important hints regarding the size of curvature and its growth along timelike curves, and can also characterize 
curvature singularities \cite{Thorpe}, while providing important information about the nature of singularities. For example, in the case of 
Schwarzschild metric, which can be obtained from (\ref{S-dS metric}) by setting $a=\infty$ (for an unambiguous definition of the notion of limit 
applied to spacetimes see Ref. \cite{Geroch69}), the Kretschmann invariant (i.e., the Riemann tensor squared) is such that 
\begin{equation}
R^{\alpha \beta \gamma \delta} R_{\alpha \beta \gamma \delta} = \dfrac{48m^2}{r^6}, \label{Schwarzschild kretsch}
\end{equation}
in agreement with the fact that in all coordinate systems the real singularity is located only at $r=0$ and not also at $r=2M$ (i.e., the event horizon). 

In order to study the features underlying the Riemann curvature of the spacetime described by the metric (\ref{ultrarelativistic boosted metric}), 
we therefore have decided to plot the Kretschmann invariant at different values of boost velocity $v$ and study the geodesic equation (dots denote derivatives with respect to the affine parameter)
\begin{equation}
\ddot{Y}^{\mu}(s) + \Gamma^{\mu}_{\; \nu \lambda} \dot{Y}^{\nu}(s) \dot{Y}^{\lambda}(s)=0, \label{geodesic eq}
\end{equation} 
$s$ being the affine parameter of the geodesic having parametric equation $Y^{\mu}=Y^{\mu}(s)$.

From the analysis of the Kretschmann invariant, we found that it is not defined unless the inequality 
(hereafter, numerical values of $Y$ coordinates have downstairs indices, to be consistent with the notation of previous sections)
\begin{equation}
(Y_{1})^2+(Y_{2})^2+(Y_{3})^2+(Y_{4})^2 > a^2, \label{singularity kretschmann}
\end{equation}
is satisfied \cite{BEST}. Then, we see that the hyperboloid constraint, i.e., condition (\ref{hyperboloid constrain}), allows us to define a 3-sphere of 
radius $a$ where the Kretschmann invariant  is not defined. This peculiar feature of our ``boosted spacetime geometry'' is indeed obvious 
if we look at formulas (\ref{g11})--(\ref{g34}), since there the quantities $\sigma$ and $\sqrt{\sigma}$ always occur at the denominator of 
the expressions of $g_{\mu \nu}$, a condition which means that the metric tensor is defined only if the inequality 
(\ref{singularity kretschmann}) holds. Moreover, it is possible to derive Eq. (\ref{singularity kretschmann}) in  the regime $v<1$ from the analysis of the Kretschmann invariant of the Schwarzschild-de Sitter metric. In fact, the Kretschmann invariant associated to (\ref{S-dS metric}) reads as \cite{BEST}
\begin{equation}
R^{\alpha \beta \gamma \delta} R_{\alpha \beta \gamma \delta} = 24 \left( \dfrac{1}{a^4}+\dfrac{2m^2}{r^6} \right), \label{S-dS kretsch}
\end{equation}
which reduces to (\ref{Schwarzschild kretsch}) in the limit $a=\infty$. Therefore, if we consider only bounded values of $a$, it follows immediately from (\ref{S-dS kretsch}) that the Schwarzschild-de Sitter metric (\ref{S-dS metric}) has an unique singularity located at $r=0$. Equation (\ref{r^2_1}) clearly shows that the condition $r=0$ leads to
\begin{equation}
\sqrt{\gamma^2 (v \sqrt{\sigma}+Y_1)^2+(Y_2)^2+(Y_3)^2}=0,
\end{equation}
which, being defined by the sum of squared quantities, in turns implies that
\begin{equation}
\begin{dcases}
& v \sqrt{\sigma}+Y_1=0, \\
& Y_2=0, \\
& Y_3 =0.
\end{dcases} 
  \label{r=0 in terms of Y}
\end{equation}
Thus, because of the presence of the term $\sqrt{\sigma}$, the condition $r=0$ is equivalent to (\ref{r=0 in terms of Y}), provided that $\sigma \geq 0$. If we now bear in mind that Eqs. (\ref{g11})--(\ref{g34}) prevent $\sigma$ from vanishing, we can conclude that the only possible choice is $\sigma >0$, which is equivalent to (\ref{singularity kretschmann}). In other words, the presence of the 3-sphere where the Kretschmann invariant is not defined follows directly from the condition $r=0$ which makes the curvature invariant (\ref{S-dS kretsch}) diverge \cite{BEST}. This fact can be interpreted as a hint indicating that this 3-sphere could represent a singularity of our ``boosted geometry''. Furthermore, it is interesting to note that the locus $r=0$ corresponds to the original position of the point source of the Schwarzschild-de Sitter metric which we have been boosted to become the two null sources (cf. Eqs. (\ref{Podolsy2.14a}) and (\ref{Podolsy2.14b})) of (\ref{ultrarelativistic boosted metric}). Eventually, if we interpret $Y_0$ as the time coordinate (see (\ref{boost 1})), we can view (\ref{singularity kretschmann}) as a condition on time.

\begin{figure} 
\centering
\includegraphics[scale=1.2]{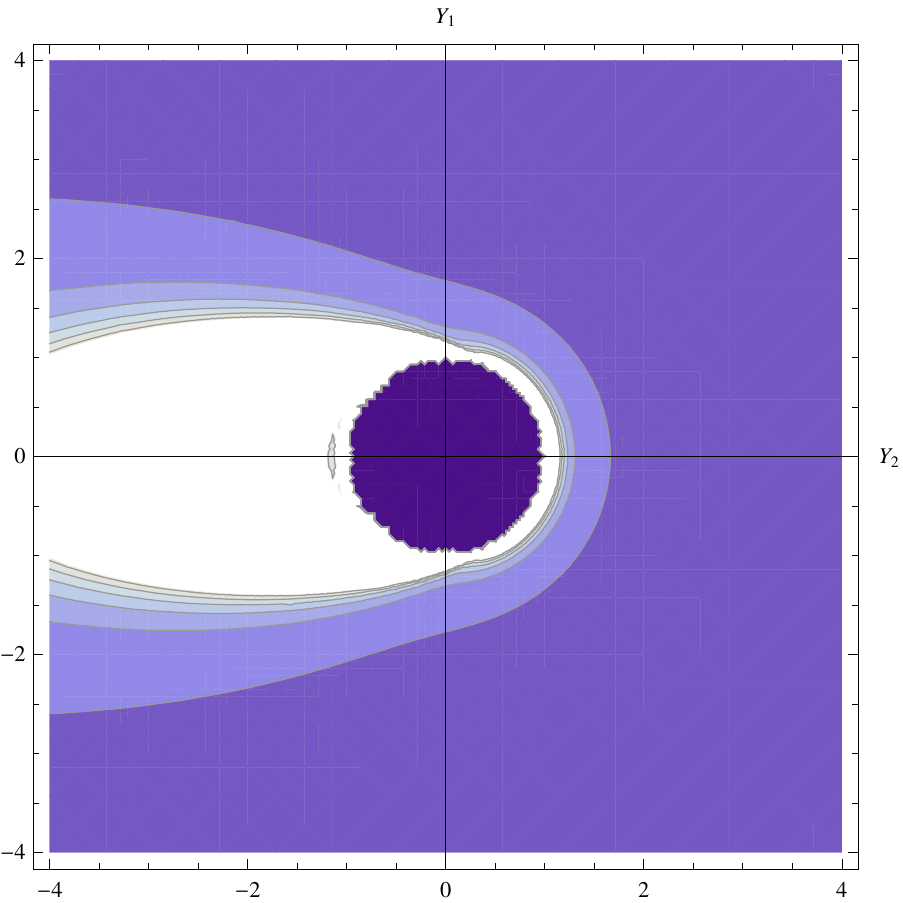}
\caption[Contour plot of the Kretschmann invariant of the boosted Schwarzschild-de Sitter metric]{Contour plot of the Kretschmann invariant numerically obtained with the following values of parameters: $a=1$, $m=0.1$, $Y_3 = Y_4 =0$ and $v=0.99$. The dark purple zone represents the circle of radius $a$ where the Kretschmann invariant is not defined.}
\label{contourplot}
\end{figure}

In the $Y_1-Y_2$ plane this 3-sphere becomes the circle with center at $Y_1 = Y_2 =0$ and radius $a$ depicted in Fig. \ref{contourplot}, which 
represents a contour plot of the Kretschmann invariant, i.e., a plot where each different color corresponds to different values of the Kretschmann 
invariant. It is thus possible to appreciate how the values assumed by the Kretschmann invariant increase as we approach this circle \cite{BEST}. 

\subsection{Boosted horizon}

\begin{figure}
\centering
\includegraphics[scale=1.2]{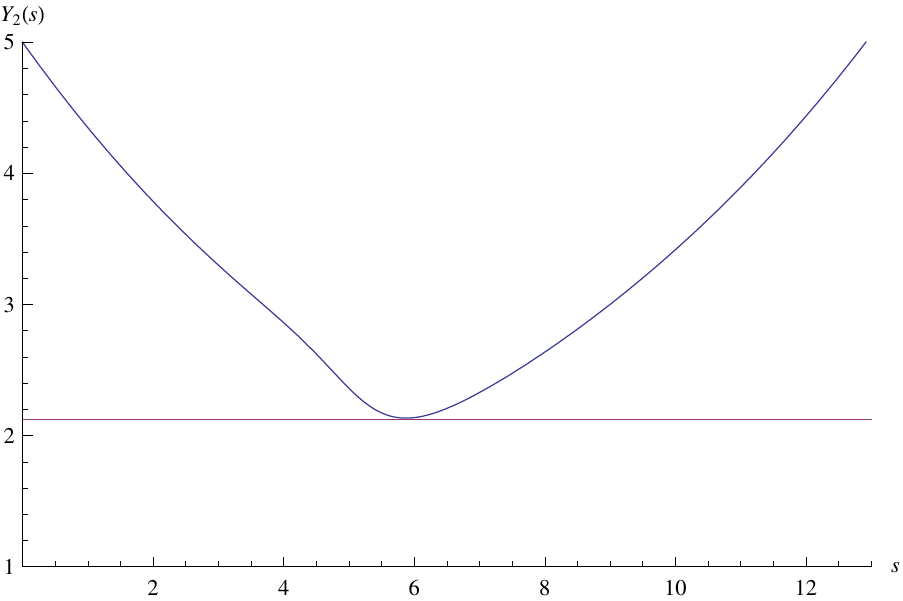}
\caption[Numerical solution of the geodesic equation I]{Numerical solution of Eq. (\ref{geodesic eq}) for the function $Y_{2}(s)$ obtained in the $Y_1-Y_2$ plane and with initial 
conditions $Y_{1}(0)=Y_3(0)=Y_4(0)=0$, $Y_2(0)=5$, $Y^{\prime}_{1}(0)=Y^{\prime}_{3}(0)=Y^{\prime}_{4}(0)=0$ and 
$Y^{\prime}_{2}(0)=-0.7$. The values of parameters are: $a=1$, $m=0.1$ and $v=0.9$. It is possible to see an ``antigravity effect'', 
since the function $Y_2(s)$ is pushed away from the ``boosted horizon'', which is represented by the horizontal line located at $Y_2=2.12$.}
\label{geod1}
\end{figure}

\begin{figure}
\centering
\includegraphics[scale=1.2]{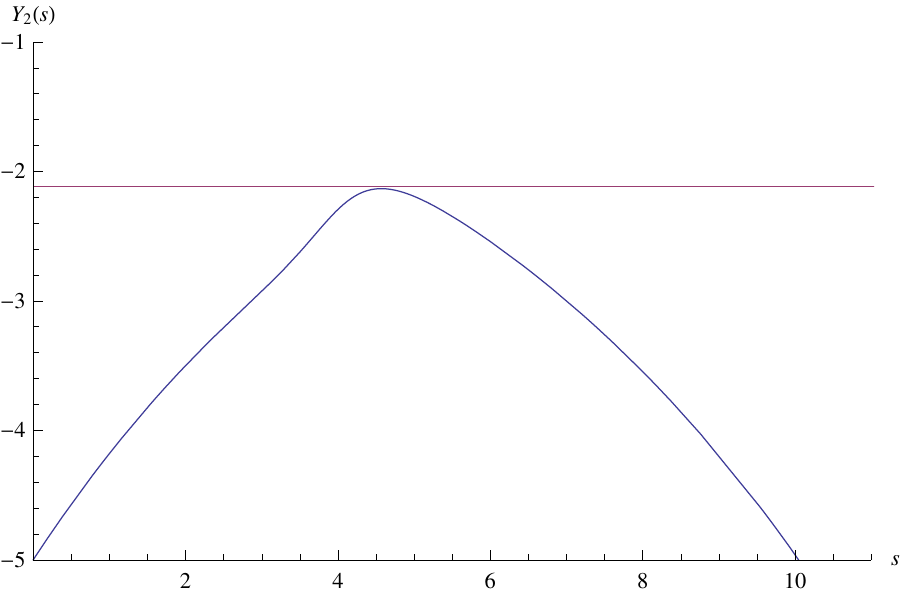}
\caption[Numerical solution of the geodesic equation II]{Numerical solution of Eq. (\ref{geodesic eq}) for the function $Y_2(s)$ obtained in the $Y_1-Y_2$ plane and with initial 
conditions $Y_1(0)=Y_3(0)=Y_4(0)=0$, $Y_2(0)=-5$, $Y^{\prime}_{1}(0)=Y^{\prime}_{3}(0)=Y^{\prime}_{4}(0)=0$ and 
$Y^{\prime}_{2}(0)=0.9$. The values of parameters are: $a=1$, $m=0.1$ and $v=0.9$.  The function $Y_2(s)$ initially moves toward the 
``boosted horizon'', i.e. the horizontal line at $Y_2=-2.12$, but then it is pushed away.}
\label{geod2}
\end{figure}

A really interesting feature characterizing ``boosted geometries'' is represented by the presence of a sort of barrier surrounding the  
3-sphere where the Kretschmann invariant is not defined, which we may call ``boosted horizon'', in the sense that all geodesics, despite maintaining their completeness condition, are surprisingly pushed away from it \cite{BEST}\footnote{More precisely, one defines an ``event horizon'' as the boundary of the causal past of future null infinity \cite{Haw-Ell}. In the ultrarelativistic regime we cannot say if this concept is still valid and hence we talk about 
``boosted horizon'' as the surface of spacetime surrounding the 3-sphere of radius $a$ where all geodesics, despite being 
complete, are pushed away.}. We have also discovered \cite{BEST} that the extension of the ``boosted horizon'' depends solely on the boost velocity $v$, as we 
will shortly see. Since we have found that all geodesics are complete, according to standard definitions of general relativity outlined in Sec. \ref{b-completeness} the ``boosted horizon'' is not a singularity but, as we will show, it seems to be a sort of elastic wall which is hit by all particles before they get away.  
We have observed this effect numerically, by varying initial conditions of (\ref{geodesic eq}) and the boost velocity $v$, so as to 
reproduce different physical situations.  Figures \ref{geod1} and \ref{geod2} indeed represent one among the many situations analyzed which witness this ``antigravity'' effect. These figures show in fact a particle initially lying on the $Y_1=0$ line of Fig. \ref{contourplot} and having an 
initial velocity directed toward the region where the Kretschmann invariant is not defined. Strikingly, the solution ``refuses'' to be 
attracted by the 3-sphere but, regardless of its initial velocity, the particle always arrives at a certain point and then it goes away from it, 
as if an elastic wall was present. We propose to call this elastic wall ``boosted horizon'' \cite{BEST}. The position of such a ``boosted horizon'' 
is independent of the initial velocity of the particle, but depends only on the boost velocity $v$. In fact, bearing in mind Fig. 
\ref{contourplot}, both for particles coming from ``above'' (i.e., particles initially lying on the positive half-line $Y_2>0$, 
$Y_1=0$ and with $Y^{\prime}_{2}(0) <0$) and for those coming from ``below'' (i.e., particles initially lying on the negative half-line 
$Y_2<0$, $Y_1=0$ and with $Y^{\prime}_{2}(0)>0$), the position of the ``boosted horizon'' does not change, as Tab. \ref{boosted horizon position} shows.
\begin{table}
\centering
\renewcommand\arraystretch{1.0}
\begin{tabular}{ |l | c|}
\hline
boost velocity &  ``boosted horizon'' location \\
& ($Y_2$ coordinate) \\
\hline
0.9995 & $\pm$ 1.02\\
0.9992 & $\pm$ 1.02\\
0.9991 & $\pm$ 1.02\\
0.999 & $\pm$ 1.02 \\
0.99  & $\pm$ 1.41 \\
0.9 & $\pm$ 2.12 \\
0.8 & $\pm$ 2.33 \\
0.7 & $\pm$ 2.43 \\
0.6 & $\pm$ 2.48 \\
0.5 & $\pm$ 2.48\\
0.4 & $\pm$ 2.42 \\
0.3 & $\pm$ 2.42 \\
0.2 & $\pm$ 2.34 \\
0.1 & $\pm$ 2.19 \\
0.01 & $\pm$ 1.52 \\
0.00155 & $\pm$ 1.00 \\
0.001 & $\pm$ 0.88 \\
0.0001 & $\pm$ 0.27 \\
\hline
\end{tabular}
\caption[Location of the ``boosted horizon'' as a function of the boost velocity $v$]{Location of the ``boosted horizon'' as a function of the boost velocity $v$. The positive sign refers to particles coming from ``above'' and the negative to those coming from ``below''. The values of parameters are $a=1$ and $m=0.1$.}
\label{boosted horizon position}
\end{table}
We have numerically checked, for each line of Tab. \ref{boosted horizon position}, 
that the minimum distance of the particle from the boundary of the $3$-sphere is always bigger than
its radius $a$, independently of the particle initial velocity. This means that the ``boosted horizon'' is always outside the
$3$-sphere. For example, we find that, when the boost velocity $v=0.5$, the minimum distance 
$d_{m}=3.1$ when $a=1$, and it decreases monotonically as $v$ increases or decreases, reaching a minimum value
of order $1.05 \div 1.10$ \cite{BEST}. 

\begin{figure}
\centering
\includegraphics[scale=1.2]{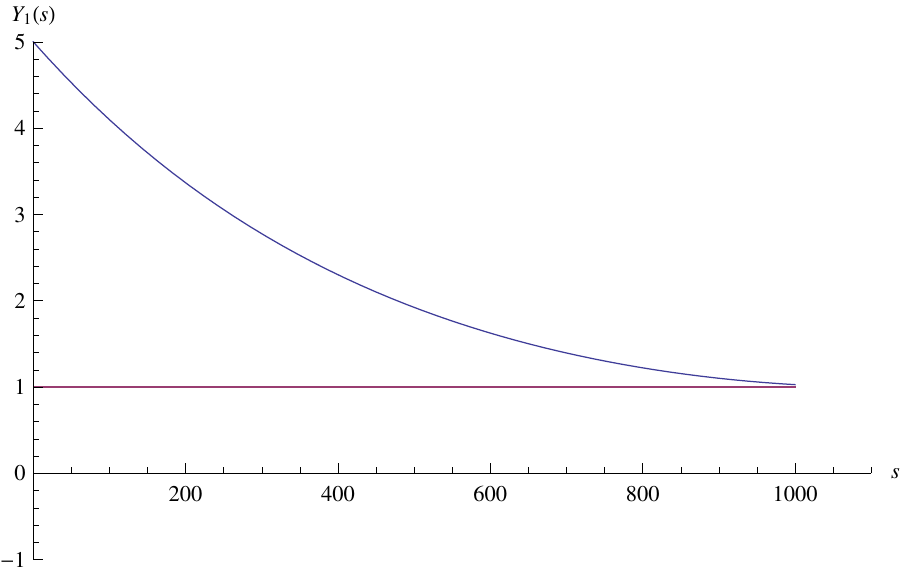}
\caption[Numerical solution of the geodesic equation III]{Numerical solution of Eq. (\ref{geodesic eq}) for the function $Y_1(s)$ obtained in the $Y_1-Y_2$ plane and with initial 
conditions $Y_1(0)=5$, $Y_2(0)=Y_3(0)=Y_4(0)=0$, $Y^{\prime}_1(0)=-0.01$, $Y^{\prime}_2(0)=Y^{\prime}_3(0)=Y^{\prime}_4(0)=0$. 
The values of parameters are: $a=1$, $m=0.1$ and $v=0.99$. The particle manages to hit the $3$-sphere, which is represented by the 
horizontal line $Y_1=1$.}
\label{geod3}
\end{figure}

The situation becomes somewhat intriguing when the particle lies initially on the $Y_2=0$ line (see Fig. \ref{contourplot}). 
In fact, in the cases in which the particle lies initially on the positive half-line $Y_1>0$, $Y_2=0$, it always manages to hit the 
3-sphere where the Kretschmann invariant is not defined, even if its initial velocity is extremely low, as we can see from Fig. 
\ref{geod3}. Once the particle has reached the 3-sphere, its geodesic is not defined anymore and hence, according to the analysis of Sec. \ref{b-completeness}, we can conclude that the 3-sphere of equation $(Y_{1})^2+(Y_{2})^2+(Y_{3})^2+(Y_{4})^2 = a^2$ defines 
a ``scalar curvature singularity'' for our ``boosted geometry'', as we have guessed before \cite{BEST}. 

\begin{figure} 
\centering
\includegraphics[scale=1.2]{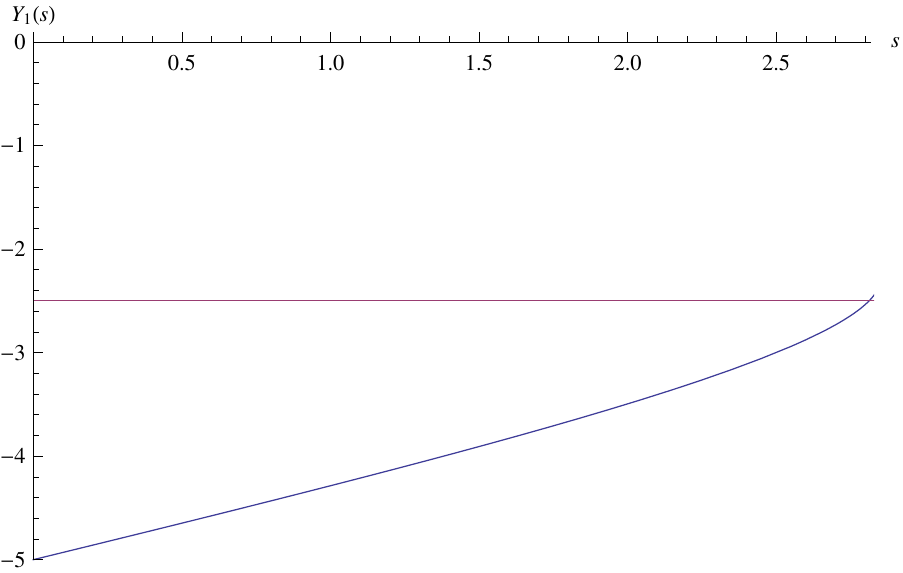}
\caption[Numerical solution of the geodesic equation IV]{Numerical solution of Eq. (\ref{geodesic eq}) for the function $Y_1(s)$ obtained in the $Y_1-Y_2$ plane and with initial conditions $Y_1(0)=-5$, $Y_2(0)=Y_3(0)=Y_4(0)=0$, $Y^{\prime}_1(0)=0.7$, $Y^{\prime}_2(0)=Y^{\prime}_3(0)=Y^{\prime}_4(0)=0$. 
The values of parameters are: $a=1$, $m=0.1$ and $v=0.99$. The particle does not manage to hit the $3$-sphere but disappears in 
correspondence of the $Y_1=-2.5$ line. }
\label{geod4}
\end{figure}

\begin{figure}
\centering
\includegraphics[scale=1.2]{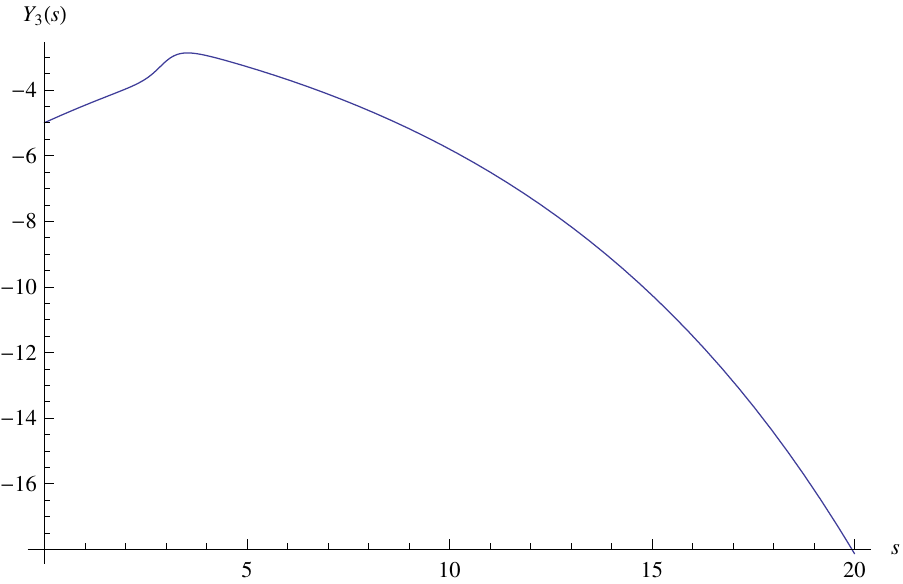}
\caption[Numerical solution of the geodesic equation V]{Numerical solution of Eq. (\ref{geodesic eq}) for the function $Y_3(s)$ obtained in the $Y_3-Y_4$ plane and with initial 
conditions $Y_1(0)=Y_2(0)=0$, $Y_3(0)=Y_4(0)=-5$, $Y^{\prime}_1(0)=Y^{\prime}_2(0)=0$, $Y^{\prime}_3(0)
=Y^{\prime}_4(0)=0.566$. The values of parameters are: $a=1$, $m=0.1$ and $v=0.99$. The ``antigravity effect'' is once again evident.}
\label{geod5}
\end{figure}

\begin{figure} 
\centering
\includegraphics[scale=1.2]{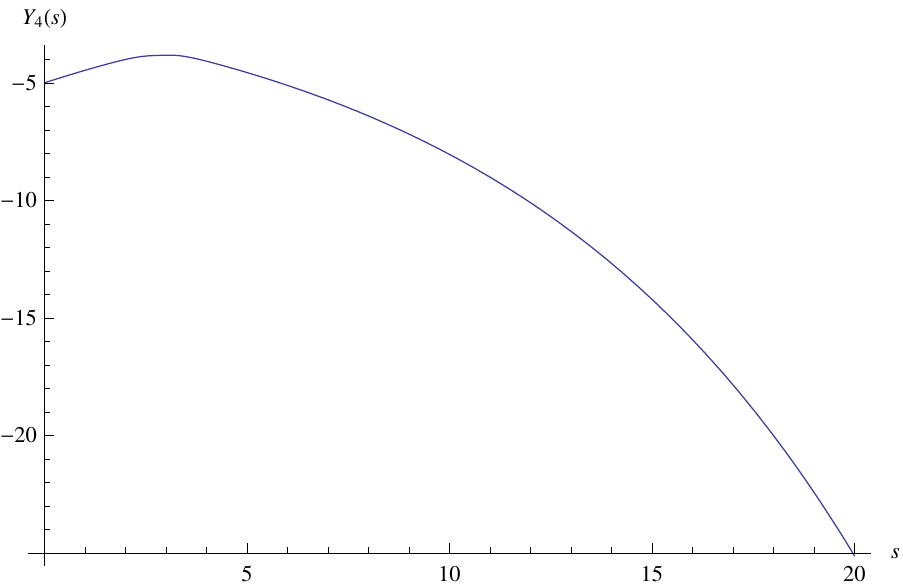}
\caption[Numerical solution of the geodesic equation VI]{Numerical solution of Eq. (\ref{geodesic eq}) for the function $Y_4(s)$ obtained in the $Y_3-Y_4$ plane and with initial 
conditions $Y_1(0)=Y_2(0)=0$, $Y_3(0)=Y_4(0)=-5$, $Y^{\prime}_1(0)=Y^{\prime}_2(0)=0$, $Y^{\prime}_3(0)
=Y^{\prime}_4(0)=0.566$. The values of parameters are: $a=1$, $m=0.1$ and $v=0.99$. The ``antigravity effect'' is once again evident.}
\label{geod6}
\end{figure}

When the particle lies initially on the negative half-line $Y_1<0$, $Y_2=0$, its geodesic is not defined even before it reaches the 
$3$-sphere (see Fig. \ref{geod4}). This means that another ``scalar curvature singularity'' exists. Its position depends only on the boost 
velocity $v$ and not on the particle initial velocity. In any case, numerical analysis shows that this kind of singularities exists only if the particle lies initially on the $Y_2=0$ line \cite{BEST}. 
We have repeated the same analysis also by putting $Y_1=Y_2=0$ in the relations defining the curvature, i.e., in the $Y_3-Y_4$ 
plane, and we have found the same ``antigravity effect'' of the previous cases, as shown in Figs. \ref{geod5} and \ref{geod6}, which 
represent some examples among the many situations numerically analyzed. Interestingly, in this case we have  found no 
``scalar curvature singularities'' \cite{BEST}.

In the ultrarelativistic regime ($v=0.9999$) the ``antigravity effects'' are still present but, as is clear from Tab. \ref{boosted horizon position}, the position of the boosted horizon tends to that of the singularity 3-sphere \cite{BEST}. 

\section{The coordinate shift method}

An important question arises while dealing with Secs. \ref{The boosting procedure_Sec} and \ref{Riemann curvature of the boosted Schwarzschild-de Sitter spacetime_Sec}, i.e., how to cope with the Riemann curvature tensor when it has terms proportional to $\delta^2$. In fact, from (\ref{ultrarelativistic boosted metric}) it is easy to understand that the Riemann tensor has got terms involving the products of two Dirac's $\delta$ distributions (a formal method to cope with multiplication of distributions can be found in Ref. \cite{Colombeau}). This means that the ``boosted Riemann tensor'' of our ``boosted geometry'' is in principle not defined. Anyway, we will be able to show that the $\delta^2$ terms appearing in the ``boosted Riemann tensor'' vanish in a distributional sense. Unlike the (rather simple) example discussed in Ref. \cite{AS1971}, we will achieve this point in a more difficult way, since the high difficulty of metric (\ref{ultrarelativistic boosted metric}) makes it quite impossible to write down explicitly all the boosted Riemann tensor components, as pointed out before. For this reason in this section we will make use of an equivalent method to describe the gravitational shock-wave of a massless particle, i.e., the coordinate shift method \cite{Dray,Sfetsos95} (or, equivalently, the scissors-and-paste method introduced by Penrose \cite{Penrose72}). The equivalence of this method and the boosting procedure has been demonstrated by the authors of Ref. \cite{Dray}, where it is explicitly shown that with the new approach it is possible to recover the results of Aichelburg and Sexl. By exploiting this equivalence between the two methods, we will show in which sense the $\delta^2$ terms appearing in the Riemann tensor of metric (\ref{ultrarelativistic boosted metric}) can be seen as vanishing, leading to a well defined spacetime function (in the sense of distributions) \cite{BEST}. 

Therefore, this section has two purposes: on one hand it elucidates the features of the new method, on the other hand it proposes a recipe for the problem concerning the presence of products of two distributions in the Riemann tensor.  

\subsection{Formal aspects}

As we know, the sources of gravitational shock-waves are massless particles moving at the speed of light. Thus, we could consider particles moving along a null surface such as the event horizon in the case of black holes. Therefore, another way to introduce a gravitational shock-wave is through a coordinate shift which reflects this peculiarity. This method can be applied both to vacuum solutions of Einstein equations \cite{Dray} and in presence of matter fields and non-vanishing cosmological constant \cite{Sfetsos95}.

Following Refs. \cite{Dray,Sfetsos95}, we start with a background geometry having line element   
\begin{equation}
{\rm d}s^2 = 2 A(u,v) {\rm d}u {\rm d}v +g(u,v) h_{ij}(x) {\rm d}x^i{\rm d}x^j, \label{Sfetsosmetric}
\end{equation}
with $i,j=1,2$ (hereafter $v$ is a spacetime coordinate, unlike the previous sections where it indicates the boost velocity). We also assume the presence of some matter fields whose  non-vanishing components of the energy-momentum tensor are given by
\begin{equation}
T=2 \; T_{uv}(u,v,x)  \; {\rm d}u {\rm d}v + T_{uu}(u,v,x)  \; {\rm d}u^2 +  T_{vv}(u,v,x)  \; {\rm d}v^2 + T_{ij}(u,v,x)  \; {\rm d}x^i {\rm d}x^j.
\end{equation} 
 Consider a massless particle located at $u=0$ and moving with the speed of light in the $v$-direction. The coordinate shift method consists in making the ansatz according to which for $u<0$ the spacetime is still described by (\ref{Sfetsosmetric}), whereas for $u>0$ we suppose that the background geometry (\ref{Sfetsosmetric}) (back-)reacts in such a way that $v$ is shifted as $v \rightarrow v + f(x)$, where $f(x)$ is a (shift) function to be determined. Therefore, the resulting line element reads as
\begin{equation}
{\rm d}s^2 = 2 A(u,v+\Theta f) {\rm d}u\left( {\rm d}v +\Theta f_{,i} {\rm d}x^i  \right) +g(u,v+\Theta f) h_{ij}(x) {\rm d}x^i{\rm d}x^j, \label{shiftedmetric}
\end{equation} 
where $\Theta=\Theta(u)$ is the Heaviside step function and
\begin{equation}
\begin{split}
T=& 2 \; T_{uv}(u,v+\Theta f,x)  \; {\rm d}u ({\rm d}v+\Theta f_{,i} {\rm d}x^i )+ T_{uu}(u,v+\Theta f,x)  \; {\rm d}u^2  \\ 
&+ T_{vv}(u,v+\Theta f,x)  \; ( {\rm d}v+\Theta f_{,i} {\rm d}x^i )^2 + T_{ij}(u,v+\Theta f,x)  \; {\rm d}x^i {\rm d}x^j.
\end{split}
\end{equation} 
With the notation
\begin{equation}
\hat{u}=u, \;\;\;\;\; \hat{v}=v+f(x) \Theta(u), \;\;\;\;\; \hat{x}^i=x^i, 
\end{equation}
the metric (\ref{shiftedmetric}) assumes the handy form
\begin{equation}
\begin{split}
{\rm d}s^2 &= 2 \hat{A}\; {\rm d}\hat{u}\left( {\rm d}\hat{v} -\delta(\hat{u}) \hat{f} {\rm d} \hat{u}\right)+ \hat{g}\; \hat{h}_{ij}(x) \;{\rm d}\hat{x}^i{\rm d}\hat{x}^j \\
&= 2 \hat{A}\; {\rm d}\hat{u} {\rm d}\hat{v} +\hat{F}\; {\rm d}\hat{u}^2 +\hat{g}\; \hat{h}_{ij}(x) \;{\rm d}\hat{x}^i{\rm d}\hat{x}^j, \label{handymetric}
 \end{split}
\end{equation}
and the energy-momentum tensor becomes
\begin{equation}
T= 2 \left(\hat{T}_{\hat{u}\hat{v}} -\hat{T}_{\hat{v}\hat{v}} \; \hat{f} \hat{\delta} \right)  {\rm d}\hat{u} {\rm d}\hat{v}  + \left(\hat{T}_{\hat{u}\hat{u}}+  \hat{T}_{\hat{v}\hat{v}} \; \hat{f}^2 \hat{\delta}^2 - 2 \hat{T}_{\hat{u}\hat{v}} \; \hat{f} \hat{\delta} \right) {\rm d}\hat{u}^2  
 + \hat{T}_{\hat{v}\hat{v}} {\rm d}\hat{v}^2 + \hat{T}_{ij} {\rm d}\hat{x}^i{\rm d}\hat{x}^j, \label{shifted_en-mom_tensor}
\end{equation}
with $\hat{F}=F(\hat{u},\hat{v},\hat{x})=-2 \; \hat{A} \; \hat{f} \; \hat{\delta}$ and where the hats indicate that the corresponding quantities are evaluated at $\hat{u}$, $\hat{v}$, $\hat{x}$ and $\hat{\delta}=\delta(\hat{u})$ is the $\delta$ distribution. We  now demand that the metric (\ref{handymetric}) satisfies Einstein equation where the energy-momentum tensor is given by Eq. (\ref{shifted_en-mom_tensor}) plus the the energy-momentum tensor of the massless particle located at the origin of the transverse $x$-space and at $u=0$ and moving at the speed of light in the $v$-direction
\begin{equation}
T^p=T^p_{\; uu} {\rm d}u^2=\hat{T}^p_{\; \hat{uu}} {\rm d}\hat{u}^2=-4p\; \hat{A}^2 \hat{\delta}^{(2)}(\hat{x}) \hat{\delta}(\hat{u}){\rm d}\hat{u}^2,
\end{equation}
where $p$ is the particle momentum. If we suppose that the parts of field equations that do not involve the function $f$ are automatically satisfied, we find, by examining the terms linear in $f\; \delta$, that the necessary and sufficient conditions for being able to introduce a gravitational shock-wave via a coordinate shift amount to demand that at $u=0$ there exist the additional conditions (hereafter we drop the hat symbol to simplify the notation)
\begin{equation}
g_{,v}=A_{,v}=T_{vv}=0, \label{g,v-A,v}
\end{equation}
\begin{equation}
\triangle_{h_{ij}}f-\dfrac{g_{,uv}}{A}f=32 \pi \; p \; g \; A \; \delta^{(2)}(x), \label{f(x)_eq}
\end{equation}
where 
\begin{equation}
\triangle_{h_{ij}}=\dfrac{1}{\sqrt{h}} \partial_i \sqrt{h}h^{ij} \partial_j,
\end{equation}
is the Laplacian with respect to the 2-metric $h_{ij}$. 

A crucial point is represented by the presence of $\delta^2$ type terms both in Riemann and in Ricci tensors.
We have found that the only Riemann tensor components of the metric (\ref{handymetric}) depending on $\hat{\delta}^2=\delta^2(\hat{u})$ are given by (dropping like before the hat symbol) \cite{BEST}
\begin{equation}
R^{v}_{\; uvu}=2 \left(A_{,uv}-\dfrac{A_{,u}A_{,v}}{A} \right)f \delta + 2 \left(\dfrac{A_{,vv}}{A}-\dfrac{A^2_{,v}}{A^2} \right)f^2 \delta^2,
\label{BESTD1}
\end{equation} 
\begin{equation}
R^{v}_{\;ux^iu}=\left(2\dfrac{A_{,v}}{A}-\dfrac{g_{,v}}{g}\right)f_{,x^i}f\delta^2, \; \; \; \;\; \; \;  (i=1,2),
\end{equation}
\begin{equation}
R^{x^i}_{\;ux^iu}=\left(\dfrac{g_{,v}}{g}\dfrac{A_{,v}}{A} \right)f^2 \delta^2 + \dots {\rm (terms \; at \; most \; linear \; in \; } \delta), \; \; \; \;\; \; \,  (i=1,2).
\label{BESTD3}
\end{equation}
Therefore the only Ricci tensor component having $\delta^2$ terms is
\begin{equation}
\begin{split}
R_{uu} &=\sum_{\rho}R^{\rho}_{\; u\rho u}=R^{v}_{\; uvu}+ R^{x^1}_{\; ux^1u}+ R^{x^2}_{\; ux^2u} \\
&= 2 \left(\dfrac{A_{,vv}}{A}-\dfrac{A^2_{,v}}{A^2}+\dfrac{g_{,v}}{g}\dfrac{A_{,v}}{A} \right)f^2 \delta^2 + \dots {\rm (terms \; at \; most \; linear \; in \; } \delta).
\end{split}
\end{equation}
These terms must vanish in a distributional sense, otherwise the Riemann and Ricci tensors are not defined. Anyway, by considering the conditions (\ref{g,v-A,v}), it is easy to show that the quantities $\dfrac{A_{,vv}}{A}$, $\dfrac{A^2_{,v}}{A^2}$, $\dfrac{g_{,v}}{g}$, $\dfrac{A_{,v}}{A}$ appearing both in Riemann and in Ricci tensors are of order ${\rm O}(u)$ or ${\rm O}(u^2)$. Since all quantities involving $\delta$ terms should be intended as distributions to be integrated over smooth functions, we can conclude that all these $\delta^2$ terms give vanishing contribution and hence both Riemann and Ricci tensors turn out to be under control as functions (in a distributional sense) of spacetime coordinates $(u,v,x^1,x^2)$ \cite{BEST}. 
The geodesic equations for the metric (\ref{handymetric}) obtained by varying the coordinates $v$ and $x^i$ are
\begin{equation}
\ddot{u}+ \dfrac{A_{,u}}{A} \dot{u}^2 -\dfrac{g_{,v}}{2 A} h_{ij} \dot{x}^i \dot{x}^j+f \; \dfrac{A_{,v}}{A} \delta \; \dot{u}^2=0,
\end{equation}
\begin{equation}
\ddot{x}^i + \Gamma^i_{\; jk} \dot{x}^j \dot{x}^k+\dfrac{g_{,u}}{g} \dot{u} \dot{x}^i +\dfrac{g_{,v}}{g} \dot{v} \dot{x}^i+\dfrac{A}{g} \; \delta \; f_{,i} h^{ij} \dot{u}^2=0,
\end{equation}
where $\Gamma^i_{\; jk}$ denote the Christoffel symbols (see Appendix A of Ref. \cite{Sfetsos95} for their lengthy expression); the geodesic equation obtained from the variation of $u$ is 
\begin{equation}
\begin{split}
& \ddot{v}+ \dfrac{A_{,v}}{A} \dot{v}^2-\dfrac{g_{,u}}{2 A} h_{ij} \dot{x}^i \dot{x}^j + \left(   f \; \dfrac{A_{,u}}{A} \dot{u}^2 -2 f \; \dfrac{A_{,v}}{A} \dot{u} \dot{v}-2 f_{,i} \dot{u} \dot{x}^i   -\dfrac{g_{,v}}{ A} \; f \; h_{ij} \dot{x}^i \dot{x}^j  \right) \delta  \\
& -f  \delta^{\prime} \dot{u}^2 + 2 f^2 \; \delta^2 \; \dfrac{A_{,v}}{A} \dot{u}^2 =0.
\end{split}
\end{equation}
On performing the integration of the geodesic equations, it is possible to understand how the original background geometry (\ref{Sfetsosmetric}) is affected by the presence of a massless particle moving in the $v$-direction at $u=0$. In fact, as the geodesic trajectory crosses the null surface $u=0$ there is a shift in its $v$-component expressed by the relation
\begin{equation}
\Delta v \equiv v\vert_{u=0^+} - v\vert_{u=0^-}=f(x), \label{v_discontinuity}
\end{equation}
and a refraction effect in the transverse $x$-plane expressed by the refraction function
\begin{equation}
R^i(x) \equiv \left. \dfrac{{\rm d}x^i}{{\rm d}u} \right\vert_{u=0^-} - \left. \dfrac{{\rm d}x^i}{{\rm d}u} \right\vert_{u=0^+} = \left. \frac{A}{g} \right\vert_{u=0} \; f_{,i} h^{ij}, \label{refraction_function}
\end{equation}
which measures the change of the angle that the trajectory forms with the $u=0$ surface after having crossed it. Therefore, when a trajectory crosses the $u=0$ null surface its $v$ component suffers from a discontinuity which, according to (\ref{v_discontinuity}), equals $f(x)$, while the other components remain continuous. Moreover, Eq. (\ref{refraction_function}) expresses the fact that the directional derivatives of $f(x)$ give information about how much the $x^i$ components change direction along $u$ while crossing the surface $u=0$. 

\subsection{de Sitter and Schwarzschild-de Sitter backgrounds}

As we know, our ``boosted geometry" is characterized by a spherical gravitational shock-wave evolving in de Sitter background. Therefore, within the pattern of the coordinate shift method we need to employ the metric (\ref{dS metric}). In this case, the computation is quite easy and hence we briefly expose the results. 

The line element (\ref{dS metric}) can be written in the equivalent form
\begin{equation}
{\rm d}s^2=-\tilde{\lambda} {\rm d}t^2 + \dfrac{{\rm d}r^2}{\tilde{\lambda}}+ r^2 {\rm d}\Omega^2,
\label{Sfetsos3.1}
\end{equation}
${\rm d}\Omega^2={\rm d}\theta^2 + \sin^2 \theta {\rm d} \, \phi^2$ being the metric on the unit 2-sphere. Bearing in mind Eq. (\ref{Sfetsos3.1}), in order to bring (\ref{dS metric}) in the form (\ref{Sfetsosmetric}), we should introduce the function \cite{Sfetsos95}
\begin{equation}
F:r \rightarrow  F(r)= {\rm exp} \left[ \dfrac{1}{a}\int {\rm d}r \; \tilde{\lambda}^{-1}\right],
\label{Sfetsos3.4}
\end{equation} 
and the independent variables
\begin{equation}
\begin{split}
& u= {\rm e}^{t/a} \; F(r), \\
& v={\rm e}^{-t/a} \; F(r). 
\end{split}
\label{coord_u,v}
\end{equation}
Moreover, we have \cite{Sfetsos95}
\begin{equation}
A(u,v)=\dfrac{1}{2} a^2 \dfrac{\tilde{\lambda}}{F(r)^2},
\end{equation}
\begin{equation}
g(u,v)=r^2.
\label{Sfetsos3.6}
\end{equation}
Therefore, from the above relations it follows that in the case of de Sitter metric
\begin{equation}
F(r)= {\rm exp} \left[ a\int {\rm d}r \; \dfrac{1}{a^2-r^2} \right]=\sqrt{\dfrac{a-r}{a+r}},
\end{equation} 
along with
\begin{equation}
\begin{split}
& A(u,v)= \dfrac{\left(1-{r^{2}\over a^{2}}\right)a^2}{2}\; F^{-2}=\dfrac{1}{2}(r+a)^2, \\
& g(u,v)=r^2.
\end{split}
\end{equation}

Having obtained the following relations for the derivatives \cite{BEST}: 
\begin{equation}
 \dfrac{{\rm d}}{{\rm d}u}= \dfrac{1}{2} \left(\dfrac{a\;{\rm e}^{-t/a}}{F(r)}\dfrac{{\rm d}}{{\rm d}t}+\dfrac{{\rm e}^{-t/a}}{F^\prime(r)}\dfrac{{\rm d}}{{\rm d}r} \right), 
 \label{BEST4.35}
\end{equation} 
\begin{equation}
  \dfrac{{\rm d}}{{\rm d}v}= \dfrac{1}{2} \left(\dfrac{-a\;{\rm e}^{t/a}}{F(r)}\dfrac{{\rm d}}{{\rm d}t}+\dfrac{{\rm e}^{t/a}}{F^\prime(r)}\dfrac{{\rm d}}{{\rm d}r} \right), 
  \label{BEST4.36}
\end{equation} 
it is easy to show that conditions (\ref{g,v-A,v}) are satisfied at the null hypersurface $u=0$ (i.e., $r=a$). 

At this stage, it is possible to show that the partial differential equation (\ref{f(x)_eq}) satisfied by the shift function $f(\theta)$ becomes \cite{Dray,Sfetsos95}
\begin{equation}
\triangle_{(2)}f-c\; f=2 \pi k \delta(\xi-1) \delta (\phi), \label{f-SdS_eq}
\end{equation}
with
\begin{equation}
\triangle_{(2)}= \partial_\xi (1-\xi^2) \partial_\xi + \dfrac{\partial^2_\phi}{(1-\xi^2)}, \; \; \; \; \; \; \; \; \;  (\xi=\cos \theta), 
\end{equation}
being the Laplacian on the unit 2-sphere, and $k$ and $c$ being  real constants. This equation represents the usual Legendre equation of order $n$ ($n$ being a solution of $n(n+1)+c=0$) with a Dirac's $\delta$ appearing on the right-hand side. Therefore, its solutions depend strongly on the values assumed by the constant $c$ and can be given in terms of Legendre polynomials as
\begin{equation}
f(\theta;c)= - k \sum_{l=0}^{+ \infty} \dfrac{\left( l+\dfrac{1}{2}\right)}{\left[ l(l+1)+c\right]}P_l(\cos \theta), \; \; \; \; \; \;  c \in \mathbb{R}-\{-N(N+1), \; N=0,1,... \}. \label{f_solution}
\end{equation}  
In the case of de Sitter background, we have
\begin{equation}
\begin{split}
& c=-2, \\
& k= 32 p a^4,
\end{split}
\label{Sfetsos4.4}
\end{equation}
and hence the solution (\ref{f_solution}) assumes the form \cite{Sfetsos95}
\begin{equation}
f(\theta)= 32 p a^4  \left[ 1 - \dfrac{1}{2} \cos \theta \log \left(\dfrac{1+ \cos \theta}{1-\cos \theta} \right) \right] \Theta \left(\dfrac{\pi}{2}- \theta\right),
\label{Sfetsos4.5}
\end{equation}
where the $\Theta$-function restricts the solution to the upper hemisphere. The shift function (\ref{Sfetsos4.5}) is such that it goes to minus infinity at $\theta=0$ and then monotonically increases until it reaches the value $32 p a^4$ at $\theta=\dfrac{\pi}{2}$. Thus, there is an angle ($\theta_0 \simeq 33.52\degree$) where it vanishes. The corresponding refraction function (cf. Eq. (\ref{refraction_function})) is given by \cite{Sfetsos95}
\begin{equation}
R(\theta)=64 p a^4 \left\{ \left[ \dfrac{\cos \theta}{\sin 2 \theta} + \dfrac{1}{2} \sin \theta \log  \left(\dfrac{1+ \cos \theta}{1-\cos \theta} \right) \right] \Theta \left(\dfrac{\pi}{2}- \theta\right) -  \delta \left( \theta - \dfrac{\pi}{2} \right)\right\}.
\end{equation} 
The first term is a monotonically decreasing function of $\theta$ which varies from plus infinity to zero as we move from the northern pole to the equator. However, exactly there the second term gives an infinite contribution. 

An important remark should be mentioned at this point. Since in this chapter, following Refs. \cite{BEST,ES2007}, we have employed the exact form of Schwarzschild-de Sitter metric and not the approximated one (cf. Eqs. (\ref{S-dS metric}) and (\ref{S-dS approx})), it is equally important to describe what happens if we would adopt a Schwarzschild-de Sitter background. Our exact approach in fact can be considered as an ultimate case of de Sitter background, because, unlike the authors of Ref. \cite{HT1993}, we have not regarded the mass parameter $m$ as a perturbation of de Sitter metric. However, this does not mean that our point of view has changed the background where the shock-wave evolves (it is still de Sitter, Eqs. (\ref{boost 1})--(\ref{boost 2}) being ruled by the de Sitter group $O(1,4)$), but we simply believe that it is ``morally'' necessary, also for possible future purposes, a discussion involving Schwarzschild-de Sitter background geometry. 

In the case of Schwarzschild-de Sitter metric, the calculations are more complex and hence we give some more details than de Sitter geometry. We will follow Ref. \cite{BEST}. From Eqs. (\ref{Sfetsos3.1})--(\ref{Sfetsos3.6}) jointly with (\ref{S-dS metric}) and (\ref{Sfetsosmetric}) we have that
\begin{equation}
F(r)= {\rm exp} \left[ \dfrac{1}{a} \int {\rm d}r \; \dfrac{r a^2}{(r a^2-r^3-2 m a^2)} \right],
\end{equation} 
\begin{equation}
A(u,v)=\dfrac{\left(1-{2m \over r}-{r^{2}\over a^{2}}\right)a^2}{2}\; F^{-2}, \label{A(u,v)} 
\end{equation}
\begin{equation}
g(u,v)=r^2.
\end{equation}
By performing the integration, we have found that \cite{BEST}
\begin{equation}
F(r)= {\rm exp} \left[ a \dfrac{r_1(r_3-r_2)\log(r-r_1)+r_2(r_1-r_3)\log(r-r_2)+r_3(r_2-r_1)\log(r-r_3)}{(r_1-r_2)(r_1-r_3)(r_2-r_3)} \right], \label{F(r)_1}
\end{equation}
$r_1$,$r_2$ and $r_3$ being the three roots of the cubic equation
\begin{equation}
r^3-ra^2+2ma^2=0, \label{cubic_eq}
\end{equation}
whose values are given by \cite{BEST}
\begin{equation}
r_1 = \dfrac{1}{3^{1/3}}\left(  \dfrac{a^2}{\Upsilon} + \dfrac{\Upsilon}{3^{1/3}} \right), \label{root_1} 
\end{equation}
\begin{equation}
 r_{2,3} =\dfrac{1}{2}\dfrac{1}{3^{1/3}}\left(- \dfrac{\left(1 \pm i \sqrt{3} \; \right)a^2}{\Upsilon} - \dfrac{ \left(1\mp i \sqrt{3} \; \right) \Upsilon}{3^{1/3}} \right), \label{root_2and3}
\end{equation}
where $\Upsilon$ is defined as
\begin{equation}
\Upsilon \equiv \left( -9a^2m+\sqrt{3}\sqrt{27a^4m^2-a^6} \right)^{1/3}.
\end{equation}
In other words, Eqs. (\ref{root_1}) and (\ref{root_2and3}) describe the three null surfaces where the metric (\ref{S-dS metric}) blows up, and hence the three horizons that characterize this geometry. With  the hypothesis $a/m >\sqrt{27}$ (which is respected by the choice $a=1$ and $m=0.1$ adopted in the previous sections) the discriminant of (\ref{cubic_eq}) becomes negative and then (\ref{root_1}) and (\ref{root_2and3}) turn out to be real roots. This condition allows us to write the quantities (\ref{root_1}) and (\ref{root_2and3}) in trigonometric form. We obtain \cite{BEST}
\begin{equation}
 r_1= \dfrac{2a}{\sqrt{3}} \cos \left( \dfrac{\varphi}{3} \right), 
\end{equation}
\begin{equation}
 r_{2,3} = -\dfrac{2a}{\sqrt{3}} \cos \left( \dfrac{\varphi \mp \pi}{3} \right)=-\dfrac{a}{\sqrt{3}} \left( \cos \dfrac{\varphi}{3} \pm \sqrt{3} \sin \dfrac{\varphi}{3} \right) \label{trigonometric_r3},
\end{equation}
where $\cos \varphi = \sqrt{27} m/a$. Note also that the roots (\ref{root_1}) and (\ref{root_2and3}) are characterized by the fact that $r_1+r_2+r_3=0$ and $r_1r_2r_3=-2ma^2$. Now, we can write (\ref{F(r)_1}) as
\begin{equation}
F(r)=\prod_{i=1}^{3}(r-r_i)^{k_i}, \label{F(r)}
\end{equation}
where the three constants $k_i \; (i=1,2,3)$ are given by 
\begin{equation}
 k_1=\dfrac{a r_1 (r_3-r_2)}{k_r} ,
\end{equation}
\begin{equation}
k_2=\dfrac{a r_2 (r_1-r_3)}{k_r} ,
\end{equation}
\begin{equation}
 k_3=\dfrac{a r_3 (r_2-r_1)}{k_r} ,
\end{equation}
with $k_r=(r_1-r_2)(r_1-r_3)(r_2-r_3)$. Therefore, bearing in mind (\ref{coord_u,v}) and (\ref{A(u,v)})  we have that
\begin{equation}
 A(u,v)=-\dfrac{1}{2r} \prod_{i=1}^{3}(r-r_i)^{1-2k_i}, 
\end{equation}
\begin{equation}
 u={\rm e}^{t/a} \prod_{i=1}^{3}(r-r_i)^{k_i}, 
\end{equation}
\begin{equation}
 v={\rm e}^{-t/a} \prod_{i=1}^{3}(r-r_i)^{k_i},
\end{equation}
and in particular we can satisfy the condition $u=0$ by choosing $r=r_i \; (i=1,2 ,3)$. Next, we have to show that conditions (\ref{g,v-A,v}) are satisfied. Bearing in mind Eqs. (\ref{BEST4.35}) and (\ref{BEST4.36}), we find that \cite{BEST}
\begin{equation}
g_{,v}= {\rm e}^{\frac{t}{a} } \; \frac{r (r-r_1)^{1-k_1} (r-r_2)^{1-k_2} (r-r_3)^{1-k_3}}{k_1 (r-r_2) (r-r_3)+k_2
   (r-r_1) (r-r_3)+k_3 (r-r_1) (r-r_2)},
\end{equation}
so that
\begin{equation}
\lim_{u \rightarrow 0} g_{,v}=0,  \; \; \; {\rm if\; and\; only\; if} \;  k_i <1. \label{g,v}
\end{equation}
Furthermore \cite{BEST},
\begin{equation}
A_{,v}={\rm e}^{\frac{t}{a}} \; \frac{  (r-r_1)^{1-3 k_1} (r-r_2)^{1-3 k_2} (r-r_3)^{1-3k_3}} {4 r^2 \left[ k_1
   (r-r_2) (r-r_3)+(r-r_1) (k_2 (r-r_3)+k_3 (r-r_2) \right]^2} \; \mathcal{F},
\end{equation}
where $\mathcal{F}=\mathcal{F}(r,r_i,k_i)$ is a function of $r$, the roots (\ref{root_1}) and (\ref{root_2and3}), and the constants $k_i$ (which in turns tend to a constant when $r \rightarrow r_i$), whose particular form is not of any special interest. We can then conclude that 
\begin{equation}
\lim_{u \rightarrow 0} A_{,v}=0,  \; \; \;{\rm if\; and\; only\; if}\;  k_i <1/3. \label{A,v}
\end{equation}
By virtue of Eqs. (\ref{g,v}) and (\ref{A,v}) we can say that conditions (\ref{g,v-A,v}) are satisfied provided that \cite{BEST}
\begin{equation}
k_i < 1/3, \; \; \; \; \; \; \; \; \; (i=1,2,3).
\end{equation}

In the case of Schwarzschild-de Sitter black hole, (\ref{f-SdS_eq}) depends on the ratio $a/m$ and thus possesses two branches of solutions for the constants $c$ and $k$. In the branch where the null hypersurface is described by a positive value of $r$ we have that \cite{BEST}
\begin{equation}
c=\dfrac{(r_1-r_3)(r_3-r_2)}{a^2}=2 \sin \left(\dfrac{\varphi}{3} \right) \left[ \sqrt{3} \cos \left(\dfrac{\varphi}{3} \right) - \sin \left(\dfrac{\varphi}{3} \right) \right],
\end{equation}
while the constant $k$ is always positive, with precise value which is not of particular interest. The inequality $a/m >\sqrt{27}$ is equivalent to the obvious condition $\cos \varphi <1$, moreover the null hypersurface $u=0$ where the massless particle is placed corresponds to $r=r_3$ (see Eq. (\ref{trigonometric_r3})). The condition $r_3 > 0$ implies that (for positive values of $m$ and $a$)
\begin{equation}
\varphi \in ( \pi/ 2, 3/2 \, \pi],
\end{equation}
so that
\begin{equation}
c \in (-2,0) \cup (0,1)  \; \; \; \; {\rm if} \; \; \; \varphi \in (\pi/2, \pi) \cup (\pi, \dfrac{3}{2} \pi ). 
\end{equation}
The boundary cases $c=-2$ ($\varphi = \dfrac{3}{2} \pi$) and $c=1$ ($\varphi = \pi/2$) correspond to de Sitter spacetime (cf. Eq. (\ref{Sfetsos4.4})) and Schwarzschild black hole, respectively, whereas the case $c=0$ ($\varphi= \pi$) is similar to the extremal Reissner-Nordstr\"{o}m charged black hole. As we pointed out before, the shift function $f(\theta)$ is given by Eq. (\ref{f_solution}). For $\dfrac{1}{4}\leq c <1$, an integral representation of the solution is given by \cite{Sfetsos95}
\begin{equation}
\begin{split}
f(\theta;c)&=\dfrac{-k}{\sqrt{2}} \int \limits_{0}^{+ \infty} {\rm d}s \; \cos (\sqrt{c-1/4} \;s) \dfrac{1}{\sqrt{\cosh s -\cos \theta}}  \\
&=\dfrac{- k \pi }{2 \cosh (\sqrt{c-1/4}\;\pi)} F (1/2-i\sqrt{c-1/4},1/2+i\sqrt{c-1/4};1;\cos^2 \dfrac{\theta}{2}),
\end{split}
\end{equation}
where $F(a,b;c;z)$ is the Gaussian or ordinary hypergeometric function (already encountered at the end of Sec. \ref{Alternative_route_5_Sec}). For $0 <c \leq \dfrac{1}{4}$ the solution is given by replacing $\sqrt{c-1/4}$ by $i\sqrt{1/4-c}$ and the trigonometric functions by hyperbolic ones, and vice versa. In both cases the shift function blows up at the point of the unit 2-sphere where the particle is located, i.e., at the northern pole $\theta=0$. Moreover, it is everywhere negative and for fixed $c$ it is a monotonically increasing function of $\theta \in [0, \pi]$, approaching a non-vanishing constant at $\theta=\pi$. For fixed $\theta$ it also monotonically increases as a function of $c \in (0,1)$. The refraction function (\ref{refraction_function}) is given by
\begin{equation}
R(\theta;c) = \left. \left( \dfrac{A}{g} \right)\right\vert_{u=0} \partial_{\theta} f(\theta;c).
\end{equation} 
It is a monotonically decreasing function of $\theta$ such that $\lim \limits_{\theta \to 0} R(\theta;c) = + \infty $ and $\lim \limits_{\theta \to \pi} R(\theta;c) = 0$. Thus, both the shift function and the refraction function blow up at $\theta=0$ and reach their minimum magnitudes at the southern pole $\theta=\pi$, where the refraction phenomenon disappears even if a particle trajectory is still discontinuous since $f(\pi;c) \neq 0$. For $-2<c<0$, the shift function is given by the integral representation 
\begin{equation}
f(\theta;c)=\dfrac{-k}{2c}-k \int \limits_{0}^{+ \infty} {\rm d}s \; \cosh (\sqrt{1/4-c} \;s) \left( \dfrac{1/\sqrt{2}}{\sqrt{\cosh s -\cos \theta}}-{\rm e}^{-s/2} \right). 
\end{equation}
The solution again blows up at $\theta=0$ and it monotonically increases as we move from $\theta=0$ to $\theta=\pi$. Moreover, it changes from negative to positive values at an angle $\theta_0$ that depends on the value assumed by the constant $c$ and reaches its minimum at $\theta=0$. On the other hand, the refraction function is a monotonically decreasing function of $\theta$ \cite{BEST}. 

As we can see, the conditions found in this section via the coordinate shift method are not in contrast with the results obtained through the boosting procedure of the previous sections. We have shown in fact that the ``boosted horizon'' gives rise to a sort of ``antigravity effect'' which, in light of the results displayed in this section, can be read as the refraction phenomenon described by the function (\ref{refraction_function}). It represents an important point the fact that these effects take place in a non-singular region of spacetime, i.e., the ``boosted horizon'' (for the boosting picture) and at the null hypersurface $u=0$ (in the coordinate shift method). Moreover, the presence of the singularity 3-sphere where the Kretschmann invariant is not defined could be probably related to the discontinuity of the $v$ component defined by Eq. (\ref{v_discontinuity}). The fact that in the ultrarelativistic regime the ``boosted horizon'' and the singularity 3-sphere positions' get blurred (as shown in Tab. \ref{boosted horizon position}) represents a clue in favour of this hypothesis. To make clearer the equivalence between the boost and the coordinate shift methods, one should be able to relate the coordinates $(u,v,x^1,x^2)$ exploited in this section with the boosted coordinates $(Y_1,Y_2,Y_3,Y_4)$ occurring in the four-dimensional metric components (\ref{g11})--(\ref{g34}). This can be done with the help of the results enlightened in Sec. \ref{coordinate-transformation}. In fact, as we said before, Eqs. (\ref{r^2_1}), (\ref{theta-Y_1})--(\ref{phi-Y_2}), represent the relations which link $(t,r,\theta,\phi)$ to $(Y_1,Y_2,Y_3,Y_4)$. By exploiting these outcomes, it is possible to express $(u,v,\theta,\phi)$ as functions of $(Y_1,Y_2,Y_3,Y_4)$. 

Therefore, by employing the equivalence between the two frameworks and the relations linking the two sets of coordinates, it is possible to relate all the results obtained through the coordinate shift method to those achieved through the boosting procedure. This means that the considerations made within the coordinate shift method about how handling the $\delta^2$ terms in the Riemann tensor (see Eqs. (\ref{BESTD1})--(\ref{BESTD3})) are valid also if we use the boost picture. Thus, the severe singularities of the Riemann tensor associated with the metric (\ref{ultrarelativistic boosted metric}) can be considered to be under control \cite{BEST}. 

\chapter*{Conclusions and open problems\markboth{Conclusions and open problems}{}}
\addcontentsline{toc}{chapter}{Conclusions and open problems}

The most astonishing result of the first part of this thesis is surely represented by the fact that, thanks to the modern Satellite/Lunar Laser Ranging technique, our effective field theory pattern produces testable (of the order of few millimetres) low-energy quantum gravity effects in a close and familiar system like the one made up of the Earth and the Moon. This represents a novel feature in the context of quantum gravity, since all other quantum frameworks of gravitation (e.g., string theory, loop quantum gravity, $f(R)$-theories, and so forth) are unable to produce detectable results, even in the large-scale structure of the universe. 

In chapter 1, we have first outlined the features of effective field theories and then we have applied such a framework to the quantization of general relativity, deriving in particular the Feynman rules for the gravitational field (Eqs. (\ref{PhdA1})--(\ref{2scalars-2gravitons vertex})). By considering only the non-analytical contribution resulting form the propagation of massless particles and their low-energy couplings in Feynman diagrams, we have achieved the expressions (\ref{1.2b})--(\ref{1.4b}) defining the quantum corrected Newtonian potential. The resulting quantum theory is not affected by ultraviolet divergences, provided that the full Lagrangian of gravity is endowed with a never ending set of higher-derivative terms compatible with the symmetries and with the general covariant criterion underlying general relativity. Anyway, the low-energy regime is ruled only by the Einstein-Hilbert sector of the theory. We have seen that within this domain three types of potentials are expected, depending on the definition adopted: one-particle reducible, scattering and bound-states potential. All calculations carried out in this manuscript have been performed by taking into account the aforementioned choices.  

In the second chapter we have applied the effective field theory point of view to the restricted three-body problem of celestial mechanics involving the Earth and the Moon as the primaries. Our contribution has been precisely a systematic investigation of the ultimate consequences of such a pattern. We have first derived the sufficient conditions (\ref{4.2a}), (\ref{4.8a}), (\ref{4.10a}), (\ref{4.12a}), (\ref{4.14a}), (\ref{4.16a}), and (\ref{4.18a}), which in an original way imply that some changes of qualitative features are unavoidable with respect to Newtonian theory, regardless of the choice of signs made in (\ref{1.2b})--(\ref{1.4b}),
although five out of seven sufficient conditions are fulfilled with the choice of scattering potential. Moreover, we have shown that the coordinates of non-collinear Lagrangian points are found by solving (both numerically and analytically by means of the pattern developed by Tschirnhaus, Bring, Jerrard, and Birkeland) the algebraic equations of fifth degree (\ref{5.4a}) and (\ref{5.11a}), and the resulting corrections on corresponding Newtonian values, obtained for the first time in the class of effective theories of gravity, are given in Tab. \ref{noncoll_corrections_tab}. On the other hand, the position of collinear libration points are governed by the ninth degree algebraic equations (\ref{3.14b}) and (\ref{3.25b}), quantum corrections being reported in Tab. \ref{coll_corrections_tab}. After a digression on the subject of variational equations, first-order stability for the five equilibrium points of the Earth-Moon system has been studied. We have proved therein that, provided the scattering potential is employed, $L_1$, $L_2$, and $L_3$ are still unstable, while $L_4$ and $L_5$ continue to be stable to first order also in the quantum corrected regime. Furthermore, displaced orbits have been evaluated in the quantum corrected domain, when the condition for the existence of such orbits is
affected by terms resulting from a solar sail model. We have found that, even when the quantum corrected potential (\ref{1.2b}) is adopted, displaced periodic orbits are of elliptical shape (see Figs. \ref{8b.pdf} and \ref{10b.pdf}) at all Lagrangian points, as in Newtonian theory.

Chapter 3 is dedicated to the full three-body problem and the restricted four-body problem in effective field theories of gravity. A central role is obviously fulfilled by the Earth and the Moon, like for the previous chapter. The aim of this chapter consists in making more realistic the model outlined in this thesis, because we hope that it could be part of some future space mission aimed at testing it in the future. As far as the full three-body problem is concerned, Eqs. (\ref{6.21f}) and (\ref{6.26f}) for the evaluation of solutions of the variational equations are our main original result. We have arrived at a broad framework that presents formidable technical difficulties, which is not the same as {\it solving} our equations. In fact, in the algorithm proposed the repeated application of a $2 \times 2$ matrix of first-order linear differential operators occurs. In addiction, we have seen how the extreme smallness of Planck length jointly with Poincar\'e theorem on periodic solutions lead to the existence of periodic orbits even at quantum level. The restricted four-body problem has been analysed in order to study the effects of the Sun in the Earth-Moon system both in the classical and in the quantum corrected context. In fact, we have demonstrated that also in the quantum regime the presence of the Sun makes the planetoid ultimately escape from the triangular libration points, which therefore can be considered as ``stable'' equilibrium points only during the length of observations. Unless we consider solar radiation pressure, from Eqs. (\ref{4aq})--(\ref{4cq}) we have obtained a plot describing the spacecraft motion about $L_4$ (Fig. \ref{3c.pdf}), which is slightly modified if compared with the corresponding classical one (Fig. \ref{1c.pdf}). If we instead take into account the solar radiation pressure, the differences between classical and quantum theory become more evident. The presence of solar pressure in the classical case, in fact, makes just the planetoid go away from the Lagrangian points $L_4$ more rapidly (see Fig. \ref{5c.pdf}), but in the quantum case, before escaping away from the libration point $L_4$, the planetoid is characterized by a less chaotic and irregular motion, as is clear from Fig. \ref{6c.pdf}. This feature remains true also if we consider several initial velocities for the planetoid (Figs. \ref{7c.pdf} and \ref{8c.pdf}). In particular, we have shown that the reduction of the envelope of the planetoid motion becomes more evident in the quantum case. After that, we have calculated the impulse needed for the stability of the spacecraft at $L_4$ both in the classical and in the quantum regime. These values, as witnessed by Eqs. (\ref{3.10c}) and (\ref{3.10cbis}), are a little bit different and therefore they suggest sending two satellites at $L_4$ and $L_5$, respectively, and checking which is the impulse truly needed for stability, in order to find out which is, between the classical and the quantum one, the best theory suited to describe these phenomena.

The fourth chapter deals with a theory involving quantum corrections to Einstein gravity, rather than to Newtonian model. First of all, we have performed a comparison between Newtonian gravity and general relativity, since of course the latter is the most successful theory describing gravitational interactions, at least in the Solar System. By evaluating the points where the gradient of the potential (\ref{4.9c}) vanishes, we have solved the algebraic equation describing the position of Lagrangian points. The distances of non-collinear Lagrangian points from the primaries are given in terms of the solutions of Eqs. (\ref{4.14c}) and (\ref{4.15c}) (or equivalently Eqs. (\ref{4.16c}) and (\ref{4.17c})) and are summarized in Tab. \ref{noncoll_GRcorrections_tab}. As far as collinear Lagrangian points are concerned, we have to focus on Eqs. (\ref{4.38c}) and (\ref{4.50c}) and on  Tab. \ref{coll_GRcorrections_tab}. After that, we have outlined the features of the new quantum theory whose underlying classical theory is represented by general relativity. By applying the map (\ref{5.8c}) and (\ref{5.10c}) to the Lagrangian (\ref{5.6c})  that general relativity provides for the restricted three-body problem, we have ended up with the quantum corrected Lagrangian (\ref{5.11c}) which, by means of 
Euler-Lagrange equations (\ref{5.12c}) together with the conditions $\ddot{\xi}=\ddot{\eta}=\ddot{\zeta}=\dot{\xi}=\dot{\eta}=\dot{\zeta}=\zeta=0$, has led us to the corrections of Tab. \ref{new_quantum_corrections_tab}. The possibility of mapping the effective potential of Newtonian gravity into an effective potential similar to the one of general relativity (cf. (\ref{5.13c}) and (\ref{5.14c})) adds evidence in favour of the choice of $\kappa_{1}$ and $\kappa_{2}$ appropriate for bound-states potential. If we bear in mind that such a pattern leads also to a correct evaluation of the perihelion shift of Mercury, we can conclude that bound-states potential could be the best choice in the context of quantum corrected phenomena occurring in celestial mechanics. 
 \begin{table}
\centering
\caption[General relativity corrections on the position of Newtonian Lagrangian points for the Sun-Earth system]{General relativity corrections on the position of Newtonian Lagrangian points for the Sun-Earth system obtained by solving Eqs. (\ref{4.16c}), (\ref{4.17c}), (\ref{4.38c}), and (\ref{4.50c}). The differences involved refer to the distances of the Sun from the planetoid}
\renewcommand\arraystretch{2.0}
\begin{tabular}{|c|c|}
\hline
\multicolumn{2}{|c|}{General relativity corrections on the Sun-Earth system}\\
\hline
\; \;  $L_i$ \; \;  &  Corrections \\
\cline{1-2}
$ L_1$ & $r_{GR}-r_{cl}=  4.8 \; {\rm m}$ \\ 
\hline
$ L_2$ & $r_{GR}-r_{cl}=  -5.0  \; {\rm m}$ \\ 
\hline
$ L_3$ & $r_{GR}-r_{cl}=  -0.3 \;  {\rm cm}$ \\ 
\hline
$ L_{4,5}$ & $r_{GR}-r_{cl}=-0.3  \;  {\rm cm}$ \\ 
\hline
\end{tabular} 
\label{Sun-Earth_GR-Newton}
\end{table} 
\begin{table}
\centering
\caption[General relativity corrections on the position of Newtonian Lagrangian points for the Sun-Jupiter system]{General relativity corrections on the position of Newtonian Lagrangian points for the Sun-Jupiter system obtained by solving Eqs. (\ref{4.16c}), (\ref{4.17c}), (\ref{4.38c}), and (\ref{4.50c}). The differences involved refer to the distances of the Sun from the planetoid}
\renewcommand\arraystretch{2.0}
\begin{tabular}{|c|c|}
\hline
\multicolumn{2}{|c|}{General relativity corrections on the Sun-Jupiter system}\\
\hline
\; \;  $L_i$ \; \;  &  Corrections \\
\cline{1-2}
$ L_1$ & $r_{GR}-r_{cl}=  30 \; {\rm m}$ \\ 
\hline
$ L_2$ & $r_{GR}-r_{cl}=   -38 \; {\rm m}$ \\ 
\hline
$ L_3$ & $r_{GR}-r_{cl}=   -1\;  {\rm m}$ \\ 
\hline
$ L_{4,5}$ & $r_{GR}-r_{cl}= -1 \;  {\rm m}$ \\ 
\hline
\end{tabular} 
\label{Sun-Jupiter_GR-Newton}
\end{table} 

Following the model developed in chapter 4, relativistic corrections to Newtonian Lagrangian points have been evaluated also for the Sun-Earth and the Sun-Jupiter systems (Tabs. \ref{Sun-Earth_GR-Newton} and \ref{Sun-Jupiter_GR-Newton}). In particular, the values reported in Tab. \ref{Sun-Jupiter_GR-Newton} are in {\it modulus} the same as the ones obtained by Yamada and Hasada. Nevertheless, also the quantum corrected model outlined in this thesis could be applied to such systems, but we feel that we first need to deal with the delicate point regarding all the possible perturbations occurring therein. In fact, as we said before theoretical predictions presented in this thesis are testable in light of modern advances in Lunar/Laser Ranging technique, but several perturbations, of gravitational and non-gravitational nature, may (slightly) modify such outcomes. Thus, if one wants to test the tiny corrections provided both by effective field theories of gravity and general relativity, it is necessary to perform a theoretical investigation of all conceivable perturbations of the Earth-Moon-satellite system. Therefore, it will become important to describe the solar system dynamics in general relativity. This will represent the aim of the NEWREFLECTIONS experiment. However, at this stage two fundamental questions could be asked:
\begin{enumerate}
\item Which is the best theory between effective field theories and general
relativity, if one wants to describe celestial mechanics phenomena?
\item Can we claim that the theoretical pattern developed in this thesis may represent a test bed between effective theories and general relativity (at least within the Solar System)? In other words, will it be possible, on experimental ground based on the corrections evaluated in this manuscript, to determine whether the effective field theory approach to general relativity is valid?
\end{enumerate}

The second part of this thesis deals with the high-energy regime of quantum gravity. Here we have numerically evaluated, for the first time in the literature, the Riemann curvature of a boosted spacetime in the ultrarelativistic limit $v \rightarrow 1$, starting from the Schwarzschild-de Sitter spacetime metric (\ref{S-dS metric}). We have exploited the fact that a de Sitter space can be seen as a four-dimensional hyperboloid embedded in a flat five-dimensional spacetime satisfying the constraint (\ref{Z hyperboloid constraint}). After that, we have introduced the boosting procedure through the relations (\ref{boost 1})--(\ref{boost 2}) which make it possible to obtain the boosted Schwarzschild-de Sitter metric (\ref{boosted metric}), whose ultrarelativistic limit is represented by (\ref{ultrarelativistic boosted metric}). By exploiting the hyperboloid constraint (\ref{Z hyperboloid constraint}) we have then expressed (\ref{boosted metric}) in the manifestly four-dimensional form (\ref{g11})--(\ref{g34}). By virtue of (\ref{Z hyperboloid constraint}), the metric components (\ref{g11})--(\ref{g34}) are defined only if $\sigma >0$, $\sigma$ being defined by relation (\ref{Y0}). This fact is strictly related to inequality (\ref{singularity kretschmann}). In fact, $\left \{ \partial / \partial Y_\mu \right \}$ being a coordinate basis, we have numerically computed the Riemann curvature tensor by using the usual relations of general relativity, and to better understand the features of curvature we have studied both the Kretschmann invariant and the geodesic equation (\ref{geodesic eq}). We have indeed found that the Kretschmann invariant 
is not defined unless (\ref{singularity kretschmann}) holds and thus we have just concluded that there exists a 3-sphere of radius $a$ 
where the spacetime  possesses a ``scalar curvature singularity''. In fact, from the numerical analysis of the geodesic equation, we have 
found that if the particle lies initially on the positive half-line $Y_1>0$, $Y_2=0$ of Fig. \ref{contourplot} it  always reaches the 
3-sphere (Fig. \ref{geod3}). After that, its geodesic is no longer defined and hence we can conclude that the 3-sphere of equation 
$(Y_{1})^2+(Y_{2})^2+(Y_{3})^2+(Y_{4})^2 = a^2$ defines a ``scalar curvature singularity'' for the ``boosted geometry'' under investigation. When 
the particle lies initially on the negative half-line $Y_1<0$, $Y_2=0$, its geodesic is not defined even before it manages to reach the 
3-sphere (see Fig. \ref{geod4}): there exists another ``scalar curvature singularity'' whose position depends solely on the boost velocity $v$. We have also discovered that ``boosted geometries'' are characterized by the presence of a sort of elastic wall surrounding the singularity 3-sphere 
whose coordinates depend only on the boost velocity (see Tab. \ref{boosted horizon position}). All geodesics indeed, despite being complete, 
are always pushed away from there, as Figs. \ref{geod1} and \ref{geod2} show. We propose to call this barrier ``boosted horizon'' because, 
as in the case of Schwarzschild geometry, it is not a singularity of spacetime, but it is related to a sort of ``antigravity effect'' 
that should rule ``boosted geometries''. As we know, boosted geometries are characterized by the fact that both the spacetime metric and the Riemann curvature tensor assume a distributional nature in the ultrarelativistic regime. This regime is still governed by ``antigravity effects'', with the peculiarity that ``boosted horizon'' and singularity 3-sphere tend to overlap. 

Eventually, we have analysed the geometry of the metric (\ref{S-dS metric}) through the coordinate shift method. We have proved that this new picture is equivalent to the boosting procedure and we have demonstrated how it solves the issues related to the presence of $\delta^2$ terms in the Riemann tensor. In particular, the ``antigravity effects'' emerged at the ``boosted horizon'' have been ascribed to the refraction phenomenon described by the function (\ref{refraction_function}). Moreover, the fact that in the ultrarelativistic regime the ``boosted horizon'' position's tends to that of the singularity 3-sphere could be related to the fact that, in the coordinate shift method picture, when the particle crosses the null surface located at $u=0$ it suffers a discontinuity in its $v$-component (Eq. (\ref{v_discontinuity})) while the $x^i$ components are refracted according to (\ref{refraction_function}). This is a really delicate point as, unlike the singularity 3-sphere, both the null hypersurface $u=0$ (coordinate shift method) and our ``boosted horizon'' (boosted picture) do not define a spacetime singularity, and we feel that some more efforts should be produced in this direction. The equivalence between the two methods, which can be formally made manifest for our ``boosted geometry'' by Eqs. (\ref{r^2_1}), (\ref{theta-Y_1})--(\ref{phi-Y_2}), has enabled us to conclude that the Riemann tensor associated with metric (\ref{ultrarelativistic boosted metric}) is defined and has a behavior under control.

We suppose that ``antigravity effects'' may result from the cosmological constant $\Lambda = 3/a^2 >0$ occurring in the Schwarzschild-de Sitter metric 
(\ref{S-dS metric}) (a positive $\Lambda$ represents a repulsive interaction), while ``scalar curvature singularities'' 
might be related to the presence of a more exotic object, i.e., a firewall, which can be a possible solution to an apparent 
inconsistency in black hole complementarity.\footnote{See Refs. \cite{AMPS,Braunstein2013,Braunstein2015,STU,SHW}.}

\begin{appendices}
\addtocontents{toc}{\protect\setcounter{tocdepth}{0}}

\chapter{Summary of Feynman rules for quantum gravity} \label{Appendix_Feynman_rules} 

In this appendix we list the Feynman rules employed in this thesis. We make use of the de Donder gauge and the flat Minkowski background. In the case of gravity-scalar interacting vertices (\ref{2scalar-1graviton_vertex}) and (\ref{2scalars-2gravitons vertex}) we always use the convention on four-momentum conservation $p^{\prime}-p=q$.

\subsection*{Scalar propagator}

The massive scalar propagator is represented by the Feynman propagator and it reads as
\begin{equation}
\Delta_{F}(q) = \dfrac{{\rm i}}{q^2 -m^2 + {\rm i} \epsilon}.
\label{PhdA1}
\end{equation} 
\begin{figure} [htbp] 
\centering
\includegraphics[scale=0.7]{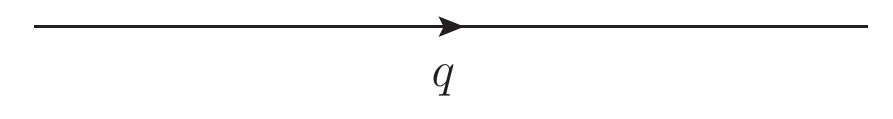}
\caption[The scalar propagator]{The propagator for massive scalar particles.} 
\end{figure}

\subsection*{Graviton propagator}

The graviton propagator is given by
\begin{equation}
D_{\mu \nu \rho \sigma} (k)= {\rm i} \dfrac{\mathcal{P}_{\mu \nu \rho \sigma} }{k^2 +  {\rm i} \epsilon}, 
\end{equation}
with
\begin{equation}
\mathcal{P}_{\mu \nu \rho \sigma}=\dfrac{1}{2} (\eta_{\mu \rho} \eta_{\nu \sigma} + \eta_{\mu \sigma} \eta_{\nu \rho}-\eta_{\mu \nu} \eta_{\rho \sigma}).
\end{equation}
\begin{figure} [htbp] 
\centering
\includegraphics[scale=0.7]{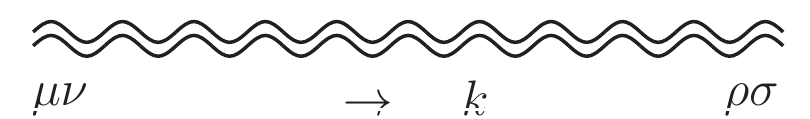}
\caption[The graviton propagator]{The graviton propagator.} 
\end{figure}

\subsection*{Three-graviton vertex}

The three-graviton vertex has the form \cite{D94b}
\begin{equation}
\begin{split}
\tau^{\mu \nu}_{\; \; \; \alpha \beta \gamma \delta}(k,q) &= - \dfrac{{\rm i} \chi}{2} \biggl( \mathcal{P}_{\alpha \beta \gamma \delta} \Bigl [k^\mu k^\nu + (k-q)^\mu (k-q)^\nu + q^\mu q^\nu - \dfrac{3}{2} \eta^{\mu \nu} q^2 \Bigr]\\
& + 2 q_\lambda q_\sigma \Bigl[ I^{\lambda \sigma}_{\; \; \; \alpha \beta} I^{\mu \nu}_{\; \; \; \gamma \delta}+ I^{\lambda \sigma}_{\; \; \; \gamma \delta} I^{\mu \nu}_{\; \; \; \alpha \beta}-I^{\lambda \mu}_{\; \; \; \alpha \beta} I^{\sigma \nu}_{\; \; \; \gamma \delta}-I^{ \sigma \nu}_{\; \; \; \alpha \beta} I^{\lambda \mu }_{\; \; \; \gamma \delta} \Bigr] \\
& + \Bigl[ q_\lambda q^\mu \Bigl( \eta_{\alpha \beta} I^{ \lambda \nu}_{\; \; \; \gamma \delta} + \eta_{\gamma \delta} I^{ \lambda \nu}_{\; \; \; \alpha \beta} \Bigr) +  q_\lambda q^\nu \Bigl( \eta_{\alpha \beta} I^{ \lambda \mu}_{\; \; \; \gamma \delta} + \eta_{\gamma \delta} I^{ \lambda \mu}_{\; \; \; \alpha \beta} \Bigr) \\
& -q^2  \Bigl( \eta_{\alpha \beta} I^{  \mu \nu}_{\; \; \; \gamma \delta} + \eta_{\gamma \delta} I^{  \mu \nu}_{\; \; \; \alpha \beta} \Bigr) - \eta^{\mu \nu} q^\lambda q^\sigma \Bigl (\eta_{\alpha \beta} I_{\gamma \delta \lambda \sigma} + \eta_{\gamma \delta} I_{\alpha \beta \lambda \sigma} \Bigr) \Bigr] \\
& + \Bigl[ 2 q^\lambda \Bigl( I^{\sigma \nu}_{\; \; \; \alpha \beta} I_{\gamma \delta \lambda \sigma} (k-q)^\mu + I^{  \sigma \mu}_{\; \; \; \alpha \beta} I_{\gamma \delta \lambda \sigma} (k-q)^\nu   \\
& - I^{\sigma \nu}_{\; \; \; \gamma \delta} I_{\alpha \beta \lambda \sigma} k^\mu -  I^{\sigma \mu}_{\; \; \; \gamma \delta} I_{\alpha \beta \lambda \sigma} k^\nu \Bigr) + q^2 \Bigl( I^{\sigma \mu}_{\; \; \; \alpha \beta} I_{ \gamma \delta \sigma}^{\; \; \; \; \; \; \nu} +  I^{\sigma \mu}_{\; \; \; \gamma \delta} I_{ \alpha \beta \sigma}^{\; \; \; \; \; \; \nu} \Bigr) \\
& + \eta^{\mu \nu} q^\lambda q_\sigma \Bigl( I_{\alpha \beta \lambda \rho} I^{\rho \sigma}_{\; \; \; \gamma \delta}+ I_{\gamma \delta \lambda \rho} I^{\rho \sigma}_{\; \; \; \alpha \beta} \Bigr) \Bigr] \\
& + \Bigl \{ \Bigl( k^2 + (k-q)^2 \Bigr) \Bigl( I^{ \sigma \mu}_{\; \; \; \alpha \beta} I_{ \gamma \delta \sigma}^{\; \; \; \; \; \; \nu} + I^{ \sigma \nu}_{\; \; \; \alpha \beta} I_{ \gamma \delta \sigma}^{\; \; \; \; \; \; \mu} - \dfrac{1}{2} \eta^{\mu \nu} \mathcal{P}_{\alpha \beta \gamma \delta} \Bigr) \\
& - \Bigl(k^2 \; \eta_{\gamma \delta}  I^{ \mu \nu}_{\; \; \; \alpha \beta} + (k-q)^2 \eta_{\alpha \beta}  I^{ \mu \nu}_{\; \; \; \gamma \delta} \Bigr) \Bigr \} \biggr),
\end{split} \label{three-graviton_vertex}
\end{equation}
where
\begin{equation}
I_{\alpha \beta \gamma \delta}= \dfrac{1}{2} (\eta_{\alpha \gamma} \eta_{\beta \delta}+\eta_{\alpha \delta} \eta_{\beta \gamma} ). \label{I_tens}
\end{equation}
Note that the graviton with Lorentz indices $\mu \nu$ represents a background graviton, which therefore has not to be used within any loop.
\begin{figure} [htbp] 
\centering
\includegraphics[scale=0.7]{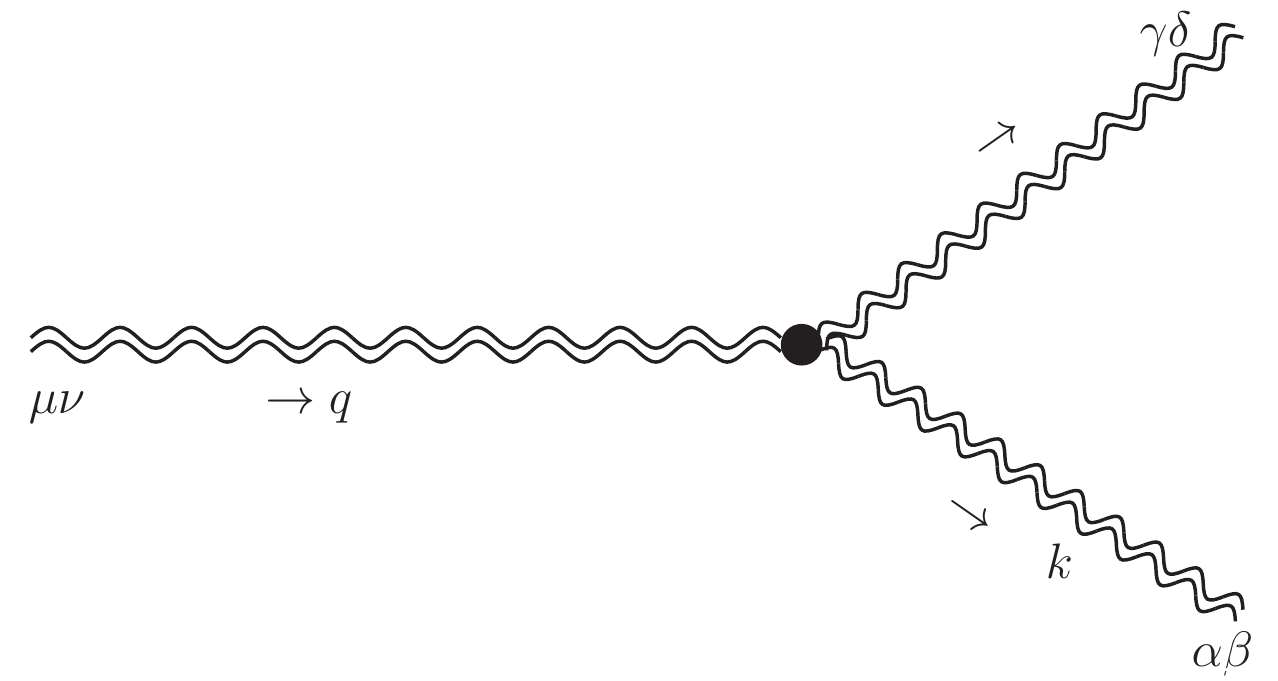}
\caption[The three-graviton vertex]{The three-graviton vertex.} 
\end{figure}

\subsection*{Two scalar-one graviton vertex}

The expression for the two scalar-one graviton vertex is \cite{D94b}
\begin{equation}
\tau_{\mu \nu} (p,p^\prime,m)=\dfrac{-{\rm i} \chi}{2} \left[ p_\mu p^{\prime}_{\nu} + p_\nu p^{\prime}_{\mu} - \eta_{\mu \nu} \left(p \cdot p^\prime - m^2 \right) \right]. \label{2scalar-1graviton_vertex} 
\end{equation}
\begin{figure} [htbp] 
\centering
\includegraphics[scale=0.7]{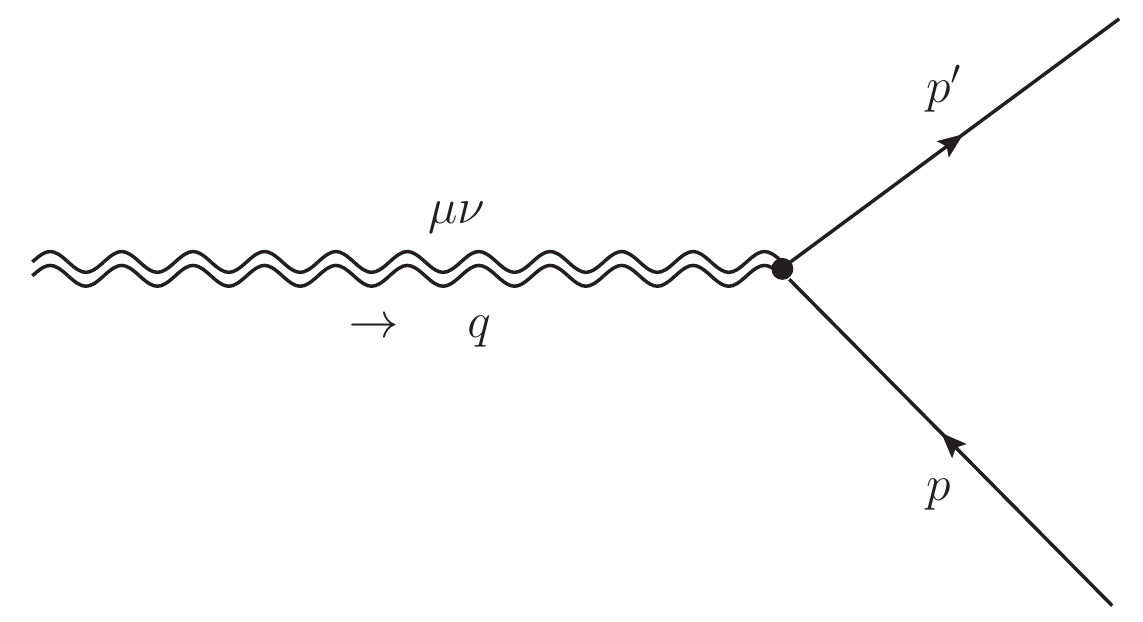}
\caption[The two scalar-one graviton vertex]{The two scalar-one graviton vertex.}
\end{figure}

\subsection*{Two scalar-two graviton vertex}

This vertex can be expressed as \cite{D94b}
\begin{equation}
\begin{split}
\tau_{\eta \lambda\rho \sigma }(p, p^\prime, m)& = \dfrac{{\rm i} \chi^2}{2} \biggl [ I_{\eta \lambda \alpha \delta} I^{\delta}_{\; \beta \rho \sigma} \left(p^\alpha p^{\prime \beta}+ p^\beta p^{\prime \alpha} \right) \\ 
& -\dfrac{1}{2} \left(\eta_{\eta  \lambda} I_{\rho \sigma \alpha \beta} + \eta_{ \rho \sigma} I_{\eta \lambda \alpha \beta} \right)  p^\beta p^{\prime \alpha} \\
& -\dfrac{1}{2} \left( I_{\eta \lambda\rho \sigma} - \dfrac{1}{2} \eta_{\eta \lambda} \eta_{\rho \sigma}  \right) \left( p \cdot p^\prime - m^2 \right) \biggr].
\end{split} \label{2scalars-2gravitons vertex}
\end{equation}
\begin{figure} [htbp] 
\centering
\includegraphics[scale=0.7]{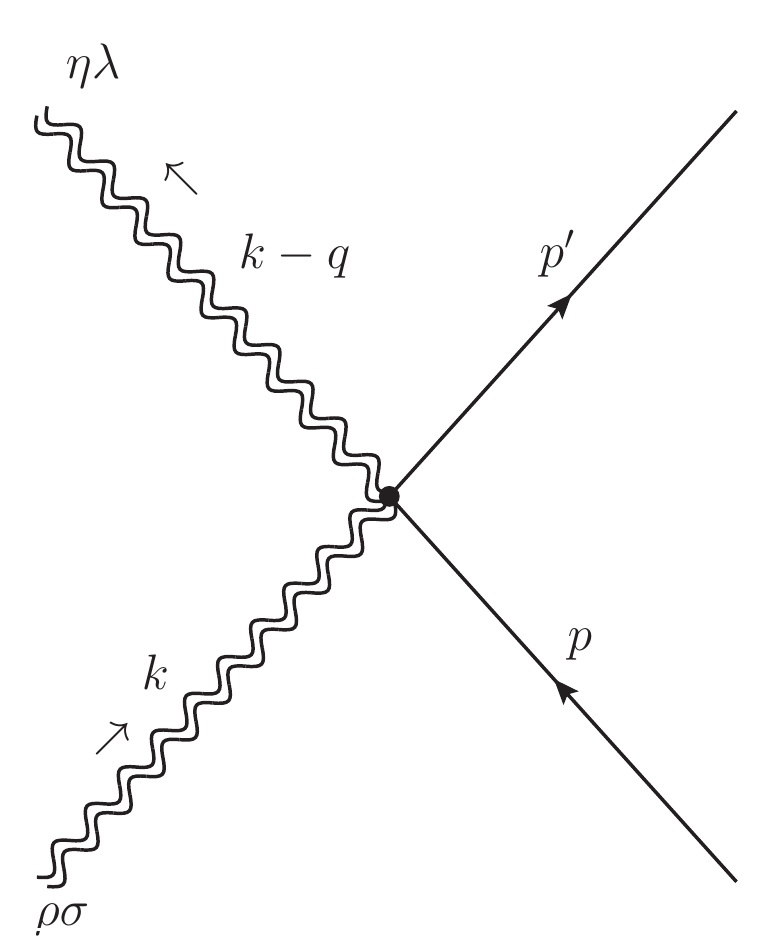}
\caption[The two scalar-two graviton vertex]{The two scalar-two graviton vertex.}
\end{figure}

\chapter{Useful integrals} \label{Appendix_Useful_integrals}

All the integrals needed to calculate the Feynman diagrams presented in this thesis are displayed in this appendix \cite{D94b,D03}. 

\section*{Fourier transforms}

Fourier transformations have been applied to Feynman diagram calculations in order to recover the potential $V_Q(r)$ of Eq. (\ref{1.2b}). The following Fourier integrals are useful:
\begin{equation}
\int \dfrac{{\rm d}^3 q}{(2 \pi)^3}\; \E^{\I {\bf q} \cdot {\bf r}} = \delta^{(3)} ({\bf r}),
\end{equation}
\begin{equation}
\int \dfrac{{\rm d}^3 q}{(2 \pi)^3}\; \E^{\I {\bf q} \cdot {\bf r}} \dfrac{1}{\vert  {\bf q}\vert^2} =\dfrac{1}{4 \pi r}, 
\label{Fourier1}
\end{equation}
\begin{equation}
\int \dfrac{{\rm d}^3 q}{(2 \pi)^3}\; \E^{\I {\bf q} \cdot {\bf r}} \dfrac{1}{\vert  {\bf q}\vert} =\dfrac{1}{2 \pi^2 r^2}, 
\end{equation}
\begin{equation}
\int \dfrac{{\rm d}^3 q}{(2 \pi)^3}\; \E^{\I {\bf q} \cdot {\bf r}} \log \left(  \vert {\bf q} \vert^2 \right) =-\dfrac{1}{2 \pi r^3}.
\end{equation}

\section*{Integrals needed in the calculation of Feynman diagrams }

The evaluation of the various diagrams presented in this thesis can be carried on by employing the integrals listed here \cite{D94,D94b,D03}. We start with the conventions used. In the derivation of the one-particle reducible potential, Figs. \ref{D94b.pdf} and \ref{D94c.pdf}  are characterized by two external momenta which are indicated with $k$ (initial or ingoing) and $k^\prime$ (final or outgoing). Their on-shell condition reads as 
\begin{equation}
k^2=k^{\prime^2}=m^2,
\end{equation}
and they are such that the momentum conservation can be written in the form
\begin{equation}
k-k^\prime=q, 
\end{equation}
$q$ being the graviton transferred momentum.  Therefore, we have
\begin{equation}
\begin{split}
(k-k^\prime)^2 & =2m^2-2 k^\prime \cdot k = q^2 \Rightarrow \\
&  k^\prime \cdot k = m^2 - \dfrac{q^2}{2}.
\end{split}
\end{equation}
Moreover, 
\begin{equation}
k \cdot q = k \cdot \left( k - k^\prime \right) = \dfrac{q^2}{2},
\end{equation}
\begin{equation}
k^\prime \cdot q = k^\prime \cdot \left( k - k^\prime \right) = \dfrac{-q^2}{2},
\end{equation}
and hence
\begin{equation}
k \cdot q = - k^\prime \cdot q = \dfrac{q^2}{2}.
\end{equation}
In the cases of the scattering and bound-state potential, the external momenta are $k_1$, $k_3$ (ingoing) and $k_2$, $k_4$ (outgoing) and the mass-shell condition is
\begin{equation}
\begin{split}
& (k_{1})^2=(k_{2})^2=(m_{1})^2, \\
& (k_{3})^2=(k_{4})^2=(m_{2})^2,
\end{split}
\end{equation}
whereas the momentum conservation is given by
\begin{equation}
k_1-k_2=k_4-k_3=q. \label{momentum_conservation}
\end{equation}
From these relations it follows that
\begin{equation}
\begin{split}
\left(k_1-k_2\right)^2 & =(k_{1})^2+(k_{2})^2 - 2 k_1 \cdot k_2 = q^2 \Rightarrow \\
& k_1 \cdot k_2 = (m_1)^2-\dfrac{q^2}{2}, \label{momentum_scalar_product1}
\end{split}
\end{equation}
and similarly
\begin{equation}
k_3 \cdot k_4 = (m_2)^2-\dfrac{q^2}{2},
\end{equation}
whereas by exploiting the knowledge of the Mandelstam variables $s=(k_1+k_3)^2=(k_4+k_2)^2$ and $u=(k_3-k_2)^2=(k_1-k_4)^2$ we also have that
\begin{equation}
k_1 \cdot k_3 = k_2 \cdot k_4,
\end{equation}
\begin{equation}
k_1 \cdot k_4 = k_2 \cdot k_3. \label{momentum_scalar_product2}
\end{equation}
Furthermore, the following identities turn out to be very useful:
\begin{equation}
k_1 \cdot q = k_1 \cdot \left( k_1 - k_2 \right)= \dfrac{q^2}{2},
\end{equation}
\begin{equation}
k_2 \cdot q = k_2 \cdot \left( k_1 - k_2 \right)=- \dfrac{q^2}{2},
\end{equation}
\begin{equation}
k_3 \cdot q = k_3 \cdot \left( k_4 - k_3 \right)=- \dfrac{q^2}{2},
\end{equation}
\begin{equation}
k_4 \cdot q = k_4 \cdot \left( k_4 - k_3 \right)= \dfrac{q^2}{2},
\end{equation}
which means that we have
\begin{equation}
k_1 \cdot q = - k_2 \cdot q = - k_3 \cdot q = k_4 \cdot q = \dfrac{q^2}{2}.
\end{equation}

The integrals used in the calculations of the Feynman diagrams are
\begin{equation}
\mathfrak{H} = \int \dfrac{{\rm d}^4 l}{(2 \pi)^4} \dfrac{1}{l^2 \left(q \pm l \right)^2} = \dfrac{\I}{32 \pi^2} \left( -2 \tilde{L} \right) + \dots,
\end{equation}
\begin{equation}
\mathfrak{H}_{\mu} = \int \dfrac{{\rm d}^4 l}{(2 \pi)^4} \dfrac{l_{\mu}}{l^2 \left(q \pm l \right)^2} = \dfrac{\I}{32 \pi^2} q_{\mu} \left( \pm \tilde{L} \right) + \dots,
\end{equation}
\begin{equation}
\mathfrak{H}_{\mu \nu} = \int \dfrac{{\rm d}^4 l}{(2 \pi)^4} \dfrac{l_{\mu} \l_{\nu}}{l^2 \left(q \pm l \right)^2} = \dfrac{\I}{32 \pi^2} \left[ q_{\mu} q_{\nu}\left( -\dfrac{2}{3} \tilde{L} \right)  - q^2 \eta_{\mu \nu}\left( -\dfrac{1}{6} \tilde{L} \right)   \right]+ \dots,
\end{equation}
together with
\begin{equation}
\mathfrak{I} = \int \dfrac{{\rm d}^4 l}{(2 \pi)^4} \dfrac{1}{l^2 \left(q \pm l \right)^2 \left[ \left( k+l \right)^2 - m^2 \right]} = \dfrac{\I}{32 \pi^2 m^2} \left( -\tilde{L} - \tilde{S} \right) + \dots, 
\end{equation}
\begin{equation}
\begin{split}
\mathfrak{I}_{\mu} & = \int \dfrac{{\rm d}^4 l}{(2 \pi)^4} \dfrac{l_{\mu}}{l^2 \left(q \pm l \right)^2 \left[ \left( k+l \right)^2 - m^2 \right]} \\
& = \dfrac{\I}{32 \pi^2 m^2} \left\{ -k_\mu \left[ \left( 1+\dfrac{1}{2}\dfrac{q^2}{m^2} \right)\tilde{L} +\dfrac{1}{4}\dfrac{q^2}{m^2} \tilde{S} \right] \pm q_\mu \left( \tilde{L} + \dfrac{1}{2} \tilde{S} \right) \right\} + \dots, 
\end{split}
\end{equation}
\begin{equation}
\begin{split}
\mathfrak{I}_{\mu \nu} & = \int \dfrac{{\rm d}^4 l}{(2 \pi)^4} \dfrac{l_{\mu} \l_\nu }{l^2 \left(q \pm l \right)^2 \left[ \left( k+l \right)^2 - m^2 \right]} \\
&  = \dfrac{\I}{32 \pi^2 m^2} \Biggl \{ -q_\mu q_\nu \left(\tilde{L} +\dfrac{3}{8} \tilde{S} \right) + \dfrac{q^2 \eta_{\mu \nu}}{2} \left(\dfrac{1}{2}\tilde{L} + \dfrac{1}{4} \tilde{S} \right) \\
& - k_\mu k_\nu \left(\dfrac{q^2}{2 m^2} \right) \left( \tilde{L} + \dfrac{1}{4} \tilde{S} \right) \pm \left( q_\mu k_\nu + q_\nu k_\mu \right) \left[ \dfrac{1}{2} \left( 1+ \dfrac{q^2}{m^2} \right)\tilde{L} + \dfrac{3}{16} \dfrac{q^2}{m^2} \tilde{S} \right] \Biggr \} + \dots, 
\end{split}
\end{equation}
\begin{equation}
\begin{split}
\mathfrak{I}_{\mu \nu \alpha} & = \int \dfrac{{\rm d}^4 l}{(2 \pi)^4} \dfrac{l_{\mu} \l_\nu l_\alpha }{l^2 \left(q \pm l \right)^2 \left[ \left( k+l \right)^2 - m^2 \right]} \\
& = \dfrac{\I}{32 \pi^2 m^2} \Biggl \{ \pm q_\mu q_\nu q_\alpha \left(\tilde{L} + \dfrac{5}{16}\tilde{S} \right) \mp \dfrac{q^2}{2}  \left( \eta_{\mu \nu} q_{\alpha} + \eta_{\mu \alpha} q_{\nu} + \eta_{\nu \alpha} q_{\mu} \right)  \left( \dfrac{1}{3} \tilde{L} + \dfrac{1}{8} \tilde{S} \right) \\
& + \dfrac{q^2}{m^2} \left[ -\dfrac{1}{6} \, k_\mu  k_\nu k_\alpha \pm \left( q_\mu k_\nu k_\alpha + q_\nu k_\mu k_\alpha + q_\alpha k_\mu k_\nu \right) \left( \dfrac{1}{3} \tilde{L} + \dfrac{1}{16} \tilde{S}  \right)\right] \\
& - \left( q_\mu q_\nu k_\alpha + q_\mu q_\alpha k_\nu + q_\nu q_\alpha k_\mu \right) \left[ \left( \dfrac{1}{3} + \dfrac{1}{2} \dfrac{q^2}{m^2} \right) \tilde{L} + \dfrac{5}{32} \dfrac{q^2}{m^2} \tilde{S} \right] \\
& + \dfrac{1}{12} q^2 \left( \eta_{\mu \nu} k_\alpha + \eta_{\mu \alpha} k_\nu + \eta_{\nu \alpha} k_\mu \right) \tilde{L} \Biggr \}+ \dots, 
\end{split}
\end{equation}
where we have set $\tilde{L}=\log \left(-q^2 \right)$, $\tilde{S} = \left( \pi^2 m \right) / \sqrt{-q^2}$ and $k$ indicates the (on-shell) external momentum satisfying $k \cdot q = q^2/2$. This fact means that for the diagrams involved in the one-particle reducible potential $k$ coincides with the only ingoing momentum, whereas for the scattering and bound-state potentials $k$ can be either $k_1$ or $k_4$. Ellipses denote that both higher-order non-analytic contributions and analytic terms have been dropped. In some cases the integrals are used with $k$ replaced by some (on-shell) $-\tilde{k}$, provided that $\tilde{k} \cdot q = -q^2/2$. This point is crucial because, bearing in mind the above relations, it means that the results can be obtained through the replacement
\begin{equation}
k \rightarrow -\tilde{k} = 
 \Biggl\{ \begin{array}{l}
- k^{\prime}, \; \; \; \; \; \; \; \; \; \; \; \; \,  {\rm (one-particle \;  reducible)}, \\
-k_2,-k_3, \; \; \; \; {\rm (scattering \; or \; bound-states)}. \end{array} 
\end{equation} 
For the box and crossed-box diagrams (Figs. \ref{box.pdf} and \ref{crossedbox.pdf}), we exploit the integrals\footnote{The exact expression of $\mathfrak{G}$ can be found in Ref. \cite{ErratumD03}.}
\begin{equation}
\begin{split}
\mathfrak{G} & = \int \dfrac{{\rm d}^4 l}{(2 \pi)^4} \dfrac{1}{l^2 \left( q+l \right)^2 \left[ \left(k_1+l \right)^2 -\left(m_1 \right)^2 \right]  \left[ \left(k_3-l \right)^2 -\left(m_2 \right)^2 \right]} \\
& =\dfrac{\I}{16 \pi^2 m_1 m_2 q^2} \left[ \left(1- \dfrac{\mathfrak{w}}{3 m_1 m_2 } \right)\tilde{L} - \dfrac{\I \pi m_1 m_2}{\left(m_1+m_2 \right) \mathfrak{p}}\right] + \dots,
\end{split}
\end{equation}
\begin{equation}
\begin{split}
\mathfrak{G}^\prime & = \int \dfrac{{\rm d}^4 l}{(2 \pi)^4} \dfrac{1}{l^2 \left( q+l \right)^2 \left[ \left(k_1+l \right)^2 -\left(m_1 \right)^2 \right]  \left[ \left(k_4+l \right)^2 -\left(m_2 \right)^2 \right]} \\
& =\dfrac{\I}{16 \pi^2 m_1 m_2 q^2} \left[ \left(-1+ \dfrac{\mathfrak{W}}{3 m_1 m_2 } \right)\tilde{L} \right] + \dots,
\end{split}
\end{equation}
where again we have written down only the lowest-order non-analytical terms and where we have defined
\begin{equation}
\mathfrak{w}= k_1 \cdot k_3 - m_1 m_2, 
\end{equation}
\begin{equation}
\mathfrak{W}= k_1 \cdot k_4 - m_1 m_2,
\end{equation}
and
\begin{equation}
\mathfrak{p}= \sqrt{\dfrac{\left[s-\left(m_1+m_2\right)^2 \right]\left[s-\left(m_1-m_2\right)^2 \right]}{4s}},
\end{equation}
being the mass center momentum. 

For the above integrals the non-analytic terms satisfy various constraints that can be verified on-shell, such as
\begin{equation}
q^\mu \mathfrak{H}_\mu = \mp \dfrac{q^2}{2} \mathfrak{H} + \dots, 
\end{equation}
\begin{equation}
q^\mu \mathfrak{H}_{\mu \nu} =\mp \dfrac{q^2}{2} \mathfrak{H}_\nu + \dots, 
\end{equation}
 \begin{equation}
\eta^{\mu \nu} \mathfrak{H}_{\mu \nu} = 0 + \dots,
 \end{equation}
\begin{equation}
q^\mu \mathfrak{I}_\mu = \mp \dfrac{q^2}{2} \mathfrak{I} + \dots, 
\end{equation}
\begin{equation}
q^\mu \mathfrak{I}_{\mu \nu} = \mp \dfrac{q^2}{2} \mathfrak{I}_\nu + \dots, 
\end{equation}
\begin{equation}
q^\mu \mathfrak{I}_{\mu \nu \alpha} = \mp \dfrac{q^2}{2} \mathfrak{I}_{\nu \alpha} + \dots, 
\end{equation}
 \begin{equation}
\eta^{\mu \nu} \mathfrak{I}_{\mu \nu} =\eta^{\mu \nu} \mathfrak{I}_{\mu \nu \alpha}  = 0 + \dots,
 \end{equation}
\begin{equation}
k^\mu \mathfrak{I}_\mu = \pm \dfrac{1}{2} \mathfrak{H} + \dots, 
\end{equation}
\begin{equation}
k^\mu \mathfrak{I}_{\mu \nu} = \pm \dfrac{1}{2} \mathfrak{H}_{\nu} + \dots, 
\end{equation}
\begin{equation}
k^\mu \mathfrak{I}_{\mu \nu \alpha } = \pm \dfrac{1}{2} \mathfrak{H}_{\nu \alpha} + \dots.
\end{equation}

Since we care only about non-analytic contributions, other on-shell identities can be used in order to simplify the integrals. For example, the on-shell relations
\begin{equation}
l \cdot q = \dfrac{\left(q+l \right)^2 -q^2 -l^2}{2},
\end{equation}
\begin{equation}
l \cdot k_1 = \dfrac{\left(k_1+l \right)^2 -(m_1)^2 -l^2}{2},
\end{equation}
\begin{equation}
l \cdot k_3 =- \dfrac{\left(k_3-l \right)^2 -(m_2)^2 -l^2}{2},
\end{equation}
turn out to be very practical. In fact, by way of illustration, consider
\begin{equation}
\begin{split}
& \int \dfrac{{\rm d}^4 l}{(2 \pi)^4}   \dfrac{l \cdot q}{l^2 \left( q+l \right)^2 \left[ \left(k_1+l \right)^2 -\left(m_1 \right)^2 \right]  \left[ \left(k_3-l \right)^2 -\left(m_2 \right)^2 \right]} \\
& =\dfrac{1}{2} \int \dfrac{{\rm d}^4 l}{(2 \pi)^4}   \dfrac{\left(q+l \right)^2 -q^2 -l^2}{l^2 \left( q+l \right)^2 \left[ \left(k_1+l \right)^2 -\left(m_1 \right)^2 \right]  \left[ \left(k_3-l \right)^2 -\left(m_2 \right)^2 \right]},
\end{split}
\end{equation} 
which, by recalling that the integrals with the factors $l^2$ and $(q\pm l)^2$ at the numerator yields no non-analytical terms, can be reduced as 
\begin{equation}
\begin{split}
& \int \dfrac{{\rm d}^4 l}{(2 \pi)^4}   \dfrac{l \cdot q}{l^2 \left( q+l \right)^2 \left[ \left(k_1+l \right)^2 -\left(m_1 \right)^2 \right]  \left[ \left(k_3-l \right)^2 -\left(m_2 \right)^2 \right]} \\
& \rightarrow - \dfrac{q^2}{2} \int \dfrac{{\rm d}^4 l}{(2 \pi)^4}   \dfrac{1}{l^2 \left( q+l \right)^2 \left[ \left(k_1+l \right)^2 -\left(m_1 \right)^2 \right]  \left[ \left(k_3-l \right)^2 -\left(m_2 \right)^2 \right]} =- \dfrac{q^2}{2} \mathfrak{G}.
\end{split}
\end{equation} 
Along the above lines, a significant situation can be found in all those cases in which it is possible to perform a contraction between a loop momentum and an external momentum, removing in this way one of the propagators and leaving a much simpler loop integral, i.e., 
\begin{equation}
\begin{split}
& \int \dfrac{{\rm d}^4 l}{(2 \pi)^4}   \dfrac{l \cdot k_1}{l^2 \left( q+l \right)^2 \left[ \left(k_1+l \right)^2 -\left(m_1 \right)^2 \right]  \left[ \left(k_3-l \right)^2 -\left(m_2 \right)^2 \right]} \\
& =\dfrac{1}{2} \int \dfrac{{\rm d}^4 l}{(2 \pi)^4}   \dfrac{ \left(k_1+l \right)^2 -(m_1)^2 -l^2  }{l^2 \left( q+l \right)^2 \left[ \left(k_1+l \right)^2 -\left(m_1 \right)^2 \right]  \left[ \left(k_3-l \right)^2 -\left(m_2 \right)^2 \right]} \\
& \dfrac{1}{2}  \int \dfrac{{\rm d}^4 l}{(2 \pi)^4} \left\{  \dfrac{ 1  }{l^2 \left( q+l \right)^2  \left[ \left(k_3-l \right)^2 -\left(m_2 \right)^2 \right]} -  \dfrac{ 1  }{\left( q+l \right)^2 \left[ \left(k_1+l \right)^2 -\left(m_1 \right)^2 \right]  \left[ \left(k_3-l \right)^2 -\left(m_2 \right)^2 \right]} \right\} \\
& = \dfrac{1}{2} \mathfrak{I} -\dfrac{1}{2} \int \dfrac{{\rm d}^4 l^\prime}{(2 \pi)^4} \dfrac{ 1  }{l^{\prime^2} \left[ \left(k_2+l^\prime \right)^2 -\left(m_1 \right)^2 \right]  \left[ \left(k_4-l^\prime \right)^2 -\left(m_2 \right)^2 \right]} \\
& \rightarrow \dfrac{1}{2} \mathfrak{I},
\end{split}
\end{equation} 
or equivalently
\begin{equation}
\begin{split}
& \int \dfrac{{\rm d}^4 l}{(2 \pi)^4}   \dfrac{l \cdot k_3}{l^2 \left( q+l \right)^2 \left[ \left(k_1+l \right)^2 -\left(m_1 \right)^2 \right]  \left[ \left(k_3-l \right)^2 -\left(m_2 \right)^2 \right]} \\
& =-\dfrac{1}{2} \int \dfrac{{\rm d}^4 l}{(2 \pi)^4}   \dfrac{ \left(k_3-l \right)^2 -(m_2)^2 -l^2  }{l^2 \left( q+l \right)^2 \left[ \left(k_1+l \right)^2 -\left(m_1 \right)^2 \right]  \left[ \left(k_3-l \right)^2 -\left(m_2 \right)^2 \right]} \\
& -\dfrac{1}{2}  \int \dfrac{{\rm d}^4 l}{(2 \pi)^4} \left\{  \dfrac{ 1  }{l^2 \left( q+l \right)^2  \left[ \left(k_1+l \right)^2 -\left(m_1 \right)^2 \right]} -  \dfrac{ 1  }{\left( q+l \right)^2 \left[ \left(k_1+l \right)^2 -\left(m_1 \right)^2 \right]  \left[ \left(k_3-l \right)^2 -\left(m_2 \right)^2 \right]} \right\} \\
& = -\dfrac{1}{2} \mathfrak{I} +\dfrac{1}{2} \int \dfrac{{\rm d}^4 l^\prime}{(2 \pi)^4} \dfrac{ 1  }{l^{\prime^2} \left[ \left(k_2+l^\prime \right)^2 -\left(m_1 \right)^2 \right]  \left[ \left(k_4-l^\prime \right)^2 -\left(m_2 \right)^2 \right]} \\
& \rightarrow -\dfrac{1}{2} \mathfrak{I},
\end{split}
\end{equation} 
where we have introduced the shift $l^\prime=l+q$ and we have exploited the momentum conservation (\ref{momentum_conservation}) so that we can write $k_1+l = k_2 + l^\prime$ and $k_3-l=k_4-l^\prime$. These reductions can be used in the box and crossed-box diagrams in order to simplify the calculations.  

\chapter{Asymptotic expansions}\label{Asymptotic exp_App}

In this appendix we will briefly describe the formal aspects of asymptotic expansions by following both Dieudonn\'e and Poincar\'e approaches \cite{Pseries,Dieudonne}.

First of all, we recall the difference between the O-notation and the o-notation.

If $f(s)$ and $g(s)$ are functions of a  complex variable $s$, defined on the arbitrary subset $S$ of $\mathbb{C}$, we write
\begin{equation}
f(s)={\rm O}(g(s)),  \; \; (s\, \in S) \Leftrightarrow \exists \, c>0\, : \; \; \; \vert f(s)\vert \leq c \vert g(s)\vert \; \; \; \forall \, s \in S.
\end{equation}
When we deal with real functions $f(x)$ and $g(x)$, such an estimation involves their limiting behavior when the arguments tend towards infinity or a particular value. In such cases we have
\begin{equation}
f(x)={\rm O}(g(x)),  \; \; (x \to \infty) \Leftrightarrow \exists \, M>0, x_0\, \in \mathbb{R} \, : \; \; \; \vert f(x)\vert \leq M \vert g(x)\vert \; \; \; \forall \, x \geq x_0,
\end{equation}
\begin{equation}
f(x)={\rm O}(g(x)),  \; \; (x \to l) \Leftrightarrow \exists \, M, \delta >0\, : \; \; \; \vert f(x)\vert \leq M \vert g(x)\vert \; \; \; {\rm for}\;  \vert x-l\vert < \delta,
\end{equation}
respectively. In many contexts, the assumption that we are interested in the growth rate as the variable $x$ goes to a particular value or to infinity is left unstated, and one writes more simply $f(x)={\rm O}(g(x))$. 

Moreover, we set $f(x)={\rm o}(g(x))$ and we say that $f(x)$ is of smaller order than $g(x)$, if $g(x)$ is a non-vanishing function and in addiction it grows faster than $f(x)$. Formally, we write
\begin{equation}
f(x)={\rm o}(g(x)) \; \; (x \to x_0) \Leftrightarrow g(x) \neq 0 \; \; {\rm for} \; x \to x_0 \;\; \;  {\rm and} \;\;  \lim_{x \to x_0} \dfrac{f(x)}{g(x)}=0,
\end{equation}
or 
\begin{equation}
f(x)={\rm o}(g(x)) \; \; (x \to \infty) \Leftrightarrow  g(x) \neq 0 \; \; {\rm for \; large \; values \; of} \; x \;\; \;  {\rm and} \;\;  \lim_{x \to \infty} \dfrac{f(x)}{g(x)}=0.
\end{equation}
In particular, if the last condition holds, it is equivalent to having an O-estimate $f(x) = {\rm O}(g(x))$ with a constant $c$ that can be chosen arbitrarily small (but positive) and a range $x \geq x_0(c)$ depending on $c$. Thus, an o-estimate is stronger than the corresponding O-estimate.

A closely related notation is that of asymptotic equivalence
\begin{equation}
f(x) \sim g(x) \; \; (x \to \infty)  \Leftrightarrow g(x) \neq 0 \; \; {\rm for} \; x \to x_0 \;\; \;  {\rm and} \;\;  \lim_{x \to x_0} \dfrac{f(x)}{g(x)}=1.
\end{equation}
In such a case we say that $f(x)$ is asymptotically equivalent to $g(x)$ as $x \to \infty$. 

At this stage, we are ready to give the definition of asymptotic expansion in the Dieudonn\'e sense \cite{Dieudonne}. In general, one starts by considering the set $\mathcal{E}$ of functions of the form
\begin{equation}
g: x \rightarrow g(x) \equiv x^{\mu} \left( \log x \right)^\nu {\rm e}^{P(x)},
\end{equation}
$\mu$, $\nu$ being real non-vanishing constants and 
\begin{equation}
P(x)= \sum_{j=1}^{k} c_j x^{\gamma_j},
\end{equation}
where $c_j$ are real constants of arbitrary sign, while
\begin{equation}
\gamma_1 > \gamma_2 > \dots > \gamma_k >0.
\end{equation}
By definition, given a function $f$, its asymptotic expansion with $k$ terms with respect to the set $\mathcal{E}$ is meant to be the sum
\begin{equation}
\Sigma_k \equiv \sum_{j=1}^{k} b_j g_j,
\end{equation}
where $b_j$ are non-vanishing constants and $g_j$ are functions belonging to the set $\mathcal{E}$ such that
\begin{equation}
g_{j+1}={\rm o}(g_j), \; \; \; \;\;\;\; \forall\, j : 1\leq j \leq k-1.
\end{equation}
One then writes
\begin{equation}
f= \sum_{j=1}^{k} b_j g_j + {\rm o}(g_k).
\end{equation}
The difference $f - \Sigma_k$ is called the remainder of the asymptotic expansion. In the physics-oriented literature, it is commonly adopted a kind of notation for which the last formula is written with the equality symbol replaced by the $\sim$ symbol. Thence, as you can see, by definition an asymptotic expansion has only finitely many terms (unlike a series, which has infinitely many terms) and hence talking about convergence (or lack of) is meaningless.

In Poincar\'e approach, the concept of asymptotic expansion assumes a completely different meaning \cite{Pseries}. In fact, Poincar\'e was interested in divergent series both in astronomy and in the context of differential equations. For this purpose, his definition involves from the very beginning a divergent series. In Ref. \cite{Pseries} in fact Poincar\'e begins by discussing the peculiar properties of Stirling series:
\begin{equation}
\log \Gamma(x+1)= \dfrac{1}{2} \log (2 \pi) + \left(x+\dfrac{1}{2}\right)\log(x)-x+\dfrac{B_1}{1 \cdot 2} \dfrac{1}{x}-\dfrac{B_2}{3 \cdot 4} \dfrac{1}{x^2}+\dfrac{B_3}{5 \cdot 6} \dfrac{1}{x^3}- \dots,
\end{equation}
$\Gamma(x)$ being the Euler gamma function. Poincar\'e pointed out that this series is always diverging, but one can use it at large $x$. In fact, what happens is that, after decreasing very rapidly, the terms become unboundedly large. Nevertheless, if we take the smallest term, the corresponding error in the evaluation of $\log \Gamma(x+1)$ is very small.

Thus, bearing in mind the above considerations, consider the divergent series
\begin{equation}
\sum_{n=0}^{+ \infty}A_n x^{-n}= A_0 + \dfrac{A_1}{x}+\dfrac{A_2}{x^2}+ \dots + \dfrac{A_n}{x^n} +\dfrac{A_{n+1}}{x^{n+1}} + \dots,
\label{Acta1}
\end{equation}  
which is such that the sum of its first $n+1$ terms is $S_n$. The series (\ref{Acta1}) represents asymptotically the function 
\begin{equation}
f(x)= S_n + {\rm O}(x^{-(n+2)})
\end{equation}
if 
\begin{equation}
\lim_{x \to \infty} x^n \vert f(x)-S_n \vert =0,
\end{equation} 
and one writes
\begin{equation}
f(x) \sim \sum_{n=0}^{+ \infty} A_n x^{-n}, \; \; \; \; \; ({\rm as}\; x \to \infty ).
\end{equation}
In fact, if $x$ is sufficiently large, then 
\begin{equation}
x^n \vert f(x)-S_n \vert < \epsilon,
\end{equation} 
$\epsilon$ being a very small constant, and hence the error
\begin{equation}
\vert f(x)-S_n \vert = \dfrac{\epsilon}{x^n},
\label{Acta2}
\end{equation}
committed on the function $f(x)$ while considering only the first $n+1$ terms of the series will be very small. 

Therefore, it is possible to realize how profoundly different the two definitions described above are. 

\chapter{Notes on the system $\dot{{\bold{x}}}=\bold{A}\bold{x}$} \label{Fundamental_App}

In this appendix we will describe some details concerning the resolution of the system of differential equations $\dot{{\bold{x}}}=\bold{A}\bold{x}$ by introducing the concepts of fundamental matrix and Jordan normal form \cite{Eastham,Bronson}.

\section*{Fundamental matrix} 

Consider the autonomous system of $n$ first-order linear homogeneous ordinary differential equations
\begin{equation}
\dot{{\bold{x}}}(t)=\bold{A}\bold{x}(t), \label{Fund1}
\end{equation}
$\bold{A}$ being a constant $n \times n$ matrix. The unknown of (\ref{Fund1}) is the column vector
\begin{equation}
\bold{x}(t)=
\left( 
\begin{matrix}
 x_1(t)\cr
 x_2(t)\cr
\vdots \cr
 x_n(t)\cr
\end{matrix}
\right),
\end{equation}
 so that the general solution reads as
 \begin{equation}
\bold{x}(t)= {\rm e}^{ t \bold{A}} \bold{C}= {\rm e}^{t \bold{A}} 
\left( 
\begin{matrix}
 C_1\cr
 C_2\cr
\vdots \cr
 C_n\cr
\end{matrix}
\right), \label{Fund3}
\end{equation}
 $C_1, C_2, \dots, C_n$ being arbitrary constants. Furthermore, the solution of the initial value problem
\begin{equation}
\begin{dcases}
& \dot{{\bold{x}}}(t)=\bold{A}\bold{x}(t), \\
& \bold{x}(t_0)=\bold{x}_0, \\
\end{dcases} \\ [2em]
\label{Fund5}
\end{equation}
 is represented by
 \begin{equation}
 \bold{x}(t)= {\rm e}^{(t-t_0)\bold{A}}\,\bold{x}_0,
 \label{Fund6}
 \end{equation}
 the exponential function of a square matrix being defined by
 \begin{equation}
 {\rm e}^{ t \bold{A}}= \sum_{k=0}^{\infty} \dfrac{t^k}{k!}\bold{A}^k=\mathbb{1}+ t \bold{A}+\left(\dfrac{t^2}{2!}\right)\bold{A}^2 + \dots.
 \end{equation}
 When the matrix occurring in Eq. (\ref{Fund1}) is a time dependent matrix $\bold{A}(t)$, the solution of
 \begin{equation}
\dot{{\bold{x}}}(t)=\bold{A}(t)\bold{x}(t), \label{Fund2}
\end{equation}
 can be expressed only in an approximate way by
 \begin{equation}
 \bold{x}(t) = \bold{R}(t)\bold{C}, \label{Fund10}
 \end{equation}
where $\bold{C}$ is the matrix of constant coefficients of Eq. (\ref{Fund3}), while $\bold{R}(t)$ is expressed by the infinite series
\begin{equation}
\bold{R}(t)= \sum_{k=0}^{\infty} \bold{D}_k(t), \label{Fund4}
\end{equation}
 where
 \begin{equation}
\bold{D}_0=\mathbb{1},
\end{equation}
 and 
 \begin{equation}
\begin{dcases}
& \dot{{\bold{D}}}_{i+1}=\bold{A}\bold{D}_i, \\
& \bold{D}_{i+1}(0)=\bold{0}. \\
\end{dcases} \\ [2em]
\end{equation}
Obviously, when the matrix $\bold{A}$ is constant, Eq. (\ref{Fund4}) becomes
\begin{equation}
\bold{R}= \mathbb{1} + t \bold{A} + \left(\dfrac{t^2}{2!}\right) \bold{A}^2 +  \left(\dfrac{t^3}{3!}\right) \bold{A}^3 + \dots = {\rm e}^{t \bold{A}},
\end{equation}
and we recover the solution (\ref{Fund3}).  

One of the approaches towards the solution of the system of differential equations (\ref{Fund1}) or (\ref{Fund2}) consists in finding its so-called fundamental matrix. Consider first the time-independent case (\ref{Fund1}). The general solution of such a system has the structure
\begin{equation}
\bold{x}(t)= C_1 \bold{x}_1(t) + C_2 \bold{x}_2(t) + \dots C_n \bold{x}_n(t), \; \; \; \; \; \; C_i \in \mathbb{R},  
\end{equation}
$\bold{x}_1(t), \bold{x}_2(t),\dots,\bold{x}_n(t)$ being $n$ linearly independent solutions. Then, the solution of (\ref{Fund1}) can be written as
\begin{equation}
\bold{x}(t)= \left[
\begin{matrix}
\bold{x}_1(t) & \bold{x}_2(t) &\dots & \bold{x}_n(t) \cr
\end{matrix}
\right] \bold{C},
\label{Fund20}
\end{equation}
whereas for the Cauchy problem (\ref{Fund5}) $\bold{C}$ is ruled by the matrix equation
\begin{equation}
\begin{split}
& \bold{x}(0)= \left[
\begin{matrix}
\bold{x}_1(0) & \bold{x}_2(0) &\dots & \bold{x}_n(0) \cr
\end{matrix}
\right] \bold{C}= \bold{x}_0 \\
 & \Rightarrow  \bold{C} = \left[ \begin{matrix}
\bold{x}_1(0) & \bold{x}_2(0) &\dots & \bold{x}_n(0) \cr
\end{matrix}
\right]^{-1} \bold{x}_0,
\label{Fund9}
\end{split}
\end{equation}
where we have set $t_0=0$ without loss of generality. Therefore, the solution of (\ref{Fund5}) is given by
\begin{equation}
\bold{x}(t)= \left[
\begin{matrix}
\bold{x}_1(t) & \bold{x}_2(t) &\dots & \bold{x}_n(t) \cr
\end{matrix}
\right]
 \left[ \begin{matrix}
\bold{x}_1(0) & \bold{x}_2(0) &\dots & \bold{x}_n(0) \cr
\end{matrix}
\right]^{-1} \bold{x}_0.
\label{Fund7}
\end{equation}
A comparison between (\ref{Fund6}) and (\ref{Fund7}) leads to (recall that we have set $t_0=0$)
\begin{equation}
{\rm e}^{t \bold{A}}=
\left[
\begin{matrix}
\bold{x}_1(t) & \bold{x}_2(t) &\dots & \bold{x}_n(t) \cr
\end{matrix}
\right]
\left[ \begin{matrix}
\bold{x}_1(0) & \bold{x}_2(0) &\dots & \bold{x}_n(0) \cr
\end{matrix}
\right]^{-1} .
\label{Fund8}
\end{equation}
We can then appreciate how in this method of evaluating the solution of (\ref{Fund1}) (or (\ref{Fund5})) the matrix $\left[
\begin{matrix}
\bold{x}_1(t) & \bold{x}_2(t) &\dots & \bold{x}_n(t) \cr
\end{matrix}
\right]$ plays an essential role. Therefore, we can give the following definition:
\newtheorem*{Fundamental1}{Definition}
\begin{Fundamental1}
If $\bold{x}_1(t), \bold{x}_2(t), \dots, \bold{x}_n(t)$ represent $n$ linearly independent solutions of the $n$-dimensional linear homogeneous system (\ref{Fund1}), then we call 
\begin{equation}
\bold{F}(t) \equiv \left[
\begin{matrix}
\bold{x}_1(t) & \bold{x}_2(t) &\dots & \bold{x}_n(t) \cr
\end{matrix}
\right],
\end{equation}
fundamental matrix solution. Moreover, the matrix $\bold{F}(t)$ is called principal fundamental matrix solution if there exists a $t_0$ such that $\bold{F}(t_0)$ is the identity. 
\end{Fundamental1}
In other words, the fundamental matrix $\bold{F}(t)$ is the $n \times n$ matrix-valued function whose columns are $n$ linearly independent solutions of (\ref{Fund1}). Its elements are such that the entry $x_{ij}$ indicates the $i$-th component of the $j$-th linearly independent vector. The exponential map (\ref{Fund8}) then is given by
\begin{equation}
{\rm e}^{t \bold{A}}= \bold{F}(t) \bold{F}(0)^{-1}. \label{Fund11}
\end{equation}
Moreover, bearing in mind Eq. (\ref{Fund20}), the fundamental matrix solution of (\ref{Fund1}) can be written as
\begin{equation}
\bold{x}(t) = \bold{F}(t) \bold{C}, \label{Fund18}
\end{equation}
while for the initial value problem (\ref{Fund5}) we have (see Eq. (\ref{Fund9}))
\begin{equation}
\bold{x}(t) = \bold{F}(t) \bold{F}(0)^{-1}\bold{x}_0.
\end{equation}
Note that $\bold{F}(0)^{-1}$ exists because the determinant $\vert \bold{F}(0) \vert $ represents the value at $t=0$ of the Wronskian of $\bold{x}_1(t), \bold{x}_2(t), \dots, \bold{x}_n(t)$, which is non-vanishing because the $n$ solutions are linearly independent. In fact, the fundamental matrix is always invertible for any value of $t$. Furthermore, it clearly satisfies
\begin{equation}
\dot{\bold{F}}(t) = \bold{A} \bold{F}(t), \label{Fund25}
\end{equation}
and it also true that, since any solution of (\ref{Fund1}) can be expressed as a linear combination, with constant coefficients, of $n$ linearly independent solutions, any other fundamental matrix can be written as $\bold{F}(t) \bold{L}$, where $\bold{L}$ is a non-singular $n\times n$ constant matrix. In particular, the matrix $\bold{R}(t)$ defined by Eq. (\ref{Fund4}) (the same as Eq. (\ref{Pars2321})) is a principal fundamental matrix solution, because, besides having columns representing linearly independent solutions, it is also characterized by the fact that there exists an instant of time $t_0$ such that $\bold{R}(t_0)=\mathbb{1}$ (in our case $t_0=0$). In fact, also for the time-dependent case (\ref{Fund2}), we can define the fundamental matrix exactly in the same way as for the time-independent one, except that for the former the map (\ref{Fund11}) is no longer valid. Thence, the fundamental matrix solution of (\ref{Fund2}) is represented by (\ref{Fund10}). 
\begin{description}
\item[{\it Example.}] Consider the system
\begin{equation}
\dot{\bold{x}}(t)= \left(
\begin{matrix}
1 & 2 \cr
2 & 1 \cr
\end{matrix}
\right) \bold{x}(t).
\label{Fund12}
\end{equation}
Then, we set
\begin{equation}
\bold{A}=\left( \begin{matrix}
1 & 2 \cr
2 & 1 \cr
\end{matrix}
\right).
\end{equation}
The characteristic equation will be given by
\begin{equation}
\det \left( \bold{A}-\mathbb{1} \lambda \right)= (\lambda-3)(\lambda+1)=0,
\end{equation}
so that the eigenvalues of $\bold{A}$ are
\begin{equation}
\begin{split}
& \lambda_1=3,\\
 & \lambda_2=-1,
\end{split}
\end{equation}
while the corresponding eigenvectors are found by solving through Gaussian elimination the system
\begin{equation}
\left( \bold{A}-\mathbb{1} \lambda \right) \bold{v}=0, \; \; \; \; \; \; \; \; \; \; \; \;  ( \bold{v} \neq 0),
\end{equation}
yielding
\begin{equation}
\bold{v}_1= \left(
\begin{matrix}
1 \cr
1
\end{matrix}
\right),
\end{equation}
\begin{equation}
\bold{v}_2= \left(
\begin{matrix}
1 \cr
-1
\end{matrix}
\right),
\end{equation}
for $\lambda_1$ and $\lambda_2$, respectively. Therefore, the solution vectors of (\ref{Fund12}) become
\begin{equation}
\begin{split}
& \bold{u}_1= {\rm e}^{3t} \left(
\begin{matrix}
1 \cr
1
\end{matrix}
\right), \\
& \bold{u}_2= {\rm e}^{-t} \left(
\begin{matrix}
1 \cr
-1
\end{matrix}
\right),
\end{split}
\end{equation}
whereas the fundamental matrix assumes the form
\begin{equation}
\bold{F}(t)= \left(
\begin{matrix}
\bold{u}_1 & \bold{u}_2 \cr
\end{matrix}
\right)= \left( 
\begin{matrix}
{\rm e}^{3t} & {\rm e}^{-t} \cr
{\rm e}^{3t} & -{\rm e}^{-t} \cr
\end{matrix}
\right).
\end{equation}
Then, the general solution of (\ref{Fund12}) then is
\begin{equation}
\bold{x}(t)= C_1 \bold{u}_1+C_2 \bold{u}_2 =\bold{F}(t) \left(
\begin{matrix}
C_1 \cr
C_2 \cr
\end{matrix}
\right),
\end{equation}
$C_1$, $C_2$ being arbitrary constants. If we need to find $C_1$, $C_2$ satisfying some initial condition $\bold{x}(t_0)=\bold{x}_0$, we have to solve the matrix equation
\begin{equation}
\bold{x}(t_0)=\bold{F}(t_0) \left(
\begin{matrix}
C_1 \cr
C_2 \cr
\end{matrix}
\right)= \left(
\begin{matrix}
x_{0_1} \cr
x_{0_2} \cr
\end{matrix}
\right),
\end{equation}
which gives
\begin{equation}
 \left(
\begin{matrix}
C_1 \cr
C_2 \cr
\end{matrix} \right) = \bold{F}(t_0)^{-1} \left(
\begin{matrix}
x_{0_1} \cr
x_{0_2} \cr
\end{matrix}
\right),
\end{equation}
and hence the solution becomes
\begin{equation}
\bold{x}(t)=\bold{F}(t)\bold{F}(t_0)^{-1} \left(
\begin{matrix}
x_{0_1} \cr
x_{0_2} \cr
\end{matrix}
\right).
\end{equation}
\end{description}

\section*{Jordan normal form}

If $\bold{A}$ is a diagonalizable matrix, i.e., there exists a non-singular matrix $\bold{L}$ such that $\bold{A}=\bold{L}\bold{D}\bold{L}^{-1}$, $\bold{D}$ being the diagonal matrix having the eigenvalues $\lambda_1,\lambda_2,\dots,\lambda_n$ of $\bold{A}$ as its entries, then it is easy to compute its exponential map, since
\begin{equation}
{\rm e}^{t \bold{A}}={\rm e}^{t \bold{L}\bold{D}\bold{L}^{-1}}= \bold{L} \, {\rm e}^{t \bold{D}} \, \bold{L}^{-1}= \bold{L} \left( \begin{matrix}
{\rm e}^{t \lambda_1} & & & \cr
& {\rm e}^{t \lambda_2} & & \cr
& & \ddots & \cr
& & &   {\rm e}^{t \lambda_n} \cr
\end{matrix} \right)\bold{L}^{-1}.
\end{equation}
We know from theorems of linear algebra that  there exist cases in which a matrix turns out to be not diagonalizable. In fact only real symmetric matrices are always (orthogonally) similar to a diagonal matrix having real eigenvalues and eigenvectors (corresponding to distinct eigenvalues) which are orthogonal. In all those circumstances in which the matrix $\bold{A}$ is not diagonalizable, we need to employ another approach to find the exponential map ${\rm e}^{t \bold{A}}$ if we want to solve the system of constant coefficients linear differential equations (\ref{Fund1}). This method involves the use of the so-called Jordan normal form (or Jordan canonical form\footnote{It is named after the French mathematician Camille Jordan (1838-1922).}) of $\bold{A}$, i.e., a special shape that can be assumed by a matrix under similarity transformations. The idea underlying such a pattern is represented by an important theorem of algebra, i.e., the Schur triangulation theorem, stating that if $\bold{A}$ is a real symmetric matrix and its characteristic polynomial ${\rm P}_{\bold{A}}(\lambda)$ factors completely, then $\bold{A}$ is orthogonally similar to an upper triangular matrix (called Schur form). Furthermore, the eigenvalues of an upper triangular matrix correspond to the entries on its diagonal. However, one may wonders if this is the best result that can be achieved. The answer is ``no'', because the ``closest-to-diagonal'' matrix that can be obtained is just the Jordan normal form, which is a particular upper triangular matrix having each non-zero off-diagonal entry equals to one and collocated immediately above the main diagonal (called the super-diagonal), and with identical diagonal entries to the left and below them.

To introduce the Jordan normal form of a generic (real) matrix, say $\bold{N}$, we first have to define the ``bricks'' forming such a matrix. They are called Jordan blocks. A Jordan block $\bold{J}_{h,p}$ is a $p \times p$ upper triangular matrix of the form
\begin{equation}
\bold{J}_{h,p}=\left( \begin{matrix}
h & 1 & 0 & \dots & 0 & 0 \cr
0 & h & 1 & \dots & 0 & 0 \cr
\vdots & \vdots & \vdots & \ddots & \vdots &\vdots  \cr
 0 & 0 & 0 &\dots  & 0 & h \cr 
\end{matrix}
\right).
\label{Jordan19}
\end{equation}
We say that $h$ is the eigenvalue associated with $\bold{J}_{h,p}$. Thus, a Jordan block is composed of vanishing elements everywhere except for the diagonal, which is filled with a fixed eigenvalue $h$ \footnote{We are supposing that $h \in \mathbb{R}$.}, and for the super-diagonal, which is composed of ones. Moreover, any Jordan block $\bold{J}_{h,p}$ is characterized  by a characteristic polynomial ${\rm P}_{\bold{J}_{h,p}}(\lambda)$ and a minimal polynomial ${\rm M}_{\bold{J}_{h,p}}(\lambda)$ given by
\begin{equation}
{\rm P}_{\bold{J}_{h,p}}(\lambda)=(-1)^{p}(\lambda-h)^{p},
\end{equation}
\begin{equation}
{\rm M}_{\bold{J}_{h,p}}(\lambda)=(\lambda-h)^{p},
\end{equation}
respectively. In other words, for a Jordan block the characteristic and minimal polynomials differ (possibly) only for the sign. Recall that a matrix $\bold{N}$ always satisfies its characteristic polynomial, i.e., ${\rm P}_{\bold{N}}(\bold{N})=0$ (Cayley-Hamilton theorem). Moreover, we define the minimal polynomial ${\rm M}_{\bold{N}}(\lambda)$ of $\bold{N}$ as the unique monic polynomial (i.e., an invariant polynomial having the leading coefficient equals to one) of least degree satisfying ${\rm M}_{\bold{N}}({\bold{N}})=0$. Such a polynomial is a factor of ${\rm P}_{\bold{N}}(\lambda)$ and contains each of the linear factors of ${\rm P}_{\bold{N}}(\lambda)$. Furthermore, similar matrices have the same minimal and characteristic polynomials and hence, in particular, have the same eigenvalues. 

Now let us introduce the concept of generalized eigenvectors. A column vector $\bold{X}_q$ represents a generalized eigenvectors of rank $q$ of a matrix $\bold{N}$ corresponding to the eigenvalue $\lambda$ if
\begin{equation}
\left( \bold{N} - \lambda \mathbb{1}  \right)^q \, \bold{X}_q = \bold{0},
\end{equation}
but meanwhile 
\begin{equation}
\left( \bold{N} - \lambda \mathbb{1} \right)^{q-1} \, \bold{X}_q\neq \bold{0}.
\end{equation}
Consider the Jordan block $\bold{J}_{h,p}$. Since $(\bold{J}_{h,p}-h \mathbb{1})$ turns out to be a matrix where the only non-vanishing elements are on the super-diagonal and are equal to one, $(\bold{J}_{h,p}-h \mathbb{1})$ has rank $p-1$ and hence $\bold{J}_{h,p}$ has only $p-(p-1)=1$ linearly independent eigenvector. 

Suppose that $\bold{N}$ is a $p \times p$ matrix which is similar to the Jordan block $\bold{J}_{h,p}$. Then, since similar matrices have the same characteristic polynomials, we have that ${\rm P}_{\bold{N}}(\lambda)={\rm P}_{\bold{J}_{h,p}}(\lambda)=(h-\lambda)^p$, so that $\lambda=h$ is an eigenvalue of $\bold{N}$ having algebraic multiplicity equals to $p$. Let $\bold{L}$ be the non-singular matrix 
\begin{equation}
\bold{L} =\left( \begin{matrix}
\bold{X}_1 & \bold{X}_2 & \dots & \bold{X}_p
\end{matrix}  
\right)
\end{equation}
($\bold{X}_i$ being column vectors) such that $\bold{L}^{-1}\bold{N}\bold{L}=\bold{J}_{h,p}$. Thence, from the condition $\bold{N}\bold{L}=\bold{L}\bold{J}_{h,p}$ we get the set of matrix equations
\begin{equation}
\begin{dcases}
\bold{N}  \bold{X}_1 & = h \bold{X}_1, \\
 \bold{N}  \bold{X}_2 & =\bold{X}_1+ h \bold{X}_2, \\
& \vdots \\
\bold{N}  \bold{X}_p & =\bold{X}_{p-1}+ h \bold{X}_p, \\
\end{dcases} \\ [2em]
\end{equation}
which can be re-arranged in order to obtain the so-called Jordan chain
\begin{equation}
\begin{dcases}
\left(\bold{N} -h \mathbb{1}\right) \bold{X}_1 & =  \bold{0}, \\
\left(\bold{N} -h \mathbb{1}\right) \bold{X}_2 & = \bold{X}_1, \\
& \vdots \\
\left(\bold{N} -h \mathbb{1}\right) \bold{X}_p & = \bold{X}_{p-1}.
\end{dcases} \\ [2em]
\label{Jordan1}
\end{equation}
As one can see, $\bold{X}_1$ is an eigenvector of $\bold{N}$ corresponding to the eigenvalue $h$, while the other $p-1$ vectors $(\bold{X}_2, \bold{X}_3, \dots, \bold{X}_p)$ are the generalized eigenvectors of $\bold{N}$. In particular, $\bold{X}_p$ is referred to as the generator or leading vector of the Jordan chain. The rank of a generalized eigenvector $\bold{X}_k$ (with $2 \leq  k \leq p$) is $k$ and $(\bold{X}_1, \bold{X}_2, \dots, \bold{X}_p)$ is named Jordan basis. Therefore, given the eigenvalue $h$, its corresponding Jordan block gives rise to a Jordan chain whose generator vector $\bold{X}_p$ is a generalized eigenvector of rank $p$, where $p$ represent the size of the Jordan block. Furthermore, the vector $\bold{X}_1=\left(\bold{N} -h \mathbb{1}\right)^{p-1}\bold{X}_p$ is an eigenvector of $\bold{N}$ corresponding to the eigenvalue $h$. Thence, we can conclude that a $p \times p$ matrix $\bold{N}$ is similar to a Jordan block $\bold{J}_{h,p}$ if there exists a Jordan basis consisting of one eigenvector and $p-1$ generalized eigenvectors. All of them have to satisfy the system (\ref{Jordan1}). 
\begin{description}
\item[{\it Example.}]
Let $\bold{N}=\left(\begin{matrix}
3 & 1 \cr
-1 & 1 \cr
\end{matrix} \right)$. Then ${\rm P}_{\bold{N}}(\lambda) = (\lambda-2)^2$ and $\lambda_1=2$ is an eigenvalue of $\bold{N}$ having algebraic multiplicity equals to two. Since its geometric multiplicity is one, the matrix is not diagonalizable and
\begin{equation}
\bold{X}_1= \left( \begin{matrix}
1 \cr
-1 \cr
\end{matrix}
\right),
\end{equation} 
represents the only linearly independent eigenvector. Let us attempt to find a vector $\bold{X}_2$ such that $(\bold{X}_1,\bold{X}_2)$ forms a Jordan basis. Bearing in mind Eq. (\ref{Jordan1}), we need to solve the matrix equation
\begin{equation}
\left(\bold{N} -\lambda_1 \mathbb{1}\right) \bold{X}_2=  \bold{X}_1,
\end{equation} 
which gives
\begin{equation}
\bold{X}_2= \left(\begin{matrix}
1 \cr
0 \cr
\end{matrix} \right).
\end{equation}
Therefore, if $\bold{L}=\left( \begin{matrix}
\bold{X}_1 & \bold{X}_2 \cr
\end{matrix} \right)=\left(\begin{matrix}
1 & 1 \cr
-1 & 0 \cr
\end{matrix} \right)$, then
\begin{equation}
\bold{L}^{-1} \bold{N} \bold{L} = \left( \begin{matrix}
2 & 1 \cr
0 & 2
\end{matrix}
\right)=\bold{J}_{2,2},
\end{equation}
which is a $2 \times 2$ Jordan block having $\lambda_1=2$ as its eigenvalue. 
\end{description}
We know from Schur triangulation theorem that if the characteristic polynomial of a matrix $\bold{A}$ factors completely, than $\bold{A}$ is similar to an upper triangular matrix. Nevertheless, this matrix is not unique. On the other side, the above-mentioned  Jordan normal form of $\bold{A}$ is both unique (in a sense that we will describe in a while) and, as we pointed out before, represents the ``closest-to-diagonal'' matrix that can be obtained by similarity transformations. These results are summarized by the following theorem:
\newtheorem*{Jordan}{Theorem}
\begin{Jordan}
Let $\bold{A}$ be an $n \times n$ matrix whose characteristic polynomial ${\rm P}_{\bold{A}}(\lambda)$ factors completely. Then $\bold{A}$ can be transformed in such a way that it turns out to be similar to a particular upper triangular matrix $\bold{J}$ having form
\begin{equation}
\bold{J}= \left( \begin{matrix}
\bold{J}_1 & 0 & \dots & 0 \cr
0 & \bold{J}_2 & \dots & 0 \cr
\vdots & \vdots & \ddots & \vdots \cr
0 & 0 & \dots & \bold{J}_k 
\end{matrix}\right),
\label{Jordan2}
\end{equation}
where $\bold{J}_1, \bold{J}_2, \dots, \bold{J}_k$ are Jordan blocks. The matrix $\bold{J}$ is called Jordan normal form of $\bold{A}$ and it is unique up to permutations of the blocks $\bold{J}_1, \bold{J}_2, \dots, \bold{J}_k$, which can occur in any order.
\end{Jordan}
The above theorem indicates that it is possible to transform the Schur form of $\bold{A}$ into the more convenient Jordan normal form $\bold{J}$, but, on the other side, it represents only an existence theorem which gives no information about the form of $\bold{J}$. Therefore, the proprieties of $\bold{J}$ can only be guessed by analysing the original matrix $\bold{A}$. To fix ideas, let $\bold{A}$ be the an $n \times n$ matrix having a completely factorized characteristic polynomial, say ${\rm P}_{\bold{A}}(\lambda)=(h_1-\lambda)^{p_1}(h_2-\lambda)^{p_2}\dots(h_s-\lambda)^{p_s}$, $h_1,h_2,\dots,h_s$ being distinct eigenvalues. Furthermore, let ${\rm M}_{\bold{A}}(\lambda)=(\lambda-h_1)^{l_1}(\lambda-h_2)^{l_2}\dots(\lambda-h_s)^{l_s}$ (with $1 \leq l_i \leq p_i$, for $i=1,2,\dots,s$). Let $\bold{J}$ be the Jordan normal form of $\bold{A}$ with Jordan blocks given by $\bold{J}_1, \bold{J}_2, \dots, \bold{J}_k$. Then $\bold{J}$ can be constructed by taking into account the following proprieties: 
\begin{itemize}
\item[-] Since $\bold{J}$ and $\bold{A}$ are similar, the eigenvalues of $\bold{A}$ will appear on the diagonal of $\bold{J}$. Therefore, the sum of the orders of the blocks in which $h_i$ occurs on the diagonal is $p_i$, i.e., an eigenvalue of $\bold{A}$ having algebraic multiplicity $p_i$ will appear $p_i$ times on the diagonal of its Jordan normal form. 
\item[-] Since there is one block for each linearly independent eigenvector of $\bold{A}$, the number of blocks associated with the eigenvalue $h_i$ corresponds to the geometric multiplicity of $h_i$. 
\item[-] The order of the largest block related to $h_i$ is the exponent $l_i$ of $(\lambda-h_i)$ in ${\rm M}_{\bold{A}}(\lambda)$. In other words, the size of the biggest Jordan block associated to a certain eigenvalue is ruled by the minimal polynomial.  
\end{itemize}
An important remark must be mentioned at this point. In fact, while the Jordan normal form determines the minimal polynomial, the converse is not true. This leads to the notion of elementary divisors. The elementary divisors of a square matrix $\bold{A}$ are the characteristic polynomials of its Jordan blocks. The factors of the minimal polynomial are the elementary divisors of the largest degree corresponding to distinct eigenvalues. The degree of an elementary divisor is the size of the corresponding Jordan block, therefore the dimension of the corresponding invariant subspace. Thus, we can interpret the diagonalization from another point of you, since it is possible to prove that if all elementary divisors are linear, then $\bold{J}$ is a diagonal matrix and hence the matrix $\bold{A}$ is diagonalizable. 
\begin{description}
\item[{\it Example.}] Assume that $\bold{A}$ is a matrix such that
\begin{equation}
\begin{split}
& {\rm P}_{\bold{A}}(\lambda)= \left(1-\lambda\right)^3 (2-\lambda)^2,\\
& {\rm M}_{\bold{A}}(\lambda)=\left(\lambda-1\right)^2 (\lambda-2).
\end{split}
\end{equation}
Then, on the diagonal of the Jordan normal form $\bold{J}$ of $\bold{A}$ $\lambda_1=1$ will appear three times and $\lambda_2=2$ twice, because their algebraic multiplicity is $p_1=3$ and $p_2=2$, respectively. Moreover, from the analysis of the minimal polynomial we realize that the order of the largest block associated with $\lambda_1$ is two, while for $\lambda_2$ is one. Therefore, $\bold{J}$ has the form
\begin{equation}
\bold{J}= \left[ 
\renewcommand{\arraystretch}{1.5}
\begin{matrix}
\left[
\begin{matrix}
1 & 1 \cr
0 & 1 \cr
\end{matrix}
\right] & 0 & 0 & 0 \cr
0 & [1] &0  & 0 \cr
0 &0  & [2] &0 \cr
0 & 0 & 0 &   [2] \cr
\end{matrix}
 \right].
\end{equation}
Note how the sum of the orders of the blocks having $\lambda_1$ and $\lambda_2$ as eigenvalues is $p_1$ and $p_2$, respectively. 
\end{description}
As we have said at the beginning of this section, the Jordan normal form of a matrix is useful in solving the set of constant coefficients differential equations (\ref{Fund1}) in all those cases in which $\bold{A}$ turns out to be not diagonalizable. Let $\bold{J}$ be the Jordan normal form of $\bold{A}$. Therefore, from the similarity condition involving $\bold{A}$ and $\bold{J}$, i.e., $\bold{A}=\bold{L}\bold{J}\bold{L}^{-1}$, the system (\ref{Fund1}) can be written as
\begin{equation}
\dot{\bold{y}}= \bold{J} \bold{y}, \label{Jordan3}
\end{equation}
where we have set
\begin{equation}
\bold{y}(t)=\bold{L}^{-1}\bold{x}(t). \label{Jordan6}
\end{equation}
Assume that each Jordan block $\bold{J}_i$ ($i=1,2,\dots,k$) of $\bold{J}$ is an $n_i \times n_i$ matrix (cf.  (\ref{Jordan2})). By writing $\bold{y}(t)$ as 
\begin{equation}
\bold{y}(t)=  \left( \begin{matrix} 
\bold{y}_1(t) \cr
\bold{y}_2(t) \cr 
\vdots \cr
\bold{y}_k(t) \cr
\end{matrix} \right), 
\label{Jordan4}
\end{equation}
where each entry $\bold{y}_i(t)$ (with $i=1,2,\dots,k$) represents an $n_i \times 1$ matrix, Eq. (\ref{Jordan3}) becomes
\begin{equation}
\dot{\bold{y}} = \left( \begin{matrix} 
\dot{\bold{y}}_1(t) \cr
\dot{\bold{y}}_2(t) \cr 
\vdots \cr
\dot{\bold{y}}_k(t) \cr
\end{matrix} \right) = 
\left( \begin{matrix}
\bold{J}_1 & 0 & \dots & 0 \cr
0 & \bold{J}_2 & \dots & 0 \cr
\vdots & \vdots & \ddots & \vdots \cr
0 & 0 & \dots & \bold{J}_k 
\end{matrix}\right) \left( \begin{matrix} 
\bold{y}_1(t) \cr
\bold{y}_2(t) \cr 
\vdots \cr
\bold{y}_k(t) \cr
\end{matrix} \right) = \left( \begin{matrix} 
 \bold{J}_1 \, \bold{y}_1(t)\cr
 \bold{J}_2 \, \bold{y}_2(t) \cr 
\vdots \cr
  \bold{J}_k \,  \bold{y}_k(t)\cr
\end{matrix} \right).
\end{equation}
Therefore, we need to solve $k$ systems of the form
\begin{equation}
\dot{\bold{y}}_i(t)= \bold{J}_i \,  \bold{y}_i(t), \; \; \; \; (i=1,2,\dots,k). \label{Jordan5}
\end{equation}
In other words, we have a system like the one in Eq. (\ref{Jordan5}) for each block $\bold{J}_i$ (with $i=1,2,\dots,k$) and the problem of solving a system of differential equations has now be reduced to solve a system associated to a single Jordan block. Given the solution $\bold{y}_i(t)$ of (\ref{Jordan5}), we construct the matrix (\ref{Jordan4}) and hence the solution of (\ref{Fund1}) can be obtained by inverting (\ref{Jordan6}), i.e., $\bold{x}(t)=\bold{L}\,\bold{y}(y)$. As an example, consider the case in which one of the Jordan block $\bold{J}_i$ appearing in (\ref{Jordan5}) is some $\bold{J}_{h,p}$. In this case, Eq. (\ref{Jordan5}) yields
\begin{equation}
\begin{dcases}
\dot{y}_1 & =h y_1 +y_2, \\
 & \vdots \\
\dot{y}_{p-1} & =h y_{p-1} +y_p, \\
\dot{y}_{p} & =h y_{p},  \\
\end{dcases} \\ [2em]
\end{equation}
which represents nothing more than a system of first-order linear differential equations having constant coefficients. It can be solved by starting from the bottom and working up: first of all, we solve the last equation for $y_p$ and substitute it into the second-last one and solve for $y_{p-1}$ and so forth. This algorithm involves solving differential equations of the form
\begin{equation}
\dot{y}+f(t)y=g(t), \label{Jordan6bis}
\end{equation}
in the special case in which the function $f(t)$ assumes a constant value, i.e.,
\begin{equation}
f(t)=-h.
\end{equation}
Since the solution of the general case (\ref{Jordan6bis}) is given by
\begin{equation}
y(t) =\dfrac{1}{q(t)} \int g(t) q(t) {\rm d}t + \dfrac{K}{q(t)},
\end{equation}
$K$ being an arbitrary constant and $q(t)$ the integrating factor defined as
\begin{equation}
q(t) \equiv {\rm e}^{\int f(t) {\rm d}t},
\end{equation}
in the case of constant coefficients we have simply
\begin{equation}
y(t)={\rm e}^{h t} \int {\rm e}^{-h t} g(t) {\rm d}t.
\end{equation}

\chapter{The tetrad formalism} \label{Tetrad}

In most situations a curvature calculation that relies upon Christoffel symbols is extremely lengthy and not obviously feasible or 
readable. However, the tetrad formalism is known to simplify such a task, at least when the metric does not possess distributional singularities. Thus, this appendix 
is devoted to some effort we made to express the highly singular ultrarelativistic boosted metric (\ref{ultrarelativistic boosted metric}) 
in terms of tetrads. 

As in the case of the boosted metric (\ref{boosted metric}), starting from the ultrarelativistic metric (\ref{ultrarelativistic boosted metric}) 
we can arrive at its manifestly four-dimensional form by exploiting (\ref{Y0}) and (\ref{dYo}) and hence we can eventually write the covariant 
metric components in the concise form \cite{BEST}
\begin{equation}
g_{kk}=1-{Y_{k}^{2}\over \sigma(Y_{j})}
+\left({Y_{k}^{2}\over \sigma(Y_{j})}+\delta_{1k}\right)
f(Y_{4})\delta\Bigr(Y_{1}+\sqrt{\sigma(Y_{j})}\Bigr), \;\; \; \; 
\forall k=1,2,3,4,
\label{gkk ultrarelativ}
\end{equation}
\begin{equation}
g_{1k}=-{Y_{1}Y_{k}\over \sigma(Y_{j})}
+\left({Y_{1}\over \sqrt{\sigma(Y_{j})}}+1 \right)
{Y_{k}\over \sqrt{\sigma(Y_{j})}}f(Y_{4})
\delta \Bigr(Y_{1}+\sqrt{\sigma(Y_{j})}\Bigr), \;\; \; \; 
\forall k=2,3,4,
\label{g1k ultrarelativ}
\end{equation}
\begin{equation}
g_{2k}=-{Y_{2}Y_{k}\over \sigma(Y_{j})}
+{Y_{2}Y_{k}\over \sigma(Y_{j})}f(Y_{4})
\delta \Bigr(Y_{1}+\sqrt{\sigma(Y_{j})}\Bigr), \;\; \; \; 
\forall k=3,4,
\label{g2k ultrarelativ}
\end{equation}
\begin{equation}
g_{34}=-{Y_{3}Y_{4}\over \sigma(Y_{j})}
+{Y_{3}Y_{4}\over \sigma(Y_{j})}
f(Y_{4})\delta \Bigr(Y_{1}+\sqrt{\sigma(Y_{j})}\Bigr),
\label{g34 ultrarelativ}
\end{equation}
where
\begin{equation}
f(Y_{4}) \equiv 4p \left[-2+{Y_{4}\over a}
\log \left({{a+Y_{4}}\over {a-Y_{4}}}\right)\right].
\label{f(Y4)}
\end{equation}
Since all components of this metric are non-vanishing, at this
stage we still assume the existence of tetrad covectors $e_{\; \mu}^{a}$ such that the covariant form of the metric reads as
\begin{equation}
g_{\mu \nu}=e_{\; \mu}^{a} e_{\; \nu}^{b}\eta_{ab},
\label{metric tetrad}
\end{equation}
$a,b$ being Lorentz-frame indices, and $\eta_{ab}$ being the 
familiar Minkowski metric ${\rm diag}(-1,1,1,1)$. By comparison
of the formulae (\ref{gkk ultrarelativ})--(\ref{g34 ultrarelativ}) with (\ref{metric tetrad}) we find that one can set \cite{BEST}
\begin{equation}
e_{\; k}^{0}={Y_{k}\over \sqrt{\sigma(Y_{j})}}, \; \; \; \; 
\forall k=1,2,3,4,
\label{e0k}
\end{equation}
while the other components of the singular, distribution-valued
limit of tetrad covectors solve the following non-linear 
algebraic system \cite{BEST}: 
\begin{equation}
\Bigr(e_{\; k}^{1}\Bigr)^{2}+\Bigr(e_{\; k}^{2}\Bigr)^{2}
+\Bigr(e_{\; k}^{3}\Bigr)^{2}
=1+\left({Y_{k}^{2}\over \sigma(Y_{j})}+\delta_{1k} \right)
f(Y_{4})\delta \Bigr(Y_{1}+\sqrt{\sigma(Y_{j})}\Bigr), \; \; \; \;
\forall k=1,2,3,4,
\label{alg tetrad 1}
\end{equation}
\begin{equation}
\sum_{i=1}^{3}e_{\; 1}^{i} e_{\; k}^{i}
=\left({Y_{1}\over \sqrt{\sigma(Y_{j})}}+1 \right)
{Y_{k}\over \sqrt{\sigma(Y_{j})}}f(Y_{4})
\delta \Bigr(Y_{1}+\sqrt{\sigma(Y_{j})}\Bigr), \; \; \; \;
\forall k=2,3,4,
\label{alg tetrad 2}
\end{equation}
\begin{equation}
\sum_{i=1}^{3} e_{\; 2}^{i}e_{\; k}^{i}
={Y_{2}Y_{k}\over \sigma(Y_{j})}f(Y_{4})
\delta \Bigr(Y_{1}+\sqrt{\sigma(Y_{j})}\Bigr), \; \; \; \;
\forall k=3,4,
\label{alg tetrad 3}
\end{equation}
\begin{equation}
\sum_{i=1}^{3}e_{\; 3}^{i}e_{\; 4}^{i}
={Y_{3}Y_{4}\over \sigma(Y_{j})}
f(Y_{4})\delta \Bigr(Y_{1}+\sqrt{\sigma(Y_{j})}\Bigr).
\label{alg tetrad 4}
\end{equation}
Since the system  (\ref{alg tetrad 1})--(\ref{alg tetrad 4}) consists of ten equations for the twelve unknown tetrad covectors, it is 
possible to find at least a particular solution. Now, once we get such a solution, the procedure should be as follows. As we know from 
general relativity, whenever the spacetime manifold is parallelizable, we can always introduce a set of Lorentz frames \cite{Cartan}, so that the spin-connection $1$-form $\omega^{ab}=\omega_{\mu}^{\;ab}dx^{\mu}$ obtained from requiring
that the torsion $2$-form should vanish has components \cite{DW}
\begin{equation}
\omega_{\mu}^{\; ab}= {1\over 2}e^{a\nu}\Bigr(e_{\; \nu,\mu}^{b}
-e_{\; \mu,\nu}^{b}\Bigr)
-{1\over 2}e^{b \nu}\Bigr(e_{\; \nu,\mu}^{a}
-e_{\; \mu,\nu}^{a}\Bigr)
+{1\over 2}e^{a \nu}e^{b \sigma}\Bigr(e_{\; \nu,\sigma}^{c}
-e_{\; \sigma, \nu}^{c}\Bigr)e_{c \mu},
\label{spin connection}
\end{equation}
where
\begin{equation}
e^{a \nu}=\eta^{ab}e_{\; b}^{\nu}, \;
e_{c \mu}=e_{\; \mu}^{a}\eta_{ac},
\end{equation} 
the tetrad vectors $e_{\; a}^{\mu}$ being computable by
comparison from the relation
\begin{equation}
{\rm d}x^{\mu}=e_{\; a}^{\mu}e^{a},
\end{equation}
which holds by virtue of the definition of tetrad $1$-forms
\begin{equation}
e^{a} \equiv e_{\; \mu}^{a}{\rm d}x^{\mu},
\end{equation}
jointly with \cite{DW}
\begin{equation}
e_{\; a}^{\rho}e_{\; \mu}^{a}=\delta_{\; \mu}^{\rho}.
\end{equation}
At this stage, we should be able to perform the curvature calculation
bearing in mind that the Riemann curvature is described by the $2$-form
\begin{equation}
R^{ab}={1\over 2}R_{\; \; \mu \nu}^{ab}{\rm d}x^{\mu} \wedge {\rm d}x^{\nu},
\label{curvature two form}
\end{equation}
where the components are given by
\begin{equation}
R_{\; \; \mu \nu}^{ab}=\Bigr(\omega_{\; \;\, \nu,\mu}^{ab}
-\omega_{\; \; \, \mu,\nu}^{ab}\Bigr)
+\eta_{cd}\Bigr(\omega_{\; \; \,\mu}^{bd} \, \omega_{\; \; \,\nu}^{ca}
-\omega_{\; \; \, \mu}^{ad} \, \omega_{\; \; \, \nu}^{cb}\Bigr).
\end{equation}
By virtue results of chapter \ref{Boosted_chapter}, the singular limit of the curvature $2$-form is a non-trivial mathematical object, since
it involves the Dirac's $\delta$ distribution, its powers and its derivatives. Finally, the Riemann curvature tensor $R_{\; \nu \rho \sigma}^{\mu}$ can be obtained from the identity
\begin{equation}
R_{\; \nu \rho \sigma}^{\mu} \; e_{\; \mu}^{a}
=R_{\; b \rho \sigma}^{a} \; e_{\; \nu}^{b}.
\end{equation}

\end{appendices}

\end{document}